\documentclass{article}
\usepackage{amssymb}
\usepackage{amsmath}

\input{tcilatex}
\begin{document}

\begin{center}
{\Huge Unitary Manipulation of a Single Atom in Time and Space --- The
Spatially-Selective and Internal- State-Selective Triggering Pulses\vspace{%
0.2in}\vspace{1.2in}}

{\huge Xijia Miao\footnote{{\large Email: miaoxijia@yahoo.com}}\vspace{0.2in}%
}

{\huge Somerville, Massachusetts\vspace{0.2in}}

{\huge Date: July 2010\newpage }

{\Large Abstract}
\end{center}

A spatially-selective and internal-state-selective triggering pulse is an
important component to realize the unitary dynamical state-locking process
in a single-atom quantum system (Arxiv: quant-ph/0607144), while a unitary
dynamical state-locking process and its inverse process may be used to
realize the reversible and unitary halting protocol and construct the
unstructured quantum search process based on the tensor-product
Hilbert-space symmetric structure and the unitary quantum dynamics. In the
previous paper (Arxiv: quant-ph/0708.2129) the ideal
internal-state-selective triggering pulses that are not spatially selective
are constructed explicitly in a single-atom quantum system.
Spatially-selective unitary operations, excitations, and processes tend to
be necessary components of a quantum computational process in an atomic
physical system, because a real-world quantum computer cannot have an
infinite dimensional size in space. In particular, the unitary manipulation
of a single atom in time and space in the unitary dynamical state-locking
process involves generally the spatially-selective operations, excitations,
and/or processes. Therefore, an ideal internal-state-selective triggering
pulse must be replaced with a spatially-selective and
internal-state-selective triggering pulse. In this paper it is shown how a
spatially-selective and internal-state-selective triggering pulse is
constructed generally in the quantum system of a single motional atom.
Rigorous theoretical calculation and error estimation, which are based on
the Trotter-Suzuki decomposition method and the
multiple-Gaussian-wave-packet ($MGWP$) expansion method, are carried out for
the time evolution process of a single atom in the double-well potential
field and in the presence of a spatially-selective and
internal-state-selective triggering pulse. It proves that the
spatially-selective and internal-state-selective triggering pulse is
feasible, that is, the possible errors generated by the triggering pulse are
shown to be controllable. This strict theoretical calculation also shows
that the spatially-selective and internal-state-selective triggering pulse
may be different from its ideal counterpart, but the difference between
their final states of the time evolution process is controllable. These
computational results are achieved with the help of the Gaussian wave-packet
motional states and the space-dependent quadratic Hamiltonians of a single
atom. The methods and techniques developed in the paper are useful for
studying the quantum-computing speedup mechanism of the unitary quantum
dynamics in the quantum system of a single atom.\newline
\newline
\newline
\newline

\qquad \qquad \qquad \qquad \qquad\ \ \ \ {\LARGE Contents\newline
}\newline
{\large Abstract \ \ \ \ \ \ \ \ \ \ \ \ \ \ \ \ \ \ \ \ \ \ \ \ \ \ \ \ \ \
\ \ \ \ \ \ \ \ \ \ \ \ \ \ \ \ \ \ \ \ \ \ \ \ \ \ \ \ \ \ \ \ \ \ \ \ \ \
\ \ \ \ \ \ 2\newline
1 Introduction \ \ \ \ \ \ \ \ \ \ \ \ \ \ \ \ \ \ \ \ \ \ \ \ \ \ \ \ \ \ \
\ \ \ \ \ \ \ \ \ \ \ \ \ \ \ \ \ \ \ \ \ \ \ \ \ \ \ \ \ \ \ \ \ \ \ \ 3%
\newline
2 The Hamiltonians to describe a spatially-selective }

{\large and internal-state-selective triggering pulse\ \ \ \ \ \ \ \ \ \ \ \
\ \ \ \ \ \ \ \ \ \ \ \ \ \ \ \ \ \ \ \ \ \ \ \ \ \ \ 8\newline
3 Imperfections of the harmonic potential well\ in the }

{\large double-well potential field\ \ \ \ \ \ \ \ \ \ \ \ \ \ \ \ \ \ \ \ \
\ \ \ \ \ \ \ \ \ \ \ \ \ \ \ \ \ \ \ \ \ \ \ \ \ \ \ \ \ \ \ \ \ \ \ \ \ \
\ \ \ \ \ \ \ \ \ 16}\newline
{\large 4 Imperfections for the spatially selective laser light beams\ \ \ \
\ \ \ \ \ \ \ \ \ \ \ \ 43}\newline
{\large 4.1 The upper bound of the norm }${\large ||}O_{p}^{V}(x,r,\tau
)\Psi _{00}(x,r,t_{0})||${\large \ \ \ \ \ \ \ \ \ \ \ \ \ \ \ \ \ \ \ \ \ 51%
}\newline
{\large 4.2 The upper bound of the norm }$||M_{1}^{V}(x,r,t_{1},t_{3})||$%
{\large \ \ \ \ \ \ \ \ \ \ \ \ \ \ \ \ \ \ \ \ \ \ \ \ \ \ \ \ \ \ \ 55}%
\newline
{\large 4.3 The upper bound of the norm }$||M_{2}^{V}(x,r,t_{1},t_{3})||$%
{\large \ \ \ \ \ \ \ \ \ \ \ \ \ \ \ \ \ \ \ \ \ \ \ \ \ \ \ \ \ \ \ 65}%
\newline
{\large 4.3.1 The upper bound of the norm }$||M_{21}^{V}(x,r,t_{1},t_{3})||$%
{\large \ \ \ \ \ \ \ \ \ \ \ \ \ \ \ \ \ \ \ \ \ \ \ \ \ \ \ 66}\newline
{\large 4.3.2 The upper bound of the norm }$||M_{22}^{V}(x,r,t_{1},t_{3})||$%
{\large \ \ \ \ \ \ \ \ \ \ \ \ \ \ \ \ \ \ \ \ \ \ \ \ \ \ \ 77}\newline
{\large 4.3.2.1 The upper bounds for the norms }$NORM(k,\lambda ,\mu )$%
{\Large \ }

{\large with }$\mu =CS${\large \ and }$S${\large \ \ \ \ \ \ \ \ \ \ \ \ \ \
\ \ \ \ \ \ \ \ \ \ \ \ \ \ \ \ \ \ \ \ \ \ \ \ \ \ \ \ \ \ \ \ \ \ \ \ \ \
\ \ \ \ \ \ \ \ \ \ \ \ \ \ \ \ \ \ \ \ \ \ \ \ \ 82}\newline
{\large 4.3.2.2 The upper bounds for the norms }$NORM(k,\lambda ,C0)${\large %
\ \ \ \ \ \ \ \ \ \ \ \ \ \ \ \ \ \ 127}\newline
{\large 4.4 The upper bound of the norm }$||E_{r}^{0}(x,r,t_{0}+\tau )||$%
{\large \ \ \ \ \ \ \ \ \ \ \ \ \ \ \ \ \ \ \ \ \ \ \ \ \ \ \ \ \ \ 144}%
\newline
{\large 5 The Lamb-Dicke limit\ \ \ \ \ \ \ \ \ \ \ \ \ \ \ \ \ \ \ \ \ \ \
\ \ \ \ \ \ \ \ \ \ \ \ \ \ \ \ \ \ \ \ \ \ \ \ \ \ \ \ \ \ \ \ \ \ \ \ \ \
\ \ \ \ \ \ \ \ \ \ \ \ \ \ 153}\newline
{\large 6 Generating the spatially-selective and internal-state-}

{\large selective triggering pulses \ \ \ \ \ \ \ \ \ \ \ \ \ \ \ \ \ \ \ \
\ \ \ \ \ \ \ \ \ \ \ \ \ \ \ \ \ \ \ \ \ \ \ \ \ \ \ \ \ \ \ \ \ \ \ \ \ \
\ \ \ \ \ \ 168}\newline
{\large 6.1 Theoretical calculation of the experimental basic }

{\large pulse sequence \ \ \ \ \ \ \ \ \ \ \ \ \ \ \ \ \ \ \ \ \ \ \ \ \ \ \
\ \ \ \ \ \ \ \ \ \ \ \ \ \ \ \ \ \ \ \ \ \ \ \ \ \ \ \ \ \ \ \ \ \ \ \ \ \
\ \ \ \ \ \ \ \ \ \ \ \ \ \ \ \ 178}\newline
{\large 6.2 Theoretical analysis of the SSISS triggering pulse} \ \ {\large %
\ \ \ \ \ \ \ \ \ \ \ \ \ \ \ \ \ \ \ \ \ \ 205}\newline
{\large 7 Discussion and conclusion \ \ \ \ \ \ \ \ \ \ \ \ \ \ \ \ \ \ \ \
\ \ \ \ \ \ \ \ \ \ \ \ \ \ \ \ \ \ \ \ \ \ \ \ \ \ \ \ \ \ \ \ \ \ \ \ \ \
\ \ \ \ \ \ \ 215}\newline
{\large Acknowledgment \ \ \ \ \ \ \ \ \ \ \ \ \ \ \ \ \ \ \ \ \ \ \ \ \ \ \
\ \ \ \ \ \ \ \ \ \ \ \ \ \ \ \ \ \ \ \ \ \ \ \ \ \ \ \ \ \ \ \ \ \ \ \ \ \
\ \ \ \ \ \ \ \ \ \ \ \ \ \ \ \ \ \ \ 217}\newline
{\large References\ \ \ \ \ \ \ \ \ \ \ \ \ \ \ \ \ \ \ \ \ \ \ \ \ \ \ \ \
\ \ \ \ \ \ \ \ \ \ \ \ \ \ \ \ \ \ \ \ \ \ \ \ \ \ \ \ \ \ \ \ \ \ \ \ \ \
\ \ \ \ \ \ \ \ \ \ \ \ \ \ \ \ \ \ \ \ \ \ \ \ \ \ \ 217}\newline
\newline
\newline
\newline
{\Large 1 Introduction}

There are a large number of works [1 -- 6] that are involved in manipulating
coherently the atomic systems in center-of-mass (COM) motion in many
research areas including the atomic laser cooling and trapping, quantum
coherence interference, quantum-state engineering, and quantum computation
and quantum information. An atomic system is one of the most promising
quantum systems to realize a large-scale quantum computation. It has some
advantages. A single atom may have the COM motion and internal (electronic
or nuclear spin) motion at the same time. Both the atomic motions may be
manipulated independently by the external electromagnetic field and/or
potential field. Therefore, an atomic system provides a quantum computation
with both the manipulating freedom degrees of the atomic COM motion and
internal motion. In quantum computation the atomic internal states are
mainly used as quantum bits. Manipulating and controlling coherently the
atomic internal motion are naturally necessary to realize a universal
quantum computation in an atomic system. In the past decades a large number
of works have been devoted to realizing a universal quantum computation in a
variety of atomic systems (See, for example, Refs. [3, 5, 6]). On the other
hand, there are also a large number of works [3, 4, 7 -- 16] related to
coherent preparation, control, and manipulation of quantum states and
especially the COM motional states of a single atom which may or may not be
confined in an external potential field. Coherent preparation and control of
quantum states of quantum systems including a single-atom quantum system are
generally important in quantum computation.

A quantum system of a single atom does not own those properties that only a
multiple-particle quantum system can have and it does not yet have the
quantum-computing resource (i.e., the symmetric structure of Hilbert space
of a composite quantum system). Therefore, such a simple quantum system is
particularly useful for studying the quantum-computing speedup mechanism of
the unitary quantum dynamics [47]. This implies that unitarily manipulating
in time and space a single atom is important in quantum computation as
spatially-selective unitary operations, excitations, and processes of a
single atom are generally necessary to realize a quantum computation in
atomic systems including a single-atom system. The present work is closely
related to the research subject of unitarily manipulating a single atom in
time and space. It has been shown [14 -- 16] that in order to realize the
reversible and unitary halting protocol and the unitary dynamical
state-locking process\footnote{%
It should be pointed out that both the reversible and unitary halting
protocol and the unitary dynamical state-locking process make sense only
when they consume the computational resource.}in a single-atom quantum
system it is necessary to manipulate unitarily a single atom in time and
space, that is, it is necessary to manipulate unitarily the COM motion,
internal motion, and their interaction of a single atom. Such a unitary
manipulation may be used to realize the quantum-state-level mutual
cooperation [14] between the COM motion and internal motion of a single
atom, while the quantum-state-level mutual cooperation is necessary to
realize in a single-atom quantum system the \textit{uni}tary \textit{dy}%
namical \textit{s}tate-\textit{lock}ing (UNIDYSLOCK) process and its inverse
process which may be used to construct the unstructured quantum search
process based on the tensor-product Hilbert-space symmetric structure and
the unitary quantum dynamics [47]. For convenience, hereafter the atom is
named the halting-qubit atom which is used to realize the reversible and
unitary halting protocol and the UNIDYSLOCK process. A spatially-selective
and internal-state-selective triggering pulse which is an important
component of the UNIDYSLOCK process also is realized in the quantum system
of the halting-qubit atom.

Coherently manipulating a spin system (in a spin ensemble) is very popular
in nuclear magnetic resonance spectroscopy, magnetic resonance imaging, and
electronic spin resonance spectroscopy. Many coherently manipulating methods
and techniques well developed in the NMR spectroscopy [17, 18] could be used
as well in quantum computation in quantum systems that are not involved in
any detailed COM motion. They also could be helpful to realize the unitary
manipulation in an atomic system in quantum computation which is often
involved in the COM motion, although manipulating coherently an atomic
system has mainly used the laser-light-based manipulating and controlling
methods and techniques [1, 3, 19, 42]. As an example, the decoupling
techniques for nuclear spin interactions in the NMR\ spectroscopy [17] could
be used to decouple the interaction between the atoms, as suggested in Ref.
[14], while the multi-pulse decoupling techniques may realize the
decoherence quantum control [20]. An atomic system in COM motion is
generally more complicated than a quantum spin system without COM motion in
coherent manipulation. It is far more complicated to realize the desired
coherent manipulation in an atomic system in COM motion. It is the atomic
COM motion that makes the coherent manipulation complicated. Many useful
coherent manipulation methods and techniques [17, 18, 19, 42] could become
ineffective in an atomic system in COM motion. Whether or not these methods
and techniques are effective is dependent to quite a large extent upon the
atomic COM motional states. Therefore, it is necessary to consider
explicitly the atomic COM motional states in the coherent manipulation in an
atomic system.

Coherent manipulation of an atomic system uses generally the electromagnetic
wave fields (mainly the laser light beams [1]) and external potential fields
including the externally applied electric field and magnetic field [3].
Sometimes the electromagnetic wave field also may be used to generate the
potential field such as the optical lattice and hence may act as the
externally potential field. If in coherent manipulation the atomic internal
motion is not explicitly involved, then it is usually convenient to use the
external potential fields to manipulate an atomic system. However, once the
atomic internal motion needs to be considered explicitly, it is more
convenient to use the electromagnetic wave fields. The UNIDYSLOCK process
needs to manipulate unitarily the COM motion, internal motion, and the
interaction between the two motions of the halting-qubit atom [14].
Therefore, both the electromagnetic wave fields and external potential
fields are necessary tools to realize the UNIDYSLOCK process.

When atoms are cooled down to a low temperature or even an ultralow
temperature, quantum effect of their COM motion becomes obvious [1, 3]. The
atomic motions including the internal motion and COM motion have to be
described by the quantum-mechanical principles [7 -- 12, 21]. Thus, for a
quantum computation on the cold atomic systems both the atomic internal
motion and COM motion have to be treated on the basis of the
quantum-mechanical principles [22]. Though a heavy quantum particle such as
an atom is quite similar to a classical particle, its quantum behavior is
completely the same as that one of a light quantum particle like an
electron. The time evolution process of a single atom in motion still obeys
the unitary quantum dynamics [22]. In quantum mechanics the COM motion of a
heavy quantum particle is usually described by a wave-packet state [22, 23].
Wave-packet COM motional states of a quantum particle such as an atom are
generally the theoretical and experimental basis to realize
spatially-selective manipulation and control. As shown in Ref. [14], the
spatially-selective unitary operations, excitations, and/or processes are
necessary components of the UNIDYSLOCK process. The latter could be realized
only when one can realize precisely and exactly the quantum-state-level
mutual cooperation between the COM motion and internal motion of the
halting-qubit atom. Then this requires that the wave-packet motional states
of the halting-qubit atom be chosen suitably, so that they can be
manipulated and controlled precisely and exactly in the entire UNIDYSLOCK
process. On the other hand, precise and exact manipulation and control
require that the time evolution process of the halting-qubit atom be
calculated exactly or evaluated strictly with the quantum-mechanical
principles. These requirements lead to that not only the wave-packet
motional states of the halting-qubit atom but also the electromagnetic wave
fields and/or the external potential fields that are used to manipulate the
halting-qubit atom need to be chosen suitably in the UNIDYSLOCK process.

It has been proposed [14 -- 16] that Gaussian wave-packet states are chosen
as the basic wave-packet motional states of the halting-qubit atom and
quadratic potential fields as the basic external potential fields to
construct the UNIDYSLOCK process, while the externally applied
electromagnetic wave fields must ensure that Gaussian wave-packet motional
states of the halting-qubit atom keep their Gaussian shape unchanged before
and after the unitary manipulation with the electromagnetic wave fields and
moreover, the possible errors generated by the manipulation are
controllable. This strategy also was used as a guidance to design the
(internal-)state-selective triggering pulses that are not spatially
selective [15] and will be used as well below to construct the
spatially-selective and internal-state-selective triggering pulses. A
Gaussian wave-packet state is one of the simplest and most basic wave-packet
states in quantum mechanics [22], while a quadratic potential field is one
of the simplest and most basic potential fields. Gaussian wave-packet states
(the coherent states [24a, 24b] and squeezed states [24c] of a harmonic
oscillator in coordinate space are also the special Gaussian wave-packet
states) have been studied extensively and have an extensive application in a
variety of research areas [22 -- 27]. The ground state of a harmonic
oscillator also is a\ Gaussian wave-packet state. Therefore, the Gaussian
wave-packet motional state of a single atom usually may be prepared easily
in experiment. On the other hand, a quantum particle in an external
quadratic potential field is a solvable quantum physical system [28 -- 33].
Its unitary propagator can be exactly obtained [28 -- 33]. Then the time
evolution process for the Gaussian wave-packet state of a single atom can be
exactly calculated in the presence of a general quadratic potential field.
Moreover, the Gaussian wave-packet state evolves into another in a general
quadratic potential field [26 -- 33, 22], meaning that Gaussian shape of the
Gaussian wave-packet state keeps unchanged during the time evolution
process. These basic properties of both the Gaussian wave-packet states and
the quadratic potential fields show that it could be most suitable to choose
both the Gaussian wave-packet states and the quadratic potential fields to
realize the UNIDYSLOCK process in the quantum system of the halting-qubit
atom. Of course, here it does not rule out the possibility to choose
non-Gaussian wave-packet states and other potential fields than the
quadratic potential fields to construct the UNIDYSLOCK process.

A \textit{S}patially-\textit{S}elective and \textit{I}nternal-\textit{S}tate-%
\textit{S}elective ($SSISS$) triggering pulse was first proposed in
constructing the reversible and unitary halting protocol [14]. It is an
important component to realize the UNIDYSLOCK process and may be used to
construct the unstructured quantum search process based on the
tensor-product Hilbert-space symmetric structure and the unitary quantum
dynamics. It was first suggested in Ref. [15] to\ construct the
state-selective triggering pulses that are not spatially selective by using
the coherent average method [46] based on the Trotter-Suzuki decomposition
method [34] or the average Hamiltonian theory [35] and the Magnus expansion
[36, 37]. Here the Trotter-Suzuki decomposition method may be conveniently
used to construct a higher-order state-selective triggering pulse. A
state-selective triggering pulse is used to manipulate unitarily the COM
motion of the halting-qubit atom in an external potential field such as a
harmonic potential field [14, 15]. The manipulation is different from a
conventional one in that it is dependent on the atomic internal motion. Only
when the halting-qubit atom is in some chosen internal states, can the
state-selective triggering pulse take a real action on the atom. Another
special point for the manipulation is that a Gaussian wave-packet motional
state of the halting-qubit atom may keep its Gaussian shape unchanged at the
end of the state-selective triggering pulse. Thus, a state-selective
triggering pulse must have the quantum-state-level mutual cooperation
between the atomic internal motion and COM motion [14]. The
internal-state-dependent manipulation for the atomic COM motion is generally
necessary to realize the quantum-state-level mutual cooperation (or
interaction) between the atomic internal motion and COM motion [14]. The
interaction could be simply achieved by applying suitable electromagnetic
wave fields such as the laser light fields to the chosen internal states of
a motional atom in an external potential field such as a harmonic potential
field [3, 7 -- 12]. In a general case, a sequence of the electromagnetic
field pulses and the potential field pulses are needed to manipulate
cooperatively the motional atom so that the interaction can be achieved.
Unlike a conventional coherent manipulation such as the laser light cooling,
here every manipulation is required to be unitary mainly due to that the
sequence could be used to construct the UNIDYSLOCK process. A
state-selective triggering pulse could simply consist of a sequence of the
external electromagnetic pulses such as the laser light pulses that are
selectively applied to the specific internal states of the halting-qubit
atom in an external potential field [14]. It was proposed to use a pair of
phase-modulation double-wave-number laser light beams to generate the pulse
sequences of the state-selective triggering pulses [15]. This scheme could
be thought of as an instance of the coherent double-photon techniques [38,
3, 9b, 10a, 10b, 12b] or the NMR double-resonance techniques [17]. The
advantage of the phase-modulation double-wave-number laser light pulses is
that the atomic Hamiltonian generated by the pair of laser light pulses are
time-independent in the rotating frame [15]. This leads to that it is much
easier to investigate strictly the time evolution process of the
halting-qubit atom in the presence of the state-selective triggering pulse.

The state-selective triggering pulses [15] are constructed in an ideal
harmonic potential field and with the electromagnetic field pulses that
extend over the whole coordinate space. They are the ideal state-selective
triggering pulses. Actually, both an electromagnetic field and an external
potential field can not be extended infinitely in space when they are used
to manipulate an atom in quantum computation. In practice an electromagnetic
field and/or external potential field tend to be applied to an atom in a
finite and chosen spatial region. Therefore, a question arises whether or
not such spatially-selective electromagnetic fields and potential fields are
suitable to construct a state-selective triggering pulse. A state-selective
triggering pulse that is constructed with the spatially-selective
electromagnetic fields and potential fields is called a spatially-selective
and internal-state-selective (or $SSISS$ briefly) triggering pulse. Then it
is needed to answer the question whether or not a $SSISS$ triggering pulse
can work well as expected to realize the UNIDYSLOCK process in the
double-well potential field [14]. The present work is devoted to solving the
problem: how to construct the $SSISS$ triggering pulses that can be used to
realize the UNIDYSLOCK process in the double-well potential field. If a $%
SSISS$ triggering pulse can achieve almost the same result that its
corresponding ideal state-selective triggering pulse achieves and moreover,
the difference between the two state-selective triggering pulses is
controllable, then such a $SSISS$ triggering pulse may replace an ideal
triggering pulse to realize the UNIDYSLOCK process. In the paper it is shown
that such a $SSISS$ triggering pulse can be constructed explicitly in the
double-well potential field. The time evolution process of the halting-qubit
atom in the presence of the $SSISS$ triggering pulse is calculated strictly
and all the possible errors generated during the process also are estimated
rigorously. It proves that the $SSISS$ triggering pulse is feasible, that
is, all these possible errors are controllable, and that the difference
between the $SSISS$ triggering pulse and its ideal counterpart is
controllable.\newline
\newline
\newline
{\Large 2 The Hamiltonians to describe a spatially-seletctive and
internal-state-selective triggering pulse}

In the paper [14] the UNIDYSLOCK process is realized in the double-well
potential field, where the left-hand ($LH$) potential well is approximately
a harmonic potential well, while the right-hand ($RH$) one is a square
potential well. When the dimensional size of the square potential well is
sufficiently large compared with the wave-packet spread of the halting-qubit
atom, the double-well potential field may be described reasonably by the
potential energy:%
\begin{equation}
V(x)=\left\{ 
\begin{array}{c}
\frac{1}{2}m\omega ^{2}x^{2},\text{ }-\infty <x<x_{L} \\ 
L_{h},\text{ }x_{L}<x<x_{L}+L \\ 
0,\text{ \ \ \ }x_{L}+L<x<+\infty%
\end{array}%
\right.  \tag{2.1}
\end{equation}%
where $x_{L}$ is the joint coordinate position between the $LH$ potential
well and the intermediate potential barrier, $L$ and $L_{h}$ are width and
height of the intermediate potential barrier, respectively, and usually the
height $L_{h}\geq m\omega ^{2}x_{L}^{2}/2,$ $L>0$, and $x_{L}>0.$ Hereafter $%
V(x)$ also is called the double-well potential field. It will be seen in the
error estimation below that the joint (coordinate) position $x_{L}$ is an
important control parameter. It is well known that the harmonic potential
field of an ideal harmonic oscillator is defined by the potential energy $%
V(x)=m\omega ^{2}x^{2}/2,$ where $\omega $ is the oscillatory frequency of
the harmonic oscillator and the coordinate position $x$ runs over the whole
one-dimensional space: $-\infty <x<+\infty .$ Thus, when the joint position $%
x_{L}\rightarrow +\infty ,$ the $LH$ potential well is an ideal harmonic
potential well. In general, the $LH$ potential well may be approximately
considered as an ideal harmonic potential well for the halting-qubit atom
when the potential energy value $V(x_{L})=m\omega ^{2}x_{L}^{2}/2$ at the
joint position $x_{L}$ is much greater than the mean motional energy of the
atom within the $LH$ potential well. The joint position $x_{L}$ can not be
taken as $x_{L}=+\infty $ due to a finite dimensional size of a real-world
quantum computer. In practice, the $LH$ potential well may deviate from an
ideal harmonic potential well. This imperfection could affect the motional
states of the atom in the $LH$ potential well, but this effect could be
generally small when the joint position $x_{L}$ is large. On the other hand,
the height of the intermediate potential barrier could be taken as an
infinitely large value in the theoretical treatment, that is, $L_{h}=+\infty 
$. This is due to that the height may be controlled at will from the outside
without having an effect on the dimensional size of a quantum computer. In
the case $L_{h}=+\infty $ the intermediate potential barrier generates two
hard potential walls, one is at the joint position $x_{L}$ if the
halting-qubit atom is in the $LH$ potential well and another at the position 
$x_{L}+L$ if the atom is in the $RH$ potential well. Such hard potential
walls may block effectively the halting-qubit atom to tunnel through the
intermediate potential barrier from one potential well to another. Though in
the case $L_{h}=+\infty $ the double-well potential field $V(x)$\ is
simpler, it could not be convenient to calculate rigorously the time
evolution process of the atom in the $LH$ potential well, especially when
there is an external electromagnetic field applying to the atom. It could be
more convenient to calculate strictly the effect of the imperfection of the $%
LH$ harmonic potential field on the atomic motional state if the height of
the intermediate potential barrier is taken as a finite and sufficiently
large value, e.g., $L_{h}=m\omega ^{2}x_{L}^{2}/2>>E_{0},$ here $E_{0}$ is
the atomic mean motional energy. Obviously, the effect of the imperfection
due to the tunneling effect in the case $L_{h}=m\omega ^{2}x_{L}^{2}/2$
generally is larger than that one in the case $L_{h}=+\infty .$ However,
when the joint position $x_{L}$ is large enough, both the cases should not
have a significant difference.

When the halting-qubit atom is in the $LH$ harmonic potential well, it may
be more convenient to write the Hamiltonian of the atom in the double-well
potential field $V(x)$ of (2.1) as 
\begin{equation}
H=H_{0}+V_{1}(x,t),\text{ }-\infty <x<+\infty ,  \tag{2.2}
\end{equation}%
where the main Hamiltonian $H_{0}$ is not spatially selective and the
perturbation term $V_{1}(x,t)$ is. The perturbation term $V_{1}(x,t)$ may be
used to measure the deviation of the $LH$ harmonic potential well from the
ideal harmonic potential well. Below two different cases are considered
explicitly. For the first case, there is not any external electromagnetic
field applying to the atom in the $LH$ potential well. In this case the main
Hamiltonian $H_{0}$ in (2.2) is taken as the Hamiltonian $H_{0}^{ho}$ of the
ideal harmonic oscillator --- the halting-qubit atom in the ideal
(left-hand) harmonic potential well, 
\begin{equation}
H_{0}^{ho}=\frac{1}{2m}p^{2}+\frac{1}{2}m\omega ^{2}x^{2},\text{ }-\infty
<x<+\infty .  \tag{2.3}
\end{equation}%
Because the atomic internal states are not involved, here the atomic
internal energy is not considered explicitly. In this case the perturbation
term $V_{1}(x,t)$ in (2.2) is time-independent. For convenience it is
denoted as $V_{1}^{ho}(x)$ and is given by%
\begin{equation}
V_{1}^{ho}(x)=\left\{ 
\begin{array}{c}
0,\text{ \ }-\infty <x<x_{L} \\ 
\frac{1}{2}m\omega ^{2}(x_{L}^{2}-x^{2}),\text{ \ }x_{L}<x<x_{L}+L \\ 
-\frac{1}{2}m\omega ^{2}x^{2},\text{ \ }x_{L}+L<x<+\infty%
\end{array}%
\right.  \tag{2.4}
\end{equation}%
where the height of the intermediate potential barrier is set to $%
L_{h}=m\omega ^{2}x_{L}^{2}/2$ in (2.1). The perturbation term $%
V_{1}^{ho}(x) $ takes a non-zero value only in the spatial region $%
x_{L}<x<+\infty .$ Thus, it is spatially selective. For the second case,
there is an external electromagnetic field or driving force field that is
applied to the atom within the $LH$ potential well. Then in this case the
main Hamiltonian $H_{0} $ is modified to the form 
\begin{equation}
H_{0}=H_{0}^{ho}+H_{a}+H_{1}(x,t),\text{ }-\infty <x<+\infty ,  \tag{2.5}
\end{equation}%
where $H_{a}$ is the atomic internal Hamiltonian and $H_{1}(x,t)$ is the
interaction between the atom and the external electromagnetic or potential
field, which may be the electric dipole interaction, etc., and the
corresponding perturbation term $V_{1}(x,t)$ is written as 
\begin{equation}
V_{1}(x,t)=\left\{ 
\begin{array}{c}
0,\text{ \ }-\infty <x<x_{L} \\ 
\frac{1}{2}m\omega ^{2}(x_{L}^{2}-x^{2})-H_{1}(x,t),\text{ \ }x_{L}<x<x_{L}+L
\\ 
-\frac{1}{2}m\omega ^{2}x^{2}-H_{1}(x,t),\text{ \ }x_{L}+L<x<+\infty%
\end{array}%
\right.  \tag{2.6}
\end{equation}%
It is clear that $V_{1}(x,t)$ is a spatially-selective perturbation term.
The present $SSISS$ triggering pulses are described by the two Hamiltonians
which are given by (2.2): the first Hamiltonian consists of the main
Hamiltonian $H_{0}^{ho}$ of (2.3) and the perturbation term $V_{1}^{ho}(x)$
of (2.4) and the second comprises the main Hamiltonian of (2.5) and the
perturbation term $V_{1}(x,t)$ of (2.6). The interaction $H_{1}(x,t)$ in
(2.5) and (2.6) will be given later. Both the perturbation terms $%
V_{1}^{ho}(x)$ and $V_{1}(x,t)$ are incontinuous in coordinate space and
their incontinuous points are $x=x_{L}$ and $x=x_{L}+L.$ These incontinuous
points could not affect the first-order approximation calculation for the
time evolution process of the halting-qubit atom, as shown below. However,
they could affect the rigor theoretical calculation for the time evolution
process.

Suppose that the propagators in the coordinate representation corresponding
to the Hamiltonians $H_{0}$ and $H$ in (2.2) are given by $%
G_{0}(x_{b},t_{b};x_{a},t_{a})$ and $G(x_{b},t_{b};x_{a},t_{a}),$
respectively. The main propagator $G_{0}(x_{b},t_{b};x_{a},t_{a})$ may be
exactly calculated by using the Feynman$^{\prime }s$ path integral technique
[28, 33] if the main Hamiltonian $H_{0}$ is a quadratic Hamiltonian. For
example, it can be calculated exactly if $H_{0}$ is the Hamiltonian $%
H_{0}^{ho}$ of (2.3) of the harmonic oscillator. However, it is generally
difficult to calculate exactly the total propagator $%
G(x_{b},t_{b};x_{a},t_{a})$ even if the main propagator $%
G_{0}(x_{b},t_{b};x_{a},t_{a})$ can be exactly obtained. When the
halting-qubit atom is in the $LH$ potential well and its motional energy is
much smaller than the height of the intermediate potential barrier, the
perturbation term $V_{1}(x,t)$ has generally a small effect on the atom.
Then in this case one may expand the total propagator $%
G(x_{b},t_{b};x_{a},t_{a})$ in terms of the perturbation term $V_{1}(x,t).$
The expansion for the propagator $G(x_{b},t_{b};x_{a},t_{a})$ may be
obtained by iterating the recursive relation [22]: 
\begin{equation*}
G(x_{b},t_{b};x_{a},t_{a})=G_{0}(x_{b},t_{b};x_{a},t_{a})
\end{equation*}%
\begin{equation}
+\frac{1}{i\hslash }\int_{t_{a}}^{t_{b}}dt_{c}\int
dx_{c}G_{0}(x_{b},t_{b};x_{c},t_{c})V_{1}(x_{c},t_{c})G(x_{c},t_{c};x_{a},t_{a}).
\tag{2.7}
\end{equation}%
When the joint position $x_{L}$ is infinitely large, the propagator $%
G(x_{b},t_{b};x_{a},t_{a})$ approaches $G_{0}(x_{b},t_{b};x_{a},t_{a})$,%
\begin{equation}
\lim_{x_{L}\rightarrow +\infty
}G(x_{b},t_{b};x_{a},t_{a})=G_{0}(x_{b},t_{b};x_{a},t_{a}).  \tag{2.8}
\end{equation}%
Therefore, if now the joint position $x_{L}$ is very large (that means that
the potential energy $L_{h}=m\omega ^{2}x_{L}^{2}/2$ is much larger than the
atomic mean motional energy), then one may take reasonably the first-order
approximation of the equation (2.7) as the propagator of the atom in the $LH$
potential well [22], 
\begin{equation*}
G(x_{b},t_{b};x_{a},t_{a})=G_{0}(x_{b},t_{b};x_{a},t_{a})
\end{equation*}%
\begin{equation}
+\frac{1}{i\hslash }\int_{t_{a}}^{t_{b}}dt_{c}\int
dx_{c}G_{0}(x_{b},t_{b};x_{c},t_{c})V_{1}(x_{c},t_{c})G_{0}(x_{c},t_{c};x_{a},t_{a}).
\tag{2.9}
\end{equation}%
Generally, one should use the exact propagator $G(x_{b},t_{b};x_{a},t_{a})$
to calculate the atomic motional state in the time evolution process of the
atom in the double-well potential field and in the presence of the
electromagnetic wave field. However, since the exact propagator $%
G(x_{b},t_{b};x_{a},t_{a})$ is usually hard to obtain, the propagator (2.9)\
of the first-order approximation may be used conveniently to calculate
approximately the time evolution process. It could be reasonable to
calculate the time evolution process if the halting-qubit atom in the $LH$
potential well is in a Gaussian wave-packet motional state approximately and
satisfies the condition that the atomic mean motional energy is sufficiently
smaller than the height of the intermediate potential barrier. This is due
to that $(i)$ there is the asymptotic form (2.8) for the exact propagator
and $(ii)$ there is the Gaussian wave-packet property that amplitude of a
Gaussian wave-packet state at a coordinate position decays exponentially
with the square deviation of the position from the COM position of the
Gaussian wave-packet state.

In many cases one needs to carry out a rigor theoretical calculation (or a
strict error estimation in an approximation calculation) for the time
evolution process of a quantum system. This is generally necessary for a
quantum computational process. In particular, it is absolutely necessary to
perform a rigor theoretical calculation and error estimation for a $SSISS$
triggering pulse. In many cases it may be more convenient to use other
quantum-mechanical theoretical methods than the above
coordinate-representation propagator $G(x_{b},t_{b};x_{a},t_{a})$ of (2.7)
to carry out a rigor theoretical calculation or other challenging tasks such
as constructing a $SSISS$ triggering pulse. The coherent average methods
[46] based on the Trotter-Suzuki decomposition method [34, 39], the Magnus
expansion [36, 37] and the average Hamiltonian theory [35] have been
proposed to construct the state-selective triggering pulses [15]. These
theoretical methods also could be used to construct a $SSISS$\textit{\ }%
triggering pulse. The present $SSISS$ triggering pulses are generated as
follows: an ideal state-selective triggering pulse is first constructed by
these coherent average methods, as proposed in Ref. [15], and then it is
converted into its corresponding $SSISS$ triggering pulse that could be
directly realized in experiment. An ideal state-selective triggering pulse
[15] consists of a sequence of the phase-modulation double-wave-number laser
light beams that are applied to the\ halting-qubit atom in an external
harmonic potential field. Both the laser light beams and the harmonic
potential field are perfect. They extend over the whole one-dimensional
coordinate space $(-\infty ,+\infty )$. However, the present $SSISS$
triggering pulses are generated by the phase-modulation double-wave-number
laser light beams that are spatial-selectively applied to the halting-qubit
atom within the $LH$ potential well $(-\infty ,x_{L})$ in the double-well
potential field of (2.1). Such spatially-selective laser light beams are not
the ideal ones that are used in an ideal state-selective triggering pulse
and the $LH$ potential field of the double-well potential field is not yet a
perfect harmonic potential field. Is such a $SSISS$ triggering pulse useful?
There are some problems to be addressed. Take a simple instance. An ideal
state-selective triggering pulse [15] needs to use the inverse unitary
propagators of the halting-qubit atom in a harmonic potential field. Now its
corresponding $SSISS$ triggering pulse also needs to use the inverse unitary
propagators of the halting-qubit atom in the double-well potential field.
While there is not any problem to obtain an exact inverse propagator of the
halting-qubit atom in a harmonic potential field that is realizable in
experiment, this is a large problem to generate in experiment an exact
inverse propagator of the halting-qubit atom in the double-well potential
field of (2.1). As shown later in the section 6, this problem may be solved
suitably. Here a spatially-dependent quadratic Hamiltonian including a
harmonic-oscillator Hamiltonian plays a crucial role in solving the problem,
largely because both the unitary propagator of a general quadratic
Hamiltonian and its inverse can be generated easily in theory and experiment
[15, 26 - 33]. On the other hand, these theoretical methods including the
Trotter-Suzuki decomposition method [34, 39], the Magnus expansion [36, 37]
and the average Hamiltonian theory [35] can not ensure that all the possible
errors generated by a $SSISS$ triggering pulse are controllable or the $%
SSISS $ triggering pulse can be convergent to its ideal counterpart. Though
there are the convergent criteria of the Trotter-Suzuki decomposition method
[40] and the Magnus expansion and the average Hamiltonian theory [41], these
criteria usually are insufficient or unavailable for an atomic system in COM
motion. Whether or not a $SSISS$ triggering pulse is convergent is largely
dependent on the COM motional states of the halting-qubit atom. Therefore,
one must pay attention to the fact that a $SSISS$ triggering pulse could not
be always useful to realize a UNIDYSLOCK process. It is necessary to carry
out a rigor theoretical calculation or a strict error estimation for the
time evolution process of the halting-qubit atom in the presence of a $SSISS$
triggering pulse, so that one can find out in what conditions the $SSISS$
triggering pulse is convergent and may be used to realize a UNIDYSLOCK
process.

It is more convenient to use a continuous double-well potential field
corresponding to the incontinuous one of (2.1) to do a rigor theoretical
calculation or a rigor error estimation for a $SSISS$ triggering pulse. The
double-well potential field of (2.1) is incontinuous, but it may be made
smooth in a suitable manner. Here one may use the smooth step function $%
\Theta (x,\varepsilon )$ defined below to make the potential function $V(x)$
of (2.1) smooth in coordinate space. With the help of the conventional
incontinuous step function $\Theta (x)$ the perturbation term $V_{1}^{ho}(x)$
of (2.4) is first rewritten as%
\begin{equation}
V_{1}^{ho}(x)=\frac{1}{2}m\omega ^{2}x_{L}^{2}\Theta (x-x_{L})\Theta
(x_{L}+L-x)-\frac{1}{2}m\omega ^{2}x^{2}\Theta (x-x_{L}),  \tag{2.10}
\end{equation}%
where the step function $\Theta (x)=1$ if $x>0$ and $\Theta (x)=0$ if $x<0.$
Its smooth form $V_{1}^{ho}(x,\varepsilon )$ then may be obtained by
replacing the incontinuous step function $\Theta (x)$ in (2.10) with the
continuous step function $\Theta (x,\varepsilon )$. The continuous step
function $\Theta (x,\varepsilon )$ with a small parameter $\varepsilon $ $%
(0<\varepsilon <<1)$ is defined by%
\begin{equation}
\Theta (x,\varepsilon )=\frac{1}{\varepsilon \sqrt{\pi }}\int_{-\infty
}^{x}\exp (-y^{2}/\varepsilon ^{2})dy,\text{ }\Theta (x)=\lim_{\varepsilon
\rightarrow 0}\Theta (x,\varepsilon ).  \tag{2.11}
\end{equation}%
The smooth function $\Theta (x,\varepsilon )$ is a Gaussian integral. It is
a continuous and monotonously increasing function. It satisfies $0<\Theta
(x,\varepsilon )<1$ for $-\infty <x<+\infty .$ There exists an
arbitrary-order coordinate derivative for the smooth function $\Theta
(x,\varepsilon )$ and the smooth perturbation term $V_{1}^{ho}(x,\varepsilon
)$ of (2.10). These properties for the function $\Theta (x,\varepsilon )$
are very helpful for carrying out a strict error estimation for the $SSISS$
triggering pulse below. Similarly, the incontinuous perturbation term $%
V_{1}(x,t)$ of (2.6) may be made smooth. It may be first rewritten as%
\begin{equation*}
V_{1}(x,t)=\frac{1}{2}m\omega ^{2}x_{L}^{2}\Theta (x-x_{L})\Theta (x_{L}+L-x)
\end{equation*}%
\begin{equation}
-\frac{1}{2}m\omega ^{2}x^{2}\Theta (x-x_{L})-H_{1}(x,t)\Theta (x-x_{L}). 
\tag{2.12}
\end{equation}%
Then its continuous form $V_{1}(x,t,\varepsilon )$ may be obtained by
replacing the step function $\Theta (x)$ in (2.12) with the smooth one $%
\Theta (x,\varepsilon ).$ Below the two smooth perturbation terms $%
V_{1}^{ho}(x,\varepsilon )$ of (2.10) and $V_{1}(x,t,\varepsilon )$ of
(2.12) will be used extensively in the strict error estimation for a $SSISS$
triggering pulse. Generally, the error upper bound obtained in the error
estimation using the smooth perturbation terms is dependent on the small
parameter $\varepsilon .$ The relevant work of the strict error estimation
is described mainly in the section 3 and 4 below.

In the present $SSISS$ triggering pulses there is the interaction between a
pair of the spatially-selective phase-modulation double-wave-number laser
light pulses and the halting-qubit atom in the $LH$ potential well. This
interaction is composed of the two parts. One part is the interaction
between the halting-qubit atom and the phase-modulation double-wave-number
laser light pulses, that is, the term $H_{1}(x,t)$ in (2.5). It may be found
from Ref. [15]. It is not spatially selective. Another part is
spatially-selective. It is contained into the spatially-selective
perturbation term $V_{1}(x,t)$ of (2.6). The total Hamiltonian in the
rotating frame is still given by (2.2) for the halting-qubit atom in the $LH$
potential well and in the presence of the spatially-selective
phase-modulation double-wave-number laser light pulses, here the main
Hamiltonian $H_{0}$ of (2.5) and the spatially-selective perturbation term $%
V_{1}(x,t)$ of (2.6) are explicitly obtained below. Suppose that a pair of
ideal laser light beams $\{E_{Ll}(t),$ $k_{l},$ $\omega _{l}\}$ with $l=0$
and $1$ are selectively applied to the two internal states $\{|g_{0}\rangle
, $ $|e\rangle \}$ of the halting-qubit atom in the $LH$ potential well. The
two laser light beams may be either counterpropagating if their wave number
values $k_{0}$ and $k_{1}$ have the opposite signs or copropagating if $%
k_{0} $ and $k_{1}$ have the same sign. They may have different frequencies $%
\{\omega _{l}\}$ and amplitudes $\{E_{Ll}(t)\}.$ Then in the laboratory
frame the electric dipole interaction $H_{L1}(x,t)$ between the
halting-qubit atom and the pair of laser light beams may be written as [15,
42, 7, 21]%
\begin{equation*}
H_{L1}(x,t)=\hslash \Omega _{0}(t)\{I^{+}\exp [i(k_{0}x-\omega _{0}t-\varphi
_{0}(t))]
\end{equation*}%
\begin{equation*}
+I^{-}\exp [-i(k_{0}x-\omega _{0}t-\varphi _{0}(t))]\}
\end{equation*}%
\begin{equation*}
+\hslash \Omega _{1}(t)\{I^{+}\exp [i(k_{1}x-\omega _{1}t-\varphi _{1}(t))]
\end{equation*}%
\begin{equation}
+I^{-}\exp [-i(k_{1}x-\omega _{1}t-\varphi _{1}(t))]\}  \tag{2.13}
\end{equation}%
where $\Omega _{l}(t)$ and $\varphi _{l}(t)$ are the Rabi frequency and
phase of the laser light beam $\{E_{Ll}(t),$ $k_{l},$ $\omega _{l}\}$,
respectively, and the atomic internal-state operators are defined by $%
I^{+}=|e\rangle \langle g_{0}|$ and $I^{-}=|g_{0}\rangle \langle e|.$ Note
that there is the rotating wave approximation for the interaction $%
H_{L1}(x,t)$ of (2.13) if the pair of laser light beams are linearly
polarized. However, there is not the rotating wave approximation for the
interaction $H_{L1}(x,t)$ if the pair of laser light beams are circularly
polarized [42]. For convenience hereafter only consider that the pair of
laser light beams are circularly polarized. Now the phase-modulation
double-wave-number laser light pulses are defined as a pair of ideal laser
light pulses whose Rabi frequencies and phases satisfy the following match
conditions [15]: 
\begin{equation}
\Omega _{0}(t)=\Omega _{1}(t),\text{ }\varphi _{0}(t)=\alpha +\gamma ,\text{ 
}\varphi _{1}(t)=(\omega _{0}-\omega _{1})t-\alpha +\gamma .  \tag{2.14}
\end{equation}%
Here the phases $\alpha $ and $\gamma $ may be set suitably in experiment.
The match conditions (2.14)\ lead directly to that the electric dipole
interaction of (2.13) is reduced to the form [15] 
\begin{equation*}
H_{L1}(x,t)=2\hslash \Omega _{0}(t)\exp (-i\omega _{0}t)I^{+}\exp [i\frac{1}{%
2}(k_{0}+k_{1})x-i\gamma ]\cos (\frac{1}{2}\Delta kx-\alpha )
\end{equation*}%
\begin{equation*}
+2\hslash \Omega _{0}(t)\exp (i\omega _{0}t)I^{-}\exp [-i\frac{1}{2}%
(k_{0}+k_{1})x+i\gamma ]\cos (\frac{1}{2}\Delta kx-\alpha ).
\end{equation*}%
Here the wave-number difference $\Delta k=k_{0}-k_{1}.$ This interaction is
similar to that one when the halting-qubit atom is irradiated by a
single-frequency laser light beam. Now setting the Rabi frequency $\Omega
_{0}(t)=\Omega _{0},$ namely that the Rabi frequency is taken as a
time-independent value. Then in the rotating frame the electric dipole
interaction $H_{L1}(x,t)$ is changed to the interaction $H_{1}(x,t)$ which
is equal to $H_{I}(x,\alpha ,\gamma )$ [15]: 
\begin{equation*}
H_{I}(x,\alpha ,\gamma )=2\hslash \Omega _{0}I^{+}\exp [i\frac{1}{2}%
(k_{0}+k_{1})x-i\gamma ]\cos (\frac{1}{2}\Delta kx-\alpha )
\end{equation*}%
\begin{equation}
+2\hslash \Omega _{0}I^{-}\exp [-i\frac{1}{2}(k_{0}+k_{1})x+i\gamma ]\cos (%
\frac{1}{2}\Delta kx-\alpha ),  \tag{2.15}
\end{equation}%
and the atomic internal Hamiltonian $H_{a}$ is given by $H_{a}=\hslash
(\omega _{a}-\omega _{0})I_{z}.$ Here the resonant frequency $\omega
_{a}=(E_{e}-E_{0})/\hslash ,$ where $E_{0}$ and $E_{e}$ are the energies of
the atomic internal states $|g_{0}\rangle $ and $|e\rangle ,$ respectively.
The spin operators $I_{\mu }$ $(\mu =x,y,z)$ of the atomic
two-internal-state subspace $\{|g_{0}\rangle ,|e\rangle \}$ are defined by $%
I_{x}=(|e\rangle \langle g_{0}|+|g_{0}\rangle \langle e|)/2,$ $%
I_{y}=(|e\rangle \langle g_{0}|-|g_{0}\rangle \langle e|)/(2i),$ and $%
I_{z}=(|e\rangle \langle e|-|g_{0}\rangle \langle g_{0}|)/2.$ They satisfy
the conventional spin angular momentum commutation relations: $%
[I_{x},I_{y}]=iI_{z},$ $etc.$ It is clear that the electric dipole
interaction $H_{I}(x,\alpha ,\gamma )$ of (2.15) is time-independent. When
the on-resonance condition $\omega _{a}=\omega _{0}$ is met, the atomic
internal Hamiltonian $H_{a}=0.$ The present $SSISS$ triggering pulses are
constructed (in a spatially-selective form) with the \textit{pha}se-\textit{m%
}odulation \textit{do}uble-\textit{w}ave-\textit{n}umber ($PHAMDOWN$) laser
light pulses $\{E_{Ll}(t),$ $k_{l},$ $\omega _{l}\}$ with $l=0$ and $1$ that
satisfy the following match conditions:%
\begin{equation*}
\Omega _{1}(t)=\Omega _{0}(t)=\Omega _{0},\text{ }\omega _{0}=\omega _{a},
\end{equation*}%
\begin{equation}
\varphi _{0}(t)=\alpha +\gamma ,\text{ }\varphi _{1}(t)=(\omega _{0}-\omega
_{1})t-\alpha +\gamma .  \tag{2.16}
\end{equation}%
The pair of $PHAMDOWN$ laser light pulses lead directly to that in the main
Hamiltonian $H_{0}$ of (2.5) the atomic internal Hamiltonian $H_{a}=0$ and
the electric dipole interaction $H_{1}(x,t)=H_{I}(x,\alpha ,\gamma )$,
indicating that both the main Hamiltonian $H_{0}$ of (2.5) and the
perturbation term $V_{1}(x,t)$ of (2.6) are time-independent. Now suppose
that the pair of $PHAMDOWN$ laser light pulses are applied
spatial-selectively to the halting-qubit atom within the $LH$ potential
well. Then in the rotating frame the time evolution process of the
halting-qubit atom may be described by the time-independent total
Hamiltonian of (2.2) whose time-independent main Hamiltonian and
perturbation term are given by (2.5) and (2.6), respectively. Here $H_{a}=0$
and $H_{1}(x,t)=H_{I}(x,\alpha ,\gamma )$ in the main Hamiltonian $H_{0}$ of
(2.5) and $H_{1}(x,t)=H_{I}(x,\alpha ,\gamma )$ in the perturbation term $%
V_{1}(x,t)$ of (2.6). The time-independent Hamiltonian of (2.2) leads to
that it is easier to carry out a strict error estimation on the basis of the
Trotter-Suzuki decomposition method [34, 39] for the present $SSISS$
triggering pulses.

It is known from Ref. [14] that the initial COM motional state of the
halting-qubit atom under the action of a $SSISS$ triggering pulse is a
Gaussian wave-packet state, i.e., the ground motional state of the
halting-qubit atom in the $LH$ harmonic potential well. A \textit{G}aussian 
\textit{W}ave-\textit{P}acket ($GWP$) state in one dimension may be written
in a standard form%
\begin{equation*}
\Psi (x,t_{0})=\exp (i\varphi _{0})[\frac{(\Delta x)^{2}}{2\pi }]^{1/4}\sqrt{%
\frac{1}{(\Delta x)^{2}+i(\frac{\hslash T_{0}}{2m})}}
\end{equation*}%
\begin{equation}
\times \exp \{-\frac{1}{4}\frac{(x-x_{0})^{2}}{(\Delta x)^{2}+i(\frac{%
\hslash T_{0}}{2m})}\}\exp \{ip_{0}x/\hslash \}.  \tag{2.17}
\end{equation}%
This standard form may be seen in the textbook \textit{quantum mechanics}
[22] (See: the section 12 of Chapt. 3) and also in the review paper [27]. It
will be used mainly in the present work. The standard $GWP$ state is
completely characterized by these parameters including the COM position $%
x_{0},$ the mean motional momentum $p_{0}$, and the wave-packet complex
linewidth $W(t_{0})=(\Delta x)^{2}+i\hslash T_{0}/(2m)$ as well as the
global phase $\varphi _{0}.$ These four parameters $\{x_{0},$ $p_{0},$ $%
(\Delta x)^{2},$ $T_{0}\}$ are the characteristic parameters of a $GWP$
state. All these characteristic parameters for a $GWP$ state used in the
present work may or may not be time-dependent. That a COM motional state has
a Gaussian shape means that the probability density in space of the motional
state is a Gaussian function which has the standard form $G(x)=\frac{1}{%
\varepsilon _{0}\sqrt{\pi }}\exp [-\frac{(x-x_{0})^{2}}{\varepsilon _{0}^{2}}%
]$ with the COM position $x_{0}$ and wave-packet spread $\varepsilon _{0}.$
For the $GWP$ state of (2.17) the wave-packet spread $\varepsilon _{0}=\sqrt{%
2[(\Delta x)^{2}+(\frac{\hslash T_{0}}{2m\Delta x})^{2}]}$ and $%
|W(t_{0})|^{2}=\frac{1}{2}(\Delta x)^{2}\varepsilon _{0}^{2}.$ Sometimes
these four parameters $\{x_{0},$ $p_{0},$ $W(t_{0}),$ $\varepsilon _{0}\}$
also are called the characteristic parameters of a $GWP$ state. There is an
important property for a $GWP$ motional state that amplitude (or probability
density) of a $GWP$ motional state at a coordinate point in space decays
exponentially with the square deviation of the coordinate point from the COM
position of the $GWP$ state. Because of this important property the $GWP$
motional states of the halting-qubit atom play a crucial role in carrying
out the rigor error estimation and proving rigorously that the $SSISS$
triggering pulse is convergent in the present work.\newline
\newline
\newline
{\Large 3 Imperfections of the harmonic potential well in the double-well
potential field}

First of all, the theoretical calculation and error estimation are carried
out for the time evolution process of the halting-qubit atom in the $LH$
potential well in the absence of any external electromagnetic field. Since
there is not any external electromagnetic field, the time evolution process
of the halting-qubit atom is described simply by the Hamiltonian of (2.2)
with the perturbation term of (2.4). Because the double-well potential field 
$V(x)$ is internal-state-independent, the propagator to describe the atomic
internal motion can only generate a global phase factor for the final atomic
state of the time evolution process if the initial atomic internal state is
an eigenstate of the atomic internal Hamiltonian. For convenience, suppose
that the halting-qubit atom is in an internal eigenstate $|g_{0}\rangle $.
Then here the atomic internal Hamiltonian may not be considered explicitly.
Suppose that at the initial time $t_{0}$ the atom is in a $GWP$ motional
state $\Psi _{0}(x,t_{0})$ like the state (2.17). Then at the initial time
the atom is in the product state $\Psi _{0}(x,r,t_{0})=\Psi
_{0}(x,t_{0})|g_{0}\rangle $. Below the first-order approximation propagator
(2.9) is first used to calculate the error originating from the imperfection
of the $LH$ harmonic potential well. Then a strict error estimation for the
imperfection is carried out on the basis of the Trotter-Suzuki decomposition
method [39]. Now by using the first-order approximation propagator (2.9) it
can turn out that at a time $t_{b}\geq t_{0}$ the atom is in the motional
state: 
\begin{equation}
\Psi (x_{b},t_{b})=\Psi _{0}(x_{b},t_{b})+E_{r}(x_{b},t_{b})  \tag{3.1}
\end{equation}%
where the motional state $\Psi _{0}(x,t_{c})$ ($t_{0}\leq t_{c}\leq t_{b}$)
is given by%
\begin{equation}
\Psi _{0}(x,t_{c})=\int dx_{a}G_{0}^{ho}(x,t_{c};x_{a},t_{0})\Psi
_{0}(x_{a},t_{0}).  \tag{3.2}
\end{equation}%
The motional state $\Psi _{0}(x,t_{c})$ also is a $GWP$ state, because $%
G_{0}^{ho}(x,t_{c};x_{a},t_{a})$ is the unitary propagator of the ideal
harmonic oscillator with the Hamiltonian $H_{0}^{ho}$ of (2.3) and a
harmonic-oscillator unitary propagator does not change the Gaussian shape of
a $GWP$ state [22, 26]. The error $E_{r}(x_{b},t_{b})$ in (3.1) measures the
deviation of the final state $\Psi (x_{b},t_{b})$ from the $GWP$ state $\Psi
_{0}(x_{b},t_{b}).$ It can be written as, in the first-order approximation,%
\begin{equation}
E_{r}(x_{b},t_{b})=\frac{1}{i\hslash }\int_{t_{0}}^{t_{b}}dt_{c}\int
dxG_{0}^{ho}(x_{b},t_{b};x,t_{c})V_{1}^{ho}(x,t_{c})\Psi _{0}(x,t_{c}). 
\tag{3.3}
\end{equation}%
For convenience the error $E_{r}(x_{b},t_{b})$ also may be thought of as an
error state or wavefunction with no normalization. The final state $\Psi
(x_{b},t_{b})$ is not an ideal $GWP$ state, but it approaches the $GWP$
state $\Psi _{0}(x_{b},t_{b})$ when the joint position $x_{L}$ in the
double-well potential field of (2.1) approaches the infinite point $+\infty $%
, 
\begin{equation*}
\underset{x_{L}\rightarrow +\infty }{\lim }\Psi (x_{b},t_{b})=\Psi
_{0}(x_{b},t_{b})\text{ and }\underset{x_{L}\rightarrow +\infty }{\lim }%
E_{r}(x_{b},t_{b})=0.
\end{equation*}%
Therefore, the error $E_{r}(x_{b},t_{b})$ measures the imperfection of the $%
LH$ harmonic potential well. Since $G_{0}^{ho}(x,t_{c};x_{a},t_{a})$ is a
unitary propagator in coordinate representation and the perturbation term $%
V_{1}^{ho}(x)$ given by (2.4)\ does not change the atomic internal states,
it follows from (3.3) that the upper bound of the probability $%
||E_{r}(x_{b},t_{b})||^{2}$ may be determined from%
\begin{equation}
||E_{r}(x_{b},t_{b})||\leq \frac{1}{\hslash }%
\int_{t_{0}}^{t_{b}}dt_{c}||V_{1}^{ho}(x,t_{c})\Psi _{0}(x,t_{c})|| 
\tag{3.4}
\end{equation}%
where the perturbation term $V_{1}^{ho}(x,t_{c})=V_{1}^{ho}(x)$ and the
probability of the motional state $V_{1}^{ho}(x,t_{c})\Psi _{0}(x,t_{c})$
with no normalization is defined by, according to quantum mechanics [22], 
\begin{equation}
||V_{1}^{ho}(x,t_{c})\Psi _{0}(x,t_{c})||^{2}=\int
dx|V_{1}^{ho}(x,t_{c})\Psi _{0}(x,t_{c})|^{2}.  \tag{3.5}
\end{equation}%
Here the norm $||V_{1}^{ho}(x,t_{c})\Psi _{0}(x,t_{c})||$ is a spectral
norm. Conventionally $||X||$ is denoted as norm of the vector $X$ if $X$ is
a vector or a quantum state and it also is denoted as norm of the operator $%
X $ if $X$ is an operator or a matrix [43]. Throughout the paper only the
spectral norm $(||X||_{2})$ [43] is used unless stated otherwise. That is, $%
||X||$ represents the spectral norm of the quantum state or operator $X$ in
the paper. Now suppose that the $GWP$ state $\Psi _{0}(x,t_{c})$ $(t_{0}\leq
t_{c}\leq t_{b})$ in (3.3) at the time $t_{c}$ has the characteristic
parameters $\{x_{c}(t_{c}),$ $p_{c}(t_{c}),$ $W(t_{c}),$ $\varepsilon
(t_{c})\},$ that is, it has the COM position $x_{c}(t_{c}),$ momentum $%
p_{c}(t_{c}),$ complex linewidth $W(t_{c}),$ and wave-packet spread $%
\varepsilon (t_{c}).$ By inserting the perturbation term $V_{1}^{ho}(x)$ of
(2.4) and the $GWP$ state $\Psi _{0}(x,t_{c})$ into (3.5) it turns out that
the probability $||V_{1}^{ho}(x,t_{c})\Psi _{0}(x,t_{c})||^{2}$ is bounded by%
\begin{equation}
||V_{1}^{ho}(x,t_{c})\Psi _{0}(x,t_{c})||^{2}<(\frac{1}{2}m\omega
^{2})^{2}I_{0}(x_{L},\varepsilon (t_{c}),x_{c}(t_{c}))  \tag{3.6}
\end{equation}%
where the integral $I_{0}(x_{L},\varepsilon (t_{c}),x_{c}(t_{c}))$ at any
time $t_{c}$ of the time region $t_{0}\leq t_{c}\leq t_{b}$ is defined by%
\begin{equation}
I_{0}(x_{L},\varepsilon (t_{c}),x_{c}(t_{c}))=\int_{x_{L}}^{+\infty
}dxx^{4}|\Psi _{0}(x,t_{c})|^{2}.  \tag{3.7}
\end{equation}%
It is clear that the integral is positive. It also can turn out that the
integral $I_{0}(x_{L},\varepsilon (t_{c}),x_{c}(t_{c}))$ can be explicitly
expressed as%
\begin{equation*}
I_{0}(x_{L},\varepsilon (t_{c}),x_{c}(t_{c}))=\frac{1}{4}\frac{1}{\sqrt{\pi }%
}\{3\varepsilon (t_{c})^{4}+12\varepsilon
(t_{c})^{2}x_{c}(t_{c})^{2}+4x_{c}(t_{c})^{4}\}
\end{equation*}%
\begin{equation*}
\times \int_{y_{M}}^{+\infty }dy\exp (-y^{2})+\frac{1}{4}\frac{1}{\sqrt{\pi }%
}\{y_{M}(3+2y_{M}^{2})\varepsilon (t_{c})^{4}+8(1+y_{M}^{2})\varepsilon
(t_{c})^{3}x_{c}(t_{c})
\end{equation*}%
\begin{equation}
+12y_{M}\varepsilon (t_{c})^{2}x_{c}(t_{c})^{2}+8\varepsilon
(t_{c})x_{c}(t_{c})^{3}\}\exp (-y_{M}^{2})  \tag{3.8}
\end{equation}%
where the parameter $y_{M}$ is defined by%
\begin{equation*}
y_{M}\equiv y_{M}(x_{L},t_{c})=(x_{L}-x_{c}(t_{c}))/\varepsilon (t_{c}).
\end{equation*}%
The parameter $y_{M}$ is an important parameter in the error estimation
throughout the paper. It is called the deviation-to-spread ratio (or the D/S
ratio in short) of the $GWP$ state $\Psi _{0}(x,t_{c})$. Since $%
x_{L}-x_{c}(t_{c})$ is the deviation of the joint position $x_{L}$ from the
COM position $x_{c}(t_{c})$ and $\varepsilon (t_{c})$ the wave-packet spread
of the $GWP$ state $\Psi _{0}(x,t_{c})$, the parameter $y_{M}$ is the ratio
between the deviation and the wave-packet spread. Notice that the joint
position $x_{L}$ is always greater than the COM position $x_{c}(t_{c})$\ due
to that the halting-qubit atom is within the $LH$ potential well for any
time $t_{c}\in \lbrack t_{0},t_{b}]$. Then the parameter $y_{M}>0.$ In order
to evaluate the upper bound of the integral $I_{0}(x_{L},\varepsilon
(t_{c}),x_{c}(t_{c}))$ one needs to use the following inequality for the
error function $\func{erf}c(x)$ [44]:%
\begin{equation*}
\int_{y_{M}}^{+\infty }dy\exp (-y^{2})\leq \frac{\exp (-y_{M}^{2})}{y_{M}+%
\sqrt{y_{M}^{2}+4/\pi }},\text{ for }y_{M}\geq 0.
\end{equation*}%
According to this inequality it can prove that the integral $%
I_{0}(x_{L},\varepsilon (t_{c}),x_{c}(t_{c}))$ of (3.8) is bounded by%
\begin{equation}
I_{0}(x_{L},\varepsilon (t_{c}),x_{c}(t_{c}))\leq P(x_{L},\varepsilon
(t_{c}),x_{c}(t_{c}))\exp [-y_{M}(x_{L},t_{c})^{2}]  \tag{3.9}
\end{equation}%
where the positive function $P(x_{L},\varepsilon (t_{c}),x_{c}(t_{c}))$ is
defined by%
\begin{equation*}
P(x_{L},\varepsilon (t_{c}),x_{c}(t_{c}))=\frac{1}{4}\frac{1}{\sqrt{\pi }}%
\frac{3\varepsilon (t_{c})^{4}+12\varepsilon
(t_{c})^{2}x_{c}(t_{c})^{2}+4x_{c}(t_{c})^{4}}{y_{M}+\sqrt{y_{M}^{2}+4/\pi }}
\end{equation*}%
\begin{equation*}
+\frac{1}{4}\frac{1}{\sqrt{\pi }}\{y_{M}(3+2y_{M}^{2})\varepsilon
(t_{c})^{4}+8(1+y_{M}^{2})\varepsilon (t_{c})^{3}x_{c}(t_{c})
\end{equation*}%
\begin{equation}
+12y_{M}\varepsilon (t_{c})^{2}x_{c}(t_{c})^{2}+8\varepsilon
(t_{c})x_{c}(t_{c})^{3}\}.  \tag{3.10}
\end{equation}%
Denote $y_{M}(x_{L},t_{c}^{\ast })$ and $P(x_{L},t_{c}^{\ast })$ as the
minimum value of the deviation-to-spread ratio $y_{M}$ and the maximum value
of the function $P(x_{L},\varepsilon (t_{c}),x_{c}(t_{c}))$ in the time
region $t_{0}\leq t_{c}\leq t_{b}$ for a given joint position $x_{L},$
respectively,%
\begin{equation*}
P(x_{L},t_{c}^{\ast })=\max_{t_{0}\leq t_{c}\leq t_{b}}\{P(x_{L},\varepsilon
(t_{c}),x_{c}(t_{c}))\},
\end{equation*}%
\begin{equation*}
y_{M}(x_{L},t_{c}^{\ast })=\min_{t_{0}\leq t_{c}\leq
t_{b}}\{y_{M}(x_{L},t_{c})\}.
\end{equation*}%
Then the integral $I_{0}(x_{L},\varepsilon (t_{c}),x_{c}(t_{c}))$ for any
time $t_{c}$ in the time region $[t_{0},t_{b}]$ is bounded by 
\begin{equation}
I_{0}(x_{L},\varepsilon (t_{c}),x_{c}(t_{c}))\leq P(x_{L},t_{c}^{\ast })\exp
[-y_{M}(x_{L},t_{c}^{\ast })^{2}].  \tag{3.11}
\end{equation}%
Therefore, it follows from the inequalities (3.4), (3.6), and (3.11) that
the error $E_{r}(x_{b},t_{b})$ is bounded by 
\begin{equation}
||E_{r}(x_{b},t_{b})||<\frac{1}{\hslash }(t_{b}-t_{0})(\frac{1}{2}m\omega
^{2})\sqrt{P(x_{L},t_{c}^{\ast })}\exp [-y_{M}(x_{L},t_{c}^{\ast })^{2}/2]. 
\tag{3.12}
\end{equation}%
Obviously, the upper bound (3.12)\ of the error $E_{r}(x_{b},t_{b})$ is
proportional to the exponentially-decaying Gaussian factor $\exp
[-y_{M}(x_{L},t_{c}^{\ast })^{2}/2]$ or it decays exponentially with the
square deviation-to-spread ratio $y_{M}(x_{L},t_{c}^{\ast })^{2}.$ This
point is particularly important. Denote $x_{M}(t_{c}^{\ast })$ and $%
\varepsilon _{M}(t_{c}^{\ast })$ as the maximum values of the COM position $%
x_{c}(t_{c})$ and the wave-packet spread $\varepsilon (t_{c})$ of the $GWP$
state $\Psi _{0}(x_{c},t_{c})$ of (3.2) in the time region $t_{0}\leq
t_{c}\leq t_{b},$ respectively, 
\begin{equation*}
x_{M}(t_{c}^{\ast })=\max_{t_{0}\leq t_{c}\leq t_{b}}\{x_{c}(t_{c})\}\text{
and }\varepsilon _{M}(t_{c}^{\ast })=\max_{t_{0}\leq t_{c}\leq
t_{b}}\{\varepsilon (t_{c})\}.
\end{equation*}%
Then one has 
\begin{equation*}
y_{M}(x_{L},t_{c})\geq y_{M}(x_{L},t_{c}^{\ast })\geq
(x_{L}-x_{M}(t_{c}^{\ast }))/\varepsilon _{M}(t_{c}^{\ast })>0.
\end{equation*}%
As can be seen below, the maximum values $x_{M}(t_{c}^{\ast })$ and $%
\varepsilon _{M}(t_{c}^{\ast })$ are really bounded through the motional
energy of the initial $GWP$ state $\Psi _{0}(x,t_{0}).$ On the other hand,
the function $P(x_{L},\varepsilon (t_{c}),x_{c}(t_{c}))$ of (3.10) increases
polynomially with the joint position $x_{L},$ the COM position $%
x_{c}(t_{c}), $ and the wave-packet spread $\varepsilon (t_{c})$
approximately. Actually, if the joint position $x_{L}$ or the
deviation-to-spread ratio $y_{M}$ is large enough, then the function $%
P(x_{L},\varepsilon (t_{c}),x_{c}(t_{c}))$ is a cubic polynomial in the
joint position $x_{L}$ approximately. Therefore, if the joint position $%
x_{L} $ is large enough such that the deviation-to-spread ratio $%
y_{M}(x_{L},t_{c}^{\ast })>>1,$ then it follows from (3.12) that the
absolute error $||E_{r}(x_{b},t_{b})||$ whose upper bound is proportional to
the exponentially-decaying factor $\exp [-y_{M}(x_{L},t_{c}^{\ast })^{2}/2]$
is so small that it can be neglected.

Below it turns out that the integral $I_{0}(x_{L},\varepsilon
(t_{c}),x_{c}(t_{c}))$ of (3.8) and the function $P(x_{L},\varepsilon
(t_{c}),x_{c}(t_{c}))$ of (3.10)\ are bounded if the motional energy $E_{ho}$
of the halting-qubit atom is given in advance. Both $I_{0}(x_{L},\varepsilon
(t_{c}),x_{c}(t_{c}))$ and $P(x_{L},\varepsilon (t_{c}),x_{c}(t_{c}))$
depend on these parameters $x_{c}(t_{c}),$ $\varepsilon (t_{c}),$ and $%
x_{L}. $ Thus, one needs only to prove that both $|x_{c}(t_{c})|$ and $%
\varepsilon (t_{c})$ have their own upper and lower bounds for a given
motional energy $E_{ho}.$ The equation (3.2) shows that the $GWP$ state $%
\Psi _{0}(x,t_{c})$ is generated by applying the propagator $%
G_{0}^{ho}(x_{c},t_{c};x_{a},t_{a})$ of the harmonic oscillator to the
initial $GWP$ state $\Psi _{0}(x,t_{0}).$ Since the Hamiltonian $H_{0}^{ho}$
of the harmonic oscillator is time-independent, the harmonic oscillator in
the $GWP$ state $\Psi _{0}(x,t_{c})$ obeys the energy conservation law over
the time region $[t_{0},t_{b}]$. Based on the energy conservation law one
may determine approximately the upper and lower bounds of these
characteristic parameters of the $GWP$ state $\Psi _{0}(x,t_{c})$ in the
time region $t_{0}\leq t_{c}\leq t_{b}$ if the motional energy $E_{ho}$ of
the halting-qubit atom is given in advance. This method will be discussed in
detail in the section 5 later. Since $\Psi _{0}(x,t_{c})$ is a $GWP$
motional state with the characteristic parameters $\{x_{c}(t_{c}),$ $%
p_{c}(t_{c}),$ $W(t_{c}),$ $\varepsilon (t_{c})\}$, a straight calculation
shows that the harmonic oscillator in the state $\Psi _{0}(x,t_{c})$ has the
motional energy:%
\begin{equation}
E_{ho}(t_{c})=\frac{p_{c}(t_{c})^{2}}{2m}+\frac{1}{2}m\omega
^{2}x_{c}(t_{c})^{2}+\frac{1}{4}m\omega ^{2}\varepsilon (t_{c})^{2}+\frac{1}{%
4}\frac{\hslash ^{2}}{2m(\Delta x)^{2}},  \tag{3.13a}
\end{equation}%
where $(\Delta x)^{2}=2|W(t_{c})|^{2}/\varepsilon (t_{c})^{2}$. The motional
energy $E_{ho}\equiv E_{ho}(t_{0})$ may be determined conveniently from the
initial $GWP$ state $\Psi _{0}(x,t_{0})$ according to the energy equation
(3.13a) with the setting $t_{c}=t_{0}$.\ The energy conservation law $%
E_{ho}(t_{c})=E_{ho}$ confines the maximum values for the COM position $%
|x_{c}(t_{c})|$ and the wave-packet spread $\varepsilon (t_{c})$ when the
harmonic oscillator evolves from the initial state $\Psi _{0}(x,t_{0})$ to
the state $\Psi _{0}(x,t_{c})$ at any time $t_{c}$ in the time region $%
[t_{0},t_{b}]$. It follows from (3.13a) that the maximum values for the COM
position $|x_{c}(t_{c})|$ and the wave-packet spread $\varepsilon (t_{c})$
in the time region $[t_{0},t_{b}]$ are always smaller than $\sqrt{%
2E_{ho}/(m\omega ^{2})}$ and $\sqrt{4E_{ho}/(m\omega ^{2})},$ respectively,
and moreover, the minimum value of the wave-packet spread $\varepsilon
(t_{c})$ is always greater than $\sqrt{\hslash ^{2}/(4mE_{ho})},$ 
\begin{equation}
0\leq x_{M}(t_{c}^{\ast })\leq |x_{c}(t_{c})|_{\max }\leq \sqrt{%
2E_{ho}/(m\omega ^{2})},  \tag{3.13b}
\end{equation}%
\begin{equation}
\sqrt{\hslash ^{2}/(4mE_{ho})}\leq \varepsilon (t_{c})\leq \varepsilon
_{M}(t_{c}^{\ast })\leq \sqrt{4E_{ho}/(m\omega ^{2})}.  \tag{3.13c}
\end{equation}%
These show that both $|x_{c}(t_{c})|$ and $\varepsilon (t_{c})$ have their
own upper and lower bounds for a given motional energy $E_{ho}.$ Thus, the
deviation-to-spread ratio $y_{M}(x_{L},t_{c})$ has the lower bound $%
y_{M}(x_{L},t_{c}^{\ast })$ and the function $P(x_{L},\varepsilon
(t_{c}),x_{c}(t_{c}))$\ has the upper bound $P(x_{L},t_{c}^{\ast })$ in the
time region $[t_{0},t_{b}]$ for a given joint position $x_{L}$ if the
motional energy $E_{ho}$ of the initial state $\Psi _{0}(x,t_{0})$ is given
in advance.

More generally, at the initial time $t_{0}$ the halting-qubit atom may be in
a superposition state $\Psi _{0}(x,r,t_{0})$. For example, suppose that $%
\Psi _{0}(x,r,t_{0})=\Psi _{0}^{g}(x,t_{0})|g_{0}\rangle +\Psi
_{0}^{e}(x,t_{0})|e\rangle ,$ where $|g_{0}\rangle $ and $|e\rangle $ are
the two orthogonal internal states and the motional state $\Psi
_{0}^{a}(x,t_{0})$ with $a=g$ or $e$ may be a superposition of $n_{a}$\ $GWP$
states. Because the harmonic-oscillator propagator $%
G_{0}^{ho}(x,t_{c};x_{a},t_{a})$ does not change the atomic internal states,
it follows from (3.2) that the atomic product state $\Psi _{0}(x,r,t_{c})$
at any time $t_{c}$ ($t_{0}\leq t_{c}\leq t_{b}$) also can be formally
written as%
\begin{equation}
\Psi _{0}(x,r,t_{c})=\Psi _{0}^{g}(x,t_{c})|g_{0}\rangle +\Psi
_{0}^{e}(x,t_{c})|e\rangle .  \tag{3.14}
\end{equation}%
Here the motional state $\Psi _{0}^{a}(x,t_{c})$ is still a superposition of 
$n_{a}$ $GWP$ states: 
\begin{equation}
\Psi _{0}^{a}(x,t_{c})=\sum_{k=1}^{n_{a}}A_{k}^{a}(t_{c})\Psi
_{0k}^{a}(x,t_{c}),  \tag{3.15}
\end{equation}%
where $A_{k}^{a}(t_{c})$ is amplitude of the $k-$th normalized $GWP$ state $%
\Psi _{0k}^{a}(x,t_{c})$. The normalization condition for the state $\Psi
_{0}(x,r,t_{c})$ is given by $||\Psi _{0}^{g}(x,t_{c})||^{2}+||\Psi
_{0}^{e}(x,t_{c})||^{2}=1.$ Now the equation (3.1) may be rewritten as%
\begin{equation}
\Psi (x_{b},r,t_{b})=\Psi _{0}^{g}(x_{b},t_{b})|g_{0}\rangle +\Psi
_{0}^{e}(x_{b},t_{b})|e\rangle +E_{r}(x_{b},r,t_{b}).  \tag{3.16}
\end{equation}%
Here the total error term $E_{r}(x_{b},r,t_{b})=E_{r}^{g}(x_{b},t_{b})|g_{0}%
\rangle +E_{r}^{e}(x_{b},t_{b})|e\rangle $ is bounded by 
\begin{equation*}
||E_{r}(x_{b},r,t_{b})||=\sqrt{%
||E_{r}^{g}(x_{b},t_{b})||^{2}+||E_{r}^{e}(x_{b},t_{b})||^{2}}
\end{equation*}%
\begin{equation}
\leq \frac{1}{\hslash }\int_{t_{0}}^{t_{b}}dt_{c}\{||V_{1}^{ho}(x,t_{c})\Psi
_{0}^{g}(x,t_{c})||+||V_{1}^{ho}(x,t_{c})\Psi _{0}^{e}(x,t_{c})||\}, 
\tag{3.17}
\end{equation}%
where the orthogonal relation between the two internal states $|g_{0}\rangle 
$ and $|e\rangle $ is already used and so is the inequality (3.4). Now the
norm $||V_{1}^{ho}(x,t_{c})\Psi _{0}^{a}(x,t_{c})||$ may be calculated by
using the superposition state of (3.15). First of all, it satisfies the
inequality [43]: 
\begin{equation}
||V_{1}^{ho}(x,t_{c})\Psi _{0}^{a}(x,t_{c})||\leq
\sum_{k=1}^{n_{a}}|A_{k}^{a}(t_{c})|\times ||V_{1}^{ho}(x,t_{c})\Psi
_{0k}^{a}(x,t_{c})||.  \tag{3.18}
\end{equation}%
Then the upper bound of each norm $||V_{1}^{ho}(x,t_{c})\Psi
_{0k}^{a}(x,t_{c})||$ may be calculated in the same way as the above.
According to the inequality (3.6) the upper bound of the probability $%
||V_{1}^{ho}(x,t_{c})\Psi _{0k}^{a}(x,t_{c})||^{2}$ may be determined from,
in the first-order approximation,%
\begin{equation}
||V_{1}^{ho}(x,t_{c})\Psi _{0k}^{a}(x,t_{c})||^{2}<(\frac{1}{2}m\omega
^{2})^{2}I_{0k}^{a}(x_{L},\varepsilon _{0k}^{a}(t_{c}),x_{ck}^{a}(t_{c})) 
\tag{3.19}
\end{equation}%
where, just like the integral $I_{0}(x_{L},\varepsilon (t_{c}),x_{c}(t_{c}))$
in (3.6), the integral $I_{0k}^{a}(x_{L},$ $\varepsilon
_{0k}^{a}(t_{c}),x_{ck}^{a}(t_{c}))$ is still defined by (3.7) and its
explicit expression is still given by (3.8) in which the parameter settings
are given by $\varepsilon (t_{c})=\varepsilon _{0k}^{a}(t_{c}),$ $%
x_{c}(t_{c})=x_{ck}^{a}(t_{c}),$ and $%
y_{M}=y_{0k}^{a}(x_{L},t_{c})=(x_{L}-x_{ck}^{a}(t_{c}))/\varepsilon
_{0k}^{a}(t_{c})$. Here $x_{ck}^{a}(t_{c})$ and $\varepsilon
_{0k}^{a}(t_{c}) $ are the COM position and wave-packet spread of the $k-$th 
$GWP$ state $\Psi _{0k}^{a}(x,t_{c})$, respectively. Moreover, just like the
integral $I_{0}(x_{L},\varepsilon (t_{c}),x_{c}(t_{c}))$ whose upper bound
is determined \ from \ the inequality (3.11), the integral $%
I_{0k}^{a}(x_{L},\varepsilon _{0k}^{a}(t_{c}),x_{ck}^{a}(t_{c}))$ has its
own upper bound in the time region $[t_{0},t_{b}],$ which may be determined
from%
\begin{equation}
I_{0k}^{a}(x_{L},\varepsilon _{0k}^{a}(t_{c}),x_{ck}^{a}(t_{c}))\leq
P_{0k}^{a}(x_{L},\varepsilon _{0k}^{a}(t_{c}),x_{ck}^{a}(t_{c}))\exp
\{-y_{0k}^{a}(x_{L},t_{c}^{\ast })^{2}\}.  \tag{3.20}
\end{equation}%
Here the positive function $P_{0k}^{a}(x_{L},\varepsilon
_{0k}^{a}(t_{c}),x_{ck}^{a}(t_{c}))$ is still defined by (3.10) in which the
parameter settings are given by $\varepsilon (t_{c})=\varepsilon
_{0k}^{a}(t_{c}),$ $x_{c}(t_{c})=x_{ck}^{a}(t_{c}),$ and $%
y_{M}=y_{0k}^{a}(x_{L},t_{c}),$ while $y_{0k}^{a}(x_{L},t_{c}^{\ast })$ is
the minimum value of the deviation-to-spread ratio $y_{0k}^{a}(x_{L},t_{c})$
in the time region $[t_{0},t_{b}]$. Now denote that $y_{M}^{a}(x_{L})=%
\min_{1\leq k\leq n_{a}}\{y_{0k}^{a}(x_{L},t_{c}^{\ast })\}.$ Then these
three inequalities (3.18), (3.19), and (3.20) together lead to that the norm 
$||V_{1}^{ho}(x,t_{c})\Psi _{0}^{a}(x,t_{c})||$ is bounded by%
\begin{equation}
||V_{1}^{ho}(x,t_{c})\Psi _{0}^{a}(x,t_{c})||<(\frac{1}{2}m\omega
^{2})Q_{a}(x_{L},t_{c})\exp \{-y_{M}^{a}(x_{L})^{2}/2\},  \tag{3.21}
\end{equation}%
where the positive function $Q_{a}(x_{L},t_{c})$ is given by%
\begin{equation}
Q_{a}(x_{L},t_{c})=\sum_{k=1}^{n_{a}}|A_{k}^{a}(t_{c})|\sqrt{%
P_{0k}^{a}(x_{L},\varepsilon _{0k}^{a}(t_{c}),x_{ck}^{a}(t_{c}))}. 
\tag{3.22}
\end{equation}%
Denote $Q_{a}(x_{L},t_{c}^{\ast })=\max_{t_{0}\leq t_{c}\leq
t_{b}}\{Q_{a}(x_{L},t_{c})\}.$ Then it follows from (3.17) and (3.21) that
the error $E_{r}(x_{b},r,t_{b})$ in (3.16) is bounded by%
\begin{equation*}
||E_{r}(x_{b},r,t_{b})||<\frac{1}{\hslash }(t_{b}-t_{0})(\frac{1}{2}m\omega
^{2})Q_{g}(x_{L},t_{c}^{\ast })\exp \{-y_{M}^{g}(x_{L})^{2}/2\}
\end{equation*}%
\begin{equation}
+\frac{1}{\hslash }(t_{b}-t_{0})(\frac{1}{2}m\omega
^{2})Q_{e}(x_{L},t_{c}^{\ast })\exp \{-y_{M}^{e}(x_{L})^{2}/2\}.  \tag{3.23}
\end{equation}%
The upper bound (3.23) is proportional to the exponentially-decaying factors 
$\{\exp \{-y_{M}^{a}(x_{L})^{2}/2\}\}$. Suppose now that $n_{a}$ is a finite
positive integer for the index $a=g$ and $e$. Then the inequality (3.23)
shows that when the joint position $x_{L}$ is large enough such that the
minimum deviation-to-spread ratio $\min
\{y_{M}^{g}(x_{L}),y_{M}^{e}(x_{L})\}>>1,$ the error $E_{r}(x_{b},r,t_{b})$
can be neglected. Hence it follows from (3.16) that the final state $\Psi
(x_{b},r,t_{b})$ is equal to the state $\Psi _{0}(x_{b},r,t_{b})$ of (3.14)
after the error $E_{r}(x_{b},r,t_{b})$ is neglected. This indicates that the
imperfection of the $LH$ harmonic potential well does not have a significant
effect on the time evolution process of the halting-qubit atom even when the
atom is initially in a superposition of a finite number of the $GWP$ states.

In the above paragraphs the first-order approximation propagator (2.9) has
been used to investigate the imperfection of the $LH$ harmonic potential
well. This is a quite convenient approximation method to treat theoretically
the spatially selective unitary operation, process, and excitation. The
problem is that one does not know how exact it is for the result obtained by
the approximation method. Here an exact theoretical treatment is carried out
for the imperfection of the $LH$ harmonic potential well. This exact
theoretical treatment is based on the Trotter-Suzuki decomposition method
[39]. It shows in what conditions the first-order approximation propagator
(2.9) may be used reasonably. The exact theoretical treatment provides a
correction to the first-order approximation calculation based on the formula
(2.9). It can turn out below that this correction decays exponentially with
the square deviation-to-spread ratios of the $GWP$ states of the
halting-qubit atom. This shows that the result obtained from the first-order
approximation calculation above is really quite general that upper bound of
the error originating from the imperfection of the $LH$ harmonic potential
well decays exponentially with the square deviation-to-spread ratios. The
exact theoretical treatment is described below. Suppose that the Hamiltonian
of the halting-qubit atom in the $LH$ potential well is time-independent and
it may be generally written as $H=H_{0}+V_{1}.$ Here $H_{0}$ is usually the
main Hamiltonian to describe the non-space-selective excitation, operation,
or process, while $V_{1}$ is the perturbation term that could originate from
the spatially-selective excitation, operation, or process. Then it can turn
out that the propagator $\exp [-i(H_{0}+V_{1})t/\hslash ]$ satisfies the
operator identity [39]: 
\begin{equation*}
\exp [-i(H_{0}+V_{1})t/\hslash ]=\exp (-iV_{1}t/\hslash )\exp
(-iH_{0}t/\hslash )
\end{equation*}%
\begin{equation*}
-\frac{1}{\hslash ^{2}}\int_{0}^{t}d\lambda \int_{0}^{\lambda }d\lambda
^{\prime }\{\exp [-i(H_{0}+V_{1})(t-\lambda )/\hslash ]\exp [-iV_{1}(\lambda
-\lambda ^{\prime })/\hslash ]
\end{equation*}%
\begin{equation}
\times \lbrack H_{0},V_{1}]\exp (-iV_{1}\lambda ^{\prime }/\hslash )\exp
(-iH_{0}\lambda /\hslash )\}.  \tag{3.24}
\end{equation}%
For convenience, this operator identity is called the Trotter-Suzuki
decomposition formula. It may be used to investigate strictly the
imperfection of the $LH$ harmonic potential well. As shown in the previous
section 2, a $SSISS$ triggering pulse generated with the spatially-selective 
$PHAMDOWN$ laser light pulses is described completely by the two
time-independent Hamiltonians of (2.2). One of the two Hamiltonians is given
by $H=H_{0}^{ho}+V_{1}^{ho}(x)$ with the main Hamiltonian $H_{0}^{ho}$ of
(2.3) and the perturbation term $V_{1}^{ho}(x)$ of (2.4). This Hamiltonian
has nothing to do with the spatially-selective $PHAMDOWN$ laser light
pulses. Here the time evolution process of the halting-qubit atom in the $LH$
harmonic potential well is treated strictly with the aid of the operator
identity (3.24). It is governed by the time-independent Hamiltonian $%
H_{0}^{ho}+V_{1}^{ho}(x)$. The unitary propagator corresponding to this
time-independent Hamiltonian may be exactly given by (3.24). Suppose now
that the halting-qubit atom is in the initial $GWP$ product state $\Psi
_{00}(x,r,t_{0}).$ By applying this exact propagator to this initial state
one obtains%
\begin{equation*}
\Psi (x,r,t_{0}+t)\overset{\text{def}}{\equiv }\exp
[-i(H_{0}+V_{1})t/\hslash ]\Psi _{00}(x,r,t_{0})
\end{equation*}%
\begin{equation}
=\Psi _{0}(x,r,t_{0}+t)+E_{r}^{(1)}(x,r,t_{0}+t)+E_{r}^{(2)}(x,r,t_{0}+t), 
\tag{3.25}
\end{equation}%
where the symbol "$\overset{\text{def}}{\equiv }$ " means that $\Psi
(x,r,t_{0}+t)$ is defined as the final product state, the desired product
state is written as%
\begin{equation}
\Psi _{0}(x,r,t_{0}+t)=\exp (-iH_{0}t/\hslash )\Psi _{00}(x,r,t_{0}), 
\tag{3.26}
\end{equation}%
and the two error terms are given exactly by%
\begin{equation}
E_{r}^{(1)}(x,r,t_{0}+t)=-[1-\exp (-iV_{1}t/\hslash )]\exp (-iH_{0}t/\hslash
)\Psi _{00}(x,r,t_{0})  \tag{3.27a}
\end{equation}%
and%
\begin{equation*}
E_{r}^{(2)}(x,r,t_{0}+t)=-\frac{1}{\hslash ^{2}}\int_{0}^{t}d\lambda
\int_{0}^{\lambda }d\lambda ^{\prime }\{\exp [-i(H_{0}+V_{1})(t-\lambda
)/\hslash ]
\end{equation*}%
\begin{equation}
\times \exp [-iV_{1}(\lambda -\lambda ^{\prime })/\hslash ][H_{0},V_{1}]\exp
(-iV_{1}\lambda ^{\prime }/\hslash )\exp (-iH_{0}\lambda /\hslash )\Psi
_{00}(x,r,t_{0})\}.  \tag{3.27b}
\end{equation}%
Here the main Hamiltonian $H_{0}=H_{0}^{ho}$ and the perturbation term $%
V_{1}=V_{1}^{ho}(x).$ The two formulae (3.27) may be used to investigate the
imperfection of the $LH$ harmonic potential well. The equation (3.25)
describes completely the time evolution process of the halting-qubit atom in
the double-well potential field $V(x)$ of (2.1). As shown in (2.4), the
time-independent perturbation term $V_{1}^{ho}(x)$ is closely related to the
double-well potential field. The double-well potential field may be treated
as an external potential-field pulse in the time evolution process. It is
applied to the halting-qubit atom at some initial time $t_{0}$ and then is
turned off at a later time $t_{0}+t$. This process is really a unitary
operation in space and time. The double-well potential field is zero in the
space region $(x_{L}+L,+\infty ),$ meaning that it is applied to the atom
only in the chosen spatial region $(-\infty ,x_{L}+L)$ of the whole
coordinate space $(-\infty ,+\infty )$. Thus, this unitary operation is
spatially selective. Similarly, the perturbation term $V_{1}^{ho}(x)$ of
(2.4) also is spatially selective. In order to calculate exactly the time
evolution process by using the exact formula (3.25) one needs to use the
smooth perturbation term $V_{1}.$ The perturbation term $V_{1}^{ho}(x)$ of
(2.4) is not smooth. But it can be made smooth according to (2.10). Thus, in
(3.25) the perturbation term $V_{1}$ may take the smooth perturbation term $%
V_{1}^{ho}(x,\varepsilon )$ of (2.10). Now let $H_{0}=H_{0}^{ho}$ and $%
V_{1}=V_{1}^{ho}(x,\varepsilon )$ in (3.25)--(3.27). Then one can find that
the motional state $\Psi _{0}(x,t_{0}+t)$ of (3.26) with $t_{0}+t=t_{b}$ is
just the desired state $\Psi _{0}(x_{b},t_{b})$ of (3.2) with $t_{c}=t_{b}.$
Thus, both the first-order approximation propagator of (2.9) and the exact
propagator of (3.24) obtain the same desired state. But their difference is
reflected by their own error terms. The first-order approximation propagator
generates the error $E_{r}(x_{b},t_{b})$ in (3.1), while the exact
propagator leads to the exact error expression $%
E_{r}^{(1)}(x,r,t_{0}+t)+E_{r}^{(2)}(x,r,t_{0}+t)$ in (3.25). It can turn
out below that the first error $E_{r}^{(1)}(x,r,t_{0}+t)$ can result in the
same error upper bound as the error $E_{r}(x_{b},t_{b})$. Thus, the error $%
E_{r}^{(1)}(x,r,t_{0}+t)$ may be considered as the first-order approximation
error. The upper bound of the error $E_{r}(x_{b},t_{b})$ is determined
directly from (3.12). On the other hand, it follows from (3.27a) that the
upper bound of the error $E_{r}^{(1)}(x,r,t_{0}+\tau )$ satisfies the
inequality: \newline
\begin{equation}
||E_{r}^{(1)}(x,r,t_{0}+\tau )||\leq (\frac{\tau }{\hslash }%
)||V_{1}^{ho}(x,\varepsilon )\exp [-iH_{0}^{ho}\tau /\hslash ]\Psi
_{00}(x,r,t_{0})||.  \tag{3.28}
\end{equation}%
This inequality holds for any time interval $\tau \geq 0.$ Now by inserting
the continuous perturbation term $V_{1}^{ho}(x,\varepsilon )$ of (2.10) into
(3.28) one obtains%
\begin{equation*}
||V_{1}^{ho}(x,\varepsilon )\exp [-iH_{0}^{ho}\tau /\hslash ]\Psi
_{00}(x,r,t_{0})||^{2}
\end{equation*}%
\begin{equation}
<(\frac{1}{2}m\omega ^{2})^{2}\int_{-\infty }^{\infty }dx\{x^{4}|\Psi
_{0}(x,r,t_{0}+\tau )|^{2}\Theta (x-x_{L},\varepsilon )\}.  \tag{3.29}
\end{equation}%
This inequality with setting $t_{0}+\tau =t_{c}$ approaches (3.6) when the
smooth step function $\Theta (x-x_{L},\varepsilon )$ approaches the
incontinuous one $\Theta (x-x_{L}),$ that is, $\Theta (x-x_{L},\varepsilon
)\rightarrow \Theta (x-x_{L}).$ Thus, using the continuous perturbation term 
$V_{1}^{ho}(x,\varepsilon )$ or incontinuous term $V_{1}^{ho}(x)$ to
calculate the integral on the $RH$ side of (3.29) will not generate a
significant difference if the positive parameter $\varepsilon <<1$. This
leads to that both the inequalities (3.4) and (3.28) can generate the same
upper bound as $\varepsilon \rightarrow 0,$ indicating that the error $%
E_{r}^{(1)}(x,r,t_{0}+t)$ has the same error upper bound as the error $%
E_{r}(x_{b},t_{b})$. Here one needs to set the time interval $\tau
=t_{b}-t_{0}$ in (3.28) and take the maximum norms on both the $RH$ sides of
(3.4) and (3.29) over the full time region $[t_{0},$ $t_{0}+\tau ].$ In
particular, the maximum norm on the $RH$ side of (3.28) is over the full
time region $[t_{0},$ $t_{0}+\tau ]$ instead of at the final time $%
t_{0}+\tau $ only. Thus, the upper bound of the error $%
E_{r}^{(1)}(x,r,t_{0}+\tau )$ also may be calculated by using (3.4) with the
perturbation term $V_{1}^{ho}(x,\varepsilon )$ or by using (3.12) directly.
A rigorous calculation for the upper bound of the integral on the $RH$ side
of (3.29) is not difficult even if $\varepsilon $ is not very small, as can
be seen below.

One may consider the second error $E_{r}^{(2)}(x,r,t_{0}+t)$ in (3.25) as a
correction to the error $E_{r}^{(1)}(x,r,t_{0}+t)$ of the first-order
approximation. It can be found from (3.27b) that the upper bound of the
error $E_{r}^{(2)}(x,r,t_{0}+t)$ is determined from%
\begin{equation*}
||E_{r}^{(2)}(x,r,t_{0}+t)||\leq \frac{1}{\hslash ^{2}}\int_{0}^{t}d\lambda
\int_{0}^{\lambda }d\lambda ^{\prime
}\{||[H_{0}^{ho},V_{1}^{ho}(x,\varepsilon )]
\end{equation*}%
\begin{equation}
\times \exp (-iV_{1}^{ho}(x,\varepsilon )\lambda ^{\prime }/\hslash )\exp
(-iH_{0}^{ho}\lambda /\hslash )\Psi _{00}(x,r,t_{0})||\}.  \tag{3.30}
\end{equation}%
This upper bound is dependent on the parameter $\varepsilon .$ It is
proportional to the commutator $[H_{0}^{ho},V_{1}^{ho}(x,\varepsilon )],$
which contains the first- and second-order coordinate derivatives of the
smooth perturbation term $V_{1}^{ho}(x,\varepsilon )$. In order to calculate
the commutator one needs to calculate the coordinate derivatives of the
perturbation term $V_{1}^{ho}(x,\varepsilon ).$ This is the reason why the
incontinuous perturbation term $V_{1}^{ho}(x)$\ of (2.4) needs to be made
smooth in the present strict error estimation. It is not difficult to
calculate the commutator $[H_{0}^{ho},V_{1}^{ho}(x)]$ on the $RH$ side of
(3.30) by using the continuous perturbation term $V_{1}^{ho}(x,\varepsilon )$
of (2.10). It is clear that once the upper bound of the integrand on the $RH$
side of (3.30) is determined, one can easily determine the upper bound of
the error $E_{r}^{(2)}(x,r,t_{0}+t)$ from (3.30). Now by using the
Hamiltonian $H_{0}^{ho}$ of (2.3) and $V_{1}^{ho}(x,\varepsilon )$ of (2.10)
it can prove that the integrand on the $RH$ side of (3.30) is bounded by%
\begin{equation*}
2m||[H_{0}^{ho},V_{1}^{ho}(x,\varepsilon )]\exp (-iV_{1}^{ho}(x,\varepsilon
)\lambda ^{\prime }/\hslash )\exp (-iH_{0}^{ho}\lambda /\hslash )\Psi
_{00}(x,r,t_{0})||
\end{equation*}%
\begin{equation*}
\leq \hslash ^{2}||[\frac{\partial ^{2}}{\partial x^{2}}V_{1}^{ho}(x,%
\varepsilon )]\Psi _{0}(x,r,t_{0}+\lambda )||+2\hslash \lambda ^{\prime }||[%
\frac{\partial }{\partial x}V_{1}^{ho}(x,\varepsilon )]^{2}\Psi
_{0}(x,r,t_{0}+\lambda )||
\end{equation*}%
\begin{equation}
+2\hslash ^{2}||[\frac{\partial }{\partial x}V_{1}^{ho}(x,\varepsilon )][%
\frac{\partial }{\partial x}\Psi _{0}(x,r,t_{0}+\lambda )]||  \tag{3.31}
\end{equation}%
where $\Psi _{0}(x,r,t_{0}+\lambda )=\exp (-iH_{0}^{ho}\lambda /\hslash
)\Psi _{00}(x,r,t_{0})$ is a $GWP$ product state and the time $\lambda $ is
in $[0,$ $t]$. The next step is to prove that each one of the three norms on
the $RH$ side of (3.31) decays exponentially with the square
deviation-to-spread ratio of the state $\Psi _{0}(x,r,t)$. It is not
difficult to calculate the upper bounds of the three norms, but the
calculation is still quite cumbersome. For simplicity, consider that the
atomic initial state is a single $GWP$ product state $\Psi
_{00}(x,r,t_{0})=\Psi _{00}(x,t_{0})|\psi (r)\rangle ,$ where $\Psi
_{00}(x,t_{0})$ is a single $GWP$ motional state and $|\psi (r)\rangle $ an
atomic internal eigenstate. Suppose that the $GWP$ motional state $\Psi
_{0}(x,t_{0}+\lambda )$ has the characteristic parameters $%
\{x_{c}(t_{0}+\lambda ),$ $p_{c}(t_{0}+\lambda ),$ $W_{c}(t_{0}+\lambda ),$ $%
\varepsilon _{c}(t_{0}+\lambda )\}$, which may depend on the time $\lambda $
with $0\leq \lambda \leq t.$ Below using these parameter and initial-state
settings a detailed calculation is carried out for the upper bounds of these
three norms.

In order to calculate conveniently the upper bounds of these three norms
here introduce an important function: the smooth $\delta -$function $\delta
(x,\varepsilon ).$\ The smooth $\delta -$function $\delta (x,\varepsilon )$
is defined by 
\begin{equation}
\delta (x,\varepsilon )=\frac{\partial }{\partial x}\Theta (x,\varepsilon )=(%
\frac{1}{\varepsilon \sqrt{\pi }})\exp (-\frac{x^{2}}{\varepsilon ^{2}}). 
\tag{3.32}
\end{equation}%
Here the spread parameter $\varepsilon $ satisfies $0<\varepsilon <<1.$ The
smooth $\delta -$function is a standard Gaussian function. It is equal to
the first-order coordinate derivative of the smooth step function $\Theta
(x,\varepsilon )$ that is defined in (2.11). It satisfies $0<\delta
(x,\varepsilon )\leq \frac{1}{\varepsilon \sqrt{\pi }}$ for $-\infty
<x<+\infty .$ The relation between the smooth $\delta -$function and the
conventional $\delta -$function $\delta (x)$ is described by $\delta
(x,\varepsilon )\rightarrow \delta (x)$ when $\varepsilon \rightarrow 0.$
Both the smooth $\delta -$function $\delta (x,\varepsilon )$ and the smooth
step function $\Theta (x,\varepsilon )$ have been used extensively below.

First of all, calculate the third norm on the $RH$ side of (3.31) as it is
simpler. It is easy to calculate exactly both the first-order coordinate
derivatives of $V_{1}^{ho}(x,\varepsilon )$ and $\Psi _{0}(x,t_{0}+\lambda )$
in the norm. It can turn out that the norm is bounded by%
\begin{equation*}
||[\frac{\partial }{\partial x}V_{1}^{ho}(x,\varepsilon )]\frac{\partial }{%
\partial x}\Psi _{0}(x,r,t_{0}+\lambda )||
\end{equation*}%
\begin{equation*}
\leq \sum_{l=0}^{1}A_{l}^{a}||x^{l}\delta (x-x_{L}-L,\varepsilon )\Psi
_{0}(x,r,t_{0}+\lambda )||
\end{equation*}%
\begin{equation*}
+\sum_{l=0}^{3}A_{l}^{b}||x^{l}\delta (x-x_{L},\varepsilon )\Psi
_{0}(x,r,t_{0}+\lambda )||
\end{equation*}%
\begin{equation}
+\sum_{l=0}^{2}A_{l}^{c}||x^{l}\Theta (x-x_{L},\varepsilon )\Psi
_{0}(x,r,t_{0}+\lambda )||,  \tag{3.33a}
\end{equation}%
where the non-negative parameters $\{A_{l}^{\alpha }\}$ are given by%
\begin{equation*}
A_{0}^{c}=0,\text{ }A_{1}^{c}=m\omega ^{2}(\frac{1}{2}|\frac{%
x_{c}(t_{0}+\lambda )}{W_{c}(t_{0}+\lambda )}|+|p_{c}(t_{0}+\lambda
)/\hslash |),
\end{equation*}%
\begin{equation*}
A_{2}^{c}=\frac{1}{2}\frac{m\omega ^{2}}{|W_{c}(t_{0}+\lambda )|},
\end{equation*}%
and the rest parameters can be obtained from the two parameters $A_{1}^{c}$
and $A_{2}^{c},$ that is, $A_{0}^{a}=A_{0}^{b}=x_{L}^{2}A_{1}^{c}/2,$ $%
A_{1}^{a}=A_{1}^{b}=x_{L}^{2}A_{2}^{c}/2,$ $A_{2}^{b}=A_{1}^{c}/2,$ and $%
A_{3}^{b}=A_{2}^{c}/2.$ By calculating directly the second-order coordinate
derivative $\frac{\partial ^{2}}{\partial x^{2}}V_{1}^{ho}(x,\varepsilon )$
and then using the derivative to calculate the first norm on the $RH$ side
of (3.31) one can prove that the norm is bounded by%
\begin{equation*}
||[\frac{\partial ^{2}}{\partial x^{2}}V_{1}^{ho}(x,\varepsilon )]\Psi
_{0}(x,r,t_{0}+\lambda )||\leq (m\omega ^{2})||\Theta (x-x_{L},\varepsilon
)\Psi _{0}(x,r,t_{0}+\lambda )||
\end{equation*}%
\begin{equation*}
+\sum_{l=0}^{3}B_{l}^{a}||x^{l}\delta (x-x_{L},\varepsilon )\Psi
_{0}(x,r,t_{0}+\lambda )||
\end{equation*}%
\begin{equation*}
+\sum_{l=0}^{1}B_{l}^{b}||x^{l}\delta (x-x_{L}-L,\varepsilon )\Psi
_{0}(x,r,t_{0}+\lambda )||
\end{equation*}%
\begin{equation}
+(m\omega ^{2}x_{L}^{2})||\delta (x-x_{L},\varepsilon )\delta
(x-x_{L}-L,\varepsilon )\Psi _{0}(x,r,t_{0}+\lambda )||  \tag{3.33b}
\end{equation}%
where the non-negative parameters $\{B_{l}^{\alpha }\}$ are given by $%
B_{0}^{a}=x_{L}(\frac{m\omega ^{2}x_{L}^{2}}{\varepsilon ^{2}}),$ $%
B_{1}^{a}=(\frac{m\omega ^{2}x_{L}^{2}}{\varepsilon ^{2}})+2m\omega ^{2},$ $%
B_{2}^{a}=x_{L}(\frac{m\omega ^{2}}{\varepsilon ^{2}}),$ $B_{3}^{a}=(\frac{%
m\omega ^{2}}{\varepsilon ^{2}});$ $B_{0}^{b}=(x_{L}+L)(\frac{m\omega
^{2}x_{L}^{2}}{\varepsilon ^{2}}),$ $B_{1}^{0}=(\frac{m\omega ^{2}x_{L}^{2}}{%
\varepsilon ^{2}}).$ Similarly, one can prove that the second norm on the $%
RH $ side of (3.31) is bounded by%
\begin{equation*}
||[\frac{\partial }{\partial x}V_{1}^{ho}(x,\varepsilon )]^{2}\Psi
_{0}(x,r,t_{0}+\lambda )||
\end{equation*}%
\begin{equation*}
\leq (m\omega ^{2})^{2}||x^{2}\Theta (x-x_{L},\varepsilon )\Psi
_{0}(x,r,t_{0}+\lambda )||
\end{equation*}%
\begin{equation*}
+\sum_{l=0}^{1}D_{l}^{a}||x^{2l+1}\delta (x-x_{L},\varepsilon )\Psi
_{0}(x,r,t_{0}+\lambda )||
\end{equation*}%
\begin{equation*}
+\sum_{l=0}^{2}D_{l}^{b}||x^{2l}\delta (x-x_{L},\varepsilon )^{2}\Psi
_{0}(x,r,t_{0}+\lambda )||
\end{equation*}%
\begin{equation*}
+x_{L}^{2}(m\omega ^{2})^{2}||x\delta (x-x_{L}-L,\varepsilon )\Psi
_{0}(x,r,t_{0}+\lambda )||
\end{equation*}%
\begin{equation*}
+\frac{1}{4}(m\omega ^{2}x_{L}^{2})^{2}||\delta (x-x_{L}-L,\varepsilon
)^{2}\Psi _{0}(x,r,t_{0}+\lambda )||
\end{equation*}%
\begin{equation}
+\sum_{l=0}^{1}D_{l}^{c}||x^{2l}\delta (x-x_{L}-L,\varepsilon )\delta
(x-x_{L},\varepsilon )\Psi _{0}(x,r,t_{0}+\lambda )||  \tag{3.33c}
\end{equation}%
where the non-negative parameters $\{D_{l}^{\alpha }\}$ are given by $%
D_{0}^{a}=(m\omega ^{2})^{2}x_{L}^{2},$ $D_{1}^{a}=(m\omega ^{2})^{2};$ $%
D_{0}^{b}=\frac{1}{4}(m\omega ^{2}x_{L}^{2})^{2},$ $D_{1}^{b}=\frac{1}{2}%
x_{L}^{2}(m\omega ^{2})^{2},$ $D_{2}^{b}=\frac{1}{4}(m\omega ^{2})^{2};$ $%
D_{0}^{c}=\frac{1}{2}(m\omega ^{2}x_{L}^{2})^{2},$ $D_{1}^{c}=\frac{1}{2}%
x_{L}^{2}(m\omega ^{2})^{2}.$ Since the functions $\delta
(x-x_{L},\varepsilon )^{2}$ and $\delta (x-x_{L}-L,\varepsilon )^{2}$ are
proportional to $\varepsilon ^{-2},$ the third and fifth norms are
dominating terms on the $RH$ side of (3.33c) when the positive parameter $%
\varepsilon <<1$. Though the function $\delta (x-x_{L}-L,\varepsilon )\delta
(x-x_{L},\varepsilon )$ is also proportional to $\varepsilon ^{-2},$ it can
prove below that the last terms with this function on the $RH$ sides of both
(3.33b) and (3.33c) can be neglected when the distance $L>>\varepsilon .$

The upper bounds of the three norms on the $RH$ side of (3.31) are
determined from the $RH$ sides of the three inequalities of (3.33),
respectively. The $RH$ side of (3.33a) consists of the two types of basic
norms. The first type is defined by%
\begin{equation*}
NBAS1=||x^{l}\delta (x-x_{c},\varepsilon _{c})\Psi _{0}(x,r,t_{0}+\lambda )||
\end{equation*}%
and the second by 
\begin{equation*}
NBAS2=||x^{l}\Theta (x-x_{L},\varepsilon )\Psi _{0}(x,r,t_{0}+\lambda )||,
\end{equation*}%
where the finite integer $l=0,$ $1,$ $2,$ $...,$ and the other parameters
have the same physical meanings as stated before. The two types of basic
norms are very important in the strict error estimation below. It can be
found below that the upper bounds of many complex norms can be expressed as
a linear combination of the two types of basic norms. For example, it will
be shown below that the $RH$ sides of the two inequalities (3.33b) and
(3.33c) also consist of the two types of basic norms. Here one needs to
prove that the last norms with the function $\delta (x-x_{L}-L,\varepsilon
)\delta (x-x_{L},\varepsilon )$ on the $RH$ sides of (3.33b) and (3.33c) can
be reduced to the first type of basic norms $\{NBAS1\}.$ One also needs to
prove that the third and fifth norms with the functions $\delta
(x-x_{L},\varepsilon )^{2}$ and $\delta (x-x_{L}-L,\varepsilon )^{2}$,
respectively, on the $RH$ side of (3.33c) can be reduced to the first type
of basic norms. There is an identity for the product of a pair of Gaussian
functions: 
\begin{equation}
\exp [-\frac{(x-z_{1})^{2}}{\varepsilon _{1}^{2}}]\exp [-\frac{(x-z_{2})^{2}%
}{\varepsilon _{2}^{2}}]=\exp \{-\frac{(z_{1}-z_{2})^{2}}{(\varepsilon
_{1}^{2}+\varepsilon _{2}^{2})}\}\exp \{-\frac{[x-x_{0}]^{2}}{\varepsilon
_{0}^{2}}\},  \tag{3.34}
\end{equation}%
where the parameters $x_{0}$ and $\varepsilon _{0}^{2}$ are given by 
\begin{equation*}
\varepsilon _{0}^{2}=\frac{\varepsilon _{1}^{2}\varepsilon _{2}^{2}}{%
(\varepsilon _{1}^{2}+\varepsilon _{2}^{2})},\text{ }x_{0}=\frac{%
(\varepsilon _{2}^{2}z_{1}+\varepsilon _{1}^{2}z_{2})}{(\varepsilon
_{1}^{2}+\varepsilon _{2}^{2})}.
\end{equation*}%
By this identity and the definition (3.32) of the smooth $\delta -$function
it can turn out that 
\begin{equation}
\delta (x-x_{L},\varepsilon )\delta (x-x_{L}-L,\varepsilon )=\frac{1}{%
\varepsilon \sqrt{2\pi }}\exp (-\frac{1}{2}\frac{L^{2}}{\varepsilon ^{2}}%
)\delta (x-x_{L}-L/2,\varepsilon /\sqrt{2}).  \tag{3.35}
\end{equation}%
This formula shows that the product of a pair of the smooth $\delta -$%
functions with different COM positions can be expressed as the product of
another smooth $\delta -$function and a Gaussian factor. This shows that the
function $\delta (x-x_{L},\varepsilon )\delta (x-x_{L}-L,\varepsilon )$ is
proportional to the exponentially-decaying factor $\exp
[-L^{2}/(2\varepsilon ^{2})]$, indicating that it is much smaller than a
usual smooth $\delta -$function and can be neglected when the length $L$
satisfies $L>>\varepsilon >0.$ In particular, when the length $L=0$, the
formula (3.35) is reduced to the form 
\begin{equation}
\delta (x-x_{L},\varepsilon )^{2}=\frac{1}{\varepsilon \sqrt{2\pi }}\delta
(x-x_{L},\varepsilon /\sqrt{2}).  \tag{3.36}
\end{equation}%
There is not an extra exponentially-decaying factor for the function $\delta
(x-x_{L},\varepsilon )^{2},$ indicating that the function $\delta
(x-x_{L},\varepsilon )^{2}$ is really a usual smooth $\delta -$function up
to a constant $(\varepsilon \sqrt{2\pi })^{-1}$. The formula (3.35) shows
that the last norms on the $RH$ sides of (3.33b) and (3.33c) can be reduced
to the first type of basic norms and at the same time an extra
exponentially-decaying factor is generated for each norm. This extra
exponentially-decaying factor results in that these last norms can be
neglected when the length $L>>\varepsilon .$ The relation (3.36) directly
shows that the third and fifth norms on the $RH$ side of (3.33c) can be
reduced to the first type of basic norms. Now these three inequalities
(3.33) show that the upper bound of each one of the three norms on the $RH$
side of (3.31) can be expressed as a linear combination of the two types of
the basic norms $\{NBAS1\}$ and $\{NBAS2\}$. Therefore, in order to prove
that the error norm $||E_{r}^{(2)}(x,r,t_{0}+t)||$ in (3.30) decays
exponentially with the square deviation-to-spread ratios one needs only to
calculate explicitly the two types of the basic norms $\{NBAS1\}$ and $%
\{NBAS2\}$ and shows that they decay exponentially with the square
deviation-to-spread ratios, respectively.

Below the two basic norms $NBAS1$ and $NBAS2$ are explicitly calculated.
First of all, consider the first basic norm $NBAS1$. In the basic norm the
smooth $\delta -$function $\delta (x-x_{c},\varepsilon _{c})$ is defined by
(3.32). For those basic norms $\{NBAS1\}$ that appear on the $RH$ sides of
the three inequalities (3.33) the COM position $x_{c}$ may be taken as $%
x_{L} $ or $x_{L}+L$ and the spread $\varepsilon _{c}$ taken as $\varepsilon 
$ or $\varepsilon /\sqrt{2}$ in the smooth $\delta -$function $\delta
(x-x_{c},\varepsilon _{c}).$ Now by using directly the $GWP$ state $\Psi
_{0}(x,r,t_{0}+\lambda )$ one can prove that the upper bound of the basic
norm $NBAS1$ can be calculated through the relation:%
\begin{equation*}
||x^{l}\delta (x-x_{c},\varepsilon _{c})\Psi _{0}(x,r,t_{0}+\lambda )||^{2}=(%
\frac{1}{\varepsilon _{c}\sqrt{\pi }})^{2}\frac{1}{\varepsilon
_{c}(t_{0}+\lambda )\sqrt{\pi }}
\end{equation*}%
\begin{equation}
\times \exp \{-\frac{[x_{c}-x_{c}(t_{0}+\lambda )]^{2}}{\varepsilon
_{c}^{2}/2+\varepsilon _{c}(t_{0}+\lambda )^{2}}\}I_{2l}(x_{0},\varepsilon
_{0})  \tag{3.37}
\end{equation}%
where the product identity (3.34) of a pair of Gaussian functions is already
used, the Gaussian integral $I_{k}(x_{0},\varepsilon _{0})$ (here $k=2l$) is
defined by%
\begin{equation}
I_{k}(x_{0},\varepsilon _{0})=\int_{-\infty }^{\infty }dx\{x^{k}\exp [-\frac{%
(x-x_{0})^{2}}{\varepsilon _{0}^{2}}]\},  \tag{3.38a}
\end{equation}%
and the parameters $x_{0}$ and $\varepsilon _{0}^{2}$ are given by%
\begin{equation*}
\varepsilon _{0}^{2}=\frac{\varepsilon _{c}^{2}\varepsilon
_{c}(t_{0}+\lambda )^{2}}{\varepsilon _{c}^{2}+2\varepsilon
_{c}(t_{0}+\lambda )^{2}},\text{ }x_{0}=\frac{2x_{c}\varepsilon
_{c}(t_{0}+\lambda )^{2}+\varepsilon _{c}^{2}x_{c}(t_{0}+\lambda )}{%
\varepsilon _{c}^{2}+2\varepsilon _{c}(t_{0}+\lambda )^{2}}.
\end{equation*}%
It can prove that the Gaussian integral $I_{2l}(x_{0},\varepsilon _{0})$ is
a $2l-$order polynomial in parameter $x_{0}.$ Denote $y=(x-x_{0})/%
\varepsilon _{0}.$ The binomial expansion for $x^{k}$ with the variable $y$
is given by%
\begin{equation*}
x^{k}=(y\varepsilon _{0}+x_{0})^{k}=\sum_{j=0}^{k}\left( 
\begin{array}{c}
k \\ 
j%
\end{array}%
\right) x_{0}^{k-j}\varepsilon _{0}^{j}y^{j}.
\end{equation*}%
Then by inserting this expansion into the Gaussian integral $%
I_{k}(x_{0},\varepsilon _{0})$ one obtains, with the help of the Gaussian
integral formula [44a] (see also (5.11) in the section 5 below),%
\begin{equation}
I_{k}(x_{0},\varepsilon _{0})=\varepsilon _{0}\sqrt{\pi }%
\sum_{j=0,2,4,...}^{k}\left( 
\begin{array}{c}
k \\ 
j%
\end{array}%
\right) \frac{(j-1)!!}{\sqrt{2^{j}}}x_{0}^{k-j}\varepsilon _{0}^{j}. 
\tag{3.38b}
\end{equation}%
Indeed, the Gaussian integral $I_{2l}(x_{0},\varepsilon _{0})$ is a $2l-$%
order polynomial in the COM position $x_{0}$ or a $(2l+1)-$order polynomial
in the spread $\varepsilon _{0}.$ Then it follows from (3.37) and (3.38)
that the first basic norm $NBAS1$ is proportional to the following
exponentially-decaying factor, 
\begin{equation*}
||x^{l}\delta (x-x_{c},\varepsilon _{c})\Psi _{0}(x,r,t_{0}+\lambda
)||\varpropto \exp \{-\frac{1}{2}Y_{eff}(x_{c},\varepsilon _{eff})^{2}\}.
\end{equation*}%
Here the effective deviation-to-spread ratio $Y_{eff}(x_{c},\varepsilon
_{eff})$ is defined by%
\begin{equation}
Y_{eff}(x_{c},\varepsilon _{eff})^{2}=\frac{[x_{c}-x_{c}(t_{0}+\lambda )]^{2}%
}{\varepsilon _{eff}^{2}}  \tag{3.39}
\end{equation}%
and the effective wave-packet spread $\varepsilon _{eff}=\varepsilon
_{eff}(t_{0}+\lambda )$ by 
\begin{equation*}
\varepsilon _{eff}^{2}=\varepsilon _{c}^{2}/2+\varepsilon _{c}(t_{0}+\lambda
)^{2}.
\end{equation*}%
On the other hand, the deviation-to-spread ratio $y_{M}(x_{c},t_{0}+\lambda
) $ of the $GWP$ state $\Psi _{0}(x,r,t_{0}+\lambda )$ is determined from $%
y_{M}(x_{c},t_{0}+\lambda )^{2}=[x_{c}-x_{c}(t_{0}+\lambda
)]^{2}/\varepsilon _{c}(t_{0}+\lambda )^{2}.$ When $\varepsilon
_{c}^{2}/2<<\varepsilon _{c}(t_{0}+\lambda )^{2},$ the square effective
wave-packet spread $\varepsilon _{eff}(t_{0}+\lambda )^{2}$ is close to $%
\varepsilon _{c}(t_{0}+\lambda )^{2}$ and the square effective
deviation-to-spread ratio $Y_{eff}(x_{c},\varepsilon _{eff})^{2}$ close to $%
y_{M}(x_{c},t_{0}+\lambda )^{2}$. This means that when $\varepsilon
_{c}^{2}/2<<\varepsilon _{c}(t_{0}+\lambda )^{2},$ the effective wave-packet
spread and the effective deviation-to-spread ratio are approximately equal
to those of the $GWP$ state $\Psi _{0}(x,r,t_{0}+\lambda ),$ respectively.
Now the first basic norm $NBAS1$ decays exponentially with the square
effective deviation-to-spread ratio. This means that it decays exponentially
with the square deviation-to-spread ratio of the $GWP$ state $\Psi
_{0}(x,r,t_{0}+\lambda )$ approximately when $\varepsilon
_{c}^{2}/2<<\varepsilon _{c}(t_{0}+\lambda )^{2}$.

Now the second basic norm $NBAS2$ is calculated explicitly. By using
directly the $GWP$ state $\Psi _{0}(x,r,t_{0}+\lambda )$ and the smooth step
function $\Theta (x-x_{L},\varepsilon )$ of (2.11) one can prove that the
upper bound for the second basic norm may be calculated by%
\begin{equation*}
||x^{l}\Theta (x-x_{L},\varepsilon )\Psi _{0}(x,r,t_{0}+\lambda )||^{2}\leq (%
\frac{1}{\varepsilon \sqrt{\pi }})\frac{1}{\varepsilon _{c}(t_{0}+\lambda )%
\sqrt{\pi }}
\end{equation*}%
\begin{equation}
\times \int_{-\infty }^{\infty }dx\{x^{2l}\int_{-\infty }^{x-x_{L}}\exp (-%
\frac{z^{2}}{\varepsilon ^{2}})dz\exp \{-\frac{[x-x_{c}(t_{0}+\lambda )]^{2}%
}{\varepsilon _{c}(t_{0}+\lambda )^{2}}\}\}.  \tag{3.40}
\end{equation}%
Actually, this basic norm also may be calculated approximately by using
directly the usual step function $\Theta (x-x_{L})$ as $\Theta
(x-x_{L},\varepsilon )\rightarrow \Theta (x-x_{L})$ when $\varepsilon
\rightarrow 0.$ Below it is calculated strictly with a two-dimensional
integration method in polar coordinate system. This strict calculation is
based on the basic integral: 
\begin{equation}
J(\varepsilon )=\int_{-\infty }^{\infty }dx\{\exp [-\frac{(x-x_{cm})^{2}}{%
\varepsilon _{cm}^{2}}]\int_{-\infty }^{x-x_{L}}\exp (-\frac{z^{2}}{%
\varepsilon ^{2}})dz\}.  \tag{3.41}
\end{equation}%
By making the variable transformation $y=-z+(x-x_{L})$ the integral $%
J(\varepsilon )$ is changed to the two-dimensional form%
\begin{equation*}
J(\varepsilon )=\int_{-\infty }^{\infty }dx\int_{0}^{\infty }dy\{\exp \{-%
\frac{(x-x_{cm})^{2}}{\varepsilon _{cm}^{2}}-\frac{[y-(x-x_{L})]^{2}}{%
\varepsilon ^{2}}\}\}
\end{equation*}%
where the integral region is the up-half $xy-$plane. This integral may be
treated more easily in the polar coordinate system $(r,$ $\theta )$:%
\begin{equation*}
x=r\cos \theta ,\text{ }y=r\sin \theta ,\text{ }dxdy=rdrd\theta .
\end{equation*}%
In the polar coordinate system the integral may be written as%
\begin{equation}
J(\varepsilon )=\int_{0}^{\pi }d\theta \int_{0}^{\infty }rdr\exp \{-[\frac{%
r^{2}}{\varepsilon _{0}(\theta )^{2}}+\frac{2r_{0}(\theta )}{\varepsilon
_{0}(\theta )^{2}}r+(\frac{x_{cm}^{2}}{\varepsilon _{cm}^{2}}+\frac{x_{L}^{2}%
}{\varepsilon ^{2}})]\}  \tag{3.42}
\end{equation}%
where the functions $\varepsilon _{0}(\theta )$ and $r_{0}(\theta )$ are
given by 
\begin{equation}
\varepsilon _{0}(\theta )^{-2}=[\frac{\cos ^{2}\theta }{\varepsilon _{cm}^{2}%
}+\frac{(\sin \theta -\cos \theta )^{2}}{\varepsilon ^{2}}],  \tag{3.43}
\end{equation}%
\begin{equation}
r_{0}(\theta )=[\frac{\cos ^{2}\theta }{\varepsilon _{cm}^{2}}+\frac{(\sin
\theta -\cos \theta )^{2}}{\varepsilon ^{2}}]^{-1}[\frac{-x_{cm}\cos \theta 
}{\varepsilon _{cm}^{2}}+\frac{x_{L}(\sin \theta -\cos \theta )}{\varepsilon
^{2}}].  \tag{3.44}
\end{equation}%
It is clear that $\varepsilon _{0}(\theta )^{-2}>0,$ but $r_{0}(\theta )$
may be positive, zero, and negative. The integral (3.42) may be further
simplified by making the integration by parts on the variable $r$ and then
it may be written as a sum of the two integrals: 
\begin{equation}
J(\varepsilon )=J_{1}(\varepsilon )+J_{2}(\varepsilon ),  \tag{3.45a}
\end{equation}%
where the two integrals $J_{1}(\varepsilon )$ and $J_{2}(\varepsilon )$ are
given by 
\begin{equation}
J_{1}(\varepsilon )=\frac{1}{2}\exp \{-[\frac{x_{cm}^{2}}{\varepsilon
_{cm}^{2}}+\frac{x_{L}^{2}}{\varepsilon ^{2}}]\}\int_{0}^{\pi }d\theta
\varepsilon _{0}(\theta )^{2},  \tag{3.45b}
\end{equation}%
\begin{equation*}
J_{2}(\varepsilon )=-\exp \{-[\frac{x_{cm}^{2}}{\varepsilon _{cm}^{2}}+\frac{%
x_{L}^{2}}{\varepsilon ^{2}}]\}
\end{equation*}%
\begin{equation}
\times \int_{0}^{\pi }d\theta \{r_{0}(\theta )\exp [\frac{r_{0}(\theta )^{2}%
}{\varepsilon _{0}(\theta )^{2}}]\int_{0}^{\infty }dr\exp \{-\frac{%
[r+r_{0}(\theta )]^{2}}{\varepsilon _{0}(\theta )^{2}}\}\}.  \tag{3.45c}
\end{equation}%
Below calculate the upper bounds of the two integrals. It can turn out that
the function $\varepsilon _{0}(\theta )^{2}$ of (3.43) is bounded. Denote $%
\varepsilon _{0}^{M}(\theta _{u})^{2}$ and $\varepsilon _{0}^{m}(\theta
_{l})^{2}$ as the maximum and minimum values of the function $\varepsilon
_{0}(\theta )^{2}$ in the region $0\leq \theta \leq \pi ,$ respectively,
that is, $\varepsilon _{0}^{m}(\theta _{l})^{2}\leq $ $\varepsilon
_{0}(\theta )^{2}\leq $ $\varepsilon _{0}^{M}(\theta _{u})^{2}.$ Then
according to (3.43) it can turn out that $\varepsilon _{0}^{M}(\theta
_{u})^{2}$ and $\varepsilon _{0}^{m}(\theta _{l})^{2}$ may be obtained from%
\begin{equation*}
\varepsilon _{0}^{M}(\theta _{u})^{2}=\max \{\varepsilon _{0}^{+}(\theta
^{\ast })^{2},\text{ }\varepsilon _{0}^{-}(\theta ^{\ast })^{2},\text{ }%
\frac{\varepsilon ^{2}\varepsilon _{cm}^{2}}{\varepsilon ^{2}+\varepsilon
_{cm}^{2}}\},
\end{equation*}%
\begin{equation*}
\varepsilon _{0}^{m}(\theta _{l})^{2}=\min \{\varepsilon _{0}^{+}(\theta
^{\ast })^{2},\text{ }\varepsilon _{0}^{-}(\theta ^{\ast })^{2},\text{ }%
\frac{\varepsilon ^{2}\varepsilon _{cm}^{2}}{\varepsilon ^{2}+\varepsilon
_{cm}^{2}}\},
\end{equation*}%
where $\varepsilon _{0}^{\pm }(\theta ^{\ast })^{2}$ are given by 
\begin{equation*}
\varepsilon _{0}^{\pm }(\theta ^{\ast })^{2}=\frac{\varepsilon
^{2}\varepsilon _{cm}^{2}\{2+(\frac{\varepsilon ^{2}}{\varepsilon _{cm}^{2}}%
)[(\frac{\varepsilon ^{2}}{2\varepsilon _{cm}^{2}})\pm \sqrt{1+(\frac{%
\varepsilon ^{2}}{2\varepsilon _{cm}^{2}})^{2}}]\}}{\varepsilon
^{2}+\varepsilon _{cm}^{2}\{(\frac{\varepsilon ^{2}}{2\varepsilon _{cm}^{2}}%
)\pm \sqrt{1+(\frac{\varepsilon ^{2}}{2\varepsilon _{cm}^{2}})^{2}}-1\}^{2}}.
\end{equation*}%
Therefore, the first integral $J_{1}(\varepsilon )$ of (3.45b) is bounded by%
\begin{equation}
J_{1}(\varepsilon )\leq \frac{\pi }{2}\exp \{-[\frac{x_{cm}^{2}}{\varepsilon
_{cm}^{2}}+\frac{x_{L}^{2}}{\varepsilon ^{2}}]\}\varepsilon _{0}^{M}(\theta
_{u})^{2}.  \tag{3.46}
\end{equation}%
The second integral $J_{2}(\varepsilon )$ of (3.45c) is more complex.
According to (3.44) it can turn out that $r_{0}(\theta )\leq 0$ if $0\leq
\theta \leq \theta _{m}$ and $r_{0}(\theta )>0$ if $\theta _{m}<\theta \leq
\pi ,$ where the angle $\theta _{m}$ is given by%
\begin{equation}
\theta _{m}=\tan ^{-1}\{(\frac{x_{cm}}{\varepsilon _{cm}^{2}}+\frac{x_{L}}{%
\varepsilon ^{2}})/(\frac{x_{L}}{\varepsilon ^{2}})\}<\pi /2.  \tag{3.47}
\end{equation}%
Then the function $r_{0}(\theta )$ may be written as 
\begin{equation}
r_{0}(\theta )=\left\{ 
\begin{array}{c}
|r_{0}(\theta )|,\text{ if }\theta _{m}<\theta \leq \pi \\ 
-|r_{0}(\theta )|,\text{ if }0\leq \theta \leq \theta _{m}<\pi /2%
\end{array}%
\right. .  \tag{3.48}
\end{equation}%
With the help of (3.48) the integral $J_{2}(\varepsilon )$ may be written as%
\begin{equation*}
J_{2}(\varepsilon )=\exp \{-[\frac{x_{c}^{2}}{\varepsilon _{c}^{2}}+\frac{%
x_{L}^{2}}{\varepsilon ^{2}}]\}\int_{0}^{\theta _{m}}d\theta \{J_{-}(\theta
)|r_{0}(\theta )|\exp [\frac{r_{0}(\theta )^{2}}{\varepsilon _{0}(\theta
)^{2}}]\}
\end{equation*}%
\begin{equation}
-\exp \{-[\frac{x_{c}^{2}}{\varepsilon _{c}^{2}}+\frac{x_{L}^{2}}{%
\varepsilon ^{2}}]\}\int_{\theta _{m}}^{\pi }d\theta \{J_{+}(\theta
)|r_{0}(\theta )|\exp [\frac{r_{0}(\theta )^{2}}{\varepsilon _{0}(\theta
)^{2}}]\}  \tag{3.49}
\end{equation}%
where the integral $J_{\pm }(\theta )$ is given by%
\begin{equation*}
J_{\pm }(\theta )=\int_{0}^{\infty }dr\exp \{-\frac{[r\pm |r_{0}(\theta
)|]^{2}}{\varepsilon _{0}(\theta )^{2}}\}.
\end{equation*}%
Since the second term is negative on the $RH$ side of (3.49), one can find
that the integral $J_{2}(\varepsilon )$ is bounded by%
\begin{equation}
J_{2}(\varepsilon )<\exp \{-[\frac{x_{c}^{2}}{\varepsilon _{c}^{2}}+\frac{%
x_{L}^{2}}{\varepsilon ^{2}}]\}\int_{0}^{\theta _{m}}d\theta \{J_{-}(\theta
)|r_{0}(\theta )|\exp [\frac{r_{0}(\theta )^{2}}{\varepsilon _{0}(\theta
)^{2}}]\}.  \tag{3.50}
\end{equation}%
Notice that the integral on the variable $r$ on the $RH$ side of (3.50) has
the upper and lower bounds: 
\begin{equation*}
\frac{1}{2}\varepsilon _{0}(\theta )\sqrt{\pi }\leq J_{-}(\theta
)<\varepsilon _{0}(\theta )\sqrt{\pi },
\end{equation*}%
and it can be found from (3.44) that the absolute function $|r_{0}(\theta )|$
is bounded by%
\begin{equation*}
|r_{0}(\theta )|<\varepsilon _{0}^{M}(\theta _{u})^{2}(|\frac{x_{cm}}{%
\varepsilon _{cm}^{2}}+\frac{x_{L}}{\varepsilon ^{2}}|+|\frac{x_{L}}{%
\varepsilon ^{2}}|).
\end{equation*}%
Then it follows from (3.50) that the integral $J_{2}(\varepsilon )$ is
bounded by%
\begin{equation}
J_{2}(\varepsilon )<\sqrt{\pi }\theta _{m}\varepsilon _{0}^{M}(\theta
_{u})^{3}(|\frac{x_{cm}}{\varepsilon _{cm}^{2}}+\frac{x_{L}}{\varepsilon ^{2}%
}|+|\frac{x_{L}}{\varepsilon ^{2}}|)\exp \{-[\frac{x_{cm}^{2}}{\varepsilon
_{cm}^{2}}+\frac{x_{L}^{2}}{\varepsilon ^{2}}]+[\frac{r_{0}(\theta )^{2}}{%
\varepsilon _{0}(\theta )^{2}}]_{\max }\}  \tag{3.51}
\end{equation}%
where $[\frac{r_{0}(\theta )^{2}}{\varepsilon _{0}(\theta )^{2}}]_{\max }$
is the maximum value of the ratio $r_{0}(\theta )^{2}/\varepsilon
_{0}(\theta )^{2}$ in the region $0\leq \theta \leq \theta _{m}$. It can
turn out that the maximum value is given by%
\begin{equation*}
\lbrack \frac{r_{0}(\theta )^{2}}{\varepsilon _{0}(\theta )^{2}}]_{\max
}=\max_{0\leq \theta \leq \theta _{m}}\{\frac{r_{0}(\theta )^{2}}{%
\varepsilon _{0}(\theta )^{2}}\}=\frac{(\frac{x_{L}}{\varepsilon ^{2}}+\frac{%
x_{cm}}{\varepsilon _{cm}^{2}})^{2}}{(\frac{1}{\varepsilon _{cm}^{2}}+\frac{1%
}{\varepsilon ^{2}})}.
\end{equation*}%
Now the upper bounds of both the integrals $J_{1}(\varepsilon )$ and $%
J_{2}(\varepsilon )$ are determined from (3.46) and (3.51), respectively.
Therefore, it follows from (3.46) and (3.51) that the integral $%
J(\varepsilon )$ of (3.45a) is bounded by%
\begin{equation*}
J(\varepsilon )<\frac{\pi }{2}\varepsilon _{0}^{M}(\theta _{u})^{2}\exp \{-[%
\frac{x_{cm}^{2}}{\varepsilon _{cm}^{2}}+\frac{x_{L}^{2}}{\varepsilon ^{2}}%
]\}
\end{equation*}%
\begin{equation}
+\sqrt{\pi }\theta _{m}(|\frac{x_{cm}}{\varepsilon _{cm}^{2}}+\frac{x_{L}}{%
\varepsilon ^{2}}|+|\frac{x_{L}}{\varepsilon ^{2}}|)\varepsilon
_{0}^{M}(\theta _{u})^{3}\exp \{-\frac{(x_{L}-x_{cm})^{2}}{(\varepsilon
^{2}+\varepsilon _{cm}^{2})}\}.  \tag{3.52}
\end{equation}%
The first term on the $RH$ side of (3.52) is proportional to the
exponentially-decaying factor $\exp \{-[\frac{x_{c}^{2}}{\varepsilon _{c}^{2}%
}+\frac{x_{L}^{2}}{\varepsilon ^{2}}]\},$ while the second term is
proportional to another exponentially-decaying factor $\exp \{-\frac{%
(x_{c}-x_{L})^{2}}{(\varepsilon ^{2}+\varepsilon _{c}^{2})}\}.$ Notice that $%
x_{cm}$ is taken as the COM position $x_{c}(t_{0}+\lambda )$ of the $GWP$
state $\Psi _{0}(x,r,t_{0}+\lambda )$ in (3.40) in the calculation of the
second basic norm. Thus, $x_{cm}$ is time-dependent and $x_{L}>|x_{cm}|.$
Then $0<(x_{L}-x_{cm})<x_{L}$ for $x_{cm}>0.$ In particular, the minimum
value of $(x_{L}-x_{cm})$ may be much less than $x_{L}.$ Therefore, it tends
to be%
\begin{equation*}
\lbrack \frac{(x_{L}-x_{cm})^{2}}{(\varepsilon ^{2}+\varepsilon _{cm}^{2})}%
]_{\min }<<\frac{x_{cm}^{2}}{\varepsilon _{cm}^{2}}+\frac{x_{L}^{2}}{%
\varepsilon ^{2}}.
\end{equation*}%
This means that the second term on the $RH$ side of (3.52) is the dominating
term, while the first term can be neglected. Thus, the upper bound of the
integral $J(\varepsilon )$ may be approximately determined from%
\begin{equation}
J(\varepsilon )<\sqrt{\pi }\theta _{m}(|\frac{x_{cm}}{\varepsilon _{cm}^{2}}+%
\frac{x_{L}}{\varepsilon ^{2}}|+|\frac{x_{L}}{\varepsilon ^{2}}|)\varepsilon
_{0}^{M}(\theta _{u})^{3}\exp \{-[\frac{(x_{L}-x_{cm})^{2}}{(\varepsilon
^{2}+\varepsilon _{cm}^{2})}]_{\min }\}.  \tag{3.53}
\end{equation}%
When $\varepsilon <<\varepsilon _{cm},$ the ratio $(x_{L}-x_{cm})^{2}/(%
\varepsilon ^{2}+\varepsilon _{cm}^{2})$ is approximately equal to the
square deviation-to-spread ratio, $(x_{L}-x_{cm})^{2}/\varepsilon _{cm}^{2},$
of the $GWP$ state with the COM position $x_{cm}$ and the wave-packet spread 
$\varepsilon _{cm}.$ Thus, the upper bound of the integral $J(\varepsilon )$
decays exponentially with the square deviation-to-spread ratio of the $GWP$
state when $\varepsilon <<\varepsilon _{cm}$.

Now both the equation (3.37) that is used to determine the upper bound of
the first basic norm $NBAS1$ and the inequality (3.53) that is used to
determine the upper bound of the integral $J(\varepsilon )$ together may be
used to determine the upper bound of the second basic norm $NBAS2$. First of
all, the inequality (3.40) that is used to determine the upper bound of the
second basic norm may be rewritten as%
\begin{equation}
||x^{l}\Theta (x-x_{L},\varepsilon )\Psi _{0}(x,r,t_{0}+\lambda )||^{2}\leq (%
\frac{1}{\varepsilon \sqrt{\pi }})\frac{J_{2l}(x_{c}(t_{0}+\lambda
),\varepsilon _{c}(t_{0}+\lambda ),\varepsilon )}{\varepsilon
_{c}(t_{0}+\lambda )\sqrt{\pi }}.  \tag{3.54}
\end{equation}%
Here the integral $J_{2l}(x_{cm},\varepsilon _{cm},\varepsilon )$ is defined
as%
\begin{equation}
J_{2l}(x_{cm},\varepsilon _{cm},\varepsilon )=\int_{-\infty }^{\infty
}dx\{x^{2l}\exp [-\frac{(x-x_{cm})^{2}}{\varepsilon _{cm}^{2}}]\int_{-\infty
}^{x-x_{L}}\exp (-\frac{z^{2}}{\varepsilon ^{2}})dz\}.  \tag{3.55a}
\end{equation}%
This integral may be reduced to the integral $J(\varepsilon )$. In fact,
when $l=0$, the integral $J_{0}(x_{cm},\varepsilon _{cm},\varepsilon
)=J(\varepsilon ).$ Thus, the upper bound of the integral $%
J_{0}(x_{cm},\varepsilon _{cm},\varepsilon )$ may be determined from (3.53).
For any other cases $l>0$ the situation is not so simple. The integral $%
J_{2l}(x_{cm},\varepsilon _{cm},\varepsilon )$ may be reduced by using the
binomial expansion: $x^{2l}=[(x-x_{cm})+x_{cm}]^{2l}=\sum_{k=0}^{2l}\left( 
\begin{array}{c}
2l \\ 
k%
\end{array}%
\right) (x_{cm})^{2l-k}(x-x_{cm})^{k}$. The reduced result may be written as%
\begin{equation}
J_{2l}(x_{cm},\varepsilon _{cm},\varepsilon )=\sum_{k=0}^{2l}\left( 
\begin{array}{c}
2l \\ 
k%
\end{array}%
\right) (x_{cm})^{2l-k}J_{k}^{0}(x_{cm},\varepsilon _{cm},\varepsilon ) 
\tag{3.55b}
\end{equation}%
where the integral $J_{k}^{0}(x_{cm},\varepsilon _{cm},\varepsilon )$ is
defined as 
\begin{equation}
J_{k}^{0}(x_{cm},\varepsilon _{cm},\varepsilon )=\int_{-\infty }^{\infty
}dx\{(x-x_{cm})^{k}\exp [-\frac{(x-x_{cm})^{2}}{\varepsilon _{cm}^{2}}%
]\int_{-\infty }^{x-x_{L}}\exp (-\frac{z^{2}}{\varepsilon ^{2}})dz\}. 
\tag{3.56}
\end{equation}%
Obviously, the integral $J_{0}^{0}(x_{cm},\varepsilon _{cm},\varepsilon
)=J(\varepsilon )$ and its upper bound may be determined from (3.53). By
calculating directly the integral $J_{1}^{0}(x_{cm},\varepsilon
_{cm},\varepsilon )$ one obtains%
\begin{equation*}
J_{1}^{0}(x_{cm},\varepsilon _{cm},\varepsilon )=\frac{\sqrt{\pi }}{2}\frac{%
\varepsilon ^{2}\varepsilon _{cm}^{4}}{(\varepsilon _{cm}^{2}+\varepsilon
^{2})}\exp \{-\frac{(x_{L}-x_{cm})^{2}}{(\varepsilon _{cm}^{2}+\varepsilon
^{2})}\}.
\end{equation*}%
Both the integrals $J_{0}^{0}(x_{cm},\varepsilon _{cm},\varepsilon )$ and $%
J_{1}^{0}(x_{cm},\varepsilon _{cm},\varepsilon )$ are basic. Now one may set
up the recursive relation for the integral $J_{k}^{0}(x_{cm},\varepsilon
_{cm},\varepsilon ).$ Then this recursive relation is further used to
calculate the integral $J_{2l}(x_{cm},\varepsilon _{cm},\varepsilon ).$ In
general, by using the integration by parts and the product formula (3.34) of
a pair of Gaussian functions it can turn out that the integral $%
J_{k}^{0}(x_{cm},\varepsilon _{cm},\varepsilon )$ $(k\geq 2)$ satisfies the
recursive relation: 
\begin{equation*}
J_{k}^{0}(x_{cm},\varepsilon _{cm},\varepsilon )=\frac{1}{2}(k-1)\varepsilon
_{cm}^{2}J_{k-2}^{0}(x_{cm},\varepsilon _{cm},\varepsilon )
\end{equation*}%
\begin{equation}
+\frac{1}{2}\varepsilon _{cm}^{2}\exp \{-\frac{(x_{L}-x_{cm})^{2}}{%
(\varepsilon _{cm}^{2}+\varepsilon ^{2})}\}I_{k-1}(\frac{\varepsilon
_{cm}^{2}(x_{L}-x_{cm})}{(\varepsilon _{cm}^{2}+\varepsilon ^{2})},\frac{%
\varepsilon \varepsilon _{cm}}{\sqrt{\varepsilon _{cm}^{2}+\varepsilon ^{2}}}%
)  \tag{3.57}
\end{equation}%
where the integral function $I_{k}(x_{0},\varepsilon _{0})$ is determined
from (3.38). The recursive relation (3.57) shows that an extra term, i.e.,
the second term on the $RH$ side of (3.57), is generated after the integral $%
J_{k}^{0}(x_{cm},\varepsilon _{cm},\varepsilon )$ is reduced to $%
J_{k-2}^{0}(x_{cm},\varepsilon _{cm},\varepsilon )$. It is known from
(3.38b) that the integral $I_{k}(x_{0},\varepsilon _{0})$ is a $k-$order
polynomial in $x_{0},$ and it is also a $(k+1)-$order polynomial in $%
\varepsilon _{0}$ if $k$ is even or a $k-$order polynomial in $\varepsilon
_{0}$ if $k$ is odd. Then this extra term is proportional to the
exponentially-decaying factor $\exp \{-\frac{(x_{L}-x_{cm})^{2}}{%
(\varepsilon _{cm}^{2}+\varepsilon ^{2})}\}.$ The recursive formula (3.57)
may be used repeatedly. This results in that any integral $%
J_{2j}^{0}(x_{cm},\varepsilon _{cm},\varepsilon )$ for $j=1,2,...,l$ may be
expressed as a linear combination of the integral $J_{0}^{0}(x_{cm},%
\varepsilon _{cm},\varepsilon )$ and the polynomials $\{I_{k}(x_{0}^{\prime
},\varepsilon _{0}^{\prime })\},$ and any integral $J_{2j+1}^{0}(x_{cm},%
\varepsilon _{cm},\varepsilon )$ for $j=1,2,...,l-1$ may be expressed as a
linear combination of the integral $J_{1}^{0}(x_{cm},\varepsilon
_{cm},\varepsilon )$ and the polynomials $\{I_{k}(x_{0}^{\prime
},\varepsilon _{0}^{\prime })\}.$ Here the parameter $x_{0}^{\prime
}=\varepsilon _{cm}^{2}(x_{L}-x_{cm})/(\varepsilon _{cm}^{2}+\varepsilon
^{2})$ and $\varepsilon _{0}^{\prime }=\varepsilon _{cm}\varepsilon /\sqrt{%
\varepsilon _{cm}^{2}+\varepsilon ^{2}}.$ Then by substituting these
expressions of the integrals $\{J_{k}^{0}(x_{cm},\varepsilon
_{cm},\varepsilon )\}$ into (3.55b) one obtains%
\begin{equation*}
J_{2l}(x_{cm},\varepsilon _{cm},\varepsilon )=J_{0}^{0}(x_{cm},\varepsilon
_{cm},\varepsilon )\sum_{j=0}^{l}\left( 
\begin{array}{c}
2l \\ 
2j%
\end{array}%
\right) (2j-1)!!(x_{cm})^{2l-2j}(\frac{1}{2}\varepsilon _{cm}^{2})^{j}
\end{equation*}%
\begin{equation*}
+J_{1}^{0}(x_{cm},\varepsilon _{cm},\varepsilon )\sum_{k=0}^{l-1}\left( 
\begin{array}{c}
2l \\ 
2k+1%
\end{array}%
\right) (2k)!!(x_{cm})^{2l-2k-1}(\frac{1}{2}\varepsilon _{cm}^{2})^{k}
\end{equation*}%
\begin{equation}
+\exp \{-\frac{(x_{L}-x_{cm})^{2}}{(\varepsilon _{cm}^{2}+\varepsilon ^{2})}%
\}F_{l}(x_{0}^{\prime },\varepsilon _{0}^{\prime })  \tag{3.58}
\end{equation}%
where the function $F_{l}(x_{0}^{\prime },\varepsilon _{0}^{\prime })$ is
given by%
\begin{equation*}
F_{l}(x_{0}^{\prime },\varepsilon _{0}^{\prime
})=\sum_{j=1}^{l}\sum_{m=0}^{j-1}\left( 
\begin{array}{c}
2l \\ 
2j%
\end{array}%
\right) \frac{(2j-1)!!}{(2m+1)!!}(x_{cm})^{2l-2j}(\frac{1}{2}\varepsilon
_{cm}^{2})^{j-m}I_{2m+1}(x_{0}^{\prime },\varepsilon _{0}^{\prime })
\end{equation*}%
\begin{equation}
+\sum_{k=1}^{l-1}\sum_{m=1}^{k}\left( 
\begin{array}{c}
2l \\ 
2k+1%
\end{array}%
\right) \frac{(2k)!!}{(2m)!!}(x_{cm})^{2l-2k-1}(\frac{1}{2}\varepsilon
_{cm}^{2})^{k-m+1}I_{2m}(x_{0}^{\prime },\varepsilon _{0}^{\prime }). 
\tag{3.59}
\end{equation}%
The integral $J_{2l}(x_{cm},\varepsilon _{cm},\varepsilon )$ consists of the
three terms. The first term is proportional to $J_{0}^{0}(x_{cm},\varepsilon
_{cm},\varepsilon )$ and a $2l-$order polynomial in the COM position $%
x_{cm}. $ It is known from (3.52) and (3.53) that the integral $%
J_{0}^{0}(x_{cm},\varepsilon _{cm},\varepsilon ),$ i.e., the integral $%
J(\varepsilon )$ in (3.52), is proportional to the exponentially-decaying
factor $\exp \{-\frac{(x_{L}-x_{cm})^{2}}{(\varepsilon _{cm}^{2}+\varepsilon
^{2})}\}.$ Then the first term in (3.58) is proportional to the
exponentially-decaying factor $\exp \{-\frac{(x_{L}-x_{cm})^{2}}{%
(\varepsilon _{cm}^{2}+\varepsilon ^{2})}\}.$ The second term in (3.58) is
proportional to the integral $J_{1}^{0}(x_{cm},\varepsilon _{cm},\varepsilon
)$ and a $(2l-1)-$order polynomial in $x_{cm}.$ Here $J_{1}^{0}(x_{cm},%
\varepsilon _{cm},\varepsilon )$ is proportional to the
exponentially-decaying factor $\exp \{-\frac{(x_{L}-x_{cm})^{2}}{%
(\varepsilon _{cm}^{2}+\varepsilon ^{2})}\}.$ Hence the second term is
proportional to the exponentially-decaying factor $\exp \{-\frac{%
(x_{L}-x_{cm})^{2}}{(\varepsilon _{cm}^{2}+\varepsilon ^{2})}\}.$ The last
term in (3.58) is proportional to the exponentially-decaying factor $\exp \{-%
\frac{(x_{L}-x_{cm})^{2}}{(\varepsilon _{cm}^{2}+\varepsilon ^{2})}\} $ and
the polynomial $F_{l}(x_{0}^{\prime },\varepsilon _{0}^{\prime })$ in $%
x_{cm} $ which is given by (3.59). Since $I_{2m+1}(x_{0}^{\prime
},\varepsilon _{0}^{\prime })$ is a $(2m+1)-$order polynomial in $%
x_{0}^{\prime },$ the first term of $F_{l}(x_{0}^{\prime },\varepsilon
_{0}^{\prime })$ in (3.59) is a $(2l-1)-$order polynomial in $x_{cm}$ or $%
x_{L}.$ Similarly, the second term of $F_{l}(x_{0}^{\prime },\varepsilon
_{0}^{\prime })$ in (3.59) is also a $(2l-1)-$order polynomial in $x_{cm}$
or a $(2l-2)-$order polynomial in $x_{L}.$ Thus, the polynomial $%
F_{l}(x_{0}^{\prime },\varepsilon _{0}^{\prime })$ in $x_{cm}$ or $x_{L}$
has an order of $2l-1$. The above analysis shows that the integral $%
J_{2l}(x_{cm},\varepsilon _{cm},\varepsilon )$ is proportional to the
exponentially-decaying factor $\exp \{-\frac{(x_{L}-x_{cm})^{2}}{%
(\varepsilon _{cm}^{2}+\varepsilon ^{2})}\}.$

Now let $x_{cm}$ and $\varepsilon _{cm}$ in the integral $%
J_{2l}(x_{cm},\varepsilon _{cm},\varepsilon )$ of (3.55a) be the COM
position $x_{c}(t_{0}+\lambda )$ and wave-packet spread $\varepsilon
_{c}(t_{0}+\lambda )$ of the $GWP$ state $\Psi _{0}(x,r,$ $t_{0}+\lambda ),$
respectively. Then the integral $J_{2l}(x_{cm},\varepsilon _{cm},\varepsilon
)$ is just the integral $J_{2l}(x_{c}(t_{0}+\lambda ),\varepsilon
_{c}(t_{0}+\lambda ),\varepsilon )$ on the $RH$ side of (3.54). As shown in
(3.54), the upper bound of the second basic norm $NBAS2$ is proportional to
square root of the integral $J_{2l}(x_{c}(t_{0}+\lambda ),\varepsilon
_{c}(t_{0}+\lambda ),\varepsilon ).$ Now the integral $J_{2l}(x_{cm},%
\varepsilon _{cm},\varepsilon )$ is proportional to the
exponentially-decaying factor $\exp \{-\frac{(x_{L}-x_{cm})^{2}}{%
(\varepsilon _{cm}^{2}+\varepsilon ^{2})}\}.$ By inserting the parameters $%
x_{cm}=x_{c}(t_{0}+\lambda )$ and $\varepsilon _{cm}=\varepsilon
_{c}(t_{0}+\lambda )$ into the exponentially-decaying factor one can find
that the integral $J_{2l}(x_{c}(t_{0}+\lambda ),\varepsilon
_{c}(t_{0}+\lambda ),\varepsilon )$ is proportional to the
exponentially-decaying factor $\exp \{-Y_{eff}(x_{L},\varepsilon
_{eff})^{2}/2\},$ where the effective deviation-to-spread ratio $%
Y_{eff}(x_{L},\varepsilon _{eff})$ is defined by (3.39) with the effective
wave-packet spread $\varepsilon _{eff}$ given by $\varepsilon
_{eff}^{2}=\varepsilon ^{2}+\varepsilon _{c}(t_{0}+\lambda )^{2}$. Then it
follows from (3.54) that the second basic norm $NBAS2$ has an upper bound
that is proportional to the exponentially-decaying factor $\exp
\{-Y_{eff}(x_{L},\varepsilon _{eff})^{2}/2\}:$ 
\begin{equation}
||x^{l}\Theta (x-x_{L},\varepsilon )\Psi _{0}(x,r,t_{0}+\lambda
)||_{u}\varpropto \exp \{-\frac{1}{2}\frac{[x_{L}-x_{c}(t_{0}+\lambda )]^{2}%
}{\varepsilon _{eff}^{2}}\},  \tag{3.60}
\end{equation}%
here the subscript $^{\prime }u^{\prime }$ stands for the upper bound of the
basic norm $NBAS2$. This means that when $\varepsilon _{c}(t_{0}+\lambda
)^{2}>>\varepsilon ^{2},$ the upper bound of the basic norm $NBAS2$ decays
exponentially with the square deviation-to-spread ratio $y_{M}(x_{L},$ $%
t_{0}+\lambda )^{2}$ of the $GWP$ state $\Psi _{0}(x,r,t_{0}+\lambda ).$
This is the desired result. As shown above, a similar result also may be
obtained if one uses the first-order approximation formula (2.9) with the
incontinuous step function $\Theta (x-x_{L})$ to calculate the basic norm $%
NBAS2$. However, that result is approximate. The current result is rigorous.
It confirms further and also corrects the first-order approximation result.

Once it proves that both the basic norms $NBAS1$ and $NBAS2$ decay
exponentially with the square deviation-to-spread ratio or more exactly with
the square effective deviation-to-spread ratio of the $GWP$ state $\Psi
_{0}(x,r,t_{0}+\lambda ),$ it is easy to prove that the upper bounds of the
errors $E_{r}^{(1)}(x,r,$ $t_{0}+t)$ of (3.27a) and $%
E_{r}^{(2)}(x,r,t_{0}+t) $ of (3.27b) decay exponentially with the square
deviation-to-spread ratio of the $GWP$ state $\Psi _{0}(x,r,t_{0}+\lambda ).$
Actually, the inequalities (3.28) and (3.29) show that the upper bound of
the error $E_{r}^{(1)}(x,r,t_{0}+t)$ is proportional to the maximum value of
the integral $J_{4}(x_{cm},\varepsilon _{cm},\varepsilon )$ in the time
region $[t_{0},$ $t_{0}+\tau ]$, here $x_{cm}$ and $\varepsilon _{cm}$ are
the COM position $x_{c}(t_{0}+\tau )$ and wave-packet spread $\varepsilon
_{c}(t_{0}+\tau )$ of the $GWP$ state $\Psi _{0}(x,r,t_{0}+\tau ),$
respectively. Thus, the error $E_{r}^{(1)}(x,r,t_{0}+t)$ decays
exponentially with the square deviation-to-spread ratio of the $GWP$ state $%
\Psi _{0}(x,r,t_{0}+\tau ).$ It is known from (3.30) and (3.31) that the
upper bound of the error $E_{r}^{(2)}(x,r,t_{0}+t)$ may be determined from
the three norms on the $RH$ side of (3.31). The upper bound of the last norm
on the $RH$ side of (3.31) is determined from (3.33a), while the upper
bounds of the first two norms are determined from (3.33b) and (3.33c),
respectively. The $RH$ side of (3.33a) consists of nine basic norms $%
\{NBSA1\}$ and $\{NBSA2\}.$ Similarly, those of (3.33b) and (3.33c) have
eight and ten basic norms, respectively. Therefore, the upper bound of the
integrand on the $RH$ side of (3.30) decays exponentially with the square
deviation-to-spread ratio of the $GWP$ state $\Psi _{0}(x,r,t_{0}+\lambda )$
or%
\begin{equation*}
||[H_{0}^{ho},V_{1}^{ho}(x,\varepsilon )]\exp (-iV_{1}^{ho}(x,\varepsilon
)\lambda ^{\prime }/\hslash )\exp (-iH_{0}^{ho}\lambda /\hslash )\Psi
_{00}(x,r,t_{0})||_{u}
\end{equation*}%
\begin{equation}
\varpropto \exp \{-\frac{1}{2}[\frac{[x_{L}-x_{c}(t_{0}+\lambda )]^{2}}{%
\varepsilon _{c}(t_{0}+\lambda )^{2}}]_{\min }\},  \tag{3.61}
\end{equation}%
here suppose that $\varepsilon _{c}(t_{0}+\lambda )^{2}>>\varepsilon ^{2}$
so that the effective wave-packet spread $\varepsilon _{eff}$ can be
approximately replaced with $\varepsilon _{c}(t_{0}+\lambda ).$ This means
that the error $E_{r}^{(2)}(x,r,t_{0}+t)$ decays exponentially with the
square deviation-to-spread ratio of the $GWP$ state $\Psi
_{0}(x,r,t_{0}+\lambda ).$ Now both the errors $E_{r}^{(1)}(x,r,t_{0}+t)$
and $E_{r}^{(2)}(x,r,t_{0}+t)$ in (3.25) are shown rigorously to have the
upper bounds that decay exponentially with the square deviation-to-spread
ratio of the $GWP$ state $\Psi _{0}(x,r,t_{0}+\lambda ).$ Therefore, when
the joint position $x_{L}$ is taken as a large enough value, both the errors 
$E_{r}^{(1)}(x,r,t_{0}+t)$ and $E_{r}^{(2)}(x,r,t_{0}+t)$ are negligible,
leading directly to that $\Psi (x,r,t_{0}+t)=\Psi _{0}(x,r,t_{0}+t)$ in
(3.25). This shows that the spatially-selective effect of the perturbation
term $V_{1}^{ho}(x,\varepsilon )$ has a negligible contribution to the time
evolution process of the halting-qubit atom in the $LH$ potential well (or
the imperfection of the $LH$ harmonic potential field has a negligible
effect on the time evolution process) when the joint position $x_{L}$ is
large enough. This is the desired result in the section.

The above rigorous error estimation considers the initial state $\Psi
_{00}(x,r,t_{0})$ to be a single $GWP$ state. As shown in the section 5
below, a Gaussian superposition state is often used as the initial state $%
\Psi _{00}(x,r,t_{0})$ in an error estimation. This general case is also
easy to consider in the above error estimation. Suppose that the initial
state $\Psi _{00}(x,r,t_{0})$ is a Gaussian superposition state: $\Psi
_{00}(x,r,t_{0})=\sum_{k=1}^{m}A_{k}\Psi _{0k}(x,r,t_{0}).$ Here the
amplitude $A_{k}$ may be a complex coefficient. By inserting this initial
product state into (3.25) one finds that the desired product state is
written as%
\begin{equation}
\Psi _{0}(x,r,t_{0}+t)=\sum_{k=1}^{m}A_{k}\exp (-iH_{0}t/\hslash )\Psi
_{0k}(x,r,t_{0}),  \tag{3.62}
\end{equation}%
while the two error terms are given exactly by%
\begin{equation}
E_{r}^{(l)}(x,r,t_{0}+t)=\sum_{k=1}^{m}A_{k}E_{rk}^{(l)}(x,r,t_{0}+t),\text{ 
}l=1,2,  \tag{3.63}
\end{equation}%
and their upper bounds may be determined from 
\begin{equation}
||E_{r}^{(l)}(x,r,t_{0}+t)||\leq \sum_{k=1}^{m}|A_{k}|\times
||E_{rk}^{(l)}(x,r,t_{0}+t)||,\text{ }l=1,2,  \tag{3.64}
\end{equation}%
where the error $E_{rk}^{(1)}(x,r,t_{0}+t)$ is still given by (3.27a) with
the product state $\Psi _{00}(x,r,t_{0})=\Psi _{0k}(x,r,t_{0}),$ while the
error $E_{rk}^{(2)}(x,r,t_{0}+t)$ may be expressed as (3.27b) with the
product state $\Psi _{00}(x,r,t_{0})=\Psi _{0k}(x,r,t_{0}).$ Therefore, the
upper bound of each error $E_{rk}^{(l)}(x,r,t_{0}+t)$ for $l=1$, $2$ and $%
k=1 $, $2$, ..., $m$ can be strictly calculated according to the above
error-estimation method. Then according to (3.64) all these upper bounds are
summed up with the weights $\{|A_{k}|\}$ to generate the error upper bounds $%
\{||E_{r}^{(l)}(x,r,t_{0}+t)||_{u}\}$. Thus, the desired result is still
obtained that the total error upper bound $%
||E_{r}^{(1)}(x,r,t_{0}+t)||_{u}+||E_{r}^{(2)}(x,r,t_{0}+t)||_{u}$ in (3.25)
decays exponentially with the square deviation-to-spread ratios of the $GWP$
states $\{\Psi _{0k}(x,r,t_{0}+\lambda )\}$, here the state $\Psi
_{0k}(x,r,t_{0}+\lambda )=\exp (-iH_{0}^{ho}\lambda /\hslash )\Psi
_{0k}(x,r,t_{0}).$

There are some applications of the rigorous theoretical calculation method
for the time evolution process (3.25). As an example, the theoretical
calculation method may be used to investigate strictly the
spatially-selective effect of the laser light pulse, when the
spatially-selective laser light pulse is used to prepare the one-qubit
quantum gates of the halting-qubit atom in the $LH$ harmonic potential well.
The spatially-selective effect needs to be evaluated strictly for the
preparation of the one-qubit quantum gates of the unitary operations $%
U_{h}^{c}$ and $V_{h}^{c}$ of the reversible and unitary halting protocol
[14]. Suppose that the halting-qubit atom motions in the $LH$ potential well
only along one-dimensional direction $x$, while the spatially-selective
laser light pulse is externally applied to the two internal states $%
\{|g_{0}\rangle ,|e\rangle \}$ of the halting-qubit atom along the direction 
$y$. This means that the motion of the halting-qubit atom is strongly
constrained along the direction $y$ such that it is not affected
significantly by the laser light pulse. It could not be difficult to achieve
this point in experiment [3]. Below it is shown that the laser light pulse
could not yet affect significantly the atomic COM motion along the direction 
$x$. The spatially-selective excitation region for the laser light pulse
along the direction $x$ is within the $LH$ potential well $(-\infty ,x_{L})$%
. In general, the excitation spatial region of the laser light pulse is far
more narrow than the spatial region $(-\infty ,x_{L}),$ but in practice it
must be chosen in such a way that the laser light beam can cover
sufficiently the effective spatial region of the wave-packet motional state
of the halting-qubit atom in motion. As a typical instance, here still
assume that the excitation spatial region of the laser light pulse is $%
(-\infty ,x_{L}).$ For a practical case the theoretical treatment is also
easy and the obtained result also is similar to that one obtained in this
typical instance if the halting-qubit atom is in a $GWP$ motional state. It
is clear that the laser light pulse is spatially-selective in the direction $%
x$. Here the laser light beam may be circularly-polarized. Then in the
rotating frame the Hamiltonian to describe the spatially-selective
excitation induced by the laser light pulse in the two-level system $%
\{|g_{0}\rangle ,|e\rangle \}$ of the halting-qubit atom may be written as%
\begin{equation}
H=H_{0}^{ho}+V_{1}^{ho}(x)+\hslash (\omega _{a}-\omega _{0})I_{z}+\hslash
\Omega (x)I_{x}.  \tag{3.65}
\end{equation}%
Here $H_{0}^{ho}$ and $V_{1}^{ho}(x)$ are still given by (2.3) and (2.4),
respectively. The spin operators $I_{\mu }$ $(\mu =x,y,z)$ are defined in
the previous section 2. This Hamiltonian is time-independent. Suppose
further that in the spatially-selective excitation the amplitude $\Omega (x)$
of the spatially-selective laser light pulse is given simply by $\Omega
(x)=\omega _{1}$ if $-\infty <x\leq x_{L}$ and $\Omega (x)=0$ if $x>x_{L}$.
A more practical spatially-selective excitation could be defined by $\Omega
(x)=\omega _{1}$ if $-x_{L}\leq x\leq x_{L}$ and $\Omega (x)=0$ if $x>x_{L}$
and $x<-x_{L}$. The Hamiltonian may be simply rewritten as $%
H=H_{0}+H_{1}(x), $ where the main Hamiltonian $H_{0}=H_{0}^{ho}+\hslash
(\omega _{a}-\omega _{0})I_{z}+\hslash \omega _{1}I_{x}$ and the
spatially-selective perturbation term $H_{1}(x)=V_{1}^{ho}(x)+\hslash
\lbrack \Omega (x)-\omega _{1}]I_{x}.$ Obviously, the main Hamiltonian $%
H_{0} $ is not spatially selective, while the perturbation term $H_{1}(x)$
is. The main Hamiltonian $H_{0}$ is responsible for the preparation of the
one-qubit quantum gates of the halting-qubit atom. The on-resonance
condition $\omega _{0}=\omega _{a}$ leads to that the main Hamiltonian is
written as $H_{0}=H_{0}^{ho}+\hslash \omega _{1}I_{x}.$ Note that $%
[H_{0}^{ho},$ $I_{x}]=0$. Thus, in on-resonance condition the internal-state
excitation of the halting-qubit atom by the spatially-selective laser light
pulse is independent on the atomic COM motion. On the other hand, the
perturbation term $H_{1}(x)$ could affect the preparation. It originates
from the imperfections of both the $LH$ harmonic potential field and the
spatially-selective laser light pulse. Both $H_{0}$ and $H_{1}(x)$ are
time-independent. Now the error estimation method above can be used as well
to evaluate strictly the effect of the perturbation term $H_{1}(x)$ on the
preparation. Here one needs to consider explicitly the atomic internal
states. This point is slightly different from the above error estimation.
Then by using the above rigorous error estimation method one can prove that
the error generated by the imperfections decays exponentially with the
square deviation-to-spread ratios of the relevant $GWP$ states of the
halting-qubit atom. These show that the laser light pulse does not have a
significant effect on the atomic COM motion in the direction $x$.

Another important application for the rigor theoretical calculation method
above is to use the method to investigate the spatially-selective excitation
of a single $GWP$ state in a Gaussian superposition state of an atom freely
moving along one-dimensional direction $x$. Such a spatially-selective
excitation for a freely-moving atom may be achieved by the STIRAP-based
decelerating (or accelerating) process [16]. Here a simpler
spatially-selective excitation method is proposed for it. Consider the
simple case that the Gaussian superposition state consists of two $GWP$
states, 
\begin{equation}
\Psi (x,r,t_{0})=a_{1}(t_{0})\Psi _{01}(x,t_{0})|g_{0}\rangle
+a_{2}(t_{0})\Psi _{02}(x,t_{0})|g_{0}\rangle ,  \tag{3.66}
\end{equation}%
where both the $GWP$ motional states $\Psi _{01}(x,t_{0})$ and $\Psi
_{02}(x,t_{0})$ have different COM positions in the one-dimensional
coordinate space. Here suppose that the two $GWP$ motional states are well
distinguished in space so that a spatially-selective excitation for one of
them can be carried out. The characteristic parameters of the $GWP$ state $%
\Psi _{0l}(x,t_{0})$ are denoted as $\{x_{c}^{l}(t_{0}),$ $p_{c}^{l}(t_{0}),$
$W_{c}^{l}(t_{0}),$ $\varepsilon _{c}^{l}(t_{0})\}$ for $l=1$ and $2$.
Assume that the $GWP$ state $\Psi _{0l}(x,t_{0})$ with $l=1$ or $2$ locates
in the effective spatial region $[x_{c}^{l}(t_{0})-D_{l}^{L},$ $%
x_{c}^{l}(t_{0})+D_{l}^{R}]$ with the effective spatial width $%
D_{l}^{L}+D_{l}^{R}.$ Both the effective spatial regions of the motional
states $\{\Psi _{0l}(x,t_{0})\}$ do not overlap with each other in the
one-dimensional space. Now a laser light beam with the propagating direction 
$y$ is applied to the halting-qubit atom which is moving freely along the
direction $x$. Just like before, here also consider the one-dimensional case
that the atomic motion along the direction $y$ is strongly constrained such
that it is not affected significantly by the laser light pulse. This laser
light beam is space-selective in the direction $x$ and also internal-state
selective. That is, it is selectively applied to the atomic two-level system 
$\{|g_{0}\rangle ,|e\rangle \}$ within the effective spatial region $%
[x_{c}^{1}(t_{0})-D_{1}^{L},$ $x_{c}^{1}(t_{0})+D_{1}^{R}].$ A similar
spatially-selective excitation has been used extensively in quantum coherent
interference experiments in cold atomic systems (See, for example, Refs.
[1]). Bear in mind that the present quantum system under study is a
pure-state quantum system of an individual atom instead of a quantum
ensemble and the present spatially-selective excitation is unitary.
Therefore, different from the conventional one in a cold atomic quantum
ensemble, the present spatially-selective excitation is nontrivial. In such
a spatially-selective excitation only the internal state of the first $GWP$
product state $\Psi _{01}(x,t_{0})|g_{0}\rangle $ is effectively excited in
the superposition state $\Psi (x,r,t_{0})$, while that one of the second
product state $\Psi _{02}(x,t_{0})|g_{0}\rangle $ keeps almost unchanged.
This spatially-selective excitation may be expressed as%
\begin{equation}
\Psi (x,r,t_{0})\rightarrow a_{1}(t_{f})\Psi _{01}(x,t_{f})|e\rangle
+a_{2}(t_{f})\Psi _{02}(x,t_{f})|g_{0}\rangle .  \tag{3.67}
\end{equation}%
Now the Hamiltonian of the atom in the spatially-selective excitation
process may be written as%
\begin{equation}
H=\frac{1}{2m}p^{2}+\hslash (\omega _{a}-\omega _{0})I_{z}+\hslash \Omega
(x)I_{x}.  \tag{3.68}
\end{equation}%
Here the intensity of the spatially-selective laser light beam is simply
given by%
\begin{equation}
\Omega (x)=\left\{ 
\begin{array}{c}
\omega _{1}\text{ if }x\in \lbrack x_{c}^{1}(t_{0})-D_{1}^{L},\text{ }%
x_{c}^{1}(t_{0})+D_{1}^{R}] \\ 
0\text{ if }x\notin \lbrack x_{c}^{1}(t_{0})-D_{1}^{L},\text{ }%
x_{c}^{1}(t_{0})+D_{1}^{R}]%
\end{array}%
\right.  \tag{3.69}
\end{equation}%
This Hamiltonian is time-independent. Then the time evolution process of the
halting-qubit atom during the spatially-selective excitation is expressed as%
\begin{equation*}
\Psi (x,r,t_{f})=\exp [-iH(t_{f}-t_{0})]\Psi (x,r,t_{0})
\end{equation*}%
\begin{equation}
=\sum_{l=1,2}a_{l}(t_{0})\exp [-iH(t_{f}-t_{0})]\Psi
_{0l}(x,t_{0})|g_{0}\rangle .  \tag{3.70}
\end{equation}%
This time evolution process can be calculated by separately calculating the
time evolution process: 
\begin{equation}
\Psi _{l}(x,r,t_{f})=\exp [-iH(t_{f}-t_{0})]\Psi _{0l}(x,t_{0})|g_{0}\rangle
.  \tag{3.71}
\end{equation}%
Now the Hamiltonian $H$ still may be written as $H=H_{0}+H_{1}(x),$ here the
main Hamiltonian $H_{0}$ is non-space-selective, while the perturbation term 
$H_{1}(x)$ is spatially selective. According to (3.25) the time evolution
process (3.71) may be written as%
\begin{equation}
\Psi _{l}(x,r,t_{f})=\Psi
_{0l}(x,r,t_{f})+E_{rl}^{(1)}(x,r,t_{f})+E_{rl}^{(2)}(x,r,t_{f}),  \tag{3.72}
\end{equation}%
where $\Psi _{0l}(x,r,t_{f})$ is the desired product state, which is
determined from (3.26), and $E_{rl}^{(1)}(x,r,t_{f})$ and $%
E_{rl}^{(2)}(x,r,t_{f})$ are the two errors, which are given by the two
equation (3.27), respectively. Firstly consider the first $GWP$ state $\Psi
_{01}(x,r,t_{f}).$ Notice that here the laser light beam covers the entire
effective spatial region of the first $GWP$ state $\Psi _{01}(x,t)$ over the
whole time region $t_{0}\leq t\leq t_{f}.$ Then in this case the main
Hamiltonian $H_{0}$ may be taken as $H_{0}=\frac{1}{2m}p^{2}+\hslash (\omega
_{a}-\omega _{0})I_{z}+\hslash \omega _{1}I_{x}$ and accordingly the
spatially-selective perturbation term is given by $H_{1}(x)=\hslash \lbrack
\Omega (x)-\omega _{1}]I_{x}.$ For convenience, the on-resonance condition $%
\omega _{0}=\omega _{a}$ is used in calculation below. Then the main
Hamiltonian is reduced to $H_{0}=\frac{1}{2m}p^{2}+\hslash \omega _{1}I_{x}.$
Now according to (3.26) the desired product state at the end of the
spatially-selective excitation is given by%
\begin{equation}
\Psi _{01}(x,r,t_{f})=\Psi _{01}(x,t_{f})|e\rangle ,  \tag{3.73}
\end{equation}%
where the time interval $(t_{f}-t_{0})=\pi /\omega _{1}$ and the final $GWP$
state is given by%
\begin{equation}
\Psi _{01}(x,t_{f})=-i\exp [-i\frac{1}{2m}p^{2}(t_{f}-t_{0})/\hslash ]\Psi
_{01}(x,t_{0}).  \tag{3.74}
\end{equation}%
The equation (3.74) is really the motional process of a free particle with
the initial state $-i\Psi _{01}(x,t_{0}).$ The final state $\Psi
_{01}(x,r,t_{f})$ of (3.73) shows that the initial internal state $%
|g_{0}\rangle $ of the first $GWP$ product state $\Psi _{01}(x,r,t_{0})$
indeed is excited to another internal state $|e\rangle $ by the laser light
pulse. The two error terms $E_{r1}^{(k)}(x,r,t_{f})$ for $k=1$ and $2$ in
(3.72), which are generated by the spatially-selective perturbation term $%
H_{1}(x),$ may be directly calculated by using the two equations (3.27) with
the perturbation term $V_{1}=H_{1}(x)$, respectively. The error term $%
E_{r1}^{(1)}(x,r,t_{f})$ also may be directly calculated by using the
inequality (3.28) with the time interval $\tau =t_{f}-t_{0}$ and the
replacement $V_{1}^{ho}(x,\varepsilon )\leftrightarrow H_{1}(x)$ and $%
H_{0}^{ho}\leftrightarrow H_{0}.$ Here the initial state $\Psi
_{00}(x,r,t_{0})$ is taken as $\Psi _{01}(x,t_{0})|g_{0}\rangle $ and the
spatially-selective intensity $\Omega (x)$ of the laser light beam is given
by (3.69). The calculation based on the inequality (3.28) shows that the
error $E_{r1}^{(1)}(x,r,t_{f})$ decays exponentially with the square
deviation-to-spread ratios of the $GWP$ state $\Psi _{01}^{0}(x,t)$. Here
the state $\Psi _{01}^{0}(x,t)=\exp [-i\frac{1}{2m}p^{2}(t-t_{0})/\hslash
]\Psi _{01}(x,t_{0})$ for $t_{0}\leq t\leq t_{f}.$ There are two different
deviation-to-spread ratios for the $GWP$ state $\Psi _{01}^{0}(x,t).$ They
are defined by $Y_{R}^{1}(t)=[(D_{1}^{R}+x_{c}^{1}(t_{0}))-x_{c}^{1}(t)]/%
\varepsilon _{c}^{1}(t)>0$ and $%
Y_{L}^{1}(t)=[x_{c}^{1}(t)-(x_{c}^{1}(t_{0})-D_{1}^{L})]/\varepsilon
_{c}^{1}(t)>0.$ Therefore, the dimensional-size parameters $D_{1}^{L}$ and $%
D_{1}^{R}$ for the effective spatial region $[x_{c}^{1}(t_{0})-D_{1}^{L},$ $%
x_{c}^{1}(t_{0})+D_{1}^{R}]$ must be chosen suitably such that both the
deviation-to-spread ratios satisfy $Y_{R}^{1}(t)>>1$ and $Y_{L}^{1}(t)>>1$
for $t_{0}\leq t\leq t_{f}.$ When $Y_{R}^{1}(t)>>1$ and $Y_{L}^{1}(t)>>1,$
the error $E_{r1}^{(1)}(x,r,t_{f})$ can be neglected. In order to calculate
strictly the error $E_{r1}^{(2)}(x,r,t_{f})$ one needs to first make it
smooth for the incontinuous intensity $\Omega (x)$ of the
spatially-selective laser light beam. The spatially-selective intensity of
(3.69) is first rewritten as 
\begin{equation*}
\Omega (x)=\omega _{1}\Theta (x-[x_{c}^{1}(t_{0})-D_{1}^{L}])\Theta
([x_{c}^{1}(t_{0})+D_{1}^{R}]-x).
\end{equation*}%
Then it is made smooth by using the continuous step function $\Theta
(x,\varepsilon )$ in (2.11), 
\begin{equation*}
\Omega (x,\varepsilon )=\omega _{1}\Theta
(x-[x_{c}^{1}(t_{0})-D_{1}^{L}],\varepsilon )\Theta
([x_{c}^{1}(t_{0})+D_{1}^{R}]-x,\varepsilon ).
\end{equation*}%
Now by using the smooth perturbation term $H_{1}(x,\varepsilon )=\hslash
\lbrack \Omega (x,\varepsilon )-\omega _{1}]I_{x}$ and the main Hamiltonian $%
H_{0}=\frac{1}{2m}p^{2}+\hslash \omega _{1}I_{x}$ one may exactly calculate
the commutator $[H_{0},$ $H_{1}(x,\varepsilon )].$ Then the commutator is
substituted into (3.27b) to calculate the error $E_{r1}^{(2)}(x,r,t_{f}).$
It can turn out that the upper bound of the error $E_{r1}^{(2)}(x,r,t_{f})$
consists of a few basic norms $\{NBAS1\}.$ This indicates that the error $%
E_{r1}^{(2)}(x,r,t_{f})$ decays exponentially with the square
deviation-to-spread ratios of the $GWP$ state $\Psi _{01}^{0}(x,t).$ On the
other hand, the second $GWP$ state $\Psi _{02}(x,t)$ is distant from the
laser light beam. It is not affected significantly by the laser light beam.
Then in this case the main Hamiltonian may be taken as $H_{0}=\frac{1}{2m}%
p^{2}+\hslash (\omega _{a}-\omega _{0})I_{z},$ while the spatially-selective
perturbation term is given by $H_{1}(x)=\hslash \Omega (x)I_{x}.$ In the
on-resonance condition the main Hamiltonian is given by $H_{0}=\frac{1}{2m}%
p^{2}.$ This main Hamiltonian does not affect any internal state of the
halting-qubit atom. Now the desired product state at the end of the
spatially-selective excitation is still given by (3.26). It is written as%
\begin{equation}
\Psi _{02}(x,r,t_{f})=\Psi _{02}(x,t_{f})|g_{0}\rangle ,  \tag{3.75}
\end{equation}%
where the final $GWP$ state is given by%
\begin{equation}
\Psi _{02}(x,t_{f})=\exp [-i\frac{1}{2m}p^{2}(t_{f}-t_{0})/\hslash ]\Psi
_{02}(x,t_{0}).  \tag{3.76}
\end{equation}%
The equation (3.76) is really the motional process of a free particle with
the initial state $\Psi _{02}(x,t_{0}).$ The final state $\Psi
_{02}(x,r,t_{f})$ of (3.75) shows that the initial internal state $%
|g_{0}\rangle $ of the second $GWP$ product state $\Psi _{02}(x,r,t_{0})$
indeed is not excited by the laser light pulse. Now the two error terms $%
E_{r2}^{(k)}(x,r,t_{f})$ for $k=1$ and $2$ in (3.72) in the
spatially-selective excitation may be exactly calculated by the two
equations (3.27), respectively. Here the error $E_{r2}^{(1)}(x,r,t_{f})$
also may be calculated directly by using the inequality (3.28) with the time
interval $\tau =t_{f}-t_{0}$ and the replacement $V_{1}^{ho}(x,\varepsilon
)\leftrightarrow H_{1}(x)$ and $H_{0}^{ho}\leftrightarrow H_{0}.$ There are
two cases to be considered. The first case is that the $GWP$ state $\Psi
_{02}(x,t)$ always locates on the left side of the effective spatial region $%
[x_{c}^{1}(t_{0})-D_{1}^{L},$ $x_{c}^{1}(t_{0})+D_{1}^{R}].$ The second one
is that the state $\Psi _{02}(x,t)$ always locates on the right side of the
effective spatial region. Either case generates a similar result. Thus, here
only the first case is considered. Now on the basis of the inequality (3.28)
one can prove that the error $E_{r2}^{(1)}(x,r,t_{f})$ decays exponentially
with the square deviation-to-spread ratio of the $GWP$ motional state $\Psi
_{02}(x,t)$ $(t_{0}\leq t\leq t_{f})$ which is given by (3.76) with the
replacement $t_{f}\leftrightarrow t$. Here the deviation-to-spread ratio is
defined by $Y_{L}^{2}(t)=[(x_{c}^{1}(t_{0})-D_{1}^{L})-x_{c}^{2}(t)]/%
\varepsilon _{c}^{2}(t)>0.$ Therefore, when $Y_{L}^{2}(t)>>1,$ the error $%
E_{r2}^{(1)}(x,r,t_{f})$ can be neglected. Another error $%
E_{r2}^{(2)}(x,r,t_{f})$ is calculated by using (3.27b). Here the commutator 
$[H_{0},$ $H_{1}(x,\varepsilon )]$ is calculated by using the main
Hamiltonian $H_{0}=\frac{1}{2m}p^{2}$ and the smooth perturbation term $%
H_{1}(x,\varepsilon )=\hslash \Omega (x,\varepsilon )I_{x}.$ By a strict
calculation it can turn out that the upper bound of the error $%
E_{r2}^{(2)}(x,r,t_{f})$ consists of a few basic norms $\{NBAS1\}$,
indicating that the error decays exponentially with the square
deviation-to-spread ratios of the $GWP$ motional state $\Psi _{02}(x,t).$%
\newline
\newline
\newline
{\Large 4 Imperfections for the spatially\ selective\ laser light }

{\Large beams}

When an external electromagnetic wave field is space-selectively applied to
the halting-qubit atom within the $LH$ potential well, the error estimation
method in the preceding section 3 needs to be modified, because one also
needs to consider the possible errors generated by the spatially selective
electromagnetic wave field. In theory a state-selective triggering pulse may
be generated by the ideal plane-wave electromagnetic fields of the $PHAMDOWN$
laser light beams over the whole coordinate space $(-\infty ,+\infty )$, as
shown in Ref. [15]. However, in practice the $PHAMDOWN$ laser light beams
used to generate a $SSISS$ triggering pulse are applied space-selectively to
the halting-qubit atom within the $LH$ potential well with the spatial
region $(-\infty ,x_{L})$. Therefore, such $PHAMDOWN$ laser light beams are
spatially selective. Notice that the $PHAMDOWN$ laser light beams are
applied to the halting-qubit atom along the direction parallel to the atomic
motional direction. This is different from the case that the
spatially-selective laser light beam is used to prepare the one-qubit
quantum gates in the preceding section 3. In the latter case the
spatially-selective laser light pulse is applied to the atom along the
direction perpendicular to the atomic motional direction. This latter case
is much simpler and may be strictly treated by using the error estimation
method in the section 3. There is a difference between an ideal
state-selective triggering pulse and a $SSISS$ triggering pulse generated by
the spatially-selective $PHAMDOWN$ laser light beams. This difference is
generated by the spatially-selective effect of the $PHAMDOWN$ laser light
beams and the imperfection of the $LH$ harmonic potential field. Therefore,
in addition to the imperfection of the $LH$ harmonic potential well there is
an extra imperfection of the spatially-selective $PHAMDOWN$ laser light
beams for a $SSISS$ triggering pulse. Since there are spatially selective
external potential field and electromagnetic wave field, the time evolution
process for the halting-qubit atom in the presence of a $SSISS$ triggering
pulse becomes complicated and is hard to calculate exactly. It is not yet
easy to calculate the time evolution process even when one uses the
first-order approximation propagator of (2.9). On the other hand, it could
be more convenient to use other theoretical methods rather than the
coordinate-representation propagator to calculate strictly the time
evolution process. One of these theoretical methods is the Trotter-Suzuki
decomposition method [34, 39]. This method has already been used
successfully in the preceding section 3. Now it is used to investigate
strictly the imperfection of the spatially-selective $PHAMDOWN$ laser light
beams. It could be more suited to treat strictly the time evolution process
when the time period of the process is short and the Hamiltonian to describe
the process is time-independent. For simplicity here consider the
extensively used three-order symmetric decomposition formula [39] for the
propagator $\exp [-i(H_{0}+V_{1})\tau /\hslash ]$, 
\begin{equation}
\exp [-i(H_{0}+V_{1})\tau /\hslash ]=\exp [-\frac{1}{2}iH_{0}\tau /\hslash
]\exp [-iV_{1}\tau /\hslash ]\exp [-\frac{1}{2}iH_{0}\tau /\hslash
]+O_{p}(\tau ^{3}),  \tag{4.1}
\end{equation}%
where both the operators $H_{0}$ and $V_{1}$ are time-independent and the
error operator $O_{p}(\tau ^{3})$ is given explicitly below. The operator
identity (4.1) also is called the Trotter-Suzuki decomposition formula. A
higher-order decomposition formula also may be found in Ref. [34, 39]. It
also may be used to calculate the time evolution process and usually could
achieve a better result. In order to use the decomposition formula (4.1) to
calculate the time evolution process one needs to choose suitably the main
Hamiltonian $H_{0}$ and the perturbation term $V_{1}$ for a given total
Hamiltonian $H_{0}+V_{1}$ and at the same time make the upper bound (or
norm) of the error operator $O_{p}(\tau ^{3})$ in (4.1) as small as
possible. It has been proven [39, 40] that the upper bound of the error
operator $O_{p}(\tau ^{3})$ is convergent and can be controlled by the time
interval $\tau $ if both the operators $H_{0}$ and $V_{1}$ are bounded. This
upper bound is proportional to $\tau ^{3}$ approximately. However, in an
atomic system in COM motion both the operators $H_{0}$ and $V_{1}$ may not
be bounded. Then the error operator $O_{p}(\tau ^{3})$ may not be convergent
and it may not yet be controlled by the single time interval $\tau .$ It
seems that it could not be suited to use such a decomposition formula as
(4.1) to calculate the time evolution process for a space-dependent quantum
system such as a motional atomic system, because the Hamiltonian of the
quantum system is generally dependent on unbounded coordinate and momentum
operators. Actually, whether or not such a decomposition formula as (4.1) is
useful to calculate the time evolution process of the quantum system is
dependent on the error that is generated by acting the error operator $%
O_{p}(\tau ^{3})$ in (4.1) on the initial state of the quantum system.
Suppose that the error can be controlled so that it can be neglected in the
suitable parameter settings. Then in this case the exact propagator $\exp
[-i(H_{0}+V_{1})\tau /\hslash ]$ could be approximated well by the
decomposition formula (4.1), and it is suited to use the decomposition
formula (4.1) to calculate the time evolution process even if the operators $%
H_{0}$ and $V_{1}$ are unbounded operators. As will be seen later, whether
or not the error is controllable is largely dependent the quantum state and
especially the COM motional state of the quantum system in addition to the
error operator $O_{p}(\tau ^{3})$ itself.

In order to prove the decomposition formula (4.1) useful both the propagator
and the quantum states and especially the COM motional states of the quantum
system need to be considered explicitly, and the explicit expression of the
error operator $O_{p}(\tau ^{3})$ also is needed. This is different from the
conventional theoretical treatments based on the Trotter-Suzuki
decomposition method [34, 39, 40] and the Magnus expansion and the average
Hamiltonian theory [35, 36, 37, 41], where quantum states of a physical
system usually need not be considered explicitly. It can turn out that there
is the exact three-order symmetric decomposition formula of (4.1)\ [39].
This means that there is the exact expression for the error operator $%
O_{p}(\tau ^{3})$ in (4.1). This error operator $O_{p}(\tau ^{3})$ may be
exactly and explicitly expressed as [39]%
\begin{equation*}
O_{p}(\tau ^{3})=\frac{i}{2\hslash ^{3}}\int_{0}^{\tau
}\int_{0}^{t_{1}}\int_{0}^{t_{2}}dt_{1}dt_{2}dt_{3}\{\exp [-\frac{i}{\hslash 
}(H_{0}+V_{1})(\tau -t_{1})]\exp [-\frac{i}{2\hslash }H_{0}t_{1}]
\end{equation*}%
\begin{equation*}
\times \exp [-\frac{i}{\hslash }V_{1}t_{3}][V_{1},[H_{0},V_{1}]]\exp [-\frac{%
i}{\hslash }V_{1}(t_{1}-t_{3})]\exp [-\frac{i}{2\hslash }H_{0}t_{1}]\}
\end{equation*}%
\begin{equation*}
+\frac{i}{4\hslash ^{3}}\int_{0}^{\tau
}\int_{0}^{t_{1}}\int_{0}^{t_{2}}dt_{1}dt_{2}dt_{3}\{\exp [-\frac{i}{\hslash 
}(H_{0}+V_{1})(\tau -t_{1})]\exp [-\frac{i}{2\hslash }H_{0}(t_{1}-t_{3})]
\end{equation*}%
\begin{equation}
\times \lbrack H_{0},[H_{0},V_{1}]]\exp [-\frac{i}{2\hslash }H_{0}t_{3}]\exp
[-\frac{i}{\hslash }V_{1}t_{1}]\exp [-\frac{i}{2\hslash }H_{0}t_{1}]\}. 
\tag{4.2}
\end{equation}%
Once the explicit expression for the error operator $O_{p}(\tau ^{3})$ is
obtained, it could not be hard to calculate the upper bound of the error
generated by acting the error operator $O_{p}(\tau ^{3})$ on the initial
state of the quantum system. However, this calculation is still very
complex, as can be seen below. Now denote $\Psi _{00}(x,r,t_{0})$ as the
initial state of the quantum system. Then the generated error is given by $%
E_{r}(x,r,t_{0}+\tau )=O_{p}(\tau ^{3})\Psi _{00}(x,r,t_{0}).$ With the help
of the exact error operator $O_{p}(\tau ^{3})$ of (4.2) it can turn out that
this error is bounded by%
\begin{equation*}
||E_{r}(x,r,t_{0}+\tau )||\leq \frac{1}{2\hslash ^{3}}\int_{0}^{\tau
}dt_{1}\int_{0}^{t_{1}}dt_{2}\int_{0}^{t_{2}}dt_{3}||M_{1}(x,r,t_{1},t_{3})||
\end{equation*}%
\begin{equation}
+\frac{1}{4\hslash ^{3}}\int_{0}^{\tau
}dt_{1}\int_{0}^{t_{1}}dt_{2}\int_{0}^{t_{2}}dt_{3}||M_{2}(x,r,t_{1},t_{3})||
\tag{4.3}
\end{equation}%
where the two error states $\{M_{k}(x,r,t_{1},t_{3})\}$ for $k=1$ and $2$
are respectively defined by%
\begin{equation*}
M_{1}(x,r,t_{1},t_{3})=[V_{1},[H_{0},V_{1}]]\exp [-\frac{i}{\hslash }%
V_{1}(t_{1}-t_{3})]
\end{equation*}%
\begin{equation}
\times \exp [-\frac{i}{2\hslash }H_{0}t_{1}]\Psi _{00}(x,r,t_{0}),  \tag{4.4}
\end{equation}%
\begin{equation*}
M_{2}(x,r,t_{1},t_{3})=[H_{0},[H_{0},V_{1}]]\exp [-\frac{i}{2\hslash }%
H_{0}t_{3}]\exp [-\frac{i}{\hslash }V_{1}t_{1}]
\end{equation*}%
\begin{equation}
\times \exp [-\frac{i}{2\hslash }H_{0}t_{1}]\Psi _{00}(x,r,t_{0}).  \tag{4.5}
\end{equation}%
Here the important fact has been used in obtaining the inequality (4.3) that
any unitary operator does not change the norm of a quantum state. Now define
the maximum norm $||M_{k}(x,r,t_{1},t_{3})||_{\max }$ for $k=1$ or $2$ over
the time region $0\leq t_{3},t_{1}\leq \tau $ as 
\begin{equation}
||M_{k}(x,r,t_{1},t_{3})||_{\max }=\max_{0\leq t_{3},t_{1}\leq \tau
}\{||M_{k}(x,r,t_{1},t_{3})||\}.  \tag{4.6}
\end{equation}%
Then the inequality (4.3) and the relation (4.6) show that the upper bound
for the error $E_{r}(x,r,t_{0}+\tau )$ may be determined from 
\begin{equation}
||E_{r}(x,r,t_{0}+\tau )||\leq \frac{1}{2}\frac{1}{3!}\frac{\tau ^{3}}{%
\hslash ^{3}}\{||M_{1}(x,r,t_{1},t_{3})||_{\max }+\frac{1}{2}%
||M_{2}(x,r,t_{1},t_{3})||_{\max }\}.  \tag{4.7}
\end{equation}%
This inequality indicates that if the maximum norms $%
\{||M_{k}(x,r,t_{1},t_{3})||_{\max }\}$ for $k=1$ and $2$ are bounded over
the whole coordinate space $-\infty <x<+\infty $, then the upper bound of
the error $E_{r}(x,r,t_{0}+\tau )$ may be controlled effectively by the
single time parameter $\tau .$ This upper bound is clearly proportional to $%
\tau ^{3},$ which is in agreement with the conventional cases [34, 39, 40].

An important application of the operator identity (4.1) is that the operator
identity may be used to investigate strictly the imperfections of the
spatially selective $PHAMDOWN$ laser light pulses in a $SSISS$ triggering
pulse. As shown in the section 2, when the halting-qubit atom in the $LH$
potential well is applied by the spatially selective $PHAMDOWN$ laser light
pulses, the time-independent Hamiltonian of the halting-qubit atom is given
by (2.2), i.e., $H=H_{0}^{ho}+H_{I}(x,\alpha ,\gamma )+V_{1}(x,\alpha
,\gamma ),$ where the atomic internal Hamiltonian $H_{a}=0$, the interaction 
$H_{1}(x,t)=H_{I}(x,\alpha ,\gamma )$ which is given by (2.15), and the
time-independent perturbation term $V_{1}(x,\alpha ,\gamma )\equiv
V_{1}(x,t) $ is given by (2.12) in which $H_{1}(x,t)=H_{I}(x,\alpha ,\gamma
).$ Let the main Hamiltonian $H_{0}=H_{0}^{ho}$ and the perturbation term $%
V_{1}=H_{I}(x,\alpha ,\gamma )+V_{1}(x,\alpha ,\gamma )$ in the operator
identity (4.1) so that the error upper bound on the $RH$ side of (4.3) may
be calculated conveniently. Now the decomposition operator on the $RH$ side
of (4.1) may be rewritten as%
\begin{equation*}
\exp [-\frac{i}{2\hslash }H_{0}\tau ]\exp [-\frac{i}{\hslash }V_{1}\tau
]\exp [-\frac{i}{2\hslash }H_{0}\tau ]
\end{equation*}%
\begin{equation}
=\exp [-\frac{i}{2\hslash }H_{0}^{ho}\tau ]\exp [-\frac{i}{\hslash }%
H_{I}(x,\alpha ,\gamma )\tau ]\exp [-\frac{i}{2\hslash }H_{0}^{ho}\tau
]+O_{p}^{V}(x,r,\tau )  \tag{4.8}
\end{equation}%
where the error operator $O_{p}^{V}(x,r,\tau )$ is defined by%
\begin{equation*}
O_{p}^{V}(x,r,\tau )=-\exp [-\frac{i}{2\hslash }H_{0}^{ho}\tau ]\exp [-\frac{%
i}{\hslash }H_{I}(x,\alpha ,\gamma )\tau ]
\end{equation*}%
\begin{equation}
\times \{1-\exp [-\frac{i}{\hslash }V_{1}(x,\alpha ,\gamma )\tau ]\}\exp [-%
\frac{i}{2\hslash }H_{0}^{ho}\tau ].  \tag{4.9}
\end{equation}%
Notice that the perturbation term $V_{1}(x,\alpha ,\gamma )$ measures the
imperfections of the $LH$ potential well and the spatially selective $%
PHAMDOWN$ laser light pulses. If now $V_{1}(x,\alpha ,\gamma )=0,$ then the
error operator $O_{p}^{V}(x,r,\tau )=0.$ This means that the error operator $%
O_{p}^{V}(x,r,\tau )$ originates from these imperfections. This error
operator generates an error state $O_{p}^{V}(x,r,\tau )\Psi _{00}(x,r,t_{0})$
when it acts on the initial product state $\Psi _{00}(x,r,t_{0})$ of the
halting-qubit atom. Similarly, the error states $M_{k}(x,r,t_{1},t_{3})$ for 
$k=1$ and $2,$ which are given by (4.4) and (4.5), respectively, may be
re-expressed as%
\begin{equation}
M_{k}(x,r,t_{1},t_{3})=M_{k}^{0}(x,r,t_{1},t_{3})+M_{k}^{V}(x,r,t_{1},t_{3}),
\tag{4.10}
\end{equation}%
where $\{M_{k}^{0}(x,r,t_{1},t_{3})\}$ are independent on these
imperfections and may be written as%
\begin{equation*}
M_{1}^{0}(x,r,t_{1},t_{3})=[H_{I}(x,\alpha ,\gamma
),[H_{0}^{ho},H_{I}(x,\alpha ,\gamma )]]
\end{equation*}%
\begin{equation}
\times \exp [-\frac{i}{\hslash }H_{I}(x,\alpha ,\gamma )(t_{1}-t_{3})]\exp [-%
\frac{i}{2\hslash }H_{0}^{ho}t_{1}]\Psi _{00}(x,r,t_{0})  \tag{4.11a}
\end{equation}%
and%
\begin{equation*}
M_{2}^{0}(x,r,t_{1},t_{3})=[H_{0}^{ho},[H_{0}^{ho},H_{I}(x,\alpha ,\gamma
)]]\exp [-\frac{i}{2\hslash }H_{0}^{ho}t_{3}]
\end{equation*}%
\begin{equation}
\times \exp [-\frac{i}{\hslash }H_{I}(x,\alpha ,\gamma )t_{1}]\exp [-\frac{i%
}{2\hslash }H_{0}^{ho}t_{1}]\Psi _{00}(x,r,t_{0}),  \tag{4.11b}
\end{equation}%
while $\{M_{k}^{V}(x,r,t_{1},t_{3})\}$ are generated by these imperfections
and explicitly given by%
\begin{equation*}
M_{1}^{V}(x,r,t_{1},t_{3})=-[H_{I}(x,\alpha ,\gamma
),[H_{0}^{ho},H_{I}(x,\alpha ,\gamma )]]\exp [-\frac{i}{\hslash }%
H_{I}(x,\alpha ,\gamma )(t_{1}-t_{3})]
\end{equation*}%
\begin{equation*}
\times \{1-\exp [-\frac{i}{\hslash }V_{1}(x,\alpha ,\gamma
)(t_{1}-t_{3})]\}\exp [-\frac{i}{2\hslash }H_{0}^{ho}t_{1}]\Psi
_{00}(x,r,t_{0})
\end{equation*}%
\begin{equation*}
+\{[H_{I}(x,\alpha ,\gamma ),[H_{0}^{ho},V_{1}(x,\alpha ,\gamma
)]]+[V_{1}(x,\alpha ,\gamma ),[H_{0}^{ho},H_{I}(x,\alpha ,\gamma )]]
\end{equation*}%
\begin{equation*}
+[V_{1}(x,\alpha ,\gamma ),[H_{0}^{ho},V_{1}(x,\alpha ,\gamma )]]\}\exp [-%
\frac{i}{\hslash }H_{I}(x,\alpha ,\gamma )(t_{1}-t_{3})]
\end{equation*}%
\begin{equation}
\times \exp [-\frac{i}{\hslash }V_{1}(x,\alpha ,\gamma )(t_{1}-t_{3})]\exp [-%
\frac{i}{2\hslash }H_{0}^{ho}t_{1}]\Psi _{00}(x,r,t_{0})  \tag{4.12a}
\end{equation}%
and 
\begin{equation*}
M_{2}^{V}(x,r,t_{1},t_{3})=-[H_{0}^{ho},[H_{0}^{ho},H_{I}(x,\alpha ,\gamma
)]]\exp [-\frac{i}{2\hslash }H_{0}^{ho}t_{3}]
\end{equation*}%
\begin{equation*}
\times \exp [-\frac{i}{\hslash }H_{I}(x,\alpha ,\gamma )t_{1}]\{1-\exp [-%
\frac{i}{\hslash }V_{1}(x,\alpha ,\gamma )t_{1}]\}\exp [-\frac{i}{2\hslash }%
H_{0}^{ho}t_{1}]\Psi _{00}(x,r,t_{0})
\end{equation*}%
\begin{equation*}
+[H_{0}^{ho},[H_{0}^{ho},V_{1}(x,\alpha ,\gamma )]]\exp [-\frac{i}{2\hslash }%
H_{0}^{ho}t_{3}]\exp [-\frac{i}{\hslash }H_{I}(x,\alpha ,\gamma )t_{1}]
\end{equation*}%
\begin{equation}
\times \exp [-\frac{i}{\hslash }V_{1}(x,\alpha ,\gamma )t_{1}]\exp [-\frac{i%
}{2\hslash }H_{0}^{ho}t_{1}]\Psi _{00}(x,r,t_{0}).  \tag{4.12b}
\end{equation}%
Indeed, both the error states $M_{1}^{V}(x,r,t_{1},t_{3})$ and $%
M_{2}^{V}(x,r,t_{1},t_{3})$ are equal to zero if the spatially-selective
perturbation term $V_{1}(x,\alpha ,\gamma )=0.$ These error states $%
O_{p}^{V}(x,r,\tau )\Psi _{00}(x,r,t_{0}),$ $M_{k}^{0}(x,r,t_{1},t_{3}),$
and $M_{k}^{V}(x,r,t_{1},t_{3})$ will be further used to calculate the upper
bound of the total error originating from these imperfections.

Now the time evolution process of the halting-qubit atom in the presence of
the spatially-selective $PHAMDOWN$ laser light beams may be written as,
according to the exact propagator (4.1) with $H_{0}=H_{0}^{ho}$ and $%
V_{1}=H_{I}(x,\alpha ,\gamma )+V_{1}(x,\alpha ,\gamma )$ and the formula
(4.8), 
\begin{equation*}
\Psi (x,r,t_{0}+\tau )\overset{\text{def}}{\equiv }\exp
[-i(H_{0}^{ho}+H_{I}(x,\alpha ,\gamma )+V_{1}(x,\alpha ,\gamma ))\tau
/\hslash ]\Psi _{00}(x,r,t_{0})
\end{equation*}%
\begin{equation}
=\Psi _{0}(x,r,t_{0}+\tau )+E_{r}^{0}(x,r,t_{0}+\tau
)+E_{r}^{V}(x,r,t_{0}+\tau )  \tag{4.13}
\end{equation}%
where the desired product state is written as%
\begin{equation*}
\Psi _{0}(x,r,t_{0}+\tau )=\exp [-\frac{i}{2\hslash }H_{0}^{ho}\tau ]\exp [-%
\frac{i}{\hslash }H_{I}(x,\alpha ,\gamma )\tau ]
\end{equation*}%
\begin{equation}
\times \exp [-\frac{i}{2\hslash }H_{0}^{ho}\tau ]\Psi _{00}(x,r,t_{0}). 
\tag{4.14}
\end{equation}%
This time evolution process generates the two errors $E_{r}^{0}(x,r,t_{0}+%
\tau )$ and $E_{r}^{V}(x,r,$ $t_{0}+\tau ).$ The error $E_{r}^{0}(x,r,t_{0}+%
\tau )$ is independent on the imperfections and its upper bound is still
determined from (4.3) by setting $%
M_{k}(x,r,t_{1},t_{3})=M_{k}^{0}(x,r,t_{1},t_{3})$ for $k=1$ and $2$ on the $%
RH$ side of (4.3), while the error $E_{r}^{V}(x,r,$ $t_{0}+\tau )$ is
generated only by the imperfections and its upper bound can be determined
from%
\begin{equation*}
||E_{r}^{V}(x,r,t_{0}+\tau )||\leq \frac{1}{2\hslash ^{3}}\int_{0}^{\tau
}dt_{1}\int_{0}^{t_{1}}dt_{2}%
\int_{0}^{t_{2}}dt_{3}||M_{1}^{V}(x,r,t_{1},t_{3})||
\end{equation*}%
\begin{equation}
+\frac{1}{4\hslash ^{3}}\int_{0}^{\tau
}dt_{1}\int_{0}^{t_{1}}dt_{2}%
\int_{0}^{t_{2}}dt_{3}||M_{2}^{V}(x,r,t_{1},t_{3})||+||O_{p}^{V}(x,r,\tau
)\Psi _{00}(x,r,t_{0})||.  \tag{4.15}
\end{equation}%
The imperfections of the $LH$ harmonic potential well and the
spatially-selective $PHAMDOWN$ laser light beams may be measured by the
upper bound of the error $E_{r}^{V}(x,r,t_{0}+\tau ).$ This error can be
controlled by the joint position $x_{L}$ in the double-well potential field\
of (2.1). When the joint position $x_{L}\rightarrow +\infty ,$ the error
approaches zero,%
\begin{equation*}
\lim_{x_{L}\rightarrow +\infty }||E_{r}^{V}(x,r,t_{0}+\tau )||=0.
\end{equation*}%
Thus, the error tends to be secondary when the joint position $x_{L}$ is
large, if the initial state $\Psi _{00}(x,r,t_{0})$ is a $GWP$ product
state. Then this means that the error $E_{r}^{0}(x,r,t_{0}+\tau )$ is the
main error in (4.13) when the joint position $x_{L}$ is large. Later the
equation (4.13) will be further used to calculate the time evolution process
of halting-qubit atom in the presence of the $SSISS$ triggering pulse
consisting of the spatially-selective $PHAMDOWN$ laser light pulses.

The main task in the section is to prove that the main error $%
E_{r}^{0}(x,r,t_{0}+\tau )$ in (4.13) can be controlled by the time interval 
$\tau $, while the error $E_{r}^{V}(x,r,t_{0}+\tau )$ generated only by the
imperfections decays exponentially with the square deviation-to-spread
ratios of the relevant $GWP$ motional states of the halting-qubit atom.

Because now the Hamiltonian of the halting-qubit atom contains the electric
dipole interaction $H_{1}(x,t)$ between the halting-qubit atom and the
spatially-selective $PHAMDOWN$ laser light beams, the atomic internal states
could be changed by the laser light beams in the time evolution process of
(4.13). Therefore, in addition to the atomic motional states the atomic
internal states must be considered explicitly in the time evolution process.
There are two cases that need to be considered separately during the
spatially-selective $PHAMDOWN$ laser light beams. One of which is that at
the initial time the halting-qubit atom is in the internal state which is
not really acted on by the electromagnetic fields of the spatially-selective 
$PHAMDOWN$ laser light beams. For example, if the atom is in the internal
state $|g_{1}\rangle \notin \{|g_{0}\rangle ,|e\rangle \},$\ then it is not
really affected by the $PHAMDOWN$ laser light beams, because the laser light
beams are selectively applied to the internal-state subspace $%
\{|g_{0}\rangle ,|e\rangle \}$. In this case the interaction $H_{1}(x,t)$
need not be explicitly taken into account, that is, it may be set to zero, $%
H_{1}(x,t)=0,$ in the exact propagator. Actually, in this case the
Hamiltonian $H=H_{0}^{ho}+H_{1}(x,t)+V_{1}(x,t)$ is reduced to $%
H=H_{0}^{ho}+V_{1}^{ho}(x)$ and hence one may use directly (3.25) to
calculate the final atomic product state $\Psi (x,r,t_{0}+\tau )=\Psi
(x,t_{0}+\tau )|g_{1}\rangle $ up to a global phase factor. Another case is
that at the initial time the halting-qubit atom is in an internal state of
the subspace $\{|g_{0}\rangle ,|e\rangle \}$. In this case the $PHAMDOWN$
laser light beams could induce the atom to make a transition between the two
internal states $|g_{0}\rangle $ and $|e\rangle .$ Then one must consider
explicitly the effect of the interaction $H_{1}(x,t)$ on the time evolution
process. This is the case that will be discussed in detail below. The final
atomic product state $\Psi (x,r,t_{0}+\tau )$ of (4.13) could be a
superposition of the two atomic internal states $|g_{0}\rangle $ and $%
|e\rangle $ even if at the initial time $t_{0}$ the atom is in a given
internal state $(|g_{0}\rangle $ or $|e\rangle )$ and a $GWP$ motional
state. In general, the initial product state $\Psi _{00}(x,r,t_{0})$ in
(4.13) may be a superposition product state. It may be directly used in the
error estimation. Actually, a more convenient scheme is to use the simple
initial product states $\Psi _{00}(x,r,t_{0})=\Psi
_{0}^{g}(x,t_{0})|g_{0}\rangle $ and $\Psi _{0}^{e}(x,t_{0})|e\rangle $ in
the error estimation, respectively. Here $\Psi _{0}^{a}(x,t_{0})$ with $a=g$
or $e$ may be a single $GWP$ motional state or a Gaussian superposition
motional state. The error upper bounds are first calculated explicitly by
using these different initial product states, respectively. Then these error
upper bounds are further summed up to obtain the total error upper bound.

The error estimation also needs to use the interaction $H_{I}(x,\alpha
,\gamma )$ between the halting-qubit atom and the $PHAMDOWN$ laser light
beams in the rotating frame. The interaction $H_{I}(x,\alpha ,\gamma )$ is
already given in Ref. [15] or in (2.15). For convenience it may be rewritten
as%
\begin{equation}
H_{I}(x,\alpha ,\gamma )=Q_{x}(x,\alpha ,\gamma )I_{x}+Q_{y}(x,\alpha
,\gamma )I_{y}.  \tag{4.16}
\end{equation}%
Here the time-independent amplitudes $\{Q_{x,y}(x,\alpha ,\gamma )\}$ are
explicitly given by%
\begin{equation}
Q_{x}(x,\alpha ,\gamma )=4\hslash \Omega _{0}\cos [\frac{1}{2}%
(k_{0}+k_{1})x-\gamma ]\cos (\frac{1}{2}\Delta kx-\alpha )  \tag{4.17a}
\end{equation}%
and%
\begin{equation}
Q_{y}(x,\alpha ,\gamma )=-4\hslash \Omega _{0}\sin [\frac{1}{2}%
(k_{0}+k_{1})x-\gamma ]\cos (\frac{1}{2}\Delta kx-\alpha ).  \tag{4.17b}
\end{equation}%
For the present $SSISS$ triggering pulses there are the parameter settings $%
\alpha =\pi /4$ and $\gamma =0,$ $\pi /2,$ $\pi ,$ and $3\pi /2$ in the
interaction $H_{I}(x,\alpha ,\gamma )$ [15] (See also the section 6 below).
The interaction is dependent on the coordinate $x$ over the whole coordinate
space $(-\infty <x<+\infty )$ and hence it is not spatially selective. In
order to calculate conveniently the time evolution process of (4.13) one may
decompose the propagator $\exp \{-iH_{I}(x,\alpha ,\gamma )\tau /\hslash \}$
as a product of the simple unitary operators. By making a unitary
transformation on the Hamiltonian $H_{I}(x,\alpha ,\gamma )$ of (4.16) one
obtains%
\begin{equation*}
\exp [-i\varphi (x)I_{z}]H_{I}(x,\pi /4,0)\exp [i\varphi (x)I_{z}]=\Omega
(x)I_{x},
\end{equation*}%
\begin{equation*}
\exp [-i\varphi (x)I_{z}]H_{I}(x,\pi /4,\pi /2)\exp [i\varphi
(x)I_{z}]=\Omega (x)I_{y},
\end{equation*}%
where the parameters $\Omega (x)$ and $\varphi (x)$ are given by 
\begin{equation*}
\Omega (x)=4\hslash \Omega _{0}\cos [\frac{1}{2}\Delta kx-\pi /4],\text{ }%
\varphi (x)=\frac{1}{2}(k_{0}+k_{1})x.
\end{equation*}%
Then one can further obtain 
\begin{equation}
\exp \{\pm iH_{I}(x,\pi /4,0)t/\hslash \}=\exp [i\varphi (x)I_{z}]\exp \{\pm
i\Omega (x)I_{x}t/\hslash \}\exp [-i\varphi (x)I_{z}]  \tag{4.18a}
\end{equation}%
and 
\begin{equation}
\exp \{\pm iH_{I}(x,\pi /4,\pi /2)t/\hslash \}=\exp [i\varphi (x)I_{z}]\exp
\{\pm i\Omega (x)I_{y}t/\hslash \}\exp [-i\varphi (x)I_{z}].  \tag{4.18b}
\end{equation}%
The decompositions for the other propagators $\exp \{\pm iH_{I}(x,\pi /4,\pi
)t/\hslash \}$ and $\exp \{\pm iH_{I}(x,\pi /4,3\pi /2)t/\hslash \}$ can be
further obtained from (4.18a) and (4.18b) with the help of the following
relations:%
\begin{equation*}
H_{I}(x,\pi /4,0)=-H_{I}(x,\pi /4,\pi ),\text{ }H_{I}(x,\pi /4,\pi
/2)=-H_{I}(x,\pi /4,3\pi /2).
\end{equation*}%
Actually, the decomposition formulae (4.18) also may be obtained from the
unified unitary transformation: 
\begin{equation}
\exp [-i\varphi (x,\gamma )I_{z}]H_{I}(x,\pi /4,\gamma )\exp [i\varphi
(x,\gamma )I_{z}]=\Omega (x)I_{x}  \tag{4.19}
\end{equation}%
where the phase angle $\varphi (x,\gamma )$ is given by 
\begin{equation*}
\varphi (x,\gamma )=\varphi (x)-\gamma \text{ for }\gamma =0,\text{ }\pi /2,%
\text{ }\pi ,\text{ }3\pi /2.
\end{equation*}%
It is necessary to employ the two unitary transformations (4.18) to
investigate the time evolution process of (4.13) and evaluate strictly the
errors $E_{r}^{0}(x,r,t_{0}+\tau )$ and $E_{r}^{V}(x,r,t_{0}+\tau )$ in
(4.13).

It is known from (4.3) and (4.15) that a strict error estimation for the
errors $E_{r}^{0}(x,r,t_{0}+\tau )$ and $E_{r}^{V}(x,r,t_{0}+\tau )$ in
(4.13) needs to calculate the three norms $||O_{p}^{V}(x,r,\tau )\Psi
_{00}(x,r,t_{0})||,$ $||M_{k}^{0}(x,r,t_{1},t_{3})||,$ and $%
||M_{k}^{V}(x,r,t_{1},t_{3})||$ for $k=1$, $2$. It is easy to calculate the
first norm. However, it is complex to calculate the last two norms.
Particularly, it is very complex to calculate strictly the norm $%
||M_{2}^{V}(x,r,t_{1},t_{3})||.$ Below a rigorous calculation is carried out
for the upper bound of every one of these norms. \newline
\newline
{\large 4.1 The upper bound of the norm }${\large ||}O_{p}^{V}(x,r,\tau
)\Psi _{00}(x,r,t_{0})||$

According to (4.9) it can turn out that the error $O_{p}^{V}(x,r,\tau )\Psi
_{00}(x,r,t_{0})$ is bounded by%
\begin{equation}
||O_{p}^{V}(x,r,\tau )\Psi _{00}(x,r,t_{0})||\leq (\frac{\tau }{\hslash }%
)||V_{1}(x,\alpha ,\gamma )\exp [-\frac{i}{2\hslash }H_{0}^{ho}\tau ]\Psi
_{00}(x,r,t_{0})||.  \tag{4.20}
\end{equation}%
This means that the upper bound of the error also may be calculated strictly
by using the first-order approximation propagator (2.9), as shown in the
previous section 3. In general, the initial product state $\Psi
_{00}(x,r,t_{0})$ may be a Gaussian superposition state: $\Psi
_{00}(x,r,t_{0})=\Psi _{0}^{g}(x,t_{0})|g_{0}\rangle +\Psi
_{0}^{e}(x,t_{0})|e\rangle $. Then the product state $\Psi
_{0}(x,r,t_{0}+t)=\exp [-iH_{0}^{ho}t/\hslash ]\Psi _{00}(x,r,t_{0})$ also
is a Gaussian superposition state. Moreover, the product state $\Psi
_{0}(x,r,t_{c})$ with $t_{0}\leq t_{c}\leq t_{0}+\tau /2$ may be formally
written as $\Psi _{0}(x,r,t_{c})=\Psi _{0}^{g}(x,t_{c})|g_{0}\rangle +\Psi
_{0}^{e}(x,t_{c})|e\rangle .$ Now by using the product state $\Psi
_{0}(x,r,t_{c})$ and the incontinuous perturbation term $V_{1}(x,\alpha
,\gamma )$ of (2.12) one can calculate the error state $\Psi
_{V_{1}}(x,r,t_{c})=V_{1}(x,\alpha ,\gamma )\Psi _{0}(x,r,t_{c}).$ Of
course, here one also may use the smooth perturbation term $V_{1}(x,\alpha
,\gamma ,\varepsilon )$ of (2.12) in the calculation, but there is a
negligible difference between the calculated results for both the cases if
the parameter $\varepsilon <<1$, as shown in the previous section 3. The
calculated result is: (a)\ if $-\infty <x<x_{L},$ the state $\Psi
_{V_{1}}(x,r,t_{c})=0;$ (b) if $x_{L}<x<x_{L}+L,$ the state is 
\begin{equation*}
\Psi _{V_{1}}(x,r,t_{c})=\frac{1}{2}\{m\omega ^{2}(x_{L}^{2}-x^{2})\Psi
_{0}^{g}(x,t_{c})
\end{equation*}%
\begin{equation*}
-[Q_{x}(x,\alpha ,\gamma )+iQ_{y}(x,\alpha ,\gamma )]\Psi
_{0}^{e}(x,t_{c})\}|g_{0}\rangle
\end{equation*}%
\begin{equation*}
+\frac{1}{2}\{m\omega ^{2}(x_{L}^{2}-x^{2})\Psi
_{0}^{e}(x,t_{c})-[Q_{x}(x,\alpha ,\gamma )-iQ_{y}(x,\alpha ,\gamma )]\Psi
_{0}^{g}(x,t_{c})\}|e\rangle ;
\end{equation*}%
and (c) if $x_{L}+L<x<+\infty ,$ the state is 
\begin{equation*}
\Psi _{V_{1}}(x,r,t_{c})=-\frac{1}{2}\{m\omega ^{2}x^{2}\Psi
_{0}^{g}(x,t_{c})+[Q_{x}(x,\alpha ,\gamma )+iQ_{y}(x,\alpha ,\gamma )]\Psi
_{0}^{e}(x,t_{c})\}|g_{0}\rangle
\end{equation*}%
\begin{equation*}
-\frac{1}{2}\{m\omega ^{2}x^{2}\Psi _{0}^{e}(x,t_{c})+[Q_{x}(x,\alpha
,\gamma )-iQ_{y}(x,\alpha ,\gamma )]\Psi _{0}^{g}(x,t_{c})\}|e\rangle .
\end{equation*}%
One therefore obtains explicitly the motional states $\Psi
_{V_{1}}^{g}(x,t_{c})$ and $\Psi _{V_{1}}^{e}(x,t_{c})$ from the error state 
$\Psi _{V_{1}}(x,r,t_{c}).$ Now by using the two motional states $\Psi
_{V_{1}}^{g}(x,t_{c})$ and $\Psi _{V_{1}}^{e}(x,t_{c})$ it can turn out that
the total probability of the error state $\Psi _{V_{1}}(x,r,t_{c})$ is
bounded by%
\begin{equation*}
||\Psi _{V_{1}}(x,r,t_{c})||^{2}\leq (\frac{1}{2}m\omega
^{2}x_{L}^{2})\int_{x_{L}}^{x_{L}+L}dx_{a}\{\{\frac{1}{2}m\omega
^{2}x_{L}^{2}-m\omega ^{2}x_{a}^{2}
\end{equation*}%
\begin{equation*}
+|Q_{x}(x_{a},\alpha ,\gamma )|+|Q_{y}(x_{a},\alpha ,\gamma )|\}\{|\Psi
_{0}^{g}(x_{a},t_{c})|^{2}+\Psi _{0}^{e}(x_{a},t_{c})|^{2}\}\}
\end{equation*}%
\begin{equation*}
+\int_{x_{L}}^{\infty }dx_{a}\{\{(\frac{1}{2}m\omega ^{2}x_{a}^{2})^{2}+(%
\frac{1}{2}m\omega ^{2}x_{a}^{2})(|Q_{x}(x_{a},\alpha ,\gamma
)|+|Q_{y}(x_{a},\alpha ,\gamma )|)
\end{equation*}%
\begin{equation}
+\frac{1}{4}|Q_{x}(x_{a},\alpha ,\gamma )|^{2}+\frac{1}{4}%
|Q_{y}(x_{a},\alpha ,\gamma )|^{2}\}\{|\Psi _{0}^{g}(x_{a},t_{c})|^{2}+|\Psi
_{0}^{e}(x_{a},t_{c})|^{2}\}\}.  \tag{4.21}
\end{equation}%
Suppose that the amplitudes $\{Q_{x,y}(x,\alpha ,\gamma )\}$ have the upper
bounds: 
\begin{equation}
|Q_{x}(x,\alpha ,\gamma )|\leq Q_{x}^{M},\text{ }|Q_{y}(x,\alpha ,\gamma
)|\leq Q_{y}^{M}.  \tag{4.22}
\end{equation}%
In fact, it follows from (4.17) that both the amplitudes $\{Q_{x,y}(x,\alpha
,\gamma )\}$ are bounded by $|Q_{x,y}(x,\alpha ,\gamma )|\leq 4\hslash
\Omega _{0},$ indicating that $Q_{x}^{M}=Q_{y}^{M}=4\hslash \Omega _{0}.$ By
inserting the inequalities (4.22) into (4.21) it can prove that the
probability $||\Psi _{V_{1}}(x,r,t_{c})||^{2}$ is bounded by%
\begin{equation}
||\Psi
_{V_{1}}(x,r,t_{c})||^{2}<M_{0}^{g}(x_{L},t_{c})+M_{0}^{e}(x_{L},t_{c}), 
\tag{4.23}
\end{equation}%
where the positive function $M_{0}^{a}(x_{L},t_{c})$ with the label $a=g$ or 
$e$ is defined by 
\begin{equation*}
M_{0}^{a}(x_{L},t_{c})=P_{3}(x_{L})\int_{x_{L}}^{\infty }dx|\Psi
_{0}^{a}(x,t_{c})|^{2}
\end{equation*}%
\begin{equation}
+C_{2}\int_{x_{L}}^{\infty }dxx^{2}|\Psi
_{0}^{a}(x,t_{c})|^{2}+C_{4}\int_{x_{L}+L}^{\infty }dxx^{4}|\Psi
_{0}^{a}(x,t_{c})|^{2},  \tag{4.24}
\end{equation}%
while the positive parameters $P_{3}(x_{L}),$ $C_{2},$ and $C_{4}$ are given
by 
\begin{equation*}
P_{3}(x_{L})=[\frac{1}{4}(Q_{x}^{M})^{2}+\frac{1}{4}(Q_{y}^{M})^{2}]+(\frac{1%
}{2}m\omega ^{2}x_{L}^{2})(Q_{x}^{M}+Q_{y}^{M})
\end{equation*}%
\begin{equation*}
+L(2+L/x_{L})[1+(1+L/x_{L})^{2}](\frac{1}{2}m\omega ^{2})^{2}x_{L}^{3},
\end{equation*}%
\begin{equation*}
C_{2}=(\frac{1}{2}m\omega ^{2})(Q_{x}^{M}+Q_{y}^{M}),\text{ }C_{4}=(\frac{1}{%
2}m\omega ^{2})^{2}.
\end{equation*}%
The parameter $P_{3}(x_{L})$ is a cubic polynomial in the joint position $%
x_{L}$ approximately if $x_{L}>>L.$ Actually, $M_{0}^{a}(x_{L},t_{c})$ is
the upper bound of the probability $||\Psi _{V_{1}}^{a}(x,t_{c})||^{2}.$ It
can be calculated explicitly. Suppose that the motional state $\Psi
_{0}^{a}(x,t_{c})$ with $a=g$ or $e$ is a single $GWP$ state with the COM
position $x_{0}^{a}(t_{c})$ and wave-packet spread $\varepsilon
_{0}^{a}(t_{c}).$ The state $\Psi _{0}^{a}(x,t_{c})$ may not be normalized
and its amplitude is denoted as $A_{0}^{a}(t_{c})$. Now by inserting the
state $\Psi _{0}^{a}(x,t_{c})$ into (4.24) the function $%
M_{0}^{a}(x_{L},t_{c})$ can be written as 
\begin{equation}
M_{0}^{a}(x_{L},t_{c})=|A_{0}^{a}(t_{c})|^{2}M_{0}^{a}(x_{L},\varepsilon
_{0}^{a}(t_{c}),x_{0}^{a}(t_{c}))  \tag{4.25}
\end{equation}%
where the positive function $M_{0}^{a}(x_{L},\varepsilon
_{0}^{a}(t_{c}),x_{0}^{a}(t_{c}))$ turns out to be%
\begin{equation*}
M_{0}^{a}(x_{L},\varepsilon _{0}^{a}(t_{c}),x_{0}^{a}(t_{c}))=\frac{1}{\sqrt{%
\pi }}P_{3}(x_{L})\int_{y_{M}^{a}}^{\infty }dy\exp (-y^{2})
\end{equation*}%
\begin{equation*}
+\frac{1}{\sqrt{\pi }}C_{2}\{\varepsilon _{0}^{a}(t_{c})[x_{0}^{a}(t_{c})+%
\frac{1}{2}y_{M}^{a}\varepsilon _{0}^{a}(t_{c})]\exp [-(y_{M}^{a})^{2}]
\end{equation*}%
\begin{equation}
+[\frac{1}{2}\varepsilon
_{0}^{a}(t_{c})^{2}+x_{0}^{a}(t_{c})^{2}]\int_{y_{M}^{a}}^{+\infty }dy\exp
(-y^{2})\}+C_{4}I_{0}(x_{L}+L,\varepsilon _{0}^{a}(t_{c}),x_{0}^{a}(t_{c})).
\tag{4.26}
\end{equation}%
Here the deviation-to-spread ratio $y_{M}^{a}\equiv
y_{M}^{a}(x_{L},t_{c})=(x_{L}-x_{0}^{a}(t_{c}))$ $/\varepsilon
_{0}^{a}(t_{c})>0.$ The integral $I_{0}(x_{L}+L,\varepsilon
_{0}^{a}(t_{c}),x_{0}^{a}(t_{c}))$ is still defined by (3.7) and its
explicit expression is given by (3.8), in which the parameter settings are $%
\varepsilon (t_{c})=\varepsilon _{0}^{a}(t_{c}),$ $%
x_{c}(t_{c})=x_{0}^{a}(t_{c}),$ and $y_{M}=z_{M}^{a}.$ Here the parameter $%
z_{M}^{a}\equiv (x_{L}+L-x_{0}^{a}(t_{c}))/\varepsilon _{0}^{a}(t_{c})$.
Since the error function $\int_{y_{M}^{a}}^{\infty }dy\exp (-y^{2})\leq $ $%
\exp [-(y_{M}^{a})^{2}]/[y_{M}^{a}+\sqrt{(y_{M}^{a})^{2}+4/\pi }]$ [44], it
can turn out that the function $M_{0}^{a}(x_{L},\varepsilon
_{0}^{a}(t_{c}),x_{0}^{a}(t_{c}))$\ of (4.26) satisfies%
\begin{equation}
M_{0}^{a}(x_{L},\varepsilon _{0}^{a}(t_{c}),x_{0}^{a}(t_{c}))\leq
Q_{0}^{a}(x_{L},\varepsilon _{0}^{a}(t_{c}),x_{0}^{a}(t_{c}))\exp
[-y_{M}^{a}(x_{L},t_{c})^{2}]  \tag{4.27}
\end{equation}%
where the inequality (3.9) for the integral $I_{0}(x_{L}+L,\varepsilon
_{0}^{a}(t_{c}),x_{0}^{a}(t_{c}))$ has been used and the positive function $%
Q_{0}^{a}(x_{L},\varepsilon _{0}^{a}(t_{c}),x_{0}^{a}(t_{c}))$ is defined by%
\begin{equation*}
Q_{0}^{a}(x_{L},\varepsilon _{0}^{a}(t_{c}),x_{0}^{a}(t_{c}))=\frac{1}{\sqrt{%
\pi }}\frac{P_{3}(x_{L})}{y_{M}^{a}+\sqrt{(y_{M}^{a})^{2}+4/\pi }}
\end{equation*}%
\begin{equation*}
+\frac{1}{2}\frac{1}{\sqrt{\pi }}C_{2}\{\varepsilon
_{0}^{a}(t_{c})[2x_{0}^{a}(t_{c})+y_{M}^{a}\varepsilon _{0}^{a}(t_{c})]+%
\frac{\varepsilon _{0}^{a}(t_{c})^{2}+2x_{0}^{a}(t_{c})^{2}}{y_{M}^{a}+\sqrt{%
(y_{M}^{a})^{2}+4/\pi }}\}
\end{equation*}%
\begin{equation}
+C_{4}P(x_{L}+L,\varepsilon _{0}^{a}(t_{c}),x_{0}^{a}(t_{c}))\exp
\{-(z_{M}^{a})^{2}+(y_{M}^{a})^{2}\}.  \tag{4.28}
\end{equation}%
Here the function $P(x_{L}+L,\varepsilon _{0}^{a}(t_{c}),x_{0}^{a}(t_{c}))$
is still defined by (3.10), in which the parameter settings are $\varepsilon
(t_{c})=\varepsilon _{0}^{a}(t_{c}),$ $x_{c}(t_{c})=x_{0}^{a}(t_{c}),$ and $%
y_{M}=z_{M}^{a}$. Notice that $z_{M}^{a}>y_{M}^{a}>0.$ When the joint
position $x_{L}$ is large enough, the function $Q_{0}^{a}(x_{L},\varepsilon
_{0}^{a}(t_{c}),x_{0}^{a}(t_{c}))$ is approximately a cubic polynomial in
the joint position $x_{L}$. Thus, the inequality (4.27) shows that the upper
bound of the function $M_{0}^{a}(x_{L},\varepsilon
_{0}^{a}(t_{c}),x_{0}^{a}(t_{c}))$ decays exponentially with the square
deviation-to-spread ratio $(y_{M}^{a})^{2}.$ Suppose now that 
\begin{equation*}
Q_{0}^{a}(x_{L},t_{c}^{\ast })=\max_{t_{0}\leq t_{c}\leq t_{0}+\tau
/2}\{Q_{0}^{a}(x_{L},\varepsilon _{0}^{a}(t_{c}),x_{0}^{a}(t_{c}))\},
\end{equation*}%
\begin{equation*}
|A_{0}^{a}(t_{c}^{\ast })|=\max_{t_{0}\leq t_{c}\leq t_{0}+\tau
/2}\{|A_{0}^{a}(t_{c})|\},\text{ }y_{M}^{a}(x_{L},t_{c}^{\ast
})=\min_{t_{0}\leq t_{c}\leq t_{0}+\tau /2}\{y_{M}^{a}(x_{L},t_{c})\}.
\end{equation*}%
Then the inequality (4.27) further shows that for any time $t_{c}\in \lbrack
t_{0},t_{0}+\tau /2]$ the function $M_{0}^{a}(x_{L},\varepsilon
_{0}^{a}(t_{c}),x_{0}^{a}(t_{c}))$ is bounded by 
\begin{equation}
M_{0}^{a}(x_{L},\varepsilon _{0}^{a}(t_{c}),x_{0}^{a}(t_{c}))\leq
Q_{0}^{a}(x_{L},t_{c}^{\ast })\exp [-y_{M}^{a}(x_{L},t_{c}^{\ast })^{2}]. 
\tag{4.29}
\end{equation}%
Now it follows from (4.29), (4.25), (4.23), and (4.20) that the error $%
O_{p}^{V}(x,r,\tau )$ $\times \Psi _{00}(x,r,t_{0})$ is bounded by%
\begin{equation*}
||O_{p}^{V}(x,r,\tau )\Psi _{00}(x,r,t_{0})||<\frac{\tau }{\hslash }%
|A_{0}^{g}(t_{c}^{\ast })|\sqrt{Q_{0}^{g}(x_{L},t_{c}^{\ast })}\exp
[-y_{M}^{g}(x_{L},t_{c}^{\ast })^{2}/2]
\end{equation*}%
\begin{equation}
+\frac{\tau }{\hslash }|A_{0}^{e}(t_{c}^{\ast })|\sqrt{%
Q_{0}^{e}(x_{L},t_{c}^{\ast })}\exp [-y_{M}^{e}(x_{L},t_{c}^{\ast })^{2}/2].
\tag{4.30}
\end{equation}%
This shows that the upper bound of the error $O_{p}^{V}(x,r,\tau )\Psi
_{00}(x,r,t_{0})$ is proportional to the exponentially-decaying factors $%
\{\exp [-y_{M}^{a}(x_{L},t_{c}^{\ast })^{2}/2]\}.$ Thus, this error, that is
due to the imperfections for the $LH$ harmonic potential well and the
space-selective $PHAMDOWN$ laser light beams, is negligible if the joint
position $x_{L}$ is large enough such that $\min
\{y_{M}^{g}(x_{L},t_{c}^{\ast }),y_{M}^{e}(x_{L},t_{c}^{\ast })\}>>1$.

More generally, the initial motional state $\Psi _{0}^{a}(x,t_{0})$ with $%
a=g $ or $e$ may be a Gaussian superposition motional state. Then the
motional state $\Psi _{0}^{a}(x,t_{c})$ in the function $%
M_{0}^{a}(x_{L},t_{c})$ of (4.24) is also a Gaussian superposition state. It
may be expressed as (3.15). According to the Cauchy inequality [44b] it
follows from (3.15) that the probability density $|\Psi
_{0}^{a}(x,t_{c})|^{2}$ is bounded by%
\begin{equation*}
|\Psi _{0}^{a}(x,t_{c})|^{2}\leq
\{\sum_{k=1}^{n_{a}}|A_{k}^{a}(t_{c})|^{2}\}\{\sum_{k=1}^{n_{a}}|\Psi
_{0k}^{a}(x,t_{c})|^{2}\}.
\end{equation*}%
This inequality and the formula (4.24) lead to that the function $%
M_{0}^{a}(x_{L},t_{c})$ is bounded by%
\begin{equation}
M_{0}^{a}(x_{L},t_{c})\leq
\{\sum_{k=1}^{n_{a}}|A_{k}^{a}(t_{c})|^{2}\}%
\sum_{k=1}^{n_{a}}M_{0k}^{a}(x_{L},\varepsilon
_{0k}^{a}(t_{c}),x_{0k}^{a}(t_{c}))  \tag{4.31}
\end{equation}%
where the positive function $M_{0k}^{a}(x_{L},\varepsilon
_{0k}^{a}(t_{c}),x_{0k}^{a}(t_{c}))$ is defined by%
\begin{equation*}
M_{0k}^{a}(x_{L},\varepsilon
_{0k}^{a}(t_{c}),x_{0k}^{a}(t_{c}))=P_{3}(x_{L})\int_{x_{L}}^{\infty
}dx|\Psi _{0k}^{a}(x,t_{c})|^{2}
\end{equation*}%
\begin{equation}
+C_{2}\int_{x_{L}}^{\infty }dxx^{2}|\Psi
_{0k}^{a}(x,t_{c})|^{2}+C_{4}\int_{x_{L}+L}^{\infty }dxx^{4}|\Psi
_{0k}^{a}(x,t_{c})|^{2}.  \tag{4.32}
\end{equation}%
Since now the $GWP$ state $\Psi _{0k}^{a}(x,t_{c})$ is normalized, the
function of (4.32) may be explicitly expressed as (4.26) with the parameter
settings $\varepsilon _{0}^{a}(t_{c})=\varepsilon _{0k}^{a}(t_{c})$, $%
x_{0}^{a}(t_{c})=x_{0k}^{a}(t_{c}),$ $%
z_{M}^{a}=z_{0k}^{a}=(x_{L}+L-x_{0k}^{a}(t_{c}))/\varepsilon
_{0k}^{a}(t_{c})>0$, and $y_{M}^{a}=y_{0k}^{a}=(x_{L}-x_{0k}^{a}(t_{c}))/%
\varepsilon _{0k}^{a}(t_{c})>0$. Here $x_{0k}^{a}(t_{c})$ and $\varepsilon
_{0k}^{a}(t_{c})$ are the COM position and wave-packet spread of the $k-$th
normalized $GWP$ state $\Psi _{0k}^{a}(x,t_{c}),$ respectively. Obviously,
each one of these functions $\{M_{0k}^{a}(x_{L},\varepsilon
_{0k}^{a}(t_{c}),x_{0k}^{a}(t_{c}))\}$ satisfies the following inequality,
which is similar to (4.27), 
\begin{equation}
M_{0k}^{a}(x_{L},\varepsilon _{0k}^{a}(t_{c}),x_{0k}^{a}(t_{c}))\leq
Q_{0k}^{a}(x_{L},\varepsilon _{0k}^{a}(t_{c}),x_{0k}^{a}(t_{c}))\exp
[-y_{0k}^{a}(x_{L},t_{c})^{2}]  \tag{4.33}
\end{equation}%
where the function $Q_{0k}^{a}(x_{L},\varepsilon
_{0k}^{a}(t_{c}),x_{0k}^{a}(t_{c}))$ is still defined by (4.28), in which
the parameter settings are $\varepsilon _{0}^{a}(t_{c})=\varepsilon
_{0k}^{a}(t_{c}),$ $x_{0}^{a}(t_{c})=x_{0k}^{a}(t_{c}),$ $%
y_{M}^{a}=y_{0k}^{a},$ and $z_{M}^{a}=z_{0k}^{a}$. Denote $%
Q_{0k}^{a}(x_{L},t_{c}^{\ast })$ as the maximum value of the function $%
Q_{0k}^{a}(x_{L},\varepsilon _{0k}^{a}(t_{c}),$ $x_{0k}^{a}(t_{c}))$ in the
time region $[t_{0},t_{0}+\tau /2]$ and $y_{0k}^{a}(x_{L},t_{c}^{\ast })$ as
the minimum value of the deviation-to-spread ratio $y_{0k}^{a}(x_{L},t_{c})$
in the same time region. Furthermore, denote that 
\begin{equation*}
Q_{0}^{a}(x_{L})=\max_{1\leq k\leq n_{a}}\{Q_{0k}^{a}(x_{L},t_{c}^{\ast })\}%
\text{ and }y_{M}^{a}(x_{L})=\min_{1\leq k\leq
n_{a}}\{y_{0k}^{a}(x_{L},t_{c}^{\ast })\}.
\end{equation*}%
Then there is the inequality for any index $k\in \lbrack 1,n_{a}]:$%
\begin{equation}
Q_{0k}^{a}(x_{L},\varepsilon _{0k}^{a}(t_{c}),x_{0k}^{a}(t_{c}))\exp
[-y_{0k}^{a}(x_{L},t_{c})^{2}]\leq Q_{0}^{a}(x_{L})\exp
[-y_{M}^{a}(x_{L})^{2}].  \tag{4.34}
\end{equation}%
Now it follows from (4.34), (4.33), (4.31), (4.23), and (4.20) that the
error $O_{p}^{V}(x,r,\tau )\Psi _{00}(x,r,t_{0})$ is bounded by%
\begin{equation*}
||O_{p}^{V}(x,r,\tau )\Psi _{00}(x,r,t_{0})||<\frac{\tau }{\hslash }\sqrt{%
n_{g}A_{0}^{g}Q_{0}^{g}(x_{L})}\exp [-y_{M}^{g}(x_{L})^{2}/2]
\end{equation*}%
\begin{equation}
+\frac{\tau }{\hslash }\sqrt{n_{e}A_{0}^{e}Q_{0}^{e}(x_{L})}\exp
[-y_{M}^{e}(x_{L})^{2}/2]  \tag{4.35}
\end{equation}%
where $A_{0}^{a}=\max_{t_{0}\leq t_{c}\leq t_{0}+\tau
/2}\{\sum_{k=1}^{n_{a}}|A_{k}^{a}(t_{c})|^{2}\}.$ This inequality shows that
the upper bound of the error $O_{p}^{V}(x,r,\tau )\Psi _{00}(x,r,t_{0})$ is
proportional to the exponen- tially-decaying factors $\{\exp
[-y_{M}^{a}(x_{L})^{2}/2]\}.$ Therefore, if both $n_{g}$ and $n_{e}$ are
finite positive integers, then the upper bound is close to zero and hence
the error $O_{p}^{V}(x,r,\tau )\Psi _{00}(x,r,t_{0})$ can be neglected when
the joint position $x_{L}$ is large enough such that the minimum
deviation-to-spread ratio $\min \{y_{M}^{g}(x_{L}),$ $y_{M}^{e}(x_{L})\}>>1.$%
\newline
\newline
{\large 4.2 The upper bound of the norm }$||M_{1}^{V}(x,r,t_{1},t_{3})||$

The upper bound of the error $E_{r}^{V}(x,r,t_{0}+\tau )$ consists of the
three terms on the $RH$ side of (4.15). Among these three terms the upper
bound for the last term $||O_{p}^{V}(x,r,\tau )\Psi _{00}(x,r,t_{0})||$ is
already determined from (4.30) or (4.35). The first term is an integral
whose integrand is the norm $||M_{1}^{V}(x,r,t_{1},t_{3})||/(2\hslash ^{3}).$
Now in order to prove that the term decays exponentially with the square
deviation-to-spread ratios of the relevant $GWP$ states one needs to
calculate strictly the norm $||M_{1}^{V}(x,r,t_{1},t_{3})||.$ The task in
the subsection is to calculate strictly this norm. It follows from (4.12a)
that the error state $M_{1}^{V}(x,r,t_{1},t_{3})$ may be rewritten as%
\begin{equation}
M_{1}^{V}(x,r,t_{1},t_{3})=M_{11}^{V}(x,r,t_{1},t_{3})+M_{12}^{V}(x,r,t_{1},t_{3})
\tag{4.36a}
\end{equation}%
where the two error states are given by%
\begin{equation*}
M_{11}^{V}(x,r,t_{1},t_{3})=-[H_{I}(x,\alpha ,\gamma
),[H_{0}^{ho},H_{I}(x,\alpha ,\gamma )]]\exp [-\frac{i}{\hslash }%
H_{I}(x,\alpha ,\gamma )(t_{1}-t_{3})]
\end{equation*}%
\begin{equation}
\times \{1-\exp [-\frac{i}{\hslash }V_{1}(x,\alpha ,\gamma
)(t_{1}-t_{3})]\}\Psi _{0}(x,r,t_{0}+t_{1}/2)  \tag{4.36b}
\end{equation}%
and 
\begin{equation*}
M_{12}^{V}(x,r,t_{1},t_{3})=\{[H_{I}(x,\alpha ,\gamma
),[H_{0}^{ho},V_{1}(x,\alpha ,\gamma )]]
\end{equation*}%
\begin{equation*}
+[V_{1}(x,\alpha ,\gamma ),[H_{0}^{ho},H_{I}(x,\alpha ,\gamma
)]]+[V_{1}(x,\alpha ,\gamma ),[H_{0}^{ho},V_{1}(x,\alpha ,\gamma )]]\}
\end{equation*}%
\begin{equation}
\times \exp [-\frac{i}{\hslash }H_{I}(x,\alpha ,\gamma )(t_{1}-t_{3})]\exp [-%
\frac{i}{\hslash }V_{1}(x,\alpha ,\gamma )(t_{1}-t_{3})]\Psi
_{0}(x,r,t_{0}+t_{1}/2)  \tag{4.36c}
\end{equation}%
where $\Psi _{0}(x,r,t_{0}+t_{1}/2)=\exp [-(i/\hslash
)H_{0}^{ho}t_{1}/2]\Psi _{00}(x,r,t_{0})$ is a Gaussian product state. These
commutation relations $[H_{I}(x,\alpha ,\gamma ),[H_{0}^{ho},H_{I}(x,\alpha
,\gamma )]],$ $etc.$, in (4.36) can be calculated by using the basic
commutation relations [22]: 
\begin{equation}
\lbrack f(x),p]=i\hslash \frac{\partial }{\partial x}f(x)  \tag{4.37a}
\end{equation}%
and 
\begin{equation}
\lbrack f(x),p^{2}]=2i\hslash \lbrack \frac{\partial }{\partial x}%
f(x)]p+\hslash ^{2}\frac{\partial ^{2}}{\partial x^{2}}f(x)  \tag{4.37b}
\end{equation}%
where $p=-i\hslash \partial /\partial x$ is the momentum operator and $f(x)$
is a function of the coordinate $x$. The two norms $%
||M_{11}^{V}(x,r,t_{1},t_{3})||$ and $||M_{12}^{V}(x,r,t_{1},t_{3})||$ may
be strictly calculated by using the harmonic-oscillator Hamiltonian $%
H_{0}^{ho}$ of (2.3), the interaction $H_{I}(x,\alpha ,\gamma )$ of (4.16),
and the continuous perturbation term $V_{1}(x,\alpha ,\gamma ,\varepsilon )$
of (2.12). After a simple calculation by using the basic commutation
relations (4.37) it turns out that the upper bound for the error state $%
M_{11}^{V}(x,r,t_{1},t_{3})$ is determined from%
\begin{equation*}
2m||M_{11}^{V}(x,r,t_{1},t_{3})||\leq 2\hslash ||[H_{I}(x,\alpha ,\gamma ),%
\frac{\partial }{\partial x}H_{I}(x,\alpha ,\gamma )]||
\end{equation*}%
\begin{equation*}
\times ||p\{1-\exp [-\frac{i}{\hslash }V_{1}(x,\alpha ,\gamma ,\varepsilon
)(t_{1}-t_{3})]\}\Psi _{0}(x,r,t_{0}+t_{1}/2)||
\end{equation*}%
\begin{equation*}
+\hslash ^{2}||[H_{I}(x,\alpha ,\gamma ),\frac{\partial ^{2}}{\partial x^{2}}%
H_{I}(x,\alpha ,\gamma )]-2[\frac{\partial }{\partial x}H_{I}(x,\alpha
,\gamma )]^{2}||
\end{equation*}%
\begin{equation}
\times ||\{1-\exp [-\frac{i}{\hslash }V_{1}(x,\alpha ,\gamma ,\varepsilon
)(t_{1}-t_{3})]\}\Psi _{0}(x,r,t_{0}+t_{1}/2)||.  \tag{4.38}
\end{equation}%
The $RH$ side of (4.38) contains two different types of norms. The first
type consists of the two norms that contain the product state $\Psi
_{0}(x,r,t_{0}+t_{1}/2),$ while the second type includes the two norms that
do not contain the product state but the interaction $H_{I}(x,\alpha ,\gamma
)$. Since the interaction $H_{I}(x,\alpha ,\gamma )$ of (4.16) and its $k-$%
order coordinate derivatives ($k=1$, $2$, ...) all are bounded, the two
norms that do not contain the product state are bounded. Hereafter any norm
that does not contain a quantum state like $\Psi _{0}(x,r,t_{0}+t_{1}/2)$ is
considered as a parameter. Thus, here one needs only to prove that the two
norms that contain the product state $\Psi _{0}(x,r,t_{0}+t_{1}/2)$ on the $%
RH$ side of (4.38) decay exponentially with the square deviation-to-spread
ratios of the relevant $GWP$ states. For convenience, here define an
auxiliary product state $\chi (x,r,\tau _{1},\tau _{2})$ by%
\begin{equation}
\chi (x,r,\tau _{1},\tau _{2})\overset{\text{def}}{\equiv }\{1-\exp [-\frac{i%
}{\hslash }V_{1}(x,\alpha ,\gamma ,\varepsilon )\tau _{1}]\}\Psi
_{0}(x,r,t_{0}+\tau _{2}/2).  \tag{4.39a}
\end{equation}%
Then it can turn out that the upper bound of the last norm $||\chi
(x,r,t_{1}-t_{3},t_{1})||$ on the $RH$ side of (4.38) may be determined from%
\begin{equation*}
||\chi (x,r,t_{1}-t_{3},t_{1})||\leq \frac{1}{\hslash }|t_{1}-t_{3}|\times
\{||V_{1}^{ho}(x,\varepsilon )\Psi _{0}(x,r,t_{0}+t_{1}/2)||
\end{equation*}%
\begin{equation}
+4\hslash |\Omega _{0}|\times ||\Theta (x-x_{L},\varepsilon )\Psi
_{0}(x,r,t_{0}+t_{1}/2)||\}  \tag{4.39b}
\end{equation}%
where the trigonometric inequality $|\sin \theta |\leq |\theta |$ and the
inequality $||H_{I}(x,\alpha ,\gamma )||$ $\leq 4\hslash |\Omega _{0}|$ are
already used. Consider first that the product state $\Psi
_{0}(x,r,t_{0}+t_{1}/2)$ is a single $GWP$ product state. Then the last norm
on the $RH$ side of (4.39b) is clearly one of the second type of basic norms 
$\{NBAS2\}.$ The basic norm $NBAS2$ is defined in the previous section 3.
Its upper bound may be determined from (3.54). It already proves in the
section 3 that the basic norm decays exponentially with the square
deviation-to-spread ratio of the relevant $GWP$ state. Thus, the last norm
on the $RH$ side of (4.39b) decays exponentially with the square
deviation-to-spread ratio of the $GWP$ state $\Psi _{0}(x,r,t_{0}+t_{1}/2).$
The first norm on the $RH$ side of (4.39b) also can be reduced to the basic
norms $\{NBAS2\}$, meaning that its upper bound may be further expressed as
a linear combination of the basic norms $(i.e.,$ $\{NBAS2\})$. Hereafter
such a reduction for the upper bound of a norm will be often used. By using
the continuous perturbation term $V_{1}^{ho}(x,\varepsilon )$ of (2.10) it
can turn out that this norm is bounded by%
\begin{equation*}
||V_{1}^{ho}(x,\varepsilon )\Psi _{0}(x,r,t_{0}+t_{1}/2)||\leq (\frac{1}{2}%
m\omega ^{2}x_{L}^{2})||\Theta (x-x_{L},\varepsilon )\Psi
_{0}(x,r,t_{0}+t_{1}/2)||
\end{equation*}%
\begin{equation}
+(\frac{1}{2}m\omega ^{2})||x^{2}\Theta (x-x_{L},\varepsilon )\Psi
_{0}(x,r,t_{0}+t_{1}/2)||.  \tag{4.40}
\end{equation}%
Indeed, the upper bound of the norm is a linear combination of the two basic
norms $\{NBAS2\}$. Thus, it decays exponentially with the square
deviation-to-spread ratio of the $GWP$ state $\Psi _{0}(x,r,t_{0}+t_{1}/2).$
Now each one of the two norms on the $RH$ side of (4.39b) has an upper bound
that consists of the basic norms $\{NBAS2\}.$ Thus, the inequality (4.39b)
shows that the last norm on the $RH$ side of (4.38) has an upper bound
consisting of the basic norms $\{NBAS2\}.$ Now consider that the product
state $\Psi _{0}(x,r,t_{0}+t_{1}/2)$ is a superposition of the $GWP$ product
states. Because there is the relation $||\Psi _{A}+\Psi _{B}||\leq ||\Psi
_{A}||+||\Psi _{B}||$ for a pair of states $\Psi _{A}$ and $\Psi _{B},$ by
replacing the single $GWP$ product state $\Psi _{0}(x,r,t_{0}+t_{1}/2)$ with
the $GWP$ superposition state $\Psi _{0}(x,r,t_{0}+t_{1}/2)$ in the above
theoretical analysis, one can prove that the upper bound of the last norm on
the $RH$ side of (4.38) still consists of the basic norms $\{NBAS2\}.$ Of
course, in the present case the number of the basic norms $\{NBAS2\}$ is
more than that one in the previous case.

The first norm on the $RH$ side of (4.38) may be calculated through the
product state:%
\begin{equation*}
p\chi
(x,r,t_{1}-t_{3},t_{1})=P_{111}^{V}(x,r,t_{1},t_{3})+P_{112}^{V}(x,r,t_{1},t_{3})
\end{equation*}%
\begin{equation}
+P_{113}^{V}(x,r,t_{1},t_{3}).  \tag{4.41}
\end{equation}%
Below calculate the three product states $\{P_{11j}^{V}(x,r,t_{1},t_{3})\}$
with $j=1$, $2$, and $3$. Notice that the product state $\Psi
_{0}(x,r,t_{0}+t_{1}/2)$ may be generally written as%
\begin{equation}
\Psi _{0}(x,r,t_{0}+t_{1}/2)=\Psi _{0}^{g}(x,t_{0}+t_{1}/2)|g_{0}\rangle
+\Psi _{0}^{e}(x,t_{0}+t_{1}/2)|e\rangle .  \tag{4.42}
\end{equation}%
Here $\{\Psi _{0}^{a}(x,t_{0}+t_{1}/2)\}$ with $a=g$ and $e$ are
non-normalized $GWP$ states. They satisfy the normalization condition of the
product state $\Psi _{0}(x,r,t_{0}+t_{1}/2):$%
\begin{equation*}
||\Psi _{0}^{g}(x,t_{0}+t_{1}/2)||^{2}+||\Psi
_{0}^{e}(x,t_{0}+t_{1}/2)||^{2}=1.
\end{equation*}%
First of all, the product state $\exp [-iV_{1}(x,\alpha ,\gamma ,\varepsilon
)t_{13}/\hslash ]\Psi _{0}(x,r,t_{0}+t_{1}/2)$ $(t_{13}=t_{1}-t_{3})$ is
calculated exactly with the help of the unitary transformations (4.18) or
(4.19). It is given by%
\begin{equation*}
\exp [-\frac{i}{\hslash }V_{1}(x,\alpha ,\gamma ,\varepsilon )t_{13}]\Psi
_{0}(x,r,t_{0}+t_{1}/2)=\exp [-\frac{i}{\hslash }t_{13}V_{1}^{ho}(x,%
\varepsilon )]
\end{equation*}%
\begin{equation*}
\times \{\cos [\frac{1}{2\hslash }t_{13}\Theta (x-x_{L},\varepsilon )\Omega
(x)]\Psi _{0}^{g}(x,t_{0}+t_{1}/2)|g_{0}\rangle
\end{equation*}%
\begin{equation*}
+i\sin [\frac{1}{2\hslash }t_{13}\Theta (x-x_{L},\varepsilon )\Omega
(x)]\exp [i\varphi (x,\gamma )]\Psi _{0}^{g}(x,t_{0}+t_{1}/2)|e\rangle
\end{equation*}%
\begin{equation*}
+\cos [\frac{1}{2\hslash }t_{13}\Theta (x-x_{L},\varepsilon )\Omega (x)]\Psi
_{0}^{e}(x,t_{0}+t_{1}/2)|e\rangle
\end{equation*}%
\begin{equation}
+i\sin [\frac{1}{2\hslash }t_{13}\Theta (x-x_{L},\varepsilon )\Omega
(x)]\exp [-i\varphi (x,\gamma )]\Psi _{0}^{e}(x,t_{0}+t_{1}/2)|g_{0}\rangle
\}.  \tag{4.43}
\end{equation}%
Then by substituting (4.42) and (4.43) into (4.39a) one may obtain the
product state $\chi (x,r,t_{1}-t_{3},t_{1}).$ The state $\chi
(x,r,t_{1}-t_{3},t_{1})$ is a superposition state. It contains the phase
factor $\exp [-iV_{1}^{ho}(x,\varepsilon )t_{13}/\hslash ],$ the
trigonometric functions $\{\cos \theta (x),$ $i\sin \theta (x)\exp [\pm
i\varphi (x,\gamma )]\}$ with $\theta (x)=\frac{1}{2\hslash }t_{13}\Theta
(x-x_{L},\varepsilon )\Omega (x),$ and the product states $\{\Psi
_{0}^{a}(x,t_{0}+t_{1}/2)|\psi (r)\rangle \}$ with $a=g$, $e$ and $|\psi
(r)\rangle =|g_{0}\rangle ,$ $|e\rangle .$ As shown in (4.41), the three
product states $\{P_{11j}^{V}(x,r,t_{1},t_{3})\}$ with $j=1,2,3$ are
generated when the momentum operator $p=-i\hslash \partial /\partial x$ is
applied to the product state $\chi (x,r,t_{1}-t_{3},t_{1})$. The first
product state is $P_{111}^{V}(x,r,t_{1},t_{3})$. It is generated by applying
the differential operator $(\partial /\partial x)$ only to the phase factor $%
\exp [-iV_{1}^{ho}(x,\varepsilon )t_{13}/\hslash ]$ in the state $\chi
(x,r,t_{1}-t_{3},t_{1})$. The second is $P_{112}^{V}(x,r,t_{1},t_{3}).$ It
is generated by applying the differential operator $(\partial /\partial x)$
only to the trigonometric functions $\{\cos \theta (x),$ $i\sin \theta
(x)\exp [\pm i\varphi (x,\gamma )]\}$ in the state $\chi
(x,r,t_{1}-t_{3},t_{1})$. The third is $P_{113}^{V}(x,r,t_{1},t_{3}).$ It is
generated by applying the differential operator $(\partial /\partial x)$
only to every motional state $\Psi _{0}^{a}(x,t_{0}+t_{1}/2)$ in the state $%
\chi (x,r,t_{1}-t_{3},t_{1})$. Once the three product states $%
\{P_{11j}^{V}(x,r,t_{1},t_{3})\}$ are obtained, it is easy to calculate the
upper bound of the first norm on the $RH$ side of (4.38) through the
inequality:%
\begin{equation*}
||p\chi (x,r,t_{1}-t_{3},t_{1})||\leq
||P_{111}^{V}(x,r,t_{1},t_{3})||+||P_{112}^{V}(x,r,t_{1},t_{3})||
\end{equation*}%
\begin{equation}
+||P_{113}^{V}(x,r,t_{1},t_{3})||.  \tag{4.44}
\end{equation}%
A strict calculation for the norms $\{||P_{11j}^{V}(x,r,t_{1},t_{3})||\}$
usually needs to use the trigonometric inequalities $|\sin \theta |\leq
|\theta |$ or $\sin ^{2}\theta \leq \theta ^{2},$ $|\sin \theta |\leq 1,$ $%
|\cos \theta |\leq 1,$ and the relations $0\leq \Theta (x-x_{L},\varepsilon
)\leq 1$ and $\Theta (x-x_{L},\varepsilon )^{2}\leq \Theta
(x-x_{L},\varepsilon ).$ Now it can turn out that the norm $%
||P_{111}^{V}(x,r,t_{1},t_{3})||$ is bounded by%
\begin{equation}
||P_{111}^{V}(x,r,t_{1},t_{3})||\leq 2|t_{1}-t_{3}|\times \sum_{a=g,e}||[%
\frac{\partial }{\partial x}V_{1}^{ho}(x,\varepsilon )]\Psi
_{0}^{a}(x,t_{0}+t_{1}/2)||.  \tag{4.45}
\end{equation}%
This upper bound may be further reduced by using the perturbation term $%
V_{1}^{ho}(x,\varepsilon )$ of (2.10). Here it turns out that the norm $%
||x^{l}[\frac{\partial }{\partial x}V_{1}^{ho}(x,\varepsilon )]\Psi
_{0}^{a}(x,t_{0}+t_{1}/2)||$ with $l=0$, $1$, $2$, $...$, satisfies%
\begin{equation*}
||x^{l}[\frac{\partial }{\partial x}V_{1}^{ho}(x,\varepsilon )]\Psi
_{0}^{a}(x,t_{0}+t_{1}/2)||
\end{equation*}%
\begin{equation*}
\leq (\frac{1}{2}m\omega ^{2}x_{L}^{2})||x^{l}\delta (x-x_{L},\varepsilon
)\Psi _{0}^{a}(x,t_{0}+t_{1}/2)||
\end{equation*}%
\begin{equation*}
+(\frac{1}{2}m\omega ^{2})||x^{l+2}\delta (x-x_{L},\varepsilon )\Psi
_{0}^{a}(x,t_{0}+t_{1}/2)||
\end{equation*}%
\begin{equation*}
+(\frac{1}{2}m\omega ^{2}x_{L}^{2})||x^{l}\delta (x-x_{L}-L,\varepsilon
)\Psi _{0}^{a}(x,t_{0}+t_{1}/2)||
\end{equation*}%
\begin{equation}
+(m\omega ^{2})||x^{l+1}\Theta (x-x_{L},\varepsilon )\Psi
_{0}^{a}(x,t_{0}+t_{1}/2)||.  \tag{4.46}
\end{equation}%
Now by setting the index $l=0$ one can determine the upper bound of the norm 
$||[\frac{\partial }{\partial x}V_{1}^{ho}(x,\varepsilon )]\Psi
_{0}^{a}(x,t_{0}+t_{1}/2)||$ in (4.45) from the inequality (4.46). It is
clear that the first three norms on the $RH$ side of (4.46) are the first
type of basic norms $\{NBAS1\}$, while the last one belongs to the second
type of basic norms $\{NBAS2\}$. Therefore, the upper bound of the norm on
the RH side of (4.45) consists of three basic norms $\{NBAS1\}$ and one
basic norm $NBAS2$. Both the basic norms $NBAS1$ and $NBAS2$ are defined in
the previous section 3. Their upper bounds are determined from (3.37) and
(3.54), respectively. It also is shown in the section 3 that both the basic
norms decay exponentially with the square deviation-to-spread ratios of the
relevant $GWP$ states. Then the inequality (4.45) shows that the upper bound
of the norm $||P_{111}^{V}(x,r,t_{1},t_{3})||$ consists of the basic norms $%
\{NBAS1\}$ and $\{NBAS2\}$, indicating that it decays exponentially with the
square deviation-to-spread ratios of the $GWP$ states $\Psi
_{0}^{a}(x,t_{0}+t_{1}/2)$ with $a=g$ and $e$. A strict calculation for the
upper bounds of the error states $P_{112}^{V}(x,r,t_{1},t_{3})$ and $%
P_{113}^{V}(x,r,t_{1},t_{3})$ also needs to use the inequalities $|\Omega
(x)|\leq |4\hslash \Omega _{0}|,$ $|\frac{\partial }{\partial x}\Omega
(x)|\leq 2|(\hslash \Delta k)\Omega _{0}|,$ and the relation $|\frac{%
\partial }{\partial x}\varphi (x,\gamma )|=\frac{1}{2}|k_{0}+k_{1}|$ besides
those inequalities and relations used before. It turns out that the error
state $P_{112}^{V}(x,r,t_{1},t_{3})$ is bounded by%
\begin{equation*}
||P_{112}^{V}(x,r,t_{1},t_{3})||\leq \sum_{a=g,e}\{4\hslash |\Omega
_{0}(t_{1}-t_{3})|\times ||\delta (x-x_{L},\varepsilon )\Psi
_{0}^{a}(x,t_{0}+t_{1}/2)||
\end{equation*}%
\begin{equation}
+\hslash |\Omega _{0}(t_{1}-t_{3})|(2|\Delta k|+|k_{0}+k_{1}|)||\Theta
(x-x_{L},\varepsilon )\Psi _{0}^{a}(x,t_{0}+t_{1}/2)||\}.  \tag{4.47}
\end{equation}%
The first norm on the $RH$ side of (4.47) for the index $a=g$ or $e$ is a
basic norm $NBAS1$, while the second is a basic norm $NBAS2$. This shows
that the upper bound of the error state $P_{112}^{V}(x,r,t_{1},t_{3})$
consists of the basic norms $\{NBAS1\}$ and $\{NBAS2\}$, indicating that the
error state $P_{112}^{V}(x,r,t_{1},t_{3})$ decays exponentially with the
square deviation-to-spread ratios of the $GWP$ states $\Psi
_{0}^{a}(x,t_{0}+t_{1}/2)$ with $a=g$ and $e$. Now for the error state $%
P_{113}^{V}(x,r,t_{1},t_{3})$ it can prove that%
\begin{equation*}
||P_{113}^{V}(x,r,t_{1},t_{3})||\leq \sum_{a=g,e}\{\frac{1}{2\hslash }%
(t_{1}-t_{3})^{2}||V_{1}^{ho}(x,\varepsilon )^{2}\frac{\partial }{\partial x}%
\Psi _{0}^{a}(x,t_{0}+t_{1}/2)||
\end{equation*}%
\begin{equation*}
+\{2|\Omega _{0}|(t_{1}-t_{3})^{2}+|t_{1}-t_{3}|\}||V_{1}^{ho}(x,\varepsilon
)\frac{\partial }{\partial x}\Psi _{0}^{a}(x,t_{0}+t_{1}/2)||
\end{equation*}%
\begin{equation}
+2\hslash \{[\Omega _{0}(t_{1}-t_{3})]^{2}+|\Omega
_{0}(t_{1}-t_{3})|\}||\Theta (x-x_{L},\varepsilon )\frac{\partial }{\partial
x}\Psi _{0}^{a}(x,t_{0}+t_{1}/2)||\}.  \tag{4.48}
\end{equation}%
This upper bound may be further reduced to a linear sum of the basic norms $%
\{NBAS2\}$. Suppose that the $GWP$ state $\Psi _{0}^{a}(x,t_{0}+t_{1}/2)$
has the characteristic parameters $\{x_{c}^{a}(t_{0}+t_{1}/2),$ $%
p_{c}^{a}(t_{0}+t_{1}/2),$ $W_{c}^{a}(t_{0}+t_{1}/2),$ $\varepsilon
_{c}^{a}(t_{0}+t_{1}/2)\}$. Then it is easy to calculate the first-order
derivative of the wave function $\Psi _{0}^{a}(x,t_{0}+t_{1}/2):$ 
\begin{equation}
\frac{\partial }{\partial x}\Psi _{0}^{a}(x,t_{0}+t_{1}/2)=\{-\frac{1}{2}%
\frac{x-x_{c}^{a}(t_{0}+t_{1}/2)}{W_{c}^{a}(t_{0}+t_{1}/2)}%
+ip_{c}^{a}(t_{0}+t_{1}/2)/\hslash \}\Psi _{0}^{a}(x,t_{0}+t_{1}/2). 
\tag{4.49}
\end{equation}%
By using the first-order derivative of (4.49) and the perturbation term $%
V_{1}^{ho}(x,\varepsilon )$ of (2.10) one can prove that 
\begin{equation}
||V_{1}^{ho}(x,\varepsilon )^{2}\frac{\partial }{\partial x}\Psi
_{0}^{a}(x,t_{0}+t_{1}/2)||\leq \sum_{l=0}^{5}F_{l}^{a}||x^{l}\Theta
(x-x_{L},\varepsilon )\Psi _{0}^{a}(x,t_{0}+t_{1}/2)||  \tag{4.50}
\end{equation}%
where the positive parameters $\{F_{l}^{a}\}$ are given by 
\begin{equation*}
F_{0}^{a}=\frac{1}{4}(m\omega ^{2}x_{L}^{2})^{2}\{\frac{1}{2}|\frac{%
x_{c}^{a}(t_{0}+t_{1}/2)}{W_{c}^{a}(t_{0}+t_{1}/2)}%
|+|p_{c}^{a}(t_{0}+t_{1}/2)/\hslash |\},
\end{equation*}%
\begin{equation*}
F_{1}^{a}=\frac{1}{8}\frac{(m\omega ^{2}x_{L}^{2})^{2}}{%
|W_{c}^{a}(t_{0}+t_{1}/2)|},
\end{equation*}%
and the other parameters can be obtained from the two parameters $F_{0}^{a}$
and $F_{1}^{a}$ by $F_{0}^{a}/F_{2}^{a}=x_{L}^{2}/2$ and $%
F_{2}^{a}/F_{4}^{a}=2x_{L}^{2},$ $F_{1}^{a}/F_{3}^{a}=x_{L}^{2}/2$ and $%
F_{3}^{a}/F_{5}^{a}=2x_{L}^{2}.$ Similarly, one can prove that 
\begin{equation}
||V_{1}^{ho}(x,\varepsilon )\frac{\partial }{\partial x}\Psi
_{0}^{a}(x,t_{0}+t_{1}/2)||\leq \sum_{l=0}^{3}G_{l}^{a}||x^{l}\Theta
(x-x_{L},\varepsilon )\Psi _{0}^{a}(x,t_{0}+t_{1}/2)||,  \tag{4.51}
\end{equation}%
where $G_{0}^{a}=x_{L}^{2}G_{2}^{a},$ $G_{1}^{a}=x_{L}^{2}G_{3}^{a},$ 
\begin{equation*}
G_{0}^{a}=\frac{1}{2}(m\omega ^{2}x_{L}^{2})\{\frac{1}{2}|\frac{%
x_{c}^{a}(t_{0}+t_{1}/2)}{W_{c}^{a}(t_{0}+t_{1}/2)}%
|+|p_{c}^{a}(t_{0}+t_{1}/2)/\hslash |\},
\end{equation*}%
\begin{equation*}
G_{1}^{a}=\frac{1}{4}\frac{m\omega ^{2}x_{L}^{2}}{|W_{c}^{a}(t_{0}+t_{1}/2)|}%
;
\end{equation*}%
and one also can prove that 
\begin{equation*}
||\Theta (x-x_{L},\varepsilon )\frac{\partial }{\partial x}\Psi
_{0}^{a}(x,t_{0}+t_{1}/2)||
\end{equation*}%
\begin{equation*}
\leq \frac{1}{2}\frac{1}{|W_{c}^{a}(t_{0}+t_{1}/2)|}||x\Theta
(x-x_{L},\varepsilon )\Psi _{0}^{a}(x,t_{0}+t_{1}/2)||
\end{equation*}%
\begin{equation}
+\{\frac{1}{2}|\frac{x_{c}^{a}(t_{0}+t_{1}/2)}{W_{c}^{a}(t_{0}+t_{1}/2)}%
|+|p_{c}^{a}(t_{0}+t_{1}/2)/\hslash |\}||\Theta (x-x_{L},\varepsilon )\Psi
_{0}^{a}(x,t_{0}+t_{1}/2)||.  \tag{4.52}
\end{equation}%
These three inequalities (4.50), (4.51), and (4.52) show that the upper
bounds of the three norms on the $RH$ side of (4.48) for the index $a=g$ or $%
e$ consist of six, four, and two basic norms $\{NBAS2\}$, respectively. Then
the inequality (4.48) shows that the upper bound of the error state $%
P_{113}^{V}(x,r,t_{1},t_{3})$ consists of twelve different basic norms $%
\{NBAS2\}$. This indicates that the error state decays exponentially with
the square deviation-to-spread ratios of the $GWP$ states $\{\Psi
_{0}^{a}(x,t_{0}+t_{1}/2)\}$. Now all these three norms $%
\{||P_{11j}^{V}(x,r,t_{1},t_{3})||\}$ are proven to decay exponentially with
the square deviation-to-spread ratios of the $GWP$ states $\{\Psi
_{0}^{a}(x,t_{0}+t_{1}/2)\}$. Then the inequality (4.44) shows that the
upper bound of the error state (4.41) also decays exponentially with these
square deviation-to-spread ratios. The above theoretical calculation shows
that the upper bounds for the two norms on the $RH$ side of (4.38) consist
of the basic norms $\{NBAS1\}$ and/or $\{NBAS2\}$. Then the inequality
(4.38) indicates that the upper bound of the error state $%
M_{11}^{V}(x,r,t_{1},t_{3})$ decays exponentially with the square
deviation-to-spread ratios of the $GWP$ states $\{\Psi
_{0}^{a}(x,t_{0}+t_{1}/2)\}$.

Below it proves that the upper bound of the error state $%
M_{12}^{V}(x,r,t_{1},t_{3})$ of (4.36c) decays exponentially with the square
deviation-to-spread ratios of the relevant $GWP$ states. By using the basic
commutation relations (4.37) one can calculate all the three commutators $%
[H_{I}(x,\alpha ,\gamma ),[H_{0}^{ho},V_{1}(x,\alpha ,\gamma ,\varepsilon
)]],$ $etc.,$ on the $RH$ side of (4.36c). Then according to the calculated
results one may divide the error state $M_{12}^{V}(x,r,t_{1},t_{3})$ into
the two terms. One of which contains the momentum operator $p$ and another
does not. Thus, the error state $M_{12}^{V}(x,r,t_{1},t_{3})$ may be written
as%
\begin{equation}
M_{12}^{V}(x,r,t_{1},t_{3})=M_{121}^{V}(x,r,t_{1},t_{3})+M_{122}^{V}(x,r,t_{1},t_{3})
\tag{4.53}
\end{equation}%
where the error state $M_{121}^{V}(x,r,t_{1},t_{3})$ does not contain the
momentum operator $p$ and $M_{122}^{V}(x,r,t_{1},t_{3})$ does. After a
complex calculation one can prove that the upper bound of the error state $%
M_{121}^{V}(x,r,t_{1},t_{3})$ is determined from%
\begin{equation*}
2m||M_{121}^{V}(x,r,t_{1},t_{3})||\leq 2\hslash ^{2}||[\frac{\partial }{%
\partial x}V_{1}^{ho}(x,\varepsilon )]^{2}\Psi _{0}(x,r,t_{0}+t_{1}/2)||
\end{equation*}%
\begin{equation*}
+\sum_{l=0}^{1}A_{l}^{121}||x^{l}\Theta (x-x_{L},\varepsilon )\Psi
_{0}(x,r,t_{0}+t_{1}/2)||
\end{equation*}%
\begin{equation*}
+\sum_{k=1}^{2}\sum_{l=0}^{2}B_{kl}^{121}||x^{l}\delta (x-x_{L},\varepsilon
)^{k}\Psi _{0}(x,r,t_{0}+t_{1}/2)||
\end{equation*}%
\begin{equation*}
+4\hslash ^{2}(m\omega ^{2}x_{L}^{2})||\frac{\partial }{\partial x}%
H_{I}(x,\alpha ,\gamma )||\times ||\delta (x-x_{L}-L,\varepsilon )\Psi
_{0}(x,r,t_{0}+t_{1}/2)||
\end{equation*}%
\begin{equation}
+2\hslash ^{2}(m\omega ^{2}x_{L}^{2})||H_{I}(x,\alpha ,\gamma )||\times
||\delta (x-x_{L}-L,\varepsilon )\delta (x-x_{L},\varepsilon )\Psi
_{0}(x,r,t_{0}+t_{1}/2)||,  \tag{4.54}
\end{equation}%
where use has been made of $0\leq \Theta (x-x_{L},\varepsilon )\leq 1$ and
the inequality (4.46) as well as the inequality:%
\begin{equation*}
||\delta (x-x_{L},\varepsilon )[\frac{\partial }{\partial x}%
V_{1}^{ho}(x,\varepsilon )]\Psi _{0}(x,r,t_{0}+t_{1}/2)||
\end{equation*}%
\begin{equation*}
\leq m\omega ^{2}||x\delta (x-x_{L},\varepsilon )\Psi
_{0}(x,r,t_{0}+t_{1}/2)||
\end{equation*}%
\begin{equation*}
+\frac{1}{2}m\omega ^{2}x_{L}^{2}||\delta (x-x_{L},\varepsilon )^{2}\Psi
_{0}(x,r,t_{0}+t_{1}/2)||
\end{equation*}%
\begin{equation*}
+\frac{1}{2}m\omega ^{2}||x^{2}\delta (x-x_{L},\varepsilon )^{2}\Psi
_{0}(x,r,t_{0}+t_{1}/2)||
\end{equation*}%
\begin{equation}
+\frac{1}{2}m\omega ^{2}x_{L}^{2}||\delta (x-x_{L}-L,\varepsilon )\delta
(x-x_{L},\varepsilon )\Psi _{0}(x,r,t_{0}+t_{1}/2)||.  \tag{4.55}
\end{equation}%
The inequality (4.55) also is useful later. The non-negative parameters $%
\{A_{l}^{121}\}$ and $\{B_{kl}^{121}\}$ on the $RH$ side of (4.54) are
dependent on the interaction $H_{I}(x,\alpha ,\gamma )$ of (4.16) and its
first- and second-order derivatives of coordinate $x$. All these
non-negative parameters can be calculated exactly by using the interaction $%
H_{I}(x,\alpha ,\gamma )$ and other relevant parameters. They all are
bounded and controllable because the interaction $H_{I}(x,\alpha ,\gamma )$
and its coordinate derivatives are bounded. In particular, $B_{21}^{121}=0$
in (4.54). Now investigate the upper bound of the norm $%
2m||M_{121}^{V}(x,r,t_{1},t_{3})||$ that is determined from the $RH$ side of
(4.54). There are five terms on the $RH$ side of (4.54). The first term is a
norm whose upper bound can be determined from (3.33c). It already proves in
the section 3 that the upper bound of the norm consists of one basic norm $%
NBAS2$ and nine basic norms $\{NBAS1\}$. It is clear that the second term on
the $RH$ side of (4.54) consists of two basic norms $\{NBAS2\}$. Since the
parameter $B_{21}^{121}=0,$ the third term consists of only five norms $%
\{||x^{l}\delta (x-x_{L},\varepsilon )^{k}\Psi _{0}(x,r,t_{0}+t_{1}/2)||\}$.
According to (3.36) the function $\delta (x-x_{L},\varepsilon )^{2}$ can be
reduced to a usual smooth $\delta -$function $\delta (x-x_{L},\varepsilon )$
up to a constant. This shows that the third term really consists of five
basic norms $\{NBAS1\}$. The fourth term is clearly a basic norm $NBAS1$.
According to (3.35) the product of a pair of smooth $\delta -$functions with
different COM positions can be reduced to a smooth $\delta -$function up to
an extra Gaussian factor. Then this shows that the fifth term is a basic
norm $NBAS1$. Moreover, the fifth term is much smaller than the other terms
on the $RH$ side of (4.54) and can be neglected when $L>>\varepsilon .$ The
above analysis shows that all these five terms on the $RH$ side of (4.54)
consist of the two types of basic norms $\{NBAS1\}$ and $\{NBAS2\}$.
Therefore, the upper bound of the error state $M_{121}^{V}(x,r,t_{1},t_{3})$
is composed of the two types of basic norms $\{NBAS1\}$ and $\{NBAS2\}$,
indicating that the error state $M_{121}^{V}(x,r,t_{1},t_{3})$ decays
exponentially with the square deviation-to-spread ratios of the $GWP$ states 
$\{\Psi _{0}^{a}(x,t_{0}+t_{1}/2)\}.$ Here the product state $\Psi
_{0}(x,r,t_{0}+t_{1}/2)$ in (4.54) may take the superposition state of
(4.42). Then in this case each one of the norms on the $RH$ side of (4.54)
may be further reduced to a linear sum of two norms and the basic norms $%
\{NBAS1\}$ and $\{NBAS2\}$ should be doubly counted in the upper bound of
the norm $2m||M_{121}^{V}(x,r,t_{1},t_{3})||$.

Now calculate the upper bound of the error state $%
M_{122}^{V}(x,r,t_{1},t_{3})$ in (4.53). The error state $%
M_{122}^{V}(x,r,t_{1},t_{3})$ contains the momentum operator $p.$ For
convenience here denote the product state $\Psi
_{0}(x,r,t_{0}+t_{1}/2+t_{13})$ with $t_{13}=t_{1}-t_{3}$ as%
\begin{equation*}
\Psi _{0}(x,r,t_{0}+t_{1}/2+t_{13})=\exp [-\frac{i}{\hslash }H_{I}(x,\alpha
,\gamma )t_{13}]
\end{equation*}%
\begin{equation}
\times \exp [-\frac{i}{\hslash }V_{1}(x,\alpha ,\gamma ,\varepsilon
)t_{13}]\Psi _{0}(x,r,t_{0}+t_{1}/2).  \tag{4.56}
\end{equation}%
Then it can turn out that the error state $2mM_{122}^{V}(x,r,t_{1},t_{3})$
may be expressed as%
\begin{equation*}
2mM_{122}^{V}(x,r,t_{1},t_{3})=2i\hslash \lbrack H_{I}(x,\alpha ,\gamma ),%
\frac{\partial }{\partial x}H_{I}(x,\alpha ,\gamma )]
\end{equation*}%
\begin{equation}
\times \{2\Theta (x-x_{L},\varepsilon )-\Theta (x-x_{L},\varepsilon
)^{2}\}p\Psi _{0}(x,r,t_{0}+t_{1}/2+t_{13}).  \tag{4.57}
\end{equation}%
At first one needs to calculate analytically the product state (4.56)$.$
This can be done in a similar way that the product state (4.43) is
calculated, where the unitary transformations (4.18) or (4.19) has been
used. The product state (4.56) is explicitly given by%
\begin{equation*}
\Psi _{0}(x,r,t_{0}+t_{1}/2+t_{13})=\exp [-\frac{i}{\hslash }%
t_{13}V_{1}^{ho}(x,\varepsilon )]
\end{equation*}%
\begin{equation*}
\times \{\cos \{\frac{1}{2\hslash }t_{13}[\Theta (x-x_{L},\varepsilon
)-1]\Omega (x)\}\Psi _{0}^{g}(x,t_{0}+t_{1}/2)|g_{0}\rangle
\end{equation*}%
\begin{equation*}
+i\sin \{\frac{1}{2\hslash }t_{13}[\Theta (x-x_{L},\varepsilon )-1]\Omega
(x)\}\exp [i\varphi (x,\gamma )]\Psi _{0}^{g}(x,t_{0}+t_{1}/2)|e\rangle
\end{equation*}%
\begin{equation*}
+\cos \{\frac{1}{2\hslash }t_{13}[\Theta (x-x_{L},\varepsilon )-1]\Omega
(x)\}\Psi _{0}^{e}(x,t_{0}+t_{1}/2)|e\rangle
\end{equation*}%
\begin{equation}
+i\sin \{\frac{1}{2\hslash }t_{13}[\Theta (x-x_{L},\varepsilon )-1]\Omega
(x)\}\exp [-i\varphi (x,\gamma )]\Psi _{0}^{e}(x,t_{0}+t_{1}/2)|g_{0}\rangle
\}.  \tag{4.58}
\end{equation}%
Then by applying the momentum operator $p=-i\hslash \partial /\partial x$ to
the product state (4.58) one obtains%
\begin{equation*}
p\Psi _{0}(x,r,t_{0}+t_{1}/2+t_{13})=P_{122}^{V1}(x,r,t_{1},t_{3})
\end{equation*}%
\begin{equation}
+P_{122}^{V2}(x,r,t_{1},t_{3})+P_{122}^{V3}(x,r,t_{1},t_{3}).  \tag{4.59}
\end{equation}%
Here the error state $P_{122}^{V1}(x,r,t_{1},t_{3})$ is generated by
applying the coordinate derivative operation $(\partial /\partial x)$ only
to the phase factor $\exp [-it_{13}V_{1}^{ho}(x,\varepsilon )/\hslash ]$ in
the state (4.58), the state $P_{122}^{V2}(x,r,t_{1},t_{3})$ is obtained by
applying the coordinate derivative operation only to all the trigonometric
functions and the phase factors $\exp [\pm i\varphi (x,\gamma )]$ in the
state (4.58), and the state $P_{122}^{V3}(x,r,t_{1},t_{3})$ is generated by
applying the coordinate derivative operation only to all the motional states 
$\Psi _{0}^{a}(x,t_{0}+t_{1}/2)$ with $a=g$ and $e$ in the state (4.58).
Then it follows from (4.57) and (4.59) that the error state $%
2mM_{122}^{V}(x,r,t_{1},t_{3})$ is bounded by%
\begin{equation*}
2m||M_{122}^{V}(x,r,t_{1},t_{3})||\leq 6\hslash ||[H_{I}(x,\alpha ,\gamma ),%
\frac{\partial }{\partial x}H_{I}(x,\alpha ,\gamma )]||
\end{equation*}%
\begin{equation*}
\times \{||\Theta (x-x_{L},\varepsilon
)P_{122}^{V1}(x,r,t_{1},t_{3})||+||\Theta (x-x_{L},\varepsilon
)P_{122}^{V2}(x,r,t_{1},t_{3})||
\end{equation*}%
\begin{equation}
+||\Theta (x-x_{L},\varepsilon )P_{122}^{V3}(x,r,t_{1},t_{3})||\}. 
\tag{4.60}
\end{equation}%
Now it is easy to prove that the first norm on the $RH$ side of (4.60)
satisfies the inequality:%
\begin{equation*}
||\Theta (x-x_{L},\varepsilon )P_{122}^{V1}(x,r,t_{1},t_{3})||
\end{equation*}%
\begin{equation}
\leq 2|t_{1}-t_{3}|\sum_{a=g,e}||[\frac{\partial }{\partial x}%
V_{1}^{ho}(x,\varepsilon )]\Psi _{0}^{a}(x,t_{0}+t_{1}/2)||.  \tag{4.61}
\end{equation}%
It is known from (4.46) that the norm on the $RH$ side of (4.61) with $a=g$
or $e$ has an upper bound that is composed of three basic norms $\{NBAS1\}$
and one basic norm $NBAS2$. Then the inequality (4.61) shows that the upper
bound of the first norm on the $RH$ side of (4.60) consists of the basic
norms $\{NBAS1\}$ and $\{NBAS2\},$ indicating that the norm decays
exponentially with the square deviation-to-spread ratios of the $GWP$ states 
$\{\Psi _{0}^{a}(x,t_{0}+t_{1}/2)\}.$ Similarly, one can prove that the last
two norms on the $RH$ side of (4.60) satisfy respectively the inequalities:%
\begin{equation*}
||\Theta (x-x_{L},\varepsilon )P_{122}^{V2}(x,r,t_{1},t_{3})||
\end{equation*}%
\begin{equation*}
\leq |4\hslash \Omega _{0}(t_{1}-t_{3})|\sum_{a=g,e}||\delta
(x-x_{L},\varepsilon )\Psi _{0}^{a}(x,t_{0}+t_{1}/2)||
\end{equation*}%
\begin{equation*}
+(|4(\hslash \Delta k)\Omega _{0}(t_{1}-t_{3})|+\frac{1}{2}\hslash
|k_{0}+k_{1}|)
\end{equation*}%
\begin{equation}
\times \sum_{a=g,e}||\Theta (x-x_{L},\varepsilon )\Psi
_{0}^{a}(x,t_{0}+t_{1}/2)||  \tag{4.62}
\end{equation}%
and 
\begin{equation*}
||\Theta (x-x_{L},\varepsilon )P_{122}^{V3}(x,r,t_{1},t_{3})||
\end{equation*}%
\begin{equation}
\leq 2\hslash \sum_{a=g,e}||\Theta (x-x_{L},\varepsilon )\frac{\partial }{%
\partial x}\Psi _{0}^{a}(x,t_{0}+t_{1}/2)||.  \tag{4.63}
\end{equation}%
It is clear that the $RH$ side of (4.62) consists of two basic norms $%
\{NBAS1\}$ and two basic norms $\{NBAS2\}$, indicating that the upper bound
of the second norm on the $RH$ side of (4.60) consists of the two types of
basic norms $\{NBAS1\}$ and $\{NBAS2\}$. It is known from (4.52) that the
norm on the $RH$ side of (4.63) with $a=g$ or $e$ has an upper bound that
consists of the two basic norms $\{NBAS2\}$. Then the inequality (4.63)
indicates that the upper bound of the last norm on the $RH$ side of (4.60)$\ 
$also consists of the basic norms $\{NBAS2\}$. Then these three inequalities
(4.61), (4.62), and (4.63) show that the upper bounds of the three norms on
the $RH$ side of (4.60) consist of the two types of basic norms $\{NBAS1\}$
and $\{NBAS2\}$, respectively. This directly shows that the upper bound of
the norm $2m||M_{122}^{V}(x,r,t_{1},t_{3})||$ is also composed of the two
types of basic norms, indicating that the norm $%
||M_{122}^{V}(x,r,t_{1},t_{3})||$ decays exponentially with the square
deviation-to-spread ratios of the $GWP$ states $\{\Psi
_{0}^{a}(x,t_{0}+t_{1}/2)\}$.

It follows from (4.53) that the error state $M_{12}^{V}(x,r,t_{1},t_{3})$ is
bounded by%
\begin{equation*}
||M_{12}^{V}(x,r,t_{1},t_{3})||\leq
||M_{121}^{V}(x,r,t_{1},t_{3})||+||M_{122}^{V}(x,r,t_{1},t_{3})||.
\end{equation*}%
It is also known from (4.54) and (4.60) that both the upper bounds of the
error states $M_{121}^{V}(x,r,t_{1},t_{3})$ and $%
M_{122}^{V}(x,r,t_{1},t_{3}) $ are composed of the two types of basic norms $%
\{NBAS1\}$ and $\{NBAS2\}$. Then these results show that the error state $%
M_{12}^{V}(x,r,t_{1},t_{3})$ decays exponentially with the square
deviation-to-spread ratios of the $GWP$ states $\{\Psi
_{0}^{a}(x,t_{0}+t_{1}/2)\}$.

The above theoretical calculation in the subsection shows that the upper
bounds of both the error states $M_{11}^{V}(x,r,t_{1},t_{3})$ and $%
M_{12}^{V}(x,r,t_{1},t_{3})$ decay exponentially with the square
deviation-to-spread ratios of the $GWP$ states $\{\Psi _{0}^{a}(x,$ $%
t_{0}+t_{1}/2)\}$. It follows from (4.36a) that the error state $%
M_{1}^{V}(x,r,t_{1},t_{3})$ of (4.12a) is bounded by 
\begin{equation*}
||M_{1}^{V}(x,r,t_{1},t_{3})||\leq
||M_{11}^{V}(x,r,t_{1},t_{3})||+||M_{12}^{V}(x,r,t_{1},t_{3})||.
\end{equation*}%
Then this inequality indicates that the upper bound of the error state $%
M_{1}^{V}(x,r,$ $t_{1},t_{3})$ decays exponentially with the square
deviation-to-spread ratios of the $GWP$ states $\{\Psi
_{0}^{a}(x,t_{0}+t_{1}/2)\}$ with $a=g$ and $e$. This is the desired result!%
\newline
\newline
{\large 4.3 The upper bound of the norm }$||M_{2}^{V}(x,r,t_{1},t_{3})||$

Below it proves that the error state $M_{2}^{V}(x,r,t_{1},t_{3})$ of (4.12b)
decays exponentially with the square deviation-to-spread ratios. It is much
more complex to calculate strictly the upper bound of the error state $%
M_{2}^{V}(x,r,t_{1},t_{3})$ than that one of the error state $%
M_{1}^{V}(x,r,t_{1},t_{3})$ of (4.12a). The error state $%
M_{2}^{V}(x,r,t_{1},t_{3})$ may be rewritten as%
\begin{equation}
M_{2}^{V}(x,r,t_{1},t_{3})=M_{21}^{V}(x,r,t_{1},t_{3})+M_{22}^{V}(x,r,t_{1},t_{3})
\tag{4.64}
\end{equation}%
where both the error states $M_{21}^{V}(x,r,t_{1},t_{3})$ and $%
M_{22}^{V}(x,r,t_{1},t_{3})$ are given respectively by%
\begin{equation*}
M_{21}^{V}(x,r,t_{1},t_{3})=-[H_{0}^{ho},[H_{0}^{ho},H_{I}(x,\alpha ,\gamma
)]]\exp [-\frac{i}{2\hslash }H_{0}^{ho}t_{3}]
\end{equation*}%
\begin{equation}
\times \exp [-\frac{i}{\hslash }H_{I}(x,\alpha ,\gamma )t_{1}]\{1-\exp [-%
\frac{i}{\hslash }V_{1}(x,\alpha ,\gamma ,\varepsilon )t_{1}]\}\Psi
_{0}(x,r,t_{0}+t_{1}/2),  \tag{4.65a}
\end{equation}%
and%
\begin{equation*}
M_{22}^{V}(x,r,t_{1},t_{3})=[H_{0}^{ho},[H_{0}^{ho},V_{1}(x,\alpha ,\gamma
,\varepsilon )]]\exp [-\frac{i}{2\hslash }H_{0}^{ho}t_{3}]
\end{equation*}%
\begin{equation}
\times \exp [-\frac{i}{\hslash }H_{I}(x,\alpha ,\gamma )t_{1}]\exp [-\frac{i%
}{\hslash }V_{1}(x,\alpha ,\gamma ,\varepsilon )t_{1}]\Psi
_{0}(x,r,t_{0}+t_{1}/2).  \tag{4.65b}
\end{equation}%
Then its upper bound is determined from%
\begin{equation}
||M_{2}^{V}(x,r,t_{1},t_{3})||\leq
||M_{21}^{V}(x,r,t_{1},t_{3})||+||M_{22}^{V}(x,r,t_{1},t_{3})||.  \tag{4.65c}
\end{equation}%
It is seen below that it is far more complex to calculate strictly the upper
bound of the norm $||M_{22}^{V}(x,r,t_{1},t_{3})||$ than that one of the
norm $||M_{21}^{V}(x,r,t_{1},t_{3})||.$ Thus, the upper bound of the norm $%
||M_{21}^{V}(x,r,t_{1},t_{3})||$ is first calculated strictly in the
subsection 4.3.1. Then that one of the norm $||M_{22}^{V}(x,r,t_{1},t_{3})||$
is calculated strictly in the subsection 4.3.2 below. \newline
\newline
{\large 4.3.1 The upper bound of the norm }$||M_{21}^{V}(x,r,t_{1},t_{3})||$

By using the basic commutation relations (4.37) the commutator in (4.65a)
can be exactly calculated and it is given by%
\begin{equation*}
2m[H_{0}^{ho},[H_{0}^{ho},H_{I}(x,\alpha ,\gamma )]]=-\frac{2\hslash ^{2}}{m}%
[\frac{\partial ^{2}}{\partial x^{2}}H_{I}(x,\alpha ,\gamma )]p^{2}
\end{equation*}%
\begin{equation}
+i\frac{2\hslash ^{3}}{m}[\frac{\partial ^{3}}{\partial x^{3}}H_{I}(x,\alpha
,\gamma )]p+2\hslash ^{2}m\omega ^{2}[\frac{\partial }{\partial x}%
H_{I}(x,\alpha ,\gamma )]x+\frac{\hslash ^{4}}{2m}\frac{\partial ^{4}}{%
\partial x^{4}}H_{I}(x,\alpha ,\gamma ).  \tag{4.66}
\end{equation}%
Now define the error state $\Psi _{er}^{21}(x,r,t_{0}+t_{1}/2)$ by%
\begin{equation}
\Psi _{er}^{21}(x,r,t_{0}+t_{1}/2)=\exp [-\frac{i}{\hslash }H_{I}(x,\alpha
,\gamma )t_{1}]\chi (x,r,t_{1},t_{1}),  \tag{4.67}
\end{equation}%
where the product state $\chi (x,r,t_{1},t_{1})$ is defined by (4.39a). By
inserting (4.66) and (4.67) into (4.65a) one can find that the upper bound
of the error state $2mM_{21}^{V}(x,r,t_{1},t_{3})$ is determined from%
\begin{equation*}
||2mM_{21}^{V}(x,r,t_{1},t_{3})||\leq \frac{\hslash ^{4}}{2m}||\frac{%
\partial ^{4}}{\partial x^{4}}H_{I}(x,\alpha ,\gamma )||\times ||\Psi
_{er}^{21}(x,r,t_{0}+t_{1}/2)||
\end{equation*}%
\begin{equation*}
+2\hslash ^{2}m\omega ^{2}||\frac{\partial }{\partial x}H_{I}(x,\alpha
,\gamma )||\times ||x\exp [-\frac{i}{2\hslash }H_{0}^{ho}t_{3}]\Psi
_{er}^{21}(x,r,t_{0}+t_{1}/2)||
\end{equation*}%
\begin{equation*}
+\frac{2\hslash ^{3}}{m}||\frac{\partial ^{3}}{\partial x^{3}}H_{I}(x,\alpha
,\gamma )||\times ||p\exp [-\frac{i}{2\hslash }H_{0}^{ho}t_{3}]\Psi
_{er}^{21}(x,r,t_{0}+t_{1}/2)||
\end{equation*}%
\begin{equation}
+\frac{2\hslash ^{2}}{m}||\frac{\partial ^{2}}{\partial x^{2}}H_{I}(x,\alpha
,\gamma )||\times ||p^{2}\exp [-\frac{i}{2\hslash }H_{0}^{ho}t_{3}]\Psi
_{er}^{21}(x,r,t_{0}+t_{1}/2)||.  \tag{4.68}
\end{equation}%
This inequality may be further simplified by using the Heisenberg motion
equations $i\hslash \dot{p}(t)=[p(t),H_{0}^{ho}]$ and $i\hslash \dot{x}%
(t)=[x(t),H_{0}^{ho}]$ for a harmonic oscillator with the Hamiltonian $%
H_{0}^{ho}$, which have the solution:%
\begin{equation}
p(t)=p\cos (\omega t)-m\omega x\sin (\omega t),  \tag{4.69a}
\end{equation}%
\begin{equation}
x(t)=x\cos (\omega t)+\frac{p}{m\omega }\sin (\omega t),  \tag{4.69b}
\end{equation}%
where $\omega $ is the oscillatory frequency (or angular frequency) of the
harmonic oscillator. Now by substituting (4.69) into (4.68) and using the
basic commutation relation $[x,p]=i\hslash $ one obtains%
\begin{equation*}
2m||M_{21}^{V}(x,r,t_{1},t_{3})||\leq C_{p^{2}}||p^{2}\Psi
_{er}^{21}(x,r,t_{0}+t_{1}/2)||+C_{x^{2}}||x^{2}\Psi
_{er}^{21}(x,r,t_{0}+t_{1}/2)||
\end{equation*}%
\begin{equation*}
+C_{xp}||xp\Psi _{er}^{21}(x,r,t_{0}+t_{1}/2)||+C_{p}||p\Psi
_{er}^{21}(x,r,t_{0}+t_{1}/2)||
\end{equation*}%
\begin{equation}
+C_{x}||x\Psi _{er}^{21}(x,r,t_{0}+t_{1}/2)||+C_{0}||\Psi
_{er}^{21}(x,r,t_{0}+t_{1}/2)||.  \tag{4.70}
\end{equation}%
Here these six non-negative parameters $\{C_{\mu }\}$ ($\mu =p^{2},$ $x^{2},$
$xp,$ $p,$ $x,$ $0$) are dependent on the interaction $H_{I}(x,\alpha
,\gamma )$ of (4.16) and its coordinate derivatives. They can be calculated
exactly by using the interaction $H_{I}(x,\alpha ,\gamma )$ and other
relevant parameters. They all are bounded and controllable. Thus, the upper
bound of the error state $M_{21}^{V}(x,r,t_{1},t_{3})$ may be determined by
calculating the six norms on the $RH$ side of (4.70).

It is easy to calculate the three $p-$independent norms $||x^{j}\Psi
_{er}^{21}(x,r,t_{0}+t_{1}/2)||$ with $j=0,$ $1,$ and $2$ on the $RH$ side
of (4.70) as there is not the momentum operator inside these norms. Notice
that any unitary transformation does not affect the spectral norm [43], and
all these norms on the $RH$ side of (4.70) are the spectral norms. Since $%
[H_{I}(x,\alpha ,\gamma ),x]=0$, it follows from (4.67) that these three
norms may be reduced to the simple form%
\begin{equation}
||x^{j}\Psi _{er}^{21}(x,r,t_{0}+t_{1}/2)||=||x^{j}\chi (x,r,t_{1},t_{1})||.
\tag{4.71}
\end{equation}%
Now the upper bound of the norm on the $RH$ side of (4.71) may be easily
calculated in a similar way that the upper bound of the norm on the $LH$
side of (4.39b) is calculated. It turns out that the norm satisfies the
inequality:%
\begin{equation*}
||x^{j}\chi (x,r,t_{1},t_{1})||\leq \frac{1}{\hslash }t_{1}%
\{||x^{j}V_{1}^{ho}(x,\varepsilon )\Psi _{0}(x,r,t_{0}+t_{1}/2)||
\end{equation*}%
\begin{equation}
+4\hslash |\Omega _{0}|\times ||x^{j}\Theta (x-x_{L},\varepsilon )\Psi
_{0}(x,r,t_{0}+t_{1}/2)||\}.  \tag{4.72}
\end{equation}%
In fact, the inequality (4.39b) is a special case $(j=0)$ of the inequality
(4.72). Then by using the perturbation term $V_{1}^{ho}(x,\varepsilon )$ of
(2.10) it is easy to prove that 
\begin{equation*}
||x^{j}V_{1}^{ho}(x,\varepsilon )\Psi _{0}(x,r,t_{0}+t_{1}/2)||
\end{equation*}%
\begin{equation*}
\leq (\frac{1}{2}m\omega ^{2}x_{L}^{2})||x^{j}\Theta (x-x_{L},\varepsilon
)\Psi _{0}(x,r,t_{0}+t_{1}/2)||
\end{equation*}%
\begin{equation}
+(\frac{1}{2}m\omega ^{2})||x^{j+2}\Theta (x-x_{L},\varepsilon )\Psi
_{0}(x,r,t_{0}+t_{1}/2)||.  \tag{4.73}
\end{equation}%
This inequality is really a generalization of the inequality (4.40). These
three relations (4.71), (4.72), and (4.73) together show that each one of
these three $p-$independent norms on the $RH$ side of (4.70) has an upper
bound which consists of the basic norms $\{NBAS2\}$.

Besides these three $p-$independent norms there are also three norms that
are dependent on the momentum operator on the $RH$ side of (4.70). In order
to calculate these $p-$dependent norms one needs first to obtain the error
state (4.67). At first the product state $\chi (x,r,t_{1},t_{1})$ is
obtained from (4.39a). Then by applying the unitary operator $\exp
[-iH_{I}(x,\alpha ,\gamma )t_{1}/\hslash ]$ to the product state $\chi
(x,r,t_{1},t_{1})$ one may obtain the error state (4.67). Here the unitary
transformations (4.18) or (4.19) need to be used. The error state (4.67) may
be divided into the two parts: 
\begin{equation}
\Psi _{er}^{21}(x,r,t_{0}+t_{1}/2)=\Psi _{C}^{21}(x,r,t_{0}+t_{1}/2)+\Psi
_{S}^{21}(x,r,t_{0}+t_{1}/2).  \tag{4.74a}
\end{equation}%
Here the error state $\Psi _{C}^{21}(x,r,t_{0}+t_{1}/2)$ is written as%
\begin{equation}
\Psi _{C}^{21}(x,r,t_{0}+t_{1}/2)=F_{C}^{21}(x)\{\Psi
_{0}^{g}(x,t_{0}+t_{1}/2)|\tilde{g}_{0}\rangle +\Psi
_{0}^{e}(x,t_{0}+t_{1}/2)|\tilde{e}\rangle \}  \tag{4.74b}
\end{equation}%
where the function $F_{C}^{21}(x)$ is defined by%
\begin{equation}
F_{C}^{21}(x)=1-\cos [\frac{1}{\hslash }t_{1}V_{1}^{ho}(x,\varepsilon )]\cos
[\frac{1}{2\hslash }t_{1}\Theta (x-x_{L},\varepsilon )\Omega (x)]. 
\tag{4.74c}
\end{equation}%
The error state $\Psi _{S}^{21}(x,r,t_{0}+t_{1}/2)$ is given by%
\begin{equation*}
\Psi _{S}^{21}(x,r,t_{0}+t_{1}/2)=F_{S1}^{21}(x)\{\Psi
_{0}^{g}(x,t_{0}+t_{1}/2)|\tilde{g}_{0}\rangle +\Psi
_{0}^{e}(x,t_{0}+t_{1}/2)|\tilde{e}\rangle \}
\end{equation*}%
\begin{equation}
+F_{S2}^{21}(x)\{\exp [i\varphi (x,\gamma )]\Psi _{0}^{g}(x,t_{0}+t_{1}/2)|%
\tilde{e}\rangle +\exp [-i\varphi (x,\gamma )]\Psi _{0}^{e}(x,t_{0}+t_{1}/2)|%
\tilde{g}_{0}\rangle \},  \tag{4.74d}
\end{equation}%
where the two functions $F_{S1}^{21}(x)$ and $F_{S2}^{21}(x)$ are defined by%
\begin{equation}
F_{S1}^{21}(x)=i\sin [\frac{1}{\hslash }t_{1}V_{1}^{ho}(x,\varepsilon )]\cos
[\frac{1}{2\hslash }t_{1}\Theta (x-x_{L},\varepsilon )\Omega (x)], 
\tag{4.74e}
\end{equation}%
\begin{equation}
F_{S2}^{21}(x)=-i\exp [-\frac{i}{\hslash }t_{1}V_{1}^{ho}(x,\varepsilon
)]\sin [\frac{1}{2\hslash }t_{1}\Theta (x-x_{L},\varepsilon )\Omega (x)]. 
\tag{4.74f}
\end{equation}%
Here the normalization superposition states $|\tilde{g}_{0}\rangle =|\tilde{g%
}_{0}(x,t_{1})\rangle $ and $|\tilde{e}\rangle =|\tilde{e}(x,t_{1})\rangle ,$
which are generally defined by%
\begin{equation}
|\tilde{g}_{0}(x,t)\rangle =\cos [\frac{1}{2\hslash }\Omega
(x)t]|g_{0}\rangle -i\exp [i\varphi (x,\gamma )]\sin [\frac{1}{2\hslash }%
\Omega (x)t]|e\rangle ,  \tag{4.75a}
\end{equation}%
\begin{equation}
|\tilde{e}(x,t)\rangle =\cos [\frac{1}{2\hslash }\Omega (x)t]|e\rangle
-i\exp [-i\varphi (x,\gamma )]\sin [\frac{1}{2\hslash }\Omega
(x)t]|g_{0}\rangle .  \tag{4.75b}
\end{equation}%
In fact, both the states $|\tilde{g}_{0}\rangle $ and $|\tilde{e}\rangle $
are generated by applying the unitary operator $\exp [-iH_{I}(x,\pi
/4,\gamma )t_{1}/\hslash ]$ to the internal states $|g_{0}\rangle $ and $%
|e\rangle ,$ respectively. Therefore, both the states are normalized. It
follows from (4.74a) that the $p-$dependent norms on the $RH$ side of (4.70)
satisfy the inequality:%
\begin{equation}
||x^{k}p^{l}\Psi _{er}^{21}(x,r,t_{0}+t_{1}/2)||\leq ||x^{k}p^{l}\Psi
_{C}^{21}(x,r,t_{0}+t_{1}/2)||+||x^{k}p^{l}\Psi
_{S}^{21}(x,r,t_{0}+t_{1}/2)||  \tag{4.76}
\end{equation}%
where $k=0$, $1$ and $l=1,$ $2.$ Therefore, the upper bound for each one of
the three $p-$dependent norms in (4.70) may be determined by calculating
strictly those norms on the $RH$ side of (4.76).

Now the upper bound of the simpler norm $||x^{l}p\Psi
_{er}^{21}(x,r,t_{0}+t_{1}/2)||$ with $l=0$ or $1$ in (4.70) is calculated.
This calculation may be simplified by using the inequality (4.76). The norm $%
||x^{l}p\Psi _{C}^{21}(x,r,t_{0}+t_{1}/2)||$ with $l=0$ or $1$ in (4.76) is
first calculated strictly. The error state $\Psi
_{C}^{21}(x,r,t_{0}+t_{1}/2) $ of (4.74b) is the product of a trigonometric
function $F_{C}^{21}(x)$ and a superposition product state. After the
momentum operator $p=-i\hslash \partial /\partial x$ is applied to the error
state (4.74b), one obtains%
\begin{equation*}
p\Psi _{C}^{21}(x,r,t_{0}+t_{1}/2)=P_{211}^{C}(x,r,t_{0}+t_{1}/2)
\end{equation*}%
\begin{equation}
+P_{212}^{C}(x,r,t_{0}+t_{1}/2)+P_{213}^{C}(x,r,t_{0}+t_{1}/2).  \tag{4.77}
\end{equation}%
Here the first error state $P_{211}^{C}(x,r,t_{0}+t_{1}/2)$ is generated by
applying the coordinate derivative operation $(\partial /\partial x)$ only
to the function $F_{C}^{21}(x)$ in (4.74b), 
\begin{equation*}
P_{211}^{C}(x,r,t_{0}+t_{1}/2)=-i\hslash \lbrack \frac{\partial }{\partial x}%
F_{C}^{21}(x)]\{\Psi _{0}^{g}(x,t_{0}+t_{1}/2)|\tilde{g}_{0}\rangle +\Psi
_{0}^{e}(x,t_{0}+t_{1}/2)|\tilde{e}\rangle \},
\end{equation*}%
where the derivative $\frac{\partial }{\partial x}F_{C}^{21}(x)$ can be
obtained directly from the function $F_{C}^{21}(x)$ of (4.74c). The second
error state $P_{212}^{C}(x,r,t_{0}+t_{1}/2)$ is obtained by applying the
coordinate derivative operation only to the superposition states $|\tilde{g}%
_{0}\rangle $ and $|\tilde{e}\rangle $ in (4.74b), 
\begin{equation*}
P_{212}^{C}(x,r,t_{0}+t_{1}/2)=-i\hslash F_{C}^{21}(x)\{\Psi
_{0}^{g}(x,t_{0}+t_{1}/2)\frac{\partial }{\partial x}|\tilde{g}_{0}\rangle
+\Psi _{0}^{e}(x,t_{0}+t_{1}/2)\frac{\partial }{\partial x}|\tilde{e}\rangle
\}.
\end{equation*}%
Both the derivatives $\frac{\partial }{\partial x}|\tilde{g}_{0}\rangle $
and $\frac{\partial }{\partial x}|\tilde{e}\rangle $ can be directly
calculated by using (4.75), respectively. The last error state $%
P_{213}^{C}(x,r,t_{0}+t_{1}/2)$ is generated by applying the coordinate
derivative operation only to the motional states $\Psi
_{0}^{a}(x,t_{0}+t_{1}/2)$ with $a=g$ and $e$ in (4.74b), 
\begin{equation*}
P_{213}^{C}(x,r,t_{0}+t_{1}/2)=-i\hslash F_{C}^{21}(x)\{[\frac{\partial }{%
\partial x}\Psi _{0}^{g}(x,t_{0}+t_{1}/2)]|\tilde{g}_{0}\rangle
\end{equation*}%
\begin{equation*}
+[\frac{\partial }{\partial x}\Psi _{0}^{e}(x,t_{0}+t_{1}/2)]|\tilde{e}%
\rangle \}.
\end{equation*}%
Once these three error states $\{P_{21j}^{C}(x,r,t_{0}+t_{1}/2)\}$ are
obtained, one may use them to determine the upper bounds of the norms $%
||x^{l}P_{21j}^{C}(x,r,t_{0}+t_{1}/2)||$ with $l=0$, $1$ and $j=1$, $2$, $3.$
It can turn out that the norm $||x^{l}P_{211}^{C}(x,r,$ $t_{0}+t_{1}/2)||$
with $l=0$ or $1$ satisfies the inequality: 
\begin{equation*}
||x^{l}P_{211}^{C}(x,r,t_{0}+t_{1}/2)||\leq t_{1}\sum_{a=g,e}\{||x^{l}[\frac{%
\partial }{\partial x}V_{1}^{ho}(x,\varepsilon )]\Psi
_{0}^{a}(x,t_{0}+t_{1}/2)||
\end{equation*}%
\begin{equation*}
+|2\hslash \Omega _{0}|\times ||x^{l}\delta (x-x_{L},\varepsilon )\Psi
_{0}^{a}(x,t_{0}+t_{1}/2)||
\end{equation*}%
\begin{equation}
+|(\hslash \Delta k)\Omega _{0}|\times ||x^{l}\Theta (x-x_{L},\varepsilon
)\Psi _{0}^{a}(x,t_{0}+t_{1}/2)||\}.  \tag{4.78a}
\end{equation}%
Similarly, one can prove that the other norms $%
||x^{l}P_{21j}^{C}(x,r,t_{0}+t_{1}/2)||$ with $l=0$, $1$ and $j=2$, $3$
satisfy respectively the inequalities: 
\begin{equation*}
||x^{l}P_{212}^{C}(x,r,t_{0}+t_{1}/2)||\leq 2(\frac{1}{2\hslash }%
t_{1})^{2}(|(\hslash \Delta k)\Omega _{0}t_{1}|+\frac{1}{2}\hslash
|k_{0}+k_{1}|)
\end{equation*}%
\begin{equation*}
\times \sum_{a=g,e}\{||x^{l}V_{1}^{ho}(x,\varepsilon )^{2}\Psi
_{0}^{a}(x,t_{0}+t_{1}/2)||+|4\hslash \Omega _{0}|\times
||x^{l}V_{1}^{ho}(x,\varepsilon )\Psi _{0}^{a}(x,t_{0}+t_{1}/2)||
\end{equation*}%
\begin{equation}
+(2\hslash \Omega _{0})^{2}||x^{l}\Theta (x-x_{L},\varepsilon )\Psi
_{0}^{a}(x,t_{0}+t_{1}/2)||\}  \tag{4.78b}
\end{equation}%
and 
\begin{equation*}
||x^{l}P_{213}^{C}(x,r,t_{0}+t_{1}/2)||\leq 2\hslash (\frac{1}{2\hslash }%
t_{1})^{2}\sum_{a=g,e}\{||x^{l}V_{1}^{ho}(x,\varepsilon )^{2}\frac{\partial 
}{\partial x}\Psi _{0}^{a}(x,t_{0}+t_{1}/2)||
\end{equation*}%
\begin{equation*}
+|4\hslash \Omega _{0}|\times ||x^{l}V_{1}^{ho}(x,\varepsilon )\frac{%
\partial }{\partial x}\Psi _{0}^{a}(x,t_{0}+t_{1}/2)||
\end{equation*}%
\begin{equation}
+(2\hslash \Omega _{0})^{2}||x^{l}\Theta (x-x_{L},\varepsilon )\frac{%
\partial }{\partial x}\Psi _{0}^{a}(x,t_{0}+t_{1}/2)||\}.  \tag{4.78c}
\end{equation}%
It is known from (4.46) that the upper bound of the first norm on the $RH$
side of (4.78a) for the index $a=g$ or $e$ consists of three basic norms $%
\{NBAS1\}$ and one basic norm $NBAS2$, while the second and the third are
the basic norms $NBAS1$ and $NBAS2$, respectively. Therefore, the inequality
(4.78a) shows that the upper bound of the norm $%
||x^{l}P_{211}^{C}(x,r,t_{0}+t_{1}/2)||$ consists of the two types of basic
norms $\{NBAS1\}$ and $\{NBAS2\}$. It is easy to prove that the first norm
on the $RH$ side of (4.78b) for the index $a=g$ or $e$ satisfies the
inequality: 
\begin{equation*}
||x^{l}V_{1}^{ho}(x,\varepsilon )^{2}\Psi _{0}^{a}(x,t_{0}+t_{1}/2)||\leq (%
\frac{1}{2}m\omega ^{2}x_{L}^{2})^{2}||x^{l}\Theta (x-x_{L},\varepsilon
)\Psi _{0}^{a}(x,t_{0}+t_{1}/2)||
\end{equation*}%
\begin{equation*}
+\frac{1}{2}(m\omega ^{2})^{2}x_{L}^{2}||x^{l+2}\Theta (x-x_{L},\varepsilon
)\Psi _{0}^{a}(x,t_{0}+t_{1}/2)||
\end{equation*}%
\begin{equation}
+(\frac{1}{2}m\omega ^{2})^{2}||x^{l+4}\Theta (x-x_{L},\varepsilon )\Psi
_{0}^{a}(x,t_{0}+t_{1}/2)||.  \tag{4.79}
\end{equation}%
Thus, the upper bound of the norm is composed of three basic norms $%
\{NBAS2\} $. It is known from (4.73) that the upper bound of the second norm
on the $RH $ side of (4.78b) consists of two basic norms $\{NBAS2\}$. The
last norm on the $RH$ side of (4.78b) is clearly a basic norm $NBAS2$. These
results together show that the $RH$ side of (4.78b) may be reduced to a
linear combination of the basic norms $\{NBAS2\}$, indicating that the norm $%
||x^{l}P_{212}^{C}(x,r,t_{0}+t_{1}/2)||$ has an upper bound that consists of
the basic norms $\{NBAS2\}$. When the index $l=0,$ the first norm on the $RH$
side of (4.78c) for the index $a=g$ or $e$ satisfies (4.50). If every norm
inside is times $x^{l}$ on both sides of the inequality (4.50), then it can
prove that the inequality still holds, and such a modified inequality may be
used to determine the upper bound of the first norm on the $RH$ side of
(4.78c). By using this modified inequality one can prove that the first norm
on the $RH$ side of (4.78c) for the index $a=g$ or $e$ has an upper bound
consisting of six basic norms $\{NBAS2\}$. In an analogous way, by using the
modified inequality of (4.51) one can prove that the upper bound of the
second norm on the $RH$ side of (4.78c) for the index $a=g$ or $e$ consists
of four basic norms $\{NBAS2\}$. Similarly, the last norm on the $RH$ side
of (4.78c) for the index $a=g$ or $e$ has an upper bound consisting of two
basic norms $\{NBAS2\}$ according to the modified inequality of (4.52).
These results together show that the $RH$ side of (4.78c) may be reduced to
a linear combination of the basic norms $\{NBAS2\}$, indicating that the
norm $||x^{l}P_{213}^{C}(x,r,t_{0}+t_{1}/2)||$ has an upper bound that is
composed of the basic norms $\{NBAS2\}$. Now it follows from (4.77) that the
norm $||x^{l}p\Psi _{C}^{21}(x,r,t_{0}+t_{1}/2)||$ with $l=0$ or $1$
satisfies the inequality: 
\begin{equation*}
||x^{l}p\Psi _{C}^{21}(x,r,t_{0}+t_{1}/2)||\leq
||x^{l}P_{211}^{C}(x,r,t_{0}+t_{1}/2)||
\end{equation*}%
\begin{equation*}
+||x^{l}P_{212}^{C}(x,r,t_{0}+t_{1}/2)||+||x^{l}P_{213}^{C}(x,r,t_{0}+t_{1}/2)||.
\end{equation*}%
These three norms on the $RH$ side of the inequality are already proven to
have upper bounds consisting of the basic norms $\{NBAS1\}$ and/or $%
\{NBAS2\} $, respectively. This indicates that the norm $||x^{l}p\Psi
_{C}^{21}(x,r,t_{0}+t_{1}/2)||$ with $l=0$ or $1$ has an upper bound that
consists of the basic norms $\{NBAS1\}$ and $\{NBAS2\}$.

Below calculate the norm $||x^{l}p\Psi _{S}^{21}(x,r,t_{0}+t_{1}/2)||$ with $%
l=0$ or $1$ on the $RH$ side of (4.76). Similar to the error state $p\Psi
_{C}^{21}(x,r,t_{0}+t_{1}/2)$, the error state $p\Psi
_{S}^{21}(x,r,t_{0}+t_{1}/2)$ may be written in a sum of the three error
states:%
\begin{equation*}
p\Psi _{S}^{21}(x,r,t_{0}+t_{1}/2)=P_{211}^{S}(x,r,t_{0}+t_{1}/2)
\end{equation*}%
\begin{equation}
+P_{212}^{S}(x,r,t_{0}+t_{1}/2)+P_{213}^{S}(x,r,t_{0}+t_{1}/2).  \tag{4.80}
\end{equation}%
These three error states can be obtained by applying the momentum operator
to the state $\Psi _{S}^{21}(x,r,t_{0}+t_{1}/2)$ of (4.74d). Here the first
error state $P_{211}^{S}(x,r,t_{0}+t_{1}/2)$ is obtained by applying the
coordinate derivative operation $(\partial /\partial x)$ only to the two
functions $F_{S1}^{21}(x)$ and $F_{S2}^{21}(x)$ in (4.74d), here both the
functions are given by (4.74e) and (4.74f), respectively. The second state $%
P_{212}^{S}(x,r,t_{0}+t_{1}/2)$ is generated by applying the coordinate
derivative operation only to the superposition states $|\tilde{g}_{0}\rangle 
$ and $|\tilde{e}\rangle $ as well as the phase factors $\exp [\pm i\varphi
(x,\gamma )]$ in (4.74d), here $|\tilde{g}_{0}\rangle $ and $|\tilde{e}%
\rangle $ are given by (4.75a) and (4.75b), respectively. The last state $%
P_{213}^{S}(x,r,t_{0}+t_{1}/2)$ is obtained by applying the coordinate
derivative operation only to the motional states $\{\Psi
_{0}^{a}(x,t_{0}+t_{1}/2)\}$ in (4.74d). By using these three error states
it can prove that the norms $||x^{l}P_{21j}^{S}(x,r,t_{0}+t_{1})||$ with $%
j=1 $, $2$, $3$ and $l=0$, $1$ satisfy respectively the inequalities:%
\begin{equation*}
||x^{l}P_{211}^{S}(x,r,t_{0}+t_{1}/2)||\leq 2t_{1}\sum_{a=g,e}\{||x^{l}[%
\frac{\partial }{\partial x}V_{1}^{ho}(x,\varepsilon )]\Psi
_{0}^{a}(x,t_{0}+t_{1}/2)||
\end{equation*}%
\begin{equation*}
+|(\hslash \Delta k)\Omega _{0}|\times ||x^{l}\Theta (x-x_{L},\varepsilon
)\Psi _{0}^{a}(x,t_{0}+t_{1}/2)||
\end{equation*}%
\begin{equation}
+|2\hslash \Omega _{0}|\times ||x^{l}\delta (x-x_{L},\varepsilon )\Psi
_{0}^{a}(x,t_{0}+t_{1}/2)||\},  \tag{4.81a}
\end{equation}%
\begin{equation*}
||x^{l}P_{212}^{S}(x,r,t_{0}+t_{1}/2)||\leq t_{1}\{t_{1}|(\Delta k)\Omega
_{0}|+\frac{1}{2}|k_{0}+k_{1}|\}
\end{equation*}%
\begin{equation}
\times \sum_{a=g,e}\{||x^{l}V_{1}^{ho}(x,\varepsilon )\Psi
_{0}^{a}(x,t_{0}+t_{1}/2)||+|2\hslash \Omega _{0}|\times ||x^{l}\Theta
(x-x_{L},\varepsilon )\Psi _{0}^{a}(x,t_{0}+t_{1}/2)||\},  \tag{4.81b}
\end{equation}%
and%
\begin{equation*}
||x^{l}P_{213}^{S}(x,r,t_{0}+t_{1}/2)||\leq
t_{1}\sum_{a=g,e}\{||x^{l}V_{1}^{ho}(x,\varepsilon )\frac{\partial }{%
\partial x}\Psi _{0}^{a}(x,t_{0}+t_{1}/2)||
\end{equation*}%
\begin{equation}
+|2\hslash \Omega _{0}|\times ||x^{l}\Theta (x-x_{L},\varepsilon )\frac{%
\partial }{\partial x}\Psi _{0}^{a}(x,t_{0}+t_{1}/2)||\}.  \tag{4.81c}
\end{equation}%
As shown in (4.46), the first norm on the $RH$ side of (4.81a) with the
index $a=g$ or $e$ has an upper bound which consists of three basic norms $%
\{NBAS1\}$ and one basic norm $NBAS2$. It is clear that the second and the
third norm on the $RH$ side of (4.81a) with the index $a=g$ or $e$ are the
basic norms $NBAS2$ and $NBAS1$, respectively. Thus, the inequality (4.81a)
shows that the upper bound of the norm $%
||x^{l}P_{211}^{S}(x,r,t_{0}+t_{1}/2)||$\ consists of the basic norms $%
\{NBAS1\}$ and $\{NBAS2\}$. By comparing both the $RH$ sides of the two
inequalities (4.81b) and (4.78b) one can deduce that, just like the norm $%
||x^{l}P_{212}^{C}(x,r,t_{0}+t_{1}/2)||$ in (4.78b), the norm $%
||x^{l}P_{212}^{S}(x,r,t_{0}+t_{1}/2)||$ in (4.81b) has an upper bound
consisting of the basic norms $\{NBAS2\}$. Similarly, by comparing both the $%
RH$ sides of the two inequalities (4.81c) and (4.78c) one also can deduce
that the norm $||x^{l}P_{213}^{S}(x,r,t_{0}+t_{1}/2)||$ in (4.81c) has an
upper bound consisting of the basic norms $\{NBAS2\}$. Now it follows from
(4.80) that the norm $||x^{l}p\Psi _{S}^{21}(x,r,t_{0}+t_{1}/2)||$ with $l=0$%
, $1$ satisfies the inequality: 
\begin{equation*}
||x^{l}p\Psi _{S}^{21}(x,r,t_{0}+t_{1}/2)||\leq
||x^{l}P_{211}^{S}(x,r,t_{0}+t_{1}/2)||
\end{equation*}%
\begin{equation*}
+||x^{l}P_{212}^{S}(x,r,t_{0}+t_{1}/2)||+||x^{l}P_{213}^{S}(x,r,t_{0}+t_{1}/2)||.
\end{equation*}%
These three norms on the $RH$ side of the inequality are already shown to
have upper bounds consisting of the basic norms $\{NBAS1\}$ and/or $%
\{NBAS2\} $. This indicates that the upper bound of the norm $||x^{l}p\Psi
_{S}^{21}(x,r,t_{0}+t_{1}/2)||$ consists of the two types of basic norms $%
\{NBAS1\}$ and $\{NBAS2\}$.

The above theoretical calculations show that these norms $\{||x^{l}p\Psi
_{C/S}^{21}(x,r,t_{0}+t_{1}/2)||\}$ with $l=0$ and $1$ on the $RH$ side of
(4.76) have upper bounds that consist of the two types of basic norms $%
\{NBAS1\}$ and $\{NBAS2\}$. Then one can deduce from (4.76) that the two
norms $\{||x^{l}p\Psi _{er}^{21}(x,r,t_{0}+t_{1}/2)||\}$ with $l=0$ and $1$
on the $RH$ side of (4.70) each have an upper bound that is composed of the
two types of basic norms $\{NBAS1\}$ and $\{NBAS2\}$.

Now the norm $||p^{2}\Psi _{er}^{21}(x,r,t_{0}+t_{1}/2)||$ on the $RH$ side
of (4.70) is calculated strictly. This is the last norm to be calculated for
the upper bound of the norm $2m||M_{21}^{V}(x,r,t_{1},t_{3})||$ in (4.70).
It follows from (4.74a) that the error state $p^{2}\Psi
_{er}^{21}(x,r,t_{0}+t_{1}/2)$ may be written as%
\begin{equation*}
p^{2}\Psi _{er}^{21}(x,r,t_{0}+t_{1}/2)=p^{2}\Psi
_{C}^{21}(x,r,t_{0}+t_{1}/2)+p^{2}\Psi _{S}^{21}(x,r,t_{0}+t_{1}/2).
\end{equation*}%
By using the two relations (4.77) and (4.80) one obtains%
\begin{equation*}
p^{2}\Psi _{C/S}^{21}(x,r,t_{0}+t_{1}/2)=pP_{211}^{C/S}(x,r,t_{0}+t_{1}/2)
\end{equation*}%
\begin{equation}
+pP_{212}^{C/S}(x,r,t_{0}+t_{1}/2)+pP_{213}^{C/S}(x,r,t_{0}+t_{1}/2). 
\tag{4.82}
\end{equation}%
Therefore, by computing the six norms $%
\{||pP_{21j}^{C/S}(x,r,t_{0}+t_{1}/2)||\}$ with $j=1$, $2$, $3$ one can
obtain the upper bound of the norm $||p^{2}\Psi
_{er}^{21}(x,r,t_{0}+t_{1}/2)||.$ Furthermore, each one of the six norms may
be calculated in a similar way that is used to calculate the norm $||p\Psi
_{C/S}^{21}(x,r,t_{0}+t_{1}/2)||$ above. For example, the error state $%
pP_{211}^{C/S}(x,r,t_{0}+t_{1}/2)$ may be divided into the three error
states: 
\begin{equation*}
pP_{211}^{C/S}(x,r,t_{0}+t_{1}/2)=P_{2111}^{C/S}(x,r,t_{0}+t_{1}/2)
\end{equation*}%
\begin{equation}
+P_{2112}^{C/S}(x,r,t_{0}+t_{1}/2)+P_{2113}^{C/S}(x,r,t_{0}+t_{1}/2). 
\tag{4.83}
\end{equation}%
The other error states $pP_{21j}^{C/S}(x,r,t_{0}+t_{1}/2)$ with $j=2$ and $3$
also have similar expressions, respectively. These error states $%
\{P_{21kl}^{C/S}(x,r,t_{0}+t_{1}/2)\}$ with $k,$ $l=1,$ $2,$ $3$ are
obtained as follows. It is known from (4.74b) that the error state $\Psi
_{C}^{21}(x,r,t_{0}+t_{1}/2)$ is the product of the function $F_{C}^{21}(x)$
of (4.74c) and the superposition product state: $\Psi
_{0}^{g}(x,t_{0}+t_{1}/2)|\tilde{g}_{0}\rangle +\Psi
_{0}^{e}(x,t_{0}+t_{1}/2)|\tilde{e}\rangle .$ It is also known that the
error state $P_{211}^{C}(x,r,t_{0}+t_{1}/2)$ is obtained by applying the
coordinate derivative operation $(\partial /\partial x)$ only to the
function $F_{C}^{21}(x)$ in (4.74b). After the error state $%
P_{211}^{C}(x,r,t_{0}+t_{1}/2)$ is acted on again by the momentum operator $%
p=-i\hslash \partial /\partial x,$ the generated state $%
pP_{211}^{C}(x,r,t_{0}+t_{1}/2)$ is divided into the three error states on
the $RH$ side of (4.83). Thus, the first error state $%
P_{2111}^{C}(x,r,t_{0}+t_{1}/2)$ is generated by applying the second-order
coordinate derivative operation $(\partial ^{2}/\partial x^{2})$ only to the
function $F_{C}^{21}(x)$ in (4.74b), 
\begin{equation*}
P_{2111}^{C}(x,r,t_{0}+t_{1}/2)=-\hslash ^{2}[\frac{\partial ^{2}}{\partial
x^{2}}F_{C}^{21}(x)]\{\Psi _{0}^{g}(x,t_{0}+t_{1}/2)|\tilde{g}_{0}\rangle
+\Psi _{0}^{e}(x,t_{0}+t_{1}/2)|\tilde{e}\rangle \}.
\end{equation*}%
The second error state $P_{2112}^{C}(x,r,t_{0}+t_{1}/2)$ is generated by
applying the coordinate derivative operation $(\partial /\partial x)$
simultaneously to both the function $F_{C}^{21}(x)$ and the superposition
states $|\tilde{g}_{0}\rangle $ and $|\tilde{e}\rangle $ in (4.74b), 
\begin{equation*}
P_{2112}^{C}(x,r,t_{0}+t_{1}/2)=-\hslash ^{2}[\frac{\partial }{\partial x}%
F_{C}^{21}(x)]
\end{equation*}%
\begin{equation*}
\times \{\Psi _{0}^{g}(x,t_{0}+t_{1}/2)\frac{\partial }{\partial x}|\tilde{g}%
_{0}\rangle +\Psi _{0}^{e}(x,t_{0}+t_{1}/2)\frac{\partial }{\partial x}|%
\tilde{e}\rangle \}.
\end{equation*}%
The last error state $P_{2113}^{C}(x,r,t_{0}+t_{1}/2)$ is generated by
applying the coordinate derivative operation $(\partial /\partial x)$
simultaneously to both the function $F_{C}^{21}(x)$ and the motional states $%
\Psi _{0}^{a}(x,t_{0}+t_{1}/2)$ with $a=g$ and $e$ in (4.74b), 
\begin{equation*}
P_{2113}^{C}(x,r,t_{0}+t_{1}/2)=-\hslash ^{2}[\frac{\partial }{\partial x}%
F_{C}^{21}(x)]
\end{equation*}%
\begin{equation*}
\times \{[\frac{\partial }{\partial x}\Psi _{0}^{g}(x,t_{0}+t_{1}/2)]|\tilde{%
g}_{0}\rangle +[\frac{\partial }{\partial x}\Psi _{0}^{e}(x,t_{0}+t_{1}/2)]|%
\tilde{e}\rangle \}.
\end{equation*}%
In an analogous way, the other error states $%
\{P_{21kl}^{C/S}(x,r,t_{0}+t_{1}/2)\}$ can be obtained. It can be found that
among these error states there are the symmetrical relations: 
\begin{equation*}
P_{21kl}^{C/S}(x,r,t_{0}+t_{1}/2)=P_{21lk}^{C/S}(x,r,t_{0}+t_{1}/2),\text{ }%
for\text{ }k\neq l.
\end{equation*}%
Now by substituting (4.83), etc., into (4.82) and using these symmetrical
relations the error state $p^{2}\Psi _{C/S}^{12}(x,r,t_{0}+t_{1}/2)$ may be
expressed as%
\begin{equation*}
p^{2}\Psi
_{C/S}^{21}(x,r,t_{0}+t_{1}/2)=P_{2111}^{C/S}(x,r,t_{0}+t_{1}/2)+P_{2122}^{C/S}(x,r,t_{0}+t_{1}/2)
\end{equation*}%
\begin{equation*}
+P_{2133}^{C/S}(x,r,t_{0}+t_{1}/2)+2P_{2112}^{C/S}(x,r,t_{0}+t_{1}/2)
\end{equation*}%
\begin{equation}
+2P_{2113}^{C/S}(x,r,t_{0}+t_{1}/2)+2P_{2123}^{C/S}(x,r,t_{0}+t_{1}/2). 
\tag{4.84}
\end{equation}%
Here the error states $P_{2111}^{C/S}(x,r,t_{0}+t_{1}/2),$ $%
P_{2122}^{C/S}(x,r,t_{0}+t_{1}/2),$ and $P_{2112}^{C/S}(x,r,$ $%
t_{0}+t_{1}/2) $ are proportional to the motional states $\Psi
_{0}^{a}(x,t_{0}+t_{1}/2)$ with $a=g$ and $e.$ Now define $SUM1(C/S)$ as the
sum of norms of these error states,%
\begin{equation*}
SUM1(C/S)\overset{\text{def}}{\equiv }||P_{2111}^{C/S}(x,r,t_{0}+t_{1}/2)||
\end{equation*}%
\begin{equation*}
+||P_{2122}^{C/S}(x,r,t_{0}+t_{1}/2)||+2||P_{2112}^{C/S}(x,r,t_{0}+t_{1}/2)||.
\end{equation*}%
Then it can turn out that $SUM1(C/S)$ is bounded by%
\begin{equation*}
SUM1(C/S)\leq \Gamma _{1}^{C/S}\sum_{a=g,e}||[\frac{\partial ^{2}}{\partial
x^{2}}V_{1}^{ho}(x,\varepsilon )]\Psi _{0}^{a}(x,t_{0}+t_{1}/2)||
\end{equation*}%
\begin{equation*}
+\Gamma _{2}^{C/S}\sum_{a=g,e}||[\frac{\partial }{\partial x}%
V_{1}^{ho}(x,\varepsilon )]^{2}\Psi _{0}^{a}(x,t_{0}+t_{1}/2)||
\end{equation*}%
\begin{equation*}
+\Gamma _{3}^{C/S}\sum_{a=g,e}||\delta (x-x_{L},\varepsilon )[\frac{\partial 
}{\partial x}V_{1}^{ho}(x,\varepsilon )]\Psi _{0}^{a}(x,t_{0}+t_{1}/2)||
\end{equation*}%
\begin{equation*}
+\Gamma _{4}^{C/S}\sum_{a=g,e}||[\frac{\partial }{\partial x}%
V_{1}^{ho}(x,\varepsilon )]\Psi _{0}^{a}(x,t_{0}+t_{1}/2)||
\end{equation*}%
\begin{equation*}
+\sum_{a=g,e}\sum_{k=1}^{2}\sum_{l=0}^{1}J_{kl}^{C/S}||x^{l}\delta
(x-x_{L},\varepsilon )^{k}\Psi _{0}^{a}(x,t_{0}+t_{1}/2)||
\end{equation*}%
\begin{equation}
+\sum_{a=g,e}\sum_{l=0}^{1}L_{l}^{C/S}||x^{2l}\Theta (x-x_{L},\varepsilon
)\Psi _{0}^{a}(x,t_{0}+t_{1}/2)||,  \tag{4.85}
\end{equation}%
where all these parameters $\{\Gamma _{k}^{C/S}\},$ $\{J_{kl}^{C/S}\},$ and $%
\{L_{l}^{C/S}\}$ are non-negative, bounded, and controllable. Both the error
states $P_{2113}^{C/S}(x,r,t_{0}+t_{1}/2)$ and $%
P_{2123}^{C/S}(x,r,t_{0}+t_{1}/2)$ in (4.84) are proportional to the
first-order coordinate derivatives $\partial \Psi
_{0}^{a}(x,t_{0}+t_{1}/2)/\partial x$ with $a=g$ and $e.$ It can turn out
that the sum of norms of the two error states is bounded by%
\begin{equation*}
SUM2(C/S)\overset{\text{def}}{\equiv }%
2||P_{2113}^{C/S}(x,r,t_{0}+t_{1}/2)||+2||P_{2123}^{C/S}(x,r,t_{0}+t_{1}/2)||
\end{equation*}%
\begin{equation*}
\leq \Lambda _{1}^{C/S}\sum_{a=g,e}||[\frac{\partial }{\partial x}%
V_{1}^{ho}(x,\varepsilon )]\frac{\partial }{\partial x}\Psi
_{0}^{a}(x,t_{0}+t_{1}/2)||
\end{equation*}%
\begin{equation*}
+\Lambda _{2}^{C/S}\sum_{a=g,e}||\delta (x-x_{L},\varepsilon )\frac{\partial 
}{\partial x}\Psi _{0}^{a}(x,t_{0}+t_{1}/2)||
\end{equation*}%
\begin{equation}
+\sum_{a=g,e}\sum_{l=0}^{2}K_{l}^{C/S}||x^{2l}\Theta (x-x_{L},\varepsilon )%
\frac{\partial }{\partial x}\Psi _{0}^{a}(x,t_{0}+t_{1}/2)||  \tag{4.86}
\end{equation}%
where all these non-negative parameters $\{\Lambda _{k}^{C/S}\}$ and $%
\{K_{l}^{C/S}\}$ are bounded and controllable. The error state $%
P_{2133}^{C/S}(x,r,t_{0}+t_{1}/2)$ in (4.84) is proportional to the
second-order coordinate derivatives $\partial ^{2}\Psi
_{0}^{a}(x,t_{0}+t_{1}/2)/\partial x^{2}$ with $a=g$ and $e.$ It can turn
out that%
\begin{equation*}
||P_{2133}^{C}(x,r,t_{0}+t_{1}/2)||\leq \frac{1}{2}t_{1}^{2}%
\sum_{a=g,e}||V_{1}^{ho}(x,\varepsilon )^{2}\frac{\partial ^{2}}{\partial
x^{2}}\Psi _{0}^{a}(x,t_{0}+t_{1}/2)||
\end{equation*}%
\begin{equation}
+2|\hslash \Omega _{0}t_{1}|^{2}\sum_{a=g,e}||\Theta (x-x_{L},\varepsilon )%
\frac{\partial ^{2}}{\partial x^{2}}\Psi _{0}^{a}(x,t_{0}+t_{1}/2)|| 
\tag{4.87a}
\end{equation}%
and%
\begin{equation*}
||P_{2133}^{S}(x,r,t_{0}+t_{1}/2)||\leq \hslash
t_{1}\sum_{a=g,e}||V_{1}^{ho}(x,\varepsilon )\frac{\partial ^{2}}{\partial
x^{2}}\Psi _{0}^{a}(x,t_{0}+t_{1}/2)||
\end{equation*}%
\begin{equation}
+2\hslash t_{1}|\hslash \Omega _{0}|\sum_{a=g,e}||\Theta
(x-x_{L},\varepsilon )\frac{\partial ^{2}}{\partial x^{2}}\Psi
_{0}^{a}(x,t_{0}+t_{1}/2)||.  \tag{4.87b}
\end{equation}%
Now it follows from (4.84) that the error state $p^{2}\Psi
_{C/S}^{21}(x,r,t_{0}+t_{1}/2)$ is bounded by%
\begin{equation*}
||p^{2}\Psi _{C/S}^{21}(x,r,t_{0}+t_{1}/2)||\leq
SUM1(C/S)+SUM2(C/S)+||P_{2133}^{C/S}(x,r,t_{0}+t_{1}/2)||
\end{equation*}%
where the upper bounds of $SUM1(C/S),$ $SUM2(C/S),$ and $%
||P_{2133}^{C/S}(x,r,t_{0}+t_{1}/2)||$ are determined from (4.85), (4.86),
and (4.87), respectively. Below investigate in detail the upper bounds of $%
SUM1(C/S),$ $SUM2(C/S),$ and $||P_{2133}^{C/S}(x,r,$ $t_{0}+t_{1}/2)||,$
respectively.

The upper bound of $SUM1(C/S)$ consists of six sum terms on the $RH$ side of
(4.85). The first sum term contains the two norms $||[\frac{\partial ^{2}}{%
\partial x^{2}}V_{1}^{ho}(x,\varepsilon )]\Psi _{0}^{a}(x,t_{0}+t_{1}/2)||$
with $a=g$ and $e$. It is known from (3.33b) in the section 3 that each one
of the two norms has an upper bound consisting of one basic norm $NBAS2$ and
seven basic norms $\{NBAS1\}$. The second sum term consists of the two norms 
$||[\frac{\partial }{\partial x}V_{1}^{ho}(x,\varepsilon )]^{2}\Psi
_{0}^{a}(x,t_{0}+t_{1}/2)||$ with $a=g$ and $e$. It is known from (3.33c) in
the section 3 that each one of the two norms has an upper bound consisting
of one basic norm $NBAS2$ and nine basic norms $\{NBAS1\}$. The third sum
term consists of the two norms $||\delta (x-x_{L},\varepsilon )[\frac{%
\partial }{\partial x}V_{1}^{ho}(x,\varepsilon )]\Psi
_{0}^{a}(x,t_{0}+t_{1}/2)||$ with $a=g$ and $e.$ It is known from (4.55)
that each one of the two norms has an upper bound consisting of four basic
norms $\{NBAS1\}$. The fourth sum term contains the two norms $||[\frac{%
\partial }{\partial x}V_{1}^{ho}(x,\varepsilon )]\Psi
_{0}^{a}(x,t_{0}+t_{1}/2)||$ with $a=g$ and $e$. As can be seen in (4.46),
each one of the two norms has an upper bound consisting of three basic norms 
$\{NBAS1\}$ and one basic norm $NBAS2$. The fifth sum term consists of eight
basic norms $\{NBAS1\}$ at most and the sixth sum term is composed of four
basic norms $\{NBAS2\}$. Therefore, the inequality (4.85) shows that the
upper bound of $SUM1(C/S)$ consists of a finite number of the basic norms $%
\{NBAS1\}$ and $\{NBAS2\}$.

The upper bound of $SUM2(C/S)$ contains three sum terms on the $RH$ side of
(4.86). The first sum term contains the two norms $||[\frac{\partial }{%
\partial x}V_{1}^{ho}(x,\varepsilon )]\frac{\partial }{\partial x}\Psi
_{0}^{a}(x,t_{0}+t_{1}/2)||$ with $a=g$ and $e$. Formally, $\frac{\partial }{%
\partial x}\Psi _{0}^{a}(x,t_{0}+t_{1}/2)=(C_{1}^{a}x+C_{0}^{a})\Psi
_{0}^{a}(x,t_{0}+t_{1}/2),$ where the parameters $C_{1}^{a}$ and $C_{0}^{a}$
are determined from (4.49). Then it can be found that 
\begin{equation*}
||[\frac{\partial }{\partial x}V_{1}^{ho}(x,\varepsilon )]\frac{\partial }{%
\partial x}\Psi _{0}^{a}(x,t_{0}+t_{1}/2)||
\end{equation*}%
\begin{equation*}
\leq |C_{1}^{a}|\times ||x[\frac{\partial }{\partial x}V_{1}^{ho}(x,%
\varepsilon )]\Psi _{0}^{a}(x,t_{0}+t_{1}/2)||
\end{equation*}%
\begin{equation*}
+|C_{0}^{a}|\times ||[\frac{\partial }{\partial x}V_{1}^{ho}(x,\varepsilon
)]\Psi _{0}^{a}(x,t_{0}+t_{1}/2)||.
\end{equation*}%
Furthermore, it is known from (4.46) that the norm $||x^{l}[\frac{\partial }{%
\partial x}V_{1}^{ho}(x,\varepsilon )]\Psi _{0}^{a}(x,t_{0}+t_{1}/2)||$ for $%
l=0$, $1$, ..., has an upper bound consisting of three basic norms $%
\{NBAS1\} $ and one basic norm $NBAS2$. Thus, the norm $||[\frac{\partial }{%
\partial x}V_{1}^{ho}(x,\varepsilon )]\frac{\partial }{\partial x}\Psi
_{0}^{a}(x,$ $t_{0}+t_{1}/2)||$ of the first sum term has an upper bound
consisting of the basic norms $\{NBAS1\}$ and $\{NBAS2\}$. The second sum
term consists of the two norms $||\delta (x-x_{L},\varepsilon )\frac{%
\partial }{\partial x}\Psi _{0}^{a}(x,t_{0}+t_{1}/2)||$ with $a=g$ and $e$.
It is easy to prove that each one of the two norms has an upper bound
consisting of two basic norms $\{NBAS1\}.$ The third sum term consists of
six norms $\{||x^{2l}\Theta (x-x_{L},\varepsilon )\frac{\partial }{\partial x%
}\Psi _{0}^{a}(x,$ $t_{0}+t_{1}/2)||\}$ at most, each one of which has an
upper bound consisting of two basic norms $\{NBAS2\}$. Thus, the inequality
(4.86) shows that the upper bound of $SUM2(C/S)$ consists of a finite number
of the basic norms $\{NBAS1\}$ and $\{NBAS2\}$.

The upper bound of the norm $||P_{2133}^{C}(x,r,t_{0}+t_{1}/2)||$ contains
two sum terms on the $RH$ side of (4.87a). The first sum term contains the
two norms $||V_{1}^{ho}(x,\varepsilon )^{2}\frac{\partial ^{2}}{\partial
x^{2}}\Psi _{0}^{a}(x,t_{0}+t_{1}/2)||$ with $a=g$ and $e$. Here the
second-order derivative $\frac{\partial ^{2}}{\partial x^{2}}\Psi
_{0}^{a}(x,t_{0}+t_{1}/2)$ can be computed exactly with the help of (4.49).
It may be written formally as $\frac{\partial ^{2}}{\partial x^{2}}\Psi
_{0}^{a}(x,t_{0}+t_{1}/2)=(C_{2}^{a}x^{2}+C_{1}^{a}x+C_{0}^{a})\Psi
_{0}^{a}(x,t_{0}+t_{1}/2),$ where $C_{2}^{a},$ $C_{1}^{a},$ and $C_{0}^{a}$
are the complex parameters which may be obtained with the aid of (4.49).
Then it is easy to prove that%
\begin{equation*}
||V_{1}^{ho}(x,\varepsilon )^{l}\frac{\partial ^{2}}{\partial x^{2}}\Psi
_{0}^{a}(x,t_{0}+t_{1}/2)||\leq |C_{2}^{a}|\times
||x^{2}V_{1}^{ho}(x,\varepsilon )^{l}\Psi _{0}^{a}(x,t_{0}+t_{1}/2)||
\end{equation*}%
\begin{equation*}
+|C_{1}^{a}|\times ||xV_{1}^{ho}(x,\varepsilon )^{l}\Psi
_{0}^{a}(x,t_{0}+t_{1}/2)||
\end{equation*}%
\begin{equation*}
+|C_{0}^{a}|\times ||V_{1}^{ho}(x,\varepsilon )^{l}\Psi
_{0}^{a}(x,t_{0}+t_{1}/2)||.
\end{equation*}%
It is known from (4.73) that the norm $||x^{j}V_{1}^{ho}(x,\varepsilon )\Psi
_{0}(x,r,t_{0}+t_{1}/2)||$ with $j=0$, $1$, or $2$ has an upper bound
consisting of two basic norms $\{NBAS2\}$. It is also known from (4.79) that
the norm $||x^{l}V_{1}^{ho}(x,\varepsilon )^{2}\Psi
_{0}^{a}(x,t_{0}+t_{1}/2)||$ with $l=0$, $1$, or $2$ has an upper bound
consisting of three basic norms $\{NBAS2\}$. Therefore, the norm $%
||V_{1}^{ho}(x,\varepsilon )^{l}\frac{\partial ^{2}}{\partial x^{2}}\Psi
_{0}^{a}(x,t_{0}+t_{1}/2)||$ with $l=1$ or $2$ has an upper bound consisting
of the basic norms $\{NBAS2\}$. The second sum term on the $RH$ side of
(4.87a) contains the two norms $||\Theta (x-x_{L},\varepsilon )\frac{%
\partial ^{2}}{\partial x^{2}}\Psi _{0}^{a}(x,t_{0}+t_{1}/2)||$ with $a=g$
and $e,$ each one of which has evidently an upper bound consisting of three
basic norms $\{NBAS2\}$. Therefore, the inequalities (4.87a) indicates that
the norm $||P_{2133}^{C}(x,r,t_{0}+t_{1}/2)||$ has an upper bound consisting
of the basic norms $\{NBAS2\}$. Similarly, it can be shown by the inequality
(4.87b) that the norm $||P_{2133}^{S}(x,r,t_{0}+t_{1}/2)||$ has an upper
bound consisting of the basic norms $\{NBAS2\}$.

Now all these terms $SUM1(C/S),$ $SUM2(C/S),$ and $%
||P_{2133}^{C/S}(x,r,t_{0}+t_{1}/2)||$ are shown to have upper bounds
consisting of the two types of basic norms $\{NBAS1\}$ and $\{NBAS2\}$. This
shows that the norm $||p^{2}\Psi _{C/S}^{21}(x,r,t_{0}+t_{1}/2)||$ has an
upper bound consisting of the basic norms $\{NBAS1\}$ and $\{NBAS2\}.$ Then
the inequality,%
\begin{equation*}
||p^{2}\Psi _{er}^{21}(x,r,t_{0}+t_{1}/2)||\leq ||p^{2}\Psi
_{C}^{21}(x,r,t_{0}+t_{1}/2)||+||p^{2}\Psi _{S}^{21}(x,r,t_{0}+t_{1}/2)||,
\end{equation*}%
further indicates that the norm $||p^{2}\Psi _{er}^{21}(x,r,t_{0}+t_{1}/2)||$
has an upper bound consisting of the basic norms $\{NBAS1\}$ and $\{NBAS2\}.$

As a summary, all the six norms on the $RH$ side of (4.70) are shown to have
upper bounds consisting of a finite number of the basic norms $\{NBAS1\}$
and $\{NBAS2\},$ respectively. Then the inequality (4.70) indicates that the
norm $2m||M_{21}^{V}(x,r,t_{1},t_{3})||$ has an upper bound consisting of a
finite number of the basic norms $\{NBAS1\}$ and $\{NBAS2\}.$ This shows
that the norm $||M_{21}^{V}(x,r,t_{1},t_{3})||$ decays exponentially with
the square deviation-to-spread ratios of the $GWP$ states $\{\Psi
_{0}^{a}(x,t_{0}+t_{1}/2)\}.$ This is the desired result! \newline
\newline
{\large 4.3.2 The upper bound of the norm }$||M_{22}^{V}(x,r,t_{1},t_{3})||$

It is very complex to calculate strictly the upper bound of the error state $%
M_{22}^{V}(x,r,t_{1},t_{3})$. The error state $M_{22}^{V}(x,r,t_{1},t_{3})$
of (4.65b) is the product of the commutator $%
[H_{0}^{ho},[H_{0}^{ho},V_{1}(x,\alpha ,\gamma ,\varepsilon )]]$ and the
following product state:%
\begin{equation*}
\exp [-\frac{i}{\hslash }H_{I}(x,\alpha ,\gamma )t_{1}]\exp [-\frac{i}{%
\hslash }V_{1}(x,\alpha ,\gamma ,\varepsilon )t_{1}]\Psi
_{0}(x,r,t_{0}+t_{1}/2)
\end{equation*}%
\begin{equation}
\overset{\text{def}}{\equiv }\Psi _{0}^{C0}(x,r,t_{0}+t_{1}/2)+\Psi
_{0}^{CS}(x,r,t_{0}+t_{1}/2)+\Psi _{0}^{S}(x,r,t_{0}+t_{1}/2).  \tag{4.88}
\end{equation}%
Here the error state $\Psi _{0}^{\mu }(x,r,t_{0}+t_{1}/2)$ with the label $%
\mu =C0,$ $CS,$ or $S$ is determined by calculating explicitly the product
state on the $LH$ side of (4.88). This explicit calculation needs to use the
unitary transformation (4.19) and the product state $\Psi
_{0}(x,r,t_{0}+t_{1}/2)$ of (4.42). According to the calculated result these
three error states in (4.88) are defined by%
\begin{equation}
\Psi _{0}^{C0}(x,r,t_{0}+t_{1}/2)=\Psi _{0}^{g}(x,t_{0}+t_{1}/2)|\tilde{g}%
_{0}\rangle +\Psi _{0}^{e}(x,t_{0}+t_{1}/2)|\tilde{e}\rangle ,  \tag{4.89a}
\end{equation}%
\begin{equation*}
\Psi _{0}^{CS}(x,r,t_{0}+t_{1}/2)=-\{1-\cos [\frac{1}{\hslash }%
t_{1}V_{1}^{ho}(x,\varepsilon )]\cos [\frac{1}{2\hslash }\Theta
(x-x_{L},\varepsilon )\Omega (x)t_{1}]
\end{equation*}%
\begin{equation}
+i\sin [\frac{1}{\hslash }t_{1}V_{1}^{ho}(x,\varepsilon )]\cos [\frac{1}{%
2\hslash }\Theta (x-x_{L},\varepsilon )\Omega (x)t_{1}]\}\Psi
_{0}^{C0}(x,r,t_{0}+t_{1}/2),  \tag{4.89b}
\end{equation}%
and 
\begin{equation*}
\Psi _{0}^{S}(x,r,t_{0}+t_{1}/2)=i\exp [-\frac{i}{\hslash }%
t_{1}V_{1}^{ho}(x,\varepsilon )]\sin [\frac{1}{2\hslash }\Theta
(x-x_{L},\varepsilon )\Omega (x)t_{1}]
\end{equation*}%
\begin{equation}
\times \{\exp [i\varphi (x,\gamma )]\Psi _{0}^{g}(x,t_{0}+t_{1}/2)|\tilde{e}%
\rangle +\exp [-i\varphi (x,\gamma )]\Psi _{0}^{e}(x,t_{0}+t_{1}/2)|\tilde{g}%
_{0}\rangle \},  \tag{4.89c}
\end{equation}%
where the superposition states $|\tilde{g}_{0}\rangle $ and $|\tilde{e}%
\rangle $ are given by (4.75). Both the error states $\Psi
_{0}^{CS}(x,r,t_{0}+t_{1}/2)$ and $\Psi _{0}^{S}(x,r,t_{0}+t_{1}/2)$ are
spatially selective, while $\Psi _{0}^{C0}(x,r,t_{0}+t_{1}/2)$ is not. Then
according to (4.88) the error state $M_{22}^{V}(x,r,t_{1},t_{3})$ may be
divided into the three error states:%
\begin{equation}
M_{22}^{V}(x,r,t_{1},t_{3})=M_{22}^{VC0}(x,r,t_{1},t_{3})+M_{22}^{VCS}(x,r,t_{1},t_{3})+M_{22}^{VS}(x,r,t_{1},t_{3})
\tag{4.90}
\end{equation}%
where the state $M_{22}^{V\mu }(x,r,t_{1},t_{3})$ with the label $\mu =C0,$ $%
CS,$ or $S$ is given by%
\begin{equation}
M_{22}^{V\mu }(x,r,t_{1},t_{3})=[H_{0}^{ho},[H_{0}^{ho},V_{1}(x,\alpha
,\gamma ,\varepsilon )]]\exp [-\frac{i}{2\hslash }H_{0}^{ho}t_{3}]\Psi
_{0}^{\mu }(x,r,t_{0}+t_{1}/2).  \tag{4.91}
\end{equation}%
It is very complex to compute strictly the norms of these error states $%
\{M_{22}^{V\mu }(x,r,$ $t_{1},t_{3})\}.$ This is due to that there is the
propagator $\exp [-\frac{i}{2\hslash }H_{0}^{ho}t_{3}]$ between the
commutator $[H_{0}^{ho},[H_{0}^{ho},V_{1}(x,\alpha ,\gamma ,\varepsilon )]]$
and the non-Gaussian product state $\Psi _{0}^{\mu }(x,r,t_{0}+t_{1}/2)$ in
(4.91). The commutator in (4.91) is proportional to the momentum operators $%
\{p^{k}\}$ with the order $k\leq 2$ and the coordinate operators $\{x^{l}\}$
with the total order $k+l\leq 5.$ Thus, in order to calculate the upper
bound of the error state $M_{22}^{V\mu }(x,r,t_{1},t_{3})$ it is involved in
calculating the fifth-order coordinate derivative of the product state $\Psi
_{0}^{\mu }(x,r,t_{0}+t_{1}/2)$. This is a complex calculational task. On
the other hand, it is quite different in method to calculate the upper bound
of the product state $M_{22}^{VC0}(x,r,t_{1},t_{3})$ from those of the two
states $M_{22}^{VCS}(x,r,t_{1},t_{3})$ and $M_{22}^{VS}(x,r,t_{1},t_{3}).$
The upper bounds for the two error states $M_{22}^{VCS}(x,r,t_{1},t_{3})$
and $M_{22}^{VS}(x,r,t_{1},t_{3})$ may be controlled by the two states $\Psi
_{0}^{CS}(x,r,t_{0}+t)$ and $\Psi _{0}^{S}(x,r,t_{0}+t)$ themselves,
respectively, while that one of the error state $%
M_{22}^{VC0}(x,r,t_{1},t_{3})$ has to be controlled by the commutator $%
[H_{0}^{ho},[H_{0}^{ho},V_{1}(x,\alpha ,\gamma ,\varepsilon )]]$, here the
control factor that is related to a Gaussian factor comes from the
perturbation term $V_{1}(x,\alpha ,\gamma ,\varepsilon )$ in the commutator.
A Gaussian (coordinate) operator does not commute with the
harmonic-oscillator propagator $\exp [-\frac{i}{2\hslash }H_{0}^{ho}t_{3}].$
Moreover, the product state $\Psi _{0}^{\mu }(x,r,t_{0}+t_{1}/2)$ is not a
pure $GWP$ state. These lead to that it is quite complex to calculate
strictly the upper bound of the error state $M_{22}^{VC0}(x,r,t_{1},t_{3}).$
There are two methods to calculate strictly the upper bound of the error
state $M_{22}^{VC0}(x,r,t_{1},t_{3})$ based on

$(i)$ the \textit{M}ultiple \textit{G}aussian \textit{W}ave-\textit{P}acket $%
(MGWP)$ expansion (See also the next section 5)

$(ii)$ the commutation relation $[G(x-x_{c}),$ $\exp (-\frac{i}{2\hslash }%
H_{0}^{ho}t_{3})],$ here $G(x-x_{c})$ is a Gaussian coordinate operator.%
\newline
Either method is not easy! The first method is mainly used below.

In order to calculate conveniently the product state $M_{22}^{V\mu
}(x,r,t_{1},t_{3})$ the commutator $2m[H_{0}^{ho},[H_{0}^{ho},V_{1}(x,\alpha
,\gamma ,\varepsilon )]]$ in (4.91) is divided into the two operators $%
T_{1}(x,p,\varepsilon )$ and $T_{2}(x,p,\varepsilon )$:%
\begin{equation}
2m[H_{0}^{ho},[H_{0}^{ho},V_{1}(x,\alpha ,\gamma ,\varepsilon
)]]=T_{1}(x,p,\varepsilon )+T_{2}(x,p,\varepsilon )  \tag{4.92}
\end{equation}%
where the two operators are given by%
\begin{equation*}
T_{1}(x,p,\varepsilon )=-\frac{2\hslash ^{2}}{m}[\frac{\partial ^{2}}{%
\partial x^{2}}V_{1}^{ho}(x,\varepsilon )]p^{2}+i\frac{2\hslash ^{3}}{m}[%
\frac{\partial ^{3}}{\partial x^{3}}V_{1}^{ho}(x,\varepsilon )]p
\end{equation*}%
\begin{equation}
+2\hslash ^{2}m\omega ^{2}[\frac{\partial }{\partial x}V_{1}^{ho}(x,%
\varepsilon )]x+\frac{\hslash ^{4}}{2m}[\frac{\partial ^{4}}{\partial x^{4}}%
V_{1}^{ho}(x,\varepsilon )]  \tag{4.93a}
\end{equation}%
and%
\begin{equation*}
T_{2}(x,p,\varepsilon )=\frac{2\hslash ^{2}}{m}\{\frac{\partial ^{2}}{%
\partial x^{2}}[H_{I}(x,\alpha ,\gamma )\Theta (x-x_{L},\varepsilon )]\}p^{2}
\end{equation*}%
\begin{equation*}
-i\frac{2\hslash ^{3}}{m}\{\frac{\partial ^{3}}{\partial x^{3}}%
[H_{I}(x,\alpha ,\gamma )\Theta (x-x_{L},\varepsilon )]\}p
\end{equation*}%
\begin{equation}
-2\hslash ^{2}m\omega ^{2}\{\frac{\partial }{\partial x}[H_{I}(x,\alpha
,\gamma )\Theta (x-x_{L},\varepsilon )]\}x-\frac{\hslash ^{4}}{2m}\frac{%
\partial ^{4}}{\partial x^{4}}[H_{I}(x,\alpha ,\gamma )\Theta
(x-x_{L},\varepsilon )].  \tag{4.93b}
\end{equation}%
The operator $T_{1}(x,p,\varepsilon )$ is independent on the interaction $%
H_{I}(x,\alpha ,\gamma ).$ By substituting (4.92) into (4.91) one can find
that the error state $M_{22}^{V\mu }(x,r,t_{1},t_{3})$ is bounded by%
\begin{equation*}
||M_{22}^{V\mu }(x,r,t_{1},t_{3})||\leq ||T_{1}(x,p,\varepsilon )\exp [-%
\frac{i}{2\hslash }H_{0}^{ho}t_{3}]\Psi _{0}^{\mu }(x,r,t_{0}+t_{1}/2)||
\end{equation*}%
\begin{equation}
+||T_{2}(x,p,\varepsilon )\exp [-\frac{i}{2\hslash }H_{0}^{ho}t_{3}]\Psi
_{0}^{\mu }(x,r,t_{0}+t_{1}/2)||.  \tag{4.94}
\end{equation}%
Then by using this inequality one further finds from (4.90) that%
\begin{equation*}
2m||M_{22}^{V}(x,r,t_{1},t_{3})||\leq \sum_{\mu
=C0,CS,S}\{||T_{1}(x,p,\varepsilon )\exp [-\frac{i}{2\hslash }%
H_{0}^{ho}t_{3}]\Psi _{0}^{\mu }(x,r,t_{0}+t_{1}/2)||
\end{equation*}%
\begin{equation}
+||T_{2}(x,p,\varepsilon )\exp [-\frac{i}{2\hslash }H_{0}^{ho}t_{3}]\Psi
_{0}^{\mu }(x,r,t_{0}+t_{1}/2)||\}.  \tag{4.95}
\end{equation}%
Therefore, one may calculate the six norms on the $RH$ side of (4.95) to
determine the upper bound of the norm $||M_{22}^{V}(x,r,t_{1},t_{3})||.$

Now by inserting the continuous perturbation term $V_{1}^{ho}(x,\varepsilon
) $ of (2.10) into (4.93a) one obtains%
\begin{equation*}
T_{1}(x,p,\varepsilon )=\delta (x-x_{L},\varepsilon )\Theta
(x_{L}+L-x,\varepsilon )\Gamma _{1}(x,p)
\end{equation*}%
\begin{equation*}
+\delta (x-x_{L},\varepsilon )\delta (x-x_{L}-L,\varepsilon )\Gamma _{2}(x,p)
\end{equation*}%
\begin{equation*}
+\Theta (x-x_{L},\varepsilon )\delta (x_{L}+L-x,\varepsilon )\Gamma _{3}(x,p)
\end{equation*}%
\begin{equation}
+\Theta (x-x_{L},\varepsilon )\Gamma _{4}(x,p)+\delta (x-x_{L},\varepsilon
)\Gamma _{5}(x,p).  \tag{4.96}
\end{equation}%
Here $\{\Gamma _{k}(x,p)\}$ with $k=1$, $2$, ..., $5$ are polynomials in
momentum operator $p$ and coordinate operator $x$. The order of momentum
operator $p$ in all these operator polynomials $\{\Gamma _{k}(x,p)\}$ is not
more than two and the total order for each one of these polynomial operators
is not more than five. In particular, the polynomial operator $\Gamma
_{5}(x,p)$ contains the operators $x^{3}p^{2},$ $x^{4}p$, and $x^{5}$ whose
polynomial order are highest. The operator $T_{1}(x,p,\varepsilon )$ of
(4.96) leads to that the upper bound of the first norm on the $RH$ side of
(4.95) for any given index $\mu $ may be determined from%
\begin{equation*}
||T_{1}(x,p,\varepsilon )\exp [-\frac{i}{2\hslash }H_{0}^{ho}t_{3}]\Psi
_{0}^{\mu }(x,r,t_{0}+t_{1}/2)||
\end{equation*}%
\begin{equation*}
\leq ||\delta (x-x_{L},\varepsilon )\exp [-\frac{i}{2\hslash }%
H_{0}^{ho}t_{3}]\Gamma _{1}(x(t_{3}/2),p(t_{3}/2))\Psi _{0}^{\mu
}(x,r,t_{0}+t_{1}/2)||
\end{equation*}%
\begin{equation*}
+||\delta (x-x_{L},\varepsilon )\delta (x-x_{L}-L,\varepsilon )\exp [-\frac{i%
}{2\hslash }H_{0}^{ho}t_{3}]
\end{equation*}%
\begin{equation*}
\times \Gamma _{2}(x(t_{3}/2),p(t_{3}/2))\Psi _{0}^{\mu
}(x,r,t_{0}+t_{1}/2)||
\end{equation*}%
\begin{equation*}
+||\delta (x-x_{L}-L,\varepsilon )\exp [-\frac{i}{2\hslash }%
H_{0}^{ho}t_{3}]\Gamma _{3}(x(t_{3}/2),p(t_{3}/2))\Psi _{0}^{\mu
}(x,r,t_{0}+t_{1}/2)||
\end{equation*}%
\begin{equation*}
+||\Theta (x-x_{L},\varepsilon )\exp [-\frac{i}{2\hslash }%
H_{0}^{ho}t_{3}]\Gamma _{4}(x(t_{3}/2),p(t_{3}/2))\Psi _{0}^{\mu
}(x,r,t_{0}+t_{1}/2)||
\end{equation*}%
\begin{equation}
+||\delta (x-x_{L},\varepsilon )\exp [-\frac{i}{2\hslash }%
H_{0}^{ho}t_{3}]\Gamma _{5}(x(t_{3}/2),p(t_{3}/2))\Psi _{0}^{\mu
}(x,r,t_{0}+t_{1}/2)||.  \tag{4.97}
\end{equation}%
Here in the Heisenberg picture the momentum operator $p(t_{3}/2)$ and
coordinate operator $x(t_{3}/2)$ are given by (4.69), respectively, and the
operator polynomial $\Gamma _{k}(x(t_{3}/2),p(t_{3}/2))$ with $k=1,$ $2$, $%
...$, $5$ is written as%
\begin{equation}
\Gamma _{k}(x(t_{3}/2),p(t_{3}/2))=\exp [\frac{i}{2\hslash }%
H_{0}^{ho}t_{3}]\Gamma _{k}(x,p)\exp [-\frac{i}{2\hslash }H_{0}^{ho}t_{3}]. 
\tag{4.98}
\end{equation}%
On the other hand, after the operator $T_{2}(x,p,\varepsilon )$ of (4.93b)
is further simplified, it is used to calculate the second norm on the $RH$
side of (4.95) for any given index $\mu $. It can turn out that this norm is
bounded by%
\begin{equation*}
||T_{2}(x,p,\varepsilon )\exp [-\frac{i}{2\hslash }H_{0}^{ho}t_{3}]\Psi
_{0}^{\mu }(x,r,t_{0}+t_{1}/2)||
\end{equation*}%
\begin{equation*}
\leq \sum_{j=0}^{1}A_{j}^{T2}||\Theta (x-x_{L},\varepsilon )\exp [-\frac{i}{%
2\hslash }H_{0}^{ho}t_{3}]x(t_{3}/2)^{j}\Psi _{0}^{\mu }(x,r,t_{0}+t_{1}/2)||
\end{equation*}%
\begin{equation*}
+\sum_{j=1}^{2}B_{j}^{T2}||\Theta (x-x_{L},\varepsilon )\exp [-\frac{i}{%
2\hslash }H_{0}^{ho}t_{3}]p(t_{3}/2)^{j}\Psi _{0}^{\mu }(x,r,t_{0}+t_{1}/2)||
\end{equation*}%
\begin{equation}
+\sum_{k=0}^{3}\sum_{l=0}^{2}C_{kl}^{T2}||\delta (x-x_{L},\varepsilon )\exp
[-\frac{i}{2\hslash }H_{0}^{ho}t_{3}]x(t_{3}/2)^{k}p(t_{3}/2)^{l}\Psi
_{0}^{\mu }(x,r,t_{0}+t_{1}/2)||.  \tag{4.99}
\end{equation}%
Here the parameter $C_{kl}^{T2}=0$ if the subscript sum $k+l>3,$ indicating
that the total order $(k+l)$ for any operator polynomial $%
x(t_{3}/2)^{k}p(t_{3}/2)^{l}$ is not more than three on the $RH$ side of
(4.99). These non-negative parameters $\{A_{j}^{T2}\},$ $\{B_{j}^{T2}\}$,
and $\{C_{kl}^{T2}\}$ in (4.99) can be exactly calculated by using the
interaction $H_{I}(x,\alpha ,\gamma )$ and other relevant parameters. They
are bounded and controllable. The two inequalities (4.97) and (4.99) will be
used to determine the upper bounds of the norms $||M_{22}^{V\mu
}(x,r,t_{1},t_{3})||$ with the label $\mu =C0,$ $CS,$ and $S.$

There are five norms on the $RH$ side of (4.97), while there are also
thirteen norms on the $RH$ side of (4.99). Then in order to prove that these
norms $\{||M_{22}^{V\mu }(x,r,t_{1},t_{3})||\}$ decay exponentially with the
square deviation-to-spread ratios one may first prove that all these norms
on the $RH$ sides of (4.97) and (4.99) decay exponentially with the square
deviation-to-spread ratios. Note that the total polynomial order is not more
than five for each one of these five operator polynomials $\{\Gamma
_{k}(x,p)\}$ in (4.96). By substituting the unitary transformations (4.69)
into (4.98) one can find that any one of the five polynomials $\{\Gamma
_{k}(x(t_{3}/2),p(t_{3}/2))\}$ in momentum and coordinate operators on the $%
RH$ side of (4.97) also has a total order not more than five. Similarly, it
can be seen that any operator polynomial $x(t_{3}/2)^{k}p(t_{3}/2)^{l}$ in
momentum and coordinate operators on the $RH$ side of (4.99) has a total
order not more than three. Furthermore, by using the basic commutation
relation $[x,p]=i\hslash $ one always can express any one of the operator
polynomials $\{\Gamma _{k}(x(t_{3}/2),p(t_{3}/2))\}$ in (4.97) and $%
\{x(t_{3}/2)^{k}p(t_{3}/2)^{l}\}$ in (4.99) as%
\begin{equation}
Q_{\lambda }(x,p,t_{3})=F_{0}^{\lambda }(x,t_{3})p^{5}+F_{1}^{\lambda
}(x,t_{3})p^{4}+...+F_{4}^{\lambda }(x,t_{3})p+F_{5}^{\lambda }(x,t_{3}), 
\tag{4.100}
\end{equation}%
where $F_{l}^{\lambda }(x,t_{3})$ is a polynomial in coordinate $x$ and its
polynomial order is not more than the index $l$ for $l=0,1,...,5.$ The
polynomial $F_{l}^{\lambda }(x,t_{3})$ may be generally written as%
\begin{equation}
F_{l}^{\lambda }(x,t_{3})=F_{l,l}^{\lambda }(t_{3})x^{l}+F_{l,l-1}^{\lambda
}(t_{3})x^{l-1}+...+F_{l,1}^{\lambda }(t_{3})x+F_{l,0}^{\lambda }(t_{3}), 
\tag{4.101}
\end{equation}%
where all these parameters $\{F_{l,l^{\prime }}^{\lambda }(t_{3})\}$ $(0\leq
l^{\prime }\leq l)$ are bounded and controllable. For each one of the
operator polynomials $\{\Gamma _{k}(x(t_{3}/2),p(t_{3}/2))\}$ in (4.97) and $%
\{x(t_{3}/2)^{k}p(t_{3}/2)^{l}\}$ in (4.99) there may be different
polynomial $\{F_{l}^{\lambda }(x,t_{3})\}$ in (4.100). For example, any
operator polynomial $x(t_{3}/2)^{k}p(t_{3}/2)^{l}$ in (4.99) has a total
order not more than three. Then there are always $F_{0}^{\lambda
}(x,t_{3})=0 $ and $F_{1}^{\lambda }(x,t_{3})=0$ in (4.100) when any
operator polynomial $x(t_{3}/2)^{k}p(t_{3}/2)^{l}$ in (4.99) is written
formally as the operator polynomial $Q_{\lambda }(x,p,t_{3})$ of (4.100).
Now by using the operator polynomial $Q_{\lambda }(x,p,t_{3})$ every norm on
the $RH$ sides of (4.97) and (4.99) may be written as%
\begin{equation}
NORM(1,\lambda ,\mu )=||\delta (x-x_{c},\varepsilon _{c})\exp [-\frac{i}{%
2\hslash }H_{0}^{ho}t_{3}]Q_{\lambda }(x,p,t_{3})\Psi _{0}^{\mu
}(x,r,t_{0}+t_{1}/2)||  \tag{4.102}
\end{equation}%
or 
\begin{equation}
NORM(2,\lambda ,\mu )=||\Theta (x-x_{L},\varepsilon )\exp [-\frac{i}{%
2\hslash }H_{0}^{ho}t_{3}]Q_{\lambda }(x,p,t_{3})\Psi _{0}^{\mu
}(x,r,t_{0}+t_{1}/2)||.  \tag{4.103}
\end{equation}%
Here the parameters $x_{c}$ and $\varepsilon _{c}$ in the smooth $\delta -$%
function $\delta (x-x_{c},\varepsilon _{c})$ are given as follows. Except
for the second and third norms on the $RH$ side of (4.97) the parameter $%
x_{c}=x_{L}$ and $\varepsilon _{c}=\varepsilon $ for any one of the norms on
the $RH$ sides of (4.97) and (4.99). The parameter $x_{c}=x_{L}+L$ and $%
\varepsilon _{c}=\varepsilon $ for the third norm on the $RH$ side of
(4.97). With the help of the relation (3.35) the second norm on the $RH$
side of (4.97) may be rewritten as%
\begin{equation*}
\frac{1}{\varepsilon \sqrt{2\pi }}\exp (-\frac{1}{2}\frac{L^{2}}{\varepsilon
^{2}})||\delta (x-x_{L}-L/2,\varepsilon /\sqrt{2})\exp [-\frac{i}{2\hslash }%
H_{0}^{ho}t_{3}]
\end{equation*}%
\begin{equation*}
\times \Gamma _{2}(x(t_{3}/2),p(t_{3}/2))\Psi _{0}^{\mu
}(x,r,t_{0}+t_{1}/2)||.
\end{equation*}%
Thus, for this norm the parameter $x_{c}=x_{L}+L/2$ and $\varepsilon
_{c}=\varepsilon /\sqrt{2}.$ In comparison with other norms on the $RH$
sides of (4.97) and (4.99) this norm has an extra exponentially-decaying
factor $\exp (-\frac{1}{2}\frac{L^{2}}{\varepsilon ^{2}}).$ When the length $%
L>>\varepsilon ,$ this exponentially-decaying factor is so small that this
norm can be neglected with respect to other norms on the $RH$ sides of
(4.97) and (4.99). Thus, a norm such as the second norm on the $RH$ side of
(4.97) that contains at least two $\delta -$functions with different COM
positions may be neglected in calculation. It can be found that there are
four norms $\{NORM(1,\lambda ,\mu )\}$ and one norm $NORM(2,\lambda ,\mu )$
on the $RH$ side of (4.97), while there are nine norms $\{NORM(1,\lambda
,\mu )\}$ and four norms $\{NORM(2,\lambda ,\mu )\}$ on the $RH$ side of
(4.99). Because every one of the norms on the $RH$ sides of (4.97) and
(4.99) is either the norm $NORM(1,\lambda ,\mu )$ of (4.102) or $%
NORM(2,\lambda ,\mu )$ of (4.103), one needs only to calculate strictly the
two norms $NORM(1,\lambda ,\mu )$ and $NORM(2,\lambda ,\mu )$ below. \newline
\newline
{\Large 4.3.2.1 The upper bounds for the norms }$NORM(k,\lambda ,\mu )$%
{\Large \ with }$\mu =CS${\Large \ and }$S$

Because both the error states $\Psi _{0}^{\mu }(x,r,t_{0}+t_{1}/2)$ with the
index $\mu =CS$ and $S$ contain the spatially selective functions, as can be
seen below, all the norms inside containing the states $\Psi _{0}^{\mu
}(x,r,t_{0}+t_{1}/2)$ with the index $\mu =CS$ and $S$ on the $RH$ sides of
(4.97) and (4.99) may be controlled by the states $\Psi _{0}^{\mu
}(x,r,t_{0}+t_{1}/2)$ themselves. On the other hand, since the state $\Psi
_{0}^{C0}(x,r,t_{0}+t_{1}/2)$ does not contain any spatially selective
function, all these norms inside containing the state $\Psi
_{0}^{C0}(x,r,t_{0}+t_{1}/2)$ on the $RH$ sides of (4.97) and (4.99) must be
controlled by the spatially selective factor $\Theta (x-x_{L},\varepsilon )$
or $\delta (x-x_{c},\varepsilon _{c})$ inside these norms. These result in
that it is different to calculate the norms inside containing the states $%
\Psi _{0}^{\mu }(x,r,t_{0}+t_{1}/2)$ with $\mu =CS$ and $S$ from those norms
inside containing the state $\Psi _{0}^{C0}(x,r,t_{0}+t_{1}/2).$ In this
subsection it is carried out to calculate strictly the norms $NORM(1,\lambda
,\mu )$ and $NORM(2,\lambda ,\mu )$ that contain the states $\Psi _{0}^{\mu
}(x,r,t_{0}+t_{1}/2)$ with $\mu =CS$ and $S.$ These norms may be controlled
by the states $\Psi _{0}^{\mu }(x,r,t_{0}+t_{1}/2)$ with $\mu =CS$ and $S.$
It is known that the continuous step function $\Theta (x-x_{L},\varepsilon )$
and $\delta -$function $\delta (x-x_{c},\varepsilon _{c})$ are defined by
(2.11) and (3.32), respectively. It can be seen from these definitions that $%
0\leq \Theta (x-x_{L},\varepsilon )\leq 1$ and $0\leq \delta
(x-x_{c},\varepsilon _{c})\leq \frac{1}{\varepsilon _{c}\sqrt{\pi }}.$ Then
these two inequalities lead to that the two norms (4.102) and (4.103) can be
reduced to the simpler forms%
\begin{equation}
NORM(1,\lambda ,\mu )\leq \frac{1}{\varepsilon _{c}\sqrt{\pi }}||Q_{\lambda
}(x,p,t_{3})\Psi _{0}^{\mu }(x,r,t_{0}+t_{1}/2)||,\text{ }\mu =CS,\text{ }S,
\tag{4.104a}
\end{equation}%
and 
\begin{equation}
NORM(2,\lambda ,\mu )\leq ||Q_{\lambda }(x,p,t_{3})\Psi _{0}^{\mu
}(x,r,t_{0}+t_{1}/2)||,\text{ }\mu =CS,\text{ }S.  \tag{4.104b}
\end{equation}%
One sees that the harmonic-oscillator propagator $\exp [-\frac{i}{2\hslash }%
H_{0}^{ho}t_{3}]$ disappears on the $RH$ sides of (4.104). By these two
inequalities (4.104) one may calculate the upper bounds of the two norms $%
NORM(1,\lambda ,\mu )$ and $NORM(2,\lambda ,\mu ).$ Actually, in order to
determine these upper bounds of $NORM(1,\lambda ,\mu )$ and $NORM(2,\lambda
,\mu )$ one needs only to calculate the norm $||Q_{\lambda }(x,p,t_{3})\Psi
_{0}^{\mu }(x,r,$ $t_{0}+t_{1}/2)||$ with $\mu =CS,$ $S.$ Now by using the
operator polynomial $Q_{\lambda }(x,p,t_{3})$ of (4.100) it can turn out
that the norm is bounded by%
\begin{equation}
||Q_{\lambda }(x,p,t_{3})\Psi _{0}^{\mu }(x,r,t_{0}+t_{1}/2)||\leq
\sum_{l=0}^{5}||F_{5-l}^{\lambda }(x,t_{3})p^{l}\Psi _{0}^{\mu
}(x,r,t_{0}+t_{1}/2)||.  \tag{4.105}
\end{equation}%
The six norms on the $RH$ side of (4.105) can be calculated strictly and
directly. The complexity to calculate these norms consists in that one needs
to calculate the higher-order (up to the fifth order) coordinate derivatives
of the states $\{\Psi _{0}^{\mu }(x,r,t_{0}+t_{1}/2)\}.$ It follows from
(4.89b) and (4.89c) that the states $\Psi _{0}^{\mu }(x,r,t_{0}+t_{1}/2)$
with $\mu =CS$ and $S$ always can be expressed as the product of a
spatially-selective function $F_{0}^{\mu }$ and a product state $\tilde{\Psi}%
_{0}^{\mu }(x,r,t_{0}+t_{1}/2)$, 
\begin{equation}
\Psi _{0}^{\mu }(x,r,t_{0}+t_{1}/2)=F_{0}^{\mu }\tilde{\Psi}_{0}^{\mu
}(x,r,t_{0}+t_{1}/2),  \tag{4.106a}
\end{equation}%
where for the state $\Psi _{0}^{CS}(x,r,t_{0}+t_{1}/2)$ the
spatially-selective function is given by 
\begin{equation*}
F_{0}^{CS}=-\{1-\cos [\frac{1}{\hslash }t_{1}V_{1}^{ho}(x,\varepsilon )]\cos
[\frac{1}{2\hslash }\Theta (x-x_{L},\varepsilon )\Omega (x)t_{1}]
\end{equation*}%
\begin{equation}
+i\sin [\frac{1}{\hslash }t_{1}V_{1}^{ho}(x,\varepsilon )]\cos [\frac{1}{%
2\hslash }\Theta (x-x_{L},\varepsilon )\Omega (x)t_{1}]\}  \tag{4.106b}
\end{equation}%
and the product state is just $\Psi _{0}^{C0}(x,r,t_{0}+t)$ of (4.89a), 
\begin{equation}
\tilde{\Psi}_{0}^{CS}(x,r,t_{0}+t_{1}/2)=\Psi _{0}^{C0}(x,r,t_{0}+t); 
\tag{4.106c}
\end{equation}%
while for the state $\Psi _{0}^{S}(x,r,t_{0}+t_{1}/2)$ the
spatially-selective function is given by%
\begin{equation}
F_{0}^{S}=i\exp [-\frac{i}{\hslash }t_{1}V_{1}^{ho}(x,\varepsilon )]\sin [%
\frac{1}{2\hslash }\Theta (x-x_{L},\varepsilon )\Omega (x)t_{1}] 
\tag{4.106d}
\end{equation}%
and the product state is%
\begin{equation*}
\tilde{\Psi}_{0}^{S}(x,r,t_{0}+t_{1}/2)=\exp [i\varphi (x,\gamma )]\Psi
_{0}^{g}(x,t_{0}+t_{1}/2)|\tilde{e}\rangle
\end{equation*}%
\begin{equation}
+\exp [-i\varphi (x,\gamma )]\Psi _{0}^{e}(x,t_{0}+t_{1}/2)|\tilde{g}%
_{0}\rangle .  \tag{4.106e}
\end{equation}%
By applying the $m-$th power of momentum operator $p^{m}$ $(1\leq m\leq 5)$
to the state of (4.106a) one obtains%
\begin{equation*}
p^{m}\Psi _{0}^{\mu }(x,r,t_{0}+t_{1}/2)=(-i\hslash \partial /\partial
x)^{m}\{F_{0}^{\mu }\tilde{\Psi}_{0}^{\mu }(x,r,t_{0}+t_{1}/2)\}
\end{equation*}%
\begin{equation}
=(-i\hslash )^{m}\sum_{j=0}^{m}\left( 
\begin{array}{c}
m \\ 
j%
\end{array}%
\right) [\frac{\partial ^{m-j}}{\partial x^{m-j}}F_{0}^{\mu }]\times \lbrack 
\frac{\partial ^{j}}{\partial x^{j}}\tilde{\Psi}_{0}^{\mu
}(x,r,t_{0}+t_{1}/2)].  \tag{4.107}
\end{equation}%
Then by substituting (4.107) into (4.105) one may calculate any one of the
six norms on the $RH$ side of (4.105). In order to simplify further the
product state (4.107) it is needed to calculate explicitly any $j-$order $%
(0\leq j\leq m\leq 5$) coordinate derivatives of both the function $%
F_{0}^{\mu }$ and the product state $\tilde{\Psi}_{0}^{\mu
}(x,r,t_{0}+t_{1}/2)$ in (4.106a).

By using (4.106c) and (4.89a) one can directly calculate the $j-$order
coordinate derivative of the product state $\tilde{\Psi}%
_{0}^{CS}(x,r,t_{0}+t_{1}/2).$ It can be found that the derivative may be
expressed as%
\begin{equation*}
\frac{\partial ^{j}}{\partial x^{j}}\Psi
_{0}^{C0}(x,r,t_{0}+t_{1}/2)=\sum_{l=0}^{j}\left( 
\begin{array}{c}
j \\ 
l%
\end{array}%
\right) \frac{\partial ^{l}}{\partial x^{l}}|\tilde{g}_{0}\rangle \times 
\frac{\partial ^{j-l}}{\partial x^{j-l}}\Psi _{0}^{g}(x,t_{0}+t_{1}/2)
\end{equation*}%
\begin{equation}
+\sum_{l=0}^{j}\left( 
\begin{array}{c}
j \\ 
l%
\end{array}%
\right) \frac{\partial ^{l}}{\partial x^{l}}|\tilde{e}\rangle \times \frac{%
\partial ^{j-l}}{\partial x^{j-l}}\Psi _{0}^{e}(x,t_{0}+t_{1}/2). 
\tag{4.108a}
\end{equation}%
Similarly, by using (4.106e) one can directly calculate the $j-$order
coordinate derivative of the product state $\tilde{\Psi}%
_{0}^{S}(x,r,t_{0}+t_{1}/2),$ 
\begin{equation*}
\frac{\partial ^{j}}{\partial x^{j}}\tilde{\Psi}_{0}^{S}(x,r,t_{0}+t_{1}/2)=%
\sum_{l=0}^{j}\left( 
\begin{array}{c}
j \\ 
l%
\end{array}%
\right) \frac{\partial ^{l}}{\partial x^{l}}|\tilde{e}\rangle \times \frac{%
\partial ^{j-l}}{\partial x^{j-l}}\{\exp [i\varphi (x,\gamma )]\Psi
_{0}^{g}(x,t_{0}+t_{1}/2)\}
\end{equation*}%
\begin{equation}
+\sum_{l=0}^{j}\left( 
\begin{array}{c}
j \\ 
l%
\end{array}%
\right) \frac{\partial ^{l}}{\partial x^{l}}|\tilde{g}_{0}\rangle \times 
\frac{\partial ^{j-l}}{\partial x^{j-l}}\{\exp [-i\varphi (x,\gamma )]\Psi
_{0}^{e}(x,t_{0}+t_{1}/2)\}.  \tag{4.108b}
\end{equation}%
Here the two $l-$order coordinate derivatives $\frac{\partial ^{l}}{\partial
x^{l}}|\tilde{g}_{0}\rangle $ and $\frac{\partial ^{l}}{\partial x^{l}}|%
\tilde{e}\rangle $ can be calculated by using the normalization
superposition states $|\tilde{g}_{0}\rangle $ of (4.75a) and $|\tilde{e}%
\rangle $ of (4.75b), respectively. These coordinate derivatives of (4.108)
can be further simplified. On the basis of (4.49) it can turn out that the $%
j-$order ($j\geq 0$) coordinate derivatives of the wave functions $\Psi
_{0}^{a}(x,t_{0}+t_{1}/2)$ and $\{\exp [\pm i\varphi (x,\gamma )]\Psi
_{0}^{a}(x,t_{0}+t_{1}/2)\}$ are given by%
\begin{equation}
\frac{\partial ^{j}}{\partial x^{j}}\Psi
_{0}^{a}(x,t_{0}+t_{1}/2)=Q_{j}^{a}(x)\Psi _{0}^{a}(x,t_{0}+t_{1}/2), 
\tag{4.109a}
\end{equation}%
\begin{equation*}
\frac{\partial ^{j}}{\partial x^{j}}\{\exp [\pm i\varphi (x,\gamma )]\Psi
_{0}^{a}(x,t_{0}+t_{1}/2)\}
\end{equation*}%
\begin{equation}
=Q_{j}^{a}(x,\pm \varphi (x,\gamma ))\{\exp [\pm i\varphi (x,\gamma )]\Psi
_{0}^{a}(x,t_{0}+t_{1}/2)\},  \tag{4.109b}
\end{equation}%
where both $Q_{j}^{a}(x)$ and $Q_{j}^{a}(x,\pm \varphi (x,\gamma ))$ are the 
$j-$order polynomials in coordinate $x$ and particularly, $%
Q_{0}^{a}(x)=Q_{0}^{a}(x,\pm \varphi (x,\gamma ))=1.$ The polynomial $%
Q_{j}^{a}(x,\pm \varphi (x,\gamma ))$ is different from $Q_{j}^{a}(x)$ only
in that $Q_{j}^{a}(x,\pm \varphi (x,\gamma ))$ is dependent on $\varphi
(x,\gamma ),$ while $Q_{j}^{a}(x)$ is not. Now by substituting (4.109a) into
(4.108a) and (4.109b) into (4.108b) one obtains%
\begin{equation*}
\frac{\partial ^{j}}{\partial x^{j}}\Psi
_{0}^{C0}(x,r,t_{0}+t_{1}/2)=\sum_{l=0}^{j}\left( 
\begin{array}{c}
j \\ 
l%
\end{array}%
\right) (\frac{\partial ^{l}}{\partial x^{l}}|\tilde{g}_{0}\rangle
)Q_{j-l}^{g}(x)\Psi _{0}^{g}(x,t_{0}+t_{1}/2)
\end{equation*}%
\begin{equation}
+\sum_{l=0}^{j}\left( 
\begin{array}{c}
j \\ 
l%
\end{array}%
\right) (\frac{\partial ^{l}}{\partial x^{l}}|\tilde{e}\rangle
)Q_{j-l}^{e}(x)\Psi _{0}^{e}(x,t_{0}+t_{1}/2)  \tag{4.110a}
\end{equation}%
and%
\begin{equation*}
\frac{\partial ^{j}}{\partial x^{j}}\tilde{\Psi}_{0}^{S}(x,r,t_{0}+t_{1}/2)=%
\sum_{l=0}^{j}\left( 
\begin{array}{c}
j \\ 
l%
\end{array}%
\right) (\frac{\partial ^{l}}{\partial x^{l}}|\tilde{e}\rangle
)Q_{j-l}^{g}(x,+\varphi (x,\gamma ))
\end{equation*}%
\begin{equation*}
\times \exp [i\varphi (x,\gamma )]\Psi _{0}^{g}(x,t_{0}+t_{1}/2)
\end{equation*}%
\begin{equation}
+\sum_{l=0}^{j}\left( 
\begin{array}{c}
j \\ 
l%
\end{array}%
\right) (\frac{\partial ^{l}}{\partial x^{l}}|\tilde{g}_{0}\rangle
)Q_{j-l}^{e}(x,-\varphi (x,\gamma ))\exp [-i\varphi (x,\gamma )]\Psi
_{0}^{e}(x,t_{0}+t_{1}/2).  \tag{4.110b}
\end{equation}%
Both the equations (4.110) show that the derivatives $\frac{\partial ^{j}}{%
\partial x^{j}}\Psi _{0}^{C0}(x,r,t_{0}+t_{1}/2)$ and $\frac{\partial ^{j}}{%
\partial x^{j}}\tilde{\Psi}_{0}^{S}(x,r,t_{0}+t_{1}/2)$ are proportional to
the wave functions $\Psi _{0}^{a}(x,t_{0}+t_{1}/2)$ with $a=g$ and $e.$ By
these two equations these coordinate derivatives of the wave functions $\Psi
_{0}^{C0}(x,r,t_{0}+t_{1}/2)$ and $\tilde{\Psi}_{0}^{S}(x,r,t_{0}+t_{1}/2)$
can be converted into the wave functions $\{\Psi _{0}^{a}(x,t_{0}+t_{1}/2)\}$%
. This property is very useful for the rigorous error estimation below. By
substituting these coordinate derivatives into (4.107) one can find that the
product state $p^{m}\Psi _{0}^{\mu }(x,r,t_{0}+t_{1}/2)$ is really
proportional to the wave functions $\Psi _{0}^{a}(x,t_{0}+t_{1}/2)$ with $%
a=g $ and $e.$ On the other hand, the coordinate derivatives for the two
functions $F_{0}^{CS}$ of (4.106b) and $F_{0}^{S}$ of (4.106d) also can be
calculated directly. They are given later.

For convenience to use these coordinate derivatives of (4.110) in the error
estimation later it needs to prove that the norms of the two derivatives $%
\frac{\partial ^{l}}{\partial x^{l}}|\tilde{g}_{0}\rangle $ and $\frac{%
\partial ^{l}}{\partial x^{l}}|\tilde{e}\rangle $ are bounded. According to
(4.75) this can be done by proving that both the derivatives $\frac{\partial
^{l}}{\partial x^{l}}\cos [\frac{1}{2\hslash }\Omega (x)t_{1}]$ and $\frac{%
\partial ^{l}}{\partial x^{l}}\{\exp [\pm i\varphi (x,\gamma )]\sin [\frac{1%
}{2\hslash }\Omega (x)t_{1}]\}$ are bounded. It is known that the two
functions $\Omega (x)$ and $\varphi (x,\gamma )$ are given in (4.18) and
(4.19), respectively. Both the functions are bounded. Then the coordinate
derivatives for the functions $\varphi (x,\gamma )$ and $\Omega (x)$ are
given by $\frac{\partial }{\partial x}\varphi (x,\gamma )=\frac{1}{2}%
(k_{0}+k_{1})$ and $\frac{\partial ^{l}}{\partial x^{l}}\varphi (x,\gamma
)=0 $ for $l>1$ and%
\begin{equation*}
\frac{\partial ^{2l+1}}{\partial x^{2l+1}}\Omega (x)=(-1)^{l+1}(\frac{1}{2}%
\Delta k)^{2l+1}(4\hslash \Omega _{0})\sin [\frac{1}{2}\Delta kx-\pi /4],
\end{equation*}%
\begin{equation*}
\frac{\partial ^{2l}}{\partial x^{2l}}\Omega (x)=(-1)^{l}(\frac{1}{2}\Delta
k)^{2l}(4\hslash \Omega _{0})\cos [\frac{1}{2}\Delta kx-\pi /4],\text{ }%
l=0,1,....
\end{equation*}%
Here the zero-order derivative of a function is defined as the function
itself. These formulae show that the $l-$order derivatives $\frac{\partial
^{l}}{\partial x^{l}}\Omega (x)$ and $\frac{\partial ^{l}}{\partial x^{l}}%
\varphi (x,\gamma )$ ($l=0$, $1$, $2$, ...) are bounded. Now it can turn out
that the $l-$order coordinate derivative of the phase factor $\exp [\pm i%
\frac{1}{2\hslash }\Omega (x)t_{1}]$ ($l=0$, $1$, $2$, ...) may be generally
written as%
\begin{equation*}
\frac{\partial ^{l}}{\partial x^{l}}\exp [\pm i\frac{1}{2\hslash }\Omega
(x)t_{1}]=F_{\pm l}(x)\exp [\pm i\frac{1}{2\hslash }\Omega (x)t_{1}]
\end{equation*}%
where the $l-$th function $F_{\pm l}(x)$ ($l>0$) is determined from the
recursive relation: 
\begin{equation*}
F_{\pm l}(x)=\frac{\partial }{\partial x}F_{\pm (l-1)}(x)+(\pm i\frac{1}{%
2\hslash }t_{1})[\frac{\partial }{\partial x}\Omega (x)]F_{\pm (l-1)}(x),
\end{equation*}%
and particularly, the zeroth and first functions are given by $F_{0}(x)=1$
and $F_{\pm 1}(x)=(\pm i\frac{1}{2\hslash }t_{1})[\frac{\partial }{\partial x%
}\Omega (x)],$ respectively. This recursive relation shows that the $l-$th
function $F_{\pm l}(x)$ is composed of the derivatives $\{\frac{\partial ^{k}%
}{\partial x^{k}}\Omega (x)\}$ with order $k\leq l.$ Because the derivative $%
\frac{\partial ^{k}}{\partial x^{k}}\Omega (x)$ for $k=0$, $1$, $2$, ... is
bounded, the $l-$th function $F_{\pm l}(x)$ is bounded too, indicating that
the derivative $\frac{\partial ^{l}}{\partial x^{l}}\exp [\pm i\frac{1}{%
2\hslash }\Omega (x)t_{1}]$ is bounded. This further shows that both the $l-$%
order derivatives $\frac{\partial ^{l}}{\partial x^{l}}\cos [\frac{1}{%
2\hslash }\Omega (x)t_{1}]$ and $\frac{\partial ^{l}}{\partial x^{l}}\sin [%
\frac{1}{2\hslash }\Omega (x)t_{1}]$ ($l\geq 0$) are bounded. Moreover, it
is easy to prove that the $l-$order coordinate derivative $\frac{\partial
^{l}}{\partial x^{l}}\exp [\pm i\varphi (x,\gamma )]$ ($l\geq 0$) is
bounded. Then the fact that both the $l-$order derivatives $\frac{\partial
^{l}}{\partial x^{l}}\sin [\frac{1}{2\hslash }\Omega (x)t_{1}]$ and $\frac{%
\partial ^{l}}{\partial x^{l}}\exp [\pm i\varphi (x,\gamma )]$ are bounded
can lead directly to that the $l-$order derivative $\frac{\partial ^{l}}{%
\partial x^{l}}\{\exp [\pm i\varphi (x,\gamma )]$ $\times \sin [\frac{1}{%
2\hslash }\Omega (x)t_{1}]\}$ ($l\geq 0$) is bounded. Now according to
(4.75) one can find that\newline
\begin{equation*}
||\frac{\partial ^{l}}{\partial x^{l}}|\tilde{g}_{0}\rangle ||^{2}=|\frac{%
\partial ^{l}}{\partial x^{l}}\cos \{\frac{1}{2\hslash }\Omega
(x)t_{1}\}|^{2}+|\frac{\partial ^{l}}{\partial x^{l}}\{\exp [i\varphi
(x,\gamma )]\sin [\frac{1}{2\hslash }\Omega (x)t_{1}]\}|^{2},
\end{equation*}%
\begin{equation*}
||\frac{\partial ^{l}}{\partial x^{l}}|\tilde{e}\rangle ||^{2}=|\frac{%
\partial ^{l}}{\partial x^{l}}\cos \{\frac{1}{2\hslash }\Omega
(x)t_{1}\}|^{2}+|\frac{\partial ^{l}}{\partial x^{l}}\{\exp [-i\varphi
(x,\gamma )]\sin [\frac{1}{2\hslash }\Omega (x)t_{1}]\}|^{2}.
\end{equation*}%
Therefore, the two equations show that these norms $||\frac{\partial ^{l}}{%
\partial x^{l}}|\tilde{g}_{0}\rangle ||$ and $||\frac{\partial ^{l}}{%
\partial x^{l}}|\tilde{e}\rangle ||$ ($l\geq 0$) are bounded. On the other
hand, by using directly the derivatives $\frac{\partial ^{k}}{\partial x^{k}}%
\Omega (x)$ and $\frac{\partial ^{k}}{\partial x^{k}}\varphi (x,\gamma )$
with $k\geq 0$ one may calculate explicitly these derivatives $\frac{%
\partial ^{l}}{\partial x^{l}}\cos [\frac{1}{2\hslash }\Omega (x)t_{1}]$ and 
$\frac{\partial ^{l}}{\partial x^{l}}\{\exp [\pm i\varphi (x,\gamma )]\sin [%
\frac{1}{2\hslash }\Omega (x)t_{1}]\}$ for the present case $0\leq l\leq 5.$
For the other case $l>5$ these coordinate derivatives also can be directly
calculated. But they will not be used in the error estimation later. \newline
\newline
{\large (A) The spatially-selective functions}

Here the concept of a spatially-selective function is introduced, so that
with the aid of the concept one may prove conveniently that each one of the
six norms on the $RH$ side of (4.105) has an upper bound consisting of the
basic norms $\{NBAS1\}$ and/or $\{NBAS2\}.$ A spatially-selective function $%
S(x)$ is defined through its induced norm $||S(x)\Psi _{0}(x,r,t)||.$ If the
induced norm has an upper bound that decays exponentially with the square
deviation-to-spread ratio of the $GWP$ state $\Psi _{0}(x,r,t)$ or it
consists of the two types of basic norms $\{NBAS1\}$ and $\{NBAS2\},$ then
the function $S(x)$ is a spatially-selective function. Here $\Psi
_{0}(x,r,t)=\Psi _{0}(x,t)|\psi (r)\rangle ,$ in which $|\psi (r)\rangle $
is some internal state and $\Psi _{0}(x,t)$ is a normalized $GWP$ motional
state. There are some parameters in the spatially-selective function $S(x)$
that control the induced norm $||S(x)\Psi _{0}(x,r,t)||$. As a typical
example, the control parameter may be the joint position $x_{L}$ of the
double-well potential field of (2.1). The basic spatially-selective
functions (or factors) are $\Theta (x-x_{L},\varepsilon )$ and $\delta
(x-x_{c},\varepsilon _{c}).$ The two basic spatially-selective factors
satisfy $0\leq \Theta (x-x_{L},\varepsilon )\leq 1$ and $0\leq \delta
(x-x_{c},\varepsilon _{c})\leq \frac{1}{\varepsilon _{c}\sqrt{\pi }}.$ Thus,
the smooth step function $\Theta (x,\varepsilon )$ and $\delta -$function $%
\delta (x,\varepsilon )$ are always bounded functions. However, the smooth
step function $\Theta (x_{L}+L-x,\varepsilon )$ may not be considered as a
spatially-selective function in the sense that its induced norm $||\Theta
(x_{L}+L-x,\varepsilon )\Psi _{0}(x,r,t)||$ could not have an upper bound
that decays exponentially with the square deviation-to-spread ratio of the $%
GWP$ state $\Psi _{0}(x,r,t).$ The reason for it is that the halting-qubit
atom is within the $LH$ potential well and hence the COM position of the
motional state $\Psi _{0}(x,t)$ of the atom is smaller than the joint
position $x_{L}$ and the coordinate position $x_{L}+L,$ resulting in that $%
\Theta (x_{L}+L-x,\varepsilon )\thickapprox 1$ in the effective spatial
region of the state $\Psi _{0}(x,t).$ On the other hand, the first-order
coordinate derivative $\partial \Theta (x_{L}+L-x,\varepsilon )/\partial
x=-\delta (x-x_{L}-L,\varepsilon )$ is a spatially-selective function.

There are some properties for a spatially-selective function. It is clear
that a linear combination of spatially-selective functions is still a
spatially-selective function. If $\Omega _{l}(x)$ is an arbitrary bounded
function, i.e., $|\Omega _{l}(x)|\leq Const,$ then $\Omega _{l}(x)S(x)$ is a
spatially-selective function; and moreover, if $S_{l}(x)$ is the $l-$th
spatially-selective function, then the combination $\sum_{l}\Omega
_{l}(x)S_{l}(x)$ is also a spatially-selective function. For simplicity,
here consider the basic spatially-selective factors $\Theta
(x-x_{L},\varepsilon )$ and $\delta (x-x_{c},\varepsilon _{c})$ and their
derivatives. Suppose that $q(x)$ is any finite-order polynomial in
coordinate $x$. It proves in the section 3 that the induced norm $%
||q(x)S(x)\Psi _{0}(x,r,t)||$ for the basic spatially-selective factor $%
S(x)=\Theta (x-x_{L},\varepsilon )$ or $\delta (x-x_{c},\varepsilon _{c})$
has an upper bound that consists of the two types of basic norms $\{NBAS1\}$
and $\{NBAS2\}.$ Therefore, the functions $q(x)\Theta (x-x_{L},\varepsilon )$
and $q(x)\delta (x-x_{c},\varepsilon _{c})$ are spatially-selective
functions. Given the two spatially-selective functions $q_{1}(x)\Theta
(x-x_{L},\varepsilon )$ and $q_{2}(x)\delta (x-x_{c},\varepsilon _{c})$ with 
$q_{1}(x)$ and $q_{2}(x)$ being any finite-order polynomials, the product of
a pair of spatially-selective functions of the set $\{q_{1}(x)\Theta
(x-x_{L},\varepsilon ),$ $q_{2}(x)\delta (x-x_{c},\varepsilon _{c})\}$ is
also a spatially-selective function. That is, $q_{1}(x)\Theta
(x-x_{L}^{\prime },\varepsilon ^{\prime })q_{2}(x)\Theta (x-x_{L}^{\prime
\prime },\varepsilon ^{\prime \prime }),$ $q_{1}(x)\Theta (x-x_{L}^{\prime
},\varepsilon )q_{2}(x)\delta (x-x_{c},\varepsilon _{c}),$ and $%
q_{1}(x)\delta (x-x_{c},\varepsilon _{c})q_{2}(x)\delta (x-x_{c}^{\prime
},\varepsilon _{c}^{\prime })$ all are spatially-selective functions. That
the first two functions are spatially-selective functions is due to the fact
that $0\leq \Theta (x-x_{L},\varepsilon )^{k}\leq \Theta
(x-x_{L},\varepsilon )\leq 1$ for $k\geq 1$ and $x_{L}=x_{L}^{\prime }$ or $%
x_{L}^{\prime \prime },$ while that the third function is a
spatially-selective function is due to that the product of a pair of smooth $%
\delta -$functions is still a smooth $\delta -$function up to a Gaussian
factor, as can be seen in (3.35) and (3.36) in the previous section 3. These
properties of a spatially-selective function are important to prove more
complex functions as given below to be spatially-selective functions.
Besides the basic spatially-selective factors $\Theta (x-x_{L},\varepsilon )$
and $\delta (x-x_{c},\varepsilon _{c})$ there are some important
spatially-selective functions that may be useful in the error estimation
below. They are given as follows. \newline
$(i)$ The function $q(x)\Theta (x-x_{L},\varepsilon )^{k}\Theta
(x_{L}+L-x,\varepsilon )^{l}$ with the integers $k>0$ and $l\geq 0$ is a
spatially-selective function. This is because $(a)$ the continuous step
function $\Theta (x_{L}+L-x,\varepsilon )$ satisfies $0\leq \Theta
(x_{L}+L-x,\varepsilon )^{l}\leq 1$ for $l\geq 0,$ that is, $\Theta
(x_{L}+L-x,\varepsilon )^{l}$ is a bounded function; $(b)$ the
spatially-selective factor $\Theta (x-x_{L},\varepsilon )$ satisfies $0\leq
\Theta (x-x_{L},\varepsilon )^{k}\leq \Theta (x-x_{L},\varepsilon )\leq 1$
for $k\geq 1.$\newline
$(ii)$ The function $q(x)\delta (x-x_{c},\varepsilon _{c})^{l}\Theta
(y,\varepsilon )^{k}$ with $k\geq 0$, $l\geq 1$ and $y=x-x_{L}$ or $%
x_{L}+L-x $ is a spatially-selective function. According to the definition
(3.32) of a continuous $\delta -$function the function $\delta
(x-x_{c},\varepsilon _{c})^{l}$ may be written as%
\begin{equation}
\delta (x-x_{c},\varepsilon _{c})^{l}=\frac{1}{\sqrt{l}}(\frac{1}{%
\varepsilon _{c}\sqrt{\pi }})^{l-1}\delta (x-x_{c},\varepsilon _{c}/\sqrt{l}%
),\text{ }l\geq 1.  \tag{4.111}
\end{equation}%
In particular, this equation is reduced to (3.36) when $l=2$. Then $\delta
(x-x_{c},\varepsilon _{c})^{l}$ is a spatially-selective function and so is
the function $q(x)\delta (x-x_{c},\varepsilon _{c})^{l}\Theta (y,\varepsilon
)^{k}$ because $\Theta (y,\varepsilon )^{k}$ is a bounded function.\newline
$(iii)$ The function $q(x)\frac{\partial ^{l}}{\partial x^{l}}\delta
(x-x_{c},\varepsilon _{c})\Theta (y,\varepsilon )^{k}$ with $k\geq 0$, $%
l\geq 0$ and $y=x-x_{L}$ or $x_{L}+L-x$ is a spatially-selective function.
It is known from (3.32) that the continuous $\delta -$function $\delta
(x-x_{c},\varepsilon _{c})$ is a Gaussian function. Then with the help of
(4.49) its $l-$order coordinate derivative may be written as%
\begin{equation}
\frac{\partial ^{l}}{\partial x^{l}}\delta (x-x_{c},\varepsilon
_{c})=H_{l}(x,x_{c},\varepsilon _{c})\delta (x-x_{c},\varepsilon _{c}) 
\tag{4.112}
\end{equation}%
where $H_{l}(x,x_{c},\varepsilon _{c})$ is a $l-$order polynomial in
coordinate $x$ and it is also dependent on $x_{c}$ and $\varepsilon _{c}$.
In particular, $H_{0}(x,x_{c},\varepsilon _{c})=1.$ Therefore, the $l-$order
coordinate derivative of the $\delta -$function $\delta (x-x_{c},\varepsilon
_{c})$ is a spatially-selective function. Since $\Theta (y,\varepsilon )^{k}$
is a bounded function, the function $q(x)\frac{\partial ^{l}}{\partial x^{l}}%
\delta (x-x_{c},\varepsilon _{c})\Theta (y,\varepsilon )^{k}$ is clearly a
spatially-selective function. \newline
$(iv)$ The smooth perturbation term $V_{1}^{ho}(x,\varepsilon )$ and its $k-$%
th power term $V_{1}^{ho}(x,\varepsilon )^{k}$ with the integer $k\geq 1$
are spatially-selective functions. According to the smooth perturbation term
of (2.10) the $k-$th power term $V_{1}^{ho}(x,\varepsilon )^{k}$ may be
written as, by the binomial expansion, 
\begin{equation*}
V_{1}^{ho}(x,\varepsilon )^{k}=[\frac{1}{2}m\omega
^{2}]^{k}\sum_{l=0}^{k}\left( 
\begin{array}{c}
k \\ 
l%
\end{array}%
\right) (-1)^{k-l}x_{L}^{2l}x^{2(k-l)}
\end{equation*}%
\begin{equation}
\times \Theta (x-x_{L},\varepsilon )^{k}\Theta (x_{L}+L-x,\varepsilon )^{l}.
\tag{4.113}
\end{equation}%
The $RH$ side of (4.113) contains $k+1$ terms. Because all these $k+1$
functions $\{x^{2(k-l)}\Theta (x-x_{L},\varepsilon )^{k}\Theta
(x_{L}+L-x,\varepsilon )^{l}\}$ are spatially-selective functions, $%
V_{1}^{ho}(x,\varepsilon )^{k}$ is a spatially-selective function. \newline
$(v)$ The $k-$order coordinate derivative of the perturbation term $%
V_{1}^{ho}(x,\varepsilon )$ is a spatially-selective function. By using the
continuous perturbation term of (2.10) the $k-$order coordinate derivative
is calculated below. The first-order coordinate derivative is simply given by%
\begin{equation*}
\frac{\partial }{\partial x}V_{1}^{ho}(x,\varepsilon )=-\frac{1}{2}m\omega
^{2}x_{L}^{2}\Theta (x-x_{L},\varepsilon )\delta (x-x_{L}-L,\varepsilon )
\end{equation*}%
\begin{equation*}
+\frac{1}{2}m\omega ^{2}x_{L}^{2}\Theta (x_{L}+L-x,\varepsilon )\delta
(x-x_{L},\varepsilon )
\end{equation*}%
\begin{equation}
-\frac{1}{2}m\omega ^{2}x^{2}\delta (x-x_{L},\varepsilon )-m\omega
^{2}x\Theta (x-x_{L},\varepsilon ).  \tag{4.114a}
\end{equation}%
It is clear that this is a spatially-selective function. With the help of
(4.112) and (4.114a) the second-order coordinate derivative can be expressed
as%
\begin{equation*}
\frac{\partial ^{2}}{\partial x^{2}}V_{1}^{ho}(x,\varepsilon )=-(\frac{%
m\omega ^{2}x_{L}^{2}}{\varepsilon ^{2}})(x-x_{L})\Theta
(x_{L}+L-x,\varepsilon )\delta (x-x_{L},\varepsilon )
\end{equation*}%
\begin{equation*}
-m\omega ^{2}x_{L}^{2}\delta (x-x_{L},\varepsilon )\delta
(x-x_{L}-L,\varepsilon )-m\omega ^{2}\Theta (x-x_{L},\varepsilon )
\end{equation*}%
\begin{equation*}
+(\frac{m\omega ^{2}x_{L}^{2}}{\varepsilon ^{2}})(x-x_{L}-L)\Theta
(x-x_{L},\varepsilon )\delta (x-x_{L}-L,\varepsilon )
\end{equation*}%
\begin{equation}
+[(\frac{m\omega ^{2}}{\varepsilon ^{2}})(x-x_{L})x^{2}-2m\omega
^{2}x]\delta (x-x_{L},\varepsilon ).  \tag{4.114b}
\end{equation}%
Clearly this is also a spatially-selective function. In general, by using
the unified formula (See, (4.107)) to calculate the higher-order coordinate
derivatives and with the help of (4.112) the $k-$order coordinate derivative 
$(k>2)$ may be expressed as%
\begin{equation*}
\frac{\partial ^{k}}{\partial x^{k}}V_{1}^{ho}(x,\varepsilon )=-\frac{1}{2}%
m\omega ^{2}x_{L}^{2}H_{k-1}(x,x_{L}+L,\varepsilon )\Theta
(x-x_{L},\varepsilon )\delta (x-x_{L}-L,\varepsilon )
\end{equation*}%
\begin{equation*}
+\frac{1}{2}m\omega ^{2}x_{L}^{2}H_{k-1}(x,x_{L},\varepsilon )\Theta
(x_{L}+L-x,\varepsilon )\delta (x-x_{L},\varepsilon )
\end{equation*}%
\begin{equation*}
-m\omega ^{2}\{\frac{1}{2}x^{2}H_{k-1}(x,x_{L},\varepsilon
)+kxH_{k-2}(x,x_{L},\varepsilon )
\end{equation*}%
\begin{equation}
+\frac{1}{2}k(k-1)H_{k-3}(x,x_{L},\varepsilon )\}\delta (x-x_{L},\varepsilon
)+\Sigma _{k}^{1}(L,\varepsilon )  \tag{4.114c}
\end{equation}%
where the secondary term $\Sigma _{k}^{1}(L,\varepsilon )$ is given by, with
the help of (3.35), 
\begin{equation*}
\Sigma _{k}^{1}(L,\varepsilon )=-\frac{1}{2}m\omega
^{2}x_{L}^{2}\sum_{j=1}^{k-1}\left( 
\begin{array}{c}
k \\ 
j%
\end{array}%
\right) H_{j-1}(x,x_{L},\varepsilon )H_{k-j-1}(x,x_{L}-L,\varepsilon )
\end{equation*}%
\begin{equation*}
\times \frac{1}{\varepsilon \sqrt{2\pi }}\exp (-\frac{1}{2}\frac{L^{2}}{%
\varepsilon ^{2}})\delta (x-x_{L}-L/2,\varepsilon /\sqrt{2}).
\end{equation*}%
The term $\Sigma _{k}^{1}(L,\varepsilon )$ is clearly a spatially-selective
function. It is proportional to the Gaussian decaying factor $\exp (-\frac{1%
}{2}\frac{L^{2}}{\varepsilon ^{2}}).$ When $L>>\varepsilon ,$ it is so small
that it can be neglected with respect to the other terms on the $RH$ side of
(4.114c). Thus, the term $\Sigma _{k}^{1}(L,\varepsilon )$ is a secondary
term in (4.114c). If the secondary term $\Sigma _{k}^{1}(L,\varepsilon )$ is
neglected, then the $RH$ side of (4.114c) contains only three terms. Then
the equation (4.114c) shows that the $k-$order coordinate derivative $(k>2)$
of the perturbation term $V_{1}^{ho}(x,\varepsilon )$ is a
spatially-selective function. Now the three equations of (4.114) together
show that any finite-order coordinate derivative of the perturbation term $%
V_{1}^{ho}(x,\varepsilon )$ is a spatially-selective function. \newline
$(vi)$ The function $\{\frac{\partial }{\partial x}V_{1}^{ho}(x,\varepsilon
)\}^{k}$ with $k=2$, $3$, $...$ is a spatially-selective function. By using
the first-order derivative $\frac{\partial }{\partial x}V_{1}^{ho}(x,%
\varepsilon )$ of (4.114a) and the formula (4.111) it can turn out that the
function $[\frac{\partial }{\partial x}V_{1}^{ho}(x,\varepsilon )]^{k}$ ($%
k>1 $) may be expressed as%
\begin{equation*}
\lbrack \frac{\partial }{\partial x}V_{1}^{ho}(x,\varepsilon )]^{k}=(m\omega
^{2})^{k}(-x)^{k}\Theta (x-x_{L},\varepsilon )^{k}
\end{equation*}%
\begin{equation*}
+\sum_{l=1}^{k}\left( 
\begin{array}{c}
k \\ 
l%
\end{array}%
\right) \frac{1}{2^{l}\sqrt{l}}(\frac{1}{\varepsilon \sqrt{\pi }}%
)^{l-1}(m\omega ^{2})^{k}(-x)^{k-l}
\end{equation*}%
\begin{equation*}
\times \lbrack x_{L}^{2}\Theta (x_{L}+L-x,\varepsilon )-x^{2}]^{l}\Theta
(x-x_{L},\varepsilon )^{k-l}\delta (x-x_{L},\varepsilon /\sqrt{l})
\end{equation*}%
\begin{equation*}
+\sum_{l=1}^{k}\left( 
\begin{array}{c}
k \\ 
l%
\end{array}%
\right) \frac{(-1)^{l}x_{L}^{2l}}{2^{l}\sqrt{l}}(\frac{1}{\varepsilon \sqrt{%
\pi }})^{l-1}(m\omega ^{2})^{k}(-x)^{k-l}
\end{equation*}%
\begin{equation*}
\times \Theta (x-x_{L},\varepsilon )^{k}\delta (x-x_{L}-L,\varepsilon /\sqrt{%
l})
\end{equation*}%
\begin{equation}
+\sum_{l=2}^{k}\left( 
\begin{array}{c}
k \\ 
l%
\end{array}%
\right) \frac{1}{2^{l}}(m\omega ^{2})^{k}(-x)^{k-l}\Theta
(x-x_{L},\varepsilon )^{k-l}\Sigma _{l}^{2}(L,\varepsilon )  \tag{4.115a}
\end{equation}%
where the term $\Sigma _{l}^{2}(L,\varepsilon )$ is written as%
\begin{equation*}
\Sigma _{l}^{2}(L,\varepsilon )=\sum_{j=1}^{l-1}\left( 
\begin{array}{c}
l \\ 
j%
\end{array}%
\right) (-1)^{j}x_{L}^{2j}[x_{L}^{2}\Theta (x_{L}+L-x,\varepsilon
)-x^{2}]^{l-j}
\end{equation*}%
\begin{equation*}
\times \Theta (x-x_{L},\varepsilon )^{j}\delta (x-x_{L}-L,\varepsilon
)^{j}\delta (x-x_{L},\varepsilon )^{l-j}.
\end{equation*}%
The term $\Sigma _{l}^{2}(L,\varepsilon )$ contains $\delta
(x-x_{L}-L,\varepsilon )^{j}\delta (x-x_{L},\varepsilon )^{l-j}$ with $j\geq
1$ and $l-j\geq 1,$ which may be reduced to a single smooth $\delta -$%
function, according to (4.111) and (3.35), 
\begin{equation*}
\delta (x-x_{L}-L,\varepsilon )^{j}\delta (x-x_{L},\varepsilon )^{l-j}
\end{equation*}%
\begin{equation*}
=(\frac{1}{\varepsilon \sqrt{\pi }})^{l-1}\exp \{-\frac{j(l-j)}{l}\frac{%
x_{L}^{2}}{\varepsilon ^{2}}\}\delta (x-x_{L}-(\frac{j}{l})L,\varepsilon /%
\sqrt{l}).
\end{equation*}%
Therefore, the term $\Sigma _{l}^{2}(L,\varepsilon )$ is proportional to the
Gaussian decaying factors $\{\exp \{-\frac{j(l-j)}{l}\frac{x_{L}^{2}}{%
\varepsilon ^{2}}\}\}.$ Since $j\geq 1$ and $l-j\geq 1,$ there holds the
relation $\frac{j(l-j)}{l}\frac{x_{L}^{2}}{\varepsilon ^{2}}\geq \frac{(l-1)%
}{l}\frac{x_{L}^{2}}{\varepsilon ^{2}}$ for $j=1$, $2$, ..., $l-1$. When $%
L>>\varepsilon ,$ the secondary term $\Sigma _{l}^{2}(L,\varepsilon )$ is so
small that the last sum term that contains $\Sigma _{l}^{2}(L,\varepsilon )$
can be neglected with respect to the other terms on the $RH$ side of
(4.115a). Then the last sum term on the $RH$ side of (4.115a) is a secondary
term. Notice that $[x_{L}^{2}\Theta (x_{L}+L-x,\varepsilon )-x^{2}]^{m}$
with $m=l-j\geq 1$ or $m=l\geq 1$ that appears in the term $\Sigma
_{l}^{2}(L,\varepsilon )$ or the second sum term on the $RH$ side of
(4.115a) may be expanded as%
\begin{equation*}
\lbrack x_{L}^{2}\Theta (x_{L}+L-x,\varepsilon
)-x^{2}]^{m}=\sum_{k=0}^{m}\left( 
\begin{array}{c}
m \\ 
k%
\end{array}%
\right) x_{L}^{2(m-k)}(-x^{2})^{k}\Theta (x_{L}+L-x,\varepsilon )^{m-k}.
\end{equation*}%
Here $\Theta (x_{L}+L-x,\varepsilon )^{m-k}$ ($m-k\geq 0$) is a bounded
function. Thus, the term $\Sigma _{l}^{2}(L,\varepsilon )$ and the second
sum term on the $RH$ side of (4.115a) may be expressed as a sum of the
spatially-selective functions. The equation (4.115a) shows that if the last
sum term is neglected, then the function $[\frac{\partial }{\partial x}%
V_{1}^{ho}(x,\varepsilon )]^{k}$ consists of $2k+1$ terms, each one of which
is a spatially-selective function. Thus, the function $[\frac{\partial }{%
\partial x}V_{1}^{ho}(x,\varepsilon )]^{k}$ is a spatially-selective
function. \newline
$(vii)$ The function $\{\frac{\partial ^{2}}{\partial x^{2}}%
V_{1}^{ho}(x,\varepsilon )\}^{k}$ with $k=2$, $3$, $...$ is a
spatially-selective function. By starting from the function $\frac{\partial
^{2}}{\partial x^{2}}V_{1}^{ho}(x,\varepsilon )$ of (4.114b) one can
calculate strictly the function $[\frac{\partial ^{2}}{\partial x^{2}}%
V_{1}^{ho}(x,\varepsilon )]^{k}$ for $k>1.$ The result is written as%
\begin{equation*}
\lbrack \frac{\partial ^{2}}{\partial x^{2}}V_{1}^{ho}(x,\varepsilon
)]^{k}=\Sigma _{k}^{3}(L,\varepsilon )+(-m\omega ^{2})^{k}\Theta
(x-x_{L},\varepsilon )^{k}
\end{equation*}%
\begin{equation*}
+\sum_{j=1}^{k}\left( 
\begin{array}{c}
k \\ 
j%
\end{array}%
\right) (-m\omega ^{2})^{k-j}\tilde{Q}_{0}(x)^{j}\Theta (x-x_{L},\varepsilon
)^{k-j}\delta (x-x_{L},\varepsilon )^{j}
\end{equation*}%
\begin{equation}
+\sum_{j=1}^{k}\left( 
\begin{array}{c}
k \\ 
j%
\end{array}%
\right) (\frac{m\omega ^{2}x_{L}^{2}}{\varepsilon ^{2}})^{j}(-m\omega
^{2})^{k-j}(x-x_{L}-L)^{j}\Theta (x-x_{L},\varepsilon )^{k}\delta
(x-x_{L}-L,\varepsilon )^{j}.  \tag{4.115b}
\end{equation}%
Here the secondary term $\Sigma _{k}^{3}(L,\varepsilon )$ is%
\begin{equation*}
\Sigma _{k}^{3}(L,\varepsilon )=\Sigma _{k}^{31}(L,\varepsilon
)+\sum_{j=2}^{k}\left( 
\begin{array}{c}
k \\ 
j%
\end{array}%
\right) (-m\omega ^{2})^{k-j}\Theta (x-x_{L},\varepsilon )^{k-j}\Sigma
_{j}^{32}(L,\varepsilon ),
\end{equation*}%
while the secondary terms $\Sigma _{k}^{31}(L,\varepsilon )$ and $\Sigma
_{j}^{32}(L,\varepsilon )$ are given by%
\begin{equation*}
\Sigma _{k}^{31}(L,\varepsilon )=\sum_{l=1}^{k}\left( 
\begin{array}{c}
k \\ 
l%
\end{array}%
\right) \exp (-\frac{l}{2}\frac{L^{2}}{\varepsilon ^{2}})(-\frac{m\omega
^{2}x_{L}^{2}}{\varepsilon \sqrt{2\pi }})^{l}\delta (x-x_{L}-L/2,\varepsilon
/\sqrt{2})^{l}
\end{equation*}%
\begin{equation*}
\times \{\frac{m\omega ^{2}x_{L}^{2}}{\varepsilon ^{2}}(x-x_{L}-L)\Theta
(x-x_{L},\varepsilon )\delta (x-x_{L}-L,\varepsilon )
\end{equation*}%
\begin{equation*}
+\tilde{Q}_{0}(x)\delta (x-x_{L},\varepsilon )-m\omega ^{2}\Theta
(x-x_{L},\varepsilon )\}^{k-l}
\end{equation*}%
and 
\begin{equation*}
\Sigma _{j}^{32}(L,\varepsilon )=\sum_{n=1}^{j-1}\left( 
\begin{array}{c}
j \\ 
n%
\end{array}%
\right) \{(\frac{m\omega ^{2}x_{L}^{2}}{\varepsilon ^{2}})^{n}(x-x_{L}-L)^{n}%
\tilde{Q}_{0}(x)^{j-n}
\end{equation*}%
\begin{equation*}
\times \Theta (x-x_{L},\varepsilon )^{n}\delta (x-x_{L}-L,\varepsilon
)^{n}\delta (x-x_{L},\varepsilon )^{j-n}\},
\end{equation*}%
and the function $\tilde{Q}_{0}(x)$ is defined by%
\begin{equation*}
\tilde{Q}_{0}(x)=-\frac{m\omega ^{2}x_{L}^{2}}{\varepsilon ^{2}}%
(x-x_{L})\Theta (x_{L}+L-x,\varepsilon )-2m\omega ^{2}x+\frac{m\omega ^{2}}{%
\varepsilon ^{2}}x^{2}(x-x_{L}).
\end{equation*}%
Here $\Theta (x_{L}+L-x,\varepsilon )^{l}$ $(l=1,$ $2$, ...$)$ is a bounded
function. A simple calculation shows that both the terms $\Sigma
_{k}^{31}(L,\varepsilon )$ and $\Sigma _{k}^{32}(L,\varepsilon )$ are
spatially-selective functions, indicating that the term $\Sigma
_{k}^{3}(L,\varepsilon )$ is a spatially-selective function. The term $%
\Sigma _{k}^{31}(L,\varepsilon )$ is proportional to the Gaussian decaying
factors $\{\exp (-\frac{l}{2}\frac{L^{2}}{\varepsilon ^{2}})\}$ with $l\geq
1 $, while the term $\Sigma _{j}^{32}(L,\varepsilon )$ contains $\{\delta
(x-x_{L}-L,\varepsilon )^{n}\delta (x-x_{L},\varepsilon )^{j-n}\}$ with $%
n\geq 1$ and $j-n\geq 1,$ which are proportional to the Gaussian decaying
factors $\{\exp \{-\frac{n(j-n)}{j}\frac{x_{L}^{2}}{\varepsilon ^{2}}\}\}$.
Therefore, $\Sigma _{k}^{3}(L,\varepsilon )$ contains the Gaussian decaying
factors and hence it is a secondary term, similar to the term $\Sigma
_{l}^{2}(L,\varepsilon )$ in (4.115a). If the secondary term $\Sigma
_{k}^{3}(L,\varepsilon )$ is neglected in (4.115b), then $[\frac{\partial
^{2}}{\partial x^{2}}V_{1}^{ho}(x,\varepsilon )]^{k}$ consists of $2k+1$
terms, each one of which is a spatially-selective function. Thus, the
function $[\frac{\partial ^{2}}{\partial x^{2}}V_{1}^{ho}(x,\varepsilon
)]^{k}$ is a spatially-selective function. \newline
$(viii)$ The function $[\frac{\partial ^{k}}{\partial x^{k}}%
V_{1}^{ho}(x,\varepsilon )]^{n}$ with $k>2$ and $n>1$ is a
spatially-selective function. By using the function $\frac{\partial ^{k}}{%
\partial x^{k}}V_{1}^{ho}(x,\varepsilon )$ of (4.114c) with $k>2$ one may
compute strictly the function $[\frac{\partial ^{k}}{\partial x^{k}}%
V_{1}^{ho}(x,\varepsilon )]^{n}$ with $k>2$ and $n>1.$ The result is given by%
\begin{equation*}
\lbrack \frac{\partial ^{k}}{\partial x^{k}}V_{1}^{ho}(x,\varepsilon
)]^{n}=\Sigma _{nk}^{4}(L,\varepsilon )+\Sigma _{nk}^{5}(L,\varepsilon )
\end{equation*}%
\begin{equation*}
+\{\frac{1}{2}m\omega ^{2}x_{L}^{2}H_{k-1}(x,x_{L},\varepsilon )\Theta
(x_{L}+L-x,\varepsilon )
\end{equation*}%
\begin{equation*}
-\frac{1}{2}m\omega ^{2}x^{2}H_{k-1}(x,x_{L},\varepsilon )-k(m\omega
^{2})xH_{k-2}(x,x_{L},\varepsilon )
\end{equation*}%
\begin{equation*}
-\frac{1}{2}k(k-1)(m\omega ^{2})H_{k-3}(x,x_{L},\varepsilon )\}^{n}\delta
(x-x_{L},\varepsilon )^{n}
\end{equation*}%
\begin{equation}
+(-\frac{1}{2}m\omega ^{2}x_{L}^{2})^{n}\{H_{k-1}(x,x_{L}+L,\varepsilon
)\}^{n}\Theta (x-x_{L},\varepsilon )^{n}\delta (x-x_{L}-L,\varepsilon )^{n},
\tag{4.115c}
\end{equation}%
where $\Sigma _{nk}^{4}(L,\varepsilon )$ and $\Sigma _{nk}^{5}(L,\varepsilon
)$ are given by%
\begin{equation*}
\Sigma _{nk}^{4}(L,\varepsilon )=\sum_{l=0}^{n-1}\left( 
\begin{array}{c}
n \\ 
l%
\end{array}%
\right) \{\Sigma _{k}^{1}(L,\varepsilon )\}^{n-l}
\end{equation*}%
\begin{equation*}
\times \{-\frac{1}{2}m\omega ^{2}x_{L}^{2}H_{k-1}(x,x_{L}+L,\varepsilon
)\Theta (x-x_{L},\varepsilon )\delta (x-x_{L}-L,\varepsilon )
\end{equation*}%
\begin{equation*}
+\frac{1}{2}m\omega ^{2}x_{L}^{2}H_{k-1}(x,x_{L},\varepsilon )\Theta
(x_{L}+L-x,\varepsilon )\delta (x-x_{L},\varepsilon )
\end{equation*}%
\begin{equation*}
-m\omega ^{2}[\frac{1}{2}x^{2}H_{k-1}(x,x_{L},\varepsilon
)+kxH_{k-2}(x,x_{L},\varepsilon )
\end{equation*}%
\begin{equation*}
+\frac{1}{2}k(k-1)H_{k-3}(x,x_{L},\varepsilon )]\delta (x-x_{L},\varepsilon
)\}^{l}
\end{equation*}%
and%
\begin{equation*}
\Sigma _{nk}^{5}(L,\varepsilon )=\sum_{l=1}^{n-1}\left( 
\begin{array}{c}
n \\ 
l%
\end{array}%
\right) \{-\frac{1}{2}m\omega ^{2}x_{L}^{2}H_{k-1}(x,x_{L}+L,\varepsilon
)\Theta (x-x_{L},\varepsilon )\}^{n-l}
\end{equation*}%
\begin{equation*}
\times \{\frac{1}{2}m\omega ^{2}x_{L}^{2}H_{k-1}(x,x_{L},\varepsilon )\Theta
(x_{L}+L-x,\varepsilon )
\end{equation*}%
\begin{equation*}
-m\omega ^{2}[\frac{1}{2}x^{2}H_{k-1}(x,x_{L},\varepsilon
)+kxH_{k-2}(x,x_{L},\varepsilon )
\end{equation*}%
\begin{equation*}
+\frac{1}{2}k(k-1)H_{k-3}(x,x_{L},\varepsilon )]\}^{l}\delta
(x-x_{L},\varepsilon )^{l}\delta (x-x_{L}-L,\varepsilon )^{n-l}.
\end{equation*}%
Here the $k-$order polynomial $H_{k}(x,x_{c},\varepsilon )$ ($x_{c}=x_{L}$
or $x_{L}+L$) is defined by (4.112) and $\Theta (x_{L}+L-x,\varepsilon )^{l}$
($l=1$, $2$, ...) is a bounded function. A simple calculation shows that
both the terms $\Sigma _{nk}^{4}(L,\varepsilon )$ and $\Sigma
_{nk}^{5}(L,\varepsilon )$ are spatially-selective functions. The term $%
\Sigma _{nk}^{4}(L,\varepsilon )$ is controlled by the term $\Sigma
_{k}^{1}(L,\varepsilon )$ which is proportional to the Gaussian decaying
factor $\exp (-\frac{1}{2}\frac{L^{2}}{\varepsilon ^{2}}),$ as shown above.
The term $\Sigma _{nk}^{5}(L,\varepsilon )$ contains $\{\delta
(x-x_{L}-L,\varepsilon )^{n-l}\delta (x-x_{L},\varepsilon )^{l}\}$ with $%
l\geq 1$ and $n-l\geq 1$, which are proportional to the Gaussian decaying
factors $\{\exp \{-\frac{l(n-l)}{n}\frac{x_{L}^{2}}{\varepsilon ^{2}}\}\}$.
Thus, both the terms $\Sigma _{nk}^{4}(L,\varepsilon )$ and $\Sigma
_{nk}^{5}(L,\varepsilon )$ are secondary, similar to the terms $\Sigma
_{l}^{2}(L,\varepsilon )$ and $\Sigma _{k}^{3}(L,\varepsilon )$ above. If
the secondary terms $\Sigma _{nk}^{4}(L,\varepsilon )$ and $\Sigma
_{nk}^{5}(L,\varepsilon )$ are neglected in (4.115c), then $[\frac{\partial
^{k}}{\partial x^{k}}V_{1}^{ho}(x,\varepsilon )]^{n}$ consists of these
spatially-selective functions $q(x)\delta (x-x_{L},\varepsilon )^{n},$ $%
q(x)\Theta (x_{L}+L-x,\varepsilon )^{k}\delta (x-x_{L},\varepsilon )^{n}$ ($%
1\leq k\leq n$)$,$ and $q(x)\Theta (x-x_{L},\varepsilon )^{n}\delta
(x-x_{L}-L,\varepsilon )^{n}$. Thus, the function $[\frac{\partial ^{k}}{%
\partial x^{k}}V_{1}^{ho}(x,\varepsilon )]^{n}$ with $k>2$ and $n>1$ is a
spatially-selective function.

Actually, the first three types $(i)$, $(ii)$, and $(iii)$ of
spatially-selective functions are more basic. From them one can deduce the
last five groups $(iv)$, $(v)$, $(vi)$, $(vii)$, and $(viii)$ of
spatially-selective functions. With the help of $(1)$ the first three types
of spatially-selective functions, $(2)$ the formulae (3.35) and (4.111), and 
$(3)$\ the bounded function $\Theta (y,\varepsilon )^{k}$ that satisfies $%
0\leq \Theta (y,\varepsilon )^{k}\leq 1$ for $k\geq 1$ and $y=x-x_{L}$ or $%
x_{L}+L-x,$ it can turn out that the product of a pair of the
spatially-selective functions given by $(i)$, $(ii)$, ..., $(viii)$ is still
a spatially-selective function. This important property is often used in the
following error estimation. The following error estimation uses these
spatially-selective functions $[\frac{\partial ^{k}}{\partial x^{k}}%
V_{1}^{ho}(x,\varepsilon )]^{n}$ with $5\geq k\geq 0$ and $5\geq n\geq 1.$
Then the number of the spatially-selective functions\ is finite and not too
large which appear in the expansions (4.114) and (4.115) of the functions $[%
\frac{\partial ^{k}}{\partial x^{k}}V_{1}^{ho}(x,\varepsilon )]^{n}$ with $%
5\geq k\geq 0$ and $5\geq n\geq 1$.

In the following error estimation one also meets some specific
spatially-selective functions. It is known above that the amplitude $\Omega
(x)$ in (4.19) of the $PHAMDOWN$ laser light beams and its $k-$order ($k=0$, 
$1$, $2$, ...) coordinate derivative $\frac{\partial ^{k}}{\partial x^{k}}%
\Omega (x)$ are bounded. Then the function $\Theta (x-x_{L},\varepsilon
)\Omega (x)$ is a spatially-selective function. Besides the function $\Theta
(x-x_{L},\varepsilon )\Omega (x)$ some important spatially-selective
functions related to $\Theta (x-x_{L},\varepsilon )\Omega (x)$ are given as
follows. \newline
$(a)$ The function $[\Theta (x-x_{L},\varepsilon )\Omega (x)]^{k}$ with $k>0$
is a spatially-selective function. \newline
$(b)$ The $k-$order coordinate derivative $\frac{\partial ^{k}}{\partial
x^{k}}[\Theta (x-x_{L},\varepsilon )\Omega (x)]$ ($k\geq 0$) is a
spatially-selective function. Actually, the $k-$order derivative may be
expressed as%
\begin{equation}
\frac{\partial ^{k}}{\partial x^{k}}[\Theta (x-x_{L},\varepsilon )\Omega
(x)]=f_{k}(x)\Theta (x-x_{L},\varepsilon )+g_{k}(x)\delta
(x-x_{L},\varepsilon )  \tag{4.116}
\end{equation}%
where $f_{k}(x)=\frac{\partial ^{k}}{\partial x^{k}}\Omega (x)$ and $%
g_{k}(x) $ is given by 
\begin{equation*}
g_{k}(x)=\sum_{j=1}^{k}\left( 
\begin{array}{c}
k \\ 
j%
\end{array}%
\right) [\frac{\partial ^{k-j}}{\partial x^{k-j}}\Omega
(x)]H_{j-1}(x,x_{L},\varepsilon ).
\end{equation*}%
Here $H_{l}(x,x_{L},\varepsilon )$ is still defined by (4.112). Since $\frac{%
\partial ^{l}}{\partial x^{l}}\Omega (x)$ for $0\leq l\leq k$ is bounded,
the equation (4.116) shows that the $k-$order derivative $\frac{\partial ^{k}%
}{\partial x^{k}}[\Theta (x-x_{L},\varepsilon )\Omega (x)]$ is a
spatially-selective function. \newline
$(c)$ The function $\{\frac{\partial ^{k}}{\partial x^{k}}[\Theta
(x-x_{L},\varepsilon )\Omega (x)]\}^{m}$ is a spatially-selective function.
By starting from the function $\frac{\partial ^{k}}{\partial x^{k}}[\Theta
(x-x_{L},\varepsilon )\Omega (x)]$ of (4.116) and using the formula (4.111)
it can turn out that%
\begin{equation*}
\{\frac{\partial ^{k}}{\partial x^{k}}[\Theta (x-x_{L},\varepsilon )\Omega
(x)]\}^{m}=f_{k}(x)^{m}\Theta (x-x_{L},\varepsilon )^{m}
\end{equation*}%
\begin{equation*}
+\sum_{l=0}^{m-1}\frac{1}{\sqrt{m-l}}(\frac{1}{\varepsilon \sqrt{\pi }}%
)^{m-l-1}\left( 
\begin{array}{c}
m \\ 
l%
\end{array}%
\right) f_{k}(x)^{l}g_{k}(x)^{m-l}
\end{equation*}%
\begin{equation}
\times \Theta (x-x_{L},\varepsilon )^{l}\delta (x-x_{L},\varepsilon /\sqrt{%
m-l}).  \tag{4.117}
\end{equation}%
The $RH$ side of (4.117) contains $m+1$ terms, each one of which is a
spatially-selective function. This shows that the function $\{\frac{\partial
^{k}}{\partial x^{k}}[\Theta (x-x_{L},\varepsilon )\Omega (x)]\}^{m}$ is a
spatially-selective function. The following error estimation uses the
spatially-selective functions $\{\frac{\partial ^{k}}{\partial x^{k}}[\Theta
(x-x_{L},\varepsilon )\Omega (x)]\}^{m}$ with $0\leq k\leq 5$ and $1\leq
m\leq 5.$ Therefore, for these spatially-selective functions the number of
the spatially-selective functions appearing in the expansion (4.117) is
finite and not too large.

Now consider a more general norm $||q(x)S(x)\Psi _{0}(x,r,t)||$ in which the
product state $\Psi _{0}(x,r,t)$ is a Gaussian superposition product state.
Suppose that the Gaussian superposition state $\Psi _{0}(x,r,t)$ is given by
(3.14) and (3.15) with the setting $t_{c}=t$. Then it is easy to prove that
the norm $||q(x)S(x)\Psi _{0}(x,r,t)||$ is bounded by%
\begin{equation*}
||q(x)S(x)\Psi _{0}(x,r,t)||\leq \sum_{k=1}^{n_{g}}|A_{k}^{g}(t)|\newline
\times ||q(x)S(x)\Psi _{0k}^{g}(x,r,t)||
\end{equation*}%
\begin{equation*}
+\sum_{k=1}^{n_{e}}|A_{k}^{e}(t)|\times ||q(x)S(x)\Psi _{0k}^{e}(x,r,t)||,
\end{equation*}%
where $\Psi _{0k}^{g}(x,r,t)=\Psi _{0k}^{g}(x,t)|g_{0}\rangle ,$ $\Psi
_{0k}^{e}(x,r,t)=\Psi _{0k}^{e}(x,t)|e\rangle ,$ and both $\Psi
_{0k}^{g}(x,t)$ and $\Psi _{0k}^{e}(x,t)$ are normalized $GWP$ motional
states. If now any $k-$th normalized $GWP$ state $\Psi _{0k}^{a}(x,t)$ with $%
a=g$ or $e$ has a COM position $x_{ck}^{a}(t)<x_{L},$ then the norm $%
||q(x)S(x)\Psi _{0k}^{a}(x,r,t)||$ has an upper bound that consists of the
two types of basic norms $\{NBAS1\}$ and $\{NBAS2\}$. Therefore, given the
finite integers $n_{g}$ and $n_{e}$ the upper bound of the norm $%
||q(x)S(x)\Psi _{0}(x,r,t)||$ consists of the two types of basic norms $%
\{NBAS1\}$ and $\{NBAS2\}$. The norm $||q(x)S(x)\Psi _{0}(x,r,t)||$ with the
Gaussian superposition state $\Psi _{0}(x,r,t)$ is quite popular in the
following error estimation. In particular, here consider the two such norms: 
$\{||q(x)\Theta (x-x_{L},\varepsilon )\frac{\partial ^{j}}{\partial x^{j}}%
\tilde{\Psi}_{0}^{\mu }(x,r,t_{0}+t_{1}/2)||\}$ and $\{||q(x)\delta
(x-x_{c},\varepsilon _{c})\frac{\partial ^{j}}{\partial x^{j}}\tilde{\Psi}%
_{0}^{\mu }(x,r,t_{0}+t_{1}/2)||\}$ with the derivative order $j=0$, $1$, $2$%
, $...$. They are often met in the following error estimation. When the
order $j=0,$ the two norms are reduced to $||q(x)\Theta (x-x_{L},\varepsilon
)\tilde{\Psi}_{0}^{\mu }(x,r,t_{0}+t_{1}/2)||$ and $||q(x)\delta
(x-x_{c},\varepsilon _{c})\tilde{\Psi}_{0}^{\mu }(x,r,t_{0}+t_{1}/2)||,$
respectively. In this case the former norm has an upper bound consisting of
the basic norms $\{NBAS2\}$, while the latter has an upper bound consisting
of the basic norms $\{NBAS1\}$. Below consider the general case $j>0$. The
coordinate derivatives $\{\frac{\partial ^{j}}{\partial x^{j}}\tilde{\Psi}%
_{0}^{\mu }(x,r,t_{0}+t_{1}/2)\}$ may be obtained from the two equations
(4.110). By using these coordinate derivatives one can determine the upper
bounds of the two norms. For the case $\mu =CS$ the derivative $\frac{%
\partial ^{j}}{\partial x^{j}}\Psi _{0}^{C0}(x,r,t_{0}+t_{1}/2)$ is obtained
from (4.110a). Then by using the derivative it can turn out that the two
norms satisfy%
\begin{equation*}
||q(x)S(x)\frac{\partial ^{j}}{\partial x^{j}}\Psi
_{0}^{C0}(x,r,t_{0}+t_{1}/2)||
\end{equation*}%
\begin{equation*}
\leq \sum_{l=0}^{j}\left( 
\begin{array}{c}
j \\ 
l%
\end{array}%
\right) ||\frac{\partial ^{l}}{\partial x^{l}}|\tilde{g}_{0}\rangle ||\times
||q(x)Q_{j-l}^{g}(x)S(x)\Psi _{0}^{g}(x,t_{0}+t_{1}/2)||
\end{equation*}%
\begin{equation}
+\sum_{l=0}^{j}\left( 
\begin{array}{c}
j \\ 
l%
\end{array}%
\right) ||\frac{\partial ^{l}}{\partial x^{l}}|\tilde{e}\rangle ||\times
||q(x)Q_{j-l}^{e}(x)S(x)\Psi _{0}^{e}(x,t_{0}+t_{1}/2)||  \tag{4.118a}
\end{equation}%
where the spatially-selective factor $S(x)=\Theta (x-x_{L},\varepsilon )$ or 
$\delta (x-x_{c},\varepsilon _{c}).$ It already proves above that both the
norms $||\frac{\partial ^{l}}{\partial x^{l}}|\tilde{g}_{0}\rangle ||$ and $%
||\frac{\partial ^{l}}{\partial x^{l}}|\tilde{e}\rangle ||$ are bounded and
the function $Q_{j-l}^{a}(x)$ with $a=g$ or $e$ is a $(j-l)-$order
polynomial in coordinate $x$ (See (4.109a)). Hereafter the norms $||\frac{%
\partial ^{l}}{\partial x^{l}}|\tilde{g}_{0}\rangle ||$ and $||\frac{%
\partial ^{l}}{\partial x^{l}}|\tilde{e}\rangle ||$ are considered as
bounded parameters. Then it is clear that the function $%
q(x)Q_{j-l}^{a}(x)S(x)$ with $a=g$ or $e$ is a spatially-selective function.
Thus, the $RH$ side of (4.118a) is a linear combination of $2j+2$ norms,
each one of which has an upper bound consisting of the basic norms $%
\{NBAS2\},$ if $S(x)=\Theta (x-x_{L},\varepsilon ).$ It is also a linear
combination of $2j+2$ norms, each one of which has an upper bound consisting
of the basic norms $\{NBAS1\},$ if $S(x)=\delta (x-x_{c},\varepsilon _{c}).$
Then the inequality (4.118a) shows that the norm on the $LH$ side of
(4.118a) with $S(x)=\Theta (x-x_{L},\varepsilon )$ or $\delta
(x-x_{c},\varepsilon _{c})$ for the case $\mu =CS$ has an upper bound
consisting of the basic norms $\{NBAS2\}$ or $\{NBAS1\}$. Here the number of
the basic norms $\{NBAS1\}$ or $\{NBAS2\}$ depends on the order of the
polynomial $q(x)Q_{j-l}^{a}(x)$ in coordinate $x.$ If $q(x)Q_{j-l}^{a}(x)$
is a $k_{a}-$order polynomial$,$ then the number is $k_{a}+1$ at most for
each norm on the $RH$ side of (4.118a). Similarly, for the case $\mu =S$ the
derivative $\frac{\partial ^{j}}{\partial x^{j}}\tilde{\Psi}%
_{0}^{S}(x,r,t_{0}+t_{1}/2)$ is obtained from (4.110b). Then by using the
derivative one can prove that the two norms satisfy 
\begin{equation*}
||q(x)S(x)\frac{\partial ^{j}}{\partial x^{j}}\tilde{\Psi}%
_{0}^{S}(x,r,t_{0}+t_{1}/2)||
\end{equation*}%
\begin{equation*}
\leq \sum_{l=0}^{j}\left( 
\begin{array}{c}
j \\ 
l%
\end{array}%
\right) ||\frac{\partial ^{l}}{\partial x^{l}}|\tilde{e}\rangle ||\times
||q(x)Q_{j-l}^{g}(x,+\varphi (x,\gamma ))S(x)\Psi _{0}^{g}(x,t_{0}+t_{1}/2)||
\end{equation*}%
\begin{equation}
+\sum_{l=0}^{j}\left( 
\begin{array}{c}
j \\ 
l%
\end{array}%
\right) ||\frac{\partial ^{l}}{\partial x^{l}}|\tilde{g}_{0}\rangle ||\times
||q(x)Q_{j-l}^{e}(x,-\varphi (x,\gamma ))S(x)\Psi
_{0}^{e}(x,t_{0}+t_{1}/2)||.  \tag{4.118b}
\end{equation}%
Here $Q_{j-l}^{a}(x,\pm \varphi (x,\gamma ))$ with $a=g$ or $e$ is a $(j-l)-$%
order polynomial in coordinate $x$, as shown in (4.109b). Therefore, the
function $q(x)Q_{j-l}^{a}(x,\pm \varphi (x,\gamma ))S(x)$ is a
spatially-selective function. Note that both the norms $|\frac{\partial ^{l}%
}{\partial x^{l}}|\tilde{g}_{0}\rangle ||$ and $||\frac{\partial ^{l}}{%
\partial x^{l}}|\tilde{e}\rangle ||$ are bounded parameters. The $RH$ side
of (4.118b) is therefore a linear combination of $2j+2$ norms, each one of
which has an upper bound consisting of the basic norms $\{NBAS2\},$ if $%
S(x)=\Theta (x-x_{L},\varepsilon )$. It is also a linear combination of $%
2j+2 $ norms, each of which has an upper bound consisting of the basic norms 
$\{NBAS1\},$ if $S(x)=\delta (x-x_{c},\varepsilon _{c}).$ This indicates
that the norm on the $LH$ side of (4.118b) with $S(x)=\Theta
(x-x_{L},\varepsilon )$ or $\delta (x-x_{c},\varepsilon _{c})$ for the case $%
\mu =S$ has an upper bound consisting of the basic norms $\{NBAS2\}$ or $%
\{NBAS1\}$. Here the number of the basic norms $\{NBAS1\}$ or $\{NBAS2\}$
depends on the order of the polynomial $q(x)Q_{j-l}^{a}(x,\pm \varphi
(x,\gamma ))$ in coordinate $x. $ If $q(x)Q_{j-l}^{a}(x,\pm \varphi
(x,\gamma ))$ is a $k_{a}-$order polynomial$,$ then the number is $(k_{a}+1)$
at most for each norm on the $RH $ side of (4.118b).

More generally, for a general spatially-selective function $S(x)$ the norm $%
||q(x)S(x)\frac{\partial ^{j}}{\partial x^{j}}\tilde{\Psi}_{0}^{\mu
}(x,r,t_{0}+t_{1}/2)||$ with $j=0$, $1$, $2$, $...$ and $\mu =CS$ or $S$ has
an upper bound consisting of the basic norms $\{NBAS1\}$ and/or $\{NBAS2\}$.
These results will be used below to prove that the six norms on the $RH$
side of (4.105) have upper bounds consisting of the basic norms $\{NBAS1\}$
and/or $\{NBAS2\}$.\newline
\newline
{\large (B) The upper bound of the norm }$||Q_{\lambda }(x,p,t_{3})\Psi
_{0}^{\mu }(x,r,t_{0}+t_{1}/2)||${\large \ with }$\mu =CS$

Now begin to calculate the norms $\{||Q_{\lambda }(x,p,t_{3})\Psi _{0}^{\mu
}(x,r,t_{0}+t_{1}/2)||\}$ in (4.105). The case $\mu =CS$ is first
considered. The product state $\Psi _{0}^{CS}(x,r,t_{0}+t_{1}/2)$ in the
norms is expressed as (4.106a) with the label $\mu =CS$, here the function $%
F_{0}^{CS}$ is given by (4.106b) and $\tilde{\Psi}%
_{0}^{CS}(x,r,t_{0}+t_{1}/2)$ is equal to $\Psi
_{0}^{C0}(x,r,t_{0}+t_{1}/2), $ as shown in (4.106c). There are six norms on
the $RH$ side of (4.105) to be computed for the case $\mu =CS.$ It is easy
to calculate the first norm $||F_{5}^{\lambda }(x,t_{3})\Psi
_{0}^{CS}(x,r,t_{0}+t_{1}/2)||$ on the $RH$ side of (4.105) with the index $%
l=0$. It can turn out that this norm satisfies%
\begin{equation*}
||F_{5}^{\lambda }(x,t_{3})F_{0}^{CS}\Psi _{0}^{C0}(x,r,t_{0}+t_{1}/2)||
\end{equation*}%
\begin{equation*}
\leq 2(\frac{1}{4\hslash }t_{1})^{2}||F_{5}^{\lambda }(x,t_{3})[\Theta
(x-x_{L},\varepsilon )\Omega (x)]^{2}\Psi _{0}^{C0}(x,r,t_{0}+t_{1}/2)||
\end{equation*}%
\begin{equation*}
+2(\frac{1}{2\hslash }t_{1})^{2}||F_{5}^{\lambda
}(x,t_{3})V_{1}^{ho}(x,\varepsilon )^{2}\Psi _{0}^{C0}(x,r,t_{0}+t_{1}/2)||
\end{equation*}%
\begin{equation}
+(\frac{1}{\hslash }t_{1})||F_{5}^{\lambda
}(x,t_{3})V_{1}^{ho}(x,\varepsilon )\Psi _{0}^{C0}(x,r,t_{0}+t_{1}/2)|| 
\tag{4.119}
\end{equation}%
where the inequality below is already used:%
\begin{equation*}
1-\cos [\frac{1}{\hslash }t_{1}V_{1}^{ho}(x,\varepsilon )]\cos [\frac{1}{%
2\hslash }t_{1}\Theta (x-x_{L},\varepsilon )\Omega (x)]
\end{equation*}%
\begin{equation*}
\leq 2[\frac{1}{2\hslash }t_{1}V_{1}^{ho}(x,\varepsilon )]^{2}+2[\frac{1}{%
4\hslash }t_{1}\Theta (x-x_{L},\varepsilon )\Omega (x)]^{2}.
\end{equation*}%
The functions $V_{1}^{ho}(x,\varepsilon )^{l}$ with $l=1$ and $2$ are
spatially-selective functions, as shown in (4.113), and so is $[\Theta
(x-x_{L},\varepsilon )\Omega (x)]^{2}.$ As shown in (4.101), the function $%
F_{5}^{\lambda }(x,t_{3})$ is a polynomial in coordinate $x$ with order not
more than five. The norms $|||\tilde{g}_{0}\rangle ||=1$ and $|||\tilde{e}%
\rangle ||=1$ because both the superposition states $|\tilde{g}_{0}\rangle $
and $|\tilde{e}\rangle $ are normalized. The product state $\Psi
_{0}^{C0}(x,r,t_{0}+t_{1}/2)$ is given by (4.89a). Then it can turn out that
each one of the three norms on the $RH$ side of (4.119) can be reduced to a
linear combination of the basic norms $\{NBAS2\}$, indicating that the first
norm on the $RH$ side of (4.105) has an upper bound consisting of the basic
norms $\{NBAS2\}$.

It also is easy to calculate the second norm $||F_{4}^{\lambda
}(x,t_{3})p\Psi _{0}^{CS}(x,r,t_{0}+t_{1}/2)||$ on the $RH$ side of (4.105)
with $l=1$. With the help of (4.107) it can turn out that this norm is
bounded by 
\begin{equation*}
||F_{4}^{\lambda }(x,t_{3})p\Psi _{0}^{CS}(x,r,t_{0}+t_{1}/2)||\leq \hslash
||F_{4}^{\lambda }(x,t_{3})(\frac{\partial }{\partial x}F_{0}^{CS})\Psi
_{0}^{C0}(x,r,t_{0}+t_{1}/2)||
\end{equation*}%
\begin{equation}
+\hslash ||F_{4}^{\lambda }(x,t_{3})F_{0}^{CS}\frac{\partial }{\partial x}%
\Psi _{0}^{C0}(x,r,t_{0}+t_{1}/2)||.  \tag{4.120}
\end{equation}%
It is easy to calculate the first-order coordinate derivative $(\frac{%
\partial }{\partial x}F_{0}^{CS})$ by using the function $F_{0}^{CS}$ of
(4.106b). This derivative is a spatially-selective function. It may be
expressed as%
\begin{equation*}
\frac{\partial }{\partial x}F_{0}^{CS}=-(\frac{1}{\hslash }t_{1})\exp [-i%
\frac{1}{\hslash }t_{1}V_{1}^{ho}(x,\varepsilon )]
\end{equation*}%
\begin{equation*}
\times \{i[\frac{\partial }{\partial x}V_{1}^{ho}(x,\varepsilon )]\cos [%
\frac{1}{2\hslash }\Theta (x-x_{L},\varepsilon )\Omega (x)t_{1}]
\end{equation*}%
\begin{equation}
+\frac{1}{2}\frac{\partial }{\partial x}[\Theta (x-x_{L},\varepsilon )\Omega
(x)]\sin [\frac{1}{2\hslash }\Theta (x-x_{L},\varepsilon )\Omega (x)t_{1}]\}.
\tag{4.121}
\end{equation}%
It results in that the first norm on the $RH$ side of (4.120) is bounded by%
\begin{equation*}
||F_{4}^{\lambda }(x,t_{3})(\frac{\partial }{\partial x}F_{0}^{CS})\Psi
_{0}^{C0}(x,r,t_{0}+t_{1}/2)||
\end{equation*}%
\begin{equation*}
\leq (\frac{1}{\hslash }t_{1})||F_{4}^{\lambda }(x,t_{3})[\frac{\partial }{%
\partial x}V_{1}^{ho}(x,\varepsilon )]\Psi _{0}^{C0}(x,r,t_{0}+t_{1}/2)||
\end{equation*}%
\begin{equation}
+(\frac{1}{2\hslash }t_{1})||F_{4}^{\lambda }(x,t_{3})\frac{\partial }{%
\partial x}[\Theta (x-x_{L},\varepsilon )\Omega (x)]\Psi
_{0}^{C0}(x,r,t_{0}+t_{1}/2)||.  \tag{4.122}
\end{equation}%
Here $\frac{\partial }{\partial x}V_{1}^{ho}(x,\varepsilon )$ and $\frac{%
\partial }{\partial x}[\Theta (x-x_{L},\varepsilon )\Omega (x)]$ are
spatially-selective functions, as shown in (4.114a) and (4.116),
respectively. The function $F_{4}^{\lambda }(x,t_{3})$ is a polynomial in
coordinate $x$ with order not more than four, as shown in (4.101). Then each
one of the two norms on the $RH$ side of (4.122) can be reduced to a linear
combination of the basic norms $\{NBAS1\}$ and $\{NBAS2\}$. Thus, the
inequality (4.122) shows that the first norm on the $RH$ side of (4.120) has
an upper bound that can be expressed as a linear combination of the basic
norms $\{NBAS1\}$ and $\{NBAS2\}$. On the other hand, by using the function $%
F_{0}^{CS}$ of (4.106b) it can turn out that the second norm on the $RH$
side of (4.120) is bounded by%
\begin{equation*}
||F_{4}^{\lambda }(x,t_{3})F_{0}^{CS}\frac{\partial }{\partial x}\Psi
_{0}^{C0}(x,r,t_{0}+t_{1}/2)||
\end{equation*}%
\begin{equation*}
\leq 2(\frac{1}{4\hslash }t_{1})^{2}||F_{4}^{\lambda }(x,t_{3})[\Theta
(x-x_{L},\varepsilon )\Omega (x)]^{2}\frac{\partial }{\partial x}\Psi
_{0}^{C0}(x,r,t_{0}+t_{1}/2)||
\end{equation*}%
\begin{equation*}
+2(\frac{1}{2\hslash }t_{1})^{2}||F_{4}^{\lambda
}(x,t_{3})V_{1}^{ho}(x,\varepsilon )^{2}\frac{\partial }{\partial x}\Psi
_{0}^{C0}(x,r,t_{0}+t_{1}/2)||
\end{equation*}%
\begin{equation}
+(\frac{1}{\hslash }t_{1})||F_{4}^{\lambda
}(x,t_{3})V_{1}^{ho}(x,\varepsilon )\frac{\partial }{\partial x}\Psi
_{0}^{C0}(x,r,t_{0}+t_{1}/2)||.  \tag{4.123}
\end{equation}%
This inequality is similar to that one of (4.119). Actually, it is equal to
the inequality (4.119) if one makes the replacement $F_{5}^{\lambda
}(x,t_{3})\leftrightarrow F_{4}^{\lambda }(x,t_{3})$ and $\Psi
_{0}^{C0}(x,r,t_{0}+t_{1}/2)\leftrightarrow \frac{\partial }{\partial x}\Psi
_{0}^{C0}(x,r,t_{0}+t_{1}/2)$ in (4.119). Generally, the $k-$order
coordinate derivative $\frac{\partial ^{k}}{\partial x^{k}}\Psi
_{0}^{C0}(x,r,t_{0}+t_{1}/2)$ ($k=0$, $1$, $2$, ...) can be computed by the
formula (4.110a). It can be reduced to a combination of the motional states $%
\{\Psi _{0}^{a}(x,t_{0}+t_{1}/2)\}$, 
\begin{equation*}
\frac{\partial ^{k}}{\partial x^{k}}\Psi _{0}^{C0}(x,r,t_{0}+t_{1}/2)=\Psi
_{0}^{g}(x,t_{0}+t_{1}/2)\Phi _{k}^{g}(x,r,t_{0}+t_{1}/2)
\end{equation*}%
\begin{equation}
+\Psi _{0}^{e}(x,t_{0}+t_{1}/2)\Phi _{k}^{e}(x,r,t_{0}+t_{1}/2). 
\tag{4.123a}
\end{equation}%
Here $\Phi _{k}^{a}(x,r,t_{0}+t_{1}/2)$ with $a=g$ or $e$ is the
superposition of the internal states $|g_{0}\rangle $ and $|e\rangle .$ It
may be called the internal superposition state. It is generally determined
from (4.110a), 
\begin{equation}
\Phi _{k}^{a}(x,r,t_{0}+t_{1}/2)=\sum_{l=0}^{k}\left( 
\begin{array}{c}
k \\ 
l%
\end{array}%
\right) (\frac{\partial ^{l}}{\partial x^{l}}|\tilde{a}\rangle
)Q_{k-l}^{a}(x)  \tag{4.123b}
\end{equation}%
where $|\tilde{a}\rangle =|\tilde{g}_{0}\rangle $ if the index $a=g$ and $|%
\tilde{a}\rangle =|\tilde{e}\rangle $ if the index $a=e$. In particular,
when $k=0$, the equation (4.123a) is reduced to (4.89a). Then one has $\Phi
_{0}^{g}(x,r,t_{0}+t_{1}/2)=|\tilde{g}_{0}\rangle $ and $\Phi
_{0}^{e}(x,r,t_{0}+t_{1}/2)=|\tilde{e}\rangle .$ For $k=1$ the internal
superposition state is given by%
\begin{equation}
\Phi _{1}^{a}(x,r,t_{0}+t_{1}/2)=Q_{1}^{a}(x)|\tilde{a}\rangle +Q_{0}^{a}(x)(%
\frac{\partial }{\partial x}|\tilde{a}\rangle ).  \tag{4.123c}
\end{equation}%
Since the norm $||\frac{\partial ^{l}}{\partial x^{l}}|\tilde{a}\rangle ||$
is a bounded parameter and $Q_{l}^{a}(x)$ is a $l-$order polynomial in
coordinate $x$ ($l=0$, $1,$ $2$, ...), the two equations (4.123a) and
(4.123c) lead to that each one of the three norms on the $RH$ side of
(4.123) can be reduced to a linear sum of the four norms, as shown in
(4.118a). Actually, each norm on the $RH$ side of (4.123) can be reduced as
follows:%
\begin{equation*}
||F_{4}^{\lambda }(x,t_{3})S(x)\frac{\partial }{\partial x}\Psi
_{0}^{C0}(x,r,t_{0}+t_{1}/2)||
\end{equation*}%
\begin{equation*}
\leq ||(|\tilde{g}_{0}\rangle )||\times ||F_{4}^{\lambda
}(x,t_{3})Q_{1}^{g}(x)S(x)\Psi _{0}^{g}(x,t_{0}+t_{1}/2)||
\end{equation*}%
\begin{equation*}
+||(\frac{\partial }{\partial x}|\tilde{g}_{0}\rangle )||\times
||F_{4}^{\lambda }(x,t_{3})Q_{0}^{g}(x)S(x)\Psi _{0}^{g}(x,t_{0}+t_{1}/2)||
\end{equation*}%
\begin{equation*}
+||(|\tilde{e}\rangle )||\times ||F_{4}^{\lambda
}(x,t_{3})Q_{1}^{e}(x)S(x)\Psi _{0}^{e}(x,t_{0}+t_{1}/2)||
\end{equation*}%
\begin{equation}
+||(\frac{\partial }{\partial x}|\tilde{e}\rangle )||\times ||F_{4}^{\lambda
}(x,t_{3})Q_{0}^{e}(x)S(x)\Psi _{0}^{e}(x,t_{0}+t_{1}/2)||  \tag{4.123d}
\end{equation}%
where the spatially-selective function $S(x)=[\Theta (x-x_{L},\varepsilon
)\Omega (x)]^{2},$ $V_{1}^{ho}(x,\varepsilon )^{2},$ or $V_{1}^{ho}(x,%
\varepsilon ).$ Each one of the four norms on the $RH$ side of (4.123d) has
an upper bound consisting of the basic norms $\{NBAS2\}$, because the
spatially-selective function $S(x)$ contains only the spatially-selective
factor $\Theta (x-x_{L},\varepsilon )$ and $F_{4}^{\lambda
}(x,t_{3})Q_{l}^{a}(x)$ with $a=g$ or $e$ and $l=0$ or $1$ is a polynomial
in coordinate $x$ with order not more than five. Note that the norms $||%
\frac{\partial ^{l}}{\partial x^{l}}|\tilde{g}_{0}\rangle ||$ and $||\frac{%
\partial ^{l}}{\partial x^{l}}|\tilde{e}\rangle ||$ for $l=0$, $1$ are
bounded parameters. Then the inequality (4.123d) shows that each norm on the 
$RH$ side of (4.123) has an upper bound consisting of the basic norms $%
\{NBAS2\}.$ Therefore, all the three norms on the $RH$ side of (4.123) can
be reduced to the linear combinations of the basic norms $\{NBAS2\}$,
respectively. Then the inequality (4.123) indicates that the second norm on
the $RH$ side of (4.120) has an upper bound consisting of the basic norms $%
\{NBAS2\}$. It is already shown by the inequality (4.122) that the first
norm on the $RH$ side of (4.120) has an upper bound consisting of the basic
norms $\{NBAS1\}$ and $\{NBAS2\}$. Since now both the norms on the $RH$ side
of (4.120) have upper bounds consisting of the basic norms $\{NBAS1\}$
and/or $\{NBAS2\}$, the inequality (4.120) indicates that the second norm on
the $RH$ side of (4.105) has an upper bound consisting of the basic norms $%
\{NBAS1\}$ and $\{NBAS2\}$.

The third norm on the $RH$ side of (4.105) with $l=2$ is $||F_{3}^{\lambda
}(x,t_{3})p^{2}\Psi _{0}^{CS}(x,r,$ $t_{0}+t_{1}/2)||.$ With the help of
(4.107) it can turn out that the norm is bounded by%
\begin{equation*}
||F_{3}^{\lambda }(x,t_{3})p^{2}\Psi _{0}^{CS}(x,r,t_{0}+t_{1}/2)||\leq
\hslash ^{2}||F_{3}^{\lambda }(x,t_{3})(\frac{\partial ^{2}}{\partial x^{2}}%
F_{0}^{CS})\Psi _{0}^{C0}(x,r,t_{0}+t_{1}/2)||
\end{equation*}%
\begin{equation*}
+2\hslash ^{2}||F_{3}^{\lambda }(x,t_{3})(\frac{\partial }{\partial x}%
F_{0}^{CS})\frac{\partial }{\partial x}\Psi _{0}^{C0}(x,r,t_{0}+t_{1}/2)||
\end{equation*}%
\begin{equation}
+\hslash ^{2}||F_{3}^{\lambda }(x,t_{3})F_{0}^{CS}\frac{\partial ^{2}}{%
\partial x^{2}}\Psi _{0}^{C0}(x,r,t_{0}+t_{1}/2)||.  \tag{4.124}
\end{equation}%
Here the function $F_{3}^{\lambda }(x,t_{3})$ is a polynomial in coordinate $%
x$ with order not more than three, as shown in (4.101). The second norm on
the $RH$ side of (4.124) can be simply calculated. It can be calculated in a
similar way that one calculates the norm on the $LH$ side of (4.122) above.
Actually, if in the inequality (4.122) one makes the replacement $%
F_{4}^{\lambda }(x,t_{3})\leftrightarrow F_{3}^{\lambda }(x,t_{3})$\ and $%
\Psi _{0}^{C0}(x,r,t_{0}+t_{1}/2)\leftrightarrow \frac{\partial }{\partial x}%
\Psi _{0}^{C0}(x,r,t_{0}+t_{1}/2),$ then one finds that the second norm on
the $RH$ side of (4.124) obeys the inequality (4.122), 
\begin{equation*}
||F_{3}^{\lambda }(x,t_{3})(\frac{\partial }{\partial x}F_{0}^{CS})\frac{%
\partial }{\partial x}\Psi _{0}^{C0}(x,r,t_{0}+t_{1}/2)||
\end{equation*}%
\begin{equation*}
\leq (\frac{1}{\hslash }t_{1})||F_{3}^{\lambda }(x,t_{3})[\frac{\partial }{%
\partial x}V_{1}^{ho}(x,\varepsilon )]\frac{\partial }{\partial x}\Psi
_{0}^{C0}(x,r,t_{0}+t_{1}/2)||
\end{equation*}%
\begin{equation}
+(\frac{1}{2\hslash }t_{1})||F_{3}^{\lambda }(x,t_{3})\frac{\partial }{%
\partial x}[\Theta (x-x_{L},\varepsilon )\Omega (x)]\frac{\partial }{%
\partial x}\Psi _{0}^{C0}(x,r,t_{0}+t_{1}/2)||.  \tag{4.124a}
\end{equation}%
Here the derivative $\frac{\partial }{\partial x}\Psi
_{0}^{C0}(x,r,t_{0}+t_{1}/2)$ is given by the formula (4.123a) with the
internal superposition state $\Phi _{1}^{a}(x,r,t_{0}+t_{1}/2)$ of (4.123c).
Now one can find that each norm on the $RH$ side of (4.124a) obeys the
inequality (4.123d) if one makes the replacement $F_{4}^{\lambda
}(x,t_{3})\leftrightarrow F_{3}^{\lambda }(x,t_{3})$ and lets the
spatially-selective function $S(x)=\frac{\partial }{\partial x}%
V_{1}^{ho}(x,\varepsilon )$ or $\frac{\partial }{\partial x}[\Theta
(x-x_{L},\varepsilon )\Omega (x)]$ in (4.123d). Then such modified
inequality of (4.123d) leads to that each norm on the $RH$ side of (4.124a)
can be reduced to the four norms, each of which has an upper bound
consisting of the basic norms $\{NBAS1\}$ and $\{NBAS2\}$, because now the
spatially-selective function $S(x)$ contains both the spatially-selective
factors $\Theta (x-x_{L},\varepsilon )$ and $\delta (x-x_{c},\varepsilon
_{c}).$ Therefore, the inequality (4.124a) shows that the second norm on the 
$RH$ side of (4.124) has an upper bound consisting of the basic norms $%
\{NBAS1\}$ and $\{NBAS2\}$. Similarly, the third norm on the $RH$ side of
(4.124) obeys the inequality (4.119) if there is the replacement $%
F_{5}^{\lambda }(x,t_{3})\leftrightarrow F_{3}^{\lambda }(x,t_{3})$\ and $%
\Psi _{0}^{C0}(x,r,t_{0}+t_{1}/2)\leftrightarrow \frac{\partial ^{2}}{%
\partial x^{2}}\Psi _{0}^{C0}(x,r,t_{0}+t_{1}/2)$ in (4.119), 
\begin{equation*}
||F_{3}^{\lambda }(x,t_{3})F_{0}^{CS}\frac{\partial ^{2}}{\partial x^{2}}%
\Psi _{0}^{C0}(x,r,t_{0}+t_{1}/2)||
\end{equation*}%
\begin{equation*}
\leq 2(\frac{1}{4\hslash }t_{1})^{2}||F_{3}^{\lambda }(x,t_{3})[\Theta
(x-x_{L},\varepsilon )\Omega (x)]^{2}\frac{\partial ^{2}}{\partial x^{2}}%
\Psi _{0}^{C0}(x,r,t_{0}+t_{1}/2)||
\end{equation*}%
\begin{equation*}
+2(\frac{1}{2\hslash }t_{1})^{2}||F_{3}^{\lambda
}(x,t_{3})V_{1}^{ho}(x,\varepsilon )^{2}\frac{\partial ^{2}}{\partial x^{2}}%
\Psi _{0}^{C0}(x,r,t_{0}+t_{1}/2)||
\end{equation*}%
\begin{equation}
+(\frac{1}{\hslash }t_{1})||F_{3}^{\lambda
}(x,t_{3})V_{1}^{ho}(x,\varepsilon )\frac{\partial ^{2}}{\partial x^{2}}\Psi
_{0}^{C0}(x,r,t_{0}+t_{1}/2)||.  \tag{4.124b}
\end{equation}%
Here the second-order derivative $\frac{\partial ^{2}}{\partial x^{2}}\Psi
_{0}^{C0}(x,r,t_{0}+t_{1}/2)$ is given by the formula (4.123a), where the
internal superposition state $\Phi _{2}^{a}(x,r,t_{0}+t_{1}/2)$ of (4.123b)
is explicitly given by%
\begin{equation}
\Phi _{2}^{a}(x,r,t_{0}+t_{1}/2)=Q_{2}^{a}(x)|\tilde{a}\rangle +2Q_{1}^{a}(x)%
\frac{\partial }{\partial x}|\tilde{a}\rangle +Q_{0}^{a}(x)\frac{\partial
^{2}}{\partial x^{2}}|\tilde{a}\rangle .  \tag{4.124c}
\end{equation}%
Now by using the derivative $\frac{\partial ^{2}}{\partial x^{2}}\Psi
_{0}^{C0}(x,r,t_{0}+t_{1}/2)$ of (4.123a) with $\Phi
_{2}^{a}(x,r,t_{0}+t_{1}/2)$ of (4.124c) each norm on the $RH$ side of
(4.124b) can be further reduced to the form%
\begin{equation*}
||F_{3}^{\lambda }(x,t_{3})S(x)\frac{\partial ^{2}}{\partial x^{2}}\Psi
_{0}^{C0}(x,r,t_{0}+t_{1}/2)||
\end{equation*}%
\begin{equation*}
\leq \sum_{l=0}^{2}\left( 
\begin{array}{c}
2 \\ 
l%
\end{array}%
\right) ||(\frac{\partial ^{l}}{\partial x^{l}}|\tilde{g}_{0}\rangle
)||\times ||F_{3}^{\lambda }(x,t_{3})Q_{2-l}^{g}(x)S(x)\Psi
_{0}^{g}(x,t_{0}+t_{1}/2)||
\end{equation*}%
\begin{equation}
+\sum_{l=0}^{2}\left( 
\begin{array}{c}
2 \\ 
l%
\end{array}%
\right) ||(\frac{\partial ^{l}}{\partial x^{l}}|\tilde{e}\rangle )||\times
||F_{3}^{\lambda }(x,t_{3})Q_{2-l}^{e}(x)S(x)\Psi
_{0}^{e}(x,t_{0}+t_{1}/2)||,  \tag{4.124d}
\end{equation}%
where the spatially-selective function $S(x)=[\Theta (x-x_{L},\varepsilon
)\Omega (x)]^{2},$ $V_{1}^{ho}(x,\varepsilon )^{2},$ or $V_{1}^{ho}(x,%
\varepsilon ).$ This inequality is really a special instance of (4.118a).
Note that the norms $|\frac{\partial ^{l}}{\partial x^{l}}|\tilde{g}%
_{0}\rangle ||$ and $||\frac{\partial ^{l}}{\partial x^{l}}|\tilde{e}\rangle
||$ for $l=0$, $1$, $2$ are bounded parameters and $F_{3}^{\lambda
}(x,t_{3})Q_{2-l}^{a}(x)$ with $a=g$ or $e$ and $l=0$, $1$, $2$ is a
polynomial in coordinate $x$ with order not more than five. Therefore, each
norm on the $RH$ side of (4.124b) is reduced to a combination of six norms
on the $RH$ side of (4.124d), each one of which has an upper bound
consisting of the basic norms $\{NBAS2\}$, because here the
spatially-selective function $S(x)$ contains only $\Theta
(x-x_{L},\varepsilon ).$ Then the inequality (4.124b) shows that the third
norm on the $RH$ side of (4.124) has an upper bound consisting of the basic
norms $\{NBAS2\}$. Now one needs to calculate explicitly the first norm on
the $RH$ side of (4.124). Here the second-order derivative $\frac{\partial
^{2}}{\partial x^{2}}F_{0}^{CS}$ can be calculated directly by using the
function $F_{0}^{CS}$ of (4.106b). Then by using the derivative to calculate
further the norm one obtains%
\begin{equation*}
||F_{3}^{\lambda }(x,t_{3})(\frac{\partial ^{2}}{\partial x^{2}}%
F_{0}^{CS})\Psi _{0}^{C0}(x,r,t_{0}+t_{1}/2)||
\end{equation*}%
\begin{equation*}
\leq (\frac{1}{\hslash }t_{1})||F_{3}^{\lambda }(x,t_{3})[\frac{\partial ^{2}%
}{\partial x^{2}}V_{1}^{ho}(x,\varepsilon )]\Psi
_{0}^{C0}(x,r,t_{0}+t_{1}/2)||
\end{equation*}%
\begin{equation*}
+(\frac{1}{\hslash }t_{1})^{2}||F_{3}^{\lambda }(x,t_{3})[\frac{\partial }{%
\partial x}V_{1}^{ho}(x,\varepsilon )]^{2}\Psi _{0}^{C0}(x,r,t_{0}+t_{1}/2)||
\end{equation*}%
\begin{equation*}
+(\frac{1}{\hslash }t_{1})^{2}||F_{3}^{\lambda }(x,t_{3})[\frac{\partial }{%
\partial x}V_{1}^{ho}(x,\varepsilon )](\frac{\partial }{\partial x}[\Theta
(x-x_{L},\varepsilon )\Omega (x)])\Psi _{0}^{C0}(x,r,t_{0}+t_{1}/2)||
\end{equation*}%
\begin{equation*}
+(\frac{1}{2\hslash }t_{1})||F_{3}^{\lambda }(x,t_{3})(\frac{\partial ^{2}}{%
\partial x^{2}}[\Theta (x-x_{L},\varepsilon )\Omega (x)])\Psi
_{0}^{C0}(x,r,t_{0}+t_{1}/2)||
\end{equation*}%
\begin{equation}
+(\frac{1}{2\hslash }t_{1})^{2}||F_{3}^{\lambda }(x,t_{3})(\frac{\partial }{%
\partial x}[\Theta (x-x_{L},\varepsilon )\Omega (x)])^{2}\Psi
_{0}^{C0}(x,r,t_{0}+t_{1}/2)||.  \tag{4.125}
\end{equation}%
The functions $\frac{\partial ^{2}}{\partial x^{2}}V_{1}^{ho}(x,\varepsilon
) $ and $[\frac{\partial }{\partial x}V_{1}^{ho}(x,\varepsilon )]^{k}$ for $%
k=1 $ and $2$ are spatially-selective functions. This can be seen in
(4.114b), (4.114a), and (4.115a). On the other hand, it is known from
(4.116) that the function $\frac{\partial ^{2}}{\partial x^{2}}[\Theta
(x-x_{L},\varepsilon )\Omega (x)]$ is a spatially-selective function and it
follows from (4.117) that the functions $(\frac{\partial }{\partial x}%
[\Theta (x-x_{L},\varepsilon )\Omega (x)])^{k}$ with $k=1$ and $2$ are also
spatially-selective functions. The product of the two spatially-selective
functions $[\frac{\partial }{\partial x}V_{1}^{ho}(x,\varepsilon )]$ and $(%
\frac{\partial }{\partial x}[\Theta (x-x_{L},\varepsilon )\Omega (x)])$ is
also a spatially-selective function. These show that each one of the norms
on the $RH$ side of (4.125) has an upper bound that can be expressed as a
linear combination of the basic norms $\{NBAS1\}$ and $\{NBAS2\}$,
indicating that the first norm on the $RH$ side of (4.124) also has an upper
bound consisting of the basic norms $\{NBAS1\}$ and $\{NBAS2\}$. Now all the
three norms on the $RH$ side of (4.124) are proven to have upper bounds
consisting of the basic norms $\{NBAS1\}$ and/or $\{NBAS2\}$. This indicates
that the third norm on the $RH$ side of (4.105) indeed has an upper bound
consisting of the basic norms $\{NBAS1\}$ and $\{NBAS2\}.$

Now calculate the fourth norm on the $RH$ side of (4.105). It can turn out
by using (4.107) that the norm is bounded by%
\begin{equation*}
||F_{2}^{\lambda }(x,t_{3})p^{3}\Psi _{0}^{CS}(x,r,t_{0}+t_{1}/2)||\leq
\hslash ^{3}\{||F_{2}^{\lambda }(x,t_{3})(\frac{\partial ^{3}}{\partial x^{3}%
}F_{0}^{CS})\Psi _{0}^{C0}(x,r,t_{0}+t_{1}/2)||
\end{equation*}%
\begin{equation*}
+3||F_{2}^{\lambda }(x,t_{3})(\frac{\partial ^{2}}{\partial x^{2}}F_{0}^{CS})%
\frac{\partial }{\partial x}\Psi _{0}^{C0}(x,r,t_{0}+t_{1}/2)||
\end{equation*}%
\begin{equation*}
+3||F_{2}^{\lambda }(x,t_{3})(\frac{\partial }{\partial x}F_{0}^{CS})\frac{%
\partial ^{2}}{\partial x^{2}}\Psi _{0}^{C0}(x,r,t_{0}+t_{1}/2)||
\end{equation*}%
\begin{equation}
+||F_{2}^{\lambda }(x,t_{3})F_{0}^{CS}\frac{\partial ^{3}}{\partial x^{3}}%
\Psi _{0}^{C0}(x,r,t_{0}+t_{1}/2)||\}.  \tag{4.126}
\end{equation}%
Here the function $F_{2}^{\lambda }(x,t_{3})$ is a polynomial in coordinate $%
x$ with order not more than two, as shown in (4.101). One needs to calculate
all the four norms on the $RH$ side of the inequality (4.126). The last norm
on the $RH$ side of (4.126) satisfies the inequality (4.119) if one makes
the replacement $F_{5}^{\lambda }(x,t_{3})\leftrightarrow F_{2}^{\lambda
}(x,t_{3})$ and $\Psi _{0}^{C0}(x,r,t_{0}+t_{1}/2)\leftrightarrow \frac{%
\partial ^{3}}{\partial x^{3}}\Psi _{0}^{C0}(x,r,t_{0}+t_{1}/2)$ in (4.119).
Here the third-order derivative $\frac{\partial ^{3}}{\partial x^{3}}\Psi
_{0}^{C0}(x,r,t_{0}+t_{1}/2)$ is given by (4.123a), where the internal
superposition state $\Phi _{3}^{a}(x,r,t_{0}+t_{1}/2)$ of (4.123b) is given
by%
\begin{equation*}
\Phi _{3}^{a}(x,r,t_{0}+t_{1}/2)=Q_{3}^{a}(x)|\tilde{a}\rangle +3Q_{2}^{a}(x)%
\frac{\partial }{\partial x}|\tilde{a}\rangle
\end{equation*}%
\begin{equation}
+3Q_{1}^{a}(x)\frac{\partial ^{2}}{\partial x^{2}}|\tilde{a}\rangle
+Q_{0}^{a}(x)\frac{\partial ^{3}}{\partial x^{3}}|\tilde{a}\rangle . 
\tag{4.126a}
\end{equation}%
There are three norms on the $RH$ side of this modified inequality of
(4.119), each one of which satisfies the inequality (4.118a): 
\begin{equation*}
||F_{2}^{\lambda }(x,t_{3})S(x)\frac{\partial ^{3}}{\partial x^{3}}\Psi
_{0}^{C0}(x,r,t_{0}+t_{1}/2)||
\end{equation*}%
\begin{equation*}
\leq \sum_{l=0}^{3}\left( 
\begin{array}{c}
3 \\ 
l%
\end{array}%
\right) ||(\frac{\partial ^{l}}{\partial x^{l}}|\tilde{g}_{0}\rangle
)||\times ||F_{2}^{\lambda }(x,t_{3})Q_{3-l}^{g}(x)S(x)\Psi
_{0}^{g}(x,t_{0}+t_{1}/2)||
\end{equation*}%
\begin{equation}
+\sum_{l=0}^{3}\left( 
\begin{array}{c}
3 \\ 
l%
\end{array}%
\right) ||(\frac{\partial ^{l}}{\partial x^{l}}|\tilde{e}\rangle )||\times
||F_{2}^{\lambda }(x,t_{3})Q_{3-l}^{e}(x)S(x)\Psi
_{0}^{e}(x,t_{0}+t_{1}/2)||,  \tag{4.126b}
\end{equation}%
where the spatially-selective function $S(x)=[\Theta (x-x_{L},\varepsilon
)\Omega (x)]^{2},$ $V_{1}^{ho}(x,\varepsilon )^{2},$ or $V_{1}^{ho}(x,%
\varepsilon ).$ Note that the norms $||\frac{\partial ^{l}}{\partial x^{l}}|%
\tilde{g}_{0}\rangle ||$ and $||\frac{\partial ^{l}}{\partial x^{l}}|\tilde{e%
}\rangle ||$ for $0\leq l\leq 3$ are bounded parameters and $F_{2}^{\lambda
}(x,t_{3})Q_{3-l}^{a}(x)$ is a polynomial in coordinate $x$ with order not
more than five. The $RH$ side of (4.126b) has eight norms, each one of which
has an upper bound consisting of the basic norms $\{NBAS2\},$ because here
the spatially-selective function $S(x)$ contains only $\Theta
(x-x_{L},\varepsilon ).$ Then the norm on the $LH$ side of (4.126b) has an
upper bound consisting of the basic norms $\{NBAS2\}$, indicating that the
last norm on the $RH$ side of (4.126) has an upper bound consisting of the
basic norms $\{NBAS2\}$. Now the third norm on the $RH$ side of (4.126)
satisfies the inequality (4.122) if one makes the replacement $%
F_{4}^{\lambda }(x,t_{3})\leftrightarrow F_{2}^{\lambda }(x,t_{3})$ and $%
\Psi _{0}^{C0}(x,r,t_{0}+t_{1}/2)\leftrightarrow \frac{\partial ^{2}}{%
\partial x^{2}}\Psi _{0}^{C0}(x,r,t_{0}+t_{1}/2)$ in (4.122). Here the
derivative $\frac{\partial ^{2}}{\partial x^{2}}\Psi
_{0}^{C0}(x,r,t_{0}+t_{1}/2)$ is given by (4.123a) with the internal
superposition state $\Phi _{2}^{a}(x,r,t_{0}+t_{1}/2)$ of (4.124c). There
are two norms on the $RH$ side of this modified inequality of (4.122), each
one of which is given by $||F_{2}^{\lambda }(x,t_{3})S(x)\frac{\partial ^{2}%
}{\partial x^{2}}\Psi _{0}^{C0}(x,r,t_{0}+t_{1}/2)||,$ where the
spatially-selective function $S(x)=\frac{\partial }{\partial x}%
V_{1}^{ho}(x,\varepsilon )$ or $\frac{\partial }{\partial x}[\Theta
(x-x_{L},\varepsilon )\Omega (x)].$ One can find that such a norm satisfies
the inequality (4.124d) with the replacement $F_{3}^{\lambda
}(x,t_{3})\leftrightarrow F_{2}^{\lambda }(x,t_{3}).$ It follows from this
modified inequality of (4.124d) that the norm can be reduced to a
combination of six norms, each one of which has an upper bound consisting of
the basic norms $\{NBAS1\}$ and $\{NBAS2\}$ due to that the current
spatially-selective function $S(x)$ contains both $\Theta
(x-x_{L},\varepsilon )$ and $\delta (x-x_{c},\varepsilon _{c}).$ Therefore,
this shows that the third norm on the $RH$ side of (4.126) has an upper
bound consisting of the basic norms $\{NBAS1\}$ and $\{NBAS2\}$. Similarly,
the second norm on the $RH$ side of (4.126) satisfies the inequality (4.125)
if one makes the replacement $F_{3}^{\lambda }(x,t_{3})\leftrightarrow
F_{2}^{\lambda }(x,t_{3})$ and $\Psi
_{0}^{C0}(x,r,t_{0}+t_{1}/2)\leftrightarrow \frac{\partial }{\partial x}\Psi
_{0}^{C0}(x,r,t_{0}+t_{1}/2)$ in (4.125). Here the derivative $\frac{%
\partial }{\partial x}\Psi _{0}^{C0}(x,r,t_{0}+t_{1}/2)$ is given by
(4.123a) with the internal superposition state $\Phi
_{1}^{a}(x,r,t_{0}+t_{1}/2)$ of (4.123c). It can be found that there are
five norms on the $RH$ side of the modified inequality of (4.125), each one
of which may be written as $||F_{2}^{\lambda }(x,t_{3})S(x)\frac{\partial }{%
\partial x}\Psi _{0}^{C0}(x,r,t_{0}+t_{1}/2)||,$ where the
spatially-selective function $S(x)=\frac{\partial ^{2}}{\partial x^{2}}%
V_{1}^{ho}(x,\varepsilon ),$ $[\frac{\partial }{\partial x}%
V_{1}^{ho}(x,\varepsilon )]^{2},$ $[\frac{\partial }{\partial x}%
V_{1}^{ho}(x,\varepsilon )](\frac{\partial }{\partial x}[\Theta
(x-x_{L},\varepsilon )\Omega (x)]),$ $(\frac{\partial ^{2}}{\partial x^{2}}%
[\Theta (x-x_{L},\varepsilon )\Omega (x)]),$ or $(\frac{\partial }{\partial x%
}[\Theta (x-x_{L},\varepsilon )\Omega (x)])^{2}.$ One can further find that
such a norm satisfies the inequality (4.123d) with the replacement $%
F_{4}^{\lambda }(x,t_{3})\leftrightarrow F_{2}^{\lambda }(x,t_{3}).$ Then
this modified inequality of (4.123d) shows that the norm can be reduced to a
combination of four norms, each one of which has an upper bound consisting
of the basic norms $\{NBAS1\}$ and $\{NBAS2\}$ due to that the current
spatially-selective function $S(x)$ contains both $\Theta
(x-x_{L},\varepsilon )$ and $\delta (x-x_{c},\varepsilon _{c}).$ These show
that the second norm on the $RH$ side of (4.126) has an upper bound
consisting of the basic norms $\{NBAS1\}$ and $\{NBAS2\}$. Below one
calculates explicitly the first norm on the $RH$ side of (4.126). At first
the derivative $\frac{\partial ^{3}}{\partial x^{3}}F_{0}^{CS}$ is computed
explicitly by using the function $F_{0}^{CS}$ of (4.106b). Then by using the
derivative it can turn out that this norm satisfies the inequality:%
\begin{equation}
||F_{2}^{\lambda }(x,t_{3})(\frac{\partial ^{3}}{\partial x^{3}}%
F_{0}^{CS})\Psi _{0}^{C0}(x,r,t_{0}+t_{1}/2)||\leq NA4+NB4+NC4.  \tag{4.127}
\end{equation}%
All the three terms $NA4,$ $NB4,$ and $NC4$ on the $RH$ side of (4.127) are
non-negative. They are given below. The term $NA4$ is given by%
\begin{equation*}
NA4=\sum_{k,l,m,n}\beta _{A}^{4}(k,l,m,n)||F_{2}^{\lambda }(x,t_{3})[\frac{%
\partial ^{k}}{\partial x^{k}}V_{1}^{ho}(x,\varepsilon )]^{m}
\end{equation*}%
\begin{equation}
\times \lbrack \frac{\partial ^{l}}{\partial x^{l}}V_{1}^{ho}(x,\varepsilon
)]^{n}\Psi _{0}^{C0}(x,r,t_{0}+t_{1}/2)||.  \tag{4.128}
\end{equation}%
There are only three non-zero parameters $\{\beta _{A}^{4}(k,l,m,n)\}$ in
(4.128). Moreover, these three parameters are non-negative. They are given
by $\beta _{A}^{4}(3,0,1,0)=(\frac{1}{\hslash }t_{1}),$ $\beta
_{A}^{4}(1,2,1,1)=3(\frac{1}{\hslash }t_{1})^{2},$ and $\beta
_{A}^{4}(1,0,3,0)=(\frac{1}{\hslash }t_{1})^{3}.$ This shows that each one
of these parameters is proportional to $(t_{1}/\hslash )^{\mu }$ with the
integer $\mu $ in between $1$ and $3$ (i.e., $\mu \in \lbrack 1,$ $3])$ and
hence it can be controlled by the time $t_{1}.$ The term $NB4$ may be
written as%
\begin{equation*}
NB4=\sum_{k,l,m,n}\beta _{B}^{4}(k,l,m,n)||F_{2}^{\lambda }(x,t_{3})[\frac{%
\partial ^{k}}{\partial x^{k}}V_{1}^{ho}(x,\varepsilon )]^{m}
\end{equation*}%
\begin{equation}
\times \{\frac{\partial ^{l}}{\partial x^{l}}[\Theta (x-x_{L},\varepsilon
)\Omega (x)]\}^{n}\Psi _{0}^{C0}(x,r,t_{0}+t_{1}/2)||.  \tag{4.129}
\end{equation}%
There are only four non-zero parameters $\{\beta _{B}^{4}(k,l,m,n)\}$ in
(4.129). They are non-negative and given by $\beta _{B}^{4}(2,1,1,1)=\beta
_{B}^{4}(1,2,1,1)=\frac{3}{2}(\frac{1}{\hslash }t_{1})^{2}$ and $\beta
_{B}^{4}(1,1,2,1)=2\beta _{B}^{4}(1,1,1,2)=\frac{3}{2}(\frac{1}{\hslash }%
t_{1})^{3}.$ The term $NC4$ is given by%
\begin{equation*}
NC4=\sum_{k,l,m,n}\beta _{C}^{4}(k,l,m,n)||F_{2}^{\lambda }(x,t_{3})\{\frac{%
\partial ^{k}}{\partial x^{k}}[\Theta (x-x_{L},\varepsilon )\Omega (x)]\}^{m}
\end{equation*}%
\begin{equation}
\times \{\frac{\partial ^{l}}{\partial x^{l}}[\Theta (x-x_{L},\varepsilon
)\Omega (x)]\}^{n}\Psi _{0}^{C0}(x,r,t_{0}+t_{1}/2)||.  \tag{4.130}
\end{equation}%
There are only three non-zero parameters $\{\beta _{C}^{4}(k,l,m,n)\}$ in
(4.130). They are non-negative and given by $\beta _{C}^{4}(3,0,1,0)=(\frac{1%
}{2\hslash }t_{1}),$ $\beta _{C}^{4}(1,2,1,1)=3(\frac{1}{2\hslash }%
t_{1})^{2},$ $\beta _{C}^{4}(1,0,3,0)=(\frac{1}{2\hslash }t_{1})^{3}.$ It is
clear that all these indices ($k$, $l$, $m$, $n$) satisfy $0\leq k,$ $l,$ $%
m, $ $n\leq 3$, $k+l>0,$ and $m+n>0$ for those terms with non-zero
parameters in these three terms $NA4$, $NB4$, and $NC4$. This really means
that there is always a spatially-selective function inside every norm of
these three terms. It can be found from (4.128), (4.129), and (4.130) that
there appear only the two types of the spatially-selective functions $[\frac{%
\partial ^{k}}{\partial x^{k}}V_{1}^{ho}(x,\varepsilon )]^{m}$ and $\{\frac{%
\partial ^{k}}{\partial x^{k}}[\Theta (x-x_{L},\varepsilon )\Omega
(x)]\}^{m} $ in the three terms $NA4$, $NB4$, and $NC4,$ here $k>0$ and $%
m>0. $

Now investigate the three norms of the term $NA4$ on the $RH$ side of
(4.128). Each one of the three norms inside contains the spatially-selective
function $S(x)=[\frac{\partial ^{k}}{\partial x^{k}}V_{1}^{ho}(x,\varepsilon
)]^{m}[\frac{\partial ^{l}}{\partial x^{l}}V_{1}^{ho}(x,\varepsilon )]^{n},$
which is the product of the two spatially-selective functions $[\frac{%
\partial ^{k}}{\partial x^{k}}V_{1}^{ho}(x,\varepsilon )]^{m}$ and $[\frac{%
\partial ^{l}}{\partial x^{l}}V_{1}^{ho}(x,\varepsilon )]^{n},$ here $0\leq
k,$ $l,$ $m,$ $n\leq 3$ and $0<m+n\leq 3.$ In the previous subsection (A)
the three formulae (4.114) show that the functions $\frac{\partial }{%
\partial x}V_{1}^{ho}(x,\varepsilon ),$ $\frac{\partial ^{2}}{\partial x^{2}}%
V_{1}^{ho}(x,\varepsilon ),$ and $\frac{\partial ^{k}}{\partial x^{k}}%
V_{1}^{ho}(x,\varepsilon )$ with $k>2$ consist of four, five, and three
spatially-selective functions, respectively, here the relevant secondary
terms are already neglected. It also is known from the three formulae
(4.115) that the function $[\frac{\partial ^{k}}{\partial x^{k}}%
V_{1}^{ho}(x,\varepsilon )]^{m}$ with $k>0$ and $m>1$ consists of $2m+1$
spatially-selective functions at most after the relevant secondary terms are
neglected. These together show that the function $[\frac{\partial ^{k}}{%
\partial x^{k}}V_{1}^{ho}(x,\varepsilon )]^{m}$ with $k>0$ and $m\geq 1$ is
a spatially-selective function and moreover, the product function $[\frac{%
\partial ^{k}}{\partial x^{k}}V_{1}^{ho}(x,\varepsilon )]^{m}[\frac{\partial
^{l}}{\partial x^{l}}V_{1}^{ho}(x,\varepsilon )]^{n}$ may be expanded as a
sum of a few spatially-selective functions after the relevant secondary
terms are neglected, because here $0\leq k,$ $l,$ $m,$ $n\leq 3$ and $%
0<m+n\leq 3.$ This means that each one of the three norms of the term $NA4$
in (4.128) can be further reduced to a linear combination of the basic norms 
$\{NBAS1\}$ and $\{NBAS2\}$. Thus, the equation (4.128) shows that the term $%
NA4$ has an upper bound that is composed of the basic norms $\{NBAS1\}$ and $%
\{NBAS2\}$. On the other hand, it is known from (4.116) that the function $%
\frac{\partial ^{k}}{\partial x^{k}}[\Theta (x-x_{L},\varepsilon )\Omega
(x)] $ with $k>0$ consists of two spatially-selective functions, and it is
also known from (4.117) that the function $\{\frac{\partial ^{k}}{\partial
x^{k}}[\Theta (x-x_{L},\varepsilon )\Omega (x)]\}^{m}$ for $k>0$ and $m>1$
consists of $m+1$ spatially-selective functions. These together show that
the function $\{\frac{\partial ^{k}}{\partial x^{k}}[\Theta
(x-x_{L},\varepsilon )\Omega (x)]\}^{m}$ for $k\geq 1$ and $m\geq 1$ is a
spatially-selective function. Now as can be seen in (4.129), each one of the
four norms of the term $NB4$ inside contains the spatially-selective
function which is the product of the two spatially-selective functions $[%
\frac{\partial ^{k}}{\partial x^{k}}V_{1}^{ho}(x,\varepsilon )]^{m}$ and $\{%
\frac{\partial ^{l}}{\partial x^{l}}[\Theta (x-x_{L},\varepsilon )\Omega
(x)]\}^{n},$ here $0\leq k,$ $l,$ $m,$ $n\leq 3$ and $0<m+n\leq 3$. Then the
product function $[\frac{\partial ^{k}}{\partial x^{k}}V_{1}^{ho}(x,%
\varepsilon )]^{m}\{\frac{\partial ^{l}}{\partial x^{l}}[\Theta
(x-x_{L},\varepsilon )\Omega (x)]\}^{n}$ may be expanded as a sum of a few
spatially-selective functions after the relevant secondary terms are
neglected. Therefore, the equation (4.129) shows that the upper bound of the
term $NB4$ can be expressed as a linear combination of the basic norms $%
\{NBAS1\}$ and $\{NBAS2\}$. Similarly, each one of the three norms of the
term $NC4$ in (4.130) inside contains the spatially-selective function,
which is the product of the two spatially-selective functions $\{\frac{%
\partial ^{k}}{\partial x^{k}}[\Theta (x-x_{L},\varepsilon )\Omega
(x)]\}^{m} $ and $\{\frac{\partial ^{l}}{\partial x^{l}}[\Theta
(x-x_{L},\varepsilon )\Omega (x)]\}^{n}.$ This product function may be
expanded as a sum of a few spatially-selective functions, because here $%
0\leq k,$ $l,$ $m,$ $n\leq 3$ and $0<m+n\leq 3$. Then the equation (4.130)
indicates that the upper bound of the term $NC4$ may be expressed as a
linear combination of the basic norms $\{NBAS1\}$ and $\{NBAS2\}$. Now all
the three terms $NA4$, $NB4$, and $NC4$ on the $RH$ side of (4.127) are
shown to have the upper bounds which are composed of the basic norms $%
\{NBAS1\}$ and $\{NBAS2\}$. Then the inequality (4.127) shows that the upper
bound of the first norm on the $RH$ side of (4.126) consists of the basic
norms $\{NBAS1\}$ and $\{NBAS2\}$. Since now all the four norms on the $RH$
side of (4.126) are shown to have the upper bounds consisting of the basic
norms $\{NBAS1\}$ and $\{NBAS2\}$, the inequality (4.126) indicates that the
fourth norm on the $RH$ side of (4.105) indeed has an upper bound consisting
of the basic norms $\{NBAS1\}$ and $\{NBAS2\}$.

Now calculate the fifth norm on the $RH$ side of (4.105). With the help of
(4.107) it can turn out that the norm is bounded by%
\begin{equation*}
||F_{1}^{\lambda }(x,t_{3})p^{4}\Psi _{0}^{CS}(x,r,t_{0}+t_{1}/2)||\leq
\hslash ^{4}\{||F_{1}^{\lambda }(x,t_{3})(\frac{\partial ^{4}}{\partial x^{4}%
}F_{0}^{CS})\Psi _{0}^{C0}(x,r,t_{0}+t_{1}/2)||
\end{equation*}%
\begin{equation*}
+4||F_{1}^{\lambda }(x,t_{3})(\frac{\partial ^{3}}{\partial x^{3}}F_{0}^{CS})%
\frac{\partial }{\partial x}\Psi _{0}^{C0}(x,r,t_{0}+t_{1}/2)||
\end{equation*}%
\begin{equation*}
+6||F_{1}^{\lambda }(x,t_{3})(\frac{\partial ^{2}}{\partial x^{2}}F_{0}^{CS})%
\frac{\partial ^{2}}{\partial x^{2}}\Psi _{0}^{C0}(x,r,t_{0}+t_{1}/2)||
\end{equation*}%
\begin{equation*}
+4||F_{1}^{\lambda }(x,t_{3})(\frac{\partial }{\partial x}F_{0}^{CS})\frac{%
\partial ^{3}}{\partial x^{3}}\Psi _{0}^{C0}(x,r,t_{0}+t_{1}/2)||
\end{equation*}%
\begin{equation}
+||F_{1}^{\lambda }(x,t_{3})F_{0}^{CS}\frac{\partial ^{4}}{\partial x^{4}}%
\Psi _{0}^{C0}(x,r,t_{0}+t_{1}/2)||\}.  \tag{4.131}
\end{equation}%
There are five norms on the $RH$ side of (4.131) which will be calculated
one by one below. The last norm on the $RH$ side of (4.131) satisfies the
inequality (4.119) if one makes the replacement $F_{5}^{\lambda
}(x,t_{3})\leftrightarrow F_{1}^{\lambda }(x,t_{3})$ and $\Psi
_{0}^{C0}(x,r,t_{0}+t_{1}/2)\leftrightarrow \frac{\partial ^{4}}{\partial
x^{4}}\Psi _{0}^{C0}(x,r,t_{0}+t_{1}/2)$ in (4.119). Here the fourth-order
derivative $\frac{\partial ^{4}}{\partial x^{4}}\Psi
_{0}^{C0}(x,r,t_{0}+t_{1}/2)$ is given by (4.123a), where the internal
superposition state $\Phi _{4}^{a}(x,r,t_{0}+t_{1}/2)$ of (4.123b) is given
by%
\begin{equation*}
\Phi _{4}^{a}(x,r,t_{0}+t_{1}/2)=Q_{4}^{a}(x)|\tilde{a}\rangle +4Q_{3}^{a}(x)%
\frac{\partial }{\partial x}|\tilde{a}\rangle
\end{equation*}%
\begin{equation}
+6Q_{2}^{a}(x)\frac{\partial ^{2}}{\partial x^{2}}|\tilde{a}\rangle
+4Q_{1}^{a}(x)\frac{\partial ^{3}}{\partial x^{3}}|\tilde{a}\rangle
+Q_{0}^{a}(x)\frac{\partial ^{4}}{\partial x^{4}}|\tilde{a}\rangle . 
\tag{4.131a}
\end{equation}%
Then the last norm on the $RH$ side of (4.131) is reduced to a linear sum of
the three norms on the $RH$ side of this modified inequality of (4.119),
each one of which satisfies the inequality (4.118a):%
\begin{equation*}
||F_{1}^{\lambda }(x,t_{3})S(x)\frac{\partial ^{4}}{\partial x^{4}}\Psi
_{0}^{C0}(x,r,t_{0}+t_{1}/2)||
\end{equation*}%
\begin{equation*}
\leq \sum_{l=0}^{4}\left( 
\begin{array}{c}
4 \\ 
l%
\end{array}%
\right) ||(\frac{\partial ^{l}}{\partial x^{l}}|\tilde{g}_{0}\rangle
)||\times ||F_{1}^{\lambda }(x,t_{3})Q_{4-l}^{g}(x)S(x)\Psi
_{0}^{g}(x,t_{0}+t_{1}/2)||
\end{equation*}%
\begin{equation}
+\sum_{l=0}^{4}\left( 
\begin{array}{c}
4 \\ 
l%
\end{array}%
\right) ||(\frac{\partial ^{l}}{\partial x^{l}}|\tilde{e}\rangle )||\times
||F_{1}^{\lambda }(x,t_{3})Q_{4-l}^{e}(x)S(x)\Psi
_{0}^{e}(x,t_{0}+t_{1}/2)||,  \tag{4.131b}
\end{equation}%
where the spatially-selective function $S(x)=[\Theta (x-x_{L},\varepsilon
)\Omega (x)]^{2},$ $V_{1}^{ho}(x,\varepsilon )^{2},$ or $V_{1}^{ho}(x,%
\varepsilon ).$ The $RH$ side of (4.131b) has ten norms, each one of which
has an upper bound consisting of the basic norms $\{NBAS2\}$ due to that
here the spatially-selective function contains only $\Theta
(x-x_{L},\varepsilon ).$ Therefore, the last norm on the $RH$ side of
(4.131) has an upper bound consisting of the basic norms $\{NBAS2\}$. The
fourth norm on the $RH$ side of (4.131) satisfies the inequality (4.122) if
one makes the replacement $F_{4}^{\lambda }(x,t_{3})\leftrightarrow
F_{1}^{\lambda }(x,t_{3})$ and $\Psi
_{0}^{C0}(x,r,t_{0}+t_{1}/2)\leftrightarrow \frac{\partial ^{3}}{\partial
x^{3}}\Psi _{0}^{C0}(x,r,t_{0}+t_{1}/2)$ in (4.122). This leads to that this
fourth norm may be reduced to a linear sum of the two norms on the $RH$ side
of this modified inequality of (4.122), each one of which is given by $%
||F_{1}^{\lambda }(x,t_{3})S(x)\frac{\partial ^{3}}{\partial x^{3}}\Psi
_{0}^{C0}(x,r,t_{0}+t_{1}/2)||$ with $S(x)=$ $\frac{\partial }{\partial x}%
V_{1}^{ho}(x,\varepsilon )$ or $\frac{\partial }{\partial x}[\Theta
(x-x_{L},\varepsilon )\Omega (x)]$ and obeys the inequality (4.126b) with
the replacement $F_{2}^{\lambda }(x,t_{3})\leftrightarrow F_{1}^{\lambda
}(x,t_{3}).$ Then according to this modified inequality of (4.126b) each one
of the two norms may be further reduced to a linear sum of eight norms, each
one of which has an upper bound consisting of the basic norms $\{NBAS1\}$
and $\{NBAS2\}$ due to that here the spatially-selective function $S(x)$
contains both $\Theta (x-x_{L},\varepsilon )$ and $\delta
(x-x_{c},\varepsilon _{c}).$ Thus, the fourth norm on the $RH$ side of
(4.131) has an upper bound consisting of the basic norms $\{NBAS1\}$ and $%
\{NBAS2\}$. The third norm on the $RH$ side of (4.131) satisfies the
inequality (4.125) with the replacement $F_{3}^{\lambda
}(x,t_{3})\leftrightarrow F_{1}^{\lambda }(x,t_{3})$ and $\Psi
_{0}^{C0}(x,r,t_{0}+t_{1}/2)\leftrightarrow \frac{\partial ^{2}}{\partial
x^{2}}\Psi _{0}^{C0}(x,r,t_{0}+t_{1}/2).$ This leads to that this third norm
may be reduced to a linear sum of the five norms on the $RH$ side of this
modified inequality of (4.125), each one of which is given by $%
||F_{1}^{\lambda }(x,t_{3})S(x)\frac{\partial ^{2}}{\partial x^{2}}\Psi
_{0}^{C0}(x,r,t_{0}+t_{1}/2)||$ where the five spatially-selective functions 
$S(x)$ can be found directly from (4.125) and it also obeys the inequality
(4.124d) with the replacement $F_{3}^{\lambda }(x,t_{3})\leftrightarrow
F_{1}^{\lambda }(x,t_{3}).$ Then according to this modified inequality of
(4.124d) each one of these five norms may be further reduced to a linear
combination of six norms, each one of which has an upper bound consisting of
the basic norms $\{NBAS1\}$ and $\{NBAS2\}$. Therefore, the third norm on
the $RH$ side of (4.131) has an upper bound consisting of the basic norms $%
\{NBAS1\}$ and $\{NBAS2\}$. Similarly, the second norm on the $RH$ side of
(4.131) satisfies the inequality (4.127) with the replacement $%
F_{2}^{\lambda }(x,t_{3})\leftrightarrow F_{1}^{\lambda }(x,t_{3})$ and $%
\Psi _{0}^{C0}(x,r,t_{0}+t_{1}/2)\leftrightarrow \frac{\partial }{\partial x}%
\Psi _{0}^{C0}(x,r,t_{0}+t_{1}/2):$%
\begin{equation*}
||F_{1}^{\lambda }(x,t_{3})(\frac{\partial ^{3}}{\partial x^{3}}F_{0}^{CS})%
\frac{\partial }{\partial x}\Psi _{0}^{C0}(x,r,t_{0}+t_{1}/2)||\leq
NA4^{\prime }+NB4^{\prime }+NC4^{\prime }.
\end{equation*}%
Here the current three terms $NA4^{\prime },$ $NB4^{\prime },$ and $%
NC4^{\prime }$ correspond to the previous three terms $NA4$, $NB4$, and $NC4$
in (4.127), respectively. They are also calculated by the three modified
equations of (4.128), (4.129), and (4.130), respectively, in which $%
F_{2}^{\lambda }(x,t_{3})$ and $\Psi _{0}^{C0}(x,r,t_{0}+t_{1}/2)$ are
replaced by $F_{1}^{\lambda }(x,t_{3})$ and $\frac{\partial }{\partial x}%
\Psi _{0}^{C0}(x,r,t_{0}+t_{1}/2),$ respectively. Similar to the previous
term $NA4$ of (4.128), $NB4$ of (4.129), and $NC4$ of (4.130) the current
three terms $NA4^{\prime }$, $NB4^{\prime }$, and $NC4^{\prime }$ also
contain three, four, and three norms, respectively. Each one of these ten
norms may be formally written as $||F_{1}^{\lambda }(x,t_{3})S(x)\frac{%
\partial }{\partial x}\Psi _{0}^{C0}(x,r,t_{0}+t_{1}/2)||,$ where the
spatially-selective function $S(x)$ is obtained directly from (4.128),
(4.129), and (4.130), and moreover, it obeys the inequality (4.123d) with
the replacement $F_{4}^{\lambda }(x,t_{3})\leftrightarrow F_{1}^{\lambda
}(x,t_{3}).$ Thus, according to this modified inequality of (4.123d) this
norm can be reduced to a linear sum of four norms, each one of which has an
upper bound consisting of the basic norms $\{NBAS1\}$ and $\{NBAS2\}$.
Therefore, all the current three terms $NA4^{\prime },$ $NB4^{\prime },$ and 
$NC4^{\prime }$ have upper bounds consisting of the basic norms $\{NBAS1\}$
and $\{NBAS2\}$. This indicates that the second norm on the $RH$ side of
(4.131) has an upper bound that is a linear sum of the basic norms $%
\{NBAS1\} $ and $\{NBAS2\}$. Now one needs only to calculate explicitly the
first norm on the $RH$ side of (4.131). By using the function $F_{0}^{CS}$
of (4.106b) to compute the derivative $\frac{\partial ^{4}}{\partial x^{4}}%
F_{0}^{CS}$ it can turn out that this norm satisfies%
\begin{equation}
||F_{1}^{\lambda }(x,t_{3})(\frac{\partial ^{4}}{\partial x^{4}}%
F_{0}^{CS})\Psi _{0}^{C0}(x,r,t_{0}+t_{1}/2)||\leq NA5+NB5+NC5+ND5. 
\tag{4.132}
\end{equation}%
Here the four terms $NA5,$ $NB5,$ $NC5,$ and $ND5$ are given below. The term 
$NA5$ can be expressed as%
\begin{equation*}
NA5=\sum_{k,l,m,n}\beta _{A}^{5}(k,l,m,n)||F_{1}^{\lambda }(x,t_{3})[\frac{%
\partial ^{k}}{\partial x^{k}}V_{1}^{ho}(x,\varepsilon )]^{m}
\end{equation*}%
\begin{equation}
\times \lbrack \frac{\partial ^{l}}{\partial x^{l}}V_{1}^{ho}(x,\varepsilon
)]^{n}\Psi _{0}^{C0}(x,r,t_{0}+t_{1}/2)||,  \tag{4.133}
\end{equation}%
where only five parameters $\{\beta _{A}^{5}(k,l,m,n)\}$ take non-zero
values. The term $NB5$ is given by%
\begin{equation*}
NB5=\sum_{k,l,m,n}\beta _{B}^{5}(k,l,m,n)||F_{1}^{\lambda }(x,t_{3})[\frac{%
\partial ^{k}}{\partial x^{k}}V_{1}^{ho}(x,\varepsilon )]^{m}
\end{equation*}%
\begin{equation}
\times \{\frac{\partial ^{l}}{\partial x^{l}}[\Theta (x-x_{L},\varepsilon
)\Omega (x)]\}^{n}\Psi _{0}^{C0}(x,r,t_{0}+t_{1}/2)||,  \tag{4.134}
\end{equation}%
where among these parameters $\{\beta _{B}^{5}(k,l,m,n)\}$ there are only
eight non-zero parameters. The term $NC5$ is written as%
\begin{equation*}
NC5=\sum_{k,l,m,n}\beta _{C}^{5}(k,l,m,n)||F_{1}^{\lambda }(x,t_{3})\{\frac{%
\partial ^{k}}{\partial x^{k}}[\Theta (x-x_{L},\varepsilon )\Omega (x)]\}^{m}
\end{equation*}%
\begin{equation}
\times \{\frac{\partial ^{l}}{\partial x^{l}}[\Theta (x-x_{L},\varepsilon
)\Omega (x)]\}^{n}\Psi _{0}^{C0}(x,r,t_{0}+t_{1}/2)||,  \tag{4.135}
\end{equation}%
where among these parameters $\{\beta _{C}^{5}(k,l,m,n)\}$ there only five
non-zero parameters. The indices in (4.133) -- (4.135) satisfy $0\leq k,$ $%
l, $ $m,$ $n$; $0<k+l\leq 4;$ and $0<m+n\leq 4.$ All these non-zero
parameters $\{\beta _{\lambda }^{5}(k,l,m,n)\}$ with $\lambda =A,$ $B$, and $%
C$ are non-negative, each one of which is proportional to $(t_{1}/\hslash
)^{\mu }$ with the integer $\mu \in \lbrack 1,$ $4].$ Thus, these parameters
are controllable. The last term $ND5$ is given by%
\begin{equation*}
ND5=6(\frac{1}{\hslash }t_{1})^{3}||F_{1}^{\lambda }(x,t_{3})[\frac{\partial 
}{\partial x}V_{1}^{ho}(x,\varepsilon )][\frac{\partial ^{2}}{\partial x^{2}}%
V_{1}^{ho}(x,\varepsilon )]
\end{equation*}%
\begin{equation*}
\times (\frac{\partial }{\partial x}[\Theta (x-x_{L},\varepsilon )\Omega
(x)])\Psi _{0}^{C0}(x,r,t_{0}+t_{1}/2)||
\end{equation*}%
\begin{equation*}
+3(\frac{1}{\hslash }t_{1})^{3}||F_{1}^{\lambda }(x,t_{3})[\frac{\partial }{%
\partial x}V_{1}^{ho}(x,\varepsilon )](\frac{\partial }{\partial x}[\Theta
(x-x_{L},\varepsilon )\Omega (x)])
\end{equation*}%
\begin{equation}
\times (\frac{\partial ^{2}}{\partial x^{2}}[\Theta (x-x_{L},\varepsilon
)\Omega (x)])\Psi _{0}^{C0}(x,r,t_{0}+t_{1}/2)||.  \tag{4.136}
\end{equation}%
Below a detailed investigation is given to the four terms $NA5,$ $NB5,$ $%
NC5, $ and $ND5.$ There appear only the two types of the spatially-selective
functions $[\frac{\partial ^{k}}{\partial x^{k}}V_{1}^{ho}(x,\varepsilon
)]^{m}$ and $\{\frac{\partial ^{k}}{\partial x^{k}}[\Theta
(x-x_{L},\varepsilon )\Omega (x)]\}^{m}$ in the four terms $NA5,$ $NB5,$ $%
NC5,$ and $ND5,$ here $k>0$ and $m>0$. If one makes the replacement $%
F_{1}^{\lambda }(x,t_{3})\leftrightarrow F_{2}^{\lambda }(x,t_{3})$ in every
norm inside of the terms $NA5,$ $NB5,$ and $NC5,$ one will find that the
three terms $NA5,$ $NB5,$ and $NC5$ are formally similar to the three terms $%
NA4$, $NB4$, and $NC4$ on the $RH$ side of (4.127), respectively. The term $%
NA5$ is similar to the term $NA4$ of (4.128). As shown in (4.133), the term $%
NA5$ consists of five norms, each one of which inside contains the
spatially-selective function $[\frac{\partial ^{k}}{\partial x^{k}}%
V_{1}^{ho}(x,\varepsilon )]^{m}[\frac{\partial ^{l}}{\partial x^{l}}%
V_{1}^{ho}(x,\varepsilon )]^{n}.$ This spatially-selective product function
may be expanded as a sum of a few spatially-selective functions, because
here $0\leq k,$ $l,$ $m,$ $n\leq 4$ and $0<m+n\leq 4$. Thus, similar to the
term $NA4$ of (4.128), the term $NA5$ can be proven to have an upper bound
consisting of the basic norms $\{NBAS1\}$ and $\{NBAS2\}$. The term $NB5$ is
similar to the term $NB4$ of (4.129). As shown in (4.134), it consists of
eight norms, each one of which contains the spatially-selective function $[%
\frac{\partial ^{k}}{\partial x^{k}}V_{1}^{ho}(x,\varepsilon )]^{m}\{\frac{%
\partial ^{l}}{\partial x^{l}}[\Theta (x-x_{L},\varepsilon )\Omega
(x)]\}^{n}.$ This product function also may be expanded as a sum of a few
spatially-selective functions as $0\leq k,$ $l,$ $m,$ $n\leq 4$ and $%
0<m+n\leq 4$. Therefore, similar to the term $NB4$ of (4.129), the term $NB5$
has an upper bound that is composed of the basic norms $\{NBAS1\}$ and $%
\{NBAS2\}$. The term $NC5$ is similar to the term $NC4$ of (4.130). As shown
in (4.135), it has five norms, each one of which contains the
spatially-selective function $\{\frac{\partial ^{k}}{\partial x^{k}}[\Theta
(x-x_{L},\varepsilon )\Omega (x)]\}^{m}\{\frac{\partial ^{l}}{\partial x^{l}}%
[\Theta (x-x_{L},\varepsilon )\Omega (x)]\}^{n}.$ This spatially-selective
product function also can be expanded as a sum of a few spatially-selective
functions as $0\leq k,$ $l,$ $m,$ $n\leq 4$ and $0<m+n\leq 4$. Thus, just
like the term $NC4$ of (4.130), the term $NC5$ can be shown to have an upper
bound consisting of the basic norms $\{NBAS1\}$ and $\{NBAS2\}$. As shown in
(4.136), the term $ND5$ has two norms, the first one inside contains the
product of three spatially-selective functions: $[\frac{\partial }{\partial x%
}V_{1}^{ho}(x,\varepsilon )][\frac{\partial ^{2}}{\partial x^{2}}%
V_{1}^{ho}(x,\varepsilon )](\frac{\partial }{\partial x}[\Theta
(x-x_{L},\varepsilon )\Omega (x)]),$ while the second also contains the
product of three spatially-selective functions: $[\frac{\partial }{\partial x%
}V_{1}^{ho}(x,\varepsilon )](\frac{\partial }{\partial x}[\Theta
(x-x_{L},\varepsilon )\Omega (x)])(\frac{\partial ^{2}}{\partial x^{2}}%
[\Theta (x-x_{L},\varepsilon )\Omega (x)]).$ Each one of the two
(spatially-selective) product functions can be expanded as a sum of a few
spatially-selective functions. Thus, it is easy to prove that the upper
bound of the term $ND5$ may be expressed as a linear sum of the basic norms $%
\{NBAS1\}$ and $\{NBAS2\}$. Now all these four terms $NA5$, $NB5$, $NC5$,
and $ND5$ on the $RH$ side of (4.132) are shown to have the upper bounds
consisting of the basic norms $\{NBAS1\}$ and $\{NBAS2\}$. Then it follows
from the inequality (4.132) that the upper bound of the first norm on the $%
RH $ side of (4.131) is composed of the basic norms $\{NBAS1\}$ and $%
\{NBAS2\}$. Since now all the five norms on the $RH$ side of (4.131) are
shown to have the upper bounds consisting of the two types of basic norms $%
\{NBAS1\}$ and $\{NBAS2\}$, the inequality (4.131) shows that the fifth norm
on the $RH$ side of (4.105) indeed has an upper bound consisting of the
basic norms $\{NBAS1\}$ and $\{NBAS2\}$.

With the help of (4.107) it can turn out that the last norm on the $RH$ side
of (4.105) is bounded by%
\begin{equation*}
||F_{0}^{\lambda }(x,t_{3})p^{5}\Psi _{0}^{CS}(x,r,t_{0}+t_{1}/2)||\leq
\hslash ^{5}\{||F_{0}^{\lambda }(x,t_{3})(\frac{\partial ^{5}}{\partial x^{5}%
}F_{0}^{CS})\Psi _{0}^{C0}(x,r,t_{0}+t_{1}/2)||
\end{equation*}%
\begin{equation*}
+5||F_{0}^{\lambda }(x,t_{3})(\frac{\partial ^{4}}{\partial x^{4}}F_{0}^{CS})%
\frac{\partial }{\partial x}\Psi _{0}^{C0}(x,r,t_{0}+t_{1}/2)||
\end{equation*}%
\begin{equation*}
+10||F_{0}^{\lambda }(x,t_{3})(\frac{\partial ^{3}}{\partial x^{3}}%
F_{0}^{CS})\frac{\partial ^{2}}{\partial x^{2}}\Psi
_{0}^{C0}(x,r,t_{0}+t_{1}/2)||
\end{equation*}%
\begin{equation*}
+10||F_{0}^{\lambda }(x,t_{3})(\frac{\partial ^{2}}{\partial x^{2}}%
F_{0}^{CS})\frac{\partial ^{3}}{\partial x^{3}}\Psi
_{0}^{C0}(x,r,t_{0}+t_{1}/2)||
\end{equation*}%
\begin{equation*}
+5||F_{0}^{\lambda }(x,t_{3})(\frac{\partial }{\partial x}F_{0}^{CS})\frac{%
\partial ^{4}}{\partial x^{4}}\Psi _{0}^{C0}(x,r,t_{0}+t_{1}/2)||
\end{equation*}%
\begin{equation}
+||F_{0}^{\lambda }(x,t_{3})F_{0}^{CS}\frac{\partial ^{5}}{\partial x^{5}}%
\Psi _{0}^{C0}(x,r,t_{0}+t_{1}/2)||\}.  \tag{4.137}
\end{equation}%
There are six norms on the $RH$ side of (4.137) to be calculated below. The
last norm on the $RH$ side of (4.137) satisfies the inequality (4.119) with
the replacement $F_{5}^{\lambda }(x,t_{3})\leftrightarrow F_{0}^{\lambda
}(x,t_{3})$ and $\Psi _{0}^{C0}(x,r,t_{0}+t_{1}/2)\leftrightarrow \frac{%
\partial ^{5}}{\partial x^{5}}\Psi _{0}^{C0}(x,r,t_{0}+t_{1}/2).$ Here the
fifth-order derivative $\frac{\partial ^{5}}{\partial x^{5}}\Psi
_{0}^{C0}(x,r,t_{0}+t_{1}/2)$ is given by (4.123a) with the internal
superposition state $\Phi _{5}^{a}(x,r,t_{0}+t_{1}/2)$ of (4.123b) which is
given by%
\begin{equation*}
\Phi _{5}^{a}(x,r,t_{0}+t_{1}/2)=Q_{5}^{a}(x)|\tilde{a}\rangle +5Q_{4}^{a}(x)%
\frac{\partial }{\partial x}|\tilde{a}\rangle +10Q_{3}^{a}(x)\frac{\partial
^{2}}{\partial x^{2}}|\tilde{a}\rangle
\end{equation*}%
\begin{equation}
+10Q_{2}^{a}(x)\frac{\partial ^{3}}{\partial x^{3}}|\tilde{a}\rangle
+5Q_{1}^{a}(x)\frac{\partial ^{4}}{\partial x^{4}}|\tilde{a}\rangle
+Q_{0}^{a}(x)\frac{\partial ^{5}}{\partial x^{5}}|\tilde{a}\rangle . 
\tag{4.137a}
\end{equation}%
This leads to that the last norm on the $RH$ side of (4.137) may be reduced
to a linear sum of the three norms on the $RH$ side of this modified
inequality of (4.119). Each one of these three norms is given by $%
||F_{0}^{\lambda }(x,t_{3})S(x)\frac{\partial ^{5}}{\partial x^{5}}\Psi
_{0}^{C0}(x,r,$ $t_{0}+t_{1}/2)||,$ where the spatially-selective function $%
S(x)=[\Theta (x-x_{L},\varepsilon )\Omega (x)]^{2},$ $V_{1}^{ho}(x,%
\varepsilon )^{2},$ or $V_{1}^{ho}(x,\varepsilon ),$ and moreover, with the
help of (4.123a) and (4.137a) it can turn out that the norm obeys the
inequality (4.118a) with the settings $j=5$ and $q(x)=F_{0}^{\lambda
}(x,t_{3}),$ indicating that the norm may be reduced to a linear sum of the
twelve norms, each one of which has an upper bound consisting of the basic
norms $\{NBAS2\}$. Therefore, these show that the last norm on the $RH$ side
of (4.137) has an upper bound consisting of the basic norms $\{NBAS2\}$. Now
the fifth norm on the $RH$ side of (4.137) satisfies the inequality (4.122)
with the replacement $F_{4}^{\lambda }(x,t_{3})\leftrightarrow
F_{0}^{\lambda }(x,t_{3})$ and $\Psi
_{0}^{C0}(x,r,t_{0}+t_{1}/2)\leftrightarrow \frac{\partial ^{4}}{\partial
x^{4}}\Psi _{0}^{C0}(x,r,t_{0}+t_{1}/2).$ This leads to that the fifth norm
may be reduced to a linear sum of the two norms on the $RH$ side of this
modified inequality of (4.122). Each one of the two norms may be expressed
as $||F_{0}^{\lambda }(x,t_{3})S(x)\frac{\partial ^{4}}{\partial x^{4}}\Psi
_{0}^{C0}(x,r,t_{0}+t_{1}/2)||,$ where the spatially-selective function $%
S(x)=\frac{\partial }{\partial x}V_{1}^{ho}(x,\varepsilon )$ or $\frac{%
\partial }{\partial x}[\Theta (x-x_{L},\varepsilon )\Omega (x)].$ This norm
obeys the inequality (4.131b) with the replacement $F_{1}^{\lambda
}(x,t_{3})\leftrightarrow F_{0}^{\lambda }(x,t_{3}).$ Then according to this
modified inequality of (4.131b) it may be reduced to a linear sum of the ten
norms, each one of which has an upper bound consisting of the basic norms $%
\{NBAS1\}$ and $\{NBAS2\}$. Therefore, these show that the upper bound of
the fifth norm on the $RH$ side of (4.137) may be expressed as a linear sum
of the basic norms $\{NBAS1\}$ and $\{NBAS2\}$. The fourth norm on the $RH$
side of (4.137) satisfies the inequality (4.125) with the replacement $%
F_{3}^{\lambda }(x,t_{3})\leftrightarrow F_{0}^{\lambda }(x,t_{3})$ and $%
\Psi _{0}^{C0}(x,r,t_{0}+t_{1}/2)\leftrightarrow \frac{\partial ^{3}}{%
\partial x^{3}}\Psi _{0}^{C0}(x,r,t_{0}+t_{1}/2).$ According to this
modified inequality of (4.125) the fourth norm may be reduced to a linear
sum of the five norms. Each one of these five norms may be written as $%
||F_{0}^{\lambda }(x,t_{3})S(x)\frac{\partial ^{3}}{\partial x^{3}}\Psi
_{0}^{C0}(x,r,t_{0}+t_{1}/2)||,$ where the five spatially-selective
functions $S(x)$ can be obtained directly from (4.125). The norm clearly
obeys the inequality (4.126b) with the replacement $F_{2}^{\lambda
}(x,t_{3})\leftrightarrow F_{0}^{\lambda }(x,t_{3}).$ Then according to this
modified inequality of (4.126b) it may be reduced to a linear sum of the
eight norms, each one of which has an upper bound consisting of the basic
norms $\{NBAS1\}$ and $\{NBAS2\}$. Therefore, these show that the fourth
norm on the $RH$ side of (4.137) has an upper bound consisting of the basic
norms $\{NBAS1\}$ and $\{NBAS2\}$. The third norm on the $RH$ side of
(4.137) satisfies the inequality (4.127) with the replacement $%
F_{2}^{\lambda }(x,t_{3})\leftrightarrow F_{0}^{\lambda }(x,t_{3})$ and $%
\Psi _{0}^{C0}(x,r,t_{0}+t_{1}/2)\leftrightarrow \frac{\partial ^{2}}{%
\partial x^{2}}\Psi _{0}^{C0}(x,r,t_{0}+t_{1}/2),$ 
\begin{equation*}
||F_{0}^{\lambda }(x,t_{3})(\frac{\partial ^{3}}{\partial x^{3}}F_{0}^{CS})%
\frac{\partial ^{2}}{\partial x^{2}}\Psi _{0}^{C0}(x,r,t_{0}+t_{1}/2)||\leq
NA4^{\prime \prime }+NB4^{\prime \prime }+NC4^{\prime \prime }.
\end{equation*}%
Here the current three terms $NA4^{\prime \prime },$ $NB4^{\prime \prime },$
and $NC4^{\prime \prime }$ correspond to the previous three terms $NA4$, $%
NB4 $, and $NC4$ in (4.127), respectively. They also are given by the three
modified equations of (4.128), (4.129), and (4.130), respectively, in which $%
F_{2}^{\lambda }(x,t_{3})$ and $\Psi _{0}^{C0}(x,r,t_{0}+t_{1}/2)$ are
replaced by $F_{0}^{\lambda }(x,t_{3})$ and $\frac{\partial ^{2}}{\partial
x^{2}}\Psi _{0}^{C0}(x,r,t_{0}+t_{1}/2),$ respectively. They contain the
three, four, and three norms, respectively. Each one of these ten norms may
be formally written as $||F_{0}^{\lambda }(x,t_{3})S(x)\frac{\partial ^{2}}{%
\partial x^{2}}\Psi _{0}^{C0}(x,r,t_{0}+t_{1}/2)||,$ where the ten
spatially-selective product functions $S(x)$ can be obtained directly from
the equations (4.128), (4.129), and (4.130), respectively. This norm clearly
satisfies the inequality (4.124d) with the replacement $F_{3}^{\lambda
}(x,t_{3})\leftrightarrow F_{0}^{\lambda }(x,t_{3}).$ Then according to this
modified inequality of (4.124d) it may be reduced to a linear sum of the six
norms, each one of which has an upper bound consisting of the basic norms $%
\{NBAS1\}$ and $\{NBAS2\}$. Therefore, these show that the third norm on the 
$RH$ side of (4.137) has an upper bound consisting of the basic norms $%
\{NBAS1\}$ and $\{NBAS2\}$. The second norm on the $RH$ side of (4.137)
satisfies the inequality (4.132) with the replacement $F_{1}^{\lambda
}(x,t_{3})\leftrightarrow F_{0}^{\lambda }(x,t_{3})$ and $\Psi
_{0}^{C0}(x,r,t_{0}+t_{1}/2)\leftrightarrow \frac{\partial }{\partial x}\Psi
_{0}^{C0}(x,r,t_{0}+t_{1}/2)$, 
\begin{equation*}
||F_{0}^{\lambda }(x,t_{3})(\frac{\partial ^{4}}{\partial x^{4}}F_{0}^{CS})%
\frac{\partial }{\partial x}\Psi _{0}^{C0}(x,r,t_{0}+t_{1}/2)||\leq
NA5^{\prime }+NB5^{\prime }+NC5^{\prime }+ND5^{\prime }.
\end{equation*}%
Here the current four terms $NA5^{\prime },$ $NB5^{\prime },$ $NC5^{\prime
}, $ and $ND5^{\prime }$ correspond to the previous four terms $NA5$, $NB5$, 
$NC5$, and $ND5$ in (4.132), respectively. They also are given by (4.133),
(4.134), (4.135), and (4.136), respectively, in which $F_{1}^{\lambda
}(x,t_{3})$ and $\Psi _{0}^{C0}(x,r,t_{0}+t_{1}/2)$ are replaced with $%
F_{0}^{\lambda }(x,t_{3})$ and $\frac{\partial }{\partial x}\Psi
_{0}^{C0}(x,r,t_{0}+t_{1}/2)$, respectively. They contain five, eight, five,
and two norms, respectively. Each one of these twenty norms may be formally
written as $||F_{0}^{\lambda }(x,t_{3})S(x)$ $\times \frac{\partial }{%
\partial x}\Psi _{0}^{C0}(x,r,t_{0}+t_{1}/2)||,$ where the twenty
spatially-selective product functions $S(x)$ are obtained directly from
(4.133), (4.134), (4.135), and (4.136), respectively. This norm clearly
obeys the inequality (4.123d) with the replacement $F_{4}^{\lambda
}(x,t_{3})\leftrightarrow F_{0}^{\lambda }(x,t_{3}).$ Then according to this
modified inequality of (4.123d) it may be reduced to a linear sum of the
four norms, each one of which has an upper bound consisting of the basic
norms $\{NBAS1\}$ and $\{NBAS2\}.$ Therefore, these show that the second
norm on the $RH$ side of (4.137) has an upper bound consisting of the basic
norms $\{NBAS1\}$ and $\{NBAS2\}.$ Now one needs only to compute explicitly
the first norm on the $RH$ side of (4.137). Here one may compute directly
the fifth-order coordinate derivative $\frac{\partial ^{5}}{\partial x^{5}}%
F_{0}^{CS}$ by using the function $F_{0}^{CS}$ of (4.106b). Then by using
the derivative it can turn out that this norm is bounded by%
\begin{equation}
||F_{0}^{\lambda }(x,t_{3})(\frac{\partial ^{5}}{\partial x^{5}}%
F_{0}^{CS})\Psi _{0}^{C0}(x,r,t_{0}+t_{1}/2)||\leq NA6+NB6+NC6+ND6. 
\tag{4.138}
\end{equation}%
Here the term $NA6$ is given by%
\begin{equation*}
NA6=\sum_{k,l,m,n}\beta _{A}^{6}(k,l,m,n)||F_{0}^{\lambda }(x,t_{3})[\frac{%
\partial ^{k}}{\partial x^{k}}V_{1}^{ho}(x,\varepsilon )]^{m}
\end{equation*}%
\begin{equation}
\times \lbrack \frac{\partial ^{l}}{\partial x^{l}}V_{1}^{ho}(x,\varepsilon
)]^{n}\Psi _{0}^{C0}(x,r,t_{0}+t_{1}/2)||  \tag{4.139}
\end{equation}%
where among these parameters $\{\beta _{A}^{6}(k,l,m,n)\}$ there are only
seven non-zero parameters. The term $NB6$ may be expressed as%
\begin{equation*}
NB6=\sum_{k,l,m,n}\beta _{B}^{6}(k,l,m,n)||F_{0}^{\lambda }(x,t_{3})[\frac{%
\partial ^{k}}{\partial x^{k}}V_{1}^{ho}(x,\varepsilon )]^{m}
\end{equation*}%
\begin{equation}
\times \{\frac{\partial ^{l}}{\partial x^{l}}[\Theta (x-x_{L},\varepsilon
)\Omega (x)]\}^{n}\Psi _{0}^{C0}(x,r,t_{0}+t_{1}/2)||  \tag{4.140}
\end{equation}%
where among these parameters $\{\beta _{B}^{6}(k,l,m,n)\}$ there are only
fourteen non-zero parameters. The term $NC6$ is given by%
\begin{equation*}
NC6=\sum_{k,l,m,n}\beta _{C}^{6}(k,l,m,n)||F_{0}^{\lambda }(x,t_{3})\{\frac{%
\partial ^{k}}{\partial x^{k}}[\Theta (x-x_{L},\varepsilon )\Omega (x)]\}^{m}
\end{equation*}%
\begin{equation}
\times \{\frac{\partial ^{l}}{\partial x^{l}}[\Theta (x-x_{L},\varepsilon
)\Omega (x)]\}^{n}\Psi _{0}^{C0}(x,r,t_{0}+t_{1}/2)||  \tag{4.141}
\end{equation}%
where only seven parameters $\{\beta _{C}^{6}(k,l,m,n)\}$ are non-zero
parameters. All these non-zero parameters in the terms $NA6$, $NB6$, and $%
NC6 $ take non-negative values, each one of which is proportional to $%
(t_{1}/\hslash )^{\mu }$ with the integer $\mu \in \lbrack 1,$ $5].$
Therefore, they are controllable. Moreover, all these indices in the terms $%
NA6$, $NB6$, and $NC6$ satisfy $0\leq k,$ $l,$ $m,$ $n;$ $0<k+l\leq 5;$ and $%
0<m+n\leq 5.$ The term $ND6$ can be written as%
\begin{equation*}
ND6=\sum_{m=0}^{1}\beta _{D}^{60}(m)||F_{0}^{\lambda
}(x,t_{3})S_{m}^{60}(x)\Psi _{0}^{C0}(x,r,t_{0}+t_{1}/2)||
\end{equation*}%
\begin{equation*}
+\sum_{k,l,m,n}\beta _{D}^{61}(k,l,m,n)||F_{0}^{\lambda
}(x,t_{3})S_{klmn}^{61}(x)\Psi _{0}^{C0}(x,r,t_{0}+t_{1}/2)||
\end{equation*}%
\begin{equation}
+\sum_{k,l,m,n}\beta _{D}^{62}(k,l,m,n)||F_{0}^{\lambda
}(x,t_{3})S_{klmn}^{62}(x)\Psi _{0}^{C0}(x,r,t_{0}+t_{1}/2)||  \tag{4.142}
\end{equation}%
where $\beta _{D}^{60}(0)=5(\frac{1}{\hslash }t_{1})^{3}$ and $\beta
_{D}^{60}(1)=10(\frac{1}{\hslash }t_{1})^{3};$ the only three non-zero
parameters $\{\beta _{D}^{61}(k,l,m,n)\}$ are given by $\beta
_{D}^{61}(1,1,2,1)=2\beta _{D}^{61}(1,1,1,2)=15(\frac{1}{\hslash }t_{1})^{4}$
and $\beta _{D}^{61}(1,2,1,1)=15(\frac{1}{\hslash }t_{1})^{3};$ the only
three non-zero parameters $\{\beta _{D}^{62}(k,l,m,n)\}$ are $\beta
_{D}^{62}(2,1,1,1)=\frac{15}{2}(\frac{1}{\hslash }t_{1})^{3}$ and $\beta
_{D}^{62}(1,1,2,1)=2\beta _{D}^{62}(1,1,1,2)=\frac{15}{2}(\frac{1}{\hslash }%
t_{1})^{4};$ and correspondingly the eight spatially-selective product
functions are given by%
\begin{equation*}
S_{0}^{60}(x)=[\frac{\partial }{\partial x}V_{1}^{ho}(x,\varepsilon )](\frac{%
\partial }{\partial x}[\Theta (x-x_{L},\varepsilon )\Omega (x)])\{\frac{%
\partial ^{3}}{\partial x^{3}}[\Theta (x-x_{L},\varepsilon )\Omega (x)]\},
\end{equation*}%
\begin{equation*}
S_{1}^{60}(x)=[\frac{\partial }{\partial x}V_{1}^{ho}(x,\varepsilon )](\frac{%
\partial }{\partial x}[\Theta (x-x_{L},\varepsilon )\Omega (x)])[\frac{%
\partial ^{3}}{\partial x^{3}}V_{1}^{ho}(x,\varepsilon )],
\end{equation*}%
\begin{equation*}
S_{klmn}^{61}(x)=[\frac{\partial ^{2}}{\partial x^{2}}V_{1}^{ho}(x,%
\varepsilon )]\{\frac{\partial ^{k}}{\partial x^{k}}V_{1}^{ho}(x,\varepsilon
)\}^{m}\{\frac{\partial ^{l}}{\partial x^{l}}[\Theta (x-x_{L},\varepsilon
)\Omega (x)]\}^{n},
\end{equation*}%
\begin{equation*}
S_{klmn}^{62}(x)=(\frac{\partial ^{2}}{\partial x^{2}}[\Theta
(x-x_{L},\varepsilon )\Omega (x)])[\frac{\partial ^{k}}{\partial x^{k}}%
V_{1}^{ho}(x,\varepsilon )]^{m}\{\frac{\partial ^{l}}{\partial x^{l}}[\Theta
(x-x_{L},\varepsilon )\Omega (x)]\}^{n}.
\end{equation*}%
There are only two types of the spatially-selective functions $[\frac{%
\partial ^{k}}{\partial x^{k}}V_{1}^{ho}(x,\varepsilon )]^{m}$ and $\{\frac{%
\partial ^{k}}{\partial x^{k}}[\Theta (x-x_{L},\varepsilon )\Omega
(x)]\}^{m} $ appearing in these four terms $NA6$, $NB6$, $NC6$, and $ND6$.

Below investigate simply the four terms $NA6$, $NB6$, $NC6$, and $ND6.$
These four terms are formally similar to the previous four terms $NA5$, $NB5$%
, $NC5$, and $ND5$ on the $RH$ side of (4.132), respectively. The term $NA6$
of (4.139) consists of seven norms. Each one of which inside contains the
spatially-selective function which is the product of a pair of the
spatially-selective functions $[\frac{\partial ^{k}}{\partial x^{k}}%
V_{1}^{ho}(x,\varepsilon )]^{m}$ and $[\frac{\partial ^{l}}{\partial x^{l}}%
V_{1}^{ho}(x,\varepsilon )]^{n},$ here $0<k+l\leq 5$ and $0<m+n\leq 5.$
Thus, the term $NA6$ is similar to the term $NA4$ of (4.128) and $NA5$ of
(4.133). Similar to the terms $NA4$ and $NA5$, the term $NA6$ has an upper
bound consisting of the basic norms $\{NBAS1\}$ and $\{NBAS2\}$. As shown in
(4.140), the term $NB6$ has fourteen norms, each one of which contains the
spatially-selective function $[\frac{\partial ^{k}}{\partial x^{k}}%
V_{1}^{ho}(x,\varepsilon )]^{m}\{\frac{\partial ^{l}}{\partial x^{l}}[\Theta
(x-x_{L},\varepsilon )\Omega (x)]\}^{n},$ here $0<k+l\leq 5$ and $0<m+n\leq
5.$ The term $NB6$ is similar to the term $NB4$ of (4.129) and $NB5$ of
(4.134). Thus, just like the terms $NB4$ and $NB5$, the term $NB6$ has an
upper bound consisting of the basic norms $\{NBAS1\}$ and $\{NBAS2\}$. The
term $NC6$ of (4.141) has seven norms. Each one of these norms inside
contains the spatially-selective function $\{\frac{\partial ^{k}}{\partial
x^{k}}[\Theta (x-x_{L},\varepsilon )\Omega (x)]\}^{m}\{\frac{\partial ^{l}}{%
\partial x^{l}}[\Theta (x-x_{L},\varepsilon )\Omega (x)]\}^{n},$ here $%
0<k+l\leq 5$ and $0<m+n\leq 5.$ Then the term $NC6$ is similar to the term $%
NC4$ of (4.130) and $NC5$ of (4.135). Thus, similar to the terms $NC4$ and $%
NC5,$ the term $NC6$ has an upper bound consisting of the basic norms $%
\{NBAS1\}$ and $\{NBAS2\}$. The term $ND6$ of (4.142) is composed of eight
norms. Each one of these norms has a spatially-selective function that is
the product of three spatially-selective functions. There are eight
spatially-selective product functions in the term $ND6$, in which the first
two functions are given by $S_{0}^{60}(x)$ and $S_{1}^{60}(x),$ the central
three functions are $\{S_{klmn}^{61}(x)\},$ and the last three functions are
given by $\{S_{klmn}^{62}(x)\}$, where $0<k+l\leq 3$ and $0<m+n\leq 3.$ Each
one of these eight spatially-selective functions can be expanded as a sum of
a few spatially-selective functions. Thus, the term $ND6$ is similar to the
term $ND5$ of (4.136). It can turn out that the term $ND6$ has an upper
bound consisting of the basic norms $\{NBAS1\}$ and $\{NBAS2\}$. Now all the
four terms $NA6$, $NB6$, $NC6$, and $ND6$ on the RH side of (4.138) are
shown to have the upper bounds consisting of the basic norms $\{NBAS1\}$ and 
$\{NBAS2\}$. Then the inequality (4.138) shows that the upper bound of the
first norm on the $RH$ side of (4.137) can be expressed as a linear sum of
the basic norms $\{NBAS1\}$ and $\{NBAS2\}$. Since now all these six norms
on the $RH$ side of (4.137) are shown to have the upper bounds that consist
of the two types of basic norms $\{NBAS1\}$ and $\{NBAS2\}$, the inequality
(4.137) shows that the last norm on the $RH$ side of (4.105) has an upper
bound consisting of the basic norms $\{NBAS1\}$ and $\{NBAS2\}.$

As a summary, the theoretical calculation in this subsection shows that all
the six norms on the $RH$ side of (4.105) with $\mu =CS$ have the upper
bounds, each one of which can be expressed as a linear sum of a finite
number of the basic norms $\{NBAS1\}$ and/or $\{NBAS2\}$. Therefore, the
inequality (4.105) shows that the norm $||Q_{\lambda }(x,p,t_{3})\Psi
_{0}^{CS}(x,r,t_{0}+t_{1}/2)||$ has an upper bound that consists of a finite
number of the basic norms $\{NBAS1\}$ and $\{NBAS2\}$. Finally, the two
inequalities (4.104) indicate that both the norms $NORM(1,\lambda ,\mu )$
and $NORM(2,\lambda ,\mu )$ with $\mu =CS$ have the upper bounds consisting
of a finite number of the basic norms $\{NBAS1\}$ and $\{NBAS2\}$. This is
the desired result.\newline
\newline
{\large (C) The upper bound of the norm }$||Q_{\lambda }(x,p,t_{3})\Psi
_{0}^{\mu }(x,r,t_{0}+t_{1}/2)||${\large \ with }$\mu =S$

Now the norm $||Q_{\lambda }(x,p,t_{3})\Psi _{0}^{S}(x,r,t_{0}+t_{1}/2)||$
in (4.105) is computed. Here the product state $\Psi
_{0}^{S}(x,r,t_{0}+t_{1}/2)$ may be expressed as (4.106a) with the label $%
\mu =S$, in which the product state $\tilde{\Psi}_{0}^{S}(x,r,t_{0}+t_{1}/2)$
is given by (4.106e) and the function $F_{0}^{S}$ by (4.106d). This norm can
be strictly computed in a similar way that one calculates the norm $%
||Q_{\lambda }(x,p,t_{3})\Psi _{0}^{CS}(x,r,t_{0}+t_{1}/2)||$ in the
preceding subsection (B). This means that in this subsection there are also
six norms to be computed on the $RH$ side of (4.105) with the label $\mu =S$%
. The first norm on the $RH$ side of (4.105) with $l=0$ is $||F_{5}^{\lambda
}(x,t_{3})\Psi _{0}^{S}(x,r,t_{0}+t_{1}/2)||.$ With the help of the product
state (4.106a) and the function $F_{0}^{S}$ of (4.106d) it can turn out that
the norm is bounded by%
\begin{equation*}
||F_{5}^{\lambda }(x,t_{3})\Psi _{0}^{S}(x,r,t_{0}+t_{1}/2)||
\end{equation*}%
\begin{equation}
\leq |2\Omega _{0}t_{1}|\times ||F_{5}^{\lambda }(x,t_{3})\Theta
(x-x_{L},\varepsilon )\tilde{\Psi}_{0}^{S}(x,r,t_{0}+t_{1}/2)||.  \tag{4.143}
\end{equation}%
Note that the norms $|||\tilde{e}\rangle ||=1$ and $|||\tilde{g}_{0}\rangle
||=1.$ By using the product state $\tilde{\Psi}_{0}^{S}(x,r,t_{0}+t_{1}/2)$
of (4.106e) the norm on the $RH$ side of (4.143) may be reduced to a linear
sum of two basic norms $\{NBAS2\}$. Then the inequality (4.143) shows that
the first norm on the $RH$ side of (4.105) has an upper bound consisting of
the basic norms $\{NBAS2\}$.

The second norm on the $RH$ side of (4.105) with $l=1$ is calculated below.
With the help of (4.107) it can turn out that this norm is bounded by%
\begin{equation*}
||F_{4}^{\lambda }(x,t_{3})p\Psi _{0}^{S}(x,r,t_{0}+t_{1}/2)||\leq \hslash
||F_{4}^{\lambda }(x,t_{3})(\frac{\partial }{\partial x}F_{0}^{S})\tilde{\Psi%
}_{0}^{S}(x,r,t_{0}+t_{1}/2)||
\end{equation*}%
\begin{equation}
+\hslash ||F_{4}^{\lambda }(x,t_{3})F_{0}^{S}\frac{\partial }{\partial x}%
\tilde{\Psi}_{0}^{S}(x,r,t_{0}+t_{1}/2)||.  \tag{4.144}
\end{equation}%
Here the first-order coordinate derivative $(\frac{\partial }{\partial x}%
F_{0}^{S})$ can be calculated directly by using the function $F_{0}^{S}$ of
(4.106d). It is given by%
\begin{equation*}
\frac{\partial }{\partial x}F_{0}^{S}=(\frac{1}{\hslash }t_{1})\exp [-\frac{i%
}{\hslash }t_{1}V_{1}^{ho}(x,\varepsilon )]\{[\frac{\partial }{\partial x}%
V_{1}^{ho}(x,\varepsilon )]\sin [\frac{1}{2\hslash }\Theta
(x-x_{L},\varepsilon )\Omega (x)t_{1}]
\end{equation*}%
\begin{equation}
+i\frac{1}{2}(\frac{\partial }{\partial x}[\Theta (x-x_{L},\varepsilon
)\Omega (x)])\cos [\frac{1}{2\hslash }\Theta (x-x_{L},\varepsilon )\Omega
(x)t_{1}]\}.  \tag{4.145}
\end{equation}%
Then by using the derivative $(\frac{\partial }{\partial x}F_{0}^{S})$ it
can turn out that the two norms on the $RH$ side of (4.144) satisfy,
respectively,%
\begin{equation*}
||F_{4}^{\lambda }(x,t_{3})(\frac{\partial }{\partial x}F_{0}^{S})\tilde{\Psi%
}_{0}^{S}(x,r,t_{0}+t_{1}/2)||
\end{equation*}%
\begin{equation*}
\leq (\frac{1}{\hslash }t_{1})||F_{4}^{\lambda }(x,t_{3})[\frac{\partial }{%
\partial x}V_{1}^{ho}(x,\varepsilon )]\tilde{\Psi}%
_{0}^{S}(x,r,t_{0}+t_{1}/2)||
\end{equation*}%
\begin{equation}
+\frac{1}{2}(\frac{1}{\hslash }t_{1})||F_{4}^{\lambda }(x,t_{3})(\frac{%
\partial }{\partial x}[\Theta (x-x_{L},\varepsilon )\Omega (x)])\tilde{\Psi}%
_{0}^{S}(x,r,t_{0}+t_{1}/2)||  \tag{4.146}
\end{equation}%
and 
\begin{equation*}
||F_{4}^{\lambda }(x,t_{3})F_{0}^{S}\frac{\partial }{\partial x}\tilde{\Psi}%
_{0}^{S}(x,r,t_{0}+t_{1}/2)||
\end{equation*}%
\begin{equation}
\leq |2\Omega _{0}t_{1}|\times ||F_{4}^{\lambda }(x,t_{3})\Theta
(x-x_{L},\varepsilon )\frac{\partial }{\partial x}\tilde{\Psi}%
_{0}^{S}(x,r,t_{0}+t_{1}/2)||.  \tag{4.147}
\end{equation}%
Since $\frac{\partial }{\partial x}V_{1}^{ho}(x,\varepsilon )$ and $\frac{%
\partial }{\partial x}[\Theta (x-x_{L},\varepsilon )\Omega (x)]$ are
spatially-selective functions, the inequality (4.146) shows that the first
norm on the $RH$ side of (4.144) has an upper bound consisting of the basic
norms $\{NBAS1\}$ and $\{NBAS2\}$. In order to calculate the second norm on
the $RH$ side of (4.144) it is needed to calculate the first-order
derivative $\frac{\partial }{\partial x}\tilde{\Psi}%
_{0}^{S}(x,r,t_{0}+t_{1}/2)$. In general, by the formula (4.110b) the $k-$%
order coordinate derivative of the product state $\tilde{\Psi}%
_{0}^{S}(x,r,t_{0}+t_{1}/2)$ may be written as%
\begin{equation*}
\frac{\partial ^{k}}{\partial x^{k}}\tilde{\Psi}_{0}^{S}(x,r,t_{0}+t_{1}/2)=%
\exp [i\varphi (x,\gamma )]\Psi _{0}^{g}(x,t_{0}+t_{1}/2)\Gamma
_{k}^{g}(x,r,t_{0}+t_{1}/2)
\end{equation*}%
\begin{equation}
+\exp [-i\varphi (x,\gamma )]\Psi _{0}^{e}(x,t_{0}+t_{1}/2)\Gamma
_{k}^{e}(x,r,t_{0}+t_{1}/2)  \tag{4.147a}
\end{equation}%
where the internal superposition states $\Gamma _{k}^{a}(x,r,t_{0}+t_{1}/2)$
with $a=g$ and $e$ are respectively given by%
\begin{equation}
\Gamma _{k}^{g}(x,r,t_{0}+t_{1}/2)=\sum_{l=0}^{k}\left( 
\begin{array}{c}
k \\ 
l%
\end{array}%
\right) (\frac{\partial ^{l}}{\partial x^{l}}|\tilde{e}\rangle
)Q_{k-l}^{g}(x,+\varphi (x,\gamma )),  \tag{4.147b}
\end{equation}%
\begin{equation}
\Gamma _{k}^{e}(x,r,t_{0}+t_{1}/2)=\sum_{l=0}^{k}\left( 
\begin{array}{c}
k \\ 
l%
\end{array}%
\right) (\frac{\partial ^{l}}{\partial x^{l}}|\tilde{g}_{0}\rangle
)Q_{k-l}^{e}(x,-\varphi (x,\gamma )).  \tag{4.147c}
\end{equation}%
The internal superposition state $\Gamma _{k}^{a}(x,r,t_{0}+t_{1}/2)$ is
similar to $\Phi _{k}^{a}(x,r,t_{0}+t_{1}/2)$ of (4.123b). In particular,
when $k=1$, $\Gamma _{k}^{a}(x,r,t_{0}+t_{1}/2)$ with $a=g$ and $e$ are
respectively given by%
\begin{equation}
\Gamma _{1}^{g}(x,r,t_{0}+t_{1}/2)=Q_{1}^{g}(x,+\varphi (x,\gamma ))|\tilde{e%
}\rangle +Q_{0}^{g}(x,+\varphi (x,\gamma ))\frac{\partial }{\partial x}|%
\tilde{e}\rangle ,  \tag{4.147d}
\end{equation}%
\begin{equation}
\Gamma _{1}^{e}(x,r,t_{0}+t_{1}/2)=Q_{1}^{e}(x,-\varphi (x,\gamma ))|\tilde{g%
}_{0}\rangle +Q_{0}^{e}(x,-\varphi (x,\gamma ))\frac{\partial }{\partial x}|%
\tilde{g}_{0}\rangle .  \tag{4.147e}
\end{equation}%
Thus, the first-order derivative $\frac{\partial }{\partial x}\tilde{\Psi}%
_{0}^{S}(x,r,t_{0}+t_{1}/2)$ may be obtained by inserting (4.147d) and
(4.147e) into (4.147a). Now by using this first-order derivative it can turn
out that the norm on the $RH$ side of (4.147) satisfies%
\begin{equation*}
||F_{4}^{\lambda }(x,t_{3})S(x)\frac{\partial }{\partial x}\tilde{\Psi}%
_{0}^{S}(x,r,t_{0}+t_{1}/2)||
\end{equation*}%
\begin{equation*}
\leq ||(|\tilde{e}\rangle )||\times ||F_{4}^{\lambda
}(x,t_{3})Q_{1}^{g}(x,+\varphi (x,\gamma ))S(x)\Psi
_{0}^{g}(x,t_{0}+t_{1}/2)||
\end{equation*}%
\begin{equation*}
+||\frac{\partial }{\partial x}|\tilde{e}\rangle ||\times ||F_{4}^{\lambda
}(x,t_{3})Q_{0}^{g}(x,+\varphi (x,\gamma ))S(x)\Psi
_{0}^{g}(x,t_{0}+t_{1}/2)||
\end{equation*}%
\begin{equation*}
+||(|\tilde{g}_{0}\rangle )||\times ||F_{4}^{\lambda
}(x,t_{3})Q_{1}^{e}(x,-\varphi (x,\gamma ))S(x)\Psi
_{0}^{e}(x,t_{0}+t_{1}/2)||
\end{equation*}%
\begin{equation}
+||\frac{\partial }{\partial x}|\tilde{g}_{0}\rangle ||\times
||F_{4}^{\lambda }(x,t_{3})Q_{0}^{e}(x,-\varphi (x,\gamma ))S(x)\Psi
_{0}^{e}(x,t_{0}+t_{1}/2)||  \tag{4.147f}
\end{equation}%
where the spatially-selective function $S(x)=\Theta (x-x_{L},\varepsilon ).$
Note that the norms $||\frac{\partial ^{l}}{\partial x^{l}}|\tilde{g}%
_{0}\rangle ||$ and $||\frac{\partial ^{l}}{\partial x^{l}}|\tilde{e}\rangle
||$ with $l\geq 0$ are bounded parameters and $Q_{k}^{a}(x,\pm \varphi
(x,\gamma ))$ is a $k-$order polynomial in coordinate $x$. The inequality
(4.147f) is similar to (4.123d). It shows that the norm on the $RH$ side of
(4.147) is reduced to a linear sum of four norms, each one of which has an
upper bound consisting of the basic norms $\{NBAS2\}$. Thus, the inequality
(4.147) shows that the second norm on the $RH$ side of (4.144) has an upper
bound that is composed of the basic norms $\{NBAS2\}$. Now both the norms on
the $RH$ side of (4.144) have the upper bounds consisting of the basic norms 
$\{NBAS1\}$ and/or $\{NBAS2\}$, indicating that the second norm on the $RH$
side of (4.105) has an upper bound consisting of the basic norms $\{NBAS1\}$
and $\{NBAS2\}$.

Before the third norm on the $RH$ side of (4.105) is calculated in detail,
it is investigated how the theoretical calculation in the preceding
subsection (B) may be used to help the present theoretical calculation. One
can find that the derivative $(\frac{\partial }{\partial x}F_{0}^{S})$ of
(4.145) is similar to the derivative $(\frac{\partial }{\partial x}%
F_{0}^{CS})$ of (4.121). Both the derivatives may be respectively expressed
as%
\begin{equation}
\frac{\partial }{\partial x}F_{0}^{S}=\exp [-\frac{i}{\hslash }%
t_{1}V_{1}^{ho}(x,\varepsilon )]F_{S}(x)  \tag{4.148a}
\end{equation}%
and 
\begin{equation}
\frac{\partial }{\partial x}F_{0}^{CS}=\exp [-\frac{i}{\hslash }%
t_{1}V_{1}^{ho}(x,\varepsilon )]F_{CS}(x).  \tag{4.148b}
\end{equation}%
Their difference is in the two functions $F_{S}(x)$ and $F_{CS}(x)$ which
are given by%
\begin{equation*}
F_{S}(x)=(\frac{1}{\hslash }t_{1})[\frac{\partial }{\partial x}%
V_{1}^{ho}(x,\varepsilon )]\sin [\frac{1}{2\hslash }\Theta
(x-x_{L},\varepsilon )\Omega (x)t_{1}]
\end{equation*}%
\begin{equation}
+i\frac{1}{2}(\frac{1}{\hslash }t_{1})(\frac{\partial }{\partial x}[\Theta
(x-x_{L},\varepsilon )\Omega (x)])\cos [\frac{1}{2\hslash }\Theta
(x-x_{L},\varepsilon )\Omega (x)t_{1}]  \tag{4.148c}
\end{equation}%
and%
\begin{equation*}
F_{CS}(x)=-i(\frac{1}{\hslash }t_{1})[\frac{\partial }{\partial x}%
V_{1}^{ho}(x,\varepsilon )]\cos [\frac{1}{2\hslash }\Theta
(x-x_{L},\varepsilon )\Omega (x)t_{1}]
\end{equation*}%
\begin{equation}
-\frac{1}{2}(\frac{1}{\hslash }t_{1})(\frac{\partial }{\partial x}[\Theta
(x-x_{L},\varepsilon )\Omega (x)])\sin [\frac{1}{2\hslash }\Theta
(x-x_{L},\varepsilon )\Omega (x)t_{1}].  \tag{4.148d}
\end{equation}%
Now construct the two norms: 
\begin{equation*}
N_{S}^{k}=||q(x)[\frac{\partial ^{k}}{\partial x^{k}}(\frac{\partial }{%
\partial x}F_{0}^{S})]\tilde{\Psi}(x,r,t_{0}+t)||,\text{ }k\geq 0,
\end{equation*}%
and%
\begin{equation*}
N_{CS}^{k}=||q(x)[\frac{\partial ^{k}}{\partial x^{k}}(\frac{\partial }{%
\partial x}F_{0}^{CS})]\tilde{\Psi}(x,r,t_{0}+t)||,\text{ }k\geq 0.
\end{equation*}%
Here $q(x)$ may be a finite-order polynomial in coordinate $x$ or a general
bounded function and $\tilde{\Psi}(x,r,t_{0}+t)$ is a product state such as $%
\Psi _{0}^{C0}(x,r,t_{0}+t_{1}/2)$ or $\tilde{\Psi}%
_{0}^{S}(x,r,t_{0}+t_{1}/2)$ and so on. Below it proves that both the norms $%
N_{S}^{k}$ and $N_{CS}^{k}$ may have the same upper bound. It is known from
the theoretical calculation in the previous subsection (B) that the
coordinate derivative $\frac{\partial ^{k}}{\partial x^{k}}(\frac{\partial }{%
\partial x}F_{0}^{CS})$ for $0\leq k\leq 4$ may be expressed as%
\begin{equation}
\frac{\partial ^{k}}{\partial x^{k}}(\frac{\partial }{\partial x}%
F_{0}^{CS})=\exp [-i\frac{1}{\hslash }t_{1}V_{1}^{ho}(x,\varepsilon
)]\sum_{m=1}^{n_{k}}W_{m}^{k}Z_{km}^{CS}(x)S_{m}^{k}(x)  \tag{4.149}
\end{equation}%
where $S_{m}^{k}(x)$ is a spatially-selective function, $Z_{km}^{CS}(x)$ a
trigonometric function, and $W_{m}^{k}$ a non-negative weighting parameter.
Of course, the derivative $\frac{\partial ^{k}}{\partial x^{k}}(\frac{%
\partial }{\partial x}F_{0}^{CS})$ may be calculated directly by starting
from the derivative $\frac{\partial }{\partial x}F_{0}^{CS}$ of (4.148b) and
the function $F_{CS}(x)$ of (4.148d). All these spatially selective
functions $\{S_{m}^{k}(x)\}$ are different for a given order $k$. The
function $Z_{km}^{CS}(x)$ takes only one of the trigonometric functions $%
\{\eta \sin [\frac{1}{2\hslash }\Theta (x-x_{L},\varepsilon )\Omega
(x)t_{1}]\}$ and $\{\eta \cos [\frac{1}{2\hslash }\Theta
(x-x_{L},\varepsilon )\Omega (x)t_{1}]\}$ with $\eta =\pm ,$ $\pm i.$ The
weighting parameter $W_{m}^{k}$ is dependent on $(t_{1}/\hslash ).$ The term
number in (4.149) is $n_{k}=2,$ $5,$ $10,$ $20,$ $36$ for $k=0$, $1$, $2$, $%
3 $, $4,$ respectively. For example, when $k=0$, it is known from (4.148d)
that the spatially-selective function $S_{1}^{0}(x)=\frac{\partial }{%
\partial x}V_{1}^{ho}(x,\varepsilon )$ and $S_{2}^{0}(x)=\frac{\partial }{%
\partial x}[\Theta (x-x_{L},\varepsilon )\Omega (x)];$ the trigonometric
function $Z_{01}^{CS}(x)=-i\cos [\frac{1}{2\hslash }\Theta
(x-x_{L},\varepsilon )\Omega (x)t_{1}]$ and $Z_{02}^{CS}(x)=-\sin [\frac{1}{%
2\hslash }\Theta (x-x_{L},\varepsilon )\Omega (x)t_{1}];$ and the weighting
parameter $W_{1}^{0}=(\frac{1}{\hslash }t_{1})$ and $W_{2}^{0}=\frac{1}{2}(%
\frac{1}{\hslash }t_{1}).$ Now by using the derivative $\frac{\partial }{%
\partial x}F_{0}^{S}$ of (4.148a) and the function $F_{S}(x)$ of (4.148c) to
calculate explicitly the coordinate derivative $\frac{\partial ^{k}}{%
\partial x^{k}}(\frac{\partial }{\partial x}F_{0}^{S})$ with $0\leq k\leq 4$
one can find that the derivative can be expressed as%
\begin{equation}
\frac{\partial ^{k}}{\partial x^{k}}(\frac{\partial }{\partial x}%
F_{0}^{S})=\exp [-i\frac{1}{\hslash }t_{1}V_{1}^{ho}(x,\varepsilon
)]\sum_{m=1}^{n_{k}}W_{m}^{k}Z_{km}^{S}(x)S_{m}^{k}(x).  \tag{4.150}
\end{equation}%
This derivative is similar to that one of (4.149). The only difference
between the two derivatives (4.149) and (4.150) is in their trigonometric
functions $Z_{km}^{CS}(x)$ and $Z_{km}^{S}(x).$ Similar to the function $%
Z_{km}^{CS}(x)$ the function $Z_{km}^{S}(x)$ takes one of the trigonometric
functions $\{\eta \sin [\frac{1}{2\hslash }\Theta (x-x_{L},\varepsilon
)\Omega (x)t_{1}]\}$ and $\{\eta \cos [\frac{1}{2\hslash }\Theta
(x-x_{L},\varepsilon )\Omega (x)t_{1}]\}$ with $\eta =\pm ,$ $\pm i.$ But
both the functions $Z_{km}^{CS}(x)$ and $Z_{km}^{S}(x)$ may take different
trigonometric functions from $\{\eta \sin [\frac{1}{2\hslash }\Theta
(x-x_{L},\varepsilon )\Omega (x)t_{1}]\}$ and $\{\eta \cos [\frac{1}{%
2\hslash }\Theta (x-x_{L},\varepsilon )\Omega (x)t_{1}]\}$ with $\eta =\pm ,$
$\pm i,$ respectively. For example, it is known from (4.148c) that $%
Z_{01}^{S}(x)=\sin [\frac{1}{2\hslash }\Theta (x-x_{L},\varepsilon )\Omega
(x)t_{1}]$ and $Z_{02}^{S}(x)=i\cos [\frac{1}{2\hslash }\Theta
(x-x_{L},\varepsilon )\Omega (x)t_{1}],$ which are different from $%
Z_{01}^{CS}(x)$ and $Z_{02}^{CS}(x),$ respectively. It is clear that both
the functions $Z_{km}^{CS}(x)$ and $Z_{km}^{S}(x)$ satisfy the inequalities $%
|Z_{km}^{CS}(x)|\leq 1$ and $|Z_{km}^{S}(x)|\leq 1,$ respectively. In the
theoretical calculation in the previous subsection (B) the upper bound of
the norm $||F_{k}^{\lambda }(x,t_{3})(\frac{\partial ^{j}}{\partial x^{j}}%
F_{0}^{CS})\frac{\partial ^{l}}{\partial x^{l}}\Psi
_{0}^{C0}(x,r,t_{0}+t_{1}/2)||$ with $j>0$ and $l\geq 0$ is determined with
the help of the inequalities $|\sin [\frac{1}{2\hslash }\Theta
(x-x_{L},\varepsilon )\Omega (x)t_{1}]|\leq 1$ and $|\cos [\frac{1}{2\hslash 
}\Theta (x-x_{L},\varepsilon )\Omega (x)t_{1}]|\leq 1$ and the equality $%
||\exp [-i\frac{1}{\hslash }t_{1}V_{1}^{ho}(x,\varepsilon )]||=1.$ Thus, in
the theoretical calculation the trigonometric function $\sin [\frac{1}{%
2\hslash }\Theta (x-x_{L},\varepsilon )\Omega (x)t_{1}]$ is not treated as a
spatially-selective function. Then these trigonometric functions $\{\eta
\sin [\frac{1}{2\hslash }\Theta (x-x_{L},\varepsilon )\Omega (x)t_{1}]\}$
and $\{\eta \cos [\frac{1}{2\hslash }\Theta (x-x_{L},\varepsilon )\Omega
(x)t_{1}]\}$ with $\eta =\pm ,$ $\pm i$ do not have an effect on the upper
bounds of the norms $N_{CS}^{k}$ if here the norms $N_{CS}^{k}$ are just set
to those norms (corresponding to $N_{CS}^{k}$) in the theoretical
calculation in the previous subsection (B), resulting in that the functions $%
\{Z_{km}^{CS}(x)\}$ do not have an effect on the the upper bounds of the
norms $N_{CS}^{k}$. It can be seen below that the functions $%
\{Z_{km}^{S}(x)\}$ do not yet have an effect on the the upper bound of the
norm $N_{S}^{k}$.

In contrast, these trigonometric functions $\sin \theta $ and $\cos \theta $
with $\theta =[\frac{1}{2\hslash }\Theta (x-x_{L},\varepsilon )\Omega
(x)t_{1}]$ or $[\frac{1}{\hslash }t_{1}V_{1}^{ho}(x,\varepsilon )]$ must be
treated as the components of a spatially-selective function in the
theoretical calculation for the norm $||F_{k}^{\lambda }(x,t_{3})F_{0}^{CS}$ 
$\times \frac{\partial ^{l}}{\partial x^{l}}\Psi
_{0}^{C0}(x,r,t_{0}+t_{1}/2)||$ in the previous subsection (B) and also for
the norm $||F_{k}^{\lambda }(x,t_{3})F_{0}^{S}\frac{\partial ^{l}}{\partial
x^{l}}\tilde{\Psi}_{0}^{S}(x,r,t_{0}+t_{1}/2)||$ in the present subsection.
Both these norms are different from $N_{CS}^{k}$ and $N_{S}^{k}$.

By using the formulae (4.149) and (4.150) and then using the inequalities $%
|Z_{km}^{CS}(x)|\leq 1$ and $|Z_{km}^{S}(x)|\leq 1$ and the equality $||\exp
[-i\frac{1}{\hslash }t_{1}V_{1}^{ho}(x,\varepsilon )]||=1$ one may calculate
the upper bounds of the norms $N_{S}^{k}$ and $N_{CS}^{k}$ for $0\leq k\leq
4,$%
\begin{equation}
N_{S}^{k}\leq \sum_{m=1}^{n_{k}}W_{m}^{k}||q(x)S_{m}^{k}(x)\tilde{\Psi}%
(x,r,t_{0}+t)||,  \tag{4.151a}
\end{equation}%
\begin{equation}
N_{CS}^{k}\leq \sum_{m=1}^{n_{k}}W_{m}^{k}||q(x)S_{m}^{k}(x)\tilde{\Psi}%
(x,r,t_{0}+t)||.  \tag{4.151b}
\end{equation}%
These two inequalities show that both the norms $N_{S}^{k}$ and $N_{CS}^{k}$
have the same upper bound that is determined from the $RH$ side of (4.151a)
or (4.151b). Now setting the norm $N_{CS}^{k-1}=||F_{5-k}^{\lambda
}(x,t_{3})(\frac{\partial ^{k}}{\partial x^{k}}F_{0}^{CS})\tilde{\Psi}%
_{0}^{S}(x,r,t_{0}+t_{1}/2)||$ and $N_{S}^{k-1}=||F_{5-k}^{\lambda
}(x,t_{3})(\frac{\partial ^{k}}{\partial x^{k}}F_{0}^{S})\tilde{\Psi}%
_{0}^{S}(x,r,t_{0}+t_{1}/2)||$ for $1\leq k\leq 5.$ Both the norms $%
N_{CS}^{k-1}$ and $N_{S}^{k-1}$ clearly have the same upper bound, which is
determined from the $RH$ side of (4.151a) or (4.151b). This means that the
upper bound of the norm $N_{S}^{k-1}$ for $1\leq k\leq 5$ may be equal to
that one of $N_{CS}^{k-1}.$ Therefore, in order to obtain the upper bound of
the norm $N_{S}^{k-1}$ for $1\leq k\leq 5,$ which is the main task in the
present subsection, one needs only to determine the upper bound of the norm $%
N_{CS}^{k-1}.$ It is known that the upper bound of the norm $%
N_{CSC0}^{k-1}=||F_{5-k}^{\lambda }(x,t_{3})(\frac{\partial ^{k}}{\partial
x^{k}}F_{0}^{CS})\Psi _{0}^{C0}(x,r,t_{0}+t_{1}/2)||$ for $1\leq k\leq 5$ is
already obtained in the previous subsection (B). The only difference between
the two norms $N_{CS}^{k-1}$ and $N_{CSC0}^{k-1}$ is that the former inside
contains the product state $\tilde{\Psi}_{0}^{S}(x,r,t_{0}+t_{1}/2),$ while
the latter inside contains the product state $\Psi
_{0}^{C0}(x,r,t_{0}+t_{1}/2).$ While the norm $N_{CSC0}^{k-1}$ may be
reduced to a linear sum of the norms $\{||F_{5-k}^{\lambda
}(x,t_{3})S(x)\Psi _{0}^{C0}(x,r,t_{0}+t_{1}/2)||\},$ each one of which may
be further reduced to the $RH$ side of (4.118a) with $j=0$, the norm $%
N_{CS}^{k-1}$ may be reduced to a linear sum of the norms $%
\{||F_{5-k}^{\lambda }(x,t_{3})S(x)\tilde{\Psi}_{0}^{S}(x,r,t_{0}+t_{1}/2)||%
\},$ each one of which may be further reduced to the $RH$ side of (4.118b)
with $j=0$. This is their only difference in the theoretical calculation for
their upper bounds. Therefore, the inequality (or more generally the
theoretical calculation method) that is used to determine the upper bound of
the norm $N_{CSC0}^{k-1}$ in the previous subsection (B) also may be used to
determine that one of the norm $N_{CS}^{k-1}$ if one makes the replacement $%
\Psi _{0}^{C0}(x,r,t_{0}+t_{1}/2)\leftrightarrow \tilde{\Psi}%
_{0}^{S}(x,r,t_{0}+t_{1}/2)$ in the inequality (or in the theoretical
calculation method). When the upper bound of the norm $N_{CS}^{k-1}$ is
determined from this modified inequality, one really determines the upper
bound of the norm $N_{S}^{k-1}$ too as both the norms $N_{CS}^{k-1}$ and $%
N_{S}^{k-1}$ have the same upper bound. Similarly, the upper bound of the
norm $||F_{5-k-l}^{\lambda }(x,t_{3})(\frac{\partial ^{k}}{\partial x^{k}}%
F_{0}^{S})\frac{\partial ^{l}}{\partial x^{l}}\tilde{\Psi}%
_{0}^{S}(x,r,t_{0}+t_{1}/2)||$ for $1\leq k\leq 5$ and $0<l$ may be
determined from the inequality (or the theoretical calculation method) that
is used to determine the upper bound of the norm $||F_{5-k-l}^{\lambda
}(x,t_{3})(\frac{\partial ^{k}}{\partial x^{k}}F_{0}^{CS})\frac{\partial ^{l}%
}{\partial x^{l}}\Psi _{0}^{C0}(x,r,t_{0}+t_{1}/2)||$ in the previous
subsection (B) if one makes the replacement $\frac{\partial ^{l}}{\partial
x^{l}}\Psi _{0}^{C0}(x,r,t_{0}+t_{1}/2)\leftrightarrow \frac{\partial ^{l}}{%
\partial x^{l}}\tilde{\Psi}_{0}^{S}(x,r,t_{0}+t_{1}/2)$ in the inequality
(or in the theoretical calculation method). Here the former norm and its
upper bound are just what one wants to calculate in the present subsection.
Below this method will be used to calculate the norms on the $RH$ side of
(4.105) with $\mu =S.$ For convenience, hereafter without confusion the norm 
$||F_{5-k-l}^{\lambda }(x,t_{3})(\frac{\partial ^{k}}{\partial x^{k}}%
F_{0}^{S})\frac{\partial ^{l}}{\partial x^{l}}\tilde{\Psi}%
_{0}^{S}(x,r,t_{0}+t_{1}/2)||,$ the state $\tilde{\Psi}%
_{0}^{S}(x,r,t_{0}+t_{1}/2),$ etc., are denoted simply by $%
||F_{5-k-l}^{\lambda }(\frac{\partial ^{k}}{\partial x^{k}}F_{0}^{S})\frac{%
\partial ^{l}}{\partial x^{l}}\tilde{\Psi}_{0}^{S}||,$ $\tilde{\Psi}%
_{0}^{S}, $ etc., respectively. As an example, the inequality used to
determine the upper bound of the norm $||F_{3}^{\lambda }(\frac{\partial ^{2}%
}{\partial x^{2}}F_{0}^{CS})\Psi _{0}^{C0}||$ is given by (4.125). Now by
making the replacement $\Psi _{0}^{C0}\leftrightarrow \tilde{\Psi}_{0}^{S}$
in the inequality (4.125) one obtains the modified inequality of (4.125):%
\begin{equation*}
||F_{3}^{\lambda }(x,t_{3})(\frac{\partial ^{2}}{\partial x^{2}}F_{0}^{S})%
\tilde{\Psi}_{0}^{S}(x,r,t_{0}+t_{1}/2)||
\end{equation*}%
\begin{equation*}
\leq (\frac{1}{\hslash }t_{1})||F_{3}^{\lambda }(x,t_{3})[\frac{\partial ^{2}%
}{\partial x^{2}}V_{1}^{ho}(x,\varepsilon )]\tilde{\Psi}%
_{0}^{S}(x,r,t_{0}+t_{1}/2)||
\end{equation*}%
\begin{equation*}
+(\frac{1}{\hslash }t_{1})^{2}||F_{3}^{\lambda }(x,t_{3})[\frac{\partial }{%
\partial x}V_{1}^{ho}(x,\varepsilon )]^{2}\tilde{\Psi}%
_{0}^{S}(x,r,t_{0}+t_{1}/2)||
\end{equation*}%
\begin{equation*}
+(\frac{1}{\hslash }t_{1})^{2}||F_{3}^{\lambda }(x,t_{3})[\frac{\partial }{%
\partial x}V_{1}^{ho}(x,\varepsilon )](\frac{\partial }{\partial x}[\Theta
(x-x_{L},\varepsilon )\Omega (x)])\tilde{\Psi}_{0}^{S}(x,r,t_{0}+t_{1}/2)||
\end{equation*}%
\begin{equation*}
+(\frac{1}{2\hslash }t_{1})^{2}||F_{3}^{\lambda }(x,t_{3})(\frac{\partial }{%
\partial x}[\Theta (x-x_{L},\varepsilon )\Omega (x)])^{2}\tilde{\Psi}%
_{0}^{S}(x,r,t_{0}+t_{1}/2)||
\end{equation*}%
\begin{equation}
+(\frac{1}{2\hslash }t_{1})||F_{3}^{\lambda }(x,t_{3})(\frac{\partial ^{2}}{%
\partial x^{2}}[\Theta (x-x_{L},\varepsilon )\Omega (x)])\tilde{\Psi}%
_{0}^{S}(x,r,t_{0}+t_{1}/2)||.  \tag{4.152}
\end{equation}%
Here the norm $||F_{3}^{\lambda }(\frac{\partial ^{2}}{\partial x^{2}}%
F_{0}^{CS})\tilde{\Psi}_{0}^{S}||$ (i.e., $N_{CS}^{1}$) on the $LH$ side of
(4.152) is already replaced with the norm $||F_{3}^{\lambda }(\frac{\partial
^{2}}{\partial x^{2}}F_{0}^{S})\tilde{\Psi}_{0}^{S}||$ (i.e., $N_{S}^{1}$).
This is just the inequality (i.e., the modified inequality of (4.125)) that
is used to determine the upper bound of the norm $N_{S}^{1}.$

Now calculate the third norm on the $RH$ side of (4.105) with $l=2$. With
the help of (4.107) it can turn out that the norm is bounded by%
\begin{equation*}
||F_{3}^{\lambda }(x,t_{3})p^{2}\Psi _{0}^{S}(x,r,t_{0}+t_{1}/2)||\leq
\hslash ^{2}||F_{3}^{\lambda }(x,t_{3})(\frac{\partial ^{2}}{\partial x^{2}}%
F_{0}^{S})\tilde{\Psi}_{0}^{S}(x,r,t_{0}+t_{1}/2)||
\end{equation*}%
\begin{equation*}
+2\hslash ^{2}||F_{3}^{\lambda }(x,t_{3})(\frac{\partial }{\partial x}%
F_{0}^{S})\frac{\partial }{\partial x}\tilde{\Psi}%
_{0}^{S}(x,r,t_{0}+t_{1}/2)||
\end{equation*}%
\begin{equation}
+\hslash ^{2}||F_{3}^{\lambda }(x,t_{3})F_{0}^{S}\frac{\partial ^{2}}{%
\partial x^{2}}\tilde{\Psi}_{0}^{S}(x,r,t_{0}+t_{1}/2)||.  \tag{4.153}
\end{equation}%
This inequality is similar to (4.124) that is used to determine the upper
bound of the norm $||F_{3}^{\lambda }(x,t_{3})p^{2}\Psi
_{0}^{CS}(x,r,t_{0}+t_{1}/2)||$. Actually, the two inequalities of (4.124)
and (4.153) are convertible to each other if one makes the replacement $%
F_{0}^{CS}\leftrightarrow F_{0}^{S}$ and $\Psi _{0}^{C0}\leftrightarrow 
\tilde{\Psi}_{0}^{S}.$ The last norm on the $RH$ side of (4.153) is bounded
by%
\begin{equation*}
||F_{3}^{\lambda }(x,t_{3})F_{0}^{S}\frac{\partial ^{2}}{\partial x^{2}}%
\tilde{\Psi}_{0}^{S}(x,r,t_{0}+t_{1}/2)||
\end{equation*}%
\begin{equation}
\leq |2\Omega _{0}t_{1}|\times ||F_{3}^{\lambda }(x,t_{3})\Theta
(x-x_{L},\varepsilon )\frac{\partial ^{2}}{\partial x^{2}}\tilde{\Psi}%
_{0}^{S}(x,r,t_{0}+t_{1}/2)||.  \tag{4.154}
\end{equation}%
Here the derivative $\frac{\partial ^{2}}{\partial x^{2}}\tilde{\Psi}%
_{0}^{S} $ is given by (4.147a), in which the internal superposition states $%
\Gamma _{2}^{a}(x,r,t_{0}+t_{1}/2)$ are given by (4.147b) and (4.147c) with $%
k=2$, respectively. By using (4.147a) the norm on the $RH$ side of (4.154)
can be reduced to a linear sum of six norms, as shown in (4.118b), each one
of which has an upper bound consisting of the basic norms $\{NBAS2\}$. Then
the inequality (4.154) indicates that the upper bound of the last norm on
the $RH $ side of (4.153) consists of the basic norms $\{NBAS2\}$. The
second norm on the $RH$ side of (4.153) satisfies the inequality (4.146)
with the replacement $F_{4}^{\lambda }\leftrightarrow F_{3}^{\lambda }$ and $%
\tilde{\Psi}_{0}^{S}\leftrightarrow \frac{\partial }{\partial x}\tilde{\Psi}%
_{0}^{S}.$ Then according to this modified inequality of (4.146) the second
norm is reduced to a linear sum of two norms. Each one of the two norms may
be written as $||F_{3}^{\lambda }S(x)\frac{\partial }{\partial x}\tilde{\Psi}%
_{0}^{S}||,$ where the spatially-selective function $S(x)=[\frac{\partial }{%
\partial x}V_{1}^{ho}(x,\varepsilon )]$ or $(\frac{\partial }{\partial x}%
[\Theta (x-x_{L},\varepsilon )\Omega (x)]).$ It clearly obeys the inequality
(4.147f) with the replacement $F_{4}^{\lambda }\leftrightarrow
F_{3}^{\lambda },$ leading to that this norm is reduced to a linear sum of
four norms, each one of which has an upper bound consisting of the basic
norms $\{NBAS1\}$ and $\{NBAS2\}$. Thus, these show that the second norm on
the $RH$ side of (4.153) has an upper bound consisting of the basic norms $%
\{NBAS1\}$ and $\{NBAS2\}$. The first norm on the $RH$ side of (4.153)
satisfies the inequality (4.152). Note that the inequality (4.152) is
obtained from the inequality (4.125) by making the replacement $%
F_{0}^{CS}\leftrightarrow F_{0}^{S}$ and $\Psi _{0}^{C0}\leftrightarrow 
\tilde{\Psi}_{0}^{S}$ in the inequality (4.125). Then according to the
inequality (4.152) it can turn out that the upper bound of the first norm on
the $RH$ side of (4.153) may be expressed as a linear sum of the basic norms 
$\{NBAS1\}$ and $\{NBAS2\}$. This proof is completely according to that
procedure to prove that every norm on the $RH$ side of (4.125) has an upper
bound consisting of the basic norms $\{NBAS1\}$ and $\{NBAS2\}$ in the
previous subsection (B). Now all the three norms on the $RH$ side of (4.153)
are shown to have the upper bounds consisting of the basic norms $\{NBAS1\}$
and/or $\{NBAS2\}$. Then the inequality of (4.153) indicates that the third
norm on the $RH$ side of (4.105) has an upper bound consisting of the basic
norms $\{NBAS1\}$ and $\{NBAS2\}$.

The fourth norm on the $RH$ side of (4.105) is calculated below. With the
help of (4.107) it can turn out that it obeys the inequality:%
\begin{equation*}
||F_{2}^{\lambda }(x,t_{3})p^{3}\Psi _{0}^{S}(x,r,t_{0}+t_{1}/2)||\leq
\hslash ^{3}\{||F_{2}^{\lambda }(x,t_{3})(\frac{\partial ^{3}}{\partial x^{3}%
}F_{0}^{S})\tilde{\Psi}_{0}^{S}(x,r,t_{0}+t_{1}/2)||
\end{equation*}%
\begin{equation*}
+3||F_{2}^{\lambda }(x,t_{3})(\frac{\partial ^{2}}{\partial x^{2}}F_{0}^{S})%
\frac{\partial }{\partial x}\tilde{\Psi}_{0}^{S}(x,r,t_{0}+t_{1}/2)||
\end{equation*}%
\begin{equation*}
+3||F_{2}^{\lambda }(x,t_{3})(\frac{\partial }{\partial x}F_{0}^{S})\frac{%
\partial ^{2}}{\partial x^{2}}\tilde{\Psi}_{0}^{S}(x,r,t_{0}+t_{1}/2)||
\end{equation*}%
\begin{equation}
+||F_{2}^{\lambda }(x,t_{3})F_{0}^{S}\frac{\partial ^{3}}{\partial x^{3}}%
\tilde{\Psi}_{0}^{S}(x,r,t_{0}+t_{1}/2)||\}.  \tag{4.155}
\end{equation}%
This norm may be calculated in a similar way that one calculates the norm $%
||F_{2}^{\lambda }(x,t_{3})p^{3}\Psi _{0}^{CS}(x,r,t_{0}+t_{1}/2)||$ in the
previous subsection (B). Actually, it satisfies the inequality (4.126) with
the replacement $\Psi _{0}^{CS}\leftrightarrow \Psi _{0}^{S}$ on the $LH$
side and $F_{0}^{CS}\leftrightarrow F_{0}^{S}$ and $\Psi
_{0}^{C0}\leftrightarrow \tilde{\Psi}_{0}^{S}$ on the $RH$ side. There are
four norms to be calculated on the $RH$ side of (4.155). The last norm
satisfies%
\begin{equation*}
||F_{2}^{\lambda }(x,t_{3})F_{0}^{S}\frac{\partial ^{3}}{\partial x^{3}}%
\tilde{\Psi}_{0}^{S}(x,r,t_{0}+t_{1}/2)||
\end{equation*}%
\begin{equation}
\leq |2\Omega _{0}t_{1}|\times ||F_{2}^{\lambda }(x,t_{3})\Theta
(x-x_{L},\varepsilon )\frac{\partial ^{3}}{\partial x^{3}}\tilde{\Psi}%
_{0}^{S}(x,r,t_{0}+t_{1}/2)||.  \tag{4.155a}
\end{equation}%
Here the derivative $\frac{\partial ^{3}}{\partial x^{3}}\tilde{\Psi}%
_{0}^{S} $ is given by (4.147a), in which the internal superposition states $%
\Gamma _{3}^{a}(x,r,t_{0}+t_{1}/2)$ are given by (4.147b) and (4.147c) with $%
k=3$, respectively. Now by using (4.147a) the norm on the $RH$ side of
(4.155a) may be reduced to a linear sum of eight norms, as shown in
(4.118b), each one of which has an upper bound consisting of the basic norms 
$\{NBAS2\}$. Thus, the inequality (4.155a) shows that the upper bound of the
last norm on the $RH$ side of (4.155) may be written as a linear sum of the
basic norms $\{NBAS2\}$. As shown in the two inequalities (4.151), the third
norm on the $RH$ side of (4.155), i.e., $||F_{2}^{\lambda }(\frac{\partial }{%
\partial x}F_{0}^{S})\frac{\partial ^{2}}{\partial x^{2}}\tilde{\Psi}%
_{0}^{S}||,$ has the same upper bound as the norm $||F_{2}^{\lambda }(\frac{%
\partial }{\partial x}F_{0}^{CS})\frac{\partial ^{2}}{\partial x^{2}}\tilde{%
\Psi}_{0}^{S}||.$ The latter can be calculated in the way that is used to
calculate the third norm on the $RH$ side of (4.126), i.e., $%
||F_{2}^{\lambda }(\frac{\partial }{\partial x}F_{0}^{CS})\frac{\partial ^{2}%
}{\partial x^{2}}\Psi _{0}^{C0}||,$ in the previous subsection (B) if one
makes the replacement $\Psi _{0}^{C0}\leftrightarrow \tilde{\Psi}_{0}^{S}$.
While the norm $||F_{2}^{\lambda }(\frac{\partial }{\partial x}F_{0}^{CS})%
\frac{\partial ^{2}}{\partial x^{2}}\Psi _{0}^{C0}||$ may be reduced to a
linear sum of the norms $\{||F_{2}^{\lambda }S(x)\frac{\partial ^{2}}{%
\partial x^{2}}\Psi _{0}^{C0}||\},$ each one of which may be further reduced
to the $RH$ side of (4.118a) with $j=2$, the norm $||F_{2}^{\lambda }(\frac{%
\partial }{\partial x}F_{0}^{CS})\frac{\partial ^{2}}{\partial x^{2}}\tilde{%
\Psi}_{0}^{S}||$ may be reduced to a linear sum of the norms $%
\{||F_{2}^{\lambda }S(x)\frac{\partial ^{2}}{\partial x^{2}}\tilde{\Psi}%
_{0}^{S}||\},$ each one of which may be further reduced to the $RH$ side of
(4.118b) with $j=2$. This is the only difference generated by the
replacement $\Psi _{0}^{C0}\leftrightarrow \tilde{\Psi}_{0}^{S}$ in the
theoretical calculation for the upper bounds of the two norms. It turns out
in the previous subsection (B) that the norm $||F_{2}^{\lambda }(\frac{%
\partial }{\partial x}F_{0}^{CS})\frac{\partial ^{2}}{\partial x^{2}}\Psi
_{0}^{C0}||$ has an upper bound consisting of the basic norms $\{NBSA1\}$
and $\{NBAS2\}$. Then in an analogous way it can turn out that the norm $%
||F_{2}^{\lambda }(\frac{\partial }{\partial x}F_{0}^{CS})\frac{\partial ^{2}%
}{\partial x^{2}}\tilde{\Psi}_{0}^{S}||$ also has an upper bound consisting
of the basic norms $\{NBSA1\}$ and $\{NBAS2\}$. This indicates directly that
the third norm on the $RH$ side of (4.155) has an upper bound consisting of
the basic norms $\{NBSA1\}$ and $\{NBAS2\}$. Similarly, as shown in the two
inequalities (4.151), the second norm on the $RH$ side of (4.155), i.e., $%
||F_{2}^{\lambda }(\frac{\partial ^{2}}{\partial x^{2}}F_{0}^{S})\frac{%
\partial }{\partial x}\tilde{\Psi}_{0}^{S}||,$ has the same upper bound as
the norm $||F_{2}^{\lambda }(\frac{\partial ^{2}}{\partial x^{2}}F_{0}^{CS})%
\frac{\partial }{\partial x}\tilde{\Psi}_{0}^{S}||.$ Here the latter can be
calculated in the way that one calculates the second norm $||F_{2}^{\lambda
}(\frac{\partial ^{2}}{\partial x^{2}}F_{0}^{CS})\frac{\partial }{\partial x}%
\Psi _{0}^{C0}||$ on the $RH$ side of (4.126) in the previous subsection (B)
if one makes the replacement $\Psi _{0}^{C0}\leftrightarrow \tilde{\Psi}%
_{0}^{S}$. It already proves in the previous subsection (B) that the norm $%
||F_{2}^{\lambda }(\frac{\partial ^{2}}{\partial x^{2}}F_{0}^{CS})\frac{%
\partial }{\partial x}\Psi _{0}^{C0}||$ has an upper bound consisting of the
basic norms $\{NBAS1\}$ and $\{NBAS2\}$. Then in an analogous way it can
prove that the norm $||F_{2}^{\lambda }(\frac{\partial ^{2}}{\partial x^{2}}%
F_{0}^{CS})\frac{\partial }{\partial x}\tilde{\Psi}_{0}^{S}||$ also has an
upper bound consisting of the basic norms $\{NBAS1\}$ and $\{NBAS2\}$,
indicating directly that the second norm on the $RH$ side of (4.155) has an
upper bound consisting of the basic norms $\{NBAS1\}$ and $\{NBAS2\}$. The
first norm on the $RH$ side of (4.155), i.e., $||F_{2}^{\lambda }(\frac{%
\partial ^{3}}{\partial x^{3}}F_{0}^{S})\tilde{\Psi}_{0}^{S}||,$ has the
same upper bound as the norm $||F_{2}^{\lambda }(\frac{\partial ^{3}}{%
\partial x^{3}}F_{0}^{CS})\tilde{\Psi}_{0}^{S}||.$ The upper bound of the
latter norm can be determined in the way that the upper bound of the first
norm $||F_{2}^{\lambda }(\frac{\partial ^{3}}{\partial x^{3}}F_{0}^{CS})\Psi
_{0}^{C0}||$ on the $RH$ side of (4.126) is determined in the previous
subsection (B) if one makes the replacement $\Psi _{0}^{C0}\leftrightarrow 
\tilde{\Psi}_{0}^{S}$. It is known that the upper bound of the norm $%
||F_{2}^{\lambda }(\frac{\partial ^{3}}{\partial x^{3}}F_{0}^{CS})\Psi
_{0}^{C0}||$ is determined from the inequality (4.127) in the previous
subsection (B). Now by making the replacement $\Psi _{0}^{C0}\leftrightarrow 
\tilde{\Psi}_{0}^{S}$ in the inequality (4.127) one obtains the modified
inequality of (4.127):%
\begin{equation}
||F_{2}^{\lambda }(x,t_{3})(\frac{\partial ^{3}}{\partial x^{3}}F_{0}^{CS})%
\tilde{\Psi}_{0}^{S}(x,r,t_{0}+t_{1}/2)||\leq SA4+SB4+SC4.  \tag{4.155b}
\end{equation}%
Here the current three terms $SA4,$ $SB4,$ and $SC4$ correspond to the three
terms $NA4,$ $NB4,$ and $NC4$ in (4.127), respectively. They are still
determined from the equations (4.128), (4.129), and (4.130) with the
replacement $\Psi _{0}^{C0}\leftrightarrow \tilde{\Psi}_{0}^{S},$
respectively. Because the norm $||F_{2}^{\lambda }(\frac{\partial ^{3}}{%
\partial x^{3}}F_{0}^{S})\tilde{\Psi}_{0}^{S}||$ has the same upper bound as
the norm $||F_{2}^{\lambda }(\frac{\partial ^{3}}{\partial x^{3}}F_{0}^{CS})%
\tilde{\Psi}_{0}^{S}||,$ its upper bound also is determined from the
inequality (4.155b). With the same theoretical calculation method that is
used to calculate the upper bounds of the three terms $NA4,$ $NB4,$ and $NC4$
in (4.127) in the previous subsection (B) one can prove on the basis of
these three modified equations of (4.128), (4.129), and (4.130) that each
one of the current three terms $SA4,$ $SB4,$ and $SC4$ has the upper bound
consisting of the basic norms $\{NBAS1\}$ and $\{NBAS2\}$. This indicates
directly that the first norm on the $RH$ side of (4.155) has an upper bound
consisting of the basic norms $\{NBAS1\}$ and $\{NBAS2\}$. Now all the four
norms on the $RH$ side of (4.155) are shown to have the upper bounds
consisting of the basic norms $\{NBAS1\}$ and/or $\{NBAS2\}$, indicating
that the fourth norm on the $RH$ side of (4.105) has an upper bound
consisting of the basic norms $\{NBAS1\}$ and $\{NBAS2\}$.

The fifth norm on the $RH$ side of (4.105) with $l=4$ is calculated below.
With the aid of (4.107) one can prove that this norm obeys the inequality: 
\begin{equation*}
||F_{1}^{\lambda }(x,t_{3})p^{4}\Psi _{0}^{S}(x,r,t_{0}+t_{1}/2)||\leq
\hslash ^{4}\{||F_{1}^{\lambda }(x,t_{3})(\frac{\partial ^{4}}{\partial x^{4}%
}F_{0}^{S})\tilde{\Psi}_{0}^{S}(x,r,t_{0}+t_{1}/2)||
\end{equation*}%
\begin{equation*}
+4||F_{1}^{\lambda }(x,t_{3})(\frac{\partial ^{3}}{\partial x^{3}}F_{0}^{S})%
\frac{\partial }{\partial x}\tilde{\Psi}_{0}^{S}(x,r,t_{0}+t_{1}/2)||
\end{equation*}%
\begin{equation*}
+6||F_{1}^{\lambda }(x,t_{3})(\frac{\partial ^{2}}{\partial x^{2}}F_{0}^{S})%
\frac{\partial ^{2}}{\partial x^{2}}\tilde{\Psi}_{0}^{S}(x,r,t_{0}+t_{1}/2)||
\end{equation*}%
\begin{equation*}
+4||F_{1}^{\lambda }(x,t_{3})(\frac{\partial }{\partial x}F_{0}^{S})\frac{%
\partial ^{3}}{\partial x^{3}}\tilde{\Psi}_{0}^{S}(x,r,t_{0}+t_{1}/2)||
\end{equation*}%
\begin{equation}
+||F_{1}^{\lambda }(x,t_{3})F_{0}^{S}\frac{\partial ^{4}}{\partial x^{4}}%
\tilde{\Psi}_{0}^{S}(x,r,t_{0}+t_{1}/2)||\}.  \tag{4.156}
\end{equation}%
This inequality is really equal to (4.131) with the replacement $%
F_{0}^{CS}\leftrightarrow F_{0}^{S}$ and $\Psi _{0}^{C0}\leftrightarrow 
\tilde{\Psi}_{0}^{S}.$ It is clear that one needs to calculate the five
norms on the $RH$ side of (4.156). The last norm on the $RH$ side of (4.156)
obeys the inequality:%
\begin{equation*}
||F_{1}^{\lambda }(x,t_{3})F_{0}^{S}\frac{\partial ^{4}}{\partial x^{4}}%
\tilde{\Psi}_{0}^{S}(x,r,t_{0}+t_{1}/2)||
\end{equation*}%
\begin{equation}
\leq |2\Omega _{0}t_{1}|\times ||F_{1}^{\lambda }(x,t_{3})\Theta
(x-x_{L},\varepsilon )\frac{\partial ^{4}}{\partial x^{4}}\tilde{\Psi}%
_{0}^{S}(x,r,t_{0}+t_{1}/2)||.  \tag{4.156a}
\end{equation}%
Here the derivative $\frac{\partial ^{4}}{\partial x^{4}}\tilde{\Psi}%
_{0}^{S} $ is given by (4.147a), in which the internal superposition states $%
\Gamma _{4}^{a}(x,r,t_{0}+t_{1}/2)$ are given by (4.147b) and (4.147c),
respectively. By using (4.147a) the norm on the $RH$ side of (4.156a) may be
reduced to a linear sum of ten norms, as shown in (4.118b), each one of
which has an upper bound consisting of the basic norms $\{NBAS2\}$. Then the
inequality (4.156a) indicates that the last norm on the $RH$ side of (4.156)
has an upper bound consisting of the basic norms $\{NBAS2\}$. As shown in
the two inequalities (4.151), the fourth norm on the $RH$ side of (4.156)
has the same upper bound as the norm $||F_{1}^{\lambda }(\frac{\partial }{%
\partial x}F_{0}^{CS})\frac{\partial ^{3}}{\partial x^{3}}\tilde{\Psi}%
_{0}^{S}||.$ The latter can be calculated in the way that one calculates the
fourth norm on the $RH$ side of (4.131), i.e., $||F_{1}^{\lambda }(\frac{%
\partial }{\partial x}F_{0}^{CS})\frac{\partial ^{3}}{\partial x^{3}}\Psi
_{0}^{C0}||,$ in the previous subsection (B) if one makes the replacement $%
\Psi _{0}^{C0}\leftrightarrow \tilde{\Psi}_{0}^{S}.$ In the previous
subsection (B) it proves that the norm $||F_{1}^{\lambda }(\frac{\partial }{%
\partial x}F_{0}^{CS})\frac{\partial ^{3}}{\partial x^{3}}\Psi _{0}^{C0}||$
has an upper bound consisting of the basic norms $\{NBAS1\}$ and $\{NBAS2\}$%
. Then in an analogous way it can prove that the norm $||F_{1}^{\lambda }(%
\frac{\partial }{\partial x}F_{0}^{CS})\frac{\partial ^{3}}{\partial x^{3}}%
\tilde{\Psi}_{0}^{S}||$ also has an upper bound consisting of the basic
norms $\{NBAS1\}$ and $\{NBAS2\}$. This directly indicates that the fourth
norm on the $RH$ side of (4.156) has an upper bound consisting of the basic
norms $\{NBAS1\}$ and $\{NBAS2\}$. The third and second norms on the $RH$
side of (4.156) have the same upper bounds as the two norms $%
||F_{1}^{\lambda }(\frac{\partial ^{k}}{\partial x^{k}}F_{0}^{CS})\frac{%
\partial ^{4-k}}{\partial x^{4-k}}\tilde{\Psi}_{0}^{S}||$ with $k=2$ and $3,$
respectively. Now the two latter norms can be calculated in the ways that
one calculates the third and second norms on the $RH$ side of (4.131) in the
subsection (B), i.e., $||F_{1}^{\lambda }(\frac{\partial ^{k}}{\partial x^{k}%
}F_{0}^{CS})\frac{\partial ^{4-k}}{\partial x^{4-k}}\Psi _{0}^{C0}||$ with $%
k=2$ and $3,$ respectively, if one makes the replacement $\Psi
_{0}^{C0}\leftrightarrow \tilde{\Psi}_{0}^{S}.$ In the subsection (B) both
the norms $||F_{1}^{\lambda }(\frac{\partial ^{k}}{\partial x^{k}}F_{0}^{CS})%
\frac{\partial ^{4-k}}{\partial x^{4-k}}\Psi _{0}^{C0}||$ with $k=2$ and $3$
are already proven to have upper bounds consisting of the basic norms $%
\{NBAS1\}$ and $\{NBAS2\}$. Now by using the same theoretical calculation
method one can prove that both the norms $||F_{1}^{\lambda }(\frac{\partial
^{k}}{\partial x^{k}}F_{0}^{CS})\frac{\partial ^{4-k}}{\partial x^{4-k}}%
\tilde{\Psi}_{0}^{S}||$ with $k=2$ and $3$ also have the upper bounds
consisting of the basic norms $\{NBAS1\}$ and $\{NBAS2\}$. This indicates
directly that the third and second norms on the $RH$ side of (4.156), i.e., $%
||F_{1}^{\lambda }(\frac{\partial ^{k}}{\partial x^{k}}F_{0}^{S})\frac{%
\partial ^{4-k}}{\partial x^{4-k}}\tilde{\Psi}_{0}^{S}||$ with $k=2$ and $3,$
have the upper bounds consisting of the basic norms $\{NBAS1\}$ and $%
\{NBAS2\}$. Now the first norm on the $RH$ side of (4.156), i.e., $%
||F_{1}^{\lambda }(\frac{\partial ^{4}}{\partial x^{4}}F_{0}^{S})\tilde{\Psi}%
_{0}^{S}||,$ has the same upper bound as the norm $||F_{1}^{\lambda }(\frac{%
\partial ^{4}}{\partial x^{4}}F_{0}^{CS})\tilde{\Psi}_{0}^{S}||.$ The latter
satisfies the inequality (4.132) with the replacement $\Psi
_{0}^{C0}\leftrightarrow \tilde{\Psi}_{0}^{S}:$%
\begin{equation}
||F_{1}^{\lambda }(x,t_{3})(\frac{\partial ^{4}}{\partial x^{4}}F_{0}^{CS})%
\tilde{\Psi}_{0}^{S}(x,r,t_{0}+t_{1}/2)||\leq SA5+SB5+SC5+SD5.  \tag{4.156b}
\end{equation}%
Here the current four terms $SA5,$ $SB5,$ $SC5,$ and $SD5$ correspond to the
four terms $NA5,$ $NB5,$ $NC5,$ and $ND5$ in (4.132), respectively. They are
still given by the four modified equations of (4.133), (4.134), (4.135), and
(4.136), respectively, in which the state $\Psi _{0}^{C0}$ is replaced with $%
\tilde{\Psi}_{0}^{S}.$ From these four modified equations one can determine
the upper bounds of the current four terms $SA5,$ $SB5,$ $SC5,$ and $SD5$ by
using the same theoretical calculation method used in the previous
subsection (B). It can turn out that these four terms $SA5,$ $SB5,$ $SC5,$
and $SD5$ have the upper bounds consisting of the basic norms $\{NBAS1\}$
and $\{NBAS2\}$. Then the inequality (4.156b) shows directly that the first
norm on the $RH$ side of (4.156) has an upper bound consisting of the basic
norms $\{NBAS1\}$ and $\{NBAS2\}$. Now all the five norms on the $RH$ side
of (4.156) are shown to have the upper bounds consisting of the basic norms $%
\{NBAS1\}$ and/or $\{NBAS2\}$, indicating that the fifth norm on the $RH$
side of (4.105) also has an upper bound consisting of the basic norms $%
\{NBAS1\}$ and $\{NBAS2\}$.

The last norm on the $RH$ side of (4.105) is calculated below. With the help
of (4.107) it can turn out that this norm obeys the inequality:%
\begin{equation*}
||F_{0}^{\lambda }(x,t_{3})p^{5}\Psi _{0}^{S}(x,r,t_{0}+t_{1}/2)||\leq
\hslash ^{5}\{||F_{0}^{\lambda }(x,t_{3})(\frac{\partial ^{5}}{\partial x^{5}%
}F_{0}^{S})\tilde{\Psi}_{0}^{S}(x,r,t_{0}+t_{1}/2)||
\end{equation*}%
\begin{equation*}
+5||F_{0}^{\lambda }(x,t_{3})(\frac{\partial ^{4}}{\partial x^{4}}F_{0}^{S})%
\frac{\partial }{\partial x}\tilde{\Psi}_{0}^{S}(x,r,t_{0}+t_{1}/2)||
\end{equation*}%
\begin{equation*}
+10||F_{0}^{\lambda }(x,t_{3})(\frac{\partial ^{3}}{\partial x^{3}}F_{0}^{S})%
\frac{\partial ^{2}}{\partial x^{2}}\tilde{\Psi}_{0}^{S}(x,r,t_{0}+t_{1}/2)||
\end{equation*}%
\begin{equation*}
+10||F_{0}^{\lambda }(x,t_{3})(\frac{\partial ^{2}}{\partial x^{2}}F_{0}^{S})%
\frac{\partial ^{3}}{\partial x^{3}}\tilde{\Psi}_{0}^{S}(x,r,t_{0}+t_{1}/2)||
\end{equation*}%
\begin{equation*}
+5||F_{0}^{\lambda }(x,t_{3})(\frac{\partial }{\partial x}F_{0}^{S})\frac{%
\partial ^{4}}{\partial x^{4}}\tilde{\Psi}_{0}^{S}(x,r,t_{0}+t_{1}/2)||
\end{equation*}%
\begin{equation}
+||F_{0}^{\lambda }(x,t_{3})F_{0}^{S}\frac{\partial ^{5}}{\partial x^{5}}%
\tilde{\Psi}_{0}^{S}(x,r,t_{0}+t_{1}/2)||\}.  \tag{4.157}
\end{equation}%
This norm may be calculated in a similar way that the norm $||F_{0}^{\lambda
}(x,t_{3})p^{5}$ $\times \Psi _{0}^{CS}(x,r,t_{0}+t_{1}/2)||$ is calculated
on the basis of the inequality (4.137) in the subsection (B). The inequality
(4.157) is really equal to the inequality (4.137) with the replacement $%
F_{0}^{CS}\leftrightarrow F_{0}^{S}$ and $\Psi _{0}^{C0}\leftrightarrow 
\tilde{\Psi}_{0}^{S}.$ There are six norms to be calculated on the $RH$ side
of (4.157). The last norm satisfies the inequality: 
\begin{equation*}
||F_{0}^{\lambda }(x,t_{3})F_{0}^{S}\frac{\partial ^{5}}{\partial x^{5}}%
\tilde{\Psi}_{0}^{S}(x,r,t_{0}+t_{1}/2)||
\end{equation*}%
\begin{equation}
\leq |2\Omega _{0}t_{1}|\times ||F_{0}^{\lambda }(x,t_{3})\Theta
(x-x_{L},\varepsilon )\frac{\partial ^{5}}{\partial x^{5}}\tilde{\Psi}%
_{0}^{S}(x,r,t_{0}+t_{1}/2)||.  \tag{4.157a}
\end{equation}%
Here the derivative $\frac{\partial ^{5}}{\partial x^{5}}\tilde{\Psi}%
_{0}^{S} $ is given by (4.147a), in which the internal superposition states $%
\Gamma _{5}^{a}(x,r,t_{0}+t_{1}/2)$ are given by (4.147b) and (4.147c) with $%
k=5$, respectively. Now by using the derivative of (4.147a) the norm on the $%
RH$ side of (4.157a) may be reduced to a linear sum of the twelve norms, as
shown in (4.118b), each one of which has an upper bound consisting of the
basic norms $\{NBAS2\}$. Then the inequality (4.157a) indicates that the
last norm on the $RH$ side of (4.157) has an upper bound consisting of the
basic norms $\{NBAS2\}$. Now the first five norms on the $RH$ side of
(4.157) have the same upper bounds as the five norms $||F_{0}^{\lambda }(%
\frac{\partial ^{5-k}}{\partial x^{5-k}}F_{0}^{CS})\frac{\partial ^{k}}{%
\partial x^{k}}\tilde{\Psi}_{0}^{S}||$ with $k=0$, $1$, $2$, $3$, and $4$,
respectively. The upper bounds of the latter five norms can be determined in
the ways that the upper bounds of the corresponding five norms $%
||F_{0}^{\lambda }(\frac{\partial ^{5-k}}{\partial x^{5-k}}F_{0}^{CS})\frac{%
\partial ^{k}}{\partial x^{k}}\Psi _{0}^{C0}||$ with $k=0$, $1$, $2$, $3$,
and $4$ on the $RH$ side of (4.137) are determined in the subsection (B),
respectively, by making the replacement $\Psi _{0}^{C0}\leftrightarrow 
\tilde{\Psi}_{0}^{S}.$ It is already proven in the subsection (B) that all
the five norms $||F_{0}^{\lambda }(\frac{\partial ^{5-k}}{\partial x^{5-k}}%
F_{0}^{CS})\frac{\partial ^{k}}{\partial x^{k}}\Psi _{0}^{C0}||$ with $k=0$, 
$1$, $2$, $3$, and $4$ have the upper bounds consisting of the basic norms $%
\{NBAS1\}$ and $\{NBAS2\}$. Then by using the same theoretical calculation
method in the subsection (B) one may strictly calculate the upper bounds of
the corresponding five norms $||F_{0}^{\lambda }(\frac{\partial ^{5-k}}{%
\partial x^{5-k}}F_{0}^{CS})\frac{\partial ^{k}}{\partial x^{k}}\tilde{\Psi}%
_{0}^{S}||\}$ with $k=0,$ $1$, $2$, $3$, and $4$ and prove that these five
norms also have the upper bounds consisting of the basic norms $\{NBAS1\}$
and $\{NBAS2\}$. This really means that the first five norms on the $RH$
side of (4.157), i.e., $||F_{0}^{\lambda }(\frac{\partial ^{5-k}}{\partial
x^{5-k}}F_{0}^{S})\frac{\partial ^{k}}{\partial x^{k}}\tilde{\Psi}%
_{0}^{S}||\}$ with $k=0,$ $1$, $2$, $3$, and $4$, also have the upper bounds
consisting of the basic norms $\{NBAS1\}$ and $\{NBAS2\}$. As a typical
example, the first norm on the $RH$ side of (4.157) has the same upper bound
as the norm $||F_{0}^{\lambda }(\frac{\partial ^{5}}{\partial x^{5}}%
F_{0}^{CS})\tilde{\Psi}_{0}^{S}||,$ which satisfies the inequality: 
\begin{equation}
||F_{0}^{\lambda }(x,t_{3})(\frac{\partial ^{5}}{\partial x^{5}}%
F_{0}^{CS})\Psi _{0}^{S}(x,r,t_{0}+t_{1}/2)||\leq SA6+SB6+SC6+SD6. 
\tag{4.157b}
\end{equation}%
This inequality is similar to (4.138) that the first norm on the $RH$ side
of (4.137) obeys. Here the current four terms $SA6,$ $SB6,$ $SC6,$ and $SD6$
correspond to the four terms $NA6,$ $NB6,$ $NC6,$ and $ND6$ in (4.138),
respectively. They are still given by the four modified equations of
(4.139), (4.140), (4.141), and (4.142), respectively, in which the state $%
\Psi _{0}^{C0}$ is replaced with $\tilde{\Psi}_{0}^{S}.$ Then according to
the same theoretical calculation method that is used to calculate the
previous four terms $NA6,$ $NB6,$ $NC6,$ and $ND6$ on the $RH$ side of
(4.138) in the subsection (B) one may calculate the current four terms $SA6,$
$SB6,$ $SC6,$ and $SD6$ by using these four modified equations of (4.139),
(4.140), (4.141), and (4.142), respectively. It can turn out that all the
current four terms $SA6,$ $SB6,$ $SC6,$ and $SD6$ have the upper bounds
consisting of the basic norms $\{NBAS1\}$ and $\{NBAS2\}$. Then the
inequality (4.157b) really means that the first norm on the $RH$ side of
(4.157) has an upper bound consisting of the basic norms $\{NBAS1\}$ and $%
\{NBAS2\}$. Now all these six norms on the $RH$ side of (4.157) are shown to
have the upper bounds consisting of the basic norms $\{NBAS1\}$ and/or $%
\{NBAS2\}$. This indicates that the last norm on the $RH$ side of (4.105)
has an upper bound consisting of the basic norms $\{NBAS1\}$ and $\{NBAS2\}$.

As a summary, it proves in this subsection that all the six norms $%
\{||F_{5-k}^{\lambda }(x,t_{3})$ $\times p^{k}\Psi
_{0}^{S}(x,r,t_{0}+t_{1}/2)||\}$ for $0\leq k\leq 5$ on the $RH$ side of
(4.105) with $\mu =S$ have the upper bounds consisting of a finite number of
the basic norms $\{NBAS1\}$ and/or $\{NBAS2\}$. Then the inequality (4.105)
shows that the norm $||Q_{\lambda }(x,p,t_{3})\Psi
_{0}^{S}(x,r,t_{0}+t_{1}/2)||$ has an upper bound consisting of a finite
number of the basic norms $\{NBAS1\}$ and $\{NBAS2\}$. Finally, the two
inequalities (4.104) indicate that both the norms $NORM(1,\lambda ,\mu )$
and $NORM(2,\lambda ,\mu )$ with $\mu =S$ have the upper bounds consisting
of a finite number of the basic norms $\{NBAS1\}$ and $\{NBAS2\}$. This is
the desired result. \newline
\newline
{\large 4.3.2.2 The upper bounds for the norms }$NORM(k,\lambda ,C0)$

The product state $\Psi _{0}^{C0}(x,r,t_{0}+t_{1}/2)$ does not contain any
spatially-selective function. Both the norms $NORM(1,\lambda ,\mu )$ of
(4.102) and $NORM(2,\lambda ,\mu )$ of (4.103) with the label $\mu =C0$ are
controlled by the spatially-selective factors $\delta (x-x_{c},\varepsilon
_{c})$ and $\Theta (x-x_{L},\varepsilon )$ inside the two norms,
respectively. In order to obtain correctly the upper bounds of the two norms 
$NORM(1,\lambda ,C0)$ and $NORM(2,\lambda ,C0)$ one can not neglect the two
spatially-selective factors inside the two norms. This is different from the
two previous cases with $\mu =CS$ and $S$. By substituting $Q_{\lambda
}(x,p,t_{3})$ of (4.100) into (4.102) and (4.103), respectively, one obtains%
\begin{equation*}
NORM(1,\lambda ,C0)\leq \sum_{k=0}^{5}||\delta (x-x_{c},\varepsilon
_{c})\exp [-\frac{i}{2\hslash }H_{0}^{ho}t_{3}]
\end{equation*}%
\begin{equation}
\times F_{5-k}^{\lambda }(x,t_{3})p^{k}\Psi _{0}^{C0}(x,r,t_{0}+t_{1}/2)|| 
\tag{4.158}
\end{equation}%
and%
\begin{equation*}
NORM(2,\lambda ,C0)\leq \sum_{k=0}^{5}||\Theta (x-x_{L},\varepsilon )\exp [-%
\frac{i}{2\hslash }H_{0}^{ho}t_{3}]
\end{equation*}%
\begin{equation}
\times F_{5-k}^{\lambda }(x,t_{3})p^{k}\Psi _{0}^{C0}(x,r,t_{0}+t_{1}/2)||. 
\tag{4.159}
\end{equation}%
The upper bounds of the two norms $NORM(1,\lambda ,C0)$ and $NORM(2,\lambda
,C0)$ may be calculated rigorously on the basis of the two inequalities
(4.158) and (4.159), respectively. Here one may use the formula (4.108a) to
calculate exactly the product state $p^{k}\Psi _{0}^{C0}(x,r,t_{0}+t_{1}/2)$
for $0\leq k\leq 5.$ But it may be more convenient to use the following
method to calculate the product state. Now define the two $GWP$ motional
states by%
\begin{equation*}
\Psi _{0}^{g}(x,+\varphi ,t_{0}+t_{1}/2)=\exp [i\varphi (x,\gamma )]\Psi
_{0}^{g}(x,t_{0}+t_{1}/2)
\end{equation*}%
and 
\begin{equation*}
\Psi _{0}^{e}(x,-\varphi ,t_{0}+t_{1}/2)=\exp [-i\varphi (x,\gamma )]\Psi
_{0}^{e}(x,t_{0}+t_{1}/2).
\end{equation*}%
Then define the Gaussian superposition state $\Psi _{0}(x,r,\varphi
,t_{0}+t_{1}/2)$ as%
\begin{equation}
\Psi _{0}(x,r,\varphi ,t_{0}+t_{1}/2)=\Psi _{0}^{g}(x,+\varphi
,t_{0}+t_{1}/2)|e\rangle +\Psi _{0}^{e}(x,-\varphi
,t_{0}+t_{1}/2)|g_{0}\rangle .  \tag{4.160}
\end{equation}%
By using this superposition state the product state $\Psi
_{0}^{C0}(x,r,t_{0}+t_{1}/2)$ of (4.89a) may be simply re-expressed as%
\begin{equation*}
\Psi _{0}^{C0}(x,r,t_{0}+t_{1}/2)=\cos [\frac{1}{2\hslash }\Omega
(x)t_{1}]\Psi _{0}(x,r,t_{0}+t_{1}/2)
\end{equation*}%
\begin{equation}
-i\sin [\frac{1}{2\hslash }\Omega (x)t_{1}]\Psi _{0}(x,r,\varphi
,t_{0}+t_{1}/2).  \tag{4.161}
\end{equation}%
Here the product state $\Psi _{0}(x,r,t_{0}+t_{1}/2)$ is given by (4.42).
Now the $m-$order coordinate derivative of the product state $\Psi
_{0}^{C0}(x,r,t_{0}+t_{1}/2)$ may be expressed as%
\begin{equation*}
\frac{\partial ^{m}}{\partial x^{m}}\Psi
_{0}^{C0}(x,r,t_{0}+t_{1}/2)=\sum_{l=0}^{m}\left( 
\begin{array}{c}
m \\ 
l%
\end{array}%
\right) \{\frac{\partial ^{l}}{\partial x^{l}}\cos [\frac{1}{2\hslash }%
\Omega (x)t_{1}]\}
\end{equation*}%
\begin{equation*}
\times \{\frac{\partial ^{m-l}}{\partial x^{m-l}}\Psi
_{0}(x,r,t_{0}+t_{1}/2)\}
\end{equation*}%
\begin{equation}
-i\sum_{l=0}^{m}\left( 
\begin{array}{c}
m \\ 
l%
\end{array}%
\right) \{\frac{\partial ^{l}}{\partial x^{l}}\sin [\frac{1}{2\hslash }%
\Omega (x)t_{1}]\}\{\frac{\partial ^{m-l}}{\partial x^{m-l}}\Psi
_{0}(x,r,\varphi ,t_{0}+t_{1}/2)\}.  \tag{4.162}
\end{equation}%
Furthermore, it can turn out that the $l-$order coordinate derivatives $%
(0\leq l\leq 5)$ for the two trigonometric functions $\cos [\frac{1}{%
2\hslash }\Omega (x)t_{1}]$ and $\sin [\frac{1}{2\hslash }\Omega (x)t_{1}]$
may be formally written as%
\begin{equation}
\frac{\partial ^{l}}{\partial x^{l}}\cos [\frac{1}{2\hslash }\Omega
(x)t_{1}]=P_{C}^{l}(C_{X},S_{X})\cos [\frac{1}{2\hslash }\Omega
(x)t_{1}]+P_{S}^{l}(C_{X},S_{X})\sin [\frac{1}{2\hslash }\Omega (x)t_{1}], 
\tag{4.163a}
\end{equation}%
\begin{equation}
\frac{\partial ^{l}}{\partial x^{l}}\sin [\frac{1}{2\hslash }\Omega
(x)t_{1}]=Q_{C}^{l}(C_{X},S_{X})\cos [\frac{1}{2\hslash }\Omega
(x)t_{1}]+Q_{S}^{l}(C_{X},S_{X})\sin [\frac{1}{2\hslash }\Omega (x)t_{1}], 
\tag{4.163b}
\end{equation}%
where $P_{C}^{l}(C_{X},S_{X}),$ $P_{S}^{l}(C_{X},S_{X}),$ $%
Q_{C}^{l}(C_{X},S_{X}),$ and $Q_{S}^{l}(C_{X},S_{X})$ are the polynomials in
variables $C_{X}$ and $S_{X},$ and the trigonometric functions $C_{X}$ and $%
S_{X}$ are defined by%
\begin{equation*}
C_{X}=\cos [\frac{1}{2}\Delta kx-\pi /4],\text{ }S_{X}=\sin [\frac{1}{2}%
\Delta kx-\pi /4].
\end{equation*}%
In the previous subsection 4.3.2.1 it is shown that both the $l-$order
derivatives $\frac{\partial ^{l}}{\partial x^{l}}\cos [\frac{1}{2\hslash }%
\Omega (x)t_{1}]$ and $\frac{\partial ^{l}}{\partial x^{l}}\sin [\frac{1}{%
2\hslash }\Omega (x)t_{1}]$ ($l\geq 0$) are bounded. This means that the
four polynomials $P_{C}^{l}(C_{X},S_{X}),$ $P_{S}^{l}(C_{X},S_{X}),$ $%
Q_{C}^{l}(C_{X},S_{X}),$ and $Q_{S}^{l}(C_{X},S_{X})$ are bounded on the $RH$
sides of the two equations (4.163). Actually, these derivatives $\frac{%
\partial ^{l}}{\partial x^{l}}\cos [\frac{1}{2\hslash }\Omega (x)t_{1}]$ and 
$\frac{\partial ^{l}}{\partial x^{l}}\sin [\frac{1}{2\hslash }\Omega
(x)t_{1}]$ for $0\leq l\leq 5$ may be calculated directly. It can be found
that these four polynomials indeed are bounded and both the derivatives $%
\frac{\partial ^{l}}{\partial x^{l}}\cos [\frac{1}{2\hslash }\Omega
(x)t_{1}] $ and $\frac{\partial ^{l}}{\partial x^{l}}\sin [\frac{1}{2\hslash 
}\Omega (x)t_{1}]$ are also bounded, 
\begin{equation*}
|\frac{\partial ^{l}}{\partial x^{l}}\cos [\frac{1}{2\hslash }\Omega
(x)t_{1}]|\leq |P_{C}^{l}(C_{X},S_{X})|+|P_{S}^{l}(C_{X},S_{X})|,
\end{equation*}%
\begin{equation*}
|\frac{\partial ^{l}}{\partial x^{l}}\sin [\frac{1}{2\hslash }\Omega
(x)t_{1}]|\leq |Q_{C}^{l}(C_{X},S_{X})|+|Q_{S}^{l}(C_{X},S_{X})|.
\end{equation*}%
The two inequalities further show that both the norms $||\frac{\partial ^{l}%
}{\partial x^{l}}|\tilde{g}_{0}\rangle ||$ and $||\frac{\partial ^{l}}{%
\partial x^{l}}|\tilde{e}\rangle ||$ are indeed bounded for $0\leq l\leq 5$
in the two inequalities (4.118). This is in agreement with the results
obtained in the previous subsection 4.3.2.1.

There are two methods to calculate strictly the upper bounds of the norms $%
NORM(1,\lambda ,C0)$ and $NORM(2,\lambda ,C0).$ One of which is first to set
up the commutative relation $[\exp (-\frac{[x-x_{c}]^{2}}{\varepsilon
_{c}^{2}}),$ $\exp (-\frac{i}{2\hslash }H_{0}^{ho}t_{3})].$ An
exponentially-decaying factor is therefore generated. This
exponentially-decaying factor then is used to directly control the upper
bounds of the norms $NORM(1,\lambda ,C0)$ and $NORM(2,$ $\lambda ,C0).$ The
detailed work for this method will not be described here. Another method is
based on the multiple Gaussian wave-packet ($MGWP$) expansion. Here this
method is named the $MGWP$ expansion method. The $MGWP$ expansion method has
an extensive application in the research field of quantum computational
chemistry [25]. The detail for the $MGWP$ expansion method also may be seen
in the next section 5. Simply speaking, the $MGWP$ expansion is that a
quantum state may be approximated by a linear combination of many different
Gaussian wave-packet (basis) states. The key step to calculating strictly
the two norms of (4.158) and (4.159) is how to calculate strictly the
following time evolution process under the harmonic-oscillator propagator $%
\exp [-\frac{i}{2\hslash }H_{0}^{ho}t_{3}]$: 
\begin{equation*}
\Psi _{0k}^{C0}(x,r,t_{0}+t_{1}/2+t_{3}/2)=\exp [-\frac{i}{2\hslash }%
H_{0}^{ho}t_{3}]
\end{equation*}%
\begin{equation}
\times \{F_{5-k}^{\lambda }(x,t_{3})p^{k}\Psi _{0}^{C0}(x,r,t_{0}+t_{1}/2)\}.
\tag{4.164}
\end{equation}%
It is well known that it is easy to calculate exactly the time evolution
process of a harmonic oscillator with a $GWP$ state or more generally with a
Gaussian superposition state. However, it can be found from (4.161) and
(4.162) that the product states $\{p^{k}\Psi _{0}^{C0}\}$ are not pure
Gaussian superposition states and hence the product states $%
\{F_{5-k}^{\lambda }p^{k}\Psi _{0}^{C0}\}$ in (4.164) are not yet pure
Gaussian superposition states. This is the reason why it is difficult to
calculate exactly the time evolution process of (4.164). Here one may use
the $MGWP$ expansion method to calculate approximately the time evolution
process. In the $MGWP$ expansion method the product states $\{p^{k}\Psi
_{0}^{C0}\}$ may be approximately expanded as the Gaussian superposition
states, respectively. Here the truncation errors in the $MGWP$ expansion can
be controlled, as can be seen below. Here the amplitudes in the Gaussian
superposition states allow to be finite-order polynomials in coordinate $x$.
Note that in a conventional Gaussian superposition state the amplitudes are
generally complex coefficients. These Gaussian superposition states with the
amplitudes of finite-order polynomials in coordinate $x$ are the generalized
Gaussian superposition states. Just like a conventional Gaussian
superposition state, such a generalized Gaussian superposition state still
allows the time evolution process of (4.164) to be calculated exactly, as
can be shown below and in the next section 5. The key point for this is that
the unitary inverse of the harmonic-oscillator propagator $\exp [-\frac{i}{%
2\hslash }H_{0}^{ho}t_{3}]$ in (4.164) can be obtained exactly [15] and the
coordinate operator $x(t)$ in the Heisenberg picture also can be obtained
exactly. It is another key point to the present $MGWP$ expansion method.

It can be found from (4.161) and (4.162) that the two trigonometric
functions $\cos [\frac{1}{2\hslash }\Omega (x)t_{1}]$ and $\sin [\frac{1}{%
2\hslash }\Omega (x)t_{1}]$ directly result in that the exact calculation
for the time evolution process of (4.164) becomes difficult. Now according
to the $MGWP$ expansion one may first expand the two trigonometric functions 
$\cos [\frac{1}{2\hslash }\Omega (x)t_{1}]$ and $\sin [\frac{1}{2\hslash }%
\Omega (x)t_{1}]$ in (4.161) and (4.162) as the Taylor series:%
\begin{equation}
\cos [\frac{1}{2\hslash }\Omega (x)t_{1}]=\sum_{k=0}^{n}\frac{(-1)^{k}}{(2k)!%
}[\frac{1}{2\hslash }\Omega (x)t_{1}]^{2k}+R_{2n+1}[\frac{1}{2\hslash }%
\Omega (x)t_{1}],  \tag{4.165a}
\end{equation}%
\begin{equation}
\sin [\frac{1}{2\hslash }\Omega (x)t_{1}]=\sum_{k=0}^{n}\frac{(-1)^{k}}{%
(2k+1)!}[\frac{1}{2\hslash }\Omega (x)t_{1}]^{2k+1}+S_{2n+2}[\frac{1}{%
2\hslash }\Omega (x)t_{1}],  \tag{4.165b}
\end{equation}%
where the residual errors $R_{2n+1}[\frac{1}{2\hslash }\Omega (x)t_{1}]$ and 
$S_{2n+2}[\frac{1}{2\hslash }\Omega (x)t_{1}]$ are bounded by%
\begin{equation}
|R_{2n+1}[\frac{1}{2\hslash }\Omega (x)t_{1}]|\leq \frac{1}{(2n+1)!}\{\frac{1%
}{2\hslash }|\Omega (x)|t_{1}\}^{2n+1}\leq \frac{(2|\Omega _{0}|t_{1})^{2n+1}%
}{(2n+1)!},  \tag{4.165c}
\end{equation}%
\begin{equation}
|S_{2n+2}[\frac{1}{2\hslash }\Omega (x)t_{1}]|\leq \frac{1}{(2n+2)!}\{\frac{1%
}{2\hslash }\Omega (x)t_{1}\}^{2n+2}\leq \frac{(2\Omega _{0}t_{1})^{2n+2}}{%
(2n+2)!}.  \tag{4.165d}
\end{equation}%
The two residual error terms can be controlled by the truncation term number 
$n$. According to the Stirling formula $n!\thickapprox \sqrt{2\pi n}%
(n/e)^{n} $ one can find that the two residual error terms have the upper
bounds: $|R_{2n+1}[\frac{1}{2\hslash }\Omega (x)t_{1}]|_{\max }\propto (%
\frac{2e|\Omega _{0}|t_{1}}{2n+1})^{2n+1}$ and $|S_{2n+2}[\frac{1}{2\hslash }%
\Omega (x)t_{1}]|_{\max }\propto (\frac{2e|\Omega _{0}|t_{1}}{2n+2})^{2n+2}.$
Therefore, the two error terms decay exponentially with the numbers $2n+1$
and $2n+2,$ respectively, when $2n+1>2e|\Omega _{0}|t_{1}.$ Both the upper
bounds on the rightest sides of (4.165c) and (4.165d) are used to determine
the truncation term numbers $n$ on the $RH$ sides of (4.165a) and (4.165b),
respectively. By substituting the two equations (4.163) into (4.162) and
then using the two expansions (4.165a) and (4.165b) one obtains%
\begin{equation}
\frac{\partial ^{m}}{\partial x^{m}}\Psi
_{0}^{C0}(x,r,t_{0}+t_{1}/2)=D_{m}(x,r,t_{0}+t_{1}/2)+E_{rm}(x,r,t_{0}+t_{1}/2),
\tag{4.166}
\end{equation}%
where the main term $D_{m}(x,r,t_{0}+t_{1}/2)$ is written as%
\begin{equation*}
D_{m}(x,r,t_{0}+t_{1}/2)=\sum_{k=0}^{n}\frac{(-1)^{k}}{(2k)!}[\frac{1}{%
2\hslash }\Omega (x)t_{1}]^{2k}
\end{equation*}%
\begin{equation*}
\times \sum_{l=0}^{m}\left( 
\begin{array}{c}
m \\ 
l%
\end{array}%
\right) P_{C}^{l}(C_{X},S_{X})\frac{\partial ^{m-l}}{\partial x^{m-l}}\Psi
_{0}(x,r,t_{0}+t_{1}/2)
\end{equation*}%
\begin{equation*}
+\sum_{k=0}^{n}\frac{(-1)^{k}}{(2k+1)!}[\frac{1}{2\hslash }\Omega
(x)t_{1}]^{2k+1}\sum_{l=0}^{m}\left( 
\begin{array}{c}
m \\ 
l%
\end{array}%
\right) P_{S}^{l}(C_{X},S_{X})\frac{\partial ^{m-l}}{\partial x^{m-l}}\Psi
_{0}(x,r,t_{0}+t_{1}/2)
\end{equation*}%
\begin{equation*}
-i\sum_{k=0}^{n}\frac{(-1)^{k}}{(2k)!}[\frac{1}{2\hslash }\Omega
(x)t_{1}]^{2k}\sum_{l=0}^{m}\left( 
\begin{array}{c}
m \\ 
l%
\end{array}%
\right) Q_{C}^{l}(C_{X},S_{X})\frac{\partial ^{m-l}}{\partial x^{m-l}}\Psi
_{0}(x,r,\varphi ,t_{0}+t_{1}/2)
\end{equation*}%
\begin{equation*}
-i\sum_{k=0}^{n}\frac{(-1)^{k}}{(2k+1)!}[\frac{1}{2\hslash }\Omega
(x)t_{1}]^{2k+1}
\end{equation*}%
\begin{equation}
\times \sum_{l=0}^{m}\left( 
\begin{array}{c}
m \\ 
l%
\end{array}%
\right) Q_{S}^{l}(C_{X},S_{X})\frac{\partial ^{m-l}}{\partial x^{m-l}}\Psi
_{0}(x,r,\varphi ,t_{0}+t_{1}/2)  \tag{4.167}
\end{equation}%
and the residual error term $E_{rm}(x,r,t_{0}+t_{1}/2)$ is given by%
\begin{equation*}
E_{rm}(x,r,t_{0}+t_{1}/2)=R_{2n+1}[\frac{1}{2\hslash }\Omega (x)t_{1}]
\end{equation*}%
\begin{equation*}
\times \sum_{l=0}^{m}\left( 
\begin{array}{c}
m \\ 
l%
\end{array}%
\right) P_{C}^{l}(C_{X},S_{X})\frac{\partial ^{m-l}}{\partial x^{m-l}}\Psi
_{0}(x,r,t_{0}+t_{1}/2)
\end{equation*}%
\begin{equation*}
+S_{2n+2}[\frac{1}{2\hslash }\Omega (x)t_{1}]\sum_{l=0}^{m}\left( 
\begin{array}{c}
m \\ 
l%
\end{array}%
\right) P_{S}^{l}(C_{X},S_{X})\frac{\partial ^{m-l}}{\partial x^{m-l}}\Psi
_{0}(x,r,t_{0}+t_{1}/2)
\end{equation*}%
\begin{equation*}
-iR_{2n+1}[\frac{1}{2\hslash }\Omega (x)t_{1}]\sum_{l=0}^{m}\left( 
\begin{array}{c}
m \\ 
l%
\end{array}%
\right) Q_{C}^{l}(C_{X},S_{X})\frac{\partial ^{m-l}}{\partial x^{m-l}}\Psi
_{0}(x,r,\varphi ,t_{0}+t_{1}/2)
\end{equation*}%
\begin{equation}
-iS_{2n+2}[\frac{1}{2\hslash }\Omega (x)t_{1}]\sum_{l=0}^{m}\left( 
\begin{array}{c}
m \\ 
l%
\end{array}%
\right) Q_{S}^{l}(C_{X},S_{X})\frac{\partial ^{m-l}}{\partial x^{m-l}}\Psi
_{0}(x,r,\varphi ,t_{0}+t_{1}/2).  \tag{4.168}
\end{equation}%
Now by substituting (4.166) into (4.158) and (4.159), respectively, one
obtains%
\begin{equation*}
NORM(1,\lambda ,C0)\leq \frac{1}{\varepsilon _{c}\sqrt{\pi }}%
\sum_{m=0}^{5}\hslash ^{m}||F_{5-m}^{\lambda
}(x,t_{3})E_{rm}(x,r,t_{0}+t_{1}/2)||
\end{equation*}%
\begin{equation}
+\sum_{m=0}^{5}\hslash ^{m}||\delta (x-x_{c},\varepsilon _{c})\exp [-\frac{i%
}{2\hslash }H_{0}^{ho}t_{3}]F_{5-m}^{\lambda
}(x,t_{3})D_{m}(x,r,t_{0}+t_{1}/2)||,  \tag{4.169a}
\end{equation}%
and 
\begin{equation*}
NORM(2,\lambda ,C0)\leq \sum_{m=0}^{5}\hslash ^{m}||F_{5-m}^{\lambda
}(x,t_{3})E_{rm}(x,r,t_{0}+t_{1}/2)||
\end{equation*}%
\begin{equation}
+\sum_{m=0}^{5}\hslash ^{m}||\Theta (x-x_{L},\varepsilon )\exp [-\frac{i}{%
2\hslash }H_{0}^{ho}t_{3}]F_{5-m}^{\lambda
}(x,t_{3})D_{m}(x,r,t_{0}+t_{1}/2)||.  \tag{4.169b}
\end{equation}%
Here the relations $0\leq \delta (x-x_{c},\varepsilon _{c})\leq \frac{1}{%
\varepsilon _{c}\sqrt{\pi }}$ and $0\leq \Theta (x-x_{L},\varepsilon )\leq 1$
are already used when there appears the residual term $%
E_{rm}(x,r,t_{0}+t_{1}/2)$ inside the norms on the $RH$ sides of the two
inequalities (4.169). This means that those norms inside containing the
residual term $E_{rm}(x,r,t_{0}+t_{1}/2)$ are controlled by the residual
terms themselves on the $RH$ sides of the two inequalities (4.169),
respectively.

Below evaluate first the contribution of the residual error terms $%
\{E_{rm}(x,r,t_{0}+t_{1}/2)\}$ of (4.168) to the norm $NORM(1,\lambda ,C0)$
in (4.169a) and $NORM(2,\lambda ,C0)$ in (4.169b). With the help of the
residual error term $E_{rm}(x,r,t_{0}+t_{1}/2)$ of (4.168) and the upper
bounds of the residual terms $R_{2n+1}[\frac{1}{2\hslash }\Omega (x)t_{1}]$
and $S_{2n+2}[\frac{1}{2\hslash }\Omega (x)t_{1}],$ which are determined
from (4.165c) and (4.165d), respectively, it is easy to find that the first
sum term of the norms inside containing the residual error term $%
E_{rm}(x,r,t_{0}+t_{1}/2)$ on the $RH$ sides of (4.169a) and (4.169b)
satisfies%
\begin{equation*}
\sum_{m=0}^{5}\hslash ^{m}||F_{5-m}^{\lambda
}(x,t_{3})E_{rm}(x,r,t_{0}+t_{1}/2)||
\end{equation*}%
\begin{equation*}
\leq \frac{(2|\Omega _{0}|t_{1})^{2n+1}}{(2n+1)!}\sum_{m=0}^{5}%
\sum_{l=0}^{m}\left( 
\begin{array}{c}
m \\ 
l%
\end{array}%
\right) \hslash ^{m}|P_{C}^{l}(C_{X},S_{X})|_{\max }\times N_{ml}^{\lambda
}(0)
\end{equation*}%
\begin{equation*}
+\frac{(2\Omega _{0}t_{1})^{2n+2}}{(2n+2)!}\sum_{m=0}^{5}\sum_{l=0}^{m}%
\left( 
\begin{array}{c}
m \\ 
l%
\end{array}%
\right) \hslash ^{m}|P_{S}^{l}(C_{X},S_{X})|_{\max }\times N_{ml}^{\lambda
}(0)
\end{equation*}%
\begin{equation*}
+\frac{(2|\Omega _{0}|t_{1})^{2n+1}}{(2n+1)!}\sum_{m=0}^{5}\sum_{l=0}^{m}%
\left( 
\begin{array}{c}
m \\ 
l%
\end{array}%
\right) \hslash ^{m}|Q_{C}^{l}(C_{X},S_{X})|_{\max }\times N_{ml}^{\lambda
}(\varphi )
\end{equation*}%
\begin{equation}
+\frac{(2\Omega _{0}t_{1})^{2n+2}}{(2n+2)!}\sum_{m=0}^{5}\sum_{l=0}^{m}%
\left( 
\begin{array}{c}
m \\ 
l%
\end{array}%
\right) \hslash ^{m}|Q_{S}^{l}(C_{X},S_{X})|_{\max }\times N_{ml}^{\lambda
}(\varphi )  \tag{4.170}
\end{equation}%
where the norms $N_{ml}^{\lambda }(0)$ and $N_{ml}^{\lambda }(\varphi )$ are
defined by%
\begin{equation*}
N_{ml}^{\lambda }(0)=||F_{5-m}^{\lambda }(x,t_{3})[\frac{\partial ^{m-l}}{%
\partial x^{m-l}}\Psi _{0}(x,r,t_{0}+t_{1}/2)]||,
\end{equation*}%
\begin{equation*}
N_{ml}^{\lambda }(\varphi )=||F_{5-m}^{\lambda }(x,t_{3})[\frac{\partial
^{m-l}}{\partial x^{m-l}}\Psi _{0}(x,r,\varphi ,t_{0}+t_{1}/2)]||.
\end{equation*}%
Here one needs to prove that these norms $N_{ml}^{\lambda }(0)$ and $%
N_{ml}^{\lambda }(\varphi )$ are bounded. It is known above that all the
four polynomials $P_{C}^{l}(C_{X},S_{X}),$ $P_{S}^{l}(C_{X},S_{X}),$ $%
Q_{C}^{l}(C_{X},S_{X}),$ and $Q_{S}^{l}(C_{X},S_{X})$ are bounded. If now
these norms $N_{ml}^{\lambda }(0)$ and $N_{ml}^{\lambda }(\varphi )$ are
also bounded, then it follows from (4.170) that the upper bound of the norm
on the $LH$ side of (4.170) is controlled by the truncation-error upper
bounds $\frac{(2|\Omega _{0}|t_{1})^{2n+1}}{(2n+1)!}$ and $\frac{(2\Omega
_{0}t_{1})^{2n+2}}{(2n+2)!},$ which are the exponentially-decaying factors
when $2n+1>2e|\Omega _{0}|t_{1}.$ It is easy to calculate strictly the two
norms $N_{ml}^{\lambda }(0)$ and $N_{ml}^{\lambda }(\varphi )$. From the
formulae (4.109), (4.42), and (4.160) one obtains%
\begin{equation*}
\frac{\partial ^{m-l}}{\partial x^{m-l}}\Psi
_{0}(x,r,t_{0}+t_{1}/2)=Q_{m-l}^{g}(x)\Psi
_{0}^{g}(x,t_{0}+t_{1}/2)|g_{0}\rangle
\end{equation*}%
\begin{equation}
+Q_{m-l}^{e}(x)\Psi _{0}^{e}(x,t_{0}+t_{1}/2)|e\rangle  \tag{4.171a}
\end{equation}%
and 
\begin{equation*}
\frac{\partial ^{m-l}}{\partial x^{m-l}}\Psi _{0}(x,r,\varphi
,t_{0}+t_{1}/2)=Q_{m-l}^{g}(x,+\varphi (x,\gamma ))\Psi _{0}^{g}(x,+\varphi
,t_{0}+t_{1}/2)|e\rangle
\end{equation*}%
\begin{equation}
+Q_{m-l}^{e}(x,-\varphi (x,\gamma ))\Psi _{0}^{e}(x,-\varphi
,t_{0}+t_{1}/2)|g_{0}\rangle .  \tag{4.171b}
\end{equation}%
With the help of the two equations (4.171) it turns out that the two norms $%
N_{ml}^{\lambda }(0)$ and $N_{ml}^{\lambda }(\varphi )$ satisfy%
\begin{equation*}
N_{ml}^{\lambda }(0)\leq ||F_{5-m}^{\lambda }(x,t_{3})Q_{m-l}^{g}(x)\Psi
_{0}^{g}(x,t_{0}+t_{1}/2)||
\end{equation*}%
\begin{equation}
+||F_{5-m}^{\lambda }(x,t_{3})Q_{m-l}^{e}(x)\Psi _{0}^{e}(x,t_{0}+t_{1}/2)||
\tag{4.172a}
\end{equation}%
and%
\begin{equation*}
N_{ml}^{\lambda }(\varphi )\leq ||F_{5-m}^{\lambda
}(x,t_{3})Q_{m-l}^{g}(x,+\varphi (x,\gamma ))\Psi _{0}^{g}(x,+\varphi
,t_{0}+t_{1}/2)||
\end{equation*}%
\begin{equation}
+||F_{5-m}^{\lambda }(x,t_{3})Q_{m-l}^{e}(x,-\varphi (x,\gamma ))\Psi
_{0}^{e}(x,-\varphi ,t_{0}+t_{1}/2)||.  \tag{4.172b}
\end{equation}%
Here $F_{5-m}^{\lambda }$ is a polynomial in coordinate $x$ with order not
more than five$,$ while $Q_{m-l}^{g}(x),$ $Q_{m-l}^{e}(x),$ $%
Q_{m-l}^{g}(x,+\varphi (x,\gamma )),$ and $Q_{m-l}^{e}(x,-\varphi (x,\gamma
))$ all are the polynomials in coordinate $x$ with order not more than five$%
. $ Then in the two inequalities (4.172) $F_{5-m}^{\lambda }Q_{m-l}^{g}(x),$ 
$F_{5-m}^{\lambda }Q_{m-l}^{e}(x),$ $F_{5-m}^{\lambda
}Q_{m-l}^{g}(x,+\varphi (x,\gamma )),$ and $F_{5-m}^{\lambda
}Q_{m-l}^{e}(x,-\varphi (x,\gamma ))$ are\ the polynomials in coordinate $x$
with order not more than five. Notice that $\Psi _{0}^{g}(x,t_{0}+t_{1}/2),$ 
$\Psi _{0}^{e}(x,t_{0}+t_{1}/2),$ $\Psi _{0}^{g}(x,+\varphi ,t_{0}+t_{1}/2),$
and $\Psi _{0}^{e}(x,-\varphi ,t_{0}+t_{1}/2)$ all are $GWP$ motional
states. Then with the help of the Gaussian integrals [44a] it can turn out
that each one of the four norms on the $RH$ sides of the two inequalities
(4.172) is bounded. Thus, the two\ inequalities (4.172) indicate that both
the norms $N_{ml}^{\lambda }(0)$ and $N_{ml}^{\lambda }(\varphi )$ are
bounded. On the other hand, it is known from the two equations (4.163) that $%
P_{C}^{l}(C_{X},S_{X}),$ $P_{S}^{l}(C_{X},S_{X}),$ $Q_{C}^{l}(C_{X},S_{X}),$
and $Q_{S}^{l}(C_{X},S_{X})$ are the polynomials in the trigonometric
functions $C_{X}$ and $S_{X}$ with a\ total order not more than $l$ and here 
$0\leq l\leq 5.$ It is already shown above that all these four polynomials
are bounded. These results together show that all the four sum terms $%
(\sum_{m=0}^{5}\sum_{l=0}^{m})$ on the $RH$ side of (4.170) are bounded.
Therefore, the inequality (4.170) shows that the upper bound of the sum term
on the $LH$ side of (4.170) is proportional to the maximum one of the two
exponentially-decaying factors $\frac{(2|\Omega _{0}|t_{1})^{2n+1}}{(2n+1)!}$
$(\thickapprox (\frac{2e|\Omega _{0}|t_{1}}{2n+1})^{2n+1})$ and $\frac{%
(2\Omega _{0}t_{1})^{2n+2}}{(2n+2)!}$ $(\thickapprox (\frac{2e|\Omega
_{0}|t_{1}}{2n+2})^{2n+2})$ with $2n+1>2e|\Omega _{0}|t_{1}.$ This indicates
that the contribution of the residual error terms $%
\{E_{rm}(x,r,t_{0}+t_{1}/2)\}$ to the norm $NORM(1,\lambda ,C0)$ in (4.169a)
and $NORM(2,\lambda ,C0)$ in (4.169b) can be completely controlled by the
truncation term number $n$. It can be neglected when $2n+1>>2e|\Omega
_{0}|t_{1}.$

Now evaluate the contribution of the main terms $\{D_{m}(x,r,t_{0}+t_{1}/2)%
\} $ of (4.167) to the norm $NORM(1,\lambda ,C0)$ in (4.169a) and $%
NORM(2,\lambda ,C0)$ in (4.169b). It is needed to calculate strictly the
second sum terms on the $RH$ sides of (4.169a) and (4.169b), respectively.
The second sum term of $NORM(1,\lambda ,C0)$ on the $RH$ side of (4.169a)
consists of six norms, each one of which inside contains the
spatially-selective factor $\delta (x-x_{c},\varepsilon _{c}),$ while that
one of $NORM(2,\lambda ,C0)$ on the $RH$ side of (4.169b) also consists of
six norms, each one of which inside contains the spatially-selective factor $%
\Theta (x-x_{L},\varepsilon ).$ These six norms of $NORM(1,\lambda ,C0)$ can
be converted into those six norms of $NORM(2,\lambda ,C0)$, respectively, if
the spatially-selective factor $\delta (x-x_{c},\varepsilon _{c})$ is
replaced with $\Theta (x-x_{L},\varepsilon ).$ Thus, here one needs only to
calculate strictly the second sum term of $NORM(1,\lambda ,C0)$. The
obtained result is available as well for that one of $NORM(2,\lambda ,C0)$
as long as the spatially-selective factor $\delta (x-x_{c},\varepsilon _{c})$
is replaced with $\Theta (x-x_{L},\varepsilon ).$ With the aid of $%
D_{m}(x,r,t_{0}+t_{1}/2)$ of (4.167) it can turn out that the second sum
term on the $RH$ side of (4.169a) is bounded by%
\begin{equation*}
\sum_{m=0}^{5}\hslash ^{m}||\delta (x-x_{c},\varepsilon _{c})\exp [-\frac{i}{%
2\hslash }H_{0}^{ho}t_{3}]F_{5-m}^{\lambda
}(x,t_{3})D_{m}(x,r,t_{0}+t_{1}/2)||
\end{equation*}%
\begin{equation}
\leq NA7+NB7+NC7+ND7  \tag{4.173}
\end{equation}%
where the four terms $NA7,$ $NB7,$ $NC7,$ and $ND7$ all are non-negative and
they are given by%
\begin{equation*}
NA7=\sum_{a=g,e}\sum_{k=0}^{n}\sum_{m=0}^{5}\sum_{l=0}^{m}\frac{\hslash
^{m}(\Omega _{0}t_{1})^{2k}}{(2k)!}\left( 
\begin{array}{c}
m \\ 
l%
\end{array}%
\right) ||\delta (x-x_{c},\varepsilon _{c})F_{5-m}^{\lambda
}(x(t_{3}^{\prime }),t_{3})
\end{equation*}%
\begin{equation}
\times Q_{m-l}^{a}(x(t_{3}^{\prime }))\exp [-\frac{i}{2\hslash }%
H_{0}^{ho}t_{3}](2C_{X})^{2k}P_{C}^{l}(C_{X},S_{X})\Psi
_{0}^{a}(x,t_{0}+t_{1}/2)||,  \tag{4.174a}
\end{equation}%
\begin{equation*}
NB7=\sum_{a=g,e}\sum_{k=0}^{n}\sum_{m=0}^{5}\sum_{l=0}^{m}\frac{\hslash
^{m}(|\Omega _{0}|t_{1})^{2k+1}}{(2k+1)!}\left( 
\begin{array}{c}
m \\ 
l%
\end{array}%
\right) ||\delta (x-x_{c},\varepsilon _{c})F_{5-m}^{\lambda
}(x(t_{3}^{\prime }),t_{3})
\end{equation*}%
\begin{equation}
\times Q_{m-l}^{a}(x(t_{3}^{\prime }))\exp [-\frac{i}{2\hslash }%
H_{0}^{ho}t_{3}](2C_{X})^{2k+1}P_{S}^{l}(C_{X},S_{X})\Psi
_{0}^{a}(x,t_{0}+t_{1}/2)||,  \tag{4.174b}
\end{equation}%
\begin{equation*}
NC7=\sum_{a=g,e}\sum_{k=0}^{n}\sum_{m=0}^{5}\sum_{l=0}^{m}\frac{\hslash
^{m}(\Omega _{0}t_{1})^{2k}}{(2k)!}\left( 
\begin{array}{c}
m \\ 
l%
\end{array}%
\right) ||\delta (x-x_{c},\varepsilon _{c})F_{5-m}^{\lambda
}(x(t_{3}^{\prime }),t_{3})
\end{equation*}%
\begin{equation}
\times Q_{m-l}^{a}(x(t_{3}^{\prime }),\varphi _{a})\exp [-\frac{i}{2\hslash }%
H_{0}^{ho}t_{3}](2C_{X})^{2k}Q_{C}^{l}(C_{X},S_{X})\Psi _{0}^{a}(x,\varphi
_{a},t_{0}+t_{1}/2)||,  \tag{4.174c}
\end{equation}%
\begin{equation*}
ND7=\sum_{a=g,e}\sum_{k=0}^{n}\sum_{m=0}^{5}\sum_{l=0}^{m}\frac{\hslash
^{m}(|\Omega _{0}|t_{1})^{2k+1}}{(2k+1)!}\left( 
\begin{array}{c}
m \\ 
l%
\end{array}%
\right) ||\delta (x-x_{c},\varepsilon _{c})F_{5-m}^{\lambda
}(x(t_{3}^{\prime }),t_{3})
\end{equation*}%
\begin{equation}
\times Q_{m-l}^{a}(x(t_{3}^{\prime }),\varphi _{a})\exp [-\frac{i}{2\hslash }%
H_{0}^{ho}t_{3}](2C_{X})^{2k+1}Q_{S}^{l}(C_{X},S_{X})\Psi _{0}^{a}(x,\varphi
_{a},t_{0}+t_{1}/2)||.  \tag{4.174d}
\end{equation}%
Here define $\varphi _{a}=+\varphi (x,\gamma )$ if $a=g$ and $\varphi
_{a}=-\varphi (x,\gamma )$ if $a=e$. Some relations are already used to
derive these four terms of (4.174). They include $\Omega (x)=4\hslash \Omega
_{0}C_{X},$ the two relations (4.171), and the following unitary
transformations: 
\begin{equation*}
\exp [-\frac{i}{2\hslash }H_{0}^{ho}t_{3}]F_{5-m}^{\lambda
}(x,t_{3})Q_{m-l}^{a}(x)\exp [\frac{i}{2\hslash }H_{0}^{ho}t_{3}]
\end{equation*}%
\begin{equation}
=F_{5-m}^{\lambda }(x(t_{3}^{\prime }),t_{3})Q_{m-l}^{a}(x(t_{3}^{\prime })),%
\text{ }a=g\text{ or }e,  \tag{4.175a}
\end{equation}%
and%
\begin{equation*}
\exp [-\frac{i}{2\hslash }H_{0}^{ho}t_{3}]F_{5-m}^{\lambda
}(x,t_{3})Q_{m-l}^{a}(x,\varphi _{a})\exp [\frac{i}{2\hslash }%
H_{0}^{ho}t_{3}]
\end{equation*}%
\begin{equation}
=F_{5-m}^{\lambda }(x(t_{3}^{\prime }),t_{3})Q_{m-l}^{a}(x(t_{3}^{\prime
}),\varphi _{a}),\text{ }a=g\text{ or }e.  \tag{4.175b}
\end{equation}%
Here the coordinate operator $x(t_{3}^{\prime })$ is defined by%
\begin{equation}
x(t_{3}^{\prime })=\exp [-\frac{i}{2\hslash }H_{0}^{ho}t_{3}]x\exp [\frac{i}{%
2\hslash }H_{0}^{ho}t_{3}].  \tag{4.175c}
\end{equation}%
A key point to calculating strictly the upper bounds of the four terms $NA7,$
$NB7,$ $NC7,$ and $ND7$ by the four equations (4.174), respectively, is to
calculate strictly the two unitary transformations (4.175a) and (4.175b),
while a rigorous calculation for the two unitary transformations needs to
obtain the coordinate operator $x(t_{3}^{\prime })$ from (4.175c). Because
the propagator $\exp [-\frac{i}{2\hslash }H_{0}^{ho}t_{3}]$ is of the
harmonic oscillator with the Hamiltonian $H_{0}^{ho}$, its unitary inverse
is easy to obtain and it is given by $\exp [-iH_{0}^{ho}t_{3}^{\prime
}/\hslash ]$ up to a global phase factor [15], here $t_{3}^{\prime
}+t_{3}/2=2k\pi /\omega $ ($k=0,$ $1$, $2$, ...) and $\omega $ is the
oscillatory (angular) frequency of the harmonic oscillator. Then the
coordinate operator $x(t_{3}^{\prime })$ also can be expressed as $%
x(t_{3}^{\prime })=\exp [iH_{0}^{ho}t_{3}^{\prime }/\hslash ]x\exp
[-iH_{0}^{ho}t_{3}^{\prime }/\hslash ].$ This shows that $x(t_{3}^{\prime })$
is just the coordinate operator in the Heisenberg picture. Thus, the
coordinate operator $x(t_{3}^{\prime })$ can be exactly obtained from
(4.69b). More generally, if the Hamiltonian $H_{0}^{ho}$ is replaced with a
general quadratic Hamiltonian, it is also easy to obtain exactly the unitary
transformation (4.175c) due to that in this case the inverse unitary
propagator can be obtained exactly [15]. However, if $H_{0}^{ho}$ is
replaced with other spatially-dependent Hamiltonian than a quadratic
Hamiltonian, it is generally difficult to obtain exactly (4.175c) except for
some simple and special cases. Notice that $F_{5-m}^{\lambda
}Q_{m-l}^{g}(x), $ $F_{5-m}^{\lambda }Q_{m-l}^{e}(x),$ $F_{5-m}^{\lambda
}Q_{m-l}^{g}(x,+\varphi (x,\gamma )),$ and $F_{5-m}^{\lambda
}Q_{m-l}^{e}(x,-\varphi (x,\gamma ))$ all are polynomials in coordinate $x$
with order not more than five. It is known from (4.69b) that the coordinate
operator $x(t_{3}^{\prime })$ in the Heisenberg picture is a linear
combination of the coordinate operator $x$ and momentum operator $p$. Then
by using the coordinate operator $x(t_{3}^{\prime })$ of (4.69b) and the
commutative relation $[x,p]=i\hslash $ one always can write those operators
on the $RH$ sides of (4.175a) and (4.175b) as%
\begin{equation*}
F_{5-m}^{\lambda }(x(t_{3}^{\prime }),t_{3})Q_{m-l}^{a}(x(t_{3}^{\prime
}))=q_{ml,0}^{\lambda a}(x)p^{5}+q_{ml,1}^{\lambda a}(x)p^{4}
\end{equation*}%
\begin{equation}
+...+q_{ml,4}^{\lambda a}(x)p+q_{ml,5}^{\lambda a}(x),  \tag{4.176a}
\end{equation}%
\begin{equation*}
F_{5-m}^{\lambda }(x(t_{3}^{\prime }),t_{3})Q_{m-l}^{a}(x(t_{3}^{\prime
}),\varphi _{a})=q_{ml,0}^{\lambda a}(x,\varphi _{a})p^{5}+q_{ml,1}^{\lambda
a}(x,\varphi _{a})p^{4}
\end{equation*}%
\begin{equation}
+...+q_{ml,4}^{\lambda a}(x,\varphi _{a})p+q_{ml,5}^{\lambda a}(x,\varphi
_{a}).  \tag{4.176b}
\end{equation}%
Here both the operators $q_{ml,j}^{\lambda a}(x)$ and $q_{ml,j}^{\lambda
a}(x,\varphi _{a})$ ($0\leq j\leq 5$) are the polynomials in coordinate
operator $x$ and their orders are not more than the index $j$. These
relations (4.176) will be further used to calculate the four terms $NA7$, $%
NB7$, $NC7$, and $ND7$.

Now one needs to prove that the following motional states that appear in the
four terms $NA7$, $NB7$, $NC7$, and $ND7$ are Gaussian superposition
motional states: 
\begin{equation*}
\Psi
_{l,2k}^{a,P_{C}}(x,t_{0}+t_{1}/2)=(2C_{X})^{2k}P_{C}^{l}(C_{X},S_{X})\Psi
_{0}^{a}(x,t_{0}+t_{1}/2),
\end{equation*}%
\begin{equation*}
\Psi
_{l,2k+1}^{a,P_{S}}(x,t_{0}+t_{1}/2)=(2C_{X})^{2k+1}P_{S}^{l}(C_{X},S_{X})%
\Psi _{0}^{a}(x,t_{0}+t_{1}/2),
\end{equation*}%
\begin{equation*}
\Psi _{l,2k}^{a,Q_{C}}(x,\varphi
_{a},t_{0}+t_{1}/2)=(2C_{X})^{2k}Q_{C}^{l}(C_{X},S_{X})\Psi
_{0}^{a}(x,\varphi _{a},t_{0}+t_{1}/2),
\end{equation*}%
\begin{equation*}
\Psi _{l,2k+1}^{a,Q_{S}}(x,\varphi
_{a},t_{0}+t_{1}/2)=(2C_{X})^{2k+1}Q_{S}^{l}(C_{X},S_{X})\Psi
_{0}^{a}(x,\varphi _{a},t_{0}+t_{1}/2).
\end{equation*}%
Here the trigonometric functions $C_{X}$ and $S_{X}$ may be rewritten as%
\begin{equation}
C_{X}=\frac{1}{2}\{\exp [i(\frac{1}{2}\Delta kx-\pi /4)]+\exp [-i(\frac{1}{2}%
\Delta kx-\pi /4)]\}  \tag{4.177a}
\end{equation}%
and 
\begin{equation}
S_{X}=\frac{1}{2i}\{\exp [i(\frac{1}{2}\Delta kx-\pi /4)]-\exp [-i(\frac{1}{2%
}\Delta kx-\pi /4)]\}.  \tag{4.177b}
\end{equation}%
By using the binomial expansion the trigonometric function $(2C_{X})^{m}$
may be expressed as%
\begin{equation}
(2C_{X})^{m}=\sum_{j=0}^{m}\left( 
\begin{array}{c}
m \\ 
j%
\end{array}%
\right) \exp \{-i(m-2j)[\frac{1}{2}\Delta kx-\pi /4]\}.  \tag{4.178}
\end{equation}%
This means that the trigonometric function $(2C_{X})^{m}$ for any integer $%
m>0$ is a linear combination of the $m+1$ phase factors $\{\exp (\pm ikqx)\}$
with the real number $q$ and the integer $k$. Notice that a $GWP$ state
times any phase factor $\exp (\pm ikqx)$ is still a $GWP$ state. Then these
states $(2C_{X})^{m}\Psi _{0}^{a}(x,t_{0}+t_{1}/2)$ and $(2C_{X})^{m}\Psi
_{0}^{a}(x,\varphi _{a},t_{0}+t_{1}/2)$ are Gaussian superposition motional
states. Therefore, the above four motional states, $\Psi
_{l,2k}^{a,P_{C}}(x,t_{0}+t_{1}/2),$ $etc.$, are Gaussian superposition
states if the four polynomials $P_{C}^{l}(C_{X},S_{X}),$ $%
P_{S}^{l}(C_{X},S_{X}),$ $Q_{C}^{l}(C_{X},S_{X}),$ and $%
Q_{S}^{l}(C_{X},S_{X})$ $(0\leq l\leq 5)$ in variables $C_{X}$ and $S_{X}$
can be expressed as a linear combination of the phase factors $\{\exp (\pm
ikqx)\}$. By substituting these two formulae (4.177) into the four
polynomials and then using the binomial expansion one can prove that each
one of the four polynomials $P_{C}^{l}(C_{X},S_{X}),$ $%
P_{S}^{l}(C_{X},S_{X}),$ $Q_{C}^{l}(C_{X},S_{X}),$ and $%
Q_{S}^{l}(C_{X},S_{X})$ may be expressed as a linear combination of the
phase factors $\{\exp (\pm ikqx)\}$. Actually, these four polynomials can be
directly calculated by the two equations (4.163). In particular, it can be
found that $P_{C}^{0}(C_{X},S_{X})=Q_{S}^{0}(C_{X},S_{X})=1$ and $%
P_{S}^{0}(C_{X},S_{X})=Q_{C}^{0}(C_{X},S_{X})=0;$ $%
P_{S}^{1}(C_{X},S_{X})=-Q_{C}^{1}(C_{X},S_{X})=(4\hslash \Omega _{0})(\frac{1%
}{2}t_{1}/\hslash )(\frac{1}{2}\Delta k)S_{X}$ and $%
P_{C}^{1}(C_{X},S_{X})=Q_{S}^{1}(C_{X},S_{X})=0.$ Thus, for the two cases $%
l=0$ and $1$ each one of the four polynomials can be expressed as a linear
combination of the three phase factors $\{1,$ $\exp [i\frac{1}{2}\Delta kx],$
$\exp [-i\frac{1}{2}\Delta kx]\}$ at most. Here the value $1=\exp (ilqx)$
with $l=0$. It also can be found that the total polynomial order for each
one of the four polynomials, $P_{C}^{l}(C_{X},S_{X}),$ $etc.,$ is not more
than $l.$ Therefore, each one of the four polynomials may be expressed as a
linear combination of the $2l+1$ phase factors $\{1,$ $\exp [\pm i\frac{1}{2}%
\Delta kx],$ $...,$ $\exp [\pm li\frac{1}{2}\Delta kx]\}$ at most, 
\begin{equation}
P_{C/S}^{l}(C_{X},S_{X})=\sum_{j^{\prime }=-l}^{l}\alpha _{j^{\prime
}}^{C/S}\exp \{ij^{\prime }[\frac{1}{2}\Delta kx-\pi /4]\},  \tag{4.179a}
\end{equation}%
\begin{equation}
Q_{C/S}^{l}(C_{X},S_{X})=\sum_{j^{\prime }=-l}^{l}\beta _{j^{\prime
}}^{C/S}\exp \{ij^{\prime }[\frac{1}{2}\Delta kx-\pi /4]\},  \tag{4.179b}
\end{equation}%
where $\{\alpha _{j^{\prime }}^{C/S}\}$ and $\{\beta _{j^{\prime }}^{C/S}\}$
are bounded and controllable. Now in the four motional states, $\Psi
_{l,2k}^{a,P_{C}}(x,t_{0}+t_{1}/2),$ etc., the two polynomials $%
(2C_{X})^{2k}P_{C}^{l}(C_{X},S_{X})$ and $%
(2C_{X})^{2k}Q_{C}^{l}(C_{X},S_{X}) $ have a total order not more than $2k+l$
and other two polynomials $(2C_{X})^{2k+1}P_{S}^{l}(C_{X},S_{X})$ and $%
(2C_{X})^{2k+1}Q_{S}^{l}(C_{X},S_{X})$ have a total order not more than $%
2k+l+1$. Therefore, each one of the two polynomials $%
(2C_{X})^{2k}P_{C}^{l}(C_{X},S_{X})$ and $%
(2C_{X})^{2k}Q_{C}^{l}(C_{X},S_{X}) $ may be expressed as a linear
combination of the $2(2k+l)+1$ phase factors $\{1,$ $\exp [\pm i\frac{1}{2}%
\Delta kx],$ $...,$ $\exp [\pm i(2k+l)(\frac{1}{2}\Delta k)x]\}$ at most.
Similarly, each one of the two polynomials $%
(2C_{X})^{2k+1}P_{S}^{l}(C_{X},S_{X})$ and $%
(2C_{X})^{2k+1}Q_{S}^{l}(C_{X},S_{X})$ may be expressed as a linear
combination of the $2(2k+l+1)+1$ phase factors $\{1,$ $\exp [\pm i\frac{1}{2}%
\Delta kx],...,$ $\exp [\pm i(2k+l+1)(\frac{1}{2}\Delta k)x]\}$ at most.
Here the indices $k$ and $l$ satisfy $0\leq k\leq n$ and $0\leq l\leq 5$ for
the four terms $NA7$, $NB7$, $NC7$, and $ND7$, as can be seen in the four
formulae (4.174). These results show that each one of the two states $\Psi
_{l,2k}^{a,P_{C}}(x,t_{0}+t_{1}/2)$ and $\Psi _{l,2k}^{a,Q_{C}}(x,\varphi
_{a},t_{0}+t_{1}/2)$ may be expressed as a linear combination of the $%
2(2k+l)+1$ $GWP$ motional states at most, each one of which has a different
motional momentum. Similarly, each one of the two states $\Psi
_{l,2k+1}^{a,P_{S}}(x,t_{0}+t_{1}/2)$ and $\Psi
_{l,2k+1}^{a,Q_{S}}(x,\varphi _{a},t_{0}+t_{1}/2)$ may be expressed as a
linear combination of the $2(2k+l)+3$ $GWP$ motional states at most, each
one of which has a different motional momentum. Now it is convenient to use
directly the expansions (4.178) and (4.179) to express these four motional
states, $\Psi _{l,2k}^{a,P_{C}}(x,t_{0}+t_{1}/2),$ etc. Then the two
motional states $\Psi _{l,2k}^{a,P_{C}}(x,t_{0}+t_{1}/2)$ and $\Psi
_{l,2k+1}^{a,P_{S}}(x,t_{0}+t_{1}/2)$ may be respectively expressed as the
Gaussian superposition states: 
\begin{equation}
\Psi _{l,2k}^{a,P_{C}}(x,t_{0}+t_{1}/2)=\sum_{j=0}^{2k}\left( 
\begin{array}{c}
2k \\ 
j%
\end{array}%
\right) \sum_{j^{\prime }=-l}^{l}\alpha _{j^{\prime }}^{C}\Psi
_{0,2k-2j-j^{\prime }}^{a}(x,t_{0}+t_{1}/2),  \tag{4.180a}
\end{equation}%
\begin{equation}
\Psi _{l,2k+1}^{a,P_{S}}(x,t_{0}+t_{1}/2)=\sum_{j=0}^{2k+1}\left( 
\begin{array}{c}
2k+1 \\ 
j%
\end{array}%
\right) \sum_{j^{\prime }=-l}^{l}\alpha _{j^{\prime }}^{S}\Psi
_{0,2k+1-2j-j^{\prime }}^{a}(x,t_{0}+t_{1}/2),  \tag{4.180b}
\end{equation}%
where the Gaussian motional state $\Psi _{0,m}^{a}(x,t_{0}+t_{1}/2)$ is
defined by%
\begin{equation}
\Psi _{0,m}^{a}(x,t_{0}+t_{1}/2)=\exp \{-im[\frac{1}{2}\Delta kx-\pi
/4]\}\Psi _{0}^{a}(x,t_{0}+t_{1}/2).  \tag{4.181}
\end{equation}%
Similarly, the two motional states $\Psi _{l,2k}^{a,Q_{C}}(x,\varphi
_{a},t_{0}+t_{1}/2)$ and $\Psi _{l,2k+1}^{a,Q_{S}}(x,\varphi
_{a},t_{0}+t_{1}/2)$ also may be respectively expressed as the Gaussian
superposition states:%
\begin{equation}
\Psi _{l,2k}^{a,Q_{C}}(x,\varphi _{a},t_{0}+t_{1}/2)=\sum_{j=0}^{2k}\left( 
\begin{array}{c}
2k \\ 
j%
\end{array}%
\right) \sum_{j^{\prime }=-l}^{l}\beta _{j^{\prime }}^{C}\Psi
_{0,2k-2j-j^{\prime }}^{a}(x,\varphi _{a},t_{0}+t_{1}/2),  \tag{4.182a}
\end{equation}%
\begin{equation*}
\Psi _{l,2k+1}^{a,Q_{S}}(x,\varphi
_{a},t_{0}+t_{1}/2)=\sum_{j=0}^{2k+1}\left( 
\begin{array}{c}
2k+1 \\ 
j%
\end{array}%
\right)
\end{equation*}%
\begin{equation}
\times \sum_{j^{\prime }=-l}^{l}\beta _{j^{\prime }}^{S}\Psi
_{0,2k+1-2j-j^{\prime }}^{a}(x,\varphi _{a},t_{0}+t_{1}/2),  \tag{4.182b}
\end{equation}%
where the Gaussian motional state $\Psi _{0,m}^{a}(x,\varphi
_{a},t_{0}+t_{1}/2)$ is defined by%
\begin{equation}
\Psi _{0,m}^{a}(x,\varphi _{a},t_{0}+t_{1}/2)=\exp \{-im[\frac{1}{2}\Delta
kx-\pi /4]\}\Psi _{0}^{a}(x,\varphi _{a},t_{0}+t_{1}/2).  \tag{4.183}
\end{equation}%
These Gaussian motional states $\Psi _{0,m}^{a}(x,t_{0}+t_{1}/2)$ and $\Psi
_{0,m}^{a}(x,\varphi _{a},t_{0}+t_{1}/2)$ are different from their original
ones $\Psi _{0}^{a}(x,t_{0}+t_{1}/2)$ and $\Psi _{0}^{a}(x,\varphi
_{a},t_{0}+t_{1}/2)$ only in their motional momentum and phase factors,
respectively.

The time evolution process for these $GWP$ states $\Psi
_{0,m}^{a}(x,t_{0}+t_{1}/2)$ and $\Psi _{0,m}^{a}(x,\varphi
_{a},t_{0}+t_{1}/2)$ under the propagator $\exp [-\frac{i}{2\hslash }%
H_{0}^{ho}t_{3}]$ may be written as%
\begin{equation}
\Psi _{0,m}^{a}(x,t_{0}+t_{1}/2+t_{3}/2)=\exp [-\frac{i}{2\hslash }%
H_{0}^{ho}t_{3}]\Psi _{0,m}^{a}(x,t_{0}+t_{1}/2),  \tag{4.184a}
\end{equation}%
\begin{equation}
\Psi _{0,m}^{a}(x,\varphi _{a},t_{0}+t_{1}/2+t_{3}/2)=\exp [-\frac{i}{%
2\hslash }H_{0}^{ho}t_{3}]\Psi _{0,m}^{a}(x,\varphi _{a},t_{0}+t_{1}/2). 
\tag{4.184b}
\end{equation}%
Here $\Psi _{0,m}^{a}(x,t_{0}+t_{1}/2+t_{3}/2)$ and $\Psi
_{0,m}^{a}(x,\varphi _{a},t_{0}+t_{1}/2+t_{3}/2)$ are still the $GWP$ states
as the Hamiltonian $H_{0}^{ho}$ is of the harmonic oscillator. They are
further used to calculate the upper bounds of these four terms $NA7$, $NB7$, 
$NC7$, and $ND7$. Consider first the term $NA7$ of (4.174a). In order to
calculate the upper bound of the term $NA7$ one needs to calculate the time
evolution process: $\exp [-\frac{i}{2\hslash }H_{0}^{ho}t_{3}]\Psi
_{l,2k}^{a,P_{C}}(x,t_{0}+t_{1}/2).$ According to (4.180a) and (4.184a) this
time evolution process may be written as%
\begin{equation*}
\exp [-\frac{i}{2\hslash }%
H_{0}^{ho}t_{3}](2C_{X})^{2k}P_{C}^{l}(C_{X},S_{X})\Psi
_{0}^{a}(x,t_{0}+t_{1}/2)
\end{equation*}%
\begin{equation}
=\sum_{j=0}^{2k}\left( 
\begin{array}{c}
2k \\ 
j%
\end{array}%
\right) \sum_{j^{\prime }=-l}^{l}\alpha _{j^{\prime }}^{C}\Psi
_{0,2k-2j-j^{\prime }}^{a}(x,t_{0}+t_{1}/2+t_{3}/2).  \tag{4.185}
\end{equation}%
By substituting (4.185) into (4.174a) and then using the operator polynomial 
$F_{5-m}^{\lambda }(x(t_{3}^{\prime }),t_{3})Q_{m-l}^{a}(x(t_{3}^{\prime }))$
in coordinate $x$ and momentum $p$ of (4.176a) it is easy to prove that the
term $NA7$ is bounded by%
\begin{equation*}
NA7\leq
\sum_{a=g,e}\sum_{k=0}^{n}\sum_{j=0}^{2k}\sum_{m=0}^{5}\sum_{l=0}^{m}\left( 
\begin{array}{c}
m \\ 
l%
\end{array}%
\right) \left( 
\begin{array}{c}
2k \\ 
j%
\end{array}%
\right) \frac{(\Omega _{0}t_{1})^{2k}}{(2k)!}\sum_{j^{\prime
}=-l}^{l}\hslash ^{m}|\alpha _{j^{\prime }}^{C}|
\end{equation*}%
\begin{equation}
\times \sum_{\nu =0}^{5}||\delta (x-x_{c},\varepsilon _{c})q_{ml,5-\nu
}^{\lambda a}(x)p^{\nu }\Psi _{0,2k-2j-j^{\prime
}}^{a}(x,t_{0}+t_{1}/2+t_{3}/2)||.  \tag{4.186}
\end{equation}%
Here the Gaussian motional state $\Psi _{0,m}^{a}(x,t_{0}+t_{1}/2+t_{3}/2)$
is given by (4.184a). According to the formula (4.109a) one has the relation:%
\begin{equation*}
p^{\nu }\Psi _{0,m}^{a}(x,t_{0}+t_{1}/2+t_{3}/2)=q_{\nu }^{a,m}(x)\Psi
_{0,m}^{a}(x,t_{0}+t_{1}/2+t_{3}/2),
\end{equation*}%
where $q_{\nu }^{a,m}(x)$ is a $\nu -$order polynomial in coordinate $x$. By
substituting this relation into (4.186) one can see that the $RH$ side of
(4.186) can be further reduced to a linear combination of the basic norms $%
\{NBAS1\}$. Since $q_{ml,5-\nu }^{\lambda a}(x)q_{\nu }^{a,2k-2j-j^{\prime
}}(x)$ is a polynomial in coordinate $x$ with order not more than five, each
one of the norms on the $RH$ side of (4.186) may be reduced to a linear sum
of five norms $\{NBAS1\}$ at most. In the sum terms on the $RH$ side of
(4.186) the indices $\nu ,$ $j^{\prime },$ $l,$ and $m$ satisfy $0\leq \nu ,$
$|j^{\prime }|,$ $l,$ $m\leq 5$ and the index $a=g$, $e$. Then the total
number (here it is denoted as $N(\nu ,j^{\prime },l,m;a))$ of these terms
labelled by the index $a$ and these indices $\nu ,$ $j^{\prime },$ $l,$ and $%
m$ is fixed and finite for any given index values $k$ and $j$. On the other
hand, the indices $j$ and $k$ satisfy $0\leq k\leq n$ and $0\leq j\leq 2k,$
respectively. Here one needs to determine the truncation term number $n$. It
is known from (4.170) that the truncation term number $n$ satisfies $%
2n+1>>2e|\Omega _{0}|t_{1},$ so that the contribution of the residual error
terms $\{E_{rm}(x,r,t_{0}+t_{1}/2)\}$ to the norms $NORM(1,\lambda ,C0)$ and 
$NORM(2,\lambda ,C0)$ can be neglected. Given the index values $\nu ,$ $%
j^{\prime },$ $l,$ and $m$ and the index $a$ the number of the norms on the $%
RH$ side of (4.186) is proportional to $n^{2}$ approximately (less than $%
2(n+1)^{2}$). Then the total number of all these norms on the $RH$ side of
(4.186) is proportional to $n^{2}N(\nu ,j^{\prime },l,m;a)$ approximately.
Now examine the non-negative coefficient (or parameter) in front of each
norm on the $RH$ side of (4.186). It is clear that the non-negative
parameter is given by%
\begin{equation*}
C(a,k,j,m,l,j^{\prime },\nu )=\hslash ^{m}|\alpha _{j^{\prime }}^{C}|\left( 
\begin{array}{c}
m \\ 
l%
\end{array}%
\right) \left( 
\begin{array}{c}
2k \\ 
j%
\end{array}%
\right) \frac{(\Omega _{0}t_{1})^{2k}}{(2k)!}.
\end{equation*}%
Note that $0\leq \nu ,$ $|j^{\prime }|,$ $l,$ $m\leq 5$ and $a=g$, $e$; and $%
0\leq k\leq n,$ $0\leq j\leq 2k.$ This shows that when the truncation term
number $n$ is not large, the parameter $C(a,k,j,m,l,j^{\prime },\nu )$ is
not yet large. Thus, here one needs only to consider a large truncation term
number, e.g., $n>>5$. Then in this case the parameter $C(a,k,j,m,l,j^{\prime
},\nu )$ is mainly dependent on the parameter value $\left( 
\begin{array}{c}
2k \\ 
j%
\end{array}%
\right) \frac{(\Omega _{0}t_{1})^{2k}}{(2k)!}.$ Notice that $\left( 
\begin{array}{c}
2k \\ 
j%
\end{array}%
\right) \leq \left( 
\begin{array}{c}
2k \\ 
k%
\end{array}%
\right) \leq 2^{2k}\leq 2^{2n}$ for $0\leq j\leq 2k\leq 2n.$ This means that
the value $\left( 
\begin{array}{c}
2k \\ 
j%
\end{array}%
\right) \frac{(\Omega _{0}t_{1})^{2k}}{(2k)!}$ satisfies%
\begin{equation*}
\left( 
\begin{array}{c}
2k \\ 
j%
\end{array}%
\right) \frac{(\Omega _{0}t_{1})^{2k}}{(2k)!}\leq \left( 
\begin{array}{c}
2k \\ 
k%
\end{array}%
\right) \frac{(\Omega _{0}t_{1})^{2k}}{(2k)!}\leq \frac{(2\Omega
_{0}t_{1})^{2k}}{(2k)!}.
\end{equation*}%
It is known that $\frac{(2\Omega _{0}t_{1})^{2n}}{(2n)!}<<1$ when $%
2n+1>>2e|\Omega _{0}|t_{1}.$ Then the maximum value of $\frac{(2\Omega
_{0}t_{1})^{2k}}{(2k)!}$ occurs at some value $k$ with $0<k<n$ and so does
the one of $\left( 
\begin{array}{c}
2k \\ 
k%
\end{array}%
\right) \frac{(\Omega _{0}t_{1})^{2k}}{(2k)!}$. It also is known that in the
Poisson distribution [44] the probability $P_{k}(\lambda )=\frac{\lambda ^{k}%
}{k!}\exp (-\lambda )$ with $0\leq k<+\infty .$ Because $0\leq P_{k}(\lambda
)\leq 1,$ one has $\lambda ^{k}/k!\leq \exp (\lambda )$ for $0\leq k<+\infty
.$ This means that $[\frac{(2\Omega _{0}t_{1})^{2k}}{(2k)!}]_{\max }\leq
\exp (2|\Omega _{0}|t_{1}).$ Thus, the value $\left( 
\begin{array}{c}
2k \\ 
j%
\end{array}%
\right) \frac{(\Omega _{0}t_{1})^{2k}}{(2k)!}$ satisfies%
\begin{equation*}
\left( 
\begin{array}{c}
2k \\ 
j%
\end{array}%
\right) \frac{(\Omega _{0}t_{1})^{2k}}{(2k)!}\leq \lbrack \frac{(2\Omega
_{0}t_{1})^{2k}}{(2k)!}]_{\max }\leq \exp (2|\Omega _{0}|t_{1})\leq \exp
(2|\Omega _{0}|\tau ).
\end{equation*}%
Here $0\leq t_{1}\leq \tau $, as can be seen from (4.2). Note that the time
interval $\tau $ usually satisfies $\tau <<1,$ as can be seen in the next
sections 5 and 6. Then the value $\exp (2|\Omega _{0}|\tau )$ usually is not
large. Now given the value $2|\Omega _{0}|t_{1}$ the upper bound of $\left( 
\begin{array}{c}
2k \\ 
j%
\end{array}%
\right) \frac{(\Omega _{0}t_{1})^{2k}}{(2k)!}$ may be determined completely
and it is independent on the truncation term number $n$. Therefore, for
given index values $m,$ $l,$ $j^{\prime },$ and $\nu $ and the index $a$ the
upper bound of the parameter $C(a,k,j,m,l,j^{\prime },\nu )$ in front of
each norm on the $RH$ side of (4.186) may be determined completely by the
value $2|\Omega _{0}|t_{1}$ and it also is independent of the truncation
term number $n.$ Thus, it follows from the inequality (4.186) that the upper
bound of the term $NA7$ of (4.174a) may be expressed as a linear combination
of the basic norms $\{NBAS1\}$. Similarly, one can prove that the term $NB7$
of (4.174b) also has an upper bound consisting of the basic norms $\{NBAS1\}$%
. In an analogous way, by substituting the state $\Psi
_{l,2k}^{a,Q_{C}}(x,\varphi _{a},t_{0}+t_{1}/2)$ of (4.182a) into the $RH$
side of (4.174c), then using the equation (4.184b), and finally using the
expansion (4.176b) of the operator polynomial $F_{5-m}^{\lambda
}(x(t_{3}^{\prime }),t_{3})Q_{m-l}^{a}(x(t_{3}^{\prime }),\varphi _{a})$ one
can prove that the term $NC7$ of (4.174c) is bounded by%
\begin{equation*}
NC7\leq
\sum_{a=g,e}\sum_{k=0}^{n}\sum_{j=0}^{2k}\sum_{m=0}^{5}\sum_{l=0}^{m}\left( 
\begin{array}{c}
m \\ 
l%
\end{array}%
\right) \left( 
\begin{array}{c}
2k \\ 
j%
\end{array}%
\right) \frac{(\Omega _{0}t_{1})^{2k}}{(2k)!}\sum_{j^{\prime
}=-l}^{l}\hslash ^{m}|\beta _{j^{\prime }}^{C}|
\end{equation*}%
\begin{equation}
\times \sum_{\nu =0}^{5}||\delta (x-x_{c},\varepsilon _{c})q_{ml,5-\nu
}^{\lambda a}(x,\varphi _{a})p^{\nu }\Psi _{0,2k-2j-j^{\prime
}}^{a}(x,\varphi _{a},t_{0}+t_{1}/2+t_{3}/2)||.  \tag{4.187}
\end{equation}%
This inequality is similar to (4.186). Thus, the upper bound of the term $%
NC7 $ can be obtained in a similar way that one obtains the upper bound of
the term $NA7$ above. Here according to (4.109b) one has the relation:%
\begin{equation*}
p^{\nu }\Psi _{0,m}^{a}(x,\varphi _{a},t_{0}+t_{1}/2+t_{3}/2)=q_{\nu
}^{a,m}(x,\varphi _{a})\Psi _{0,m}^{a}(x,\varphi _{a},t_{0}+t_{1}/2+t_{3}/2),
\end{equation*}%
where $q_{\nu }^{a,m}(x,\varphi _{a})$ is a $\nu -$order polynomial in
coordinate $x$. By substituting this relation into the $RH$ side of (4.187)
one can prove that the $RH$ side of (4.187) may be further reduced to a
linear combination of the basic norms $\{NBAS1\}$, indicating that the term $%
NC7$ of (4.174c) has an upper bound consisting of the basic norms $\{NBAS1\}$%
. Similarly, one also can prove that the term $ND7$ of (4.174d) has an upper
bound consisting of the basic norms $\{NBAS1\}$. Now all the four terms $NA7$%
, $NB7$, $NC7$, and $ND7$ of (4.174)\ are shown to have the upper bounds
consisting of the basic norms $\{NBAS1\}$. Then the inequality (4.173) shows
that the second sum term on the $RH$ side of (4.169a) has an upper bound
consisting of the basic norms $\{NBAS1\}$. On the other hand, it is already
shown by the inequality (4.170) that the first sum term on the $RH$ side of
(4.169a) can be controlled completely and it can be neglected when $%
2n+1>>2e|\Omega _{0}|t_{1}.$ If now the first sum term can be neglected,
then the inequality (4.169a) shows that the upper bound of the norm $%
NORM(1,\lambda ,C0)$ is determined from the second sum term on the $RH$ side
of (4.169a). Since the second sum term has an upper bound consisting of the
basic norms $\{NBAS1\}$, the norm $NORM(1,\lambda ,C0)$ has an upper bound
consisting of the basic norms $\{NBSA1\}$ too.

Now calculate the upper bound of the norm $NORM(2,\lambda ,C0)$ by the
inequality (4.169b). The upper bound consists of the two sum terms on the $%
RH $ side of (4.169b). The first sum term originates from the residual error
terms $\{E_{rm}(x,r,t_{0}+t_{1}/2)\}$ and it obeys the inequality (4.170).
As shown in (4.170), the upper bound of the first sum term is proportional
to the maximum one of the two exponentially-decaying factors $\frac{%
(2|\Omega _{0}|t_{1})^{2n+1}}{(2n+1)!}$ $(\thickapprox (\frac{2e|\Omega
_{0}|t_{1}}{2n+1})^{2n+1})$ and $\frac{(2\Omega _{0}t_{1})^{2n+2}}{(2n+2)!}$ 
$(\thickapprox (\frac{2e|\Omega _{0}|t_{1}}{2n+2})^{2n+2})$ with $%
2n+1>2e|\Omega _{0}|t_{1}.$ Therefore, the first sum term can be controlled
by the truncation\ term number $n$ and it can be neglected when $%
2n+1>>2e|\Omega _{0}|t_{1}.$ On the other hand, the second sum term on the $%
RH$ side of (4.169b) obeys the inequality (4.173) if one makes replacement $%
\delta (x-x_{c},\varepsilon _{c})\leftrightarrow \Theta (x-x_{L},\varepsilon
)$ in the inequality, 
\begin{equation*}
\sum_{m=0}^{5}\hslash ^{m}||\Theta (x-x_{L},\varepsilon )\exp [-\frac{i}{%
2\hslash }H_{0}^{ho}t_{3}]F_{5-m}^{\lambda
}(x,t_{3})D_{m}(x,r,t_{0}+t_{1}/2)||
\end{equation*}%
\begin{equation}
\leq NA7^{\prime }+NB7^{\prime }+NC7^{\prime }+ND7^{\prime }.  \tag{4.188}
\end{equation}%
Here the current four terms $NA7^{\prime },$ $NB7^{\prime },$ $NC7^{\prime
}, $ and $ND7^{\prime }$ correspond to the four terms $NA7,$ $NB7,$ $NC7,$
and $ND7$ in (4.173), respectively. They are still given by the four
modified equations of (4.174a), (4.174b), (4.174c), and (4.174d),
respectively, in which $\delta (x-x_{c},\varepsilon _{c})$ is replaced with $%
\Theta (x-x_{L},\varepsilon ).$ Now according to the above theoretical
calculation method used for the four terms $NA7,$ $NB7,$ $NC7,$ and $ND7$
one can calculate strictly the upper bounds for the current four terms $%
NA7^{\prime },$ $NB7^{\prime },$ $NC7^{\prime },$ and $ND7^{\prime }.$ It
can turn out that each one of these upper bounds can be expressed as a
linear sum of the basic norms $\{NBAS2\}$. Actually, the current four terms $%
NA7^{\prime },$ $NB7^{\prime },$ $NC7^{\prime },$ and $ND7^{\prime }$ on the 
$RH$ side of (4.188) can be converted into the four terms $NA7,$ $NB7,$ $%
NC7, $ and $ND7$ on the $RH$ side of (4.173), respectively, and vice versa
if one makes the replacement $\delta (x-x_{c},\varepsilon
_{c})\leftrightarrow \Theta (x-x_{L},\varepsilon ).$ Then these upper bounds
of the current four terms also can be converted into those of the four terms
in (4.173), respectively, and vice versa if one makes the replacement $%
\delta (x-x_{c},\varepsilon _{c})\leftrightarrow \Theta (x-x_{L},\varepsilon
).$ If now the truncation term number $n$ is chosen such that $%
2n+1>>2e|\Omega _{0}|t_{1},$ then the first sum term can be neglected on the 
$RH$ side of (4.169b). In this case the upper bound of the norm $%
NORM(2,\lambda ,C0)$ is determined from the second sum term on the $RH$ side
of (4.169b). Since the second sum term has an upper bound consisting of the
basic norms $\{NBAS2\}$, the norm $NORM(2,\lambda ,C0)$ has an upper bound
consisting of the basic norms $\{NBAS2\}$ too.

In this subsection it proves that the norm $NORM(1,\lambda ,C0)$ has an
upper bound consisting of a finite number of the basic norms $\{NBAS1\},$
while the norm $NORM(2,\lambda ,C0)$ has an upper bound which consists of a
finite number of the basic norms $\{NBAS2\}$.

A summary for the subsection 4.3.2 is given below. This subsection is
devoted to a rigorous calculation of the norm $%
||M_{22}^{V}(x,r,t_{1},t_{3})||.$ It is known from (4.90) that the upper
bound of the norm $||M_{22}^{V}(x,r,t_{1},t_{3})||$ may be determined from%
\begin{equation*}
||M_{22}^{V}(x,r,t_{1},t_{3})||\leq
||M_{22}^{VC0}(x,r,t_{1},t_{3})||+||M_{22}^{VCS}(x,r,t_{1},t_{3})||
\end{equation*}%
\begin{equation*}
+||M_{22}^{VS}(x,r,t_{1},t_{3})||.
\end{equation*}%
Here the upper bound for each one of the three norms $||M_{22}^{V\mu
}(x,r,t_{1},t_{3})||$ with the label $\mu =C0,$ $CS,$ and $S$ can be
determined from (4.94). This means that the upper bound of the norm $%
||M_{22}^{V\mu }(x,r,t_{1},t_{3})||$ with the label $\mu =C0,$ $CS,$ or $S$
is determined by the two norms on the $RH$ side of (4.94). One of the two
norms inside contains the operator $T_{1}(x,p,\varepsilon ),$ while another
inside contains the operator $T_{2}(x,p,\varepsilon ).$ The upper bound of
the norm inside containing the operator $T_{1}(x,p,\varepsilon )$ may be
determined by the five norms on the $RH$ side of (4.97), while that one
inside containing the operator $T_{2}(x,p,\varepsilon )$ is determined by
the thirteen norms on the $RH$ side of (4.99). Therefore, for each one of
the three norms $\{||M_{22}^{V\mu }(x,r,t_{1},t_{3})||\}$ one needs to
calculate the five norms on the $RH$ side of (4.97) and the thirteen norms
on the $RH$ side of (4.99). It can be found that each one of these five
norms of (4.97) and these thirteen norms of (4.99) may be expressed in a
unified form as either the norm $NORM(1,\lambda ,\mu )$ of (4.102) or $%
NORM(2,\lambda ,\mu )$ of (4.103). This means that in order to calculate the
upper bound of the norm $||M_{22}^{V\mu }(x,r,t_{1},t_{3})||$ one needs only
to calculate the upper bounds of the two norms $NORM(1,\lambda ,\mu )$ and $%
NORM(2,\lambda ,\mu ).$ In the subsection 4.3.2 all these six norms $%
NORM(1,\lambda ,\mu )$ and $NORM(2,\lambda ,\mu )$ for $\mu =C0,$ $CS,$ and $%
S$ are calculated strictly. Here the upper bounds of the norms $%
NORM(1,\lambda ,\mu )$ and $NORM(2,\lambda ,\mu )$ for $\mu =CS$ and $S$ are
strictly calculated in the subsection 4.3.2.1, while those of the norms $%
NORM(1,\lambda ,\mu )$ and $NORM(2,\lambda ,\mu )$ for $\mu =C0$ are
strictly calculated in the subsection 4.3.2.2. These calculated results show
that each one of these six norms $NORM(1,\lambda ,\mu )$ and $NORM(2,\lambda
,\mu )$ for $\mu =C0,$ $CS,$ and $S$ has an upper bound that consists of a
finite number of the basic norms $\{NBAS1\}$ and/or $\{NBAS2\}$. This
further shows that each one of the three norms $\{||M_{22}^{V\mu
}(x,r,t_{1},t_{3})||\}$ has an upper bound consisting of a finite number of
the basic norms $\{NBAS1\}$ and $\{NBAS2\}$. Thus, the norm $%
||M_{22}^{V}(x,r,t_{1},t_{3})||$ has an upper bound consisting of a finite
number of the basic norms $\{NBAS1\}$ and $\{NBAS2\},$ indicating that it
decays exponentially with the square deviation-to-spread ratios of the
relevant $GWP$ states.

A summary for the subsection 4.3 is given below. In the subsection 4.3.1 it
proves that the upper bound of the norm $||M_{21}^{V}(x,r,t_{1},t_{3})||$
consists of a finite number of the basic norms $\{NBAS1\}$ and $\{NBAS2\}$,
while in the subsection 4.3.2 it turns out that the norm $%
||M_{22}^{V}(x,r,t_{1},t_{3})||$ has an upper bound consisting of a finite
number of the basic norms $\{NBAS1\}$ and $\{NBAS2\}.$ It is known from
(4.64) that the norm $||M_{2}^{V}(x,r,t_{1},t_{3})||$ is bounded by $%
||M_{2}^{V}(x,r,t_{1},t_{3})||\leq
||M_{21}^{V}(x,r,t_{1},t_{3})||+||M_{22}^{V}(x,r,t_{1},t_{3})||.$ This
indicates that the norm $||M_{2}^{V}(x,r,$ $t_{1},t_{3})||$ also has an
upper bound consisting of the basic norms $\{NBAS1\}$ and $\{NBAS2\}$ and it
decays exponentially with the square deviation-to-spread ratios of the
relevant $GWP$ states.

Finally, according to the inequality (4.15) one can prove that the error $%
E_{r}^{V}(x,r,t_{0}+\tau )$ in (4.13) is bounded by\newline
\begin{equation*}
||E_{r}^{V}(x,r,t_{0}+\tau )||\leq ||O_{p}^{V}(x,r,\tau )\Psi
_{00}(x,r,t_{0})||
\end{equation*}%
\begin{equation}
+\frac{1}{2}\frac{1}{3!}\frac{\tau ^{3}}{\hslash ^{3}}%
\{||M_{1}^{V}(x,r,t_{1},t_{3})||_{\max }+\frac{1}{2}%
||M_{2}^{V}(x,r,t_{1},t_{3})||_{\max }\}  \tag{4.189}
\end{equation}%
where $||M_{k}^{V}(x,r,t_{1},t_{3})||_{\max }=\max
\{||M_{k}^{V}(x,r,t_{1},t_{3})||\}$ for $k=1$ and $2$ in the time region $%
0\leq t_{3},$ $t_{1}\leq \tau $. It is known from (4.35) in the subsection
4.1 that the norm $||O_{p}^{V}(x,r,\tau )\Psi _{00}(x,r,t_{0})||$ decays
exponentially with the square deviation-to-spread ratios of the relevant $%
GWP $ states. In the subsection 4.2 it proves that for any times $t_{1}$ and 
$t_{3}$ in the time region $[0,$ $\tau ]$ the norm $%
||M_{1}^{V}(x,r,t_{1},t_{3})||$ has an upper bound consisting of a finite
number of the basic norms $\{NBAS1\}$ and $\{NBAS2\},$ while in the
subsection 4.3 it turns out that for any times $t_{1}$ and $t_{3}$ in the
time region $[0,$ $\tau ]$ the norm $||M_{2}^{V}(x,r,t_{1},t_{3})||$ has an
upper bound consisting of a finite number of the basic norms $\{NBAS1\}$ and 
$\{NBAS2\}.$ Thus, the maximum norms $||M_{k}^{V}(x,r,t_{1},t_{3})||_{\max }$
for $k=1$ and $2$ in (4.189) also have the upper bounds consisting of the
basic norms $\{NBAS1\}$ and $\{NBAS2\},$ indicating that they decay
exponentially with the square deviation-to-spread ratios of the relevant $%
GWP $ states. Then the inequality (4.189) shows that the error $%
E_{r}^{V}(x,r,t_{0}+\tau )$ also decays exponentially with the square
deviation-to-spread ratios of the relevant $GWP$ states. This is the
expected result for the error $E_{r}^{V}(x,r,t_{0}+\tau ).$\newline
\newline
{\large 4.4 The upper bound of the norm }$||E_{r}^{0}(x,r,t_{0}+\tau )||$

In addition to requiring that the error $E_{r}^{V}(x,r,t_{0}+\tau )$
originating from the spatially-selective effect be negligible the time
evolution process of (4.13) also requires that the error $%
E_{r}^{0}(x,r,t_{0}+\tau )$ be controllable. Below calculate strictly the
upper bound of the error $E_{r}^{0}(x,r,t_{0}+\tau ).$ It can turn out that
the error is bounded by%
\begin{equation}
||E_{r}^{0}(x,r,t_{0}+\tau )||\leq \frac{1}{2}\frac{1}{3!}\frac{\tau ^{3}}{%
\hslash ^{3}}\{||M_{1}^{0}(x,r,t_{1},t_{3})||_{\max }+\frac{1}{2}%
||M_{2}^{0}(x,r,t_{1},t_{3})||_{\max }\}.  \tag{4.190}
\end{equation}%
This inequality is obtained by substituting (4.10) into (4.3). It is a
special form of the general inequality (4.7). Here the error states $%
M_{1}^{0}(x,r,t_{1},t_{3})$ and $M_{2}^{0}(x,r,t_{1},t_{3})$ are given by
(4.11a) and (4.11b), respectively, while $||M_{k}^{0}(x,r,t_{1},$ $%
t_{3})||_{\max }$ for $k=1$ and $2$ are the maximum norms in the time region 
$0\leq t_{3},t_{1}\leq \tau ,$ that is, $||M_{k}^{0}(x,r,t_{1},t_{3})||_{%
\max }=\max_{0\leq t_{3},t_{1}\leq \tau }\{||M_{k}^{0}(x,r,t_{1},t_{3})||\}.$
The inequality (4.190) indicates that if the maximum norms $%
\{||M_{k}^{0}(x,r,t_{1},t_{3})||_{\max }\}$ for $k=1$ and $2$ are bounded
over the whole coordinate space $-\infty <x<+\infty $, then the upper bound
of the error $E_{r}^{0}(x,r,t_{0}+\tau )$ is proportional to $\tau ^{3}$
approximately and may be controlled effectively by the single time parameter 
$\tau ,$ and this is in agreement with the error term $O(\tau ^{3})$ in the
decomposition formula (4.1) of the conventional Trotter-Suzuki decomposition
method. It will prove below that the maximum norms $%
\{||M_{k}^{0}(x,r,t_{1},t_{3})||_{\max }\}$ are indeed bounded over the
whole coordinate space $-\infty <x<+\infty $. Now by using the two basic
commutation relations of (4.37) one may calculate explicitly the two
commutation relations of the error state $M_{1}^{0}(x,r,t_{1},t_{3})$ of
(4.11a) and $M_{2}^{0}(x,r,t_{1},t_{3})$ of (4.11b), respectively,%
\begin{equation}
\lbrack H_{I}(x,\alpha ,\gamma ),[H_{0}^{ho},H_{I}(x,\alpha ,\gamma
)]]=g_{1}(x)I_{z}p+g_{2}(x)I_{z}+g_{3}(x),  \tag{4.191a}
\end{equation}%
\begin{equation*}
\lbrack H_{0}^{ho},[H_{0}^{ho},H_{I}(x,\alpha ,\gamma
)]]=f_{1}(x)I_{x}p^{2}+f_{2}(x)I_{y}p^{2}+f_{3}(x)I_{x}p
\end{equation*}%
\begin{equation}
+f_{4}(x)I_{y}p+f_{5}(x)I_{x}x+f_{6}(x)I_{y}x+f_{7}(x)I_{x}+f_{8}(x)I_{y}. 
\tag{4.191b}
\end{equation}%
Here one also needs to use the harmonic-oscillator Hamiltonian $H_{0}^{ho}$
of (2.3) and the interaction term $H_{I}(x,\alpha ,\gamma )$ $(\alpha =\pi
/4)$ of (4.16). The interaction term $H_{I}(x,\alpha ,\gamma )$ is
proportional to the time-independent amplitudes $\{Q_{x,y}(x,\alpha ,\gamma
)\}$ of the $PHAMDOWN$ laser light beams, which are given by (4.17),
respectively. In the two equations (4.191) these functions $\{g_{k}(x)\}$
and $\{f_{k}(x)\}$ of coordinate $x$ are explicitly obtained from the two
functions $Q_{x,y}(x,\alpha ,\gamma )$ of (4.17). They also contain the
coordinate derivatives of the two functions $Q_{x,y}(x,\alpha ,\gamma ).$
Since the two functions $Q_{x,y}(x,\alpha ,\gamma )$ are the trigonometric
functions of coordinate $x$, as can be seen in (4.17), their coordinate
derivatives are also the trigonometric functions. Hence the upper bounds of
these functions $\{g_{k}(x)\}$ and $\{f_{k}(x)\}$ may be independent of the
coordinate $x$. Therefore, all these functions $\{g_{k}(x)\}$ and $%
\{f_{k}(x)\}$ are bounded. Denote the upper bounds of these functions $%
\{g_{k}(x)\}$ and $\{f_{k}(x)\}$ as $\{G_{k}\}$ and $\{F_{k}\}$,
respectively, 
\begin{equation*}
|g_{k}(x)|\leq G_{k},\text{ }k=1,2,3,\text{ }-\infty <x<+\infty ,
\end{equation*}%
\begin{equation*}
|f_{k}(x)|\leq F_{k},\text{ }k=1,2,...,8,\text{ }-\infty <x<+\infty .
\end{equation*}%
Here the upper-bound values $\{G_{k}\}$ and $\{F_{k}\}$ also may be
independent of the phases $\alpha $ and $\gamma $ of the functions $%
Q_{x,y}(x,\alpha ,\gamma ).$ The upper bounds $\{G_{k}\}$ are dependent only
on the parameters of the $PHAMDOWN$ laser light beams, while $\{F_{k}\}$
depend on these parameters of both the harmonic potential field and the $%
PHAMDOWN$ laser light beams. Therefore, the upper bounds $\{G_{k}\}$ and $%
\{F_{k}\}$ can be controlled by these parameters of both the harmonic
potential field and the $PHAMDOWN$ laser light beams.

Before the norms $\{||M_{k}^{0}(x,r,t_{1},t_{3})||\}$ are calculated, one
needs to calculate the product state:%
\begin{equation*}
\Psi _{0}(x,r,t_{0}+t_{1}/2+\tau _{1})=\exp [-\frac{i}{\hslash }%
H_{I}(x,\alpha ,\gamma )\tau _{1}]\Psi _{0}(x,r,t_{0}+t_{1}/2)
\end{equation*}%
where $\tau _{1}=t_{1}-t_{3}$ for the product state $%
M_{1}^{0}(x,r,t_{1},t_{3})$ of (4.11a) and $\tau _{1}=t_{1}$ for $%
M_{2}^{0}(x,r,t_{1},t_{3})$ of (4.11b). By using the unitary transformation
(4.19) this product state may be reduced to the form%
\begin{equation*}
\Psi _{0}(x,r,t_{0}+t_{1}/2+\tau _{1})=\Psi _{0}^{g}(x,t_{0}+t_{1}/2)|\tilde{%
g}_{0}(x,\tau _{1})\rangle +\Psi _{0}^{e}(x,t_{0}+t_{1}/2)|\tilde{e}(x,\tau
_{1})\rangle .
\end{equation*}%
Here both the superposition states $|\tilde{g}_{0}(x,\tau _{1})\rangle $ and 
$|\tilde{e}(x,\tau _{1})\rangle $ are generated by applying the unitary
operator $\exp [-iH_{I}(\pi /4,\gamma )\tau _{1}/\hslash ]$ to the states $%
|g_{0}\rangle $ and $|e\rangle ,$ respectively. They are explicitly given by
(4.75) with the time setting $t_{1}=\tau _{1}.$ Now by substituting the
commutation relation of (4.191a) and the product state $\Psi
_{0}(x,r,t_{0}+t_{1}/2+\tau _{1})$ with $\tau _{1}=t_{1}-t_{3}$ into (4.11a)
one can prove that the product state $M_{1}^{0}(x,r,t_{1},t_{3})$ is bounded
by%
\begin{equation*}
||M_{1}^{0}(x,r,t_{1},t_{3})||\leq ||g_{1}(x)I_{z}p\Psi
_{0}^{g}(x,t_{0}+t_{1}/2)|\tilde{g}_{0}(x,\tau _{1})\rangle ||
\end{equation*}%
\begin{equation}
+||g_{1}(x)I_{z}p\Psi _{0}^{e}(x,t_{0}+t_{1}/2)|\tilde{e}(x,\tau
_{1})\rangle ||+\frac{1}{2}G_{2}+G_{3},  \tag{4.192}
\end{equation}%
where the relations $|g_{k}(x)|\leq G_{k}$ for $k=1$, $2$, $3;$ $||I_{\mu
}||=1/2$ for $\mu =x,$ $y,$ $z;$ and the normalization $||\Psi _{0}||^{2}=1$
are already used. By using the superposition state $|\tilde{g}_{0}(x,\tau
_{1})\rangle $ of (4.75a) and $|\tilde{e}(x,\tau _{1})\rangle $ of (4.75b)
(here $\tau _{1}=(t_{1}-t_{3}))$ and the relations $|||\tilde{g}_{0}(x,\tau
_{1})\rangle ||=1$ and $|||\tilde{e}(x,\tau _{1})\rangle ||=1$ it can turn
out that the first ($a=g$) and the second norm ($a=e$) on the $RH$ side of
(4.192) are bounded by%
\begin{equation*}
||g_{1}(x)I_{z}p\Psi _{0}^{a}(x,t_{0}+t_{1}/2)|\tilde{a}\rangle ||\leq (%
\frac{1}{2}G_{1}\hslash )||\frac{\partial }{\partial x}\Psi
_{0}^{a}(x,t_{0}+t_{1}/2)||
\end{equation*}%
\begin{equation*}
+(\frac{1}{2}G_{1}\hslash )|\Omega _{0}\Delta k(t_{1}-t_{3})|\times ||\Psi
_{0}^{a}(x,t_{0}+t_{1}/2)||
\end{equation*}%
\begin{equation}
+\frac{1}{2}|k_{0}+k_{1}|(\frac{1}{2}G_{1}\hslash )||\Psi
_{0}^{a}(x,t_{0}+t_{1}/2)||  \tag{4.193}
\end{equation}%
where $|\tilde{a}\rangle =|\tilde{g}_{0}(x,\tau _{1})\rangle $ if $a=g$ and $%
|\tilde{a}\rangle =|\tilde{e}(x,\tau _{1})\rangle $ if $a=e,$ as usual.
Denote $A_{0}^{a}$ as the amplitude of the non-normalization $GWP$ state $%
\Psi _{0}^{a}(x,t_{0}+t_{1}/2)$ with $a=g$ or $e$. Then one has $||\Psi
_{0}^{a}(x,t_{0}+t_{1}/2)||^{2}=|A_{0}^{a}|^{2}\leq 1.$ It is clear that $%
|A_{0}^{g}|^{2}+|A_{0}^{e}|^{2}=1$ according to the normalization condition $%
||\Psi _{0}||^{2}=1.$ Notice that the $GWP$ state $\Psi
_{0}^{a}(x,t_{0}+t_{1}/2)$ has the characteristic parameters $%
\{x_{c}^{a}(t_{0}+t_{1}/2),$ $p_{c}^{a}(t_{0}+t_{1}/2),$ $%
W_{c}^{a}(t_{0}+t_{1}/2),$ $\varepsilon _{c}^{a}(t_{0}+t_{1}/2)\}$. Its
coordinate derivative is given by (4.49). Then it can turn out that the norm 
$||\frac{\partial }{\partial x}\Psi _{0}^{a}||$ is bounded by%
\begin{equation*}
||\frac{\partial }{\partial x}\Psi _{0}^{a}(x,t_{0}+t_{1}/2)||\leq \frac{1}{2%
}\frac{1}{|W_{c}^{a}(t_{0}+t_{1}/2)|}||x\Psi _{0}^{a}(x,t_{0}+t_{1}/2)||
\end{equation*}%
\begin{equation}
+\{\frac{1}{2}|\frac{x_{c}^{a}(t_{0}+t_{1}/2)}{W_{c}^{a}(t_{0}+t_{1}/2)}%
|+|p_{c}^{a}(t_{0}+t_{1}/2)|/\hslash \}||\Psi _{0}^{a}(x,t_{0}+t_{1}/2)||. 
\tag{4.194}
\end{equation}%
Here $||x\Psi _{0}^{a}||^{2}$ is given by%
\begin{equation}
||x\Psi _{0}^{a}(x,t_{0}+t_{1}/2)||^{2}=|A_{0}^{a}|^{2}\{\frac{1}{2}%
\varepsilon _{c}^{a}(t_{0}+t_{1}/2)^{2}+x_{c}^{a}(t_{0}+t_{1}/2)^{2}\}. 
\tag{4.195}
\end{equation}%
By substituting (4.194) into (4.193) one obtains%
\begin{equation*}
||g_{1}(x)I_{z}p\Psi _{0}^{a}(x,t_{0}+t_{1}/2)|\tilde{a}\rangle ||
\end{equation*}%
\begin{equation*}
\leq \frac{1}{4}\frac{\hslash G_{1}|A_{0}^{a}|}{|W_{c}^{a}(t_{0}+t_{1}/2)|}%
\sqrt{\frac{1}{2}\varepsilon
_{c}^{a}(t_{0}+t_{1}/2)^{2}+x_{c}^{a}(t_{0}+t_{1}/2)^{2}}
\end{equation*}%
\begin{equation*}
+|A_{0}^{a}|(\frac{1}{2}G_{1}\hslash )\{\frac{1}{2}|\frac{%
x_{c}^{a}(t_{0}+t_{1}/2)}{W_{c}^{a}(t_{0}+t_{1}/2)}%
|+|p_{c}^{a}(t_{0}+t_{1}/2)|/\hslash \}
\end{equation*}%
\begin{equation}
+|A_{0}^{a}|(\frac{1}{2}G_{1}\hslash )\{|\Omega _{0}\Delta k(t_{1}-t_{3})|+%
\frac{1}{2}|k_{0}+k_{1}|\}.  \tag{4.196}
\end{equation}%
It is known from (4.192) that the upper bound of the norm $%
||M_{1}^{0}(x,r,t_{1},t_{3})||$ consists of the term $(G_{2}/2+G_{3})$ and
the two norms $||g_{1}(x)I_{z}p\Psi _{0}^{a}|\tilde{a}\rangle ||$ with $a=g$
and $e.$ Now the upper bounds of the two norms $\{||g_{1}(x)I_{z}p\Psi
_{0}^{a}|\tilde{a}\rangle ||\}$ can be determined from (4.196). Obviously,
the $RH$ side of the inequality (4.196) is independent of any unbounded
variables like momentum and coordinate. Therefore, both the norms $%
\{||g_{1}(x)I_{z}p\Psi _{0}^{a}|\tilde{a}\rangle ||\}$ are bounded. This
results in that the norm $||M_{1}^{0}(x,r,t_{1},t_{3})||$ is also bounded in
the time region $0\leq t_{3},$ $t_{1}\leq \tau $ and over the whole
coordinate space $-\infty <x<+\infty .$ As shown in (4.192) and (4.196), the
upper bound of the norm $||M_{1}^{0}(x,r,t_{1},t_{3})||$ is mainly dependent
on the three types of parameters: $(i)$ the characteristic parameters of the 
$GWP$ motional states $\{\Psi _{0}^{a}(x,t_{0}+t_{1}/2)\}$, $(ii)$ the
characteristic parameters of the $PHAMDOWN$ laser light beams, $(iii)$ the
characteristic parameters of the harmonic oscillator. Therefore, the norm $%
||M_{1}^{0}(x,r,t_{1},t_{3})||$ can be controlled by these three types of
the characteristic parameters.

Now calculate the second norm $||M_{2}^{0}(x,r,t_{1},t_{3})||.$ At first the
commutation relation (4.191b) is substituted into (4.11b). Then by using the
unitary transformations:%
\begin{equation*}
\exp [\frac{i}{2\hslash }H_{0}^{ho}t_{3}]x\exp [-\frac{i}{2\hslash }%
H_{0}^{ho}t_{3}]=x(t_{3}/2),
\end{equation*}%
\begin{equation*}
\exp [\frac{i}{2\hslash }H_{0}^{ho}t_{3}]p\exp [-\frac{i}{2\hslash }%
H_{0}^{ho}t_{3}]=p(t_{3}/2),
\end{equation*}%
where the momentum operator $p(t_{3}/2)$ and the coordinate operator $%
x(t_{3}/2)$ in the Heisenberg picture are explicitly given by the two
equations (4.69), respectively, it can turn out that the norm $%
||M_{2}^{0}(x,r,t_{1},t_{3})||$ satisfies the inequality ($\tau _{1}=t_{1}$):%
\begin{equation*}
||M_{2}^{0}(x,r,t_{1},t_{3})||\leq \frac{1}{2}(F_{1}+F_{2})||p^{2}\Psi
_{0}(x,r,t_{0}+t_{1}/2+\tau _{1})||
\end{equation*}%
\begin{equation*}
+\frac{1}{4}m\omega (F_{1}+F_{2})||(px+xp)\Psi _{0}(x,r,t_{0}+t_{1}/2+\tau
_{1})||
\end{equation*}%
\begin{equation*}
+\frac{1}{2}(m\omega )^{2}(F_{1}+F_{2})||x^{2}\Psi
_{0}(x,r,t_{0}+t_{1}/2+\tau _{1})||
\end{equation*}%
\begin{equation*}
+\frac{1}{2}[(F_{3}+F_{4})+\frac{F_{5}+F_{6}}{m\omega }]||p\Psi
_{0}(x,r,t_{0}+t_{1}/2+\tau _{1})||
\end{equation*}%
\begin{equation*}
+\frac{1}{2}[m\omega (F_{3}+F_{4})+(F_{5}+F_{6})]||x\Psi
_{0}(x,r,t_{0}+t_{1}/2+\tau _{1})||
\end{equation*}%
\begin{equation}
+\frac{1}{2}(F_{7}+F_{8})  \tag{4.197}
\end{equation}%
where the relations $|f_{k}(x)|\leq F_{k}$ for $1\leq k\leq 8;$ $||I_{\mu
}||=1/2$ for $\mu =x,$ $y,$ $z;$ $|\sin \theta |\leq 1$ and $|\cos \theta
|\leq 1$ for any angle $\theta ;$ and the normalization $||\Psi _{0}||^{2}=1$
and $||\exp [\pm \frac{i}{2\hslash }H_{0}^{ho}t_{3}]||=1$ are already used.
There are six terms on the $RH$ side of (4.197). Among these six terms the
first five terms are proportional to their own norms inside containing the
product state $\Psi _{0}$, respectively. These norms may be calculated
strictly below. It is easy to calculate the fifth norm on the $RH$ side of
(4.197). Its upper bound may be directly determined from%
\begin{equation*}
||x\Psi _{0}(x,r,t_{0}+t_{1}/2+\tau _{1})||=||x\Psi _{0}(x,r,t_{0}+t_{1}/2)||
\end{equation*}%
\begin{equation*}
\leq ||x\Psi _{0}^{g}(x,t_{0}+t_{1}/2)||+||x\Psi _{0}^{e}(x,t_{0}+t_{1}/2)||.
\end{equation*}%
By using further (4.195) one finds that the norm is bounded by 
\begin{equation*}
||x\Psi _{0}(x,r,t_{0}+t_{1}/2+\tau _{1})||
\end{equation*}%
\begin{equation}
\leq \sum_{a=g,e}|A_{0}^{a}|\sqrt{\frac{1}{2}\varepsilon
_{c}^{a}(t_{0}+t_{1}/2)^{2}+x_{c}^{a}(t_{0}+t_{1}/2)^{2}}.  \tag{4.198}
\end{equation}%
Therefore, the upper bound of the norm is dependent on the Gaussian
characteristic parameters $x_{c}^{a}(t_{0}+t_{1}/2)$ and $\varepsilon
_{c}^{a}(t_{0}+t_{1}/2).$ Similarly, the upper bound of the third norm is
directly determined from%
\begin{equation}
||x^{2}\Psi _{0}(x,r,t_{0}+t_{1}/2+\tau _{1})||\leq \sum_{a=g,e}||x^{2}\Psi
_{0}^{a}(x,t_{0}+t_{1}/2)||  \tag{4.199}
\end{equation}%
Here the norm $||x^{2}\Psi _{0}^{a}||$ with $a=g$ or $e$ can be directly
calculated. It is determined from%
\begin{equation*}
||x^{2}\Psi _{0}^{a}(x,t_{0}+t_{1}/2)||^{2}=|A_{0}^{a}|^{2}\{\frac{3}{4}%
\varepsilon _{c}^{a}(t_{0}+t_{1}/2)^{4}
\end{equation*}%
\begin{equation}
+3\varepsilon
_{c}^{a}(t_{0}+t_{1}/2)^{2}x_{c}^{a}(t_{0}+t_{1}/2)^{2}+x_{c}^{a}(t_{0}+t_{1}/2)^{4}\}.
\tag{4.200}
\end{equation}%
Then by substituting (4.200) into the $RH$ side of (4.199) one finds that
the upper bound of the third norm is dependent on the Gaussian
characteristic parameters $x_{c}^{a}(t_{0}+t_{1}/2)$ and $\varepsilon
_{c}^{a}(t_{0}+t_{1}/2).$ The upper bound of the fourth norm may be
calculated by 
\begin{equation*}
||p\Psi _{0}(x,r,t_{0}+t_{1}/2+\tau _{1})||\leq \sum_{a=g,e}||p\Psi
_{0}^{a}(x,t_{0}+t_{1}/2)|\tilde{a}\rangle ||.
\end{equation*}%
Here the norms $||p\Psi _{0}^{a}|\tilde{a}\rangle ||$ with $a=g$ and $e$ can
be calculated in a similar way that the norms $||g_{1}(x)I_{z}p\Psi _{0}^{a}|%
\tilde{a}\rangle ||$ in (4.193) are calculated above. It turns out that the
norm $||p\Psi _{0}||$ is bounded by%
\begin{equation*}
||p\Psi _{0}(x,r,t_{0}+t_{1}/2+\tau _{1})||\leq \hslash \sum_{a=g,e}||\frac{%
\partial }{\partial x}\Psi _{0}^{a}(x,t_{0}+t_{1}/2)||
\end{equation*}%
\begin{equation}
+\hslash (|(\Omega _{0}\Delta k)t_{1}|+\frac{1}{2}%
|k_{0}+k_{1}|)(|A_{0}^{g}|+|A_{0}^{e}|)  \tag{4.201}
\end{equation}%
where the upper bound of the norm $||\frac{\partial }{\partial x}\Psi
_{0}^{a}||$ is determined from (4.194). It is known from (4.194) that the
upper bound of the norm $||\frac{\partial }{\partial x}\Psi _{0}^{a}||$ is
dependent on the four Gaussian characteristic parameters. Then the
inequality (4.201) shows that the upper bound of the fourth norm is
dependent on the three types of parameters as stated above. Now with the aid
of the commutation relation $[x,p]=i\hslash $ and $||\Psi _{0}||^{2}=1$ it
can turn out that the second norm on the $RH$ side of (4.197) is bounded by%
\begin{equation*}
||(px+xp)\Psi _{0}(x,r,t_{0}+t_{1}/2+\tau _{1})||\leq \hslash +2||xp\Psi
_{0}(x,r,t_{0}+t_{1}/2+\tau _{1})||.
\end{equation*}%
Here the norm $||xp\Psi _{0}||$ can be calculated in a similar way that the
norm $||p\Psi _{0}||$ is calculated in (4.201). Then it can turn out that
the second norm is bounded by%
\begin{equation*}
||(px+xp)\Psi _{0}(x,r,t_{0}+t_{1}/2+\tau _{1})||\leq \hslash +2\hslash
\sum_{a=g,e}||x\frac{\partial }{\partial x}\Psi _{0}^{a}(x,t_{0}+t_{1}/2)||
\end{equation*}%
\begin{equation}
+2\hslash \{|(\Omega _{0}\Delta k)t_{1}|+\frac{1}{2}\hslash
|k_{0}+k_{1}|\}\sum_{a=g,e}||x\Psi _{0}^{a}(x,t_{0}+t_{1}/2)||.  \tag{4.202}
\end{equation}%
Here the norm $||x\frac{\partial }{\partial x}\Psi _{0}^{a}||$ can be
calculated directly. It is bounded by%
\begin{equation*}
||x\frac{\partial }{\partial x}\Psi _{0}^{a}(x,t_{0}+t_{1}/2)||\leq \frac{1}{%
2}\frac{1}{|W_{c}^{a}(t_{0}+t_{1}/2)|}||x^{2}\Psi _{0}^{a}(x,t_{0}+t_{1}/2)||
\end{equation*}%
\begin{equation}
+\{\frac{1}{2}|\frac{x_{c}^{a}(t_{0}+t_{1}/2)}{W_{c}^{a}(t_{0}+t_{1}/2)}%
|+|p_{c}^{a}(t_{0}+t_{1}/2)|/\hslash \}||x\Psi _{0}^{a}(x,t_{0}+t_{1}/2)||, 
\tag{4.203}
\end{equation}%
where the two norms $||x^{k}\Psi _{0}^{a}||$ for $k=1$ and $2$ are
determined from (4.195) and (4.200), respectively. Since both the norms $%
\{||x^{k}\Psi _{0}^{a}||\}$ are dependent on the Gaussian characteristic
parameters $\varepsilon _{c}^{a}(t_{0}+t_{1}/2)$ and $%
x_{c}^{a}(t_{0}+t_{1}/2),$ the inequality (4.203) shows that the upper bound
of the norm $||x\frac{\partial }{\partial x}\Psi _{0}^{a}||$ is also
dependent on the four Gaussian characteristic parameters. Then the
inequality (4.202) further shows that the upper bound of the second norm on
the $RH$ side of (4.197) is dependent on the three types of parameters. Now
it follows from the inequality (4.202) that the second term on the $RH$ side
of (4.197) is bounded by%
\begin{equation*}
\frac{1}{4}m\omega (F_{1}+F_{2})||(px+xp)\Psi _{0}(x,r,t_{0}+t_{1}/2+\tau
_{1})||
\end{equation*}%
\begin{equation*}
\leq \frac{1}{4}m\omega (F_{1}+F_{2})\{\hslash +2\hslash \sum_{a=g,e}||x%
\frac{\partial }{\partial x}\Psi _{0}^{a}(x,t_{0}+t_{1}/2)||
\end{equation*}%
\begin{equation}
+2\hslash (|(\Omega _{0}\Delta k)t_{1}|+\frac{1}{2}\hslash
|k_{0}+k_{1}|)\sum_{a=g,e}||x\Psi _{0}^{a}(x,t_{0}+t_{1}/2)||\}.  \tag{4.204}
\end{equation}%
Since $F_{1}$ and $F_{2}$ are dependent on the characteristic parameters of
both the harmonic potential field and the $PHAMDOWN$ laser light beams, the
inequality (4.204) shows that the second term are also dependent on the
three types of parameters. The first norm is more complex than any other
norms on the $RH$ side of (4.197). It can turn out that the norm is bounded
by $(\tau _{1}=t_{1})$%
\begin{equation*}
||p^{2}\Psi _{0}(x,r,t_{0}+t_{1}/2+\tau _{1})||\leq \hslash
^{2}\sum_{a=g,e}||\Psi _{0}^{a}(x,t_{0}+t_{1}/2)\frac{\partial ^{2}}{%
\partial x^{2}}|\tilde{a}\rangle ||
\end{equation*}%
\begin{equation}
+2\hslash ^{2}\sum_{a=g,e}||[\frac{\partial }{\partial x}\Psi
_{0}^{a}(x,t_{0}+t_{1}/2)](\frac{\partial }{\partial x}|\tilde{a}\rangle
)||+\hslash ^{2}\sum_{a=g,e}||[\frac{\partial ^{2}}{\partial x^{2}}\Psi
_{0}^{a}(x,t_{0}+t_{1}/2)]|\tilde{a}\rangle ||.  \tag{4.205}
\end{equation}%
Then one needs to calculate the three norms on the $RH$ side of (4.205) for
a given index $a=g$ or $e$. For the first norm on the $RH$ side of (4.205)
one needs to calculate the second-order coordinate derivative $\frac{%
\partial ^{2}}{\partial x^{2}}|\tilde{a}\rangle $ for $a=g$ or $e$. It is
known that the two superposition states $|\tilde{g}_{0}(x,\tau _{1})\rangle $
and $|\tilde{e}(x,\tau _{1})\rangle $ are explicitly given by the two
equations (4.75)$,$ respectively. Note that $\tau _{1}=t_{1}$ in (4.205). By
using the two superposition states one can calculate the derivatives $\frac{%
\partial }{\partial x}|\tilde{a}\rangle $ and $\frac{\partial ^{2}}{\partial
x^{2}}|\tilde{a}\rangle $ for $a=g$ or $e$. Then the derivative $\frac{%
\partial ^{2}}{\partial x^{2}}|\tilde{a}\rangle $ is further used to
calculate the norm. It can turn out that the first norm on the $RH$ side of
(4.205) is bounded by%
\begin{equation*}
||\Psi _{0}^{a}(x,t_{0}+t_{1}/2)\frac{\partial ^{2}}{\partial x^{2}}|\tilde{a%
}\rangle ||\leq |A_{0}^{a}|\{\frac{1}{2}|\Delta k(\Omega _{0}\Delta
k)t_{1}|+|(\Omega _{0}\Delta k)t_{1}|^{2}
\end{equation*}%
\begin{equation}
+|(k_{0}+k_{1})(\Omega _{0}\Delta k)t_{1}|+\frac{1}{4}(k_{0}+k_{1})^{2}\}. 
\tag{4.206}
\end{equation}%
It is clear that in addition to the time parameter $t_{1}$ and the amplitude 
$|A_{0}^{a}|$ the upper bound of the norm is dependent only on the
characteristic parameters of the $PHAMDOWN$ laser light beams. By using the
coordinate derivative $\frac{\partial }{\partial x}\Psi _{0}^{a}$ of (4.49)
and the derivative $\frac{\partial }{\partial x}|\tilde{a}\rangle $ it can
turn out that the second norm on the $RH$ side of (4.205) is bounded by%
\begin{equation*}
||[\frac{\partial }{\partial x}\Psi _{0}^{a}(x,t_{0}+t_{1}/2)](\frac{%
\partial }{\partial x}|\tilde{a}\rangle )||
\end{equation*}%
\begin{equation*}
\leq \{|(\Omega _{0}\Delta k)t_{1}|+\frac{1}{2}|k_{0}+k_{1}|\}\{\frac{1}{2}%
\frac{1}{|W_{c}^{a}(t_{0}+t_{1}/2)|}||x\Psi _{0}^{a}(x,t_{0}+t_{1}/2)||
\end{equation*}%
\begin{equation}
+|A_{0}^{a}|\{\frac{1}{2}|\frac{x_{c}^{a}(t_{0}+t_{1}/2)}{%
W_{c}^{a}(t_{0}+t_{1}/2)}|+|p_{c}^{a}(t_{0}+t_{1}/2)|/\hslash \}\}. 
\tag{4.207}
\end{equation}%
Obviously, the upper bound of the norm is dependent on the three types of
parameters as stated above. On the other hand, the second-order coordinate
derivative $\frac{\partial ^{2}}{\partial x^{2}}\Psi _{0}^{a}$ can be
calculated on the basis of the first-order derivative $\frac{\partial }{%
\partial x}\Psi _{0}^{a}$ of (4.49). Then by using the derivative $\frac{%
\partial ^{2}}{\partial x^{2}}\Psi _{0}^{a}$ one can prove that the last
norm on the $RH$ side of (4.205) is bounded by%
\begin{equation*}
||[\frac{\partial ^{2}}{\partial x^{2}}\Psi _{0}^{a}(x,t_{0}+t_{1}/2)]|%
\tilde{a}\rangle ||\leq \frac{1}{4}\frac{1}{|W_{c}^{a}(t_{0}+t_{1}/2)|^{2}}%
||x^{2}\Psi _{0}^{a}(x,t_{0}+t_{1}/2)||
\end{equation*}%
\begin{equation*}
+\frac{1}{|W_{c}^{a}(t_{0}+t_{1}/2)|}\{\frac{1}{2}|\frac{%
x_{c}^{a}(t_{0}+t_{1}/2)}{W_{c}^{a}(t_{0}+t_{1}/2)}%
|+|p_{c}^{a}(t_{0}+t_{1}/2)|/\hslash \}||x\Psi _{0}^{a}(x,t_{0}+t_{1}/2)||
\end{equation*}%
\begin{equation*}
+|A_{0}^{a}|\{\frac{1}{4}\frac{x_{c}^{a}(t_{0}+t_{1}/2)^{2}}{%
|W_{c}^{a}(t_{0}+t_{1}/2)|^{2}}+p_{c}^{a}(t_{0}+t_{1}/2)^{2}/\hslash ^{2}
\end{equation*}%
\begin{equation}
+\frac{(\frac{1}{2}+|x_{c}^{a}(t_{0}+t_{1}/2)p_{c}^{a}(t_{0}+t_{1}/2)|/%
\hslash )}{|W_{c}^{a}(t_{0}+t_{1}/2)|}\}.  \tag{4.208}
\end{equation}%
It is clear that the upper bound of the norm is dependent on the four
Gaussian characteristic parameters. Now these three inequalities (4.206),
(4.207), and (4.208) show that all the three norms (for a given label $a=g$
or $e$) on the $RH$ side of (4.205) are dependent on the three types of
parameters as stated above. Then the inequality (4.205) further indicates
that the upper bound of the first norm on the $RH$ side of (4.197) is
dependent on these three types of parameters. The above theoretical
calculation therefore shows that the first five norms on the $RH$ side of
(4.197) are bounded and controllable. Since the upper bounds $\{F_{k}\}$ are
bounded, all the six terms on the $RH$ side of (4.197) are bounded and
controllable. This indicates that the upper bound of the norm $%
||M_{2}^{0}(x,r,t_{1},t_{3})||$ is independent of the unbounded variables,
i.e., the momentum $p$ and coordinate $x$ and it is bounded in the time
region $0\leq t_{3},$ $t_{1}\leq \tau $ and over the whole coordinate space $%
-\infty <x<+\infty $.

Both the upper bounds of the norms $||M_{1}^{0}(x,r,t_{1},t_{3})||$ and $%
||M_{2}^{0}(x,r,t_{1},t_{3})||$ are controlled by the three types of
parameters, that is, the four Gaussian characteristic parameters, the
characteristic parameters of the $PHAMDOWN$ laser light beams, and those of
the harmonic potential field. It is clear that one can set externally the
latter two types of the characteristic parameters. Below it is shown that
the Gaussian characteristic parameters can be controlled by the motional
energy of the harmonic oscillator. Let $E_{0}^{a}$ be the motional energy of
a harmonic oscillator with the Hamiltonian $H_{0}^{ho}$ and in the initial $%
GWP$ motional state $\Psi _{00}^{a}(x,t_{0}).$ Then the time evolution
process of the harmonic oscillator is given by $\Psi
_{0}^{a}(x,t_{0}+t_{1}/2)=\exp [-\frac{i}{2\hslash }H_{0}^{ho}t_{1}]\Psi
_{00}^{a}(x,t_{0}).$ According to the energy conservation law the motional
energy of the harmonic oscillator in the $GWP$ motional state $\Psi
_{0}^{a}(x,t_{0}+t_{1}/2)$ is still equal to $E_{0}^{a}.$ Then from the
motional energy $E_{0}^{a}$ and the energy equation of a harmonic oscillator
that is given by (3.13a) in the section 3 or (5.43) in the section 5, one
may determine the upper and lower bounds of the characteristic parameters of
the $GWP$ state $\Psi _{0}^{a}(x,t_{0}+t_{1}/2).$ As shown in the section 5
below, the upper and lower bounds for these characteristic parameters $%
|x_{c}^{a}(t_{0}+t_{1}/2)|,$ $|p_{c}^{a}(t_{0}+t_{1}/2)|,$ $\varepsilon
_{c}^{a}(t_{0}+t_{1}/2),$ and $|W_{c}^{a}(t_{0}+t_{1}/2)|$ in the time
region $0\leq t_{1}\leq \tau $ are determined from%
\begin{equation}
0\leq |x_{c}^{a}(t_{0}+t_{1}/2)|\leq |x_{c}^{a}(t_{0}+t_{1}/2)|_{\max }<%
\sqrt{\frac{2E_{0}^{a}}{m\omega ^{2}}},\text{ }  \tag{4.209a}
\end{equation}%
\begin{equation}
0\leq |p_{c}^{a}(t_{0}+t_{1}/2)|\leq |p_{c}^{a}(t_{0}+t_{1}/2)|_{\max }<%
\sqrt{2mE_{0}^{a}},  \tag{4.209b}
\end{equation}%
\begin{equation*}
\frac{\hslash ^{2}}{8mE_{0}^{a}}<(\Delta x^{a}(t_{0}+t_{1}/2))_{\min
}^{2}\leq (\Delta x^{a}(t_{0}+t_{1}/2))^{2}
\end{equation*}%
\begin{equation}
\leq \frac{1}{2}\varepsilon _{c}^{a}(t_{0}+t_{1}/2)^{2}\leq \frac{1}{2}%
\varepsilon _{c}^{a}(t_{0}+t_{1}/2)_{\max }^{2}<\frac{2E_{0}^{a}}{m\omega
^{2}},  \tag{4.209c}
\end{equation}%
\begin{equation*}
\frac{\hslash ^{2}}{8mE_{0}^{a}}<(\Delta x^{a}(t_{0}+t_{1}/2))_{\min
}^{2}\leq |W_{c}^{a}(t_{0}+t_{1}/2)|
\end{equation*}%
\begin{equation}
\leq \frac{1}{2}\varepsilon _{c}^{a}(t_{0}+t_{1}/2)_{\max }^{2}<\frac{%
2E_{0}^{a}}{m\omega ^{2}}.  \tag{4.209d}
\end{equation}%
where $\omega $ is the oscillatory (angular) frequency of the harmonic
oscillator. These inequalities are similar to those of (3.13b) and (3.13c).
They show that given the motional energy $E_{0}^{a}$ and the oscillatory
frequency $\omega $ one may approximately determine the upper bounds of the
absolute COM position $|x_{c}^{a}(t_{0}+t_{1}/2)|$ and momentum $%
|p_{c}^{a}(t_{0}+t_{1}/2)|$ and the lower and upper bounds of the
wave-packet spread $\varepsilon _{c}^{a}(t_{0}+t_{1}/2)$ and the absolute
complex linewidth $|W_{c}^{a}(t_{0}+t_{1}/2)|.$ Here it needs to be pointed
out that the upper bound of the norm $||M_{1}^{0}(x,r,t_{1},t_{3})||$ is
inversely proportional to the absolute complex linewidth $%
|W_{c}^{a}(t_{0}+t_{1}/2)|,$ while the upper bound of the norm $%
||M_{2}^{0}(x,r,t_{1},t_{3})||$ is inversely proportional to $%
|W_{c}^{a}(t_{0}+t_{1}/2)|$ and $|W_{c}^{a}(t_{0}+t_{1}/2)|^{2}.$ Therefore,
one must use the minimum value of the absolute complex linewidth $%
|W_{c}^{a}(t_{0}+t_{1}/2)|$ to calculate these upper bounds. Obviously, this
minimum value is greater than zero and it can be controlled by the motional
energy $E_{0}^{a}$, as shown in (4.209d).

In this subsection the theoretical calculation shows that both the maximum
norms $||M_{1}^{0}(x,r,t_{1},t_{3})||_{\max }$ and $%
||M_{2}^{0}(x,r,t_{1},t_{3})||_{\max }$ in (4.190) are bounded in the time
region $0\leq t_{3},$ $t_{1}\leq \tau $ and over the whole coordinate space $%
-\infty <x<+\infty $. They are really dependent upon the time interval $\tau 
$ and the three types of parameters. One type of parameters come from the $%
GWP$ motional states of the halting-qubit atom, while the other two types of
parameters come from the $PHAMDOWN$ laser light beams and the harmonic
potential field, respectively. Thus, both the maximum norms $%
||M_{1}^{0}(x,r,t_{1},t_{3})||_{\max }$ and $%
||M_{2}^{0}(x,r,t_{1},t_{3})||_{\max }$ may be controlled by these three
types of parameters and the time interval $\tau $. Then the inequality
(4.190) shows that once the three types of parameters are set for the time
evolution process of (4.13), the upper bound of the error $%
E_{r}^{0}(x,r,t_{0}+\tau )$ in (4.13) is proportional to $\tau ^{3}$\
approximately, that is, $||E_{r}^{0}(x,r,t_{0}+\tau )||\varpropto O(\tau
^{3}),$ and hence it may be effectively controlled by the single time
parameter $\tau .$ This is the expected result!

In the above theoretical calculation the initial $GWP$ product state is $%
\Psi _{00}(x,r,$ $t_{0})=\Psi _{0}^{g}(x,t_{0})|g_{0}\rangle +\Psi
_{0}^{e}(x,t_{0})|e\rangle $, where $\Psi _{0}^{a}(x,t_{0})$ with $a=g$ or $%
e $ is a single $GWP$ state with no normalization. This case is simpler. In
general, the initial state $\Psi _{00}(x,r,t_{0})$ may be a Gaussian
superposition state consisting of $m$ $GWP$ product states: $\Psi
_{00}(x,r,t_{0})=\sum_{l=1}^{m}\Psi _{0l}(x,r,t_{0}).$ Here the $l-$th state 
$\Psi _{0l}(x,r,t_{0})=\Psi _{0l}^{g}(x,t_{0})|g_{0}\rangle +\Psi
_{0l}^{e}(x,t_{0})|e\rangle $, where $\Psi _{0l}^{a}(x,t_{0})$ with $a=g$ or 
$e$ is a $GWP$ state with no normalization. Then in this general case the
total error generated by the error operator $O_{p}(\tau ^{3})$ of (4.2) is
given by%
\begin{equation*}
E_{r}(x,r,t_{0}+\tau )=O_{p}(\tau ^{3})\Psi
_{00}(x,r,t_{0})=\sum_{l=1}^{m}O_{p}(\tau ^{3})\Psi _{0l}(x,r,t_{0}).
\end{equation*}%
Therefore, the upper bound of the total error is determined from%
\begin{equation}
||E_{r}(x,r,t_{0}+\tau )||\leq \sum_{l=1}^{m}||O_{p}(\tau ^{3})\Psi
_{0l}(x,r,t_{0})||.  \tag{4.210}
\end{equation}%
Now the upper bound of the error $O_{p}(\tau ^{3})\Psi _{0l}(x,r,t_{0})$ can
be calculated separately according to the above theoretical calculation
method by using each state $\Psi _{0l}(x,r,t_{0})$ to replace the initial
state $\Psi _{00}(x,r,t_{0})$. Then by summing up all these $m$ upper bounds
one may obtain the total error upper bound.

A summary for this section is given below. This section is particularly
long. It is devoted to a rigorous calculation for the two errors $%
E_{r}^{V}(x,r,t_{0}+\tau )$ and $E_{r}^{0}(x,r,t_{0}+\tau )$ in the time
evolution process of (4.13) for the halting-qubit atom in the $LH$ potential
well and in the presence of the spatially-selective $PHAMDOWN$ laser light
beams. The error $E_{r}^{V}(x,r,t_{0}+\tau )$ originates from the
imperfection of the $LH$ harmonic potential well and the spatially-selective
effect of the $PHAMDOWN$ laser light beams. In the section it proves
strictly that this error can be neglected when the joint position $x_{L}$ is
large enough. It also proves strictly that the error $E_{r}^{0}(x,r,t_{0}+%
\tau )$ can be controllable. Here both the $GWP$ states and the unitary
propagators of the halting-qubit atom in a harmonic potential field (or more
generally a quadratic potential field) play a key role in the strict
theoretical calculation for the upper bounds of the two errors. These
results show that the final state of the time evolution process of (4.13)
may be reduced to the simpler desired final state of (4.14) when the two
errors are neglected. This indicates that the Trotter-Suzuki decomposition
formula (4.1) indeed can be used to calculate rigorously the time evolution
process of (4.13) of the halting-qubit atom even if the process is involved
in the COM motion of the atom. The time evolution process of (4.13)\ and the
Trotter-Suzuki decomposition formula of (4.1) may be further used to
calculate the time evolution process of the halting-qubit atom during the $%
SSISS$ triggering pulse in the section 6 below. \newline
\newline
{\Large 5 The Lamb-Dicke limit}

It is known in Ref. [15] that the two state-selective triggering pulses $%
P_{tr}(\delta t)=[P_{tr}(\delta t/\sqrt{n})]^{n},$ where the basic pulse
sequences $P_{tr}(\delta t/\sqrt{n})$ are given by (6.1a) and (6.1b) in the
next section, respectively, generate theoretically the same unitary
propagator, if neglecting their truncation errors. This unitary propagator is%
\begin{equation}
\exp [iQ(\delta t/\hslash )^{2}]=\exp \{-i(4\Omega _{0}\delta
t)^{2}I_{z}\cos ^{2}[\frac{1}{2}\Delta kx-\pi /4]\},  \tag{5.1}
\end{equation}%
where the wave-number difference $\Delta k=k_{0}-k_{1}$, the Hermitian
operator $Q=i[H_{I}(x,\pi /4,0),H_{I}(x,\pi /4,\pi /2)]$ with the
Hamiltonian $H_{I}(x,\alpha ,\gamma )$ of (4.16), and the atomic
internal-state operator $I_{z}=(|e\rangle \langle e|-|g_{0}\rangle \langle
g_{0}|)/2$. This unitary propagator may transfer a $GWP$ state to another
approximately if the Lamb-Dicke limit $|\Delta kx|<<1$ is met. However,
there are two error terms in the state transfer, one is the truncation error
term $O((\delta t)^{3}/\sqrt{n})$ or $O((\delta t)^{4}/n)$ due to the
Trotter-Suzuki formulae [34, 39] and another is the error due to the
Lamb-Dicke limit [7, 10a]. The truncation error term may be neglected if the
pulse duration $\delta t$ is short and the number $n$ is large enough [39].
Below it is discussed how the Lamb-Dicke limit affects the Gaussian shape of
the atomic $GWP$ state when the state is acted on by the unitary propagator $%
\exp [iQ(\delta t/\hslash )^{2}].$ Since $I_{z}|g_{1}\rangle =0$ for any
atomic internal state $|g_{1}\rangle \notin \{|g_{0}\rangle ,$ $|e\rangle
\}, $ the state-selective triggering pulse does not affect the halting-qubit
atom in the internal state $|g_{1}\rangle $. Here consider only the case
that the halting-qubit atom is really excited by the state-selective
triggering pulse, meaning that the atom is in the internal state $%
|g_{0}\rangle $ or $|e\rangle $ instead of $|g_{1}\rangle $ at the initial
time of the state-selective triggering pulse. Now suppose that the initial
atomic product state is $\Psi _{0}(x,t_{0})|g_{0}\rangle ,$ where the
motional state $\Psi _{0}(x,t_{0})$ is a $GWP$ state with the characteristic
parameter set $\{x_{c}(t_{0}),$ $p_{c}(t_{0}),$ $W(t_{0}),$ $\varepsilon
(t_{0})\}$. As shown in Ref. [15], at the end of the state-selective
triggering pulse the atomic product state is $\Psi (x,t)|g_{0}\rangle =\exp
[iQ(\delta t/\hslash )^{2}]\Psi _{0}(x,t_{0})|g_{0}\rangle $, if neglecting
the truncation error term $O((\delta t)^{3}/\sqrt{n})$ or $O((\delta
t)^{4}/n),$ and the motional state $\Psi (x,t)$ may be written as%
\begin{equation*}
\Psi (x,t)=\exp (i\varphi _{0})[\frac{(\Delta x)^{2}}{2\pi }]^{1/4}\sqrt{%
\frac{1}{W(t_{0})}}\exp \{-\frac{1}{4}\frac{(x-x_{c}(t_{0}))^{2}}{W(t_{0})}\}
\end{equation*}%
\begin{equation}
\times \exp \{ip_{c}(t_{0})x/\hslash \}\exp \{i(2\Omega _{0}\delta
t)^{2}\sin (\Delta kx)\}.  \tag{5.2}
\end{equation}%
The state $\Psi (x,t)$ is not an ideal $GWP$ state. According to the Bessel
function theory [45] the exponential function $\exp [iz\cos \theta ]$ can be
expanded in terms of the Bessel functions $\{J_{n}(z)\},$ i.e., $\exp
[iz\cos \theta ]=\sum_{n=-\infty }^{\infty }i^{n}J_{n}(z)\exp (in\theta ),$
where $J_{-n}(z)=(-1)^{n}J_{n}(z).$ Then according to this expansion the
exponential function $\exp \{i(2\Omega _{0}\delta t)^{2}\sin (\Delta kx)\}$
in (5.2) with $\sin (\Delta kx)$ $=\cos (\Delta kx-\pi /2)$ can be expanded
as 
\begin{equation}
\exp \{i(2\Omega _{0}\delta t)^{2}\sin (\Delta kx)\}=\sum_{n=-\infty
}^{\infty }J_{n}((2\Omega _{0}\delta t)^{2})\exp [in\Delta kx].  \tag{5.3}
\end{equation}%
Now inserting the expansion (5.3) into (5.2) one can find that the state $%
\Psi (x,t)$ is really a superposition of the $GWP$ states. All these $GWP$
states have the same COM position and complex linewidth but different
momentum $(p_{c}(t_{0})+n\hslash \Delta k)$ and amplitude $J_{n}((2\Omega
_{0}\delta t)^{2}).$ Though such an expansion is exact for the state $\Psi
(x,t),$ it may be more complex to use this expansion state than a single $%
GWP $ state in a UNIDYSLOCK process. On the other hand, if the Lamb-Dicke
limit $|\Delta kx|<<1$ is met, then one may approximate $\sin (\Delta
kx)\thickapprox \Delta kx$ and finds that the state $\Psi (x,t)$ is
approximately a single $GWP$ state. However, if the COM position $%
x_{c}(t_{0})$ is large, then the coordinate $x$ in the state $\Psi (x,t)$
may take a large value, resulting in that the Lamb-Dicke limit $|\Delta
kx|<<1$ can not be met. Thus, in a general case the Lamb-Dicke limit needs
to be modified. Now denote $y=x-x_{c}(t_{0}).$ Then the wave-packet state of
(5.2)\ may be rewritten as%
\begin{equation*}
\Psi (y,t)=\exp (i\varphi _{1})[\frac{(\Delta x)^{2}}{2\pi }]^{1/4}\sqrt{%
\frac{1}{W(t_{0})}}\exp \{-\frac{1}{4}\frac{y^{2}}{W(t_{0})}\}
\end{equation*}%
\begin{equation}
\times \exp \{ip_{c}(t_{0})y/\hslash \}\exp \{i[q_{s}\sin (\Delta
ky)+q_{c}\cos (\Delta ky)]\}  \tag{5.4}
\end{equation}%
where $q_{s}=(2\Omega _{0}\delta t)^{2}\cos [\Delta kx_{c}(t_{0})]$ and $%
q_{c}=(2\Omega _{0}\delta t)^{2}\sin [\Delta kx_{c}(t_{0})].$ Because the
probability density $|\Psi (y,t)|^{2}$ is proportional to the Gaussian
exponential decaying factor $\exp [-y^{2}/\varepsilon (t_{0})^{2}]$ with the
wave-packet spread $\varepsilon (t_{0})$, it decays exponentially with the
square deviation $y^{2}$. The probability density is very close to zero when
the deviation $|y|>>\varepsilon (t_{0}).$ Thus, there is an effective
spatial region $[-y_{M}\varepsilon (t_{0}),y_{M}\varepsilon (t_{0})]$ with
the deviation $y_{M}>>1$ for the wave-packet state $\Psi (y,t)$ outside
which the amplitude of the wave-packet state is so small that it may be
neglected. This shows that the major effect of the phase factor $\exp
\{i[q_{s}\sin (\Delta ky)+q_{c}\cos (\Delta ky)]\}$ on the wave-packet state 
$\Psi (y,t)$ occurs only within the effective spatial region $%
[-y_{M}\varepsilon (t_{0}),y_{M}\varepsilon (t_{0})].$ If now the Lamb-Dicke
limit $|\Delta ky|\leq |\Delta ky_{M}\varepsilon (t_{0})|<<1$ is met over
the whole effective spatial region $[-y_{M}\varepsilon
(t_{0}),y_{M}\varepsilon (t_{0})]$, one may expand $\sin (\Delta ky)=\Delta
ky+R_{3}(\Delta ky^{\ast })$ and $\cos (\Delta ky)=1-(\Delta
ky)^{2}/2+R_{4}(\Delta ky^{\ast })$ up to the second-order approximation for 
$y\in \lbrack -y_{M}\varepsilon (t_{0}),y_{M}\varepsilon (t_{0})].$ Here $%
R_{3}(\Delta ky^{\ast })$ and $R_{4}(\Delta ky^{\ast })$ are the residual
terms of the Taylor series expansions of the trigonometric functions $\sin
(\Delta ky)$ and $\cos (\Delta ky),$ respectively. Then in the Lamb-Dicke
limit the state $\Psi (y,t)$ may be approximated well by the following state:%
\begin{equation*}
\Psi _{LD}(y,t)=\exp (i\varphi _{1})[\frac{(\Delta x)^{2}}{2\pi }]^{1/4}%
\sqrt{\frac{1}{W(t_{0})}}\exp \{-\frac{1}{4}\frac{y^{2}}{W(t_{0})}\}
\end{equation*}%
\begin{equation}
\times \exp \{ip_{c}(t_{0})y/\hslash \}\exp \{i[q_{s}\Delta ky+q_{c}(1-\frac{%
1}{2}(\Delta ky)^{2})]\}.  \tag{5.5}
\end{equation}%
Notice that here the coordinate $y$ may be extended to the whole coordinate
space $-\infty <y<+\infty $ rather than confined in the effective spatial
region $[-y_{M}\varepsilon (t_{0}),y_{M}\varepsilon (t_{0})].$ This is
because the phase factor $\exp \{i[q_{s}\Delta ky+q_{c}(1-\frac{1}{2}(\Delta
ky)^{2})]\}$ does not have a significant effect on the wave-packet state $%
\Psi _{LD}(y,t)$ outside the effective spatial region $[-y_{M}\varepsilon
(t_{0}),y_{M}\varepsilon (t_{0})].$ The state $\Psi _{LD}(y,t)$ of (5.5) is
called the Lamb-Dicke-limit state. This is a pure $GWP$ state close to the
original wave-packet state $\Psi (y,t)$ of (5.4). The Lamb-Dicke-limit state 
$\Psi _{LD}(y,t)$ has a slightly different complex linewidth from the
original state (5.4).

In order to evaluate quantitatively how close the Lamb-Dicke-limit state $%
\Psi _{LD}(y,t)$ is to the original state $\Psi (y,t)$ one may express the
original state $\Psi (y,t)$ as%
\begin{equation}
\Psi (y,t)=\Psi _{LD}(y,t)+E_{r}(y,t)  \tag{5.6}
\end{equation}%
where $E_{r}(y,t)$ is the error term, which is also thought of as a motional
state with no normalization. The error term $E_{r}(y,t)$ can be obtained
from the two states $\Psi (y,t)$ and $\Psi _{LD}(y,t)$,%
\begin{equation*}
E_{r}(x,t)=2\Psi _{LD}(y,t)\sin \{[q_{s}R_{3}(\Delta ky^{\ast
})+q_{c}R_{4}(\Delta ky^{\ast })]/2\}
\end{equation*}%
\begin{equation*}
\times \{-\sin \{[q_{s}R_{3}(\Delta ky^{\ast })+q_{c}R_{4}(\Delta ky^{\ast
})]/2\}
\end{equation*}%
\begin{equation}
+i\cos \{[q_{s}R_{3}(\Delta ky^{\ast })+q_{c}R_{4}(\Delta ky^{\ast })]/2\}\}.
\tag{5.7}
\end{equation}%
Therefore, the norm $||E_{r}(y,t)||$ is given by%
\begin{equation}
||E_{r}(y,t)||=||2\Psi _{LD}(y,t)\sin \{[q_{s}R_{3}(\Delta ky^{\ast
})+q_{c}R_{4}(\Delta ky^{\ast })]/2\}||.  \tag{5.8}
\end{equation}%
Notice that $||E_{r}(y,t)||^{2}$ is just the probability of the error state $%
E_{r}(y,t)$. There are some relations to help the determination of the upper
bound of the error term $E_{r}(y,t)$. These relations include $|\sin \theta
|\leq |\theta |,$ $|R_{3}(\Delta ky^{\ast })|\leq |\Delta ky|^{3}/3!,$ and $%
|R_{4}(\Delta ky^{\ast })|\leq |\Delta ky|^{4}/4!.$ With the help of these
relations one can prove that the error term is bounded by%
\begin{equation}
||E_{r}(y,t)||\leq \{\sum_{l=0}^{2}\left( 
\begin{array}{c}
2 \\ 
l%
\end{array}%
\right) (\frac{1}{3!})^{l}(\frac{1}{4!})^{2-l}|q_{s}^{l}q_{c}^{2-l}||\Delta
k|^{8-l}I_{8-l}^{LD}(\varepsilon (t_{0}))\}^{1/2}  \tag{5.9}
\end{equation}%
where the Gaussian integral $I_{m}^{LD}(\varepsilon (t_{0}))$ is defined by%
\begin{equation*}
I_{m}^{LD}(\varepsilon (t_{0}))=\int_{-\infty }^{\infty }dy\{|\Psi
_{LD}(y,t)|^{2}|y|^{m}\}
\end{equation*}%
\begin{equation}
=\left\{ 
\begin{array}{c}
\frac{(m-1)!!}{\sqrt{2^{m}}}\varepsilon (t_{0})^{m},\text{ if }m\text{ is
even} \\ 
\frac{1}{\sqrt{\pi }}[(m-1)/2]!\varepsilon (t_{0})^{m},\text{ if }m\text{ is
odd}%
\end{array}%
\right.  \tag{5.10}
\end{equation}%
The second equality of (5.10) is obtained by using the Lamb-Dicke-limit
state of (5.5) and the Gaussian integral formula [44a]:%
\begin{equation}
\int_{0}^{\infty }y^{n}\exp [-ay^{2}]dy=\left\{ 
\begin{array}{c}
\frac{1}{2}\frac{(n-1)!!}{\sqrt{(2a)^{n}}}\sqrt{\frac{\pi }{a}},\text{ if }n%
\text{ is even} \\ 
\frac{1}{2}\frac{[(n-1)/2]!}{\sqrt{a^{n+1}}},\text{ if }n\text{ is odd}%
\end{array}%
\right.  \tag{5.11}
\end{equation}%
Note that $|q_{c}|,$ $|q_{s}|\leq (2\Omega _{0}\delta t)^{2}.$ Then the
upper bound of (5.9) is proportional to $(2\Omega _{0}\delta t)^{2},$ $%
\varepsilon (t_{0})^{3},$ and $|\Delta k|^{3}$ in the lowest-order
approximation. The error term $E_{r}(y,t)$ measures how close the
Lamb-Dicke-limit state of (5.5) to the original state of (5.4). If the upper
bound $||E_{r}(y,t)||$ is close to zero, then the two states are really
close to each other.

The Lamb-Dicke-limit state $\Psi _{LD}(y,t)$\ tells ones that at the end of
the state-selective triggering pulse the motional momentum of the
halting-qubit atom is given approximately by\ 
\begin{equation*}
p(t)=p_{c}(t_{0})+\hslash \Delta k(2\Omega _{0}\delta t)^{2}\cos [\Delta
kx_{c}(t_{0})].
\end{equation*}%
Thus, after the state-selective triggering pulse the atomic motional
momentum increment is given by $\Delta p=\hslash \Delta k(2\Omega _{0}\delta
t)^{2}\cos [\Delta kx_{c}(t_{0})].$ It is known that when the
state-selective triggering pulse starts to apply to the halting-qubit atom
in the quantum program of the reversible and unitary halting protocol [14],
the atom is in the $GWP$ state with the COM position $x_{c}(t_{0})=0$ and
momentum $p_{c}(t_{0})=0$. Then after the state-selective triggering pulse
the atomic motional momentum is just equal to $p_{tr}=p(t)=\hslash \Delta
k(2\Omega _{0}\delta t)^{2}.$ The atomic motional momentum $p_{tr}$ is
proportional to the wave-number difference $\Delta k$ and the strength
factor $(2\Omega _{0}\delta t)^{2}$ which is dependent upon the Rabi
frequency $\Omega _{0}$ of the $PHAMDOWN$ laser light beams. Now by using
the atomic motional momentum $p_{tr}$ one can find from (5.9) that the error
term $E_{r}(y,t)$ is bounded by%
\begin{equation}
||E_{r}(y,t)||\leq |(p_{tr}/\hslash )||\Delta k|^{2}\{\sum_{l=0}^{2}\left( 
\begin{array}{c}
2 \\ 
l%
\end{array}%
\right) (\frac{1}{3!})^{l}(\frac{1}{4!})^{2-l}|\Delta
k|^{2-l}I_{8-l}^{LD}(\varepsilon (t_{0}))\}^{1/2}.  \tag{5.12}
\end{equation}%
Suppose that the atomic motional momentum $p_{tr}$ at the end of the
state-selective triggering pulse is set to a given value. Then the upper
bound of the error term $E_{r}(y,t)$ on the $RH$ side of (5.12) is
approximately proportional to $|\Delta k|^{2}$ and $\varepsilon (t_{0})^{3}$
in the lowest-order approximation. Thus, one may choose small enough
wave-number difference $|\Delta k|$ and wave-packet spread $\varepsilon
(t_{0})$ such that the error norm $||E_{r}(y,t)||$ is less than the desired
value close to zero and at the same time the atomic motional momentum keeps
at the given value $p_{tr}$.

A slightly worse Lamb-Dicke-limit state also can be obtained from the
original state of (5.4) by using the expansions $\sin (\Delta ky)=\Delta
ky+R_{3}(\Delta ky^{\ast })$ and $\cos (\Delta ky)=1+R_{2}(\Delta ky^{\ast
}) $ for $y\in \lbrack -y_{M}\varepsilon (t_{0}),y_{M}\varepsilon (t_{0})]$.
This Lamb-Dicke-limit state $\Psi _{LD}(y,t)$ is still given by (5.5), but
now it has not the quadratic term $q_{c}(\Delta ky)^{2}/2.$ The original
state $\Psi (y,t)$ is still expressed as (5.6), but in (5.6) the error term $%
E_{r}(y,t)$ has the norm:%
\begin{equation*}
||E_{r}(y,t)||=||2\Psi _{LD}(y,t)\sin \{[q_{s}R_{3}(\Delta ky^{\ast
})+q_{c}R_{2}(\Delta ky^{\ast })]/2\}||.
\end{equation*}%
Notice that $|R_{3}(\Delta ky^{\ast })|\leq |\Delta ky|^{3}/3!$ and $%
|R_{2}(\Delta ky^{\ast })|\leq (\Delta ky)^{2}/2$. It can turn out that the
error term $E_{r}(y,t)$ is bounded by 
\begin{equation}
||E_{r}(y,t)||\leq |(p_{tr}/\hslash )||\Delta k|\{\sum_{l=0}^{2}\left( 
\begin{array}{c}
2 \\ 
l%
\end{array}%
\right) (\frac{1}{2!})^{l}(\frac{1}{3!})^{2-l}|\Delta
k|^{2-l}I_{6-l}^{LD}(\varepsilon (t_{0}))\}^{1/2}.  \tag{5.13}
\end{equation}%
Obviously, the upper bound of the error term is approximately proportional
to the wave-number difference $|\Delta k|$ and square wave-packet spread $%
\varepsilon (t_{0})^{2}$ in the lowest-order approximation if the atomic
motional momentum $p_{tr}$ is kept constant. This is different from the
previous case, where the upper bound of the error term in (5.12) is
proportional to $|\Delta k|^{2}$ and $\varepsilon (t_{0})^{3}.$ The error
term $E_{r}(y,t)$ in (5.13) can be neglected when the wave-number difference 
$|\Delta k|$ and the wave-packet spread $\varepsilon (t_{0})$ are small
enough. In order that the error norm $||E_{r}(y,t)||$ is less than the
desired value close to zero and at the same time the atomic motional
momentum $p_{tr}$ keeps at the constant one must set smaller values for the
wave-number difference $|\Delta k|$ and the wave-packet spread $\varepsilon
(t_{0})$ in the present case than in the previous case of (5.12). Obviously,
in the present case the state-selective triggering pulse does not have a
significant contribution to the complex linewidth of the Lamb-Dicke-limit
state. This is the special point for the present Lamb-Dicke-limit state.
Thus, sometimes this Lamb-Dicke-limit state may be more useful.

Now consider a general case for the Lamb-Dicke limit. The wave-packet state
of (5.4) may be rewritten as%
\begin{equation*}
\Psi (y,t)=\exp (i\varphi _{1})[\frac{(\Delta x)^{2}}{2\pi }]^{1/4}\sqrt{%
\frac{1}{W(t_{0})}}\exp \{-\frac{1}{4}\frac{y^{2}}{W(t_{0})}\}
\end{equation*}%
\begin{equation}
\times \exp \{ip_{c}(t_{0})y/\hslash \}\exp \{iS(y)\}.  \tag{5.14}
\end{equation}%
This is equivalent to that the phase factor $\exp \{i[q_{s}\sin (\Delta
ky)+q_{c}\cos (\Delta ky)]\}$ in the original state of (5.4) is replaced
with a general phase factor $\exp \{iS(y)\},$ where $S(y)$ is a real
continuous function in the effective spatial region $[-y_{M}\varepsilon
(t_{0}),$ $y_{M}\varepsilon (t_{0})].$ There are two methods to approximate
the state $\Psi (y,t)$ of (5.14) in terms of the $GWP$ states. The first
method is to use a single $GWP$ state to approximate the state $\Psi (y,t)$.
The second method is to use a superposition of a finite number of the $GWP$
states to approximate the state $\Psi (y,t).$ It is just the $MGWP$
expansion method. The first method is just the generalization of the
Lamb-Dicke limit mentioned above. In the effective spatial region $%
[-y_{M}\varepsilon (t_{0}),y_{M}\varepsilon (t_{0})]$ the function $S(y)$ is
expanded as%
\begin{equation}
S(y)=S(0)+S^{\prime }(0)y+S^{\prime \prime }(0)y^{2}/2!+R_{3}(y^{\ast }) 
\tag{5.15}
\end{equation}%
where the residual term $R_{n+1}(y^{\ast })$ with $n\geq 2$ is given by,
according to the Taylor series expansion [44b], 
\begin{equation*}
R_{n+1}(y^{\ast })=\frac{y^{n+1}}{(n+1)!}S^{(n+1)}(y^{\ast }),\text{ }%
y^{\ast }\in (0,y)\subset \lbrack -y_{M}\varepsilon (t_{0}),y_{M}\varepsilon
(t_{0})],
\end{equation*}%
and the upper bound of the residual term can be obtained from%
\begin{equation*}
|R_{n+1}(y^{\ast })|\leq \frac{|y^{n+1}|}{(n+1)!}|S^{(n+1)}(y^{\ast })|.
\end{equation*}%
Then in the second-order approximation the phase factor $\exp \{iS(y)\}$ is
written as $\exp \{iS(y)\}=\exp \{i[S(0)+S^{\prime }(0)y+S^{\prime \prime
}(0)y^{2}/2!]\}.$ The generalized Lamb-Dicke-limit state therefore is
defined as 
\begin{equation*}
\Psi _{LD}(y,t)=\exp (i\varphi _{1})[\frac{(\Delta x)^{2}}{2\pi }]^{1/4}%
\sqrt{\frac{1}{W(t_{0})}}\exp \{-\frac{1}{4}\frac{y^{2}}{W(t_{0})}\}
\end{equation*}%
\begin{equation}
\times \exp \{ip_{c}(t_{0})y/\hslash \}\exp \{i[S(0)+S^{\prime
}(0)y+S^{\prime \prime }(0)y^{2}/2!]\}.  \tag{5.16}
\end{equation}%
This is a pure $GWP$ state. Thus, in the second-order approximation the
original state (5.14) still can be approximated by a single $GWP$ state,
i.e., the Lamb-Dicke-limit state (5.16). If the original state of (5.14) is
still expressed as (5.6) in which the Lamb-Dicke-limit state $\Psi
_{LD}(y,t) $ is given by (5.16), then the error term $E_{r}(y,t)$ is given by%
\begin{equation*}
E_{r}(y,t)=2\Psi _{LD}(y,t)\sin \{R_{3}(y^{\ast })/2\}[-\sin \{R_{3}(y^{\ast
})/2\}+i\cos \{R_{3}(y^{\ast })/2\}].
\end{equation*}%
Thus, the error term has the probability: 
\begin{equation*}
||E_{r}(y,t)||^{2}=||2\Psi _{LD}(y,t)\sin \{R_{3}(y^{\ast })/2\}||^{2}.
\end{equation*}%
Suppose that the $(n+1)-$th derivative $S^{(n+1)}(y^{\ast })$ is bounded
within the effective spatial region $[-y_{M}\varepsilon
(t_{0}),y_{M}\varepsilon (t_{0})],$ that is, $|S^{(n+1)}(y^{\ast })|\leq
M_{n+1}(y_{1})$ for some value $y_{1}\in \lbrack -y_{M}\varepsilon
(t_{0}),+y_{M}\varepsilon (t_{0})].$ Then the upper bound of the error term
is determined from%
\begin{equation*}
||E_{r}(y,t)||^{2}=4\int dy|\Psi _{LD}(y,t)\sin \{R_{3}(y^{\ast })/2\}|^{2}
\end{equation*}%
\begin{equation*}
\leq (\frac{1}{3!})^{2}\int_{-y_{M}\varepsilon (t_{0})}^{y_{M}\varepsilon
(t_{0})}dy|\Psi _{LD}(y,t)S^{(3)}(y^{\ast })y^{3}|^{2}
\end{equation*}%
\begin{equation*}
+4\int_{y_{M}\varepsilon (t_{0})}^{\infty }dy|\Psi
_{LD}(y,t)|^{2}+4\int_{-\infty }^{-y_{M}\varepsilon (t_{0})}dy|\Psi
_{LD}(y,t)|^{2}
\end{equation*}%
\begin{equation}
\leq \frac{5}{96}M_{3}(y_{1})^{2}\varepsilon (t_{0})^{6}+\frac{8}{\sqrt{\pi }%
}\frac{\exp (-y_{M}^{2})}{y_{M}+\sqrt{y_{M}^{2}+4/\pi }}  \tag{5.17}
\end{equation}%
where the Gaussian integral (5.10) has been used and $M_{3}(y_{1})$ could be
dependent on the deviation $y_{M}$. This is a general upper bound of the
error term when the original state of (5.14) is approximated by a single $%
GWP $ state (5.16).

It has been shown above that on one extreme one may use the Lamb-Dicke-limit
state that is a single $GWP$ state to approximate the original state, on the
other extreme one also may use a superposition of an infinite number of $GWP$
states obtained from the Bessel function expansion (5.3) to express exactly
the same original state. The Lamb-Dicke-limit method is simpler as it uses
only one $GWP$ state to approximate the original state. It is known that at
the end of the state-selective triggering pulse the atomic motional state is
given by the wave-packet state (5.2). It has been shown above that this
wave-packet state may be well approximated by a single $GWP$ state, i.e.,
the Lamb-Dicke-limit state $\Psi _{LD}(y,t)$ of (5.5). However, during the
state-selective triggering pulse the atomic motional state could be quite
complicated. It may not be suited to use a single $GWP$ state, i.e., the
Lamb-Dicke-limit state to approximate such a motional state. Thus, it could
not be a good scheme to use a single $GWP$ state to approximate the
intermediate motional state of the halting-qubit atom during the
state-selective triggering pulse. Therefore, the Lamb-Dicke-limit method
could not lead to a better error estimation for the time evolution process
of the halting-qubit atom during the state-selective triggering pulse. On
the other hand, the Bessel function expansion (5.3) is exact, but it is too
complex to use conveniently in practice. A compromise is to use a finite
number of the $GWP$ states to express approximately the intermediate
motional state. It can be seen below that this scheme has an advantage over
the two extreme methods mentioned above when it is used to estimate the
errors of the time evolution process of the halting-qubit atom during the
state-selective triggering pulse. The $MGWP$ expansion method may result in
a better error estimation, meanwhile the computation for the $MGWP$
expansion is not too complex to carry out. In fact, the main advantage to
use the $MGWP$ expansion is that the time evolution process of a single atom
in a harmonic potential field and in a Gaussian superposition state can be
calculated exactly.

Suppose that the function $S(y)$ in the original state (5.14) is taken as $%
S(y)=q_{s}\sin (\Delta ky)-q_{c}[1-\cos (\Delta ky)].$ Obviously, $S(0)=0.$
Then the state (5.14) is really reduced to the state (5.4) up to a global
phase factor. Generally, the phase factor $\exp \{iS(y)\}$ of the original
state (5.14) may be expanded as%
\begin{equation}
F(S(y))\equiv \exp \{iS(y)\}=\sum_{k=0}^{n}\frac{1}{k!}%
[iS(y)]^{k}+R_{n+1}(S^{\ast }(y))  \tag{5.18}
\end{equation}%
where the residual term $R_{n+1}(S^{\ast }(y))$ is given by, according to
the Taylor series expansion [44], 
\begin{equation}
R_{n+1}(S^{\ast }(y))=\frac{S(y)^{n+1}}{(n+1)!}F^{(n+1)}(S^{\ast }(y)) 
\tag{5.19}
\end{equation}%
with the $(n+1)-$order derivative of the function $F(S(y))$ given by 
\begin{equation*}
F^{(n+1)}(S^{\ast }(y))=[\frac{d^{n+1}}{dS(y)^{n+1}}\exp
\{iS(y)\}]|_{S(y)=S^{\ast }(y)},\text{ }S^{\ast }(y)\in \lbrack 0,\text{ }%
S(y)],
\end{equation*}%
and the upper bound of the residual term can be obtained from 
\begin{equation}
|R_{n+1}(S^{\ast }(y))|\leq \frac{|S(y)^{n+1}|}{(n+1)!}|F^{(n+1)}(S^{\ast
}(y))|.  \tag{5.20}
\end{equation}%
If the function $S(y)=q_{s}\sin (\Delta ky)-q_{c}(1-\cos (\Delta ky))$ is
inserted into (5.18) and the residual term $R_{n+1}(S^{\ast }(y))$ is
neglected, one can find that the expansion (5.18) can be expressed as a
linear sum of the phase factors $\{\exp (\pm il\Delta ky),$ $l=0,1,...,n\}$.
Then by inserting the expansion (5.18) into (5.14) one can further find that
the state (5.14) is a superposition of a finite number of the $GWP$ states: 
\begin{equation*}
\Psi _{MG}^{n}(y,t)=\exp (i\varphi _{1})[\frac{(\Delta x)^{2}}{2\pi }]^{1/4}%
\sqrt{\frac{1}{W(t_{0})}}\exp \{-\frac{1}{4}\frac{y^{2}}{W(t_{0})}\}
\end{equation*}%
\begin{equation}
\times \exp \{ip_{c}(t_{0})y/\hslash \}\sum_{k=0}^{n}[iS(y)]^{k}/k!. 
\tag{5.21}
\end{equation}%
Obviously, this superposition state $\Psi _{MG}^{n}(y,t)$ is close to the
original state (5.14) if the integer $n$ is large. For convenience, the
state $\Psi _{MG}^{n}(y,t)$ is called the $n-$order $MGWP$ state of the
original state (5.14). If the state $\Psi (y,t)$ of (5.14) is expressed as $%
\Psi (y,t)=\Psi _{MG}^{n}(y,t)+E_{rMG}^{n}(y,t),$ then the error term $%
E_{rMG}^{n}(y,t)$ has the probability:%
\begin{equation}
||E_{rMG}^{n}(y,t)||^{2}=\frac{1}{\varepsilon (t_{0})\sqrt{\pi }}\int
dy\{\exp [-\frac{y^{2}}{\varepsilon (t_{0})^{2}}]|R_{n+1}(S^{\ast
}(y))|^{2}\}.  \tag{5.22}
\end{equation}%
The formula (5.22) together with (5.20) may be used to calculate the upper
bound of the error term $E_{rMG}^{n}(y,t).$ Therefore, the $MGWP$ expansion
may be used to estimate the error terms during the time evolution process of
the halting-qubit atom under the state-selective triggering pulse.

The $MGWP$ expansion method has been used to calculate the upper bounds of
the norms $NORM(1,\lambda ,C0)$ and $NORM(2,\lambda ,C0)$ in the previous
subsection 4.3.2.2. Below it is used to approximate the intermediate product
states of the halting-qubit atom during the state-selective triggering
pulse. It can be seen in the next section that during the $SSISS$ triggering
pulse there appears the atomic product state which is a superposition of the
wave-packet product states: 
\begin{equation*}
\Psi (x,r,t_{0}+\tau )=\cos \{\Omega (\tau )\cos (\frac{1}{2}\Delta kx-\pi
/4)\}\Psi _{0}(x,t_{0})|g_{0}\rangle
\end{equation*}%
\begin{equation}
+\sin \{\Omega (\tau )\cos (\frac{1}{2}\Delta kx-\pi /4)\}\exp [i\frac{1}{2}%
(k_{0}+k_{1})x]\Psi _{0}(x,t_{0})|e\rangle .  \tag{5.23}
\end{equation}%
Here the parameter $\Omega (\tau )$ usually takes a small value, $|\Omega
(\tau )|<<1,$ and hence $|\sin \{\Omega (\tau )\cos (\frac{1}{2}\Delta
kx-\pi /4)\}|<<1,$ indicating that the probability for the halting-qubit
atom in the excited internal state $|e\rangle $ is small. This case is met
in the next section. The atomic product state $\Psi (x,r,t_{0}+\tau )$ is
not a pure $GWP$ product state, but it could be approximated well by a
superposition of a finite number of the $GWP$ product states. Suppose that
the $GWP$ state $\Psi _{0}(x,t_{0})$ in (5.23) has the characteristic
parameter set $\{x_{c}(t_{0}),$ $p_{c}(t_{0}),$ $W(t_{0}),$ $\varepsilon
(t_{0})\}$. Denote $y=x-x_{c}(t_{0}).$ Then the trigonometric functions on
the $RH$ side of (5.23) may be expressed as 
\begin{equation}
\cos \{\Omega (\tau )\cos (\frac{1}{2}\Delta kx-\pi /4)\}=\cos \beta
_{c}\cos [S_{0}(y)]+\sin \beta _{c}\sin [S_{0}(y)],  \tag{5.24a}
\end{equation}%
\begin{equation}
\sin \{\Omega (\tau )\cos (\frac{1}{2}\Delta kx-\pi /4)\}=\sin \beta
_{c}\cos [S_{0}(y)]-\cos \beta _{c}\sin [S_{0}(y)],  \tag{5.24b}
\end{equation}%
where the function $S_{0}(y)$ is defined by%
\begin{equation}
S_{0}(y)=\beta _{c}(1-\cos (\frac{1}{2}\Delta ky))+\beta _{s}\sin (\frac{1}{2%
}\Delta ky)  \tag{5.25}
\end{equation}%
with the parameters%
\begin{equation*}
\beta _{c}=\Omega (\tau )\cos [\frac{1}{2}\Delta kx_{c}(t_{0})-\pi /4],\text{
}\beta _{s}=\Omega (\tau )\sin [\frac{1}{2}\Delta kx_{c}(t_{0})-\pi /4].
\end{equation*}%
Since the parameter $\Omega (\tau )$ usually satisfies $|\Omega (\tau )|<<1,$
the function $S_{0}(y)$ usually takes a small value, that is, $|S_{0}(y)|<<1$
and satisfies $S_{0}(0)=0.$ Moreover, in the Lamb-Dicke limit the function $%
S_{0}(y)$\ becomes much smaller in the effective spatial region $%
[-y_{M}\varepsilon (t_{0}),y_{M}\varepsilon (t_{0})]$. Now the trigonometric
functions $\sin [S_{0}(y)]$ and $\cos [S_{0}(y)]$ may be respectively
expanded as 
\begin{equation}
\sin [S_{0}(y)]=S_{0}(y)+R_{3}(S_{0}^{\ast }(y))  \tag{5.26a}
\end{equation}%
and 
\begin{equation}
\cos [S_{0}(y)]=1-\frac{1}{2}S_{0}(y)^{2}+R_{4}(S_{0}^{\ast }(y)). 
\tag{5.26b}
\end{equation}%
It is known from (5.18), (5.19), and (5.20) that the residual terms $%
R_{3}(S_{0}^{\ast }(y))$ and $R_{4}(S_{0}^{\ast }(y))$ for $y\in (-\infty
,+\infty )$ are respectively bounded by%
\begin{equation}
|R_{3}(S_{0}^{\ast }(y))|\leq \frac{1}{3!}|S_{0}(y)|^{3},\text{ }%
|R_{4}(S_{0}^{\ast }(y))|\leq \frac{1}{4!}|S_{0}(y)|^{4}.  \tag{5.27}
\end{equation}%
Now by using the two expansions (5.26) and the two equations (5.24) the
state (5.23)\ may be rewritten as%
\begin{equation}
\Psi (x,r,t_{0}+\tau )=\Psi _{0}(x,r,t_{0}+\tau )+E_{r}(x,r,t_{0}+\tau ) 
\tag{5.28}
\end{equation}%
where $\Psi _{0}(x,r,t_{0}+\tau )$ is the Gaussian superposition state, 
\begin{equation*}
\Psi _{0}(x,r,t_{0}+\tau )=\{[1-S_{0}(y)^{2}/2]\cos \beta _{c}+S_{0}(y)\sin
\beta _{c}\}\Psi _{0}(x,t_{0})|g_{0}\rangle
\end{equation*}%
\begin{equation}
+\{[1-S_{0}(y)^{2}/2]\sin \beta _{c}-S_{0}(y)\cos \beta _{c}\}\exp [i\frac{1%
}{2}(k_{0}+k_{1})x]\Psi _{0}(x,t_{0})|e\rangle ,  \tag{5.29}
\end{equation}%
and the error term $E_{r}(x,r,t_{0}+\tau )$ is given by 
\begin{equation*}
E_{r}(x,r,t_{0}+\tau )=\{R_{4}(S_{0}^{\ast }(y))\cos \beta
_{c}+R_{3}(S_{0}^{\ast }(y))\sin \beta _{c}\}\Psi _{0}(x,t_{0})|g_{0}\rangle
\end{equation*}%
\begin{equation}
+\{R_{4}(S_{0}^{\ast }(y))\sin \beta _{c}-R_{3}(S_{0}^{\ast }(y))\cos \beta
_{c}\}\exp [i\frac{1}{2}(k_{0}+k_{1})x]\Psi _{0}(x,t_{0})|e\rangle . 
\tag{5.30}
\end{equation}%
Actually, it can be easily found that the state $\Psi _{0}(x,r,t_{0}+\tau )$
is a superposition of the ten $GWP$ product states, while the upper bound of
the error term may be calculated by%
\begin{equation}
||E_{r}(x,r,t_{0}+\tau )||^{2}=\int dy\{|\Psi
_{0}(y,t_{0})|^{2}[|R_{3}(S_{0}^{\ast }(y))|^{2}+|R_{4}(S_{0}^{\ast
}(y))|^{2}]\}.  \tag{5.31}
\end{equation}%
The exact calculation for the error upper bound is not difficult. Notice
that $|S_{0}(y)|<<1$ and the residual term $|R_{3}(S_{0}^{\ast
}(y))|^{2}\leq |S_{0}(y)|^{6}/36$ \ and $\ |R_{4}(S_{0}^{\ast }(y))|^{2}$ $%
\leq $ $|S_{0}(y)|^{8}/256.$ The \ residual \ term $\ |R_{3}(S_{0}^{\ast
}(y))|^{2}$ $>>$ $|R_{4}(S_{0}^{\ast }(y))|^{2}$ \ and thus $%
|R_{4}(S_{0}^{\ast }(y))|^{2}$ may be neglected in (5.31). Then the
probability of the error term (5.31) may be approximately written as%
\begin{equation*}
||E_{r}(x,r,t_{0}+\tau )||^{2}\thickapprox \int dy\{|\Psi
_{0}(y,t_{0})|^{2}|R_{3}(S_{0}^{\ast }(y))|^{2}\}
\end{equation*}%
\begin{equation}
\leq \frac{1}{36}\int dy\{|\Psi _{0}(y,t_{0})|^{2}|S_{0}(y)|^{6}\}. 
\tag{5.32}
\end{equation}%
Now the function $|S_{0}(y)|^{6}$ in (5.32) may be expanded as 
\begin{equation*}
|S_{0}(y)|^{6}=\sum_{l=0}^{6}\left( 
\begin{array}{c}
6 \\ 
l%
\end{array}%
\right) [\beta _{c}(1-\cos (\frac{1}{2}\Delta ky))]^{6-l}[\beta _{s}\sin (%
\frac{1}{2}\Delta ky)]^{l}
\end{equation*}%
\begin{equation}
\leq \sum_{l=0}^{6}\left( 
\begin{array}{c}
6 \\ 
l%
\end{array}%
\right) \frac{|\beta _{s}^{l}\beta _{c}^{6-l}|}{2^{6-l}}|(\frac{1}{2}\Delta
ky)|^{12-l},  \tag{5.33}
\end{equation}%
where the inequalities $|(1-\cos (\frac{1}{2}\Delta ky))|\leq (\frac{1}{2}%
\Delta ky)^{2}/2$ and $|\sin (\frac{1}{2}\Delta ky)|\leq |(\frac{1}{2}\Delta
ky)|$ have been used. By combining (5.33) with (5.32) and then using the
Gaussian integrals (5.11) it can turn out that the error term $%
E_{r}(x,r,t_{0}+\tau )$ is bounded by\ 
\begin{equation}
||E_{r}(x,r,t_{0}+\tau )||^{2}\leq \frac{1}{36}\sum_{l=0}^{6}\left( 
\begin{array}{c}
6 \\ 
l%
\end{array}%
\right) (\frac{1}{2}\Delta k)^{12-l}\frac{|\beta _{s}^{l}\beta _{c}^{6-l}|}{%
2^{6-l}}I_{12-l}(\varepsilon (t_{0}))  \tag{5.34}
\end{equation}%
where the integral $I_{12-k}(\varepsilon (t_{0}))$ is defined by\newline
\begin{equation*}
I_{12-k}(\varepsilon (t_{0}))=\int_{-\infty }^{\infty }dy\{|\Psi
_{0}(y,t_{0})|^{2}|y|^{12-k}\}
\end{equation*}%
\begin{equation*}
=\left\{ 
\begin{array}{c}
\frac{(12-k-1)!!}{\sqrt{2^{12-k}}}\varepsilon (t_{0})^{12-k},\text{ if }12-k%
\text{ is even} \\ 
\frac{1}{\sqrt{\pi }}[(12-k-1)/2]!\varepsilon (t_{0})^{12-k},\text{ if }12-k%
\text{ is odd}%
\end{array}%
\right.
\end{equation*}%
If the wave-number difference $|\Delta k|$ is small enough, then one may
retain the lowest-order terms of the wave-number difference $|\Delta k|$ on
the $RH$ side of (5.34) and neglect any other higher-order terms. Then the
upper bound of the error term $E_{r}(x,r,t_{0}+\tau )$ may be determined
from 
\begin{equation}
||E_{r}(x,r,t_{0}+\tau )||_{u}^{2}\thickapprox \frac{5}{6144}\beta
_{s}^{6}\varepsilon (t_{0})^{6}(\Delta k)^{6}+\frac{1}{256}\frac{1}{\sqrt{%
\pi }}|\allowbreak \beta _{c}\beta _{s}^{5}|\varepsilon (t_{0})^{7}|\Delta
k|^{7}.  \tag{5.35}
\end{equation}%
The upper bound $||E_{r}(x,r,t_{0}+\tau )||_{u}$ is proportional to $|\beta
_{s}|^{3},$ $\varepsilon (t_{0})^{3},$ and $|\Delta k|^{3}$ approximately.
Thus, the error term $E_{r}(x,r,t_{0}+\tau )$ in (5.28) may be neglected
when the wave-number difference $|\Delta k|,$ the wave-packet spread $%
\varepsilon (t_{0}),$ and/or $|\beta _{s}|,$ $|\beta _{c}|$ are small
enough. Then in this case the state $\Psi (x,r,t_{0}+\tau )$ of (5.23) may
be approximated well by the superposition state $\Psi _{0}(x,r,t_{0}+\tau )$
of (5.29) that consists of the ten $GWP$ product states.

As a second example, consider the following atomic product state to be
approximated by a superposition of a finite number of the $GWP$ states, 
\begin{equation*}
\Psi (x,r,t)=\frac{1}{2}(1+i)\Psi _{0}(x,t_{0})|g_{0}\rangle
\end{equation*}%
\begin{equation*}
+\frac{1}{2}(1-i)\{\cos \beta _{c}\cos [S_{0}(y)]+\sin \beta _{c}\sin
[S_{0}(y)]\}\Psi _{0}(x,t_{0})|g_{0}\rangle
\end{equation*}%
\begin{equation*}
+\frac{1}{2}(1+i)\{\sin \beta _{c}\cos [S_{0}(y)]-\cos \beta _{c}\sin
[S_{0}(y)]\}
\end{equation*}%
\begin{equation}
\times \exp [i\frac{1}{2}(k_{0}+k_{1})x]\Psi _{0}(x,t_{0})|e\rangle 
\tag{5.36}
\end{equation}%
where the function $S_{0}(y)$ is still defined by (5.25). The state $\Psi
(x,r,t)$ with $t=t_{0}+\tau $ may be encountered in the next section. It may
be expressed as $\Psi (x,r,t)=\Psi _{0}(x,r,t)+E_{r}(x,r,t)$, where the
state $\Psi _{0}(x,r,t)$ is a superposition of the ten $GWP$ product states, 
\begin{equation*}
\Psi _{0}(x,r,t)=\frac{1}{2}(1+i)\Psi _{0}(x,t_{0})|g_{0}\rangle
\end{equation*}%
\begin{equation*}
+\frac{1}{2}(1-i)\{[1-S_{0}(y)^{2}/2]\cos \beta _{c}+S_{0}(y)\sin \beta
_{c}\}\Psi _{0}(x,t_{0})|g_{0}\rangle
\end{equation*}%
\begin{equation*}
+\frac{1}{2}(1+i)\{[1-S_{0}(y)^{2}/2]\sin \beta _{c}-S_{0}(y)\cos \beta
_{c}\}
\end{equation*}%
\begin{equation}
\times \exp [i\frac{1}{2}(k_{0}+k_{1})x]\Psi _{0}(x,t_{0})|e\rangle 
\tag{5.37}
\end{equation}%
and the error term $E_{r}(x,r,t)$ is given by%
\begin{equation*}
E_{r}(x,r,t)=\frac{1}{2}(1-i)\{R_{4}(S_{0}^{\ast }(y))\cos \beta
_{c}+R_{3}(S_{0}^{\ast }(y))\sin \beta _{c}\}\Psi _{0}(x,t_{0})|g_{0}\rangle
\end{equation*}%
\begin{equation*}
+\frac{1}{2}(1+i)\{R_{4}(S_{0}^{\ast }(y))\sin \beta _{c}-R_{3}(S_{0}^{\ast
}(y))\cos \beta _{c}\}
\end{equation*}%
\begin{equation}
\times \exp [i\frac{1}{2}(k_{0}+k_{1})x]\Psi _{0}(x,t_{0})|e\rangle . 
\tag{5.38}
\end{equation}%
It can turn out that the probability $||E_{r}(x,r,t)||^{2}$ with $%
t=t_{0}+\tau $ is equal to half the probability $||E_{r}(x,r,t_{0}+\tau
)||^{2}$ of (5.31), that is, $||E_{r}(x,r,t)||^{2}=||E_{r}(x,r,t_{0}+\tau
)||^{2}/2$. Then it follows from (5.35) that the upper bound $%
||E_{r}(x,r,t)||_{u}$ of the error term (5.38) is approximately given by%
\begin{equation}
||E_{r}(x,r,t)||_{u}^{2}\thickapprox \frac{5}{12288}\beta
_{s}^{6}\varepsilon (t_{0})^{6}(\Delta k)^{6}+\frac{1}{512}\frac{1}{\sqrt{%
\pi }}|\allowbreak \beta _{c}\beta _{s}^{5}|\varepsilon (t_{0})^{7}|\Delta
k|^{7}.  \tag{5.39}
\end{equation}%
Therefore, the state $\Psi (x,r,t)$ of (5.36) may be approximated well by
the Gaussian superposition state $\Psi _{0}(x,r,t)$ of (5.37) when the
wave-number difference $|\Delta k|,$ the wave-packet spread $\varepsilon
(t_{0}),$ and/or $|\beta _{s}|,$ $|\beta _{c}|$ are small enough, so that
the error term $E_{r}(x,r,t)$ can be neglected.

In addition to the second-order expansions (5.26a)\ and (5.26b) the
higher-order expansions (5.18) for the trigonometric functions $\sin
[S_{0}(y)]$ and $\cos [S_{0}(y)]$ also may be used to approximate the
original states (5.23) and (5.36), respectively. This means that one may use
a superposition state consisting of more than ten $GWP$ states to
approximate each one of these original states. Then in this case the errors $%
E_{r}(x,r,t_{0}+\tau )$ of (5.30) and $E_{r}(x,r,t)$ of (5.38) will have
smaller upper bounds than those of (5.35) and (5.39), respectively. However,
it can be seen that the present $MGWP$ expansion consisting of the ten $GWP$
states is good enough for the error estimation for the $SSISS$ triggering
pulse in the next section.

The $MGWP$\ expansion method has been extensively applied to calculating
rigorously the upper bounds of a variety of error terms in the paper. In the 
$MGWP$ expansion method (See also the subsection 4.3.2.2) a non-$GWP$
product state $\Psi (x,r,t_{0})$ may be generally expanded as a linear
combination of a finite number of the $GWP$ product states, 
\begin{equation}
\Psi (x,r,t_{0})=\sum_{k=1}^{m}A_{k}\Psi _{0k}(x,r,t_{0})+E_{r}(x,r,t_{0}). 
\tag{5.40}
\end{equation}%
Here $\Psi _{0k}(x,r,t_{0})$ is the $k-$th normalized $GWP$ product state
and $E_{r}(x,r,t_{0})$ the truncation error. If the amplitude $A_{k}$ is a
complex coefficient, then the superposition state on the $RH$ side of (5.40)
is a conventional Gaussian superposition state. However, in a general case
the amplitude $A_{k}$ may be a function of the coordinate $x$, that is, $%
A_{k}=A_{k}(x)$. In many cases the amplitude $A_{k}$ is a complex number or
a finite-order polynomial in coordinate $x$. Then the superposition state is
a generalized Gaussian superposition state. For convenience such a
generalized Gaussian superposition state is still called a Gaussian
superposition state. When the halting-qubit atom in the non-$GWP$ state $%
\Psi (x,r,t_{0})$ is acted on by a unitary propagator $\exp (-iH\tau ),$ its
time evolution process may be expressed as $\Psi (x,r,t_{0}+\tau )=\exp
(-iH\tau )\Psi (x,r,t_{0}).$ Then according to (5.40) the final product
state $\Psi (x,r,t_{0}+\tau )$ may be written as%
\begin{equation}
\Psi (x,r,t_{0}+\tau )=\sum_{k=1}^{m}A_{k}(x(-\tau ))\Psi
_{0k}(x,r,t_{0}+\tau )+\exp (-iH\tau )E_{r}(x,r,t_{0}),  \tag{5.41}
\end{equation}%
where the product state $\Psi _{0k}(x,r,t_{0}+\tau )=\exp (-iH\tau )\Psi
_{0k}(x,r,t_{0})$ and the operator function $A_{k}(x(-t))$ is defined by%
\begin{equation}
A_{k}(x(-\tau ))=\exp (-iH\tau )A_{k}(x)\exp (iH\tau ).  \tag{5.42}
\end{equation}%
It is clear that $A_{k}(x(-\tau ))=A_{k}$ is a complex coefficient if $A_{k}$
is a complex coefficient. A strict calculation for the final product state $%
\Psi (x,r,t_{0}+\tau )$ consists of several steps as follows: $(i)$
calculate the expansion (5.40); $(ii)$ calculate the product state $\Psi
_{0k}(x,r,t_{0}+\tau );$ $(iii)$ calculate the operator function $%
A_{k}(x(-\tau ))$ according to the unitary transformation (5.42); and $(v)$
calculate the product state $A_{k}(x(-\tau ))\Psi _{0k}(x,r,t_{0}+\tau ).$
The first step $(i)$ is usually trivial [25]. This also can be seen in the
present section, the previous subsection 4.3.2.2, and the next section. In
quantum mechanics [22] any wave function can be expanded in terms of the
complete set of the harmonic-oscillator eigenbases. Such an expansion also
could be considered as a special instance of the expansion (5.40). By the
first step $(i)$ one may obtain the truncation error $E_{r}(x,r,t_{0})$ or
its upper bound. Note that $||\exp (-iH\tau
)E_{r}(x,r,t_{0})||=||E_{r}(x,r,t_{0})||.$ This means that one really
obtains the upper bound of the error $\exp (-iH\tau )E_{r}(x,r,t_{0})$ of
the final product state (5.41). It is not easy and tends to be difficult to
perform in an exact form the last three steps. It seems that the second step 
$(ii)$, i.e., the calculation of the product state $\Psi
_{0k}(x,r,t_{0}+\tau ),$ is relatively easy among these three steps. Whether
or not this step is easy is dependent on the Hamiltonian $H$. However, the
third step $(iii)$, i.e., the calculation for the operator function $%
A_{k}(x(-\tau )),$ is generally difficult except for some simple and special
cases. Here the inverse unitary propagator usually plays a crucial role in
the calculation. Since the operator function $A_{k}(x(-\tau ))$ could
contain the momentum operator, whether or not the last step $(v)$ is easy is
dependent on the functional form of $A_{k}(x(-\tau ))$. If $A_{k}(x(-\tau ))$
is a complicated function of the momentum operator, then the last step tends
to be difficult. However, when the Hamiltonian $H$ in the propagator $\exp
(-iH\tau )$ is a general quadratic Hamiltonian, it is not difficult to carry
out exactly all the last three-step calculations, mainly because the inverse
unitary propagator of a general quadratic Hamiltonian can be obtained
exactly [15]. This is the main reason why the $MGWP$ expansion method has
been used extensively throughout the paper.

An important purpose for the $MGWP$\ expansion method to be used in the
error estimation is to determine the deviation-to-spread ratios (or their
lower bounds) for these $GWP$ states $\{\Psi _{0k}(x,r,t_{0}+\tau )\}$ in
(5.41). The deviation-to-spread ratio of a $GWP$ state is an important
control parameter in the error estimation. For example, as shown in the
previous sections 3 and 4, an error term generated by the imperfection of
the $LH$ harmonic potential well and/or the spatially-selective effect of
the $PHAMDOWN$ laser light beams decays exponentially with the square
deviation-to-spread ratios of the relevant $GWP$ states. Suppose that the
time evolution process of the halting-qubit atom from the initial non-$GWP$
state of (5.40) to the final state of (5.41) is governed by the
harmonic-oscillator Hamiltonian $H_{0}^{ho}$ which is a simple quadratic
Hamiltonian. Suppose further that the $GWP$ motional state $\Psi _{0k}(x,t)$
($t=t_{0}+\tau ,$ $0\leq \tau ,$ $1\leq k\leq m$) of the product state $\Psi
_{0k}(x,r,t)$ in (5.41) has the four characteristic parameters $%
\{x_{c}^{k}(t),$ $p_{c}^{k}(t),$ $W_{c}^{k}(t),$ $\varepsilon _{c}^{k}(t)\}$%
. As shown in the section 3, the deviation-to-spread ratio of the state $%
\Psi _{0k}(x,t)$ is defined as $%
y_{M}^{k}(x_{L},x_{c}^{k}(t))=[x_{L}-x_{c}^{k}(t)]/\varepsilon
_{c}^{k}(t)>0. $ For a different time interval $\tau \geq 0$ the motional
state $\Psi _{0k}(x,t_{0}+\tau )$ ($t=t_{0}+\tau $) has different COM
position $x_{c}^{k}(t_{0}+\tau )$ and wave-packet spread $\varepsilon
_{c}^{k}(t_{0}+\tau )$ and hence has different deviation-to-spread ratio $%
y_{M}^{k}(x_{L},x_{c}^{k}(t_{0}+\tau )).$ One method to determine the
deviation-to-spread ratio $y_{M}^{k}(x_{L},x_{c}^{k}(t))$ (or its lower
bound) is to calculate directly the time evolution process: $\Psi
_{0k}(x,t_{0}+\tau )=\exp (-iH_{0}^{ho}\tau )\Psi _{0k}(x,t_{0})\ $for $%
k=1,2,...,m$. For every initial motional state $\Psi _{0k}(x,t_{0})$ one
needs to calculate the time evolution process and obtains the characteristic
parameters of the motional state $\Psi _{0k}(x,t)$ with $t_{0}\leq
t=t_{0}+\tau \leq t_{b}.$ Here $t_{b}$ is the final time of the time
evolution process. Then by using these relevant characteristic parameters
one can calculate the deviation-to-spread ratio $%
y_{M}^{k}(x_{L},x_{c}^{k}(t))$ and finds out the minimum value of $%
y_{M}^{k}(x_{L},x_{c}^{k}(t))$ in the time region $[t_{0}$, $t_{b}]$. This
is an exact method to determine the deviation-to-spread ratio (or its lower
bound).

While the above exact method is not difficult to determine the
deviation-to-spread ratio (or its lower bound), it is a cumbersome method!
There could be simpler methods to determine the deviation-to-spread ratio
(or its lower bound) for a $GWP$ state. One of which is based on the energy
conservation law. It uses the motional energy of each $GWP$ motional state $%
\Psi _{0k}(x,t)$ to determine approximately the lower bound of the
deviation-to-spread ratio $y_{M}^{k}(x_{L},x_{c}^{k}(t)).$ This method is
simple and hence has been used in the paper, although it is not an exact
method. Below this method is described in detail.

According to quantum mechanics [22] the mean motional energy of the harmonic
oscillator in a motional state $\Psi _{0}(x,t)$ at any time $t\in \lbrack
t_{0},t_{b}]$ is given by%
\begin{equation*}
E_{ho}(t)=\langle \Psi _{0}(x,t)|H_{0}^{ho}|\Psi _{0}(x,t)\rangle =\langle
\Psi _{0}(x,t_{0})|H_{0}^{ho}[t-t_{0}]|\Psi _{0}(x,t_{0})\rangle ,
\end{equation*}%
where $H_{0}^{ho}(t_{0})\equiv H_{0}^{ho}$ is the Hamiltonian of the
harmonic oscillator and $H_{0}^{ho}[t-t_{0}]$ is the Hamiltonian operator of
the harmonic oscillator in the Heisenberg picture [22]. Because the
Hamiltonian $H_{0}^{ho}$ is time-independent, according to the Heisenberg
picture there hold the relations: 
\begin{equation*}
H_{0}^{ho}[t-t_{0}]=\exp [iH_{0}^{ho}(t-t_{0})/\hslash
]H_{0}^{ho}(t_{0})\exp [-iH_{0}^{ho}(t-t_{0})/\hslash
]=H_{0}^{ho}(t_{0})\equiv H_{0}^{ho}.
\end{equation*}%
These relations show that the motional energy $E_{ho}(t)$ is equal to $%
E_{ho}(t_{0})$ $($here $E_{ho}\equiv E_{ho}(t_{0})).$ Then the energy
equation $E_{ho}(t)=E_{ho}(t_{0})$ (or $E_{ho}(t)=E_{ho})$ for any time $%
t\in \lbrack t_{0},t_{b}]$ means that the energy conservation law holds for
the harmonic oscillator. It is known that the initial $GWP$ motional states $%
\{\Psi _{0k}(x,t_{0})\}$ in (5.40) evolve into the motional states $\{\Psi
_{0k}(x,t_{0}+\tau )\}$ in (5.41), respectively, according to the time
evolution process: $\Psi _{0k}(x,t_{0}+\tau )=\exp (-iH_{0}^{ho}\tau )\Psi
_{0k}(x,t_{0}).$ This time evolution process obeys the energy conservation
law because the Hamiltonian $H_{0}^{ho}$ is time-independent. Then the
energy conservation law may help one to determine the upper and lower bounds
of the four characteristic parameters $\{x_{c}^{k}(t),$ $p_{c}^{k}(t),$ $%
W_{c}^{k}(t),$ $\varepsilon _{c}^{k}(t)\}$ of the motional state $\Psi
_{0k}(x,t)$ with $t_{0}\leq t=t_{0}+\tau \leq t_{b}.$ Now the mean motional
energy of the harmonic oscillator with the Hamiltonian $H_{0}^{ho}$ and in
the motional state $\Psi _{0k}(x,t)$ at the time $t\in \lbrack t_{0},t_{b}]$
is given by $E_{ho}^{k}(t)=\langle \Psi _{0k}(x,t)|H_{0}^{ho}|\Psi
_{0k}(x,t)\rangle .$ Then a direct calculation by using the Hamiltonian $%
H_{0}^{ho}$ and the $GWP$ motional state $\Psi _{0k}(x,t)$ shows that the
mean motional energy $E_{ho}^{k}(t)$ is given by (See also (3.13a))%
\begin{equation}
E_{ho}^{k}(t)=\frac{p_{c}^{k}(t)^{2}}{2m}+\frac{1}{2}m\omega
^{2}x_{c}^{k}(t)^{2}+\frac{1}{4}m\omega ^{2}\varepsilon _{c}^{k}(t)^{2}+%
\frac{1}{4}\frac{\hslash ^{2}}{2m(\Delta x_{c}^{k}(t))^{2}}  \tag{5.43}
\end{equation}%
where $(\Delta x_{c}^{k}(t))^{2}=2|W_{c}^{k}(t)|^{2}/\varepsilon
_{c}^{k}(t)^{2}$. In particular, for the initial motional state $\Psi
_{0k}(x,t_{0})$ the mean motional energy $E_{ho}^{k}(t_{0})$ (or $%
E_{ho}^{k}) $ can be obtained directly by substituting the four
characteristic parameters $\{x_{c}^{k}(t_{0}),$ $p_{c}^{k}(t_{0}),$ $%
W_{c}^{k}(t_{0}),$ $\varepsilon _{c}^{k}(t_{0})\}$ of the initial motional
state $\Psi _{0k}(x,t_{0})$ into (5.43) with $t=t_{0}.$ The $RH$ side of
(5.43) consists of the four terms, each one of which is non-negative. In
particular, the wave-packet spread $\varepsilon _{c}^{k}(t)>0$ and $%
\varepsilon _{c}^{k}(t)=\{2(\Delta x_{c}^{k}(t))^{2}+2[\hslash T/(2m\Delta
x_{c}^{k}(t))]^{2}\}^{1/2}\geq \sqrt{2}|\Delta x_{c}^{k}(t)|.$ Then the
equation (5.43) shows that $E_{ho}^{k}(t)>\frac{p_{c}^{k}(t)^{2}}{2m},$ $%
E_{ho}^{k}(t)>\frac{1}{2}m\omega ^{2}x_{c}^{k}(t)^{2},$ $E_{ho}^{k}(t)>\frac{%
1}{4}m\omega ^{2}\varepsilon _{c}^{k}(t)^{2},$ and $E_{ho}^{k}(t)>\frac{1}{4}%
\frac{\hslash ^{2}}{2m(\Delta x_{c}^{k}(t))^{2}}.$ Now the energy
conservation law shows that $E_{ho}^{k}(t)=E_{ho}^{k}.$ Then it can be found
that the upper and lower bounds for $\{x_{c}^{k}(t),$ $p_{c}^{k}(t),$ $%
\varepsilon _{c}^{k}(t)\}$ are determined from%
\begin{equation}
0\leq |x_{c}^{k}(t)|\leq |x_{c}^{k}(t)|_{\max }<\sqrt{\frac{2E_{ho}^{k}}{%
m\omega ^{2}}},  \tag{5.44a}
\end{equation}%
\begin{equation}
0\leq |p_{c}^{k}(t)|\leq |p_{c}^{k}(t)|_{\max }<\sqrt{2mE_{ho}^{k}}, 
\tag{5.44b}
\end{equation}%
\begin{equation}
\sqrt{\frac{1}{4}\frac{\hslash ^{2}}{mE_{ho}^{k}}}<\sqrt{2}|\Delta
x_{c}^{k}(t)|_{\min }\leq \sqrt{2}|\Delta x_{c}^{k}(t)|\leq \varepsilon
_{c}^{k}(t)\leq \lbrack \varepsilon _{c}^{k}(t)]_{\max }<\sqrt{\frac{%
4E_{ho}^{k}}{m\omega ^{2}}}.  \tag{5.44c}
\end{equation}%
Note that $|W_{c}^{k}(t)|^{2}=\frac{1}{2}(\Delta
x_{c}^{k}(t))^{2}\varepsilon _{c}^{k}(t)^{2}.$ Then the upper and lower
bounds for the absolute complex linewidth $|W_{c}^{k}(t)|$ are determined
from%
\begin{equation}
\frac{\hslash ^{2}}{8mE_{ho}^{k}}<(\Delta x_{c}^{k}(t))_{\min }^{2}\leq
|W_{c}^{k}(t)|\leq \frac{1}{2}[\varepsilon _{c}^{k}(t)]_{\max }^{2}<\frac{%
2E_{ho}^{k}}{m\omega ^{2}}.  \tag{5.44d}
\end{equation}%
These upper and lower bounds are dependent on the motional energy $%
E_{ho}^{k} $ and the oscillatory frequency $\omega $ of the harmonic
oscillator with the Hamiltonian $H_{0}^{ho}$ except the lower bounds of $%
|x_{c}^{k}(t)|$ and $|p_{c}^{k}(t)|.$ Some of these inequalities (5.44) or
their invariants have appeared in the previous sections 3 and 4 and in the
next section. Once the upper bounds of $|x_{c}^{k}(t)|$ and $\varepsilon
_{c}^{k}(t)$ are obtained, one may calculate the lower bound of the
deviation-to-spread ratio $y_{M}^{k}(x_{L},x_{c}^{k}(t))$ in the time region 
$[t_{0},t_{b}],$%
\begin{equation*}
\lbrack y_{M}^{k}(x_{L},x_{c}^{k}(t))]_{\min }\geq
\{x_{L}-|x_{c}^{k}(t)|_{\max }\}/[\varepsilon _{c}^{k}(t)]_{\max }
\end{equation*}%
\begin{equation}
>\{x_{L}-\sqrt{\frac{2E_{ho}^{k}}{m\omega ^{2}}}\}/\sqrt{\frac{4E_{ho}^{k}}{%
m\omega ^{2}}}>0.  \tag{5.45}
\end{equation}%
There are $m$ different initial $GWP$ motional states $\{\Psi
_{0k}(x,t_{0})\}$ in (5.40), each one $(\Psi _{0k}(x,t_{0}))$ of which may
have different mean motional energy $(E_{ho}^{k})$ and obeys the same time
evolution process: $\Psi _{0k}(x,t_{0}+\tau )=\exp (-iH_{0}^{ho}\tau )\Psi
_{0k}(x,t_{0}).$ Now by inserting the mean motional energy value $E_{ho}^{k}$
into these inequalities (5.44) one may determine the upper and lower bounds
of the four characteristic parameter values $\{|x_{c}^{k}(t)|,$ $%
|p_{c}^{k}(t)|,$ $|W_{c}^{k}(t)|,$ $\varepsilon _{c}^{k}(t)\}$ in the time
region $[t_{0},$ $t_{b}]$ for each one of all these $m$ $GWP$ motional
states $\{\Psi _{0k}(x,t)\}$ in (5.41). Then by using (5.45) one may further
calculate the lower bounds of the deviation-to-spread ratios $%
\{y_{M}^{k}(x_{L},x_{c}^{k}(t))\}$ in the time region $[t_{0},t_{b}]$ by
using the upper bounds of the characteristic parameter values $%
\{|x_{c}^{k}(t)|,$ $\varepsilon _{c}^{k}(t)\}$ for all these $m$ motional
states $\{\Psi _{0k}(x,t)\}.$ It is known from the previous sections 3 and 4
that the upper bounds of these errors generated by the imperfection of the $%
LH$ potential well and/or the spatially-selective effect of the $PHAMDOWN$
laser light beams usually could be determined mainly by the smallest one
among these $m$ minimum deviation-to-spread ratios $%
\{[y_{M}^{k}(x_{L},x_{c}^{k}(t))]_{\min }\}.$ \newline
\newline
\newline
{\Large 6 Generating the spatially-selective and internal-state-selective
triggering pulses}

Two state-selective triggering pulses have been constructed in Ref. [15]
(i.e., the pulse sequences (41a) and (41b) in Ref. [15]). They are the ideal
state-selective triggering pulses and not spatially-selective. They could be
considered as the basic units to construct a higher-order state-selective
triggering pulse [15] with the help of the Trotter-Suzuki decomposition
method. Based on the two state-selective triggering pulses the
spatially-selective and internal-state-selective (or $SSISS$ briefly)
triggering pulses are constructed in this section. Then it is investigated
the possible errors of the $SSISS$ triggering pulses due to the imperfection
of the $LH$ harmonic potential field and the spatially-selective effect of
the $PHAMDOWN$ laser light beams. The theoretical methods to estimate
strictly these possible errors have already been developed and set up in the
previous three sections 3, 4, and 5. These methods also could be helpful to
construct a useful $SSISS$ triggering pulse. As shown in Ref. [15], each one
of the two state-selective triggering pulses consists of a sequence of the $%
PHAMDOWN$ laser light pulses (See also the section 2). The basic pulse
sequences $P_{tr}(\delta t/\sqrt{n})$ for the two state-selective triggering
pulses are given by [15]%
\begin{equation*}
P_{tr}(\delta t/\sqrt{n})=U_{I}(\pi /4,0,\delta t/\sqrt{n})U_{I}(\pi /4,\pi
/2,\delta t/\sqrt{n})
\end{equation*}%
\begin{equation}
\times U_{I}(\pi /4,\pi ,\delta t/\sqrt{n})U_{I}(\pi /4,3\pi /2,\delta t/%
\sqrt{n})  \tag{6.1a}
\end{equation}%
and%
\begin{equation*}
P_{tr}(\delta t/\sqrt{n})=U_{I}(\pi /4,0,\delta t/\sqrt{2n})U_{I}(\pi /4,\pi
/2,\delta t/\sqrt{2n})
\end{equation*}%
\begin{equation*}
\times \lbrack U_{I}(\pi /4,\pi ,\delta t/\sqrt{2n})U_{I}(\pi /4,3\pi
/2,\delta t/\sqrt{2n})]^{2}
\end{equation*}%
\begin{equation}
\times U_{I}(\pi /4,0,\delta t/\sqrt{2n})U_{I}(\pi /4,\pi /2,\delta t/\sqrt{%
2n}).  \tag{6.1b}
\end{equation}%
The state-selective triggering pulses then are generated by 
\begin{equation*}
P_{tr}(\delta t)=[P_{tr}(\delta t/\sqrt{n})]^{n}.
\end{equation*}%
Therefore, each one of them consists of $n$ basic pulse sequences $%
\{P_{tr}(\delta t/\sqrt{n})\}.$ It is clear that the basic pulse sequence
(6.1a) is much simpler than the sequence (6.1b). Both the state-selective
triggering pulses generate the same unitary propagator $\exp \{iQ(\delta
t/\hslash )^{2}\},$ where the Hermitian operator $Q=i[H_{I}(x,\pi /4,0),$ $%
H_{I}(x,\pi /4,\pi /2)]$ with the interaction term $H_{I}(x,\alpha ,\gamma )$
of (4.16), but there are also their own truncation errors [15]. For the
simpler one that consists of the basic pulse sequences of (6.1a) the
truncation error is $O((\delta t)^{3}/\sqrt{n}),$ while the truncation error
is $O((\delta t)^{4}/n)$ for that one consisting of the basic pulse
sequences of (6.1b). Therefore, the state-selective triggering pulse with
the sequence (6.1b) is slightly better than that one with the sequence
(6.1a). The convergence of the unitary propagator $U_{tr}(\delta t)$ of the
state-selective triggering pulse $[P_{tr}(\delta t/\sqrt{n})]^{n}$ for a
large number $n$ may be proven through the theoretical methods in the
previous sections 3, 4, and 5. This convergence could be not only dependent
on the pulse sequence $P_{tr}(\delta t)$ itself but also the wave-packet
motional states of the halting-qubit atom in the $LH$ potential well of the
double-well potential field.

Since the state-selective triggering pulse $P_{tr}(\delta t)$ consists of $n$
basic pulse sequences $\{P_{tr}(\delta t/\sqrt{n})\},$ the basic pulse
sequence $P_{tr}(\delta t/\sqrt{n})$ is first investigated in detail below.
The basic pulse sequence $P_{tr}(\delta t/\sqrt{n})$ of (6.1a) generates the
unitary propagator: 
\begin{equation}
U_{tr}(\delta t/\sqrt{n})=\exp \{iQ(\delta t/\hslash )^{2}/n\}+O((\delta t/%
\sqrt{n})^{3}).  \tag{6.2a}
\end{equation}%
Similarly, the basic pulse sequence (6.1b) generates the unitary propagator:%
\begin{equation}
U_{tr}(\delta t/\sqrt{n})=\exp \{iQ(\delta t/\hslash )^{2}/n\}+O((\delta t/%
\sqrt{n})^{4}).  \tag{6.2b}
\end{equation}%
Here, as a typical example, the simpler basic pulse sequence $P_{tr}(\delta
t/\sqrt{n})$ of (6.1a) will be studied in detail. For simplicity, in the
rotating frame the unitary propagator $U_{I}(\alpha ,\gamma ,\delta t/\sqrt{n%
})$ of the simpler basic pulse sequence (6.1a) is constructed by [15]%
\begin{equation*}
U_{I}(\alpha ,\gamma ,\delta t/\sqrt{n})\overset{\text{def}}{\equiv }\exp
\{Z(\alpha ,\gamma ,\delta t/\sqrt{n})\}
\end{equation*}%
\begin{equation}
=U_{o}^{+}(\delta t/\sqrt{4n})U_{I}(\delta t/\sqrt{n})U_{o}^{+}(\delta t/%
\sqrt{4n})  \tag{6.3}
\end{equation}%
where $U_{I}(\delta t/\sqrt{n})$ is the unitary propagator of the
halting-qubit atom in the ideal harmonic potential well and in the presence
of the ideal $PHAMDOWN$ laser light beams and $U_{o}^{+}(\delta t/\sqrt{4n})$
the inverse of the unitary propagator $U_{o}(\delta t/\sqrt{4n})$ of the
harmonic oscillator. According to the Baker-Campbell-Hausdorff (BCH) formula
[35, 36, 37, 18],%
\begin{equation}
\exp (A/2)\exp (B)\exp (A/2)=\exp \{A+B+\frac{1}{12}[B,[B,A]]-\frac{1}{24}%
[A,[A,B]]+...\},  \tag{6.4}
\end{equation}%
the operator $Z(\alpha ,\gamma ,\delta t/\sqrt{n})$ in (6.3) may be
expressed as 
\begin{equation}
Z(\alpha ,\gamma ,\delta t/\sqrt{n})=-\frac{i}{\hslash }(\delta t/\sqrt{n}%
)H_{I}(x,\alpha ,\gamma )+O((\delta t/\sqrt{n})^{3}).  \tag{6.5}
\end{equation}%
This formula (6.5) could be used better for a short time interval $\delta t/%
\sqrt{n}$ $(i.e.$ $\delta t/\sqrt{n}<<1).$ The operator $Z(\alpha ,\gamma
,\delta t/\sqrt{n})$ is equal to the first-order term on the $RH$ side of
(6.5) if the time interval $\delta t$ is small and/or the number $n$ is
large such that the higher-order term $O((\delta t/\sqrt{n})^{3})$ can be
neglected. In this case one may use reasonably the first-order approximation
of the operator $Z(\alpha ,\gamma ,\delta t/\sqrt{n})$ to obtain the unitary
propagator $U_{I}(\alpha ,\gamma ,\delta t/\sqrt{n})$ in (6.3). Then such a
first-order approximation propagator $U_{I}(\alpha ,\gamma ,\delta t/\sqrt{n}%
)$ is further substituted into (6.1a). This results in that the basic pulse
sequence $P_{tr}(\delta t/\sqrt{n})$ of (6.1a) generates the unitary
propagator:%
\begin{equation*}
U_{tr}(\delta t/\sqrt{n})=\exp \{-\frac{i}{\hslash }H_{I}(x,\pi /4,0)(\delta
t/\sqrt{n})\}\exp \{-\frac{i}{\hslash }H_{I}(x,\pi /4,\pi /2)(\delta t/\sqrt{%
n})\}
\end{equation*}%
\begin{equation}
\times \exp \{-\frac{i}{\hslash }H_{I}(x,\pi /4,\pi )(\delta t/\sqrt{n}%
)\}\exp \{-\frac{i}{\hslash }H_{I}(x,\pi /4,3\pi /2)(\delta t/\sqrt{n})\}. 
\tag{6.6}
\end{equation}%
Note that there is an error $O((\delta t/\sqrt{n})^{3})$ when the basic
pulse sequence (6.1a) is reduced to $U_{tr}(\delta t/\sqrt{n})$. The
propagator (6.6) can be further reduced to the unitary operator $\exp
\{iQ(\delta t/\hslash )^{2}/n\}$ of (6.2a) by using the BCH formula [35, 36,
37, 18]. Here an error $O((\delta t/\sqrt{n})^{3})$ also is generated, as
shown in (6.2a). Therefore, the total error generated by the reduction from
the basic pulse sequence (6.1a) to the propagator $\exp \{iQ(\delta
t/\hslash )^{2}/n\}$ of (6.2a) is still $O((\delta t/\sqrt{n})^{3}).$ On the
other hand, by inserting the rightest side of (6.3) into (6.1a) one may
obtain theoretically the explicit basic pulse sequence of (6.1a). For
convenience, hereafter this theoretical basic pulse sequence of (6.1a) is
denoted as $P_{tr}^{i}(\delta t/\sqrt{n}).$ Then it may be explicitly
written as%
\begin{equation*}
P_{tr}^{i}(\delta t/\sqrt{n})=\exp [\frac{i}{\hslash }H_{0}^{ho}(\delta t/%
\sqrt{4n})]\exp [-\frac{i}{\hslash }(H_{0}^{ho}+H_{I}(x,\pi /4,0))(\delta t/%
\sqrt{n})]
\end{equation*}%
\begin{equation*}
\times \exp [\frac{i}{\hslash }H_{0}^{ho}(\delta t/\sqrt{n})]\exp [-\frac{i}{%
\hslash }(H_{0}^{ho}+H_{I}(x,\pi /4,\pi /2))(\delta t/\sqrt{n})]
\end{equation*}%
\begin{equation*}
\times \exp [\frac{i}{\hslash }H_{0}^{ho}(\delta t/\sqrt{n})]\exp [-\frac{i}{%
\hslash }(H_{0}^{ho}+H_{I}(x,\pi /4,\pi ))(\delta t/\sqrt{n})]\exp [\frac{i}{%
\hslash }H_{0}^{ho}(\delta t/\sqrt{n})]
\end{equation*}%
\begin{equation}
\times \exp [-\frac{i}{\hslash }(H_{0}^{ho}+H_{I}(x,\pi /4,3\pi /2))(\delta
t/\sqrt{n})]\exp [\frac{i}{\hslash }H_{0}^{ho}(\delta t/\sqrt{4n})]. 
\tag{6.7a}
\end{equation}%
It must be pointed out that in theory the $PHAMDOWN$ laser light beams of
the basic pulse sequence of (6.7a) are applied to the halting-qubit atom
over the whole coordinate space $(-\infty ,+\infty )$ and the halting-qubit
atom is in the ideal harmonic potential well. Then in these ideal conditions
the basic pulse sequence $P_{tr}^{i}(\delta t/\sqrt{n})$ can generate the
desired propagator $U_{tr}(\delta t/\sqrt{n})$ of (6.6) up to the error term 
$O((\delta t/\sqrt{n})^{3})$ and hence the state-selective triggering pulse $%
P_{tr}(\delta t)=[P_{tr}^{i}(\delta t/\sqrt{n})]^{n}$ generates the desired
propagator of (5.1) in theory.

In practice the $PHAMDOWN$ laser light beams could be space-selectively
applied to the halting-qubit atom within the $LH$ potential well whose
spatial region is $(-\infty ,x_{L})$ instead of the whole coordinate space $%
(-\infty ,+\infty ).$ Such $PHAMDOWN$ laser light beams are spatially
selective. Moreover, the $LH$ harmonic potential well of the double-well
potential field is not yet an ideal harmonic potential well. If now these
spatially-selective $PHAMDOWN$ laser light beams and the double-well
potential field\ $V(x)$ of (2.1) replace the ideal $PHAMDOWN$ laser light
beams and the ideal harmonic potential field, respectively, then does the
basic pulse sequence $P_{tr}^{i}(\delta t/\sqrt{n})$ of (6.7a) still work
well? That is, does the basic pulse sequence (6.7a) still generate the
propagator $U_{tr}(\delta t/\sqrt{n})$ of (6.6)? This section is devoted to
proving that this spatially-selective basic pulse sequence with these
spatially-selective $PHAMDOWN$ laser light beams and the double-well
potential field\ of (2.1) still may generate the propagator $U_{tr}(\delta t/%
\sqrt{n})$ of (6.6). Then it is necessary to investigate how the
imperfections of these spatially-selective $PHAMDOWN$ laser light beams and
the $LH$ harmonic potential well affect this spatially-selective basic pulse
sequence (i.e., the $SSISS$ basic pulse sequence). All the possible errors
generated by the $SSISS$ basic pulse sequence also need to be calculated
strictly and proven to be negligible or controllable. These are the main
task in the present section. First of all, according to the theoretical
basic pulse sequence of (6.7a) one may construct the corresponding $SSISS$
basic pulse sequence (i.e., the experimental basic pulse sequence). This
experimental basic pulse sequence may be explicitly written as 
\begin{equation*}
P_{tr}^{r}(\delta t/\sqrt{n})=\exp [-\frac{i}{\hslash }H\delta t_{1}^{\prime
}]\exp [-\frac{i}{\hslash }(H+H_{I}^{r}(x,\pi /4,0))(\delta t/\sqrt{n})]
\end{equation*}%
\begin{equation*}
\times \exp [-\frac{i}{\hslash }H\delta t_{1}]\exp [-\frac{i}{\hslash }%
(H+H_{I}^{r}(x,\pi /4,\pi /2))(\delta t/\sqrt{n})]
\end{equation*}%
\begin{equation*}
\times \exp [-\frac{i}{\hslash }H\delta t_{1}]\exp [-\frac{i}{\hslash }%
(H+H_{I}^{r}(x,\pi /4,\pi ))(\delta t/\sqrt{n})]\exp [-\frac{i}{\hslash }%
H\delta t_{1}]
\end{equation*}%
\begin{equation}
\times \exp [-\frac{i}{\hslash }(H+H_{I}^{r}(x,\pi /4,3\pi /2))(\delta t/%
\sqrt{n})]\exp [-\frac{i}{\hslash }H\delta t_{1}^{\prime }].  \tag{6.7b}
\end{equation}%
Here $H$ is the Hamiltonian of the halting-qubit atom in the $LH$ potential
well of the double-well potential field of (2.1), while $H+H_{I}^{r}(x,\pi
/4,\gamma )$ is the Hamiltonian of the atom in the $LH$ potential well and
in the presence of the spatially-selective $PHAMDOWN$ laser light beams. The
experimental basic pulse sequence (6.7b) is space-selective and could be
used directly as a $SSISS$ triggering pulse. Such an experimental pulse
sequence as $P_{tr}^{r}(\delta t/\sqrt{n})$ could be directly realized in
experiment (See also its more practical version in the next paragraph). Both
the experimental basic pulse sequence (6.7b) and the theoretical one (6.7a)
are really the product of nine unitary propagators. One can find that the
experimental propagator $\exp [-iH\tau ^{\prime }/\hslash ]$ ($\tau ^{\prime
}=\delta t_{1}^{\prime }$ or $\delta t_{1}$) in (6.7b) corresponds to the
theoretical propagator $\exp [iH_{0}^{ho}\tau /\hslash ]$ ($\tau =\delta t/%
\sqrt{4n}$ or $\delta t/\sqrt{n}$) in (6.7a). In the theoretical pulse
sequence $P_{tr}^{i}(\delta t/\sqrt{n})$ the Hamiltonian $H_{0}^{ho}$ is of
the halting-qubit atom in the ideal harmonic potential well, while in the
experimental basic pulse sequence $P_{tr}^{r}(\delta t/\sqrt{n})$ the
Hamiltonian $H$ is of the halting-qubit atom in the double-well potential
field of (2.1). The latter may be written as $H=H_{0}^{ho}+V_{1}^{ho}(x,%
\varepsilon ),$ where the harmonic-oscillator Hamiltonian $H_{0}^{ho}$ is
given by (2.3) and the smooth spatially-selective perturbation term $%
V_{1}^{ho}(x,\varepsilon )$ by (2.10). In theory the inverse propagator $%
\exp [iH_{0}^{ho}\tau /\hslash ]$ in (6.7a) should correspond to the inverse
propagator $\exp [iH\tau /\hslash ]$ which would appear in the experimental
basic pulse sequence (6.7b). But the inverse propagator $\exp [iH\tau
/\hslash ]$ may not be realized directly in experiment for the halting-qubit
atom in the double-well potential field. Then this problem needs to be
solved for the experimental basic pulse sequence. Notice that the propagator 
$\exp [iH_{0}^{ho}\tau /\hslash ]$ is an inverse propagator of the harmonic
oscillator. As shown in Ref. [15], it may be expressed as $\exp
[iH_{0}^{ho}\tau /\hslash ]=\exp [-iH_{0}^{ho}\tau ^{\prime }/\hslash ]$
with $\omega \tau ^{\prime }=2k\pi -\omega \tau $ up to a global phase
factor. Now the propagator $\exp [-iH_{0}^{ho}\tau ^{\prime }/\hslash ]$ may
be realized directly in experiment for the harmonic oscillator. By replacing
the Hamiltonian $H_{0}^{ho}$ in the propagator $\exp [-iH_{0}^{ho}\tau
^{\prime }/\hslash ]$ with the spatially-selective Hamiltonian $%
H=H_{0}^{ho}+V_{1}^{ho}(x,\varepsilon )$ one may obtain the
spatially-selective propagator $\exp [-iH\tau ^{\prime }/\hslash ]$ with $%
\tau ^{\prime }=\delta t_{1}^{\prime }$ or $\delta t_{1}$ in (6.7b). Now
this spatially-selective propagator $\exp [-iH\tau ^{\prime }/\hslash ]$ may
be directly realized for the halting-qubit atom in the double-well potential
field. Therefore, there are the corresponding relations between the
theoretical and experimental propagators: 
\begin{equation*}
\exp [iH_{0}^{ho}(\delta t/\sqrt{4n})/\hslash ]=\exp [-iH_{0}^{ho}\delta
t_{1}^{\prime }/\hslash ]\leftrightarrow \exp [-iH\delta t_{1}^{\prime
}/\hslash ],
\end{equation*}%
and 
\begin{equation*}
\exp [iH_{0}^{ho}(\delta t/\sqrt{n})/\hslash ]=\exp [-iH_{0}^{ho}\delta
t_{1}/\hslash ]\leftrightarrow \exp [-iH\delta t_{1}/\hslash ].
\end{equation*}%
Obviously, such a replacement $\exp [iH_{0}^{ho}\tau /\hslash
]\leftrightarrow \exp [-iH\tau ^{\prime }/\hslash ]$ with $\tau =\delta t/%
\sqrt{4n}$ $\leftrightarrow \tau ^{\prime }=\delta t_{1}^{\prime }$ or $\tau
=\delta t/\sqrt{n}\leftrightarrow \tau ^{\prime }=\delta t_{1}$ must ensure
that the experimental basic pulse sequence (6.7b) has almost the same
performance as the theoretical one of (6.7a). This is a crucial precondition
for the experimental basic pulse sequence (6.7b) and its $SSISS$ triggering
pulse to be useful. Thus, it is necessary to show in what conditions this
precondition can be met in the section. The present method could be simpler
to solve the problem. Of course, a more complex method could be required to
solve the problem for the halting-qubit atom in a more complex external
potential field, but it will not be discussed here. Similarly, one also can
find that the experimental propagator $\exp [-i(H+H_{I}^{r}(x,\alpha ,\gamma
))\tau /\hslash ]$ $(\tau =\delta t/\sqrt{n})$ in (6.7b) corresponds to the
theoretical propagator $\exp [-i(H_{0}^{ho}+H_{I}(x,\alpha ,\gamma ))\tau
/\hslash ]$ in (6.7a). The Hamiltonian $H_{I}^{r}(x,\alpha ,\gamma )$ in
(6.7b) is the electric dipole interaction between the halting-qubit atom and
the spatially-selective $PHAMDOWN$ laser light beams within the $LH$
potential well $(-\infty <x<x_{L})$. Since the $PHAMDOWN$ laser light beams
are applied space-selectively to the halting-qubit atom, in (6.7b) the total
Hamiltonian $H+H_{I}^{r}(x,\alpha ,\gamma )$ is really given by $%
H+H_{I}^{r}(x,\alpha ,\gamma )=H_{0}^{ho}+H_{I}(x,\alpha ,\gamma
)+V_{1}(x,\alpha ,\gamma ,\varepsilon ),$ where the smooth
spatially-selective perturbation term $V_{1}(x,\alpha ,\gamma ,\varepsilon )$
is given by (2.12) and $H_{I}(x,\alpha ,\gamma )$ by (4.16) with $-\infty
<x<+\infty $. Obviously, both the spatially-selective perturbation terms $%
V_{1}^{ho}(x,\varepsilon )$ and $V_{1}(x,\alpha ,\gamma ,\varepsilon )$
result in that the experimental basic pulse sequence $P_{tr}^{r}(\delta t/%
\sqrt{n})$ is different from the theoretical one $P_{tr}^{i}(\delta t/\sqrt{n%
})$. Notice that the spatially-selective effect of the perturbation terms $%
V_{1}^{ho}(x,\varepsilon )$ and $V_{1}(x,\alpha ,\gamma ,\varepsilon )$ on
the halting-qubit atom in the $LH$ potential well becomes small and small as
the joint position $x_{L}$ becomes large and large. When the joint position $%
x_{L}$ approaches the infinite point $+\infty $, the experimental basic
pulse sequence approaches the theoretical one, 
\begin{equation*}
\lim_{x_{L}\rightarrow +\infty }P_{tr}^{r}(\delta t/\sqrt{n})=P_{tr}(\delta
t/\sqrt{n}).
\end{equation*}%
The effect of the imperfection of the $LH$ potential well and the
spatially-selective effect of the $PHAMDOWN$ laser light pulses of the
experimental basic pulse sequence may be evaluated strictly on the basis of
the theoretical methods developed in the previous sections 3, 4, and 5. In
the following it proves that the experimental basic pulse sequence $%
P_{tr}^{r}(\delta t/\sqrt{n})$ generates the propagator $U_{tr}(\delta t/%
\sqrt{n})$ of (6.6), which may be further reduced to the propagator $\exp
\{iQ(\delta t/\hslash )^{2}/n\}$ in (6.2a). The possible errors of the
experimental basic pulse sequence $P_{tr}^{r}(\delta t/\sqrt{n})$ are also
estimated strictly, with the help of these theoretical methods developed in
the sections 3, 4, and 5.

A more practical potential field that is used to generate the $SSISS$
triggering pulse could be defined by%
\begin{equation*}
V_{rw}(x)=\left\{ 
\begin{array}{c}
0,\text{ }-\infty <x<-x_{R}-R \\ 
R_{h},\text{ }-x_{R}-R<x<-x_{R} \\ 
\frac{1}{2}m\omega ^{2}x^{2},\text{ }-x_{R}<x<x_{L} \\ 
L_{h},\text{ }x_{L}<x<x_{L}+L \\ 
0,\text{ \ \ \ }x_{L}+L<x<+\infty%
\end{array}%
\right.
\end{equation*}%
where $x_{R}>x_{L};$ $R>>\varepsilon ,$ $L>>\varepsilon ;$ $R_{h}=\frac{1}{2}%
m\omega ^{2}x_{R}^{2},$ $L_{h}=\frac{1}{2}m\omega ^{2}x_{L}^{2}.$ Here the
halting-qubit atom locates within the central harmonic potential well with
the spatial region $(-x_{R},$ $x_{L}).$ The spatially-selective $PHAMDOWN$
laser light beams then are applied to the atom within the central harmonic
potential well $(-x_{R},$ $x_{L}).$ The potential field $V_{rw}(x)$ is more
close to the real world. It approaches the double-well potential field $V(x)$
of (2.1) when $x_{R}\rightarrow \infty .$ The experimental basic pulse
sequence such as (6.7b) also may be generated by the potential field $%
V_{rw}(x)$ and its corresponding spatially-selective $PHAMDOWN$ laser light
beams. Such an experimental basic pulse sequence is more close to the real
world and could be directly realized in experiment. The difference between
the potential field $V_{rw}(x)$ and the double-well potential field $V(x)$
generates only the difference between their corresponding
spatially-selective perturbation terms $V_{1}(x,t)$ in their total
Hamiltonians of (2.2). A similar difference also exists between their
corresponding spatially-selective $PHAMDOWN$ laser light beams. These result
in that their $SSISS$ triggering pulses do not have an essential difference,
because those errors generated by the spatially-selective effects can be
neglected, as shown in the previous sections 3 and 4. Below only the
experimental basic pulse sequence (6.7b) is calculated strictly. A similar
calculated result should be obtained if the experimental basic pulse
sequence (6.7b) is generated by the potential field $V_{rw}(x)$ and its
corresponding spatially-selective $PHAMDOWN$ laser light beams.

It is not easy to calculate the time evolution process of the halting-qubit
atom under the experimental basic pulse sequence $P_{tr}^{r}(\delta t/\sqrt{n%
})$ and carry out a strict error estimation. The nine unitary propagators in
the experimental basic pulse sequence may be divided into the two different
types of propagators: $\exp (-iH\tau ^{\prime }/\hslash )$ ($\tau ^{\prime
}=\delta t_{1}^{\prime }$ and $\delta t_{1}$) and $\exp
[-i(H+H_{I}^{r}(x,\alpha ,\gamma ))\tau /\hslash ]$ ($\tau =\delta t/\sqrt{n}
$). Therefore, one needs to calculate the two different types of time
evolution processes. These two types of time evolution processes have
already been calculated strictly in the previous sections 3 and 4,
respectively. These calculated results may be used here. The first kind of
time evolution process is governed by the propagator $\exp (-iH\tau ^{\prime
}/\hslash )$ ($\tau ^{\prime }=\delta t_{1}^{\prime }$ or $\delta t_{1}$).
It may be calculated strictly by using the formula (3.25). The second kind
is governed by the propagator $\exp [-i(H+H_{I}^{r}(x,\alpha ,\gamma ))\tau
/\hslash ]$ ($\tau =\delta t/\sqrt{n}$). It may be calculated strictly by
using the formula (4.13). Denote $\Psi _{0}(x,r,t_{0})$ as the initial
product state for the propagator $\exp (-iH\tau ^{\prime }/\hslash )$ or $%
\exp [-i(H+H_{I}^{r}(x,\alpha ,\gamma ))\tau /\hslash ].$ The initial state
could not be a pure $GWP$ state or a Gaussian superposition state, but it
may be generally expressed as%
\begin{equation}
\Psi _{0}(x,r,t_{0})=\Psi _{00}(x,r,t_{0})+E_{r}^{d}(x,r,t_{0}),  \tag{6.8}
\end{equation}%
where $\Psi _{00}(x,r,t_{0})$ is a Gaussian superposition state, while $%
E_{r}^{d}(x,r,t_{0})$ measures the difference between the initial Gaussian
superposition state $\Psi _{00}(x,r,t_{0})$ and the initial state $\Psi
_{0}(x,r,t_{0}).$ The expansion (6.8) may simplify greatly the theoretical
calculation of the experimental basic pulse sequence $P_{tr}^{r}(\delta t/%
\sqrt{n})$ and also the theoretical one $P_{tr}^{i}(\delta t/\sqrt{n}).$
Here take an example to show this. When an initial $GWP$ state is acted on
by the propagator $\exp [-i(H+H_{I}^{r}(x,\alpha ,\gamma ))\tau /\hslash ],$
it is generally converted into a non-$GWP$ state. If now this non-$GWP$
state is further acted on by the harmonic-oscillator propagator $\exp
(-iH_{0}^{ho}\tau /\hslash ),$ then it is difficult to calculate exactly the
time evolution process under the propagator $\exp (-iH_{0}^{ho}\tau /\hslash
)$. However, this non-$GWP$ state may be expanded approximately as a
superposition of a finite number $GWP$ states such as $\Psi _{00}(x,r,t_{0})$
in (6.8) according to the $MGWP$ expansion in the section 5. Here the
truncation error in the $MGWP$ expansion is just $E_{r}^{d}(x,r,t_{0})$ in
(6.8). It can be controlled. Now the time evolution process for the Gaussian
superposition state $\Psi _{00}(x,r,t_{0})$ under the propagator $\exp
(-iH_{0}^{ho}\tau /\hslash )$ can be calculated exactly.

Now calculate the time evolution process for the initial state $\Psi
_{0}(x,r,t_{0})$ of (6.8) under the first kind of propagator $\exp (-iH\tau
^{\prime }/\hslash )$ ($\tau ^{\prime }=\delta t_{1}^{\prime }$ or $\delta
t_{1}$). The time evolution process may be formally expressed as%
\begin{equation*}
\Psi (x,r,t_{0}+\tau ^{\prime })\overset{\text{def}}{\equiv }\exp
\{-i[H_{0}^{ho}+V_{1}^{ho}(x,\varepsilon )]\tau ^{\prime }/\hslash \}\Psi
_{0}(x,r,t_{0})
\end{equation*}%
\begin{equation}
=\Psi _{0}(x,r,t_{0}+\tau ^{\prime })+E_{r}^{(12)}(x,r,t_{0}+\tau ^{\prime
})+E_{r1}^{(3)}(x,r,t_{0}+\tau ^{\prime })  \tag{6.9}
\end{equation}%
where the expected product state $\Psi _{0}(x,r,t_{0}+\tau ^{\prime })$ is
written as%
\begin{equation}
\Psi _{0}(x,r,t_{0}+\tau ^{\prime })=\exp [-iH_{0}^{ho}\tau ^{\prime
}/\hslash ]\Psi _{0}(x,r,t_{0}).  \tag{6.10}
\end{equation}%
The errors $E_{r}^{(12)}(x,r,t_{0}+\tau ^{\prime })$ and $%
E_{r1}^{(3)}(x,r,t_{0}+\tau ^{\prime })$ in (6.9) are obtained below. The
expansion (6.8) for the initial state $\Psi _{0}(x,r,t_{0})$ is first
substituted into (6.9). Then according to the exact expression (3.25) in the
section 3, where $H_{0}=H_{0}^{ho}$ and $V_{1}$ is equal to $%
V_{1}^{ho}(x,\varepsilon )$ of (2.10), the time evolution process for the
initial Gaussian superposition state $\Psi _{00}(x,r,t_{0})$ under the
propagator $\exp (-iH\tau ^{\prime }/\hslash )$ may be expressed as%
\begin{equation*}
\exp \{-i[H_{0}^{ho}+V_{1}^{ho}(x,\varepsilon )]\tau ^{\prime }/\hslash
\}\Psi _{00}(x,r,t_{0})
\end{equation*}%
\begin{equation*}
=\exp [-iH_{0}^{ho}\tau ^{\prime }/\hslash ]\Psi
_{00}(x,r,t_{0})+E_{r}^{(12)}(x,r,t_{0}+\tau ^{\prime })
\end{equation*}%
\begin{equation}
=\Psi _{0}(x,r,t_{0}+\tau ^{\prime })+E_{r}^{(12)}(x,r,t_{0}+\tau ^{\prime
})-\exp [-iH_{0}^{ho}\tau ^{\prime }/\hslash ]E_{r}^{d}(x,r,t_{0}). 
\tag{6.11}
\end{equation}%
Here $\Psi _{0}(x,r,t_{0}+\tau ^{\prime })$ is just given by (6.10). As
shown in the section 3, the error $E_{r}^{(12)}(x,r,t_{0}+\tau ^{\prime })$
originates from the imperfection of the $LH$ potential well. It is generated
directly by the spatially-selective perturbation term $V_{1}^{ho}(x,%
\varepsilon ).$ If $\Psi _{00}(x,r,t_{0})$ is a single $GWP$ state, then the
error is given by, according to the exact expression (3.25),%
\begin{equation}
E_{r}^{(12)}(x,r,t_{0}+\tau ^{\prime })=E_{r}^{(1)}(x,r,t_{0}+\tau ^{\prime
})+E_{r}^{(2)}(x,r,t_{0}+\tau ^{\prime }).  \tag{6.12a}
\end{equation}%
Here $E_{r}^{(1)}(x,r,t_{0}+\tau ^{\prime })$ is the first-order
approximation error term and $E_{r}^{(2)}(x,r,$ $t_{0}+\tau ^{\prime })$ is
the correction to the first-order approximation error term. The two errors
are explicitly given by the two formulae (3.27), respectively. Their upper
bounds can be calculated strictly with the theoretical method in the section
3, and the initial state used in the calculation is the Gaussian
superposition state $\Psi _{00}(x,r,t_{0})$ in (6.8) instead of the original
state $\Psi _{0}(x,r,t_{0}).$ These two upper bounds can be determined from
the two inequalities (3.28) (or (3.12)) and (3.30), respectively. It has
been shown in the section 3 that the upper bound for each one of the two
error terms consists of a finite number of the basic norms $\{NBAS1\}$
and/or $\{NBAS2\}$. Therefore, both the error terms decay exponentially with
the square deviation-to-spread ratios of the relevant $GWP$ states. When the
joint position $x_{L}$ in the double-well potential field of (2.1) is large
enough, both the error terms are so small that they can be neglected. In
general, suppose that $\Psi _{00}(x,r,t_{0})$ is a Gaussian superposition
state consisting of $m$ $GWP$ states: $\Psi
_{00}(x,r,t_{0})=\sum_{k=1}^{m}A_{k}\Psi _{0k}(x,r,t_{0}).$ Then for each
normalized $GWP$ state, e.g., $\Psi _{0k}(x,r,t_{0}),$ one may separately
calculate the upper bounds of the two errors $E_{rk}^{(1)}(x,r,t_{0}+\tau
^{\prime })$ and $E_{rk}^{(2)}(x,r,t_{0}+\tau ^{\prime })$ by using the two
inequalities (3.28) (or (3.12)) and (3.30), respectively. Then the total
upper bound for the error $E_{r}^{(l)}(x,r,t_{0}+\tau ^{\prime })$ with $l=1$
or $2$ may be obtained by summing up all these upper bounds of the $m$
errors $\{E_{rk}^{(l)}(x,r,t_{0}+\tau ^{\prime })\}$ according to the
inequality (3.64) in the section 3. Thus, it also consists of a finite
number of the basic norms $\{NBAS1\}$ and $\{NBAS2\}$, indicating that the
total upper bound for the error $E_{r}^{(12)}(x,r,t_{0}+\tau ^{\prime })$ in
(6.12a) (or (6.9)) decays exponentially with the square deviation-to-spread
ratios of the relevant $GWP$ states. When the joint position $x_{L}$ is
large enough, the error $E_{r}^{(12)}(x,r,t_{0}+\tau ^{\prime })$ in (6.9)
also can be neglected. On the other hand, the error $E_{r1}^{(3)}(x,r,t_{0}+%
\tau ^{\prime })$ in (6.9) comes from the deviation of the initial Gaussian
superposition state $\Psi _{00}(x,r,t_{0})$ from the initial state $\Psi
_{0}(x,r,t_{0}).$ Now these relations (6.8), (6.9), and (6.11) together show
that the error can be written as%
\begin{equation*}
E_{r1}^{(3)}(x,r,t_{0}+\tau ^{\prime })=\exp
\{-i[H_{0}^{ho}+V_{1}^{ho}(x,\varepsilon )]\tau ^{\prime }/\hslash
\}E_{r}^{d}(x,r,t_{0})
\end{equation*}%
\begin{equation}
-\exp [-iH_{0}^{ho}\tau ^{\prime }/\hslash ]E_{r}^{d}(x,r,t_{0})\newline
.  \tag{6.12b}
\end{equation}%
This formula directly results in that the error is bounded by 
\begin{equation}
||E_{r1}^{(3)}(x,r,t_{0}+\tau ^{\prime })||\leq 2||E_{r}^{d}(x,r,t_{0})||. 
\tag{6.12c}
\end{equation}%
Since the deviation $E_{r}^{d}(x,r,t_{0})$ in (6.8) can be controlled, the
inequality (6.12c) means that the error $E_{r1}^{(3)}(x,r,t_{0}+\tau
^{\prime })$ also can be controlled. In particular, when the deviation $%
E_{r}^{d}(x,r,t_{0})$ is equal to zero, there is not the error $%
E_{r1}^{(3)}(x,r,t_{0}+\tau ^{\prime })$ in (6.9). Then the upper bound of
the error $E_{r}^{(12)}(x,r,t_{0}+\tau ^{\prime })$ can be calculated
strictly from the two formulae (3.27) by using directly the initial Gaussian
superposition state $\Psi _{0}(x,r,t_{0})$ of (6.8).

The second kind of time evolution process is governed by the propagator $%
\exp [-i(H+H_{I}^{r}(x,\alpha ,\gamma ))\tau /\hslash ]$ ($\tau =\delta t/%
\sqrt{n}<<1$). It may be formally written as%
\begin{equation*}
\Psi (x,r,t_{0}+\tau )\overset{\text{def}}{\equiv }\exp
[-i(H_{0}^{ho}+H_{I}(x,\alpha ,\gamma )+V_{1}(x,\alpha ,\gamma ,\varepsilon
))\tau /\hslash ]\Psi _{0}(x,r,t_{0})
\end{equation*}%
\begin{equation}
=\Psi _{0}(x,r,t_{0}+\tau )+E_{r}^{0}(x,r,t_{0}+\tau
)+E_{r}^{V}(x,r,t_{0}+\tau )+E_{r2}^{(3)}(x,r,t_{0}+\tau ),  \tag{6.13}
\end{equation}%
where the desired product state $\Psi _{0}(x,r,t_{0}+\tau )$ is written as 
\begin{equation}
\Psi _{0}(x,r,t_{0}+\tau )=\exp [-\frac{i}{2\hslash }H_{0}^{ho}\tau ]\exp [-%
\frac{i}{\hslash }H_{I}(x,\alpha ,\gamma )\tau ]\exp [-\frac{i}{2\hslash }%
H_{0}^{ho}\tau ]\Psi _{0}(x,r,t_{0}).  \tag{6.14}
\end{equation}%
There are three error terms on the $RH$ side of (6.13). These error terms
can be obtained below. The initial product state $\Psi _{0}(x,r,t_{0})$ of
(6.8) is first substituted into (6.13). Then according to the exact formulae
(4.13) and (4.14) in the previous section 4 the time evolution process for
the initial Gaussian superposition state $\Psi _{00}(x,r,t_{0})$ in (6.8)
under the propagator $\exp [-i(H+H_{I}^{r}(x,\alpha ,\gamma ))\tau /\hslash
] $ can be written as%
\begin{equation*}
\exp [-i(H_{0}^{ho}+H_{I}(x,\alpha ,\gamma )+V_{1}(x,\alpha ,\gamma
,\varepsilon ))\tau /\hslash ]\Psi _{00}(x,r,t_{0})
\end{equation*}%
\begin{equation*}
=\Psi _{0}(x,r,t_{0}+\tau )+E_{r}^{0}(x,r,t_{0}+\tau
)+E_{r}^{V}(x,r,t_{0}+\tau )
\end{equation*}%
\begin{equation}
-\exp [-\frac{i}{2\hslash }H_{0}^{ho}\tau ]\exp [-\frac{i}{\hslash }%
H_{I}(x,\alpha ,\gamma )\tau ]\exp [-\frac{i}{2\hslash }H_{0}^{ho}\tau
]E_{r}^{d}(x,r,t_{0}).  \tag{6.15}
\end{equation}%
Here the expected product state $\Psi _{0}(x,r,t_{0}+\tau )$ is just the
product state (6.14). Both the errors $E_{r}^{0}(x,r,t_{0}+\tau )$ and $%
E_{r}^{V}(x,r,t_{0}+\tau )$ in (6.15) can be calculated strictly with the
theoretical method in the section 4 and the initial product state used in
the calculation is the initial Gaussian superposition state $\Psi
_{00}(x,r,t_{0})$ in (6.8) instead of $\Psi _{0}(x,r,t_{0}).$ As shown in
the subsection 4.3, the error $E_{r}^{V}(x,r,t_{0}+\tau )$ originates from
the imperfection of the $LH$ harmonic potential well and the
spatially-selective effect of the $PHAMDOWN$ laser light beams. It is
generated directly by the spatially-selective perturbation term $%
V_{1}(x,\alpha ,\gamma ,\varepsilon ).$ Consider first the simple case that $%
\Psi _{00}(x,r,t_{0})$ is a Gaussian superposition state: $\Psi
_{00}(x,r,t_{0})=\Psi _{0}^{g}(x,t_{0})|g_{0}\rangle +\Psi
_{0}^{e}(x,t_{0})|e\rangle $, where $\Psi _{0}^{a}(x,t_{0})$ with $a=g$ or $%
e $ is a single $GWP$ state with no normalization. In this simple case it
proves in the section 4 that the upper bound of the error $%
E_{r}^{V}(x,r,t_{0}+\tau )$ consists of a finite number of the basic norms $%
\{NBAS1\}$ and $\{NBAS2\}$. Therefore, the error decays exponentially with
the square deviation-to-spread ratios of the relevant $GWP$ states. It can
be controlled by the joint position $x_{L}$. It may be neglected when the
joint position $x_{L}$ is large enough. The error $E_{r}^{0}(x,r,t_{0}+\tau
) $ originates from the truncation approximation of the decomposition
formula (4.1). It is usually the main error term in the time evolution
process of (6.15). It is independent on the imperfection of the $LH$
harmonic potential well and the spatially-selective effect of the $PHAMDOWN$
laser light beams. Its upper bound is calculated strictly in the subsection
4.4. As shown in the subsection 4.4, in addition to the time interval $\tau $
the upper bound of the error is dependent on the three types of parameters
including the four characteristic parameters of a $GWP$ state and the
characteristic parameters of both the $PHAMDOWN$ laser light beams and the
harmonic potential field. It can be determined from (4.190). It is
proportional to $\tau ^{3}$ approximately, that is, $||E_{r}^{0}(x,r,t_{0}+%
\tau )||\varpropto O(\tau ^{3}),$ when these three types of parameters are
given. Therefore, the error $E_{r}^{0}(x,r,t_{0}+\tau )$ is controllable and
it may be neglected when the time interval $\tau <<1.$ In a general case the
initial state $\Psi _{00}(x,r,t_{0})$ may consist of $m$ non-normalized $GWP$
product states: $\Psi _{00}(x,r,t_{0})=\sum_{l=1}^{m}\Psi _{0l}(x,r,t_{0}).$
Here the term number $m$ should be taken as a large enough integer so that
the truncation error $E_{r}^{d}(x,r,t_{0})$ in (6.8) can be neglected. For
example, the term number $m=10$ is usually large enough in the following
error estimation for the experimental basic pulse sequence (6.7b). For each $%
GWP$ state $\Psi _{0l}(x,r,t_{0})$ one may calculate strictly the upper
bounds of both the errors $E_{r}^{V}(x,r,t_{0}+\tau )$ and $%
E_{r}^{0}(x,r,t_{0}+\tau )$ completely according to the theoretical
calculation method in the section 4, in which the initial state $\Psi
_{00}(x,r,t_{0})$ is replaced with the state $\Psi _{0l}(x,r,t_{0}).$ Then
by summing up all these $m$ upper bounds of the error $E_{r}^{V}(x,r,t_{0}+%
\tau )$ one may obtain the total upper bound of the error $%
E_{r}^{V}(x,r,t_{0}+\tau ).$ Similarly, by summing up all these $m$ upper
bounds of the error $E_{r}^{0}(x,r,t_{0}+\tau )$ one also may obtain the
total upper bound of the error $E_{r}^{0}(x,r,t_{0}+\tau ).$ For a given
number $m$ the error $E_{r}^{V}(x,r,t_{0}+\tau )$ is still controlled by the
joint position $x_{L}$, while the error $E_{r}^{0}(x,r,t_{0}+\tau )$ is
still controlled approximately by the time interval $\tau .$ On the other
hand, the error $E_{r2}^{(3)}(x,r,t_{0}+\tau )$ in (6.13) originates from
the deviation $E_{r}^{d}(x,r,t_{0})$ of the initial Gaussian superposition
state $\Psi _{00}(x,r,t_{0})$ from the initial state $\Psi _{0}(x,r,t_{0})$
of (6.8). Now these relations (6.8), (6.13), and (6.15) together show that
the error $E_{r2}^{(3)}(x,r,t_{0}+\tau )$ is given by%
\begin{equation*}
E_{r2}^{(3)}(x,r,t_{0}+\tau )=\exp [-i(H_{0}^{ho}+H_{I}(x,\alpha ,\gamma
)+V_{1}(x,\alpha ,\gamma ,\varepsilon ))\tau /\hslash ]E_{r}^{d}(x,r,t_{0})
\end{equation*}%
\begin{equation}
-\exp [-\frac{i}{2\hslash }H_{0}^{ho}\tau ]\exp [-\frac{i}{\hslash }%
H_{I}(x,\alpha ,\gamma )\tau ]\exp [-\frac{i}{2\hslash }H_{0}^{ho}\tau
]E_{r}^{d}(x,r,t_{0}).  \tag{6.16a}
\end{equation}%
This equation directly leads to that the error is bounded by 
\begin{equation}
||E_{r2}^{(3)}(x,r,t_{0}+\tau )||\leq 2||E_{r}^{d}(x,r,t_{0})||.  \tag{6.16b}
\end{equation}%
The upper bound of the error indeed may be determined from that one of the
deviation $E_{r}^{d}(x,r,t_{0})$. Because the deviation $%
E_{r}^{d}(x,r,t_{0}) $ in (6.8) can be controlled, the error $%
E_{r2}^{(3)}(x,r,t_{0}+\tau )$ can be controlled too. Obviously, the error $%
E_{r2}^{(3)}(x,r,t_{0}+\tau )$ in (6.13) will not exist if there is not the
deviation $E_{r}^{d}(x,r,t_{0})$ in (6.8) or if the initial state $\Psi
_{0}(x,r,t_{0})$ of (6.8) is a Gaussian superposition state.

When the spatially-selective perturbation term $V_{1}(x,\alpha ,\gamma
,\varepsilon )$ disappears, the error $E_{r}^{V}(x,r,t_{0}+\tau )$ will not
exist in the time evolution process of (6.13). In this case the equation
(6.13) still describes the time evolution process of the halting-qubit atom
in the ideal harmonic potential well and in the presence of the ideal $%
PHAMDOWN$ laser light beams, and the error $E_{r}^{0}(x,r,t_{0}+\tau )$
therefore is a dominating term. Similarly, when the spatially-selective
perturbation term $V_{1}^{ho}(x,\varepsilon )$ disappears, the error $%
E_{r}^{(12)}(x,r,t_{0}+\tau ^{\prime })$ vanishes in the time evolution
process of (6.9). Then in this case the equation (6.9) still describes the
time evolution process of the halting-qubit atom in the ideal harmonic
potential well. Therefore, these equations (6.9) and (6.13) may be used to
analyze both the experimental basic pulse sequence (6.7b) and the
theoretical one (6.7a). They also may be used to estimate strictly the
possible errors of the two basic pulse sequences. Because the difference
between the experimental basic pulse sequence (6.7b) and the theoretical one
(6.7a) originates from the spatially-selective perturbation terms $%
V_{1}^{ho}(x,\varepsilon )$ and $V_{1}(x,\alpha ,\gamma ,\varepsilon ),$ it
can be expected that both the experimental and theoretical basic pulse
sequences generate the same result if both the perturbation terms $%
V_{1}^{ho}(x,\varepsilon )$ and $V_{1}(x,\alpha ,\gamma ,\varepsilon )$ have
a negligible effect on the halting-qubit atom. \newline
\newline
{\large 6.1 Theoretical calculation of the experimental basic pulse sequence}

On the basis of these equations (6.9) and (6.13) in the following it is
analyzed in detail the time evolution process of the halting-qubit atom in
the $LH$ potential well and in the presence of the experimental basic pulse
sequence $P_{tr}^{r}(\delta t/\sqrt{n})$. Suppose that at the starting time
of the experimental basic pulse sequence the halting-qubit atom is in the $%
GWP$ motional state $\Psi _{00}(x,t_{0})$ and in the atomic internal state $%
|g_{0}\rangle $. This means that the experimental basic pulse sequence is
applied to the atom in the starting product state $\Psi
_{00}(x,r,t_{0})=\Psi _{00}(x,t_{0})|g_{0}\rangle $. Here suppose that the $%
GWP$ state $\Psi _{00}(x,t_{0})$ has the characteristic parameters $%
\{x_{c}(t_{0}),p_{c}(t_{0}),W(t_{0}),\varepsilon (t_{0})\}$. Since the
experimental basic pulse sequence is a product of the nine unitary
propagators, the calculation for the time evolution process will be carried
out by the following nine consecutive steps. At the first step the first
propagator $\exp (-iH\delta t_{1}^{\prime })$ of the experimental basic
pulse sequence (from the right to the left side of the pulse sequence
(6.7b)) acts on the initial product state $\Psi _{00}(x,r,t_{0}).$ The
propagator $\exp (-iH\delta t_{1}^{\prime })$ is one of the first type of
propagators. The time evolution process of this step is described by the
equation (6.9). According to the equation (6.9) the time evolution process
from the starting time $t_{0}$ to the time $t_{0}+\delta t_{1}^{\prime }$
may be expressed as%
\begin{equation*}
\Psi _{1}(x,r,t_{0}+\delta t_{1}^{\prime })\overset{\text{def}}{\equiv }\exp
\{-i[H_{0}^{ho}+V_{1}^{ho}(x,\varepsilon )]\delta t_{1}^{\prime }/\hslash
\}\Psi _{00}(x,r,t_{0})
\end{equation*}%
\begin{equation}
=\Psi _{01}(x,r,t_{0}+\delta t_{1}^{\prime })+E_{r1}^{0}(x,r,t_{0}+\delta
t_{1}^{\prime })  \tag{6.17}
\end{equation}%
where the desired state $\Psi _{01}(x,r,t_{0}+\delta t_{1}^{\prime })$ is
written as, according to (6.10), 
\begin{equation}
\Psi _{01}(x,r,t_{0}+\delta t_{1}^{\prime })=\exp [\frac{i}{\hslash }%
H_{0}^{ho}(\delta t/\sqrt{4n})]\Psi _{00}(x,r,t_{0}),  \tag{6.18}
\end{equation}%
here the relation $\exp [iH_{0}^{ho}(\delta t/\sqrt{4n})/\hslash ]=\exp
[-iH_{0}^{ho}\delta t_{1}^{\prime }/\hslash ]$ is already used after
neglecting the global phase factor. Again according to (6.9) the error $%
E_{r1}^{0}(x,r,t_{0}+\delta t_{1}^{\prime })$ in (6.17) is written as\newline
\begin{equation}
E_{r1}^{0}(x,r,t_{0}+\delta t_{1}^{\prime })=E_{r}^{(12)}(x,r,t_{0}+\delta
t_{1}^{\prime })+E_{r1}^{(3)}(x,r,t_{0}+\delta t_{1}^{\prime }).  \tag{6.19a}
\end{equation}%
The physical meaning is clear for the equation (6.17) that the first
propagator $\exp (-iH\delta t_{1}^{\prime })$ of the experimental basic
pulse sequence (6.7b) acting on the initial state $\Psi _{00}(x,r,t_{0})$ is
approximately equal to the first propagator $\exp [(i/\hslash
)H_{0}^{ho}(\delta t/\sqrt{4n})]$ of the theoretical basic pulse sequence
(6.7a) acting on the same initial state, and their difference is the error $%
E_{r1}^{0}(x_{b},r,t_{0}+\delta t_{1}^{\prime })$. As shown in (6.19a), this
error consists of the two error terms $E_{r}^{(12)}(x,r,t_{0}+\delta
t_{1}^{\prime })$ and $E_{r1}^{(3)}(x,r,t_{0}+\delta t_{1}^{\prime }).$
Since the initial state $\Psi _{00}(x,r,t_{0})$ in (6.17) is a $GWP$ state,
there is not the deviation $E_{r}^{d}(x,r,t_{0})$ and according to (6.8) one
has $\Psi _{0}(x,r,t_{0})=$ $\Psi _{00}(x,r,t_{0})$. Hence according to
(6.12b) the error $E_{r1}^{(3)}(x,r,t_{0}+\delta t_{1}^{\prime })=0.$ Then
it follows from (6.19a) that the error $E_{r1}^{0}(x,r,t_{0}+\delta
t_{1}^{\prime })=E_{r}^{(12)}(x,r,t_{0}+\delta t_{1}^{\prime }).$
Furthermore, as shown in (6.12a), the error $E_{r}^{(12)}(x,r,t_{0}+\delta
t_{1}^{\prime })$ is a sum of the two error terms $E_{r}^{(k)}(x,r,t_{0}+%
\delta t_{1}^{\prime })$ with $k=1$ and $2$, indicating that the error $%
E_{r1}^{0}(x,r,t_{0}+\delta t_{1}^{\prime })$ is bounded by%
\begin{equation}
||E_{r1}^{0}(x,r,t_{0}+\delta t_{1}^{\prime })||\leq
||E_{r}^{(1)}(x,r,t_{0}+\delta t_{1}^{\prime
})||+||E_{r}^{(2)}(x,r,t_{0}+\delta t_{1}^{\prime })||.  \tag{6.19b}
\end{equation}%
As shown in the section 3, both the first-order approximate error $%
E_{r}^{(1)}(x,r,t_{0}+\delta t_{1}^{\prime })$ and its correction term $%
E_{r}^{(2)}(x,r,t_{0}+\delta t_{1}^{\prime })$ originate from the
imperfection of the $LH$ harmonic potential well. They can be calculated
strictly with the theoretical calculation method in the section 3. The
initial state used in the calculation is just the initial $GWP$ state $\Psi
_{00}(x,r,t_{0})$ and here $\Psi _{00}(x,r,t_{0})=$ $\Psi _{0}(x,r,t_{0}).$
It is shown in the section 3 that the upper bound of the error $%
E_{r}^{(1)}(x,r,t_{0}+\delta t_{1}^{\prime })$ is determined explicitly from
the inequality (3.12) with the final time $t_{b}=t_{0}+\delta t_{1}^{\prime
} $ and from (3.12) one can find that it is proportional to the
exponentially-decaying factor $\exp [-y_{M}(x_{L},t_{c0}^{\ast })^{2}/2].$
It has been shown also in the section 3 that the error $%
E_{r}^{(2)}(x,r,t_{0}+\delta t_{1}^{\prime })$ has an upper bound consisting
of a finite number of the basic norms $\{NBAS1\}$ and $\{NBAS2\}$,
indicating that the error decays exponentially with the square
deviation-to-spread ratios of the relevant $GWP$ states. These results
together with the inequality (6.19b) show that the error $%
E_{r1}^{0}(x,r,t_{0}+\delta t_{1}^{\prime })$ decays exponentially with the
square deviation-to-spread ratios of the relevant $GWP$ states. Thus, it can
be controlled by the joint position $x_{L}$. Generally, the error $%
E_{r1}^{0}(x,r,t_{0}+\delta t_{1}^{\prime })$ is secondary and may be
neglected when the joint position $x_{L}$ is large enough. Then one can
concludes that if the joint position $x_{L}$ is large enough, then according
to (6.17) the final product state $\Psi _{1}(x,r,t_{0}+\delta t_{1}^{\prime
})$, which is generated by acting the experimental propagator $\exp
(-iH\delta t_{1}^{\prime })$ on the starting product state, is really equal
to the desired product state $\Psi _{01}(x,r,t_{0}+\delta t_{1}^{\prime })$
of (6.18), which is generated by acting the theoretical propagator $\exp
[(i/\hslash )H_{0}^{ho}(\delta t/\sqrt{4n})]$ on the same starting product
state.

At the second step the second propagator of the experimental basic pulse
sequence is applied to the halting-qubit atom after the atom is acted on by
the first experimental propagator. In the experimental basic pulse sequence
the second, fourth, sixth, and eighth propagator are generated by the
spatially-selective $PHAMDOWN$ laser light beams. For convenience these
propagators are simply denoted as%
\begin{equation*}
U_{DWN}(\alpha ,\gamma ,\tau )\overset{\text{def}}{\equiv }\exp \{-\frac{i}{%
\hslash }[H_{0}^{ho}+H_{I}(x,\alpha ,\gamma )+V_{1}(x,\alpha ,\gamma
,\varepsilon )]\tau \}.
\end{equation*}%
Here $\tau =\delta t/\sqrt{n},$ $\alpha =\pi /4,$ and $\gamma =3\pi /2,$ $%
\pi ,$ $\pi /2,$ and $0$ corresponding to the second, fourth, sixth, and
eighth propagator, respectively. Then the second experimental propagator is
given by $U_{DWN}(\pi /4,3\pi /2,\delta t/\sqrt{n}).$ The time evolution
process of the second propagator may be expressed as%
\begin{equation}
\Psi _{2}(x,r,t_{0}+\delta t_{1}^{\prime }+\delta t/\sqrt{n})\overset{\text{%
def}}{\equiv }U_{DWN}(\pi /4,3\pi /2,\delta t/\sqrt{n})\Psi
_{1}(x,r,t_{0}+\delta t_{1}^{\prime }),  \tag{6.20a}
\end{equation}%
where $\Psi _{1}(x,r,t_{0}+\delta t_{1}^{\prime })$ is the atomic product
state (6.17) at the end of the first experimental propagator. The product
state (6.20a) may be expressed formally as%
\begin{equation}
\Psi _{2}(x,r,t_{0}+\tau _{1})=\Psi _{02}(x,r,t_{0}+\tau _{1})\newline
\newline
+E_{r2}^{0}(x,r,t_{0}+\tau _{1})+E_{r1}^{0}(x,r,t_{0}+\tau _{1}). 
\tag{6.20b}
\end{equation}%
Here denote $\tau _{1}=\delta t_{1}^{\prime }+\delta t/\sqrt{n}.$ The task
now is to determine the product state $\Psi _{02}(x,r,t_{0}+\tau _{1})$ and
the two error terms $E_{rk}^{0}(x,r,t_{0}+\tau _{1})$ with $k=1$ and $2$ in
(6.20b). First of all, by substituting the state $\Psi _{1}(x,r,t_{0}+\delta
t_{1}^{\prime })$ of (6.17) into (6.20a) one obtains%
\begin{equation*}
\Psi _{2}(x,r,t_{0}+\delta t_{1}^{\prime }+\delta t/\sqrt{n})=U_{DWN}(\pi
/4,3\pi /2,\delta t/\sqrt{n})\Psi _{01}(x,r,t_{0}+\delta t_{1}^{\prime })
\end{equation*}%
\begin{equation}
+E_{r1}^{0}(x,r,t_{0}+\tau _{1}).  \tag{6.21}
\end{equation}%
Here the error $E_{r1}^{0}(x,r,t_{0}+\delta t_{1}^{\prime })$ is given by 
\begin{equation}
E_{r1}^{0}(x,r,t_{0}+\tau _{1})=U_{DWN}(\pi /4,3\pi /2,\delta t/\sqrt{n}%
)E_{r1}^{0}(x,r,t_{0}+\delta t_{1}^{\prime }).  \tag{6.22a}
\end{equation}%
This error is just the last term on the $RH$ side of (6.20b). Notice that a
unitary propagator acting on an error state does not change the norm of the
error state. Then it follows from (6.22a) that there is the relation:%
\begin{equation}
||E_{r1}^{0}(x,r,t_{0}+\tau _{1})||=||E_{r1}^{0}(x,r,t_{0}+\delta
t_{1}^{\prime })||.  \tag{6.22b}
\end{equation}%
Hence the upper bound of the error $E_{r1}^{0}(x,r,t_{0}+\tau _{1})$ is just
equal to that one of the error $E_{r1}^{0}(x,r,t_{0}+\delta t_{1}^{\prime })$%
. It is shown above that the error $E_{r1}^{0}(x,r,t_{0}+\delta
t_{1}^{\prime })$ decays exponentially with the square deviation-to-spread
ratios of the relevant $GWP$ states. Then the same conclusion holds also for
the error $E_{r1}^{0}(x,r,t_{0}+\tau _{1}).$ The key step to calculate
strictly the time evolution process of (6.20a) is to calculate the first
term on the $RH$ side of (6.21). This calculation needs to use the time
evolution process of (6.13) as the propagator $U_{DWN}(\pi /4,3\pi /2,\delta
t/\sqrt{n})$ is one of the second type of propagators. It is known from
(6.13) that the first term on the $RH$ side of (6.21) may be written as%
\begin{equation*}
U_{DWN}(\pi /4,3\pi /2,\delta t/\sqrt{n})\Psi _{01}(x,r,t_{0}+\delta
t_{1}^{\prime })=\Psi _{02}(x,r,t_{0}+\tau _{1})\newline
\newline
\end{equation*}%
\begin{equation}
+E_{r}^{0}(x,r,t_{0}+\tau _{1})+E_{r}^{V}(x,r,t_{0}+\tau
_{1})+E_{r}^{(3)}(x,r,t_{0}+\tau _{1})  \tag{6.23}
\end{equation}%
where the desired product state is written as, according to (6.14), 
\begin{equation*}
\Psi _{02}(x,r,t_{0}+\tau _{1})=\exp [-\frac{i}{2\hslash }H_{0}^{ho}(\delta
t/\sqrt{n})]\exp [-\frac{i}{\hslash }H_{I}(x,\pi /4,3\pi /2)(\delta t/\sqrt{n%
})]
\end{equation*}%
\begin{equation}
\times \exp [-\frac{i}{2\hslash }H_{0}^{ho}(\delta t/\sqrt{n})]\Psi
_{01}(x,r,t_{0}+\delta t_{1}^{\prime }).  \tag{6.24}
\end{equation}%
The initial state $\Psi _{01}(x,r,t_{0}+\delta t_{1}^{\prime })$ in (6.23)
is given by (6.18). It is a $GWP$ state. The product state (6.24) is just
the first term on the $RH$ side of (6.20b). Now these relations (6.20b),
(6.21), and (6.23) together show that the error $E_{r2}^{0}(x,r,$ $%
t_{0}+\tau _{1})$ in (6.20b) is given by%
\begin{equation}
E_{r2}^{0}(x,r,t_{0}+\tau _{1})=E_{r}^{0}(x,r,t_{0}+\tau
_{1})+E_{r}^{V}(x,r,t_{0}+\tau _{1})+E_{r2}^{(3)}(x,r,t_{0}+\tau _{1}). 
\tag{6.25a}
\end{equation}%
Here the error $E_{r2}^{(3)}(x,r,t_{0}+\tau _{1})=0$ because the initial
state $\Psi _{01}(x,r,t_{0}+\delta t_{1}^{\prime })$ in (6.23) is a $GWP$
state. Thus, the upper bound of the error $E_{r2}^{0}(x,r,t_{0}+\tau _{1})$
is determined from 
\begin{equation}
||E_{r2}^{0}(x,r,t_{0}+\tau _{1})||\leq ||E_{r}^{0}(x,r,t_{0}+\tau
_{1})||+||E_{r}^{V}(x,r,t_{0}+\tau _{1})||.  \tag{6.25b}
\end{equation}%
Now all the three terms on the $RH$ side of (6.20b) can be determined from
(6.24), (6.25a), and (6.22a), respectively. The inequality (6.25b) shows
that the upper bound of the error $E_{r2}^{0}(x,r,t_{0}+\tau _{1})$ can be
determined from those of both the errors $E_{r}^{0}(x,r,t_{0}+\tau _{1})$
and $E_{r}^{V}(x,r,t_{0}+\tau _{1}).$ On the basis of the time evolution
process of (6.23) a strict calculation for the upper bounds of the two
errors $E_{r}^{0}(x,r,t_{0}+\tau _{1})$ and $E_{r}^{V}(x,r,t_{0}+\tau _{1})$
may be carried out according to the theoretical calculation method in the
section 4. Here one should use the initial $GWP$ state $\Psi
_{01}(x,r,t_{0}+\delta t_{1}^{\prime })$ of (6.18) and the interaction $%
H_{I}(x,\pi /4,3\pi /2)$ of (4.16) to calculate the upper bounds of the two
errors. The error $E_{r}^{V}(x,r,t_{0}+\tau _{1})$ originates from the
imperfection of the $LH$ harmonic potential well and the spatially-selective
effect of the $PHAMDOWN$ laser light beams. It has been shown in the section
4 that the upper bound of the error consists of a finite number of the basic
norms $\{NBAS1\}$ and $\{NBAS2\}$. Thus, it decays exponentially with the
square deviation-to-spread ratios of the relevant $GWP$ states. It can be
controlled by the joint position $x_{L}$. Therefore, it may be neglected
when the joint position $x_{L}$ is large enough. On the other hand, the
error $E_{r}^{0}(x,r,t_{0}+\tau _{1})$ is the dominating term in the error $%
E_{r2}^{0}(x,r,t_{0}+\tau _{1})$ in (6.25a). It is independent of the
imperfection of the $LH$ potential well and the spatially-selective effect
of the $PHAMDOWN$ laser light beams. It is due to the truncation
approximation of the decomposition formula (4.1). As shown in the subsection
4.4, the upper bound of the error $E_{r}^{0}(x,r,t_{0}+\tau _{1})$ may be
controlled by the time interval $\delta t/\sqrt{n}$ and the three types of
parameters. One type of parameters come from the $GWP$ motional states of
the halting-qubit atom, while the other two types come from the $PHAMDOWN$
laser light beams and the harmonic potential field, respectively. Thus, the
latter two types of parameters are of the experimental basic pulse sequence
(6.7b) itself. It is shown in the subsection 4.4 that the upper bound of the
error $E_{r}^{0}(x,r,t_{0}+\tau _{1})$ is proportional to $(\delta t/\sqrt{n}%
)^{3}$ approximately. Thus, the error can be controlled by the time interval 
$\delta t/\sqrt{n}$ in addition to the three types of parameters. Then the
inequality (6.25b) shows that the upper bound of the error $%
E_{r2}^{0}(x,r,t_{0}+\tau _{1})$ is proportional to $(\delta t/\sqrt{n})^{3}$
approximately. Therefore, the error $E_{r2}^{0}(x,r,t_{0}+\tau _{1})$ may be
controlled by the joint position $x_{L}$ and the time interval $\delta t/%
\sqrt{n}.$

Now by inserting (6.18) into (6.24) the state $\Psi _{02}(x,r,t_{0}+\tau
_{1})$ is reduced to the simple form%
\begin{equation}
\Psi _{02}(x,r,t_{0}+\tau _{1})\newline
=\exp [-\frac{i}{\hslash }H_{0}^{ho}(\delta t/\sqrt{4n})]\Psi
_{02}(x,r,t_{0}+\delta t/\sqrt{n})  \tag{6.26}
\end{equation}%
where the state $\Psi _{02}(x,r,t_{0}+\delta t/\sqrt{n})$ is defined by%
\begin{equation}
\Psi _{02}(x,r,t_{0}+\delta t/\sqrt{n})=\exp \{-\frac{i}{\hslash }%
H_{I}(x,\pi /4,3\pi /2)(\delta t/\sqrt{n})\}\Psi _{00}(x,r,t_{0}). 
\tag{6.27a}
\end{equation}%
The state $\Psi _{02}(x,r,t_{0}+\delta t/\sqrt{n})$ may be calculated
exactly by using the unitary transformation (4.18) of the propagator $\exp
\{-iH_{I}(x,\pi /4,3\pi /2)(\delta t/\sqrt{n})/\hslash \}.$ It may be
explicitly written as%
\begin{equation*}
\Psi _{02}(x,r,t_{0}+\delta t/\sqrt{n})=\cos \{2\Omega _{0}\delta t/\sqrt{n}%
\cos (\frac{1}{2}\Delta kx-\pi /4)\}\Psi _{00}(x,t_{0})|g_{0}\rangle
\end{equation*}%
\begin{equation}
+\sin \{2\Omega _{0}\delta t/\sqrt{n}\cos (\frac{1}{2}\Delta kx-\pi
/4)\}\exp [i\frac{1}{2}(k_{0}+k_{1})x]\Psi _{00}(x,t_{0})|e\rangle . 
\tag{6.27b}
\end{equation}%
The state $\Psi _{02}(x,r,t_{0}+\delta t/\sqrt{n})$ is just the state (5.23)
with the settings $\Omega (\tau )=2\Omega _{0}\delta t/\sqrt{n}$ and $\Psi
_{0}(x,t_{0})=\Psi _{00}(x,t_{0})$. Obviously, it is not a pure $GWP$ state.
The state (6.27b) will be further used in the following error estimation.

At the third step the third experimental propagator $\exp [-iH\delta
t_{1}/\hslash ]$ is applied to the halting-qubit atom at the end of the
second experimental propagator. The time evolution process of the
halting-qubit atom under the third propagator may be formally written as%
\begin{equation}
\Psi _{3}(x,r,t_{0}+\tau _{1}+\delta t_{1})\overset{\text{def}}{\equiv }\exp
\{-i[H_{0}^{ho}+V_{1}^{ho}(x,\varepsilon )]\delta t_{1}/\hslash \}\Psi
_{2}(x,r,t_{0}+\tau _{1}),  \tag{6.28a}
\end{equation}%
where the initial state $\Psi _{2}(x,r,t_{0}+\tau _{1})$ is given by
(6.20b). By substituting the state (6.20b) into (6.28a) the product state $%
\Psi _{3}(x,r,t_{0}+\tau _{1}+\delta t_{1})$ may be rewritten as%
\begin{equation}
\Psi _{3}(x,r,t_{0}+\tau _{1}+\delta t_{1})=\Psi _{03}(x,r,t_{0}+\tau
_{1}+\delta t_{1})\newline
\newline
+\sum_{k=1}^{3}E_{rk}^{0}(x,r,t_{0}+\tau _{1}+\delta t_{1}).  \tag{6.28b}
\end{equation}%
Here the two errors $E_{rk}^{0}(x,r,t_{0}+\tau _{1}+\delta t_{1})$ for $k=1$
and $2$ are obtained by the unitary transformation:%
\begin{equation}
E_{rk}^{0}(x,r,t_{0}+\tau _{1}+\delta t_{1})=\exp
\{-i[H_{0}^{ho}+V_{1}^{ho}(x,\varepsilon )]\delta t_{1}/\hslash
\}E_{rk}^{0}(x,r,t_{0}+\tau _{1}).  \tag{6.29}
\end{equation}%
Thus, the upper bounds of the two errors are determined respectively from%
\begin{equation}
||E_{r1}^{0}(x,r,t_{0}+\tau _{1}+\delta
t_{1})||=||E_{r1}^{0}(x,r,t_{0}+\delta t_{1}^{\prime })||,  \tag{6.30a}
\end{equation}%
\begin{equation}
||E_{r2}^{0}(x,r,t_{0}+\tau _{1}+\delta t_{1})||=||E_{r2}^{0}(x,r,t_{0}+\tau
_{1})||.  \tag{6.30b}
\end{equation}%
Here the equation (6.30a) is obtained from (6.29) after using the equation
(6.22b). The upper bounds of the two errors $E_{r1}^{0}(x,r,t_{0}+\delta
t_{1}^{\prime })$ and $E_{r2}^{0}(x,r,t_{0}+\tau _{1})$ are determined from
(6.19b) and (6.25b), respectively. Then both the equations (6.30) show that
the upper bounds of the two errors $E_{rk}^{0}(x,r,t_{0}+\tau _{1}+\delta
t_{1})$ with $k=1$ and $2$ also may be determined from (6.19b) and (6.25b),
respectively. The desired state $\Psi _{03}(x,r,t_{0}+\tau _{1}+\delta
t_{1}) $ and the error $E_{r3}^{0}(x,r,t_{0}+\tau _{1}+\delta t_{1})$ on the 
$RH$ side of (6.28b) may be obtained below on the basis of the time
evolution process of (6.9). By applying the current propagator $\exp
[-iH\delta t_{1}/\hslash ]$ to the initial state $\Psi _{02}(x,r,t_{0}+\tau
_{1})$, which is just the desired state (6.26) of the second experimental
propagator, one obtains formally%
\begin{equation*}
\exp \{-i[H_{0}^{ho}+V_{1}^{ho}(x,\varepsilon )]\delta t_{1}/\hslash \}\Psi
_{02}(x,r,t_{0}+\tau _{1})
\end{equation*}%
\begin{equation}
=\Psi _{03}(x,r,t_{0}+\tau _{1}+\delta t_{1})+E_{r3}^{0}(x,r,t_{0}+\tau
_{1}+\delta t_{1}).  \tag{6.31}
\end{equation}%
Here $\Psi _{03}(x,r,t_{0}+\tau _{1}+\delta t_{1})$ is the desired product
state, while $E_{r3}^{0}(x,r,t_{0}+\tau _{1}+\delta t_{1})$ is the
corresponding error. They can be explicitly obtained below. As shown in
(6.26), the state $\Psi _{02}(x,r,t_{0}+\tau _{1})$ is not a pure $GWP$
state, because the state (6.27b) is not. In order to calculate the time
evolution process of (6.31) one needs to expand the state $\Psi
_{02}(x,r,t_{0}+\tau _{1})$ as a Gaussian superposition state. This is
similar to the expansion (6.8). Now according to the $MGWP$ expansion in the
section 5 the state $\Psi _{02}(x,r,t_{0}+\delta t/\sqrt{n})$ of (6.27b) is
first expanded as a superposition of the $GWP$ states, 
\begin{equation}
\Psi _{02}(x,r,t_{0}+\delta t/\sqrt{n})=\Psi _{02}^{G}(x,r,t_{0}+\delta t/%
\sqrt{n})+E_{r2}(x,r,t_{0}+\delta t/\sqrt{n}),  \tag{6.32}
\end{equation}%
where $\Psi _{02}^{G}(x,r,t_{0}+\delta t/\sqrt{n})$ is a Gaussian
superposition state and $E_{r2}(x,r,t_{0}+\delta t/\sqrt{n})$ the truncation
error of the $MGWP$ expansion. Then inserting the expansion (6.32) into
(6.26) the state $\Psi _{02}(x,r,t_{0}+\tau _{1})$ is accordingly expanded as%
\begin{equation*}
\Psi _{02}(x,r,t_{0}+\tau _{1})=\Psi _{02}^{G}(x,r,t_{0}+\tau _{1})
\end{equation*}%
\begin{equation}
+\exp [-\frac{i}{\hslash }H_{0}^{ho}(\delta t/\sqrt{4n})]E_{r2}(x,r,t_{0}+%
\delta t/\sqrt{n})  \tag{6.33}
\end{equation}%
where the state $\Psi _{02}^{G}(x,r,t_{0}+\tau _{1})$ is given by%
\begin{equation}
\Psi _{02}^{G}(x,r,t_{0}+\tau _{1})=\exp [-\frac{i}{\hslash }%
H_{0}^{ho}(\delta t/\sqrt{4n})]\Psi _{02}^{G}(x,r,t_{0}+\delta t/\sqrt{n}). 
\tag{6.34}
\end{equation}%
The state $\Psi _{02}^{G}(x,r,t_{0}+\tau _{1})$ is a Gaussian superposition
state due to that the propagator $\exp [-iH_{0}^{ho}(\delta t/\sqrt{4n}%
)/\hslash ]$ is of a harmonic-oscillator. By comparing (6.33) with (6.8) one
finds that there are the corresponding relations:%
\begin{equation*}
\Psi _{0}(x,r,t_{0})\leftrightarrow \Psi _{02}(x,r,t_{0}+\tau _{1}),
\end{equation*}%
\begin{equation*}
\Psi _{00}(x,r,t_{0})\leftrightarrow \Psi _{02}^{G}(x,r,t_{0}+\tau _{1}),
\end{equation*}%
\begin{equation*}
E_{r}^{d}(x,r,t_{0})\leftrightarrow \exp [-\frac{i}{\hslash }%
H_{0}^{ho}(\delta t/\sqrt{4n})]E_{r2}(x,r,t_{0}+\delta t/\sqrt{n}).
\end{equation*}%
Now by inserting the state $\Psi _{02}(x,r,t_{0}+\tau _{1})$ of (6.33) into
(6.31) and then comparing (6.31) with (6.9) one can find from (6.31) that
the desired state is given by, similar to the state of (6.10),%
\begin{equation}
\Psi _{03}(x,r,t_{0}+\tau _{1}+\delta t_{1})=\exp [-iH_{0}^{ho}\delta
t_{1}/\hslash ]\Psi _{02}(x,r,t_{0}+\tau _{1})  \tag{6.35}
\end{equation}%
and the corresponding error is%
\begin{equation}
E_{r3}^{0}(x,r,t_{0}+\tau _{1}+\delta t_{1})=E_{r}^{(12)}(x,r,t_{0}+\tau
_{1}+\delta t_{1})+E_{r1}^{(3)}(x,r,t_{0}+\tau _{1}+\delta t_{1}). 
\tag{6.36}
\end{equation}%
Here the error $E_{r3}^{0}(x,r,t_{0}+\tau _{1}+\delta t_{1})$ consists of
the two following terms. One of which is given by, similar to the error of
(6.12a),%
\begin{equation}
E_{r}^{(12)}(x,r,t_{0}+\tau _{1}+\delta t_{1})=E_{r}^{(1)}(x,r,t_{0}+\tau
_{1}+\delta t_{1})+E_{r}^{(2)}(x,r,t_{0}+\tau _{1}+\delta t_{1}). 
\tag{6.37a}
\end{equation}%
and another by, similar to the error of (6.12b),%
\begin{equation*}
E_{r1}^{(3)}(x,r,t_{0}+\tau _{1}+\delta t_{1})=\exp
\{-i[H_{0}^{ho}+V_{1}^{ho}(x,\varepsilon )]\delta t_{1}/\hslash \}
\end{equation*}%
\begin{equation*}
\times \exp [-\frac{i}{\hslash }H_{0}^{ho}(\delta t/\sqrt{4n}%
)]E_{r2}(x,r,t_{0}+\delta t/\sqrt{n})
\end{equation*}%
\begin{equation}
-\exp [-iH_{0}^{ho}\delta t_{1}/\hslash ]\exp [-\frac{i}{\hslash }%
H_{0}^{ho}(\delta t/\sqrt{4n})]E_{r2}(x,r,t_{0}+\delta t/\sqrt{n}). 
\tag{6.37b}
\end{equation}%
The error $E_{r}^{(12)}(x,r,t_{0}+\tau _{1}+\delta t_{1})$ consists of the
two error terms $E_{r}^{(k)}(x,r,t_{0}+\tau _{1}+\delta t_{1})$ with $k=1$
and $2$. A strict calculation for the two error terms can be carried out
with the theoretical calculation method in the section 3. Here one should
use the initial Gaussian superposition state $\Psi _{02}^{G}(x,r,t_{0}+\tau
_{1})$ in the calculation. As shown in (6.34) and (6.38) below, the state $%
\Psi _{02}^{G}(x,r,t_{0}+\tau _{1})$ consists of the ten lower-order $GWP$
states. Thus, for each one of these ten $GWP$ states which acts as the
initial state in the calculation one may separately calculate the upper
bound of the error $E_{r}^{(k)}(x,r,t_{0}+\tau _{1}+\delta t_{1})$ for $k=1$
and $2$ with the theoretical calculation method in the section 3. Then by
summing up these ten upper bounds one may obtain the total upper bound of
the error $E_{r}^{(k)}(x,r,t_{0}+\tau _{1}+\delta t_{1}).$ Then the upper
bound of the error $E_{r}^{(12)}(x,r,t_{0}+\tau _{1}+\delta t_{1})$ is
determined from%
\begin{equation}
||E_{r}^{(12)}(x,r,t_{0}+\tau _{1}+\delta t_{1})||\leq
||E_{r}^{(1)}(x,r,t_{0}+\tau _{1}+\delta
t_{1})||+||E_{r}^{(2)}(x,r,t_{0}+\tau _{1}+\delta t_{1})||.  \tag{6.37c}
\end{equation}%
As shown in the section 3, both the errors $E_{r}^{(k)}(x,r,t_{0}+\tau
_{1}+\delta t_{1})$ with $k=1$ and $2$ originate from the imperfection of
the $LH$ harmonic potential well and hence so does the error $%
E_{r}^{(12)}(x,r,t_{0}+\tau _{1}+\delta t_{1})$. It has been shown in the
section 3 that the upper bounds of the two errors $E_{r}^{(k)}(x,r,t_{0}+%
\tau _{1}+\delta t_{1})$ with $k=1$ and $2$ decay exponentially with the
square deviation-to-spread ratios of the relevant $GWP$ states. This
indicates that the error $E_{r}^{(12)}(x,r,t_{0}+\tau _{1}+\delta t_{1})$
also decays exponentially with the square deviation-to-spread ratios of the
relevant $GWP$ states. Therefore, the error can be controlled by the joint
position $x_{L}$. It can be neglected when the joint position $x_{L}$ is
large enough.

The error $E_{r1}^{(3)}(x,r,t_{0}+\tau _{1}+\delta t_{1})$ in (6.36) really
originates from the deviation of the initial state $\Psi
_{02}(x,r,t_{0}+\tau _{1})$ from the initial Gaussian superposition state $%
\Psi _{02}^{G}(x,r,t_{0}+\tau _{1}).$ Its upper bound is determined from,
according to (6.37b), 
\begin{equation}
||E_{r1}^{(3)}(x,r,t_{0}+\tau _{1}+\delta t_{1})||\leq
2||E_{r2}(x,r,t_{0}+\delta t/\sqrt{n})||.  \tag{6.37d}
\end{equation}%
Here the error $E_{r2}(x,r,t_{0}+\delta t/\sqrt{n})$ is given by (6.32).
Below the upper bound of the error $E_{r2}(x,r,t_{0}+\delta t/\sqrt{n})$ is
obtained explicitly from the state $\Psi _{02}(x,r,t_{0}+\delta t/\sqrt{n})$
of (6.27b). It is known that the initial $GWP$ motional state $\Psi
_{00}(x,t_{0})$ in (6.27b) has the COM position $x_{c}(t_{0}),$ momentum $%
p_{c}(t_{0}),$ and wave-packet spread $\varepsilon (t_{0}).$ Denote that $%
y=x-x_{c}(t_{0}).$ According to the $MGWP$ expansion the state $\Psi
_{02}(x,r,t_{0}+\delta t/\sqrt{n})$ may be expanded as a superposition of
the $GWP$ states approximately. This Gaussian superposition state is just
the state $\Psi _{02}^{G}(x,r,t_{0}+\delta t/\sqrt{n})$ of (6.32), which may
consist of the ten lower-order $GWP$ states, just like the state (5.29),%
\begin{equation*}
\Psi _{02}^{G}(x,r,t_{0}+\delta t/\sqrt{n})=\{B_{0}^{g}+(\beta _{c}\cos
\beta _{c}-\sin \beta _{c})[\beta _{c}\cos (\frac{1}{2}\Delta ky)-\beta
_{s}\sin (\frac{1}{2}\Delta ky)]
\end{equation*}%
\begin{equation*}
+\frac{1}{2}\cos \beta _{c}[\beta _{c}\beta _{s}\sin (\Delta ky)-\frac{1}{2}%
(\beta _{c}^{2}-\beta _{s}^{2})\cos (\Delta ky)]\}\Psi
_{00}(x,t_{0})|g_{0}\rangle
\end{equation*}%
\begin{equation*}
+\{B_{0}^{e}+(\beta _{c}\sin \beta _{c}+\cos \beta _{c})[\beta _{c}\cos (%
\frac{1}{2}\Delta ky)-\beta _{s}\sin (\frac{1}{2}\Delta ky)]
\end{equation*}%
\begin{equation}
+\frac{1}{2}\sin \beta _{c}[\beta _{c}\beta _{s}\sin (\Delta ky)-\frac{1}{2}%
(\beta _{c}^{2}-\beta _{s}^{2})\cos (\Delta ky)]\}\exp [i\frac{1}{2}%
(k_{0}+k_{1})x]\Psi _{00}(x,t_{0})|e\rangle ,  \tag{6.38}
\end{equation}%
where the amplitude $B_{0}^{g}=\beta _{c}\sin \beta _{c}+(1-\frac{3}{4}\beta
_{c}^{2}-\frac{1}{4}\beta _{s}^{2})\cos \beta _{c}$ and $B_{0}^{e}=-\beta
_{c}\cos \beta _{c}$ $+(1-\frac{3}{4}\beta _{c}^{2}-\frac{1}{4}\beta
_{s}^{2})\sin \beta _{c};$ the coefficients $\beta _{s}$ and $\beta _{c}$
are given by $\beta _{c}=(2\Omega _{0}\delta t/\sqrt{n})$ $\times \cos [%
\frac{1}{2}\Delta kx_{c}(t_{0})-\pi /4]$ and $\beta _{s}=(2\Omega _{0}\delta
t/\sqrt{n})\sin [\frac{1}{2}\Delta kx_{c}(t_{0})-\pi /4].$ The two
coefficients $\beta _{s}$ and $\beta _{c}$ here are really equal to those in
(5.25), respectively, if setting $\Omega (\tau )=(2\Omega _{0}\delta t/\sqrt{%
n}).$ Obviously, the state $\Psi _{02}(x,r,t_{0}+\delta t/\sqrt{n})$ also
may be expanded as a superposition of more than ten lower-order $GWP$
states. But for the present error estimation it is good enough to use the
superposition state (6.38) of the ten lower-order $GWP$ states to calculate
the upper bound of the error $E_{r2}(x,r,t_{0}+\delta t/\sqrt{n}).$ Now by
inserting the Gaussian superposition state (6.38) into (6.34) one can
further obtain the Gaussian superposition state $\Psi
_{02}^{G}(x,r,t_{0}+\tau _{1})$. This Gaussian superposition state also
consists of the ten $GWP$ states. It serves as the initial state to
calculate the upper bounds of the error terms $E_{r}^{(k)}(x,r,t_{0}+\tau
_{1}+\delta t_{1})$ with $k=1$ and $2$ in the theoretical calculation method
in the section 3. The difference between the original state $\Psi
_{02}(x,r,t_{0}+\delta t/\sqrt{n})$ and its approximated Gaussian
superposition state $\Psi _{02}^{G}(x,r,t_{0}+\delta t/\sqrt{n})$ of (6.38)
is measured by the error $E_{r2}(x,r,t_{0}+\delta t/\sqrt{n})$ in (6.32).
This error can be estimated strictly. According to (5.35) in the section 5
the upper bound for the error may be determined from%
\begin{equation*}
||E_{r2}(x,r,t_{0}+\delta t/\sqrt{n})||_{u}\thickapprox (2\Omega _{0}\delta
t/\sqrt{n})^{3}\varepsilon (t_{0})^{3}|\Delta k|^{3}
\end{equation*}%
\begin{equation}
\times \{\frac{5}{6144}+\frac{1}{256}\frac{1}{\sqrt{\pi }}\varepsilon
(t_{0})|\Delta k|\}^{1/2}.  \tag{6.39}
\end{equation}%
This upper bound is proportional to $(\delta t/\sqrt{n})^{3},$ $\varepsilon
(t_{0})^{3},$ and $|\Delta k|^{3}$ approximately. When the time interval $%
\delta t/\sqrt{n}$ is short enough or the wave-number difference $|\Delta k|$
is small enough, the error $E_{r2}(x,r,t_{0}+\delta t/\sqrt{n})$ may be
neglected. As shown in the inequality (6.37d), the upper bound of the error $%
E_{r1}^{(3)}(x,r,t_{0}+\tau _{1}+\delta t_{1})$ is determined from that one
of the error $E_{r2}(x,r,t_{0}+\delta t/\sqrt{n}).$ Thus, the upper bound of
the error $E_{r1}^{(3)}(x,r,t_{0}+\tau _{1}+\delta t_{1})$ may be equal to $%
||E_{r2}(x,r,t_{0}+\delta t/\sqrt{n})||_{u}$ of (6.39). This means that when
the time interval $\delta t/\sqrt{n}$ is short enough or the wave-number
difference $|\Delta k|$ is small enough, the error $E_{r1}^{(3)}(x,r,$ $%
t_{0}+\tau _{1}+\delta t_{1})$ may be neglected. Now the upper bound of the
error $E_{r3}^{0}(x,r,t_{0}+\tau _{1}+\delta t_{1})$ in (6.28b) is obtained
from, according to (6.36),%
\begin{equation*}
||E_{r3}^{0}(x,r,t_{0}+\tau _{1}+\delta t_{1})||\leq
||E_{r}^{(12)}(x,r,t_{0}+\tau _{1}+\delta t_{1})||
\end{equation*}%
\begin{equation}
+||E_{r1}^{(3)}(x,r,t_{0}+\tau _{1}+\delta t_{1})||.  \tag{6.40}
\end{equation}%
This shows that the error $E_{r3}^{0}(x,r,t_{0}+\tau _{1}+\delta t_{1})$ can
be controlled by the joint position $x_{L}$, the time interval $\delta t/%
\sqrt{n},$ the wave-packet spread $\varepsilon (t_{0}),$ and the wave-number
difference $|\Delta k|.$

Since $\exp [-iH_{0}^{ho}\delta t_{1}/\hslash ]=\exp [iH_{0}^{ho}(\delta t/%
\sqrt{n})/\hslash ]$ up to a global phase factor, by inserting (6.26) into
(6.35) one may obtain the desired product state in (6.28b):%
\begin{equation}
\Psi _{03}(x,r,t_{0}+\tau _{1}+\delta t_{1})\newline
=\exp [\frac{i}{\hslash }H_{0}^{ho}(\delta t/\sqrt{4n})]\Psi
_{02}(x,r,t_{0}+\delta t/\sqrt{n}).  \tag{6.41a}
\end{equation}%
This state is not a pure $GWP$ state. By inserting the expansion (6.32) into
(6.41a) one obtains%
\begin{equation*}
\Psi _{03}(x,r,t_{0}+\tau _{1}+\delta t_{1})\newline
=\Psi _{03}^{G}(x,r,t_{0}+\tau _{1}+\delta t_{1})
\end{equation*}%
\begin{equation}
+\exp [\frac{i}{\hslash }H_{0}^{ho}(\delta t/\sqrt{4n})]E_{r2}(x,r,t_{0}+%
\delta t/\sqrt{n})  \tag{6.41b}
\end{equation}%
where the Gaussian superposition state $\Psi _{03}^{G}(x,r,t_{0}+\tau
_{1}+\delta t_{1})$ is defined by%
\begin{equation}
\Psi _{03}^{G}(x,r,t_{0}+\tau _{1}+\delta t_{1})=\exp [\frac{i}{\hslash }%
H_{0}^{ho}(\delta t/\sqrt{4n})]\Psi _{02}^{G}(x,r,t_{0}+\delta t/\sqrt{n}). 
\tag{6.42}
\end{equation}%
The state $\Psi _{03}^{G}(x,r,t_{0}+\tau _{1}+\delta t_{1})$ is a
superposition of the ten $GWP$ states. It is obtained from the Gaussian
superposition state $\Psi _{02}^{G}(x,r,t_{0}+\delta t/\sqrt{n})$ of (6.38)
by a unitary transformation with the inverse harmonic-oscillator propagator $%
\exp [\frac{i}{\hslash }H_{0}^{ho}(\delta t/\sqrt{4n})]$. The expansion
(6.41b) will be further used in the following error estimation.

Now the fourth experimental propagator $U_{DWN}(\pi /4,\pi ,\delta t/\sqrt{n}%
)$ is applied to the state $\Psi _{3}(x,r,t_{0}+\tau _{1}+\delta t_{1})$ of
(6.28b) at the end of the third experimental propagator. Then the atomic
product state at the end of the fourth propagator may be written as%
\begin{equation}
\Psi _{4}(x,r,t_{0}+\tau _{2}+\delta t_{1})\newline
=\Psi _{04}(x,r,t_{0}+\tau _{2}+\delta t_{1})\newline
+\sum_{k=1}^{4}E_{rk}^{0}(x,r,t_{0}+\tau _{2}+\delta t_{1}),  \tag{6.43}
\end{equation}%
where the time interval $\tau _{2}=\delta t_{1}^{\prime }+2\delta t/\sqrt{n}%
. $ The error $E_{rk}^{0}(x,r,t_{0}+\tau _{2}+\delta t_{1})$ with $k=1,$ $2,$
or $3$ on the $RH$ side of (6.43) is obtained by applying the current
propagator $U_{DWN}(\pi /4,\pi ,\delta t/\sqrt{n})$ to the error $%
E_{rk}^{0}(x,r,t_{0}+\tau _{1}+\delta t_{1})$ in (6.28b). Because a unitary
transformation does not change the upper bound of an error state, both the
errors $E_{rk}^{0}(x,r,t_{0}+\tau _{2}+\delta t_{1})$ and $%
E_{rk}^{0}(x,r,t_{0}+\tau _{1}+\delta t_{1})$ with $k=1,$ $2,$ or $3$ have
the same upper bound. This means that the upper bounds for the three errors $%
E_{rk}^{0}(x,r,t_{0}+\tau _{2}+\delta t_{1})$ with $k=1,$ $2,$ and $3$ in
(6.43) may be obtained from%
\begin{equation*}
||E_{r1}^{0}(x,r,t_{0}+\tau _{2}+\delta
t_{1})||=||E_{r1}^{0}(x,r,t_{0}+\delta t_{1}^{\prime })||,
\end{equation*}%
\begin{equation*}
||E_{r2}^{0}(x,r,t_{0}+\tau _{2}+\delta t_{1})||=||E_{r2}^{0}(x,r,t_{0}+\tau
_{1})||,
\end{equation*}%
\begin{equation*}
||E_{r3}^{0}(x,r,t_{0}+\tau _{2}+\delta t_{1})||=||E_{r3}^{0}(x,r,t_{0}+\tau
_{1}+\delta t_{1})||.
\end{equation*}%
Here the upper bounds of the errors $E_{r1}^{0}(x,r,t_{0}+\delta
t_{1}^{\prime }),$ $E_{r2}^{0}(x,r,t_{0}+\tau _{1}),$ and $%
E_{r3}^{0}(x,r,t_{0}+\tau _{1}+\delta t_{1})$ are determined from (6.19b),
(6.25b), and (6.40), respectively. Now only the desired product state $\Psi
_{04}(x,r,t_{0}+\tau _{2}+\delta t_{1})$ and the error $%
E_{r4}^{0}(x,r,t_{0}+\tau _{2}+\delta t_{1})$ need to be determined on the $%
RH$ side of (6.43). They can be obtained on the basis of the time evolution
process of (6.13). By applying the current propagator to the initial state $%
\Psi _{03}(x,r,t_{0}+\tau _{1}+\delta t_{1}),$ which is the final desired
state (6.41b) of the third propagator, and then according to (6.13) one may
obtain%
\begin{equation*}
U_{DWN}(\pi /4,\pi ,\delta t/\sqrt{n})\Psi _{03}(x,r,t_{0}+\tau _{1}+\delta
t_{1})
\end{equation*}%
\begin{equation}
=\Psi _{04}(x,r,t_{0}+\tau _{2}+\delta t_{1})+E_{r4}^{0}(x,r,t_{0}+\tau
_{2}+\delta t_{1})  \tag{6.44}
\end{equation}%
Here the desired state is first obtained according to (6.14) and then by
using (6.27a) and (6.41a) it can be reduced to the form%
\begin{equation}
\Psi _{04}(x,r,t_{0}+\tau _{2}+\delta t_{1})\newline
=\exp [-\frac{1}{2}\frac{i}{\hslash }H_{0}^{ho}(\delta t/\sqrt{n})]\Psi
_{04}(x,r,t_{0}+2\delta t/\sqrt{n}),  \tag{6.45}
\end{equation}%
where the state $\Psi _{04}(x,r,t_{0}+2\delta t/\sqrt{n})$ is given by 
\begin{equation*}
\Psi _{04}(x,r,t_{0}+2\delta t/\sqrt{n})=\exp \{-\frac{i}{\hslash }%
H_{I}(x,\pi /4,\pi )(\delta t/\sqrt{n})\}
\end{equation*}%
\begin{equation}
\times \exp \{-\frac{i}{\hslash }H_{I}(x,\pi /4,3\pi /2)(\delta t/\sqrt{n}%
)\}\Psi _{00}(x,r,t_{0}).  \tag{6.46}
\end{equation}%
The error $E_{r4}^{0}(x,r,t_{0}+\tau _{2}+\delta t_{1})$ is obtained
directly from (6.13). It is clearly bounded by%
\begin{equation*}
||E_{r4}^{0}(x,r,t_{0}+\tau _{2}+\delta t_{1})||\leq
||E_{r}^{0}(x,r,t_{0}+\tau _{2}+\delta t_{1})||
\end{equation*}%
\begin{equation}
+||E_{r}^{V}(x,r,t_{0}+\tau _{2}+\delta
t_{1})||+||E_{r2}^{(3)}(x,r,t_{0}+\tau _{2}+\delta t_{1})||.  \tag{6.47}
\end{equation}%
Here the two errors $E_{r}^{V}(x,r,t_{0}+\tau _{2}+\delta t_{1})$ and $%
E_{r}^{0}(x,r,t_{0}+\tau _{2}+\delta t_{1})$ can be strictly calculated
according to the theoretical calculation method in the section 4. This
calculation should use the interaction $H_{I}(x,\pi /4,\pi )$ and the
Gaussian superposition state $\Psi _{03}^{G}(x,r,t_{0}+\tau _{1}+\delta
t_{1})$ of (6.42) as its initial state. Note that the state $\Psi
_{03}^{G}(x,r,t_{0}+\tau _{1}+\delta t_{1})$ consists of the ten $GWP$
states. One may take each one of the ten $GWP$ states as the initial state
to calculate the upper bounds of the two errors. Then by summing up these
ten upper bounds of the error $E_{r}^{V}(x,r,t_{0}+\tau _{2}+\delta t_{1})$
one may obtain the total upper bound of the error $E_{r}^{V}(x,r,t_{0}+\tau
_{2}+\delta t_{1})$. Similarly, by summing up these ten upper bounds of the
error $E_{r}^{0}(x,r,t_{0}+\tau _{2}+\delta t_{1})$ one also may obtain the
total upper bound of the error $E_{r}^{0}(x,r,t_{0}+\tau _{2}+\delta t_{1}).$
The error $E_{r}^{V}(x,r,t_{0}+\tau _{2}+\delta t_{1})$ is generated by the
imperfection of the $LH$ potential well and the spatially-selective $%
PHAMDOWN $ laser light beams, while the error $E_{r}^{0}(x,r,t_{0}+\tau
_{2}+\delta t_{1})$ originates from the truncation approximation of the
decomposition formula (4.1). It is shown in the section 4 that the upper
bound of the error $E_{r}^{V}(x,r,t_{0}+\tau _{2}+\delta t_{1})$ decays
exponentially with the square deviation-to-spread ratios of the relevant $%
GWP $ states. This is similar to the error $E_{r}^{V}(x,r,t_{0}+\tau _{1})$
in (6.23). Therefore, the error can be controlled by the joint position $%
x_{L}$. It can be neglected when the joint position $x_{L}$ is large enough.
The error $E_{r}^{0}(x,r,t_{0}+\tau _{2}+\delta t_{1})$ is the dominating
term in the error $E_{r4}^{0}(x,r,t_{0}+\tau _{2}+\delta t_{1}).$ It is
shown also in the section 4 that the upper bound of the error $%
E_{r}^{0}(x,r,t_{0}+\tau _{2}+\delta t_{1})$ is proportional to $(\delta t/%
\sqrt{n})^{3}$ approximately and also dependent on the three types of
parameters as stated in the subsection 4.4. Hence the error can be
controlled by the time interval $(\delta t/\sqrt{n})$ and these three types
of parameters. This is similar to the error $E_{r}^{0}(x,r,t_{0}+\tau _{1})$
in (6.23). On the other hand, the error $E_{r2}^{(3)}(x,r,t_{0}+\tau
_{2}+\delta t_{1})$ is generated by the deviation of the initial state $\Psi
_{03}(x,r,t_{0}+\tau _{1}+\delta t_{1})$ in (6.44) from the initial Gaussian
superposition state $\Psi _{03}^{G}(x,r,t_{0}+\tau _{1}+\delta t_{1})$ of
(6.42). It is known from (6.41b) that this deviation is just $\exp [\frac{i}{%
\hslash }H_{0}^{ho}(\delta t/\sqrt{4n})]E_{r2}(x,r,t_{0}+\delta t/\sqrt{n}).$
Then according to (6.16) it can turn out that the error is bounded by%
\begin{equation*}
||E_{r2}^{(3)}(x,r,t_{0}+\tau _{2}+\delta t_{1})||\leq
2||E_{r2}(x,r,t_{0}+\delta t/\sqrt{n})||.
\end{equation*}%
Here the upper bound of the error $E_{r2}(x,r,t_{0}+\delta t/\sqrt{n})$ is
obtained directly from (6.39). Therefore, it follows from (6.39) that the
error $E_{r2}^{(3)}(x,r,t_{0}+\tau _{2}+\delta t_{1})$ can be controlled by
the time interval $\delta t/\sqrt{n},$ the wave-packet spread $\varepsilon
(t_{0}),$ and the wave-number difference $|\Delta k|$. Now the upper bounds
of the three error terms $E_{r}^{0}(x,r,t_{0}+\tau _{2}+\delta t_{1}),$ $%
E_{r}^{V}(x,r,t_{0}+\tau _{2}+\delta t_{1}),$ and $E_{r2}^{(3)}(x,r,t_{0}+%
\tau _{2}+\delta t_{1})$ together with the inequality (6.47) show that the
error $E_{r4}^{0}(x,r,t_{0}+\tau _{2}+\delta t_{1})$ can be controlled by
the joint position $x_{L},$ the time interval $\delta t/\sqrt{n},$ the
wave-packet spread $\varepsilon (t_{0}),$ and the wave-number difference $%
|\Delta k|$.

It is clear that both the desired state $\Psi _{04}(x,r,t_{0}+\tau
_{2}+\delta t_{1})$ of (6.45) and the state $\Psi _{04}(x,r,t_{0}+2\delta t/%
\sqrt{n})$ of (6.46) are not pure $GWP$ states. The former state can be
calculated analytically by using the unitary transformations (4.18) of the
propagators $\exp \{-iH_{I}(x,\pi /4,\gamma )(\delta t/\sqrt{n})/\hslash \}$
with $\gamma =\pi $ and $3\pi /2.$ It is explicitly given by%
\begin{equation*}
\Psi _{04}(x,r,t_{0}+2\delta t/\sqrt{n})=\frac{1}{2}(1+i)\Psi
_{00}(x,t_{0})|g_{0}\rangle
\end{equation*}%
\begin{equation*}
+\frac{1}{2}(1-i)\cos \{4\Omega _{0}(\delta t/\sqrt{n})\cos [\frac{1}{2}%
\Delta kx-\pi /4]\}\Psi _{00}(x,t_{0})|g_{0}\rangle
\end{equation*}%
\begin{equation}
+\frac{1}{2}(1+i)\sin \{4\Omega _{0}(\delta t/\sqrt{n})\cos [\frac{1}{2}%
\Delta kx-\pi /4]\}\exp [i\frac{1}{2}(k_{0}+k_{1})x]\Psi
_{00}(x,t_{0})|e\rangle .  \tag{6.48}
\end{equation}%
The state $\Psi _{04}(x,r,t_{0}+2\delta t/\sqrt{n})$\ is really equal to the
state (5.36) if one sets $\Omega (\tau )=2\Omega _{0}\delta t/\sqrt{n}$ and $%
\Psi _{0}(x,t_{0})=\Psi _{00}(x,t_{0})$ in (5.36). It may be expanded as a
Gaussian superposition state approximately according to the $MGWP$
expansion. Then by inserting the $MGWP$ expansion of (6.48) into (6.45) one
may obtain the explicit $MGWP$ expansion of the desired state $\Psi
_{04}(x,r,t_{0}+\tau _{2}+\delta t_{1}).$ Both the desired state of (6.45)
and its $MGWP$ expansion will be further used in the next-step error
estimation.

Now the fifth experimental propagator $\exp [-iH\delta t_{1}/\hslash ]$ is
applied to the halting-qubit atom at the end of the fourth experimental
propagator. Here the initial state is the final state $\Psi
_{4}(x,r,t_{0}+\tau _{2}+\delta t_{1})$ of (6.43) of the fourth propagator.
Then the atomic product state at the end of the fifth propagator may be
written as%
\begin{equation*}
\Psi _{5}(x,r,t_{0}+\tau _{2}+2\delta t_{1})\newline
=\Psi _{05}(x,r,t_{0}+\tau _{2}+2\delta t_{1})\newline
\end{equation*}%
\begin{equation}
+\sum_{k=1}^{5}E_{rk}^{0}(x,r,t_{0}+\tau _{2}+2\delta t_{1}).  \tag{6.49}
\end{equation}%
There are five error terms in the state (6.49). The first four error terms
in (6.49)\ are obtained by applying the current propagator $\exp [-iH\delta
t_{1}/\hslash ]$ to the four error terms on the $RH$ side of (6.43),
respectively. Therefore, the upper bounds of these four error terms are
equal to those of the four error terms $E_{r1}^{0}(x,r,t_{0}+\delta
t_{1}^{\prime }),$ $E_{r2}^{0}(x,r,t_{0}+\tau _{1}),$ $E_{r3}^{0}(x,r,t_{0}+%
\tau _{1}+\delta t_{1}),$ $E_{r4}^{0}(x,r,t_{0}+\tau _{2}+\delta t_{1}),$
respectively. This shows that the upper bounds of the first four error terms
in (6.49) can be directly obtained in the previous error estimation. Thus,
only the fifth error $E_{r5}^{0}(x,r,t_{0}+\tau _{2}+2\delta t_{1})$ and the
desired state $\Psi _{05}(x,r,t_{0}+\tau _{2}+2\delta t_{1})$ in (6.49) need
to be calculated explicitly. The desired state and the fifth error may be
determined below on the basis of the time evolution process of (6.9). By
applying the current propagator to the initial state $\Psi
_{04}(x,r,t_{0}+\tau _{2}+\delta t_{1}),$ which is the desired state (6.45)
of the fourth propagator, and then according to (6.9) one may obtain formally%
\begin{equation*}
\exp \{-i[H_{0}^{ho}+V_{1}^{ho}(x,\varepsilon )]\delta t_{1}/\hslash \}\Psi
_{04}(x,r,t_{0}+\tau _{2}+\delta t_{1})
\end{equation*}%
\begin{equation}
=\Psi _{05}(x,r,t_{0}+\tau _{2}+2\delta t_{1})\newline
+E_{r5}^{0}(x,r,t_{0}+\tau _{2}+2\delta t_{1})  \tag{6.50}
\end{equation}%
Here the desired state is first obtained according to (6.10) and then by
using (6.45) it can be reduced to the form%
\begin{equation}
\Psi _{05}(x,r,t_{0}+\tau _{2}+2\delta t_{1})\newline
=\exp [\frac{1}{2}\frac{i}{\hslash }H_{0}^{ho}(\delta t/\sqrt{n})]\Psi
_{04}(x,r,t_{0}+2\delta t/\sqrt{n}).  \tag{6.51}
\end{equation}%
On the other hand, according to (6.9) it can turn out that the fifth error
is bounded by%
\begin{equation*}
||E_{r5}^{0}(x,r,t_{0}+\tau _{2}+2\delta t_{1})||\leq
||E_{r}^{(12)}(x,r,t_{0}+\tau _{2}+2\delta t_{1})||
\end{equation*}%
\begin{equation}
+||E_{r1}^{(3)}(x,r,t_{0}+\tau _{2}+2\delta t_{1})||.  \tag{6.52}
\end{equation}%
As shown in (6.45) and (6.48), the state $\Psi _{04}(x,r,t_{0}+\tau
_{2}+\delta t_{1})$ is not a pure $GWP$ state. In order to calculate the
time evolution process of (6.50) one needs to expand its initial state $\Psi
_{04}(x,r,t_{0}+\tau _{2}+\delta t_{1})$ as a Gaussian superposition state.
According to the $MGWP$ expansion the state (6.48) is first expanded as a
Gaussian superposition state $\Psi _{04}^{G}(x,r,t_{0}+2\delta t/\sqrt{n}):$%
\begin{equation}
\Psi _{04}(x,r,t_{0}+2\delta t/\sqrt{n})=\Psi _{04}^{G}(x,r,t_{0}+2\delta t/%
\sqrt{n})+E_{r4}(x,r,t_{0}+2\delta t/\sqrt{n}).  \tag{6.53}
\end{equation}%
Here $E_{r4}(x,r,t_{0}+2\delta t/\sqrt{n})$ is the difference between the
two states $\Psi _{04}(x,r,t_{0}+2\delta t/\sqrt{n})$ and $\Psi
_{04}^{G}(x,r,t_{0}+2\delta t/\sqrt{n}).$ For the present error estimation
it is good enough to express the state $\Psi _{04}^{G}(x,r,t_{0}+2\delta t/%
\sqrt{n})$ as a superposition of the ten lower-order $GWP$ states. Then it
can be found that both the state $\Psi _{04}(x,r,t_{0}+2\delta t/\sqrt{n})$
and its Gaussian superposition state $\Psi _{04}^{G}(x,r,t_{0}+2\delta t/%
\sqrt{n})$ are just equal to the states (5.36) and (5.37), respectively,
with the settings $\Omega (\tau )=2\Omega _{0}\delta t/\sqrt{n}$ and $\Psi
_{0}(x,t_{0})=\Psi _{00}(x,t_{0})$. This means that the error $%
E_{r4}(x,r,t_{0}+2\delta t/\sqrt{n})$ is the same as that one of (5.38).
Therefore, the upper bound of the error may be determined by, according to
(5.39),%
\begin{equation*}
||E_{r4}(x,r,t_{0}+2\delta t/\sqrt{n})||_{u}\thickapprox (2\Omega _{0}\delta
t/\sqrt{n})^{3}\varepsilon (t_{0})^{3}|\Delta k|^{3}
\end{equation*}%
\begin{equation}
\times \{\frac{5}{12288}+\frac{1}{512}\frac{1}{\sqrt{\pi }}\varepsilon
(t_{0})|\Delta k|\}^{1/2}.  \tag{6.54}
\end{equation}%
This upper bound is proportional to $(\delta t/\sqrt{n})^{3},$ $\varepsilon
(t_{0})^{3},$ and $|\Delta k|^{3}$ approximately and hence the error can be
controlled by the time interval $(\delta t/\sqrt{n}),$ the wave-packet
spread $\varepsilon (t_{0}),$ and the wave-number difference $|\Delta k|.$
Now by substituting the expansion (6.53) into (6.45) one obtains the $MGWP$\
expansion of the initial state $\Psi _{04}(x,r,t_{0}+\tau _{2}+\delta t_{1})$
in (6.50): 
\begin{equation*}
\Psi _{04}(x,r,t_{0}+\tau _{2}+\delta t_{1})\newline
=\Psi _{04}^{G}(x,r,t_{0}+\tau _{2}+\delta t_{1})
\end{equation*}%
\begin{equation}
+\exp [-\frac{1}{2}\frac{i}{\hslash }H_{0}^{ho}(\delta t/\sqrt{n}%
)]E_{r4}(x,r,t_{0}+2\delta t/\sqrt{n})  \tag{6.55}
\end{equation}%
where the Gaussian superposition state is 
\begin{equation}
\Psi _{04}^{G}(x,r,t_{0}+\tau _{2}+\delta t_{1})=\exp [-\frac{1}{2}\frac{i}{%
\hslash }H_{0}^{ho}(\delta t/\sqrt{n})]\Psi _{04}^{G}(x,r,t_{0}+2\delta t/%
\sqrt{n}).  \tag{6.56}
\end{equation}%
The expansion (6.55) is similar to (6.8). Here the second term on the $RH$
side of (6.55) is the deviation of the state $\Psi _{04}(x,r,t_{0}+\tau
_{2}+\delta t_{1})$ from the Gaussian superposition state $\Psi
_{04}^{G}(x,r,t_{0}+\tau _{2}+\delta t_{1}).$ It corresponds to the
deviation $E_{r}^{d}(x,r,t_{0})$ in (6.8). The error term $%
E_{r1}^{(3)}(x,r,t_{0}+\tau _{2}+2\delta t_{1})$ in (6.52) originates from
this deviation. Thus, according to (6.12b) and (6.12c) it turns out that the
upper bound of the error $E_{r1}^{(3)}(x,r,t_{0}+\tau _{2}+2\delta t_{1})$
may be obtained from%
\begin{equation}
||E_{r1}^{(3)}(x,r,t_{0}+\tau _{2}+2\delta t_{1})||\leq
2||E_{r4}(x,r,t_{0}+2\delta t/\sqrt{n})||.  \tag{6.57a}
\end{equation}%
These two relations (6.54) and (6.57a) show that the upper bound of the
error $E_{r1}^{(3)}(x,r,t_{0}+\tau _{2}+2\delta t_{1})$ is proportional to $%
(\delta t/\sqrt{n})^{3},$ $\varepsilon (t_{0})^{3},$ and $|\Delta k|^{3}$
approximately. Thus, this error may be controlled by the time interval $%
(\delta t/\sqrt{n}),$ the wave-packet spread $\varepsilon (t_{0}),$ and the
wave-number difference $|\Delta k|.$ Now one may use the initial Gaussian
superposition state $\Psi _{04}^{G}(x,r,t_{0}+\tau _{2}+\delta t_{1})$ of
(6.56) to calculate strictly the error term $E_{r}^{(12)}(x,r,t_{0}+\tau
_{2}+2\delta t_{1})$ in (6.52) according to the theoretical calculation
method in the section 3. As shown in (6.12a), the error $%
E_{r}^{(12)}(x,r,t_{0}+\tau _{2}+2\delta t_{1})$ consists of the two error
terms and its upper bound is determined from%
\begin{equation*}
||E_{r}^{(12)}(x,r,t_{0}+\tau _{2}+2\delta t_{1})||\leq
||E_{r}^{(1)}(x,r,t_{0}+\tau _{2}+2\delta t_{1})||
\end{equation*}%
\begin{equation}
+||E_{r}^{(2)}(x,r,t_{0}+\tau _{2}+2\delta t_{1})||.  \tag{6.57b}
\end{equation}%
Both the error terms $E_{r}^{(k)}(x,r,t_{0}+\tau _{2}+2\delta t_{1})$ with $%
k=1$ and $2$ are generated by the imperfection of the $LH$ potential well.
Their upper bounds can be calculated strictly by using the theoretical
calculation method in the section 3. Here the initial state in the
calculation is the Gaussian superposition state $\Psi
_{04}^{G}(x,r,t_{0}+\tau _{2}+\delta t_{1})$ of (6.56), which consists of
the ten $GWP$ states. One may calculate strictly the upper bounds of the
errors $E_{r}^{(k)}(x,r,t_{0}+\tau _{2}+2\delta t_{1})$ with $k=1$ and $2$
according to the theoretical calculation method in the section 3 by taking
each one of these ten $GWP$ states as the initial state in the calculation.
Then by summing up these ten upper bounds of the error $%
E_{r}^{(1)}(x,r,t_{0}+\tau _{2}+2\delta t_{1})$ according to (3.64) one may
obtain the total upper bound of the error $E_{r}^{(1)}(x,r,t_{0}+\tau
_{2}+2\delta t_{1}).$ Similarly, by summing up these ten upper bounds of the
error $E_{r}^{(2)}(x,r,t_{0}+\tau _{2}+2\delta t_{1})$ according to (3.64)
one also may obtain the total upper bound of the error $%
E_{r}^{(2)}(x,r,t_{0}+\tau _{2}+2\delta t_{1}).$ It has been shown in the
section 3 that the upper bounds of both the errors $E_{r}^{(k)}(x,r,t_{0}+%
\tau _{2}+2\delta t_{1})$ with $k=1$ and $2$ decay exponentially with the
deviation-to-spread ratios of the relevant $GWP$ states. Hence the
inequality (6.57b) shows that the upper bound of the error $%
E_{r}^{(12)}(x,r,t_{0}+\tau _{2}+2\delta t_{1})$ also decay exponentially
with the deviation-to-spread ratios of the relevant $GWP$ states. Then the
error $E_{r}^{(12)}(x,r,t_{0}+\tau _{2}+2\delta t_{1})$ can be controlled by
the joint position $x_{L}$. Now the upper bounds of the errors $%
E_{r}^{(12)}(x,r,t_{0}+\tau _{2}+2\delta t_{1})$ and $%
E_{r1}^{(3)}(x,r,t_{0}+\tau _{2}+2\delta t_{1})$ together with the
inequality (6.52) show that the error $E_{r5}^{0}(x,r,t_{0}+\tau
_{2}+2\delta t_{1})$ can be controlled by the joint position $x_{L},$ the
time interval $(\delta t/\sqrt{n}),$ the wave-packet spread $\varepsilon
(t_{0}),$ and the wave-number difference $|\Delta k|.$

The desired state (6.51) is not a pure $GWP$ state, but it may be expanded
as a Gaussian superposition state by using the $MGWP$ expansion (6.53),%
\begin{equation*}
\Psi _{05}(x,r,t_{0}+\tau _{2}+2\delta t_{1})\newline
=\Psi _{05}^{G}(x,r,t_{0}+\tau _{2}+2\delta t_{1})\newline
\end{equation*}%
\begin{equation}
+\exp [\frac{1}{2}\frac{i}{\hslash }H_{0}^{ho}(\delta t/\sqrt{n}%
)]E_{r4}(x,r,t_{0}+2\delta t/\sqrt{n})  \tag{6.58}
\end{equation}%
where $\Psi _{05}^{G}(x,r,t_{0}+\tau _{2}+2\delta t_{1})$ is a Gaussian
superposition state: 
\begin{equation}
\Psi _{05}^{G}(x,r,t_{0}+\tau _{2}+2\delta t_{1})\newline
=\exp [\frac{1}{2}\frac{i}{\hslash }H_{0}^{ho}(\delta t/\sqrt{n})]\Psi
_{04}^{G}(x,r,t_{0}+2\delta t/\sqrt{n}).  \tag{6.59}
\end{equation}%
It is clear that the state $\Psi _{05}^{G}(x,r,t_{0}+\tau _{2}+2\delta
t_{1}) $ consists of the ten $GWP$ states. The $MGWP$ expansion (6.58) will
be further used in the error estimation below.

Now the sixth propagator $U_{DWN}(\pi /4,\pi /2,\delta t/\sqrt{n})$ is
applied to the halting-qubit atom at the end of the fifth propagator. The
initial state of the current propagator $U_{DWN}(\pi /4,\pi /2,\delta t/%
\sqrt{n})$ is the final state (6.49) of the fifth propagator. At the end of
the current propagator the atomic product state may be written as%
\begin{equation*}
\Psi _{6}(x,r,t_{0}+\tau _{3}+2\delta t_{1})\newline
=\Psi _{06}(x,r,t_{0}+\tau _{3}+2\delta t_{1})\newline
\end{equation*}%
\begin{equation}
+\sum_{k=1}^{6}E_{rk}^{0}(x,r,t_{0}+\tau _{3}+2\delta t_{1})  \tag{6.60}
\end{equation}%
where $\tau _{3}=\delta t_{1}^{\prime }+3\delta t/\sqrt{n}.$ There are six
error terms on the $RH$ side of (6.60). The first five error terms in
(6.60)\ are generated by applying the current propagator to the five error
terms on the $RH$ side of (6.49), respectively. Then their upper bounds are
equal to those of the error terms $E_{r1}^{0}(x,r,t_{0}+\delta t_{1}^{\prime
}),$ $E_{r2}^{0}(x,r,t_{0}+\tau _{1}),$ $E_{r3}^{0}(x,r,t_{0}+\tau
_{1}+\delta t_{1}),$ $E_{r4}^{0}(x,r,t_{0}+\tau _{2}+\delta t_{1}),$ $%
E_{r5}^{0}(x,r,t_{0}+\tau _{2}+2\delta t_{1}),$ respectively. Therefore,
these upper bounds are already obtained in the previous error estimation.
Now only the sixth error $E_{r6}^{0}(x,r,t_{0}+\tau _{3}+2\delta t_{1})$ and
the desired state $\Psi _{06}(x,r,t_{0}+\tau _{3}+2\delta t_{1})$ in (6.60)
need to be explicitly calculated. Both the desired state and the sixth error
can be obtained below on the basis of the time evolution process (6.13). By
applying the current propagator to the state $\Psi _{05}(x,r,t_{0}+\tau
_{2}+2\delta t_{1})$ of (6.58), which is the final desired state of the
fifth propagator, and then according to (6.13) one may formally obtain%
\begin{equation*}
U_{DWN}(\pi /4,\pi /2,\delta t/\sqrt{n})\Psi _{05}(x,r,t_{0}+\tau
_{2}+2\delta t_{1})\newline
\end{equation*}%
\begin{equation}
=\Psi _{06}(x,r,t_{0}+\tau _{3}+2\delta t_{1})\newline
+E_{r6}^{0}(x,r,t_{0}+\tau _{3}+2\delta t_{1}).  \tag{6.61}
\end{equation}%
Here the desired state is first obtained according to (6.14) and then by
using (6.51) and (6.46) it can be reduced to the form%
\begin{equation}
\Psi _{06}(x,r,t_{0}+\tau _{3}+2\delta t_{1})\newline
=\exp [-\frac{1}{2}\frac{i}{\hslash }H_{0}^{ho}(\delta t/\sqrt{n})]\Psi
_{06}(x,r,t_{0}+3\delta t/\sqrt{n}),  \tag{6.62}
\end{equation}%
where the state $\Psi _{06}(x,r,t_{0}+3\delta t/\sqrt{n})$ is defined by 
\newline
\begin{equation*}
\Psi _{06}(x,r,t_{0}+3\delta t/\sqrt{n})=\exp \{-\frac{i}{\hslash }%
H_{I}(x,\pi /4,\pi /2)(\delta t/\sqrt{n})\}
\end{equation*}%
\begin{equation*}
\times \exp \{-\frac{i}{\hslash }H_{I}(x,\pi /4,\pi )(\delta t/\sqrt{n})\}
\end{equation*}%
\begin{equation}
\times \exp \{-\frac{i}{\hslash }H_{I}(x,\pi /4,3\pi /2)(\delta t/\sqrt{n}%
)\}\Psi _{00}(x,r,t_{0}).  \tag{6.63}
\end{equation}%
On the other hand, according to (6.13) and (6.61) it can be found that the
sixth error $E_{r6}^{0}(x,r,t_{0}+\tau _{3}+2\delta t_{1})$ consists of the
three error terms and its upper bound is determined from 
\begin{equation*}
||E_{r6}^{0}(x,r,t_{0}+\tau _{3}+2\delta t_{1})||\leq
||E_{r}^{0}(x,r,t_{0}+\tau _{3}+2\delta t_{1})||
\end{equation*}%
\begin{equation}
+||E_{r}^{V}(x,r,t_{0}+\tau _{3}+2\delta
t_{1})||+||E_{r2}^{(3)}(x,r,t_{0}+\tau _{3}+2\delta t_{1})||.  \tag{6.64}
\end{equation}%
This inequality is similar to (6.47) that the error $E_{r4}^{0}(x,r,t_{0}+%
\tau _{2}+\delta t_{1})$ obeys. The upper bounds of both the errors $%
E_{r}^{0}(x,r,t_{0}+\tau _{3}+2\delta t_{1})$ and $E_{r}^{V}(x,r,t_{0}+\tau
_{3}+2\delta t_{1})$ in (6.64) can be strictly calculated according to the
theoretical calculation method in the section 4. Here one should use the
interaction $H_{I}(x,\pi /4,\pi /2)$ and take the Gaussian superposition
state $\Psi _{05}^{G}(x,r,t_{0}+\tau _{2}+2\delta t_{1})$ of (6.59) as the
initial state in the calculation. Note that the state $\Psi
_{05}^{G}(x,r,t_{0}+\tau _{2}+2\delta t_{1})$ consists of the ten $GWP$
states. Then one may separately calculate the upper bounds of the errors $%
E_{r}^{0}(x,r,t_{0}+\tau _{3}+2\delta t_{1})$ and $E_{r}^{V}(x,r,t_{0}+\tau
_{3}+2\delta t_{1})$ by taking each one of the ten $GWP$ states as the
initial state in the calculation. Then by summing up the ten upper bounds of
the error $E_{r}^{0}(x,r,t_{0}+\tau _{3}+2\delta t_{1})$ one may obtain the
total upper bound of the error $E_{r}^{0}(x,r,t_{0}+\tau _{3}+2\delta
t_{1}). $ Similarly, by summing up the ten upper bounds of the error $%
E_{r}^{V}(x,r,t_{0}+\tau _{3}+2\delta t_{1})$ one also may obtain the total
upper bound of the error $E_{r}^{V}(x,r,t_{0}+\tau _{3}+2\delta t_{1}).$ The
error $E_{r}^{V}(x,r,t_{0}+\tau _{3}+2\delta t_{1})$ is generated by the
imperfection of the $LH$ harmonic potential well and the spatially-selective
effect of the $PHAMDOWN$ laser light beams. It is shown in the section 4
that the error decays exponentially with the square deviation-to-spread
ratios of the relevant $GWP$ states. Thus, it can be controlled by the joint
position $x_{L}$. The error $E_{r}^{0}(x,r,t_{0}+\tau _{3}+2\delta t_{1})$
originates from the truncation approximation of the decomposition formula
(4.1). The strict calculation in the section 4 shows that its upper bound is
proportional to $(\delta t/\sqrt{n})^{3}$ approximately and also dependent
on the three types of parameters as stated in the subsection 4.4. Thus, it
can be controlled by the time interval $\delta t/\sqrt{n}$ and these three
types of parameters. The error $E_{r2}^{(3)}(x,r,t_{0}+\tau _{3}+2\delta
t_{1})$ in (6.64) measures the deviation of the initial state $\Psi
_{05}(x,r,t_{0}+\tau _{2}+2\delta t_{1})$ from the Gaussian superposition
state $\Psi _{05}^{G}(x,r,t_{0}+\tau _{2}+2\delta t_{1})$ of (6.59). This
deviation is given by the second term on the $RH$ side of (6.58). It
corresponds to the deviation of (6.8). Then according to (6.16) it can be
found that the error $E_{r2}^{(3)}(x,r,t_{0}+\tau _{3}+2\delta t_{1})$ is
bounded by%
\begin{equation}
||E_{r2}^{(3)}(x,r,t_{0}+\tau _{3}+2\delta t_{1})||\leq
2||E_{r4}(x,r,t_{0}+2\delta t/\sqrt{n})||.  \tag{6.65}
\end{equation}%
Here the upper bound of the error $E_{r4}(x,r,t_{0}+2\delta t/\sqrt{n})$ is
determined from (6.54). As shown in (6.54), this upper bound is proportional
to $(\delta t/\sqrt{n})^{3},$ $\varepsilon (t_{0})^{3},$ and $|\Delta k|^{3}$
approximately. Hence the inequality (6.65) shows that the error $%
E_{r2}^{(3)}(x,r,t_{0}+\tau _{3}+2\delta t_{1})$ can be controlled by the
time interval $(\delta t/\sqrt{n}),$ the wave-packet spread $\varepsilon
(t_{0}),$ and the wave-number difference $|\Delta k|.$ In the error $%
E_{r6}^{0}(x,r,t_{0}+\tau _{3}+2\delta t_{1})$ both the error terms $%
E_{r2}^{(3)}(x,r,t_{0}+\tau _{3}+2\delta t_{1})$ and $E_{r}^{V}(x,r,t_{0}+%
\tau _{3}+2\delta t_{1})$ are secondary, while the error term $%
E_{r}^{0}(x,r,t_{0}+\tau _{3}+2\delta t_{1})$ is the dominating term. Now
the upper bounds of these three error terms together with the inequality
(6.64) show that the sixth error $E_{r6}^{0}(x,r,t_{0}+\tau _{3}+2\delta
t_{1})$ in (6.60) can be controlled by the joint position $x_{L},$ the time
interval $(\delta t/\sqrt{n}),$ the wave-packet spread $\varepsilon (t_{0}),$
and the wave-number difference $|\Delta k|.$

With the help of the unitary transformations (6.18) of the propagators $\exp
\{-iH_{I}(x,\pi /4,\gamma )(\delta t/\sqrt{n})/\hslash \}$ with $\gamma =\pi
/2,$ $\pi ,$ and $3\pi /2$ the state $\Psi _{06}(x,r,$ $t_{0}+3\delta t/%
\sqrt{n})$ of (6.63) may be calculated analytically. It turns out that the
state may be exactly expressed as%
\begin{equation*}
\Psi _{06}(x,r,t_{0}+3\delta t/\sqrt{n})=\frac{1}{2}[(2+i)\cos \{2\Omega
_{0}(\delta t/\sqrt{n})\cos [\frac{1}{2}\Delta kx-\pi /4]\}
\end{equation*}%
\begin{equation*}
-i\cos \{6\Omega _{0}(\delta t/\sqrt{n})\cos [\frac{1}{2}\Delta kx-\pi
/4]\}]\Psi _{00}(x,t_{0})|g_{0}\rangle
\end{equation*}%
\begin{equation*}
-\frac{1}{2}i[\sin \{2\Omega _{0}(\delta t/\sqrt{n})\cos [\frac{1}{2}\Delta
kx-\pi /4]\}
\end{equation*}%
\begin{equation}
-\sin \{6\Omega _{0}(\delta t/\sqrt{n})\cos [\frac{1}{2}\Delta kx-\pi
/4]\}]\exp [i\frac{1}{2}(k_{0}+k_{1})x]\Psi _{00}(x,t_{0})|e\rangle . 
\tag{6.66}
\end{equation}%
This is a non-$GWP$ state. Both the states (6.62) and (6.66) will be further
used in the error estimation below.

Now the seventh propagator $\exp [-iH\delta t_{1}]$ is applied to the
halting-qubit atom at the end of the sixth propagator. Here the initial
state of the current propagator $\exp [-iH\delta t_{1}]$ is the final state
(6.60) of the sixth propagator. The atomic product state at the end of the
current propagator may be written as%
\begin{equation*}
\Psi _{7}(x,r,t_{0}+\tau _{3}+3\delta t_{1})\newline
=\Psi _{07}(x,r,t_{0}+\tau _{3}+3\delta t_{1})\newline
\end{equation*}%
\begin{equation}
+\sum_{k=1}^{7}E_{rk}^{0}(x,r,t_{0}+\tau _{3}+3\delta t_{1}).  \tag{6.67}
\end{equation}%
There are seven errors on the $RH$ side of (6.67). In analogous way to the
above error estimation one can find that the first six errors in (6.67) have
the same upper bounds as the errors $E_{r1}^{0}(x,r,t_{0}+\delta
t_{1}^{\prime }),$ $E_{r2}^{0}(x,r,t_{0}+\tau _{1}),$ $E_{r3}^{0}(x,r,t_{0}+%
\tau _{1}+\delta t_{1}),$ $E_{r4}^{0}(x,r,t_{0}+\tau _{2}+\delta t_{1}),$ $%
E_{r5}^{0}(x,r,t_{0}+\tau _{2}+2\delta t_{1}),$ and $E_{r6}^{0}(x,r,t_{0}+%
\tau _{3}+2\delta t_{1}),$ respectively. This means that the upper bounds of
these six errors can be obtained from the previous error estimation. Thus,
only the seventh error $E_{r7}^{0}(x,r,t_{0}+\tau _{3}+3\delta t_{1})$ and
the desired state $\Psi _{07}(x,r,t_{0}+\tau _{3}+3\delta t_{1})$ need to be
calculated explicitly below. The desired state and the seventh error may be
obtained below on the basis of the time evolution process (6.9). By acting
the current propagator on the initial state $\Psi _{06}(x,r,t_{0}+\tau
_{3}+2\delta t_{1})$, which is just the final desired state (6.62) of the
sixth propagator, and then according to (6.9) one may formally obtain%
\begin{equation*}
\exp \{-i[H_{0}^{ho}+V_{1}^{ho}(x,\varepsilon )]\delta t_{1}/\hslash \}\Psi
_{06}(x,r,t_{0}+\tau _{3}+2\delta t_{1})\newline
\end{equation*}%
\begin{equation}
=\Psi _{07}(x,r,t_{0}+\tau _{3}+3\delta t_{1})\newline
+E_{r7}^{0}(x,r,t_{0}+\tau _{3}+3\delta t_{1}).  \tag{6.68}
\end{equation}%
Here the desired state is written as, according to (6.10), 
\begin{equation}
\Psi _{07}(x,r,t_{0}+\tau _{3}+3\delta t_{1})=\exp [\frac{1}{2}\frac{i}{%
\hslash }H_{0}^{ho}(\delta t/\sqrt{n})]\Psi _{06}(x,r,t_{0}+3\delta t/\sqrt{n%
}).  \tag{6.69}
\end{equation}%
It follows from (6.9) and (6.68) that the seventh error consists of the two
error terms $E_{r}^{(12)}(x,r,t_{0}+\tau _{3}+3\delta t_{1})$ and $%
E_{r1}^{(3)}(x,r,t_{0}+\tau _{3}+3\delta t_{1})$ and it is bounded by%
\begin{equation*}
||E_{r7}^{0}(x,r,t_{0}+\tau _{3}+3\delta t_{1})||\leq
||E_{r}^{(12)}(x,r,t_{0}+\tau _{3}+3\delta t_{1})||
\end{equation*}%
\begin{equation}
+||E_{r1}^{(3)}(x,r,t_{0}+\tau _{3}+3\delta t_{1})||.  \tag{6.70}
\end{equation}%
Here in order to calculate the two error terms one needs to expand the non-$%
GWP$ state $\Psi _{06}(x,r,t_{0}+\tau _{3}+2\delta t_{1})$ in (6.68) as a
Gaussian superposition state. This can be done by first expanding the state $%
\Psi _{06}(x,r,t_{0}+3\delta t/\sqrt{n})$ of (6.66) as a superposition of
the $GWP$ states, according to the $MGWP$ expansion, 
\begin{equation}
\Psi _{06}(x,r,t_{0}+3\delta t/\sqrt{n})=\Psi _{06}^{G}(x,r,t_{0}+3\delta t/%
\sqrt{n})+E_{r6}(x,r,t_{0}+3\delta t/\sqrt{n}),  \tag{6.71}
\end{equation}%
where $E_{r6}(x,r,t_{0}+3\delta t/\sqrt{n})$ is the truncation error and $%
\Psi _{06}^{G}(x,r,t_{0}+3\delta t/\sqrt{n})$ is a superposition of the ten
lower-order $GWP$ states in the present error estimation. It can turn out
that the error $E_{r6}(x,r,t_{0}+3\delta t/\sqrt{n})$ has the upper bound:%
\begin{equation*}
||E_{r6}(x,r,t_{0}+3\delta t/\sqrt{n})||_{u}\thickapprox (2\Omega _{0}\delta
t/\sqrt{n})^{3}\varepsilon (t_{0})^{3}|\Delta k|^{3}
\end{equation*}%
\begin{equation}
\times \{\frac{245}{1536}+\frac{49}{64}\frac{1}{\sqrt{\pi }}\varepsilon
(t_{0})|\Delta k|\}^{1/2}.  \tag{6.72}
\end{equation}%
This upper bound is proportional to $(\delta t/\sqrt{n})^{3},$ $\varepsilon
(t_{0})^{3},$ and $|\Delta k|^{3}$ approximately. Thus, the error $%
E_{r6}(x,r,t_{0}+3\delta t/\sqrt{n})$ may be controlled by the time interval 
$(\delta t/\sqrt{n}),$ the wave-packet spread $\varepsilon (t_{0}),$ and the
wave-number difference $|\Delta k|$. Then by substituting the expansion
(6.71) into (6.62) the state $\Psi _{06}(x,r,t_{0}+\tau _{3}+2\delta t_{1})$
may be expanded as%
\begin{equation*}
\Psi _{06}(x,r,t_{0}+\tau _{3}+2\delta t_{1})=\Psi _{06}^{G}(x,r,t_{0}+\tau
_{3}+2\delta t_{1})
\end{equation*}%
\begin{equation}
+\exp [-\frac{1}{2}\frac{i}{\hslash }H_{0}^{ho}(\delta t/\sqrt{n}%
)]E_{r6}(x,r,t_{0}+3\delta t/\sqrt{n})  \tag{6.73}
\end{equation}%
where $\Psi _{06}^{G}(x,r,t_{0}+\tau _{3}+2\delta t_{1})$ is given by 
\begin{equation}
\Psi _{06}^{G}(x,r,t_{0}+\tau _{3}+2\delta t_{1})=\exp [-\frac{1}{2}\frac{i}{%
\hslash }H_{0}^{ho}(\delta t/\sqrt{n})]\Psi _{06}^{G}(x,r,t_{0}+3\delta t/%
\sqrt{n}).  \tag{6.74}
\end{equation}%
In the present error estimation the state $\Psi _{06}^{G}(x,r,t_{0}+\tau
_{3}+2\delta t_{1})$ is a superposition of the ten $GWP$ states. Now the
error term $E_{r}^{(12)}(x,r,t_{0}+\tau _{3}+3\delta t_{1})$ in (6.70),
which according to (6.12a) consists of the two error terms $%
E_{r}^{(k)}(x,r,t_{0}+\tau _{3}+3\delta t_{1})$ with $k=1$ and $2$, can be
strictly calculated according to the theoretical calculation method in the
section 3. Here one should use the superposition state $\Psi
_{06}^{G}(x,r,t_{0}+\tau _{3}+2\delta t_{1})$ of the ten $GWP$ states as the
initial state in the calculation. One may separately calculate the upper
bounds of the error terms $E_{r}^{(k)}(x,r,t_{0}+\tau _{3}+3\delta t_{1})$
with $k=1$ and $2$ by taking each one of the ten $GWP$ states as the initial
state in the theoretical calculation. Then by summing up these ten upper
bounds of the error term $E_{r}^{(k)}(x,r,t_{0}+\tau _{3}+3\delta t_{1})$
for $k=1$ or $2$ according to (3.64) one may obtain the total upper bound of
the error term. As shown in the section 3, the error terms $%
E_{r}^{(k)}(x,r,t_{0}+\tau _{3}+3\delta t_{1})$ with $k=1$ and $2$ originate
from the imperfection of the $LH$ harmonic potential well and hence so does
the error $E_{r}^{(12)}(x,r,t_{0}+\tau _{3}+3\delta t_{1})$. It is shown in
the section 3 that the error terms $E_{r}^{(k)}(x,r,t_{0}+\tau _{3}+3\delta
t_{1})$ with $k=1$ and $2$ decay exponentially with the square
deviation-to-spread ratios of the relevant $GWP$ states and hence so does
the error $E_{r}^{(12)}(x,r,t_{0}+\tau _{3}+3\delta t_{1}).$ Then the error $%
E_{r}^{(12)}(x,r,t_{0}+\tau _{3}+3\delta t_{1})$ in (6.70) can be controlled
by the joint position $x_{L}$. The second term on the $RH$ side of (6.73)
corresponds to the deviation of (6.8). Thus, as shown in (6.12b), this term
generates the error $E_{r1}^{(3)}(x,r,t_{0}+\tau _{3}+3\delta t_{1})$ in
(6.70). It follows from (6.12c) that the error is bounded by%
\begin{equation}
||E_{r1}^{(3)}(x,r,t_{0}+\tau _{3}+3\delta t_{1})||\leq
2||E_{r6}(x,r,t_{0}+3\delta t/\sqrt{n})||.  \tag{6.75}
\end{equation}%
This inequality shows that the upper bound of the error $%
E_{r1}^{(3)}(x,r,t_{0}+\tau _{3}+3\delta t_{1})$ is also determined from
(6.72). This upper bound is proportional to $(\delta t/\sqrt{n})^{3},$ $%
\varepsilon (t_{0})^{3},$ and $|\Delta k|^{3}$ approximately. Thus, the
error can be controlled by the time interval $(\delta t/\sqrt{n}),$ the
wave-packet spread $\varepsilon (t_{0}),$ and the wave-number difference $%
|\Delta k|.$ Now the upper bounds of the two errors $E_{r}^{(12)}(x,r,t_{0}+%
\tau _{3}+3\delta t_{1})$ and $E_{r1}^{(3)}(x,r,t_{0}+\tau _{3}+3\delta
t_{1})$ together with the inequality (6.70) show that the seventh error $%
E_{r7}^{0}(x,r,t_{0}+\tau _{3}+3\delta t_{1})$ in (6.67) can be controlled
by the joint position $x_{L},$ the time interval $(\delta t/\sqrt{n}),$ the
wave-packet spread $\varepsilon (t_{0}),$ and the wave-number difference $%
|\Delta k|.$

The desired state $\Psi _{07}(x,r,t_{0}+\tau _{3}+3\delta t_{1})$ of (6.69)
is not a pure $GWP$ state. According to the $MGWP$ expansion it can be
expanded as a Gaussian superposition state approximately. This can be done
easily by inserting the state (6.71) into (6.69), 
\begin{equation*}
\Psi _{07}(x,r,t_{0}+\tau _{3}+3\delta t_{1})=\Psi _{07}^{G}(x,r,t_{0}+\tau
_{3}+3\delta t_{1})
\end{equation*}%
\begin{equation}
+\exp [\frac{1}{2}\frac{i}{\hslash }H_{0}^{ho}(\delta t/\sqrt{n}%
)]E_{r6}(x,r,t_{0}+3\delta t/\sqrt{n})  \tag{6.76}
\end{equation}%
where $\Psi _{07}^{G}(x,r,t_{0}+\tau _{3}+3\delta t_{1})$ is a Gaussian
superposition state:%
\begin{equation}
\Psi _{07}^{G}(x,r,t_{0}+\tau _{3}+3\delta t_{1})=\exp [\frac{1}{2}\frac{i}{%
\hslash }H_{0}^{ho}(\delta t/\sqrt{n})]\Psi _{06}^{G}(x,r,t_{0}+3\delta t/%
\sqrt{n}).  \tag{6.77}
\end{equation}%
It is clear that the state $\Psi _{07}^{G}(x,r,t_{0}+\tau _{3}+3\delta
t_{1}) $ is a superposition of the ten $GWP$ states. Both the states (6.76)
and (6.77) will be used in the following error estimation.

Now the eighth propagator $U_{DWN}(\pi /4,0,\delta t/\sqrt{n})$ is applied
to the halting-qubit atom at the end of the seventh propagator. At this step
the initial state is the final state (6.67) of the seventh propagator. Then
the atomic product state at the end of the eighth propagator may be written
formally as%
\begin{equation*}
\Psi _{8}(x,r,t_{0}+\tau _{4}+3\delta t_{1})\newline
=\Psi _{08}(x,r,t_{0}+\tau _{4}+3\delta t_{1})\newline
\end{equation*}%
\begin{equation}
+\sum_{k=1}^{8}E_{rk}^{0}(x,r,t_{0}+\tau _{4}+3\delta t_{1})  \tag{6.78}
\end{equation}%
where $\tau _{4}=\delta t_{1}^{\prime }+4\delta t/\sqrt{n}.$ There are eight
errors on the $RH$ side of (6.78) to be calculated. But the upper bounds of
the first seven errors in (6.78) are equal to those of the errors $%
E_{r1}^{0}(x,r,t_{0}+\delta t_{1}^{\prime }),$ $E_{r2}^{0}(x,r,t_{0}+\tau
_{1}),$ $E_{r3}^{0}(x,r,t_{0}+\tau _{1}+\delta t_{1}),$ $%
E_{r4}^{0}(x,r,t_{0}+\tau _{2}+\delta t_{1}),$ $E_{r5}^{0}(x,r,t_{0}+\tau
_{2}+2\delta t_{1}),$ $E_{r6}^{0}(x,r,t_{0}+\tau _{3}+2\delta t_{1}),$ and $%
E_{r7}^{0}(x,r,t_{0}+\tau _{3}+3\delta t_{1})$, respectively. Therefore,
actually these seven upper bounds can be obtained from the previous error
estimation. Now only the eighth error $E_{r8}^{0}(x,r,t_{0}+\tau
_{4}+3\delta t_{1})$ and the desired state $\Psi _{08}(x,r,t_{0}+\tau
_{4}+3\delta t_{1})$ have not yet been calculated in (6.78). The desired
state and the eighth error can be obtained below on the basis of the time
evolution process (6.13). By applying the current propagator to the initial
state $\Psi _{07}(x,r,t_{0}+\tau _{3}+3\delta t_{1})$, which is just the
final desired state (6.69) of the seventh propagator, and then according to
(6.13) one may obtain formally%
\begin{equation*}
U_{DWN}(\pi /4,0,\delta t/\sqrt{n})\Psi _{07}(x,r,t_{0}+\tau _{3}+3\delta
t_{1})
\end{equation*}%
\begin{equation}
=\Psi _{08}(x,r,t_{0}+\tau _{4}+3\delta t_{1})+E_{r8}^{0}(x,r,t_{0}+\tau
_{4}+3\delta t_{1}).  \tag{6.79}
\end{equation}%
Here the desired state is first obtained according to (6.14) and then by
using (6.63) it can be reduced to the form%
\begin{equation}
\Psi _{08}(x,r,t_{0}+\tau _{4}+3\delta t_{1})=\exp [-\frac{1}{2}\frac{i}{%
\hslash }H_{0}^{ho}(\delta t/\sqrt{n})]\Psi _{08}(x,r,t_{0}+4\delta t/\sqrt{n%
}),  \tag{6.80}
\end{equation}%
where the state $\Psi _{08}(x,r,t_{0}+4\delta t/\sqrt{n})$ may be written as%
\begin{equation}
\Psi _{08}(x,r,t_{0}+4\delta t/\sqrt{n})=U_{tr}(\delta t/\sqrt{n})\Psi
_{00}(x,r,t_{0}),  \tag{6.81}
\end{equation}%
and the propagator $U_{tr}(\delta t/\sqrt{n})$ is just given by (6.6). By
comparing (6.79) with (6.13) it can be found that the error $%
E_{r8}^{0}(x,r,t_{0}+\tau _{4}+3\delta t_{1})$ consists of the three error
terms and its upper bound may be determined from%
\begin{equation*}
||E_{r8}^{0}(x,r,t_{0}+\tau _{4}+3\delta t_{1})||\leq
||E_{r}^{0}(x,r,t_{0}+\tau _{4}+3\delta t_{1})||
\end{equation*}%
\begin{equation}
+||E_{r}^{V}(x,r,t_{0}+\tau _{4}+3\delta
t_{1})||+||E_{r2}^{(3)}(x,r,t_{0}+\tau _{4}+3\delta t_{1})||.  \tag{6.82}
\end{equation}%
The two error terms $E_{r}^{0}(x,r,t_{0}+\tau _{4}+3\delta t_{1})$ and $%
E_{r}^{V}(x,r,t_{0}+\tau _{4}+3\delta t_{1})$ can be calculated strictly
with the theoretical calculation method in the section 4. This calculation
needs to use the interaction $H_{I}(x,\pi /4,0)$ and take the $MGWP$
expansion state $\Psi _{07}^{G}(x,r,t_{0}+\tau _{3}+3\delta t_{1})$ of
(6.77) as its initial state. Note that $\Psi _{07}^{G}(x,r,t_{0}+\tau
_{3}+3\delta t_{1})$ consists of the ten $GWP$ states in the present error
estimation. One may separately calculate the upper bounds of the two error
terms $E_{r}^{0}(x,r,t_{0}+\tau _{4}+3\delta t_{1})$ and $%
E_{r}^{V}(x,r,t_{0}+\tau _{4}+3\delta t_{1})$ by taking each one of the ten $%
GWP$ states as the initial state in the calculation. Then by summing up
these ten upper bounds for each one of the two error terms one may obtain
the total upper bound of the error term. The error $E_{r}^{V}(x,r,t_{0}+\tau
_{4}+3\delta t_{1})$ is generated by the imperfection of the $LH$ potential
well and the spatially-selective effect of the $PHAMDOWN$ laser light beams.
Its upper bound decays exponentially with the square deviation-to-spread
ratios of the relevant $GWP$ states. Thus, the error can be controlled by
the joint position $x_{L}$. The error $E_{r}^{0}(x,r,t_{0}+\tau _{4}+3\delta
t_{1})$ is the dominating term in the eighth error $E_{r8}^{0}(x,r,t_{0}+%
\tau _{4}+3\delta t_{1}).$ It originates from the truncation approximation
of the decomposition formula (4.1). Its upper bound is proportional to $%
(\delta t/\sqrt{n})^{3}$ approximately and also dependent on the three types
of parameters as stated in the subsection 4.4. Hence the error may be
controlled by the time interval $\delta t/\sqrt{n}$ and these three types of
parameters. On the other hand, the error $E_{r2}^{(3)}(x,r,t_{0}+\tau
_{4}+3\delta t_{1})$ in (6.82) is generated by the second term on the $RH$
side of (6.76), which is the truncation error of the $MGWP$ expansion of the
initial state $\Psi _{07}(x,r,t_{0}+\tau _{3}+3\delta t_{1})$ in (6.79).
Then according to (6.16) it can turn out that the upper bound of the error
can be determined from%
\begin{equation}
||E_{r2}^{(3)}(x,r,t_{0}+\tau _{4}+3\delta t_{1})||\leq
2||E_{r6}(x,r,t_{0}+3\delta t/\sqrt{n})||.  \tag{6.83}
\end{equation}%
Here the error $E_{r6}(x,r,t_{0}+3\delta t/\sqrt{n})$ has the upper bound
that is determined from (6.72). It follows from (6.72) and (6.83) that the
upper bound of the error $E_{r2}^{(3)}(x,r,t_{0}+\tau _{4}+3\delta t_{1})$
is proportional to $(\delta t/\sqrt{n})^{3},$ $\varepsilon (t_{0})^{3},$ and 
$|\Delta k|^{3}$ approximately. Thus, the error may be controlled by the
time interval $\delta t/\sqrt{n},$ the wave-packet spread $\varepsilon
(t_{0}),$ and the wave-number difference $|\Delta k|$. Now the upper bounds
of the three error terms $E_{r}^{0}(x,r,t_{0}+\tau _{4}+3\delta t_{1}),$ $%
E_{r}^{V}(x,r,t_{0}+\tau _{4}+3\delta t_{1}),$ and $E_{r2}^{(3)}(x,r,t_{0}+%
\tau _{4}+3\delta t_{1})$ together with the inequality (6.82) show that the
eighth error $E_{r8}^{0}(x,r,t_{0}+\tau _{4}+3\delta t_{1})$ can be
controlled by the joint position $x_{L},$ the time interval $\delta t/\sqrt{n%
},$ the wave-packet spread $\varepsilon (t_{0}),$ and the wave-number
difference $|\Delta k|$.

Now the last propagator $\exp [-iH\delta t_{1}^{\prime }]$ of the
experimental basic pulse sequence $P_{tr}^{r}(\delta t/\sqrt{n})$ is applied
to the halting-qubit atom at the end of the eighth propagator. At this last
step the initial state is $\Psi _{8}(x,r,t_{0}+\tau _{4}+3\delta t_{1})$ of
(6.78). Then the atomic product state at the end of the experimental basic
pulse sequence may be written as%
\begin{equation*}
\Psi _{9}(x,r,t_{0}+\tau _{4}+3\delta t_{1}+\delta t_{1}^{\prime })\newline
=\Psi _{09}(x,r,t_{0}+\tau _{4}+3\delta t_{1}+\delta t_{1}^{\prime })\newline
\end{equation*}%
\begin{equation}
+\sum_{k=1}^{9}E_{rk}^{0}(x,r,t_{0}+\tau _{4}+3\delta t_{1}+\delta
t_{1}^{\prime }).  \tag{6.84}
\end{equation}%
There are nine errors to be calculated in (6.84). But the upper bounds of
the first eight errors in (6.84) are equal to those of the errors $%
E_{r1}^{0}(x,r,t_{0}+\delta t_{1}^{\prime }),$ $E_{r2}^{0}(x,r,t_{0}+\tau
_{1}),$ $E_{r3}^{0}(x,r,t_{0}+\tau _{1}+\delta t_{1}),$ $%
E_{r4}^{0}(x,r,t_{0}+\tau _{2}+\delta t_{1}),$ $E_{r5}^{0}(x,r,t_{0}+\tau
_{2}+2\delta t_{1}),$ $E_{r6}^{0}(x,r,t_{0}+\tau _{3}+2\delta t_{1}),$ $%
E_{r7}^{0}(x,r,t_{0}+\tau _{3}+3\delta t_{1})$, and $E_{r8}^{0}(x,r,t_{0}+%
\tau _{4}+3\delta t_{1}),$ respectively. Thus, these eight upper bounds can
be really obtained from the previous error estimation. Now only the ninth
error $E_{r9}^{0}(x,r,t_{0}+\tau _{4}+3\delta t_{1}+\delta t_{1}^{\prime })$
and the desired state $\Psi _{09}(x,r,t_{0}+\tau _{4}+3\delta t_{1}+\delta
t_{1}^{\prime })$ in (6.84) need to be calculated strictly. Both the ninth
error and the desired state may be obtained below on the basis of the time
evolution process (6.9). By applying the current propagator $\exp [-iH\delta
t_{1}^{\prime }]$ to the initial state $\Psi _{08}(x,r,t_{0}+\tau
_{4}+3\delta t_{1}),$ which is the final desired state (6.80) of the eighth
propagator, and then according to (6.9) one may obtain formally%
\begin{equation*}
\exp \{-i[H_{0}^{ho}+V_{1}^{ho}(x,\varepsilon )]\delta t_{1}^{\prime
}/\hslash \}\Psi _{08}(x,r,t_{0}+\tau _{4}+3\delta t_{1})
\end{equation*}%
\begin{equation}
=\Psi _{09}(x,r,t_{0}+\tau _{4}+3\delta t_{1}+\delta t_{1}^{\prime
})+E_{r9}^{0}(x,r,t_{0}+\tau _{4}+3\delta t_{1}+\delta t_{1}^{\prime }). 
\tag{6.85}
\end{equation}%
Here the desired state is first obtained according to (6.10) and then by
further using (6.80) it can be reduced to the form%
\begin{equation}
\Psi _{09}(x,r,t_{0}+\tau _{4}+3\delta t_{1}+\delta t_{1}^{\prime
})=U_{tr}(\delta t/\sqrt{n})\Psi _{00}(x,r,t_{0}).  \tag{6.86}
\end{equation}%
One can find that the desired state (6.86) is just equal to the state $\Psi
_{08}(x,r,t_{0}+4\delta t/\sqrt{n})$ of (6.81). By comparing (6.85) with
(6.9) one can find that the ninth error consists of the two error terms and
its upper bound may be determined from 
\begin{equation*}
||E_{r9}^{0}(x,r,t_{0}+\tau _{4}+3\delta t_{1}+\delta t_{1}^{\prime })||\leq
||E_{r}^{(12)}(x,r,t_{0}+\tau _{4}+3\delta t_{1}+\delta t_{1}^{\prime })||
\end{equation*}%
\begin{equation}
+||E_{r1}^{(3)}(x,r,t_{0}+\tau _{4}+3\delta t_{1}+\delta t_{1}^{\prime })||.
\tag{6.87}
\end{equation}%
Both the error terms on the $RH$ side of (6.87) can be calculated strictly.
But here one needs first to expand the initial state $\Psi
_{08}(x,r,t_{0}+\tau _{4}+3\delta t_{1})$ in (6.85) as a superposition of
the $GWP$ states. As shown in (6.80), such an expansion may be achieved by
expanding the state $\Psi _{08}(x,r,t_{0}+4\delta t/\sqrt{n})$ of (6.81) as
a Gaussian superposition state. The state $\Psi _{08}(x,r,t_{0}+4\delta t/%
\sqrt{n})$ can be exactly calculated and it is equal to the state (6.93a)
below. Then by using the $MGWP$ expansion one may expand the state $\Psi
_{08}(x,r,t_{0}+4\delta t/\sqrt{n})$ as%
\begin{equation}
\Psi _{08}(x,r,t_{0}+4\delta t/\sqrt{n})=\Psi _{08}^{G}(x,r,t_{0}+4\delta t/%
\sqrt{n})+E_{r8}(x,r,t_{0}+4\delta t/\sqrt{n}),  \tag{6.88}
\end{equation}%
where $E_{r8}(x,r,t_{0}+4\delta t/\sqrt{n})$ is the truncation error and in
the present error estimation $\Psi _{08}^{G}(x,r,t_{0}+4\delta t/\sqrt{n})$
is a superposition of the ten lower-order $GWP$ states. It can turn out that
the upper bound of the error $E_{r8}(x,r,t_{0}+4\delta t/\sqrt{n})$ may be
determined from%
\begin{equation*}
||E_{r8}(x,r,t_{0}+4\delta t/\sqrt{n})||_{u}\thickapprox (2\Omega _{0}\delta
t/\sqrt{n})^{3}\varepsilon (t_{0})^{3}|\Delta k|^{3}
\end{equation*}%
\begin{equation}
\times \{\frac{125}{192}+\frac{25}{8}\frac{1}{\sqrt{\pi }}\varepsilon
(t_{0})|\Delta k|\}^{1/2}.  \tag{6.89}
\end{equation}%
This upper bound is approximately proportional to $(\delta t/\sqrt{n})^{3},$ 
$\varepsilon (t_{0})^{3},$ and $|\Delta k|^{3}.$ Thus, the error may be
controlled by the time interval $\delta t/\sqrt{n},$ the wave-packet spread $%
\varepsilon (t_{0}),$ and the wave-number difference $|\Delta k|$. The upper
bound of (6.89) may be further used to obtain the upper bound of the error $%
E_{r1}^{(3)}(x,r,t_{0}+\tau _{4}+3\delta t_{1}+\delta t_{1}^{\prime })$ in
(6.87). By inserting the state (6.88) into (6.80) one obtains the $MGWP$
expansion of the state $\Psi _{08}(x,r,t_{0}+\tau _{4}+3\delta t_{1})$:%
\begin{equation*}
\Psi _{08}(x,r,t_{0}+\tau _{4}+3\delta t_{1})=\Psi _{08}^{G}(x,r,t_{0}+\tau
_{4}+3\delta t_{1})
\end{equation*}%
\begin{equation}
+\exp [-\frac{1}{2}\frac{i}{\hslash }H_{0}^{ho}(\delta t/\sqrt{n}%
)]E_{r8}(x,r,t_{0}+4\delta t/\sqrt{n}).  \tag{6.90}
\end{equation}%
Here $\Psi _{08}^{G}(x,r,t_{0}+\tau _{4}+3\delta t_{1})$ is given by%
\begin{equation}
\Psi _{08}^{G}(x,r,t_{0}+\tau _{4}+3\delta t_{1})=\exp [-\frac{1}{2}\frac{i}{%
\hslash }H_{0}^{ho}(\delta t/\sqrt{n})]\Psi _{08}^{G}(x,r,t_{0}+4\delta t/%
\sqrt{n}).  \tag{6.91}
\end{equation}%
It is a superposition of the ten $GWP$ states. The second term on the $RH$
side of (6.90) corresponds to the deviation of (6.8). Then it follows from
(6.85), (6.12b), and (6.12c) that the error $E_{r1}^{(3)}(x,r,t_{0}+\tau
_{4}+3\delta t_{1}+\delta t_{1}^{\prime })$ in (6.87) is bounded by%
\begin{equation}
||E_{r1}^{(3)}(x,r,t_{0}+\tau _{4}+3\delta t_{1}+\delta t_{1}^{\prime
})||\leq 2||E_{r8}(x,r,t_{0}+4\delta t/\sqrt{n})||.  \tag{6.92}
\end{equation}%
This inequality shows that the upper bound of the error $%
E_{r1}^{(3)}(x,r,t_{0}+\tau _{4}+3\delta t_{1}+\delta t_{1}^{\prime })$ also
may be determined from (6.89). As shown in (6.89), this upper bound is
approximately proportional to $(\delta t/\sqrt{n})^{3},$ $\varepsilon
(t_{0})^{3},$ and $|\Delta k|^{3}$. Then the error may be controlled by the
time interval $\delta t/\sqrt{n},$ the wave-packet spread $\varepsilon
(t_{0}),$ and the wave-number difference $|\Delta k|$. On the other hand,
the error $E_{r}^{(12)}(x,r,t_{0}+\tau _{4}+3\delta t_{1}+\delta
t_{1}^{\prime })$ in (6.87) can be strictly calculated with the theoretical
calculation method in the section 3. As shown in (6.12a), the error $%
E_{r}^{(12)}(x,r,t_{0}+\tau _{4}+3\delta t_{1}+\delta t_{1}^{\prime })$
consists of the two error terms $E_{r}^{(k)}(x,r,t_{0}+\tau _{4}+3\delta
t_{1}+\delta t_{1}^{\prime })$ with $k=1$ and $2$, which can be strictly
calculated with the theoretical calculation method in the section 3. Here
one should use the Gaussian superposition state $\Psi
_{08}^{G}(x,r,t_{0}+\tau _{4}+3\delta t_{1})$ of (6.91) as the initial state
in the calculation. Note that the state $\Psi _{08}^{G}(x,r,t_{0}+\tau
_{4}+3\delta t_{1})$ consists of the ten $GWP$ states. One may separately
calculate the upper bounds of the error terms $E_{r}^{(k)}(x,r,t_{0}+\tau
_{4}+3\delta t_{1}+\delta t_{1}^{\prime })$ with $k=1$ and $2$ by taking
each one of the ten $GWP$ states as the initial state in the calculation.
Then by summing up the ten upper bound of the error term $%
E_{r}^{(k)}(x,r,t_{0}+\tau _{4}+3\delta t_{1}+\delta t_{1}^{\prime })$ for $%
k=1$ or $2$ according to (3.64) one may obtain the total upper bound of the
error term. Both the error terms originate from the imperfection of the $LH$
harmonic potential well. It is shown in the section 3 that they decay
exponentially with the square deviation-to-spread ratios of the relevant $%
GWP $ states. Thus, they may be controlled by the joint position $x_{L}$.
Then according to (6.12a) the error $E_{r}^{(12)}(x,r,t_{0}+\tau
_{4}+3\delta t_{1}+\delta t_{1}^{\prime })$ in (6.87) also decays
exponentially with the square deviation-to-spread ratios of the relevant $%
GWP $ states and can be controlled by the joint position $x_{L}.$ Now the
upper bounds of the errors $E_{r1}^{(3)}(x,r,t_{0}+\tau _{4}+3\delta
t_{1}+\delta t_{1}^{\prime })$ and $E_{r}^{(12)}(x,r,t_{0}+\tau _{4}+3\delta
t_{1}+\delta t_{1}^{\prime })$ together with the inequality (6.87) show that
the ninth error $E_{r9}^{0}(x,r,t_{0}+\tau _{4}+3\delta t_{1}+\delta
t_{1}^{\prime })$ in (6.84) can be controlled by the joint position $x_{L},$
the time interval $\delta t/\sqrt{n},$ the wave-packet spread $\varepsilon
(t_{0}),$ and the wave-number difference $|\Delta k|$.

The desired state $\Psi _{09}(x,r,t_{0}+\tau _{4}+3\delta t_{1}+\delta
t_{1}^{\prime })$ of (6.86) is also the desired final state of the
experimental basic pulse sequence. It is just equal to the state $\Psi
_{08}(x,r,t_{0}+4\delta t/\sqrt{n})$ of (6.81). Then the two equations
(6.84) and (6.86) together show that the experimental basic pulse sequence $%
P_{tr}^{r}(\delta t/\sqrt{n})$ indeed generates the unitary propagator $%
U_{tr}(\delta t/\sqrt{n})$ of (6.6) if all the nine errors in (6.84) can be
neglected. By using the propagator $U_{tr}(\delta t/\sqrt{n})$ of (6.6) and
the unitary transformations (4.18) of the propagators $\exp \{-iH_{I}(x,\pi
/4,\gamma )(\delta t/\sqrt{n})/\hslash \}$ with $\gamma =0,$ $\pi /2,$ $\pi
, $ and $3\pi /2$ one can obtain exactly the state $\Psi
_{08}(x,r,t_{0}+4\delta t/\sqrt{n})$ from (6.81), which is just the desired
state (6.86). Then it can be found that the desired state (6.86) may be
exactly expressed as%
\begin{equation*}
\Psi _{09}(x,r,t_{0}+\tau _{4}+3\delta t_{1}+\delta t_{1}^{\prime })=\{\frac{%
1}{4}(1+i)+\cos \{4\Omega _{0}(\delta t/\sqrt{n})\cos [\frac{1}{2}\Delta
kx-\pi /4]\}
\end{equation*}%
\begin{equation*}
-\frac{1}{4}(1+i)\cos \{8\Omega _{0}(\delta t/\sqrt{n})\cos [\frac{1}{2}%
\Delta kx-\pi /4]\}\}\Psi _{00}(x,t_{0})|g_{0}\rangle
\end{equation*}%
\begin{equation*}
+\frac{1}{2}(1-i)\{\sin \{4\Omega _{0}(\delta t/\sqrt{n})\cos [\frac{1}{2}%
\Delta kx-\pi /4]\}
\end{equation*}%
\begin{equation}
-\frac{1}{2}\sin \{8\Omega _{0}(\delta t/\sqrt{n})\cos [\frac{1}{2}\Delta
kx-\pi /4]\}\}\exp [i\frac{1}{2}(k_{0}+k_{1})x]\Psi _{00}(x,t_{0})|e\rangle .
\tag{6.93a}
\end{equation}%
On the other hand, according to the BCH\ formula (6.4) the unitary
propagator $U_{tr}(\delta t/\sqrt{n})$ also is given by (6.2a). Then by
inserting the propagator $U_{tr}(\delta t/\sqrt{n})$ of (6.2a) into (6.86)
the desired state (6.86) also can be written as 
\begin{equation*}
\Psi _{09}(x,r,t_{0}+\tau _{4}+3\delta t_{1}+\delta t_{1}^{\prime })=\exp
\{iQ(\delta t/\hslash )^{2}/n\}\Psi _{00}(x,r,t_{0})
\end{equation*}%
\begin{equation}
+O_{p}((\delta t/\sqrt{n})^{3})\Psi _{00}(x,r,t_{0})  \tag{6.93b}
\end{equation}%
where $O_{p}((\delta t/\sqrt{n})^{3})$ should be considered as an error
operator which is just equal to $O((\delta t/\sqrt{n})^{3})$ in (6.2a). The
upper bound of the error $O_{p}((\delta t/\sqrt{n})^{3})$ $\times \Psi
_{00}(x,r,t_{0})$ in (6.93b) can be calculated strictly from the exact
desired state (6.93a). It can turn out that it is proportional to $(\delta t/%
\sqrt{n})^{3}$ approximately, 
\begin{equation*}
||O_{p}((\delta t/\sqrt{n})^{3})\Psi _{00}(x,r,t_{0})||\leq \frac{5}{6}%
|(4\Omega _{0}\delta t/\sqrt{n})|^{3}
\end{equation*}%
\begin{equation}
+\frac{1}{2}(4\Omega _{0}\delta t/\sqrt{n})^{4}+\frac{1}{48}(4\Omega
_{0}\delta t/\sqrt{n})^{6}.  \tag{6.94}
\end{equation}%
Since $(\delta t/\sqrt{n})<<1,$ one may expand both the desired states
(6.93a) and (6.93b) as power series of the time interval $(\delta t/\sqrt{n}%
).$ Then by comparing the two states (6.93a) and (6.93b) with each other one
can find that the two states are really the same up to the error term $%
O((\delta t/\sqrt{n})^{3}).$ This shows that the experimental basic pulse
sequence $P_{tr}^{r}(\delta t/\sqrt{n})$ indeed generates the unitary
propagator $\exp \{iQ(\delta t/\hslash )^{2}/n\}$ in (6.2a) up to the error
term $O((\delta t/\sqrt{n})^{3}).$ This also indicates that the propagator $%
U_{tr}(\delta t/\sqrt{n})$ of (6.6) indeed can be reduced to the propagator $%
\exp \{iQ(\delta t/\hslash )^{2}/n\}$ in (6.2a) up to the error term $%
O((\delta t/\sqrt{n})^{3}).$ This is in agreement with the result obtained
by the conventional BCH formula.

It is known that the propagator $U_{tr}(\delta t/\sqrt{n})$ of (6.6) is
generated by the theoretical basic pulse sequence $P_{tr}^{i}(\delta t/\sqrt{%
n})$ of (6.7a). Of course, there is an error $O((\delta t/\sqrt{n})^{3})$
when the theoretical basic pulse sequence is reduced to $U_{tr}(\delta t/%
\sqrt{n})$. Thus, the theoretical basic pulse sequence acting on the initial
state $\Psi _{00}(x,r,t_{0})$ is equal to the propagator $U_{tr}(\delta t/%
\sqrt{n})$ acting on the same initial state which generates the desired
state (6.86) after neglecting the error $O((\delta t/\sqrt{n})^{3}).$ On the
other hand, the desired state (6.86) is generated by the experimental basic
pulse sequence $P_{tr}^{r}(\delta t/\sqrt{n})$ of (6.7b). If now all these
nine errors on the $RH$ side of (6.84) can be neglected, then both the
states (6.84) and (6.86) show that the experimental basic pulse sequence
applying to the initial state $\Psi _{00}(x,r,t_{0})$ is equal to the
propagator $U_{tr}(\delta t/\sqrt{n})$ applying to the same initial state.
These show that both the theoretical basic pulse sequence $P_{tr}^{i}(\delta
t/\sqrt{n})$ and the experimental one $P_{tr}^{r}(\delta t/\sqrt{n})$
applying to the same initial state $\Psi _{00}(x,r,t_{0})$ generate the same
desired state (6.86) if all these relevant errors can be neglected. This is
just the expected result in the section.

Below the nine errors on the $RH$ side of (6.84) is analyzed in a summary
form. For convenience, here the final product state of the experimental
basic pulse sequence may be simply written as%
\begin{equation}
P_{tr}^{r}(\delta t/\sqrt{n})\Psi _{00}(x,r,t_{0})=U_{tr}(\delta t/\sqrt{n}%
)\Psi _{00}(x,r,t_{0})+E_{rt}^{0}(x,r,t_{10})  \tag{6.95}
\end{equation}%
where the total error $E_{rt}^{0}(x,r,t_{10})$ consists of all the nine
errors on the $RH$ side of (6.84) and it is bounded by%
\begin{equation}
||E_{rt}^{0}(x,r,t_{10})||\leq \sum_{k=1}^{9}||E_{rk}^{0}(x,r,t_{0}+\delta
t_{1}^{\prime }+4\delta t/\sqrt{n}+3\delta t_{1}+\delta t_{1}^{\prime })||. 
\tag{6.96}
\end{equation}%
These nine errors may be divided into the two families, the first family
consisting of the first, third, fifth, seventh, and ninth error, and the
second comprising the second, fourth, sixth, and eighth error. The first
family are generated during the time evolution processes of the first kind
of propagators $\exp \{-i[H_{0}^{ho}+V_{1}^{ho}(x,\varepsilon )]\tau
^{\prime }\}.$ They originate from the imperfection of the $LH$ harmonic
potential well and the deviation of the initial state from the initial
Gaussian superposition state of the first kind of propagator. As shown in
the above error estimation, each error of the family is bounded by%
\begin{equation*}
||E_{rk}^{0}(x,r,t_{0}+\delta t_{1}^{\prime }+4\delta t/\sqrt{n}+3\delta
t_{1}+\delta t_{1}^{\prime })||
\end{equation*}%
\begin{equation*}
\leq ||E_{r}^{(12)}(x,r,t_{0}+t_{k})||+||E_{r1}^{(3)}(x,r,t_{0}+t_{k})||,%
\text{ }k=1,3,5,7,9,
\end{equation*}%
where the time interval $t_{k}$ is given by $t_{1}=\delta t_{1}^{\prime },$ $%
t_{3}=\tau _{1}+\delta t_{1},$ $t_{5}=\tau _{2}+2\delta t_{1},$ $t_{7}=\tau
_{3}+3\delta t_{1},$ and $t_{9}=\tau _{4}+3\delta t_{1}+\delta t_{1}^{\prime
}$ or $t_{9}=\delta t_{1}^{\prime }+4\delta t/\sqrt{n}+3\delta t_{1}+\delta
t_{1}^{\prime }.$ The error $E_{r1}^{(3)}(x,r,t_{0}+t_{k})$ is a truncation
error of the $MGWP$ expansion of the initial state. The above error
estimation shows that each one of the five errors $%
\{E_{r1}^{(3)}(x,r,t_{0}+t_{k})\}$ is bounded by%
\begin{equation*}
||E_{r1}^{(3)}(x,r,t_{0}+t_{1})||=0,
\end{equation*}%
\begin{equation*}
||E_{r1}^{(3)}(x,r,t_{0}+t_{2l+1})||\leq 2||E_{r2l}(x,r,t_{0}+l\times \delta
t/\sqrt{n})||,\text{ }l=1,2,3,4.
\end{equation*}%
The upper bound of the error $E_{r1}^{(3)}(x,r,t_{0}+t_{2l+1})$ for $l=1,$ $%
2,$ $3,$ $4$ is therefore determined from that one of the error $%
E_{r2l}(x,r,t_{0}+l\times \delta t/\sqrt{n}).$ When the initial Gaussian
superposition state consists of the ten lower-order $GWP$ states, the upper
bound of the error $E_{r2l}(x,r,t_{0}+l\times \delta t/\sqrt{n})$ is
proportional to $(\delta t/\sqrt{n})^{3},$ $\varepsilon (t_{0})^{3},$ and $%
|\Delta k|^{3}$ approximately. This can be seen from (6.39), (6.54), (6.72),
and (6.89). Therefore, the upper bound of the error $%
E_{r1}^{(3)}(x,r,t_{0}+t_{2l+1})$ also is proportional to $(\delta t/\sqrt{n}%
)^{3},$ $\varepsilon (t_{0})^{3},$ and $|\Delta k|^{3}$ approximately. Then
the error may be controlled by the time interval $(\delta t/\sqrt{n}),$ the
wave-packet spread $\varepsilon (t_{0}),$ and the wave-number difference $%
|\Delta k|.$ It may be improved greatly when the initial Gaussian
superposition state consists of more than ten lower-order $GWP$ states.
Therefore, it usually may make a secondary contribution to the error $%
E_{rk}^{0}(x,r,t_{0}+t_{9})$. On the other hand, the error $%
E_{r}^{(12)}(x,r,t_{0}+t_{k})$ for $k=1,$ $3,$ $5,$ $7,$ $9$ originates from
the imperfection of the $LH$ harmonic potential well. As shown in (6.12a),
the error $E_{r}^{(12)}(x,r,t_{0}+t_{k})$ consists of the two error terms
and it is bounded by%
\begin{equation*}
||E_{r}^{(12)}(x,r,t_{0}+t_{k})||\leq
||E_{r}^{(1)}(x,r,t_{0}+t_{k})||+||E_{r}^{(2)}(x,r,t_{0}+t_{k})||.
\end{equation*}%
It is shown above that they decay exponentially with the deviation-to-spread
ratios of the relevant $GWP$ states. They can be controlled by the joint
position $x_{L}$. When the joint position $x_{L}$ is large enough, they can
be neglected. Therefore, the first family of errors $%
\{E_{rk}^{0}(x,r,t_{0}+t_{9})\}$ can be controlled by the joint position $%
x_{L},$ the time interval $(\delta t/\sqrt{n}),$ the wave-packet spread $%
\varepsilon (t_{0}),$ and the wave-number difference $|\Delta k|.$ They
usually make a secondary contribution to the total error $%
E_{rt}^{0}(x,r,t_{10})$ in (6.95).

The second family of errors $\{E_{rk}^{0}(x,r,t_{0}+t_{9})\}$ with $k=2$, $4$%
, $6$, $8$ on the $RH$ side of (6.96) are generated during the time
evolution processes of the second kind of propagators $\{U_{DWN}(\pi
/4,\gamma ,\delta t/\sqrt{n})\}$. They originate from the imperfection of
the $LH$ harmonic potential well and the spatially-selective effect of the $%
PHAMDOWN$ laser light beams, the truncation approximation for the
decomposition formula (4.1), and the deviation of the initial state from the
initial Gaussian superposition state of the second kind of propagator. Each
error of the family is bounded by%
\begin{equation*}
||E_{rk}^{0}(x,r,t_{0}+\delta t_{1}^{\prime }+4\delta t/\sqrt{n}+3\delta
t_{1}+\delta t_{1}^{\prime })||\leq ||E_{r}^{0}(x,r,t_{0}+t_{k})||
\end{equation*}%
\begin{equation*}
+||E_{r}^{V}(x,r,t_{0}+t_{k})||+||E_{r2}^{(3)}(x,r,t_{0}+t_{k})||,\text{ }%
k=2,4,6,8,
\end{equation*}%
where the time interval $t_{k}$ is given by $t_{2}=\tau _{1},$ $t_{4}=\tau
_{2}+\delta t_{1},$ $t_{6}=\tau _{3}+2\delta t_{1},$ and $t_{8}=\tau
_{4}+3\delta t_{1}.$ The error $E_{r2}^{(3)}(x,r,t_{0}+t_{k})$ for $k=2$, $4$%
, $6$, $8$ is generated by the deviation of the initial state from the
initial Gaussian superposition state of the second kind of propagator. It is
bounded by%
\begin{equation*}
||E_{r2}^{(3)}(x,r,t_{0}+t_{2l})||\leq 2||E_{r2l}(x,r,t_{0}+l\times \delta t/%
\sqrt{n})||,\text{ }l=1,2,3,4.
\end{equation*}%
Then the upper bound of the error $E_{r2}^{(3)}(x,r,t_{0}+t_{2l})$ may be
determined from that one of the error $E_{r2l}(x,r,t_{0}+l\times \delta t/%
\sqrt{n})$ for $l=1$, $2$, $3$, $4.$ As shown above, this means that the
upper bound of the error $E_{r2}^{(3)}(x,r,t_{0}+t_{2l})$ for $l=1,$ $2,$ $%
3, $ $4$ is proportional to $(\delta t/\sqrt{n})^{3},$ $\varepsilon
(t_{0})^{3}, $ and $|\Delta k|^{3}$ approximately, when the initial Gaussian
superposition state consists of the ten $GWP$ states. The error can be
controlled by the time interval $\delta t/\sqrt{n},$ the wave-packet spread $%
\varepsilon (t_{0}),$ and the wave-number difference $|\Delta k|.$ It may be
greatly improved when the initial Gaussian superposition state consists of
more than ten $GWP$ states. Therefore, it usually may make a secondary
contribution to the error $E_{rk}^{0}(x,r,t_{0}+t_{9})$. The error $%
E_{r}^{V}(x,r,t_{0}+t_{k})$ is generated by the imperfection of the $LH$
harmonic potential well and the spatially selective effect of the $PHAMDOWN$
laser light beams. It decays exponentially with the square
deviation-to-spread ratios of the relevant $GWP$ states. Hence it may be
controlled by the joint position $x_{L}$. When the joint position $x_{L}$ is
large enough, it can be neglected. Therefore, the error $%
E_{r}^{V}(x,r,t_{0}+t_{k})$ usually makes a secondary contribution to the
error $E_{rk}^{0}(x,r,t_{0}+t_{9}).$ The error $E_{r}^{0}(x,r,t_{0}+t_{k})$
originates from the truncation approximation of the decomposition formula
(4.1). It is independent of the imperfection of the $LH$ harmonic potential
well and the spatially selective effect of the $PHAMDOWN$ laser light beams.
It is approximately proportional to $(\delta t/\sqrt{n})^{3}$ and also
dependent on the three types of parameters of the $GWP$ motional states of
the halting-qubit atom and the experimental basic pulse sequence (6.7b).
Hence it can be controlled by the time interval $\delta t/\sqrt{n}$ and
these three types of parameters. The error $E_{r}^{0}(x,r,t_{0}+t_{k})$ is
usually the dominating term in the error $E_{rk}^{0}(x,r,t_{0}+t_{9}).$ The
above error analysis shows that the second family of errors $%
\{E_{rk}^{0}(x,r,t_{0}+t_{9})\}$ with $k=2$, $4$, $6$, $8$ on the $RH$ side
of (6.96) may be controlled by the joint position $x_{L},$ the time interval 
$\delta t/\sqrt{n},$ the wave-packet spread $\varepsilon (t_{0}),$ the
wave-number difference $|\Delta k|$ and so on. They usually make a
dominating contribution to the total error $E_{rt}^{0}(x,r,t_{10})$ in
(6.95).

Here a summary is given for the above theoretical analysis. After the
initial $GWP$ product state $\Psi _{00}(x,r,t_{0})$ is acted on
consecutively by the nine propagators of the experimental basic pulse
sequence $P_{tr}^{r}(\delta t/\sqrt{n})$ (from the right to the left side in
(6.7b)), it is converted into the final product state $P_{tr}^{r}(\delta t/%
\sqrt{n})\Psi _{00}(x,r,t_{0}).$ The detailed theoretical analysis above
shows that the final product state is really equal to approximately the
desired product state $U_{tr}(\delta t/\sqrt{n})\Psi _{00}(x,r,t_{0})$ which
also is generated approximately by applying the theoretical basic pulse
sequence $P_{tr}^{i}(\delta t/\sqrt{n})$ to the same initial product state.%
\newline
\newline
\newline
{\large 6.2 Theoretical analysis of the SSISS triggering pulse}

The experimental basic pulse sequence $P_{tr}^{r}(\delta t/\sqrt{n})$ is the
basic starting point to further analyze the $SSISS$ triggering pulse $%
P_{tr}^{r}(\delta t)=[P_{tr}^{r}(\delta t/\sqrt{n})]^{n}$ below. For
convenience, here the desired state (6.86) is simply written as $\Psi
_{10}^{tr}(x,r,t_{10})=U_{tr}(\delta t/\sqrt{n})\Psi _{00}(x,r,t_{0}),$
where the time $t_{10}=t_{0}+\delta t_{r}$ and $\delta t_{r}\equiv
t_{9}=2\delta t_{1}^{\prime }+4\delta t/\sqrt{n}+3\delta t_{1}$ is the total
time period of the experimental basic pulse sequence $P_{tr}^{r}(\delta t/%
\sqrt{n}).$ It follows from (6.93b) that the desired state $\Psi
_{10}^{tr}(x,r,t_{10})$ may be rewritten as%
\begin{equation}
\Psi _{10}^{tr}(x,r,t_{10})=\Psi _{10}^{TR}(x,r,t_{10})+O_{10}((\delta t/%
\sqrt{n})^{3}).  \tag{6.97}
\end{equation}%
Here the product state $\Psi _{10}^{TR}(x,r,t_{10})$ is given by%
\begin{equation*}
\Psi _{10}^{TR}(x,r,t_{10})=\exp \{iQ(\delta t/\hslash )^{2}/n\}\Psi
_{00}(x,r,t_{0}),
\end{equation*}%
and the error $O_{10}((\delta t/\sqrt{n})^{3})$ by%
\begin{equation*}
O_{10}((\delta t/\sqrt{n})^{3})=O_{p}((\delta t/\sqrt{n})^{3})\Psi
_{00}(x,r,t_{0}).
\end{equation*}%
The upper bound of the error is proportional to $(\delta t/\sqrt{n})^{3}$
approximately, as shown in (6.94). Furthermore, it is shown in the section 5
that in the Lamb-Dicke limit the state $\Psi _{10}^{TR}(x,r,t_{10})$ may be
approximated well by a single $GWP$ state. Notice that $\Psi
_{00}(x,r,t_{0})=\Psi _{00}(x,t_{0})|g_{0}\rangle .$ As shown in (5.1) and
(5.2), the product state $\Psi _{10}^{TR}(x,r,t_{10})$ may be explicitly
written as%
\begin{equation*}
\Psi _{10}^{TR}(x,r,t_{10})=\exp \{i[(2\Omega _{0}\delta t)^{2}/n]\sin
(\Delta kx)\}\Psi _{00}(x,r,t_{0})
\end{equation*}%
\begin{equation}
=\Psi _{10}^{LD}(x,r,t_{10})+E_{r10}^{LD}(x,r,t_{10}).  \tag{6.98}
\end{equation}%
Here the Lamb-Dicke-limit state is defined by%
\begin{equation}
\Psi _{10}^{LD}(x,r,t_{10})=\exp \{i[(q_{s}/n)\Delta ky+(q_{c}/n)(1-\frac{1}{%
2}(\Delta ky)^{2})]\}\Psi _{00}(x,r,t_{0}),  \tag{6.99}
\end{equation}%
and it can turn out that the upper bound of the error $%
E_{r10}^{LD}(x,r,t_{10})$ is determined from%
\begin{equation*}
||E_{r10}^{LD}(x,r,t_{10})||_{u}\thickapprox \lbrack (2\Omega _{0}\delta
t)^{2}/n]\varepsilon (t_{0})^{3}|\Delta k|^{3}
\end{equation*}%
\begin{equation}
\times \{\frac{15}{288}+\frac{1}{12}\frac{1}{\sqrt{\pi }}\varepsilon
(t_{0})|\Delta k|\}^{1/2}.  \tag{6.100}
\end{equation}%
Obviously, this upper bound is proportional to $|\Delta k|^{3}$ and $%
\varepsilon (t_{0})^{3}$ approximately and inversely proportional to the
number $n$. Thus, the error $E_{r10}^{LD}(x,r,t_{10})$ can be controlled by
the wave-number difference $|\Delta k|,$ the wave-packet spread $\varepsilon
(t_{0}),$ and the number $n$. The Lamb-Dicke-limit state $\Psi
_{10}^{LD}(x,r,t_{10})$ is a single $GWP$ state. It has the same internal
state $|g_{0}\rangle $ as the starting $GWP$ state $\Psi _{00}(x,r,t_{0}).$
It also has the same COM position $x_{c}(t_{10})=x_{c}(t_{0})$ and
wave-packet spread $\varepsilon (t_{10})=\varepsilon (t_{0})$ as the
starting state $\Psi _{00}(x,r,t_{0}).$ But its momentum $%
p_{c}(t_{10})=p_{c}(t_{0})+(q_{s}/n)\hslash \Delta k$ is different from that
one ($p_{c}(t_{0}))$ of the starting state and its complex linewidth $%
W(t_{10})$ also slightly different from that one ($W(t_{0}))$ of the
starting state. Though by using the $MGWP$ expansion the state $\Psi
_{10}^{TR}(x,r,t_{10})$ may be expanded approximately as a superposition of
a finite number of the $GWP$ states instead of a single $GWP$ state (6.99)
so that a better approximation can be achieved, this expansion could lead to
a big trouble if a superposition of the $GWP$ states, instead of a single $%
GWP$ state, appears in the unitary STIRAP decelerating and accelerating
processes [14, 16]. Now by substituting (6.97) and (6.98)\ into (6.95) the
final product state of the experimental basic pulse sequence is reduced to
the form%
\begin{equation}
P_{tr}^{r}(\delta t/\sqrt{n})\Psi _{00}(x,r,t_{0})=\Psi
_{10}^{LD}(x,r,t_{10})+E_{rT}^{0}(x,r,t_{10})  \tag{6.101}
\end{equation}%
where the total error $E_{rT}^{0}(x,r,t_{10})$ is given by%
\begin{equation}
E_{rT}^{0}(x,r,t_{10})=E_{rt}^{0}(x,r,t_{10})+O_{10}((\delta t/\sqrt{n}%
)^{3})+E_{r10}^{LD}(x,r,t_{10}).  \tag{6.102}
\end{equation}%
This error is generated when the product state $P_{tr}^{r}(\delta t/\sqrt{n}%
)\Psi _{00}(x,r,t_{0})$ is reduced to the desired state $\Psi
_{10}^{tr}(x,r,t_{10})$ and then to the state $\Psi _{10}^{TR}(x,r,t_{10})$\
and finally to the single $GWP$ state $\Psi _{10}^{LD}(x,r,t_{10}).$ The
three error terms on the $RH$ side of (6.102) are given in (6.95), (6.97),
and (6.98), respectively.

There is only one\ experimental basic pulse sequence $P_{tr}^{r}(\delta t/%
\sqrt{n})$ to be analyzed in the above error estimation. A $SSISS$
triggering pulse usually consists of many experimental basic pulse sequences 
$\{P_{tr}^{r}(\delta t/\sqrt{n})\}$. For example, the $SSISS$ triggering
pulse may be constructed by $P_{tr}^{r}(\delta t)=[P_{tr}^{r}(\delta t/\sqrt{%
n})]^{n}.$ Then the above error estimation may be used as well for the $%
SSISS $ trigger pulse $P_{tr}^{r}(\delta t).$ The equation (6.101) shows
that at the end of the first experimental basic pulse sequence $%
P_{tr}^{r}(\delta t/\sqrt{n})$ of the $SSISS$ trigger pulse $%
P_{tr}^{r}(\delta t)$ the halting-qubit atom is in the product state $\Psi
_{10}^{LD}(x,r,t_{10})$ of (6.99) if the error $E_{rT}^{0}(x,r,t_{10})$ of
(6.102) is neglected. Obviously, this product state also is the starting
product state of the second experimental basic pulse sequence $%
P_{tr}^{r}(\delta t/\sqrt{n})$ of the $SSISS$ trigger pulse $%
P_{tr}^{r}(\delta t).$ Then it follows from (6.101) that at the end of the
second experimental basic pulse sequence the halting-qubit atom is in the
product state:%
\begin{equation*}
\lbrack P_{tr}^{r}(\delta t/\sqrt{n})]^{2}\Psi
_{00}(x,r,t_{0})=P_{tr}^{r}(\delta t/\sqrt{n})\Psi _{10}^{LD}(x,r,t_{10})
\end{equation*}%
\begin{equation}
+P_{tr}^{r}(\delta t/\sqrt{n})E_{rT}^{0}(x,r,t_{10}).  \tag{6.103}
\end{equation}%
Since the unitary propagator $P_{tr}^{r}(\delta t/\sqrt{n})$ does not change
the total probability of the error state $E_{rT}^{0}(x,r,t_{10}),$ the error 
$P_{tr}^{r}(\delta t/\sqrt{n})E_{rT}^{0}(x,r,t_{10})$ in (6.103) has really
the same upper bound as the error $E_{rT}^{0}(x,r,t_{10})$,%
\begin{equation}
||P_{tr}^{r}(\delta t/\sqrt{n}%
)E_{rT}^{0}(x,r,t_{10})||=||E_{rT}^{0}(x,r,t_{10})||.  \tag{6.104}
\end{equation}%
It can be seen in (6.103) that the product state $\Psi
_{10}^{LD}(x,r,t_{10}) $ now acts as the initial state of the second
experimental basic pulse sequence. As shown in (6.99), the product state $%
\Psi _{10}^{LD}(x,r,t_{10})$ is a single $GWP$ state. This is similar to the
starting product state $\Psi _{00}(x,r,t_{0})$ of the first experimental
basic pulse sequence. Its internal state $|g_{0}\rangle $ is also the same
as that one of the starting product state. Then the theoretical analysis in
the subsection 6.1 may be used as well to evaluate the time evolution
process for the second experimental basic pulse sequence applying to the
starting product state $\Psi _{10}^{LD}(x,r,t_{10}).$ Hence by applying the
equation (6.95) to the first term on the $RH$ side of (6.103) one obtains%
\begin{equation}
P_{tr}^{r}(\delta t/\sqrt{n})\Psi _{10}^{LD}(x,r,t_{10})=U_{tr}(\delta t/%
\sqrt{n})\Psi _{10}^{LD}(x,r,t_{10})+E_{rt}^{1}(x,r,t_{20}).  \tag{6.105}
\end{equation}%
Here the time $t_{20}=t_{10}+\delta t_{r}.$ By comparing (6.105) with (6.95)
one sees that there are the corresponding relations:\ $\Psi
_{10}^{LD}(x,r,t_{10})\leftrightarrow \Psi _{00}(x,r,t_{0})$ and $%
E_{rt}^{1}(x,r,t_{20})\leftrightarrow E_{rt}^{0}(x,r,t_{10}).$ Thus, the
error $E_{rt}^{1}(x,r,t_{20})$ here is similar to the error $%
E_{rt}^{0}(x,r,t_{10})$ in (6.95). It also consists of the nine error terms.
If the initial product state $\Psi _{00}(x,r,t_{0})$ in the nine error terms 
$\{E_{rk}^{0}(x,r,t_{0}+t_{9})\}$ of the first experimental basic pulse
sequence is replaced with the initial product state $\Psi
_{10}^{LD}(x,r,t_{10}),$ then these nine error terms $%
\{E_{rk}^{0}(x,r,t_{0}+t_{9})\}$ are changed to the corresponding nine error
terms $\{E_{rk}^{1}(x,r,t_{10}+\delta t_{r})\}$ of the second experimental
basic pulse sequence, respectively. Therefore, the nine error terms $%
\{E_{rk}^{1}(x,r,t_{10}+\delta t_{r})\}$ of the second experimental pulse
sequence can be strictly calculated with the same theoretical calculation
method that is used to calculate strictly the nine error terms $%
\{E_{rk}^{0}(x,r,t_{0}+t_{9})\}$ of the first experimental basic pulse
sequence in the subsection 6.1. Here the calculation uses the starting
product state $\Psi _{10}^{LD}(x,r,t_{10})$ instead of $\Psi
_{00}(x,r,t_{0}).$ Once the nine error terms $\{E_{rk}^{1}(x,r,t_{10}+\delta
t_{r})\}$ are obtained, the upper bound of the error $E_{rt}^{1}(x,r,t_{20})$
is determined from%
\begin{equation}
||E_{rt}^{1}(x,r,t_{20})||\leq \sum_{k=1}^{9}||E_{rk}^{1}(x,r,t_{10}+\delta
t_{r})||.  \tag{6.106}
\end{equation}%
This is similar to the inequality (6.96) of the error $%
E_{rt}^{0}(x,r,t_{10}).$ Notice that the product state $\Psi
_{10}^{LD}(x,r,t_{10})$ of (6.99) has the same COM position $%
x_{c}(t_{10})=x_{c}(t_{0})$ and wave-packet spread $\varepsilon
(t_{10})=\varepsilon (t_{0})$ as the starting product state $\Psi
_{00}(x,r,t_{0}).$ Denote $y=x-x_{c}(t_{10})=x-x_{c}(t_{0}).$ Then in the
Lamb-Dicke limit the desired product state $U_{tr}(\delta t/\sqrt{n})\Psi
_{10}^{LD}(x,r,t_{10})$ in (6.105) may be written as%
\begin{equation*}
U_{tr}(\delta t/\sqrt{n})\Psi _{10}^{LD}(x,r,t_{10})=\Psi
_{20}^{TR}(x,r,t_{20})+O_{20}((\delta t/\sqrt{n})^{3})
\end{equation*}%
\begin{equation}
=\Psi _{20}^{LD}(x,r,t_{20})+E_{r20}^{LD}(x,r,t_{20})+O_{20}((\delta t/\sqrt{%
n})^{3})  \tag{6.107}
\end{equation}%
where the product state $\Psi _{20}^{TR}(x,r,t_{20})=\exp \{iQ(\delta
t/\hslash )^{2}/n\}\Psi _{10}^{LD}(x,r,t_{10})$, the error $O_{20}((\delta t/%
\sqrt{n})^{3})=O_{p}((\delta t/\sqrt{n})^{3})\Psi _{10}^{LD}(x,r,t_{10}),$
and the Lamb-Dicke-limit state $\Psi _{20}^{LD}(x,r,t_{20})$ is given by 
\begin{equation}
\Psi _{20}^{LD}(x,r,t_{20})=\exp \{i[(q_{s}/n)\Delta ky+(q_{c}/n)(1-\frac{1}{%
2}(\Delta ky)^{2})]\}\Psi _{10}^{LD}(x,r,t_{10}).  \tag{6.108}
\end{equation}%
These equations appearing in (6.107) are similar to those equations of
(6.97) and (6.98), while the equation (6.108) is similar to (6.99). By using
(6.107) the state (6.105)\ may be reduced to the form%
\begin{equation}
P_{tr}^{r}(\delta t/\sqrt{n})\Psi _{10}^{LD}(x,r,t_{10})=\Psi
_{20}^{LD}(x,r,t_{20})+E_{rT}^{1}(x,r,t_{20}).  \tag{6.109}
\end{equation}%
Here the total error $E_{rT}^{1}(x,r,t_{20})$ is given by%
\begin{equation}
E_{rT}^{1}(x,r,t_{20})=E_{rt}^{1}(x,r,t_{20})+O_{20}((\delta t/\sqrt{n}%
)^{3})+E_{r20}^{LD}(x,r,t_{20}).  \tag{6.110}
\end{equation}%
This error is similar to the error $E_{rT}^{0}(x,r,t_{10})$ of (6.102). Now
by inserting the state (6.109) into (6.103) one further obtains%
\begin{equation*}
\lbrack P_{tr}^{r}(\delta t/\sqrt{n})]^{2}\Psi _{00}(x,r,t_{0})=\Psi
_{20}^{LD}(x,r,t_{20})
\end{equation*}%
\begin{equation}
+E_{rT}^{1}(x,r,t_{20})+P_{tr}^{r}(\delta t/\sqrt{n})E_{rT}^{0}(x,r,t_{10}).
\tag{6.111}
\end{equation}%
This is the final state after the experimental pulse sequence $%
[P_{tr}^{r}(\delta t/\sqrt{n})]^{2}$ is applied to the halting-qubit atom in
the starting product state $\Psi _{00}(x,r,t_{0}).$ It can prove that the
product state $\Psi _{20}^{LD}(x,r,t_{20})$ is still a single $GWP$ state.
By inserting the state $\Psi _{10}^{LD}(x,r,t_{10})$ of (6.99) into (6.108)
the state $\Psi _{20}^{LD}(x,r,t_{20})$ is reduced to the form%
\begin{equation}
\Psi _{20}^{LD}(x,r,t_{20})=\exp \{i2[(q_{s}/n)\Delta ky+(q_{c}/n)(1-\frac{1%
}{2}(\Delta ky)^{2})]\}\Psi _{00}(x,r,t_{0}).  \tag{6.112}
\end{equation}%
This is indeed a single $GWP$ state. Moreover, this state has the same COM
position $x_{c}(t_{20})=x(t_{0})$ and wave-packet spread $\varepsilon
(t_{20})=\varepsilon (t_{0})$ as the starting state $\Psi _{00}(x,r,t_{0}).$
Thus, the equation (6.111) shows that after the starting state $\Psi
_{00}(x,r,t_{0})$ is acted on by the experimental pulse sequence $%
[P_{tr}^{r}(\delta t/\sqrt{n})]^{2},$ it is converted into the pure $GWP$
state $\Psi _{20}^{LD}(x,r,t_{20})$ of (6.112), which has the same COM
position and wave-packet spread but different momentum and complex linewidth
from the starting state $\Psi _{00}(x,r,t_{0}),$ if all the error terms on
the $RH$ side of (6.111) are neglected. As shown in (6.112), the momentum
for the final state $\Psi _{20}^{LD}(x,r,t_{20})$ is $%
p_{c}(t_{20})=p_{c}(t_{0})+2(q_{s}/n)\hslash \Delta k$. Thus, the
experimental pulse sequence $[P_{tr}^{r}(\delta t/\sqrt{n})]^{2}$ achieves a
double increment for the motional momentum of the halting-qubit atom with
respect to the experimental basic pulse sequence $P_{tr}^{r}(\delta t/\sqrt{n%
}).$

More generally, an experimental pulse sequence $[P_{tr}^{r}(\delta t/\sqrt{n}%
)]^{n^{\prime }}$ $(n\geq n^{\prime }\geq 1)$ may be used to excite the
halting-qubit atom in the initial state $\Psi _{00}(x,r,t_{0})$. The above
theoretical analysis for the experimental pulse sequences $P_{tr}^{r}(\delta
t/\sqrt{n})$ and $[P_{tr}^{r}(\delta t/\sqrt{n})]^{2}$ needs to be
generalized to describe the time evolution process of the experimental pulse
sequence $[P_{tr}^{r}(\delta t/\sqrt{n})]^{n^{\prime }}.$ According to the
relations (6.101) and (6.111) one can deduce that the atomic product state
at the end of the experimental pulse sequence $[P_{tr}^{r}(\delta t/\sqrt{n}%
)]^{n^{\prime }}$ is given by%
\begin{equation}
\lbrack P_{tr}^{r}(\delta t/\sqrt{n})]^{n^{\prime }}\Psi
_{00}(x,r,t_{0})=\Psi _{n^{\prime }0}^{LD}(x,r,t_{n^{\prime
}0})+E_{R}(n^{\prime }).  \tag{6.113}
\end{equation}%
Here the Lamb-Dicke-limit state $\Psi _{n^{\prime }0}^{LD}(x,r,t_{n^{\prime
}0})$ is written as%
\begin{equation}
\Psi _{n^{\prime }0}^{LD}(x,r,t_{n^{\prime }0})=\exp \{in^{\prime
}[(q_{s}/n)\Delta ky+(q_{c}/n)(1-\frac{1}{2}(\Delta ky)^{2})]\}\Psi
_{00}(x,r,t_{0}),  \tag{6.114}
\end{equation}%
and the total error $E_{R}(n^{\prime })$ is given by 
\begin{equation}
E_{R}(n^{\prime })=\sum_{k=1}^{n^{\prime }}[P_{tr}^{r}(\delta t/\sqrt{n}%
)]^{k-1}E_{rT}^{n^{\prime }-k}(x,r,t_{n^{\prime }-k+1,0})  \tag{6.115}
\end{equation}%
where the error term $E_{rT}^{m}(x,r,t_{m+1,0})$ with $m=0,1,...,n^{\prime
}-1$ is given by, as can be deduced from the relations (6.102) and (6.110), 
\begin{equation}
E_{rT}^{m}(x,r,t_{m+1,0})=E_{rt}^{m}(x,r,t_{m+1,0})+O_{m+1,0}((\delta t/%
\sqrt{n})^{3})+E_{r,m+1,0}^{LD}(x,r,t_{m+1,0}).  \tag{6.116}
\end{equation}%
In particular, when $n^{\prime }=n,$ the $SSISS$ triggering pulse $%
P_{tr}^{r}(\delta t)=[P_{tr}^{r}(\delta t/\sqrt{n})]^{n}$ generates the
Lamb-Dicke-limit state$,$%
\begin{equation}
\Psi _{n0}^{LD}(x,r,t_{n0})=\exp \{i[q_{s}\Delta ky+q_{c}(1-\frac{1}{2}%
(\Delta ky)^{2})]\}\Psi _{00}(x,r,t_{0}),  \tag{6.117}
\end{equation}%
and the total error $E_{R}(n)$ that is given by (6.115) with $n^{\prime }=n$%
. Obviously, the Lamb-Dicke-limit state $\Psi _{n^{\prime
}0}^{LD}(x,r,t_{n^{\prime }0})$ for $n^{\prime }=1,2,...,n$ is a pure $GWP$
state. It has the same internal state $|g_{0}\rangle $ as the initial state $%
\Psi _{00}(x,r,t_{0})$ and it also has the same COM position $%
x_{c}(t_{n^{\prime }0})=x_{c}(t_{0})$ and wave-packet spread $\varepsilon
(t_{n^{\prime }0})=\varepsilon (t_{0})$ as the initial state, but it has
different motional momentum $p_{c}(t_{n^{\prime }0})$, complex linewidth $%
W(t_{n^{\prime }0}),$ and global phase factor from the initial state. In
particular, the motional momentum of the final $GWP$ state $\Psi
_{n0}^{LD}(x,r,t_{n0})$ of the $SSISS$ triggering pulse $P_{tr}^{r}(\delta
t) $ is given by%
\begin{equation*}
p_{c}(t_{n0})=p_{c}(t_{0})+q_{s}\hslash \Delta k,
\end{equation*}%
where $q_{s}$ is given in (5.4) in the section 5. Actually, one can find
that the state $\Psi _{n0}^{LD}(x,r,t_{n0})$ of (6.117) is just equal to the
Lamb-Dicke-limit state $\Psi _{LD}(y,r,t)=\Psi _{LD}(y,t)|g_{0}\rangle $ in
which the motional state $\Psi _{LD}(y,t)$ is given by (5.5). It is known in
the section 5 that the original motional state $\Psi (y,t)$ of (5.4) of the
Lamb-Dicke-limit state $\Psi _{LD}(y,t)$ of (5.5) is generated by acting the
unitary propagator $\exp [iQ(\delta t/\hslash )^{2}]$ of (5.1) on the
starting state $\Psi _{0}(x,r,t_{0})$ $($here $\Psi _{0}(x,r,t_{0})$ is just
equal to $\Psi _{00}(x,r,t_{0})),$ while the propagator $\exp [iQ(\delta
t/\hslash )^{2}]$ of (5.1) is generated approximately by the theoretical
state-selective triggering pulse $[P_{tr}^{i}(\delta t/\sqrt{n})]^{n}$ whose
basic pulse sequence $P_{tr}^{i}(\delta t/\sqrt{n})$ is given by (6.7a).
Then the equation (6.113) with $n^{\prime }=n$ shows that, just like the
theoretical state-selective triggering pulse $[P_{tr}^{i}(\delta t/\sqrt{n}%
)]^{n}$, the $SSISS$ triggering pulse $[P_{tr}^{r}(\delta t/\sqrt{n})]^{n}$
indeed generates the desired atomic motional state $\Psi _{LD}(y,t)$ of
(5.5) if the total error $E_{R}(n)$ on the $RH$ side of (6.113) can be
neglected.

Below investigate the total error $E_{R}(n)$ in (6.113) with $n^{\prime }=n$%
. Note that any unitary operator acting on a state does not change the norm
of the state. Thus, both the errors $[P_{tr}^{r}(\delta t/\sqrt{n}%
)]^{k-1}E_{rT}^{n-k}(x,r,t_{n-k+1,0})$ and $E_{rT}^{n-k}(x,r,t_{n-k+1,0})$
for $k=1,2,...,n$ on the $RH$ side of (6.115) have the same norm:%
\begin{equation*}
||[P_{tr}^{r}(\delta t/\sqrt{n}%
)]^{k-1}E_{rT}^{n-k}(x,r,t_{n-k+1,0})||=||E_{rT}^{n-k}(x,r,t_{n-k+1,0})||.
\end{equation*}%
Now by using these relations it follows from (6.115) and (6.116) that the
total error $E_{R}(n)$ is bounded by%
\begin{equation}
||E_{R}(n)||\leq
\sum_{l=1}^{n}||E_{rt}^{l-1}(x,r,t_{l,0})||+\sum_{l=1}^{n}O_{l,0}((\delta t/%
\sqrt{n})^{3})+\sum_{l=1}^{n}||E_{r,l,0}^{LD}(x,r,t_{l,0})||,  \tag{6.118}
\end{equation}%
where $O_{l,0}((\delta t/\sqrt{n})^{3})$ should be referred to as the upper
bound of the error $O_{l,0}((\delta t/\sqrt{n})^{3})$ in (6.116) without any
confusion. The total error $E_{R}(n)$ is divided into the three types of
errors. The first type of errors are the $n$ errors $%
\{E_{rt}^{l-1}(x,r,t_{l,0})\}$. Each one of which consists of the nine error
terms. The first error $E_{rt}^{0}(x,r,t_{10})$ consists of the nine error
terms $\{E_{rk}^{0}(x,r,t_{0}+t_{9})\}$ in (6.84). These nine error terms
are already calculated in detail in the subsection 6.1. The upper bound of
the error $E_{rt}^{0}(x,r,t_{10})$ is determined from (6.96). The second
error $E_{rt}^{1}(x,r,t_{20})$ also consists of the nine error terms $%
\{E_{rk}^{1}(x,r,t_{10}+\delta t_{r})\}.$ Its upper bound is determined from
(6.106). Generally, the $l-$th error $E_{rt}^{l-1}(x,r,t_{l,0})$ with $%
l=1,2,...,n$ consists of the nine error terms $\{E_{rk}^{l-1}(x,r,t_{l-1,0}+%
\delta t_{r}\}.$ Its upper bound is determined from%
\begin{equation}
||E_{rt}^{l-1}(x,r,t_{l,0})||\leq
\sum_{k=1}^{9}||E_{rk}^{l-1}(x,r,t_{l-1,0}+\delta t_{r})||.  \tag{6.119}
\end{equation}%
Here $t_{l,0}=t_{l-1,0}+\delta t_{r}$ for $l=1,2,...,n$ and $t_{0,0}\equiv
t_{0}.$ This inequality is a generalization of (6.96) and (6.106). Since the 
$l-$th experimental basic pulse sequence $(1\leq l\leq n)$ is the same as
the first one in the $SSISS$ triggering pulse $P_{tr}^{r}(\delta
t)=[P_{tr}^{r}(\delta t/\sqrt{n})]^{n},$ the difference between the $l-$th
error $E_{rt}^{l-1}(x,r,t_{l,0})$ ($1<l\leq n$) and the first error $%
E_{rt}^{0}(x,r,t_{10})$ originates from different initial states of the $l-$%
th experimental basic pulse sequence and the first one. The nine error terms 
$\{E_{rk}^{l-1}(x,r,t_{l-1,0}+\delta t_{r})\}$ of the $l-$th error $%
E_{rt}^{l-1}(x,r,t_{l,0})$ can be calculated strictly with the same
theoretical calculation method that is used in the subsection 6.1 to
calculate the nine error terms $\{E_{rk}^{0}(x,r,t_{0}+t_{9})\}$ of the
first error $E_{rt}^{0}(x,r,t_{10})$. Here the initial state in the
calculation should be taken as the $GWP$ state $\Psi
_{l-1,0}^{LD}(x,r,t_{l-1,0}).$ In particular, here denote $\Psi
_{0,0}^{LD}(x,r,t_{0,0})=\Psi _{00}(x,r,t_{0}).$ Therefore, if in the
previous theoretical calculation for the nine error terms $%
\{E_{rk}^{0}(x,r,t_{0}+t_{9})\}$ in (6.84) one replaces the initial state $%
\Psi _{00}(x,r,t_{0})$ with the initial state $\Psi
_{l-1,0}^{LD}(x,r,t_{l-1,0}),$ then one may obtain the nine error terms $%
\{E_{rk}^{l-1}(x,r,t_{l-1,0}+\delta t_{r})\}$ of the $l-$th error $%
E_{rt}^{l-1}(x,r,t_{l,0})$. Now investigate how the initial states $\{\Psi
_{l-1,0}^{LD}(x,r,t_{l-1,0})\}$ affect their corresponding errors $%
\{E_{rt}^{l-1}(x,r,t_{l,0})\}$, respectively. Since a $GWP$ state is
completely characterized by its four characteristic parameters, one needs
only to investigate how each one of these error terms is affected by the
four characteristic parameters of its initial $GWP$ state $\Psi
_{l-1,0}^{LD}(x,r,t_{l-1,0})$. Just like the nine error terms $%
\{E_{rk}^{0}(x,r,t_{0}+t_{9})\}$ of the first error $E_{rt}^{0}(x,r,t_{10}),$
the nine error terms $\{E_{rk}^{l-1}(x,r,t_{l-1,0}+t_{9})\}$ of the $l-$th
error $E_{rt}^{l-1}(x,r,t_{l,0})$ $(1\leq l\leq n)$ have the upper bounds:%
\begin{equation*}
||E_{rk}^{l-1}(x,r,t_{l-1,0}+t_{9})||\leq
||E_{r}^{(12)}(x,r,t_{l-1,0}+t_{k})||
\end{equation*}%
\begin{equation}
+||E_{r1}^{(3)}(x,r,t_{l-1,0}+t_{k})||,\text{ }k=1,3,5,7,9,  \tag{6.120a}
\end{equation}%
and 
\begin{equation*}
||E_{rk}^{l-1}(x,r,t_{l-1,0}+t_{9})||\leq ||E_{r}^{0}(x,r,t_{l-1,0}+t_{k})||
\end{equation*}%
\begin{equation}
+||E_{r}^{V}(x,r,t_{l-1,0}+t_{k})||+||E_{r2}^{(3)}(x,r,t_{l-1,0}+t_{k})||,%
\text{ }k=2,4,6,8.  \tag{6.120b}
\end{equation}%
Then it follows from (6.119) and (6.120) that the total upper bound of the $%
n $ errors $\{E_{rt}^{l-1}(x,r,t_{l,0})\}$ is determined from%
\begin{equation*}
\sum_{l=1}^{n}||E_{rt}^{l-1}(x,r,t_{l,0})||\leq
\sum_{l=1}^{n}\sum_{k=1,3,5,7,9}^{{}}||E_{r}^{(12)}(x,r,t_{l-1,0}+t_{k})||
\end{equation*}%
\begin{equation*}
+\sum_{l=1}^{n}\sum_{k=1,3,5,7,9}^{{}}||E_{r1}^{(3)}(x,r,t_{l-1,0}+t_{k})||+%
\sum_{l=1}^{n}\sum_{k=2,4,6,8}^{{}}||E_{r}^{V}(x,r,t_{l-1,0}+t_{k})||
\end{equation*}%
\begin{equation}
+\sum_{l=1}^{n}\sum_{k=2,4,6,8}^{{}}||E_{r}^{0}(x,r,t_{l-1,0}+t_{k})||+%
\sum_{l=1}^{n}\sum_{k=2,4,6,8}^{{}}||E_{r2}^{(3)}(x,r,t_{l-1,0}+t_{k})||. 
\tag{6.121}
\end{equation}%
This upper bound consists of five groups of the error norms on the $RH$ side
of (6.121). The first group consists of $5n$ error norms $%
\{||E_{r}^{(12)}(x,r,t_{l-1,0}+t_{k})||\}.$ These error terms $%
\{E_{r}^{(12)}(x,r,t_{l-1,0}+t_{k})\}$ of the group originate from the
imperfection of the $LH$ harmonic potential well. Each one of which can be
strictly calculated with the theoretical calculation method in the section
3. As shown in the subsection 6.1, it can be controlled by the joint
position $x_{L}$ as its upper bound decays exponentially with the
deviation-to-spread ratios of the relevant $GWP$ states. When the joint
position $x_{L}$ is large enough, each error norm of the first group can be
neglected. Thus, this group of the error norms are usually secondary on the $%
RH$ side of (6.121). The third group consists of $4n$ error norms $%
\{||E_{r}^{V}(x,r,t_{l-1,0}+t_{k})||\}.$ The error terms $%
\{E_{r}^{V}(x,r,t_{l-1,0}+t_{k})\}$ of the group originate from the
imperfection of the $LH$ harmonic potential well and the spatially-selective
effect of the $PHAMDOWN$ laser light beams. They can be calculated strictly
with the theoretical calculation method in the section 4. As shown in the
subsection 6.1, their upper bounds decays exponentially with the square
deviation-to-spread ratios of the relevant $GWP$ states. Then the sum of
these $4n$ error norms still decays exponentially with the square
deviation-to-spread ratios of the relevant $GWP$ states. Thus, the group can
be controlled by the joint position $x_{L}$. When the joint position $x_{L}$
is large enough, it is negligible. Therefore, the third group of error norms
are usually secondary on the $RH$ side of (6.121).

It is known from (6.114) that the $l-$th initial $GWP$ state $\Psi
_{l-1,0}^{LD}(x,r,t_{l-1,0})$ for $l=2,3,...,n$ has the same COM position
and wave-packet spread as the first initial state $\Psi _{00}(x,r,t_{0})$,
and it has different motional momentum and slightly different complex
linewidth from the first initial state. Then, as shown in the energy
equation (5.43), the differences of the motional energy among these $n$
initial states $\{\Psi _{l-1,0}^{LD}(x,r,t_{l-1,0})\}$ come mainly from the
motional momentum differences among these initial states. These motional
energy differences can lead to the differences of the minimum
deviation-to-spread ratios, as shown in (5.45) in the section 5. Therefore,
when one calculates the upper bounds of these error terms $%
\{E_{r}^{V}(x,r,t_{l-1,0}+t_{k})\}$ and $\{E_{r}^{(12)}(x,r,t_{l-1,0}+t_{k})%
\},$ one needs to take into account these different motional momentums or
energies of these $n$ initial states $\{\Psi _{l-1,0}^{LD}(x,r,t_{l-1,0})\}.$

The second group on the $RH$ side of (6.121) consists of $5n$ error norms $%
\{||E_{r1}^{(3)}(x,r,t_{l-1,0}+t_{k})||\}.$ As shown in the subsection 6.1,
these error terms $\{E_{r1}^{(3)}(x,r,t_{l-1,0}+t_{k})\}$ ($k\geq 1$) of the
group originate from the deviation of the initial states from their initial
Gaussian superposition states in (6.8), which are used to calculate the
errors terms $\{E_{r}^{(12)}(x,r,t_{l-1,0}+t_{k})\}$ in (6.121).
Particularly in the group these error norms $%
||E_{r1}^{(3)}(x,r,t_{l-1,0}+t_{1})||=0$ for $l=1,2,...,n$ because their
initial states $\Psi _{l-1,0}^{LD}(x,r,t_{l-1,0})$ are single $GWP$ states.
The fifth group is similar to the second group. It consists of $4n$ error
norms $\{||E_{r2}^{(3)}(x,r,t_{l-1,0}+t_{k})||\}$. These error terms $%
\{E_{r2}^{(3)}(x,r,t_{l-1,0}+t_{k})\}$ of the group originate from the
deviation of the initial states from their initial Gaussian superposition
states in (6.8), which are used to calculate the errors terms $%
\{E_{r}^{0}(x,r,$ $t_{l-1,0}+t_{k})\}$ and $\{E_{r}^{V}(x,r,t_{l-1,0}+t_{k})%
\}$ in (6.121). All these error terms $\{E_{r1}^{(3)}(x,r,$ $%
t_{l-1,0}+t_{k})\}$ and $\{E_{r2}^{(3)}(x,r,t_{l-1,0}+t_{k})\}$ can be
calculated strictly, as shown in the subsection 6.1. Their upper bounds are
determined from those of the errors $\{E_{r2j}(x,r,t_{l-1,0}+j\times \delta
t/\sqrt{n})\}$ with $j=1,$ $2,$ $3,$ $4$ and $l=1,$ $2,$ $...,$ $n.$ That
is, there are the relations ($l=1,2,...,n$):%
\begin{equation*}
||E_{r1}^{(3)}(x,r,t_{l-1,0}+t_{1})||=0,
\end{equation*}%
\begin{equation*}
||E_{r1}^{(3)}(x,r,t_{l-1,0}+t_{2j+1})||\leq
2||E_{r2j}(x,r,t_{l-1,0}+j\times \delta t/\sqrt{n})||,\text{ }j=1,2,3,4;
\end{equation*}%
and 
\begin{equation*}
||E_{r2}^{(3)}(x,r,t_{l-1,0}+t_{2j})||\leq 2||E_{r2j}(x,r,t_{l-1,0}+j\times
\delta t/\sqrt{n})||,\text{ }j=1,2,3,4.
\end{equation*}%
The upper bounds of the four errors $E_{r2j}(x,r,t_{l-1,0}+j\times \delta t/%
\sqrt{n})$ with $j=1,$ $2,$ $3,$ $4$ and for a given $l$ are still obtained
from (6.39), (6.54), (6.72), and (6.89), respectively. Each one of which is
proportional to $(\delta t/\sqrt{n})^{3},$ $|\Delta k|^{3},$ and $%
\varepsilon (t_{l-1,0})^{3}$ approximately. Here $\varepsilon (t_{l-1,0})$
is the wave-packet spread of the initial $GWP$ state $\Psi
_{l-1,0}^{LD}(x,r,t_{l-1,0})$ of the $l-$th experimental basic pulse
sequence. Notice that the initial state $\Psi _{l-1,0}^{LD}(x,r,t_{l-1,0})$
has the same wave-packet spread as the initial state $\Psi _{00}(x,r,t_{0})$%
, that is, $\varepsilon (t_{l-1,0})=\varepsilon (t_{0}).$ Then it follows
from (6.39), (6.54), (6.72), and (6.89) that these $n$ norms $%
\{||E_{r2j}(x,r,t_{l-1,0}+j\times \delta t/\sqrt{n})||\}$ for a given $j$
and for $l=1,$ $2,$ $...,$ $n$ have really the same upper bound:%
\begin{equation*}
||E_{r2j}(x,r,t_{n-1,0}+j\times \delta t/\sqrt{n}%
)||_{u}=||E_{r2j}(x,r,t_{n-2,0}+j\times \delta t/\sqrt{n})||_{u}
\end{equation*}%
\begin{equation*}
=...=||E_{r2j}(x,r,t_{0}+j\times \delta t/\sqrt{n})||_{u}.
\end{equation*}%
These result in that the sum of the norms of the second and fifth groups on
the $RH$ side of (6.121) is bounded by%
\begin{equation*}
\sum_{l=1}^{n}\sum_{k=1,3,5,7,9}^{{}}||E_{r1}^{(3)}(x,r,t_{l-1,0}+t_{k})||+%
\sum_{l=1}^{n}\sum_{k=2,4,6,8}^{{}}||E_{r2}^{(3)}(x,r,t_{l-1,0}+t_{k})||.
\end{equation*}%
\begin{equation*}
\leq 4\sum_{j=1,2,3,4}^{{}}n||E_{r2j}(x,r,t_{0}+j\times \delta t/\sqrt{n}%
)||_{u}.
\end{equation*}%
This upper bound is proportional to $(\delta t)^{3}/\sqrt{n},$ $|\Delta
k|^{3},$ and $\varepsilon (t_{0})^{3}$ approximately because the upper bound 
$||E_{r2j}(x,r,t_{0}+j\times \delta t/\sqrt{n})||_{u}$ is proportional to $%
(\delta t/\sqrt{n})^{3},$ $|\Delta k|^{3},$ and $\varepsilon (t_{l-1,0})^{3}$
approximately. Then it may be neglected when the number $n$ is large enough
and/or the wave-number difference $|\Delta k|$ is small enough. Because all
these errors $\{E_{r2j}(x,r,t_{l-1,0}+j\times \delta t/\sqrt{n})\}$ may be
greatly improved if each one of the initial Gaussian superposition states
contains more than ten $GWP$ states, both the second and fifth groups of the
error norms usually may be secondary on the $RH$ side of (6.121).

The fourth group consists of $4n$ error norms $%
\{||E_{r}^{0}(x,r,t_{l-1,0}+t_{k})||\}$. These error terms $%
\{E_{r}^{0}(x,r,t_{l-1,0}+t_{k})\}$ of the group are generated by the
truncation approximation of the decomposition formula (4.1). Their upper
bounds may be calculated strictly according to the theoretical calculation
method in the section 4. Their norms are usually the dominating terms on the 
$RH$ side of (6.121). As shown in the subsection 6.1, each one of their
upper bounds is approximately proportional to $(\delta t/\sqrt{n})^{3}$ and
also dependent on the three types of parameters of the relevant $GWP$ states
of the halting-qubit atom and the experimental basic pulse sequence (6.7b).
Then the sum of these $4n$ error norms $\{||E_{r}^{0}(x,r,t_{l-1,0}+t_{k})||%
\}$ is approximately proportional to $(\delta t)^{3}/\sqrt{n}$ and also
dependent on these three types of parameters. Thus, the fourth group of
error norms on the $RH$ side of (6.121) may be controlled by the number $n$,
the time interval $\delta t,$ and these three types of parameters. Note that
the initial $GWP$ states $\{\Psi _{l-1,0}^{LD}(x,r,t_{l-1,0})\}$ may have
different motional momentums and energies. As shown in the four inequalities
(5.44), different motional energies may lead to different upper bounds
and/or lower bounds of the Gaussian characteristic parameters. These facts
need to be considered when one calculates the upper bounds of the errors $%
\{E_{r}^{0}(x,r,t_{l-1,0}+t_{k})\}.$

The above theoretical analysis for the five groups of error norms on the $RH$
side of (6.121) shows that the first sum term $%
\sum_{l=1}^{n}||E_{rt}^{l-1}(x,r,t_{l,0})||$ on the $RH$ side of (6.118) can
be controlled by the joint position $x_{L}$, the number $n$ and the time
interval $\delta t$, the wave-number difference $|\Delta k|$ and the
wave-packet spread $\varepsilon (t_{0})$ and so on.

In the total error $E_{R}(n)$ of (6.118) the second type of errors consist
of the $n$ error terms $\{O_{l,0}((\delta t/\sqrt{n})^{3})\}.$ Each one of
these error terms is generated due to the reduction of the propagator $%
U_{tr}(\delta t/\sqrt{n})$ to the propagator $\exp \{iQ(\delta t/\hslash
)^{2}/n\},$ as can be seen in (6.93b). As shown in (6.94), its upper bound
is proportional to $(\delta t/\sqrt{n})^{3}$ approximately.$\ $Thus, the
upper bound for the sum of the $n$ error norms $\{O_{l,0}((\delta t/\sqrt{n}%
)^{3})\}$ is proportional to $(\delta t)^{3}/\sqrt{n}$ approximately. Then
the second sum term $\sum_{l=1}^{n}O_{l,0}((\delta t/\sqrt{n})^{3})$ on the $%
RH$ side of (6.118) may have a negligible contribution to the total error $%
E_{R}(n)$ if the number $n$ is large enough and/or the time interval $\delta
t$ is short enough.

The third type of errors of the total error $E_{R}(n)$ consist of the $n$
error terms $\{E_{r,l,0}^{LD}(x,r,t_{l,0})\}.$ It is known from (6.100) that
the upper bound of the error term $E_{r10}^{LD}(x,r,t_{10})$ is proportional
to $|\Delta k|^{3}$ and $\varepsilon (t_{0})^{3}$ approximately and
inversely proportional to the number $n$. Similarly, the upper bound of the
error term $E_{rl,0}^{LD}(x,r,t_{l,0})$ $(l=1,2,...,n)$ is proportional to $%
|\Delta k|^{3}$ and the wave-packet spread $\varepsilon (t_{l-1,0})^{3}$ of
the initial state $\Psi _{l-1,0}^{LD}(x,r,$ $t_{l-1,0})$ approximately and
inversely proportional to the number $n$. Notice that the initial state $%
\Psi _{l-1,0}^{LD}(x,r,t_{l-1,0})$ has the same wave-packet spread as the
first initial state $\Psi _{00}(x,r,t_{0}).$ That is, $\varepsilon
(t_{l-1,0})=\varepsilon (t_{0})$ for $l=1$, $2$, $...$, $n$. Then according
to (6.100) all these errors $\{E_{rl,0}^{LD}(x,r,t_{l,0})\}$ for $l=1$, $2$, 
$...$, $n$ have the same upper bound:%
\begin{equation*}
||E_{r,n,0}^{LD}(x,r,t_{n,0})||_{u}=...=||E_{r2,0}^{LD}(x,r,t_{2,0})||_{u}=||E_{r1,0}^{LD}(x,r,t_{1,0})||_{u}.
\end{equation*}%
Thus, the upper bound of sum of the $n$ error norms $%
\{||E_{rl,0}^{LD}(x,r,t_{l,0})||\}$ satisfies the inequality: 
\begin{equation*}
\sum_{l=1}^{n}||E_{r,l,0}^{LD}(x,r,t_{l,0})||\leq
n||E_{r1,0}^{LD}(x,r,t_{1,0})||_{u}.
\end{equation*}%
It can be found from (6.100) that the upper bound $%
n||E_{r10}^{LD}(x,r,t_{10})||_{u}$ is independent of the number $n$ and it
is proportional to $(2\Omega _{0}\delta t)^{2}\varepsilon (t_{0})^{3}|\Delta
k|^{3}$ or $|p_{tr}/\hslash |\varepsilon (t_{0})^{3}|\Delta k|^{2}$
approximately, here the final atomic motional momentum value $%
|p_{tr}|=(2\Omega _{0}\delta t)^{2}\hslash |\Delta k|$ is usually set to a
given value for the $SSISS$ triggering pulse $[P_{tr}^{r}(\delta t/\sqrt{n}%
)]^{n}$. Thus, the last sum term $%
\sum_{l=1}^{n}||E_{r,l,0}^{LD}(x,r,t_{l,0})||$ on the $RH$ side of (6.118)
can be controlled by the wave-number difference $|\Delta k|$ and the
wave-packet spread $\varepsilon (t_{0}).$

Now by setting suitably the joint position $x_{L}$ $(>0)$ in the double-well
potential field of (2.1), the number $n$ $(>>1)$ and the time interval $%
\delta t$, the wave-packet spread $\varepsilon (t_{0}),$ and the wave-number
difference $|\Delta k|$ and so on one may control the upper bound of the
total error $E_{R}(n)$ of (6.118). When the upper bound of the total error $%
E_{R}(n)$ is controlled to be so small that the total error $E_{R}(n)$\ can
be neglected on the $RH$ side of (6.113), the equation (6.113) with $%
n^{\prime }=n$ is reduced to the form%
\begin{equation}
\lbrack P_{tr}^{r}(\delta t/\sqrt{n})]^{n}\Psi _{00}(x,r,t_{0})=\Psi
_{n0}^{LD}(x,r,t_{n0}).  \tag{6.122}
\end{equation}%
This means that the final state of the $SSISS$ triggering pulse $%
[P_{tr}^{r}(\delta t/\sqrt{n})]^{n}$ is just the $GWP$ state $\Psi
_{n0}^{LD}(x,r,t_{n0}).$ This is just the desired result of the $SSISS$
triggering pulse.

In this section the experimental basic pulse sequence $P_{tr}^{r}(\delta t/%
\sqrt{n})$ and the $SSISS$ triggering pulse $[P_{tr}^{r}(\delta t/\sqrt{n}%
)]^{n}$ are constructed explicitly and are calculated strictly on the basis
of the theoretical methods in the previous three sections 3, 4, and 5. This
theoretical calculation shows that the $SSISS$ triggering pulse may be
different from its ideal non-spatially-selective counterpart [15], but the
difference of their final states can be controlled when they are applied to
the halting-qubit atom in a $GWP$ state, respectively. It also proves
rigorously that both the $SSISS$ triggering pulse $[P_{tr}^{r}(\delta t/%
\sqrt{n})]^{n}$ and its ideal non-spatially-selective counterpart are
convergent, that is, all the possible errors that are generated by the $%
SSISS $ triggering pulse and its ideal counterpart are controllable when the
two triggering pulses are applied to the halting-qubit atom in a $GWP$
state, respectively. These are the most important results of the present
theoretical calculation. \newline
\newline
\newline
{\Large 7 Discussion and conclusion}

The $SSISS$ triggering pulses are constructed explicitly in the paper. A $%
SSISS$ triggering pulse consists of the spatially-selective $PHAMDOWN$ laser
light beams that are applied to the halting-qubit atom in the (approximate)
harmonic potential well of the double-well potential field or other more
practical potential field like $V_{rw}(x)$ in the section 6. The present $%
SSISS$ triggering pulse $[P_{tr}^{r}(\delta t/\sqrt{n})]^{n}$ and the
experimental basic pulse sequence $P_{tr}^{r}(\delta t/\sqrt{n})$ are
simple. A better or much better $SSISS$ triggering pulse also could be
constructed explicitly. Such a $SSISS$ triggering pulse is generally much
more complex than the present one. A slightly better $SSISS$ triggering
pulse may be generated by using the theoretical basic pulse sequence (6.1b).
Actually, as pointed out in Ref. [15], a higher-order state-selective
triggering pulse may be constructed by using the Trotter-Suzuki
decomposition method [34], while here the basic pulse sequences of (6.1) may
act as the building blocks to construct the triggering pulse. An ideal
state-selective triggering pulse could become no useful in a real physical
system. It is necessary to solve the problem how to convert an ideal
state-selective triggering pulse into a real one (i.e., a $SSISS$ triggering
pulse) that can be realized directly in a real physical system. This problem
may be solved with the theoretical method in the section 6. According to the
theoretical method one may generate a $SSISS$ triggering pulse that
corresponds to the ideal state-selective triggering pulse. A $SSISS$
triggering pulse could be useful and could be realized directly in the real
physical system. The present work solves completely the problem how to
construct a $SSISS$ triggering pulse.

It is not difficult to construct a $SSISS$ triggering pulse in theory, but
it is a challenge to prove that the $SSISS$ triggering pulse is convergent!
Only a convergent $SSISS$ triggering pulse is useful in practice. Therefore,
it is necessary to make a strict error estimation for a $SSISS$ triggering
pulse. This means that all the possible errors that are generated during the 
$SSISS$ triggering pulse need to be estimated rigorously. The present
theoretical calculation in the error estimation is based on the
Trotter-Suzuki decomposition method, while the multiple Gaussian wave-packet
expansion method, i.e., the $MGWP$ expansion method, has been used
extensively in the theoretical calculation. The theory and method can be
used as well to prove a higher-order $SSISS$ triggering pulse to be
convergent. It tends to be very complex to prove a higher-order $SSISS$
triggering pulse to be convergent, although it is not difficult in theory.
The radical reason for it is that the unitary quantum dynamics in time and
space for a quantum system in COM motion is generally complicated except for
some simple and special cases. This can be seen in the present work. The
present theoretical calculation shows that the upper bounds of these
possible errors generated by the $SSISS$ triggering pulse are dependent on
not only the characteristic parameters of the $SSISS$ triggering pulse
itself but also the characteristic parameters of the $GWP$ motional states
of the halting-qubit atom. Therefore, whether or not a $SSISS$ triggering
pulse is convergent is not only dependent on the $SSISS$ triggering pulse
itself but also largely dependent on the COM motional states of the
halting-qubit atom that is acted on by the $SSISS$ triggering pulse.

It proves in the paper that the present $SSISS$ triggering pulse is
convergent. All possible errors generated by the $SSISS$ triggering pulse
are strictly calculated and shown to be controllable. This strict error
estimation is necessary because by the strict error estimation one may set
suitably the relevant parameters for the $SSISS$ triggering pulse. It also
is shown in the paper that the $SSISS$ triggering pulse may achieve the same
result as its ideal counterpart if dropping the negligible errors, when both
the triggering pulses are applied to the halting-qubit atom in a $GWP$
state, respectively. This further shows that the spatially-selective effect
of the $SSISS$ triggering pulse is negligible. This point is important.
Spatially selective excitations, operations, or processes are necessary
components for a quantum computer based on an atomic physical system,
because a quantum computer cannot have an infinite dimensional size in
space. The present work shows that these spatially selective excitations,
operations, or processes are completely realizable in a single-atom system
in an external potential field if the atomic motional state, the externally
applied electromagnetic wave field and potential field are chosen suitably.
Here one of the best schemes is that the atomic motional state is chosen as
a $GWP$ motional state, while the external potential field is chosen as a
general quadratic potential field.

The present $SSISS$ triggering pulse could be used to construct the
reversible and unitary halting protocol and the UNIDYSLOCK process and its
inverse process which may be further used to construct the unstructured
quantum search process based on the tensor-product Hilbert-space symmetric
structure and the unitary quantum dynamics [47]. The methods and techniques
developed in the present work also could be used to realize the spatially
selective excitations, operations, or processes even in a multiple-atom
system in an external potential field. They are also useful for studying the
quantum-computing speedup mechanism of the unitary quantum dynamics [47] in
the quantum system of a single atom. \newline
\newline
\newline
{\Large Acknowledgment}

The author thanks Dr. Klaas Bergmann for his comment on the paper$^{\prime }$%
s title.\newline
\newline
\newline
{\Large References}\newline
1. (a) S. Chu, \textit{Nobel Lecture: The manipulation of neutral particles}%
, Rev. Mod. Phys. 70, 685 (1998), (b) C. N. Cohen-Tannoudji, \textit{Nobel
Lecture: Manipulating atoms with photons}, Rev. Mod. Phys. 70, 707 (1998),
(c) W. D. Phillips, \textit{Nobel Lecture: Laser cooling and trapping of
neutral atoms}, Rev. Mod. Phys. 70, 721 (1998); and references therein 
\newline
2. (a)\ M. Kasevich and S. Chu, \textit{Measurement of the gravitational
acceleration of an atom with a light-pulse atom interferometer, }Appl. Phys.
B 54, 321 (1992); (b)\ P. Marte, P. Zoller, and J. L. Hall, \textit{Coherent
atomic mirrors and beam splitters by adiabatic passage in multilevel systems}%
, Phys. Rev. A 44, R4118 (1991)\newline
3. D. J. Wineland, C. Monroe, W. M. Itano, D. Leibfried, B. E. King, D. M.
Meekhof, \textit{Experimental issues in coherent quantum-state manipulation
of trapped atomic ions}, J. Res. NIST, 103, 259 (1998); and references
therein \newline
4. D. Leibfried, R. Blatt, C. Monroe, and D. Wineland, \textit{Quantum
dynamics of single trapped ion}, Rev. Mod. Phys. 75, 281 (2003)\newline
5. J. I. Cirac and P. Zoller, \textit{Quantum computations with cold trapped
ions}, Phys. Rev. Lett. 74, 4091 (1995)\newline
6. (a) G. K. Brennen, C. M. Caves, P. S. Jessen, and I. H. Deutsch, \textit{%
Quantum logic gates in optical lattices}, Phys. Rev. Lett. 82, 1060 (1999);
(b)\ D. Jaksch, H. J. Briegel, J. I. Cirac, C. W. Gardiner, and P. Zoller, 
\textit{Entanglement of atoms via cold controlled collisions}, Phys. Rev.
Lett. 82, 1975 (1999); (c) T. Calarco, E. A. Hinds, D. Jaksch, J.
Schmiedmayer, J. I. Cirac, and P. Zoller, \textit{Quantum gates with neutral
atoms: Controlling collisional interactions in time dependent traps}, Phys.
Rev. A 61, 022304 (2000); (d) D. Hayes, P. S. Julienne, and I. H. Deutsch, 
\textit{Quantum logic via exchange blockade in ultracold collisions}, Phys.
Rev. lett. 98, 070501 (2007); (e) M. Anderlini, P. J. Lee, B. L. Brown, J.
Sebby-Strabley, W. D. Phillips, and J. V. Porto, \textit{Controlled exchange
interaction between pairs of neutral atoms in an optical lattice}, Nature
448, 452 (2007); (f) P. Treutlein, T. Steinmetz, Y. Colombe, B. Lev, P. Homme%
$LH$off, J. Reichel, M. Greiner, O. Mandel, A. Widera, T. Rom, I. Bloch, and
T. W. H\"{a}nsch, \textit{Quantum information processing in optical lattices
and magnetic microtraps}, http://arxiv.org/abs/quant-ph/0605163 (2006)%
\newline
7. C. A. Blockley, D. F. Walls, and H. Risken, \textit{Quantum collapses and
revivals in a quantized trap}, Europhys. Lett. 17, 509 (1992)\newline
8. (a) J. I. Cirac, A. S. Parkins, R. Blatt, and P. Zoller, \textit{"Dark"
squeezed states of the motion of a trapped ion}, Phys. Rev. Lett. 70, 556
(1993); (b) J. I. Cirac, R. Blatt, A. S. Parkin, and P. Zoller, \textit{%
Preparation of Fock states by observation of quantum jumps in an ion trap},
Phys. Rev. Lett. 70, 762 (1993)\newline
9. (a) W. Vogel and R. L. de Matos Filho, \textit{Nonlinear Jaynes-Cummings
dynamics of a trapped ion}, Phys. Rev. A 52, 4214 (1995); (b) S.
Wallentowitz and W. Vogel, \textit{Reconstruction of the quantum mechanical
state of a trapped ion}, Phys. Rev. Lett. 75, 2932 (1995)\newline
10. (a)\ D. M. Meekhof, C. Monroe, B. E. King, W. M. Itano, and D. J.
Wineland, \textit{Generation of nonclassical motional states of a trapped
atom}, Phys. Rev. Lett. 76, 1796 (1996); Erratum, 77, 2346 (1996); (b) C.
Monroe, D. M. Meekhof, B. E. King, and D. J. Wineland, \textit{A }$^{\prime
\prime }$\textit{Schr\"{o}dinger cat}$^{\prime \prime }$\textit{\
superposition state of an atom,} Science 272, 1131 (1996); (c) D. Leibfried,
D. M. Meekhof, B. E. King, C. Monroe, W. M. Itano, and D. J. Wineland, 
\textit{Experimental determination of the motional quantum state of a
trapped atom}, Phys. Rev. Lett. 77, 4281 (1996) \newline
11. C. D$^{\prime }$Hellon and G. J. Milburn, \textit{Reconstructing the
vibrational state of a trapped ion}, Phys. Rev. A 54, R25 (1996) \newline
12. (a) S. C. Gou, J. Steinbach, and P. L. Knight, \textit{Dark pair
coherent states of the motion of a trapped ion}, Phys. Rev. A 54, R1014
(1996); (b) J. Steinbach, J. Twamley, and P. L. Knight, \textit{Engineering
two-mode interactions in ion traps}, Phys. Rev. A 56, 4815 (1997)\newline
13. (a) C. K. Law and J. H. Eberly, \textit{Arbitrary control of a quantum
electromagnetic field}, Phys. Rev. Lett. 76, 1055 (1996); (b) A. S. Parkins,
P. Marte, P. Zoller, and H. J. Kimble, \textit{Synthesis of arbitrary
quantum states via adiabatic transfer of Zeeman coherence}, Phys. Rev. Lett.
71, 3095 (1993)\newline
14. X. Miao, \textit{The basic principles to construct a generalized
state-locking pulse field and simulate efficiently the reversible and
unitary halting protocol of a universal quantum computer},
http://arxiv.org/abs/quant-ph/0607144 (2006) \newline
15. X. Miao, \textit{Unitarily manipulating in time and space a Gaussian
wave-packet motional state of a single atom in a quadratic potential field},
http: //arxiv.org /abs/quant-ph/0708.2129 (2007)\newline
16. X. Miao, \textit{The STIRAP-based \ unitary \ decelerating \ and \
accelerating \ processes \ of a single free atom},
http://arxiv.org/abs/quant-ph/0707.0063 (2007)\newline
17. R. Freeman, \textit{Spin Choreography}, Spektrum, Oxford, 1997\newline
18. R. R. Ernst, G. Bodenhausen, and A. Wokaun, \textit{Principles of
Nuclear Magnetic Resonance in One and Two Dimensions}, Oxford University
Press, Oxford, 1987\newline
19. K. Bergmann, H. Theuer, and B. W. Shore, \textit{Coherent population
transfer among quantum states of atoms and molecules}, Rev. Mod. Phys. 70,
1003 (1998); and references therein\newline
20. L. Viola, \textit{Advances in decoherence control}, J. Mod. Opt. 51,
2357 (2004) or http://arxiv.org/abs/quant-ph/0404038 (2004)\newline
21. E. T. Jaynes and F. W. Cummings, \textit{Comparison of quantum and
semiclassical radiation theories with application to the beam maser}, Proc.
Inst. Electr. Eng. 51, 89 (1963)\newline
22. L. I. Schiff, \textit{Quantum mechanics}, 3rd, McGraw-Hill book company,
New York, 1968\newline
23.\ (a) M. L. Goldberger and K. M. Watson, \textit{Collision theory},
Chapt. 3, Wiley, New York, 1964; (b)\ R. G. Newton, \textit{Scattering
theory of waves and particles}, Chapt. 6, McGraw-Hill, New York, 1966 
\newline
24. (a)\ R. J. Glauber, \textit{Coherent and incoherent states of the
radiation field}, Phys. Rev. 131, 2766 (1963); (b) J. R. Klauder and E. C.
G. Sudarshan, \textit{Foundamentals of quantum optics}, Chapt. 7, W. A.
Benjamin, New York, 1968; (c)\ D. Stoler, \textit{Equivalence classes of
minimum uncertainty packets}, Phys. Rev. D 1, 3217 (1970); \textit{%
Equivalence classes of minimum-uncertainty packets. II}, Phys. Rev. D 4,
1925 (1971) \newline
25. (a) S. F. Boys, \textit{Electronic wavefunctions I. A general method of
calculation for stationary states of any molecular systems}, Proc. Roy. Soc.
London A 200, 542 (1950); (b) W. J. Hehre, R. F. Stewart, and J. A. Pople, 
\textit{Self-consistent molecular-orbital method. I. Use of Gaussian
expansions of Slater-type atomic orbitals}, J. Chem. Phys. 51, 2657 (1969) 
\newline
26.\ (a) E. J. Heller, \textit{Time-dependent approach to semiclassical
dynamics}, J. Chem. Phys. 62, 1544 (1975); (b) D. Huber and E. J. Heller, 
\textit{Generalized Gaussian wave packet dynamics}, J. Chem. Phys. 87, 5302
(1987) \newline
27. R. G. Littlejohn, \textit{The semiclassical evolution of wave packets},
Phys. Rep. 138, 193 (1986)\newline
28. (a) R. P. Feynman, \textit{Space-time approach to non-relativistic
quantum mechanics}, Rev. Mod. Phys. 20, 367 (1948); (b) R. P. Feynmann and
A. R. Hibbs, \textit{Quantum mechanics and path integrals}, McGraw-Hill, New
York, 1965 \newline
29. K. Husimi, \textit{Miscellanea in elementary quantum mechanics II},
Prog. Theor. Phys. 9, 381 (1953)\newline
30. E. H. Kerner, \textit{Note on the forced and damped oscillator in
quantum mechanics}, Can. J. Phys. 36, 371 (1957)\newline
31. P. Carruthers and M. M. Nieto, \textit{Coherent states and the forced
quantum oscillator}, Am. J. Phys. 33, 537 (1965)\newline
32. (a) M. M. Mizrahi, \textit{Phase space path integrals, without limiting
procedures}, J. Math. Phys. 19, 298 (1978); (b) D. C. Khandekar and S. V.
Lawande, \textit{Exact solution of a time-dependent quantal harmonic
oscillator with damping and a perturbative force}, J. Math. Phys. 20, 1870
(1979); (c) B. K. Cheng, \textit{The propagator of the time-dependent forced
harmonic oscillator with time-dependent damping}, J. Math. Phys. 27, 217
(1986); (d) J. T. Marshall and J. T. Pell, \textit{Path-integral evaluation
of the space-time propagator for quadratic Hamiltonian systems}, J. Math.
Phys. 20, 1297 (1979) \newline
33. L. S. Schulman, \textit{Techniques and applications of path integration}%
, Dover, New York, 2005\newline
34. M. Suzuki, \textit{Fractal decomposition of exponential operators with
applications to many-body theories and Monte Carlo simulations}, Phys. Lett.
A 146, 319 (1990); \textit{General theory of higher-order decomposition of
exponential operators and symplectic integrators}, Phys. Lett. A 165, 387
(1992) \newline
35. (a) U. Haeberlen and J. Waugh, \textit{Coherent averaging effects in
magnetic resonance}, Phys. Rev. 175, 453 (1968); (b) U. Haeberlen, \textit{%
High resolution NMR in solids}, Adv. Magn. Reson. Suppl. 1, 1976 \newline
36. W. Magnus, \textit{On the exponential solution of differential equations
for a linear operator}, Commun. Pure Appl. Math. 7, 649 (1954) \newline
37. R. M. Wilcox, \textit{Exponential operators and parameter
differentiation in quantum physics}, J. Math. Phys. 8, 962 (1967) \newline
38. R. G. Brewer and E. L. Hahn, \textit{Coherent two-photon process:
Transient and steady-state cases}, Phys. Rev. A 11, 1641 (1975)\newline
39. M. Suzuki, \textit{Decomposition formulas of exponential operators and
Lie exponentials with some applications to quantum mechanics and statistical
physics}, J. Math. Phys. 26, 601 (1985)\newline
40. (a) M. Suzuki, \textit{Convergence of general decompositions of
exponential operators}, Commun. Math. Phys. 163, 491 (1994); (b) M. Suzuki
and T. Yamauchi, \textit{Convergence of unitary and complex decomposition of
exponential operators}, J. Math. Phys. 34, 4892 (1993) \newline
41. (a)\ M. Matti Maricq, \textit{Application of average Hamiltonian theory
to the NMR of solids}, Phys. Rev. B 25, 6622 (1982); (b)\ M. Matti Maricq, 
\textit{Convergence of the Magnus expansion for time dependent two level
systems}, J. Chem. Phys. 86, 5647 (1987); (c) X. Miao, \textit{An explicit
criterion for existence of the Magnus solution for a coupled spin system
under a time-dependent radiofrequency pulse}, Phys. Lett. A 271, 296 (2000)
(the error correction is seen at http://arxiv.org/abs/quant-ph/1204.4872v1
(2012))\newline
42. L. Allen and J. H. Eberly, \textit{Optical resonance and two-level atoms}%
, Dover, New York, 1987\newline
43. J. H. Wilkinson, \textit{The algebraic eigenvalue problem}, Chapt. One,
Oxford university press, 1965\newline
44. (a) E. W. Weisstein, \textit{CRC concise encyclopeda of mathematics},
2nd, A CRC Press Company, New York, 2003, PP. 934 (Erfc(x)\ function) and
PP. 1162 (Gaussian integrals); (b) M. Abramowitz and C. A. Stegun (Eds.), 
\textit{Handbook of mathematical functions with formulas, graphs, and
mathematical tables}, 3rd printing, National bureau of standards, 1965%
\newline
45. G. N. Watson, \textit{A treatise on the theory of Bessel functions},
Chapt. II, 2nd ed., Cambridge university press, 1944\newline
46. E. L. Hahn, \textit{Spin echoes}, Phys. Rev. 80, 580 (1950)\newline
47. X. Miao, \textit{The universal quantum driving force to speed up a
quantum computation --- The unitary quantum dynamics},
http://arxiv.org/abs/quant-ph/ 1105.3573 (2011)\newline

\end{document}